\pdfoutput=1
\documentclass[cernpreprint,texlive=2016,UKenglish,texmf]{atlasdoc}
 
\usepackage[subcaption,backend=biber]{atlaspackage}
\usepackage{atlasbiblatex}
 
\usepackage{atlasphysics}
 
\graphicspath{{logos/}}
 
\usepackage{ANA-PERF-2014-02-PAPER-defs}
\usepackage{rotating}
\usepackage{comment}
\usepackage{multirow}
\usepackage{afterpage}
 
\newcounter{listctr}
\newcounter{subctr}

\AtlasTitle{\Large Determination of jet calibration and energy resolution in proton--proton collisions at $\sqrt{s}$ = 8~TeV using the ATLAS detector}
 
\AtlasJournalRef{Eur. Phys. J. C 80 (2020) 1104}
\AtlasDOI{10.1140/epjc/s10052-020-08477-8}
 
\AtlasAbstract{
The jet energy scale, jet energy resolution, and their systematic uncertainties are measured for jets reconstructed with the ATLAS detector in 2012 using proton--proton data produced at a centre-of-mass energy of 8~TeV with an integrated luminosity of 20 fb$^{-1}$.
Jets are reconstructed from clusters of energy depositions in the ATLAS calorimeters using the anti-$k_t$ algorithm. 
A jet calibration scheme is applied in multiple steps, each addressing specific effects including mitigation of contributions from additional proton--proton collisions, loss of energy in dead material, calorimeter non-compensation, angular biases and other global jet effects.
The final calibration step uses several \textit{in situ} techniques and corrects for residual effects not captured by the initial calibration.
These analyses measure both the jet energy scale and resolution by exploiting the transverse momentum balance in $\gamma$+jet, $Z$+jet, dijet, and multijet events. A statistical combination of these measurements is performed. 
In the central detector region, the derived calibration has a precision better than 1\% for jets with transverse momentum
150 GeV $< p_{\mathrm{T}}<$ 1500 GeV, and the relative energy resolution is $(8.4\pm 0.6)$\% for $p_{\mathrm{T}}=$ 100 GeV and $(23\pm 2)$\% for $p_{\mathrm{T}}=$ 20 GeV.
The calibration scheme for jets with radius parameter $R=1.0$, for which jets receive a dedicated calibration of the jet mass, is also discussed.
}
 
\author{The ATLAS Collaboration}
 
\AtlasRefCode{PERF-2014-02}
 
\PreprintIdNumber{CERN-EP-2019-057}

\AtlasJournal{Eur.\ Phys.\ J.\ C}

\hypersetup{pdftitle={ATLAS document},pdfauthor={The ATLAS Collaboration}}
 
\begin{document}
 
\maketitle
 
\tableofcontents
 
\clearpage
 
% The next lines are included from the .//sections/intro.tex input file
\section{Introduction}
\label{sec:intro}
 
Collimated sprays of energetic hadrons, known as jets, are the dominant final-state objects of high-energy proton--proton (\pp{}) interactions at
the Large Hadron Collider (\LHC) located at CERN.
They are key ingredients for many physics measurements and for searches for new phenomena.
This paper describes the reconstruction of jets in the ATLAS detector~\cite{PERF-2007-01} using 2012 data.
Jets are reconstructed using the \antikt{}~\cite{Cacciari:2008gp} jet algorithm,
where the inputs to the jet algorithm are typically energy depositions in the \ATLAS{} calorimeters 
that have been grouped into ``topological clusters''~\cite{PERF-2014-07}.
Jet radius parameter values of $R=0.4$, $R=0.6$, and $R=1.0$ are considered.
The first two values are typically used for jets initiated by gluons or quarks, except top quarks.
The last choice of $R=1.0$ is used for jets containing the hadronic decays of massive particles, such as \Wboson/\Zboson/Higgs bosons and top quarks.
The same jet algorithm can also be used to form jets from other inputs, such as inner-detector 
tracks associated with charged particles or simulated stable particles from the Monte Carlo event record.
 
Calorimeter jets, which are reconstructed from calorimeter energy depositions, are calibrated to the energy scale of jets created with the same
jet clustering algorithm from stable interacting particles.
This calibration accounts for the following effects:
\begin{itemize}
\item \textbf{Calorimeter non-compensation:} different energy scales for hadronic and electromagnetic showers.
\item \textbf{Dead material:} energy lost in inactive areas of the detector.
\item \textbf{Leakage:} showers reaching the outer edge of the calorimeters.
\item \textbf{Out-of-calorimeter jet:} energy contributions which are included in the stable particle jet but which are not included in the reconstructed jet.
\item \textbf{Energy depositions below noise thresholds:} energy from particles that do not form calorimeter clusters or have energy depositions not included in these clusters due to the noise suppression in the cluster formation algorithm.
\item \textbf{\Pileup{}:} energy deposition in jets is affected by the presence of multiple \pp{} collisions in the same \pp{} bunch crossing as well as residual signals from other bunch crossings.
\end{itemize}

A first estimate of the jet energy scale (JES) uncertainty of $5\%\mbox{--}9$\% was based on information available prior to \pp{} collision data and initial analysis of early data taken in 2010~\cite{STDM-2010-03}.
An improved jet calibration with an uncertainty evaluated to be about $2.5\%$ for jets with pseudorapidity\footnote{
ATLAS uses a right-handed coordinate system with its origin at the nominal interaction point (IP) in the centre of the detector and the $z$-axis along the beam pipe. The $x$-axis points from the IP to the centre of the LHC ring, and the $y$-axis points upward. Cylindrical coordinates $(r,\phi)$ are used in the transverse plane, $\phi$ being the azimuthal angle around the $z$-axis.
The pseudorapidity $\eta$ is an approximation of rapidity $y\equiv 0.5\ln\left[(E+p_z)/(E-p_z)\right]$ in the high-energy limit and is defined in terms of the polar angle $\theta$ as $\eta\equiv-\ln\tan(\theta/2)$.}
$|\eta|<0.8$ over a wide range of transverse momenta (\pt{}) was achieved with the full 2010 dataset using  test-beam measurements, single-hadron response measurements, and \insitu{} techniques~\cite{PERF-2011-03}.
A much larger dataset, recorded during the 2011 data-taking period, improved the precision of JES measurements to $1\%\mbox{--}3\%$ for jets
with $\pt>40$~\GeV{} within $|\eta|<2.5$ using a statistical combination of several \insitu{} techniques~\cite{PERF-2012-01}.
 
This paper describes the derivation of the ATLAS jet calibration and jet energy resolution using the full 2012 \pp{} collision dataset,
which is more than four times larger than the 2011 dataset used for the previous calibration~\cite{PERF-2012-01}.
Due to the increased instantaneous luminosity, the beam conditions in 2012 were more challenging than those in 2011, and
the ability to mitigate the effects of additional \pp{} interactions is of major importance for robust performance, especially for jets with low \pT{}.
The jet calibration is derived using a combination of methods based both on Monte Carlo (MC) simulation and on \insitu{} techniques.
The jet energy resolution (JER), which previously was studied using events with dijet topologies~\cite{PERF-2011-04},
is determined using a combination of several \insitu{} JER measurements for the first time.
A subset of these jet calibration techniques were subsequently used for $R=0.4$ jets recorded during the 2015 data-taking period~\cite{PERF-2016-04}, and for $R=1.0$ jets recorded during the 2015-2016 data-taking period~\cite{JETM-2018-02}.
 
The outline of the paper is as follows.
Section~\ref{sec:atlas} describes the ATLAS detector and the dataset used.
The MC simulation framework is presented in Section~\ref{sec:mc}, and the jet reconstruction and calibration strategy is summarized in Section~\ref{sec:jets}.
Section~\ref{sec:gsc} describes the global sequential calibration method, which exploits information from the tracking system (including the muon chambers)
and the topology of the energy depositions in the calorimeter to improve the JES uncertainties and the JER.
The \insitu{} techniques based on a \pt{} balance are described in Sections~\ref{sec:dijet} to \ref{sec:multijet}.
First, the intercalibration between the central and forward detector, using events with dijet-like topologies, is presented in Section~\ref{sec:dijet}.
The methods based on the \pt{} balance between a jet and a well-calibrated photon or $Z$ boson are discussed in Section~\ref{sec:vjets},
while the study of the balance between a high-\pt{} jet and a system of several low-\pt{} jets is presented in Section~\ref{sec:multijet}.
The combination of the JES \insitu{} results and the corresponding uncertainties are discussed in Section~\ref{sec:JEScomb},
while the \insitu{} combination and the results for the JER are presented in Section~\ref{sec:JERcomb}.
% End of text imported from the .//sections/intro.tex input file
 
% The next lines are included from the .//sections/atlas.tex input file
\section{The ATLAS detector and data-taking conditions}
\label{sec:atlas}
 
The \ATLAS{} detector
consists of an inner tracking detector,
sampling electromagnetic and hadronic calorimeters, and muon chambers in a toroidal magnetic field.
A detailed description of the ATLAS detector is in Ref.~\cite{PERF-2007-01}.
 
The inner detector (ID) has complete azimuthal coverage and spans the
pseudorapidity range of $|\eta|<2.5$.
It consists of three subdetectors: a high-granularity silicon pixel detector, a silicon microstrip detector, and a transition
radiation tracking detector. 
These are placed inside a solenoid that provides a uniform magnetic field of 2~T.
The ID reconstructs tracks from charged particles and determines their transverse momenta from the curvature in the magnetic field.
 
Jets are reconstructed from energy deposited in the ATLAS calorimeter system. Electromagnetic calorimetry is provided by
high-granularity liquid argon (\LAr) sampling calorimeters, using lead as an absorber,
which are split into barrel ($|\eta|<1.475$) and endcap
($1.375<|\eta|<3.2$) regions, where the endcap is further subdivided into outer and inner wheels. The
hadronic calorimeter is divided into the barrel ($|\eta|<0.8$) and
extended barrel ($0.8<|\eta|<1.7$) regions, which are instrumented with tile scintillator/steel modules, and the endcap
region ($1.5<|\eta|<3.2$), which uses \LAr/copper
modules. The forward calorimeter region
($3.1<|\eta|<4.9$) is instrumented with \LAr/copper and \LAr/tungsten modules
to provide electromagnetic and hadronic energy measurements,
respectively. The electromagnetic and hadronic calorimeters are segmented into layers, allowing a determination of the longitudinal
profiles of showers. The electromagnetic barrel, the electromagnetic endcap outer wheel, and tile calorimeters consist of three layers.
The electromagnetic endcap inner wheel consists of two layers. The
hadronic endcap calorimeter consists of four layers.
The forward calorimeter has one electromagnetic and two hadronic layers.
There is also an additional thin LAr presampler, covering $|\eta| <1.8$, dedicated to correcting for energy loss in material upstream of the calorimeters.
 
The muon spectrometer surrounds the \ATLAS{} calorimeter. A system of three large
air-core toroids with eight coils each, a barrel and two endcaps, generates a magnetic field in the
pseudorapidity range $|\eta| < 2.7$.
The muon spectrometer measures muon tracks
with three layers of precision tracking chambers and is instrumented with separate trigger
chambers.

Events are retained for analysis using a trigger system~\cite{TRIG-2011-02}
consisting of a hardware-based level-1~trigger followed by
a software-based high-level trigger with two
levels: level-2 and subsequently the event filter.
Jets are identified using a sliding-window algorithm at level-1 that takes coarse-granularity
calorimeter towers as input.  This is refined with an improved jet
reconstruction based on trigger towers at level-2 and on
calorimeter cells in the event filter~\cite{TRIG-2012-01}.
 
% End of text imported from the .//sections/atlas.tex input file
% The next lines are included from the .//sections/data.tex input file
 
The dataset consists of \pp{} collisions recorded from April
to December 2012 at a centre-of-mass energy ($\sqrt{s}$) of 8~\TeV.
All ATLAS subdetectors were required to be operational and events were rejected if any data quality issues were present, resulting in a usable dataset with a total integrated luminosity of 20~\ifb{}.
The LHC beams were operated with proton bunches organized in bunch trains,
with bunch crossing intervals (bunch spacing) of 50~ns.
The average number of \pp{} interactions per bunch crossing, denoted $\avgmu$, was typically between 10 and 30~\cite{DAPR-2013-01}.

The typical electron drift time within the ATLAS \LAr{} calorimeters is 450~ns~\cite{Aad:2010zh}.
Thus, it is not possible to read out the full detector signal from one event before the next event occurs.
To mitigate this issue, a bipolar shaper~\cite{Abreu:2010zzc} is applied to the output, creating signals with a pulse sufficiently short to be read between bunch crossings.
After bipolar shaping, the average energy induced by \pileup{} interactions should be zero in the ideal situation of sufficiently long bunch trains with the same luminosity in each pair of colliding bunches.
A bunch-crossing identification number dependent offset correction is applied to account for the finite train length such that the average energy induced by pileup is zero for every crossing.
However, fluctuations in \pileup{} activity, both from in-time and out-of-time collisions, contribute to the calorimeter energy read out of the collision of interest.
Multiple methods to suppress the effects of \pileup{} are discussed in subsequent sections.
 
% End of text imported from the .//sections/data.tex input file
 
% The next lines are included from the .//sections/mc.tex input file
\section{Simulation of jets in the ATLAS detector}
\label{sec:mc}
 
Monte Carlo event generators simulate the type, energy, and direction of particles produced in $pp$ collisions.
Table~\ref{tab:mc} presents a summary of the various event generators used to determine the ATLAS jet calibration.
A detailed overview of the MC event generators used in ATLAS analyses can be found in Ref.~\cite{Buckley:2011ms}.
 
\begin{table}[!ht]
\caption{ Summary of the simulated samples used to derive the jet calibration and to assess systematic uncertainties.
\label{tab:mc}}
\begin{center}
\begin{tabular}{l|l|l|l}
\hline
\hline
Process & Event generator & PDF set & MPI/shower tune set   \\
\hline
Dijet \&  & \textsc{Pythia} 8.160 & CT10~\cite{Lai:2010vv} & AU2~\cite{ATL-PHYS-PUB-2011-008} \\
multijet & \herwigpp{} 2.5.2 &  CTEQ6L1~\cite{Nadolsky:2008zw} & EE3 MRST LO**~\cite{Gieseke:2012ft} \\  
& \powheg+\textsc{Pythia} 8.175 & CT10  & AU2  \\
& \powheg+\herwig{} 6.520.2 & CT10 & AUET2~\cite{ATL-PHYS-PUB-2011-014}  \\
& \sherpa{} 1.4.5   & CT10 & \sherpa{}-default~\cite{Sjostrand:1987su}  \\ \hline
\Zjet{} & \powheg+\pythia   & CT10 & AU2  \\
& \sherpa{}            & CT10 & \sherpa{}-default \\ \hline
\gammajet{} & \pythia  & CTEQ6L1 & AU2 \\
& \herwigpp{} & CTEQ6L1 & UE-EE-3~\cite{Gieseke:2012ft} \\ \hline
\Pileup{} & \pythia{} & MSTW2008LO~\cite{Martin:2009iq} & AM2~\cite{ATL-PHYS-PUB-2012-003} \\
\hline
\hline
\end{tabular}
\end{center}
\end{table}
 
The baseline simulation samples used to obtain the MC-based jet calibration were produced using \textsc{Pythia} version 8.160~\cite{Sjostrand:2007gs}.
\textsc{Pythia} uses a $2\to 2$ matrix element interfaced with a parton distribution function (PDF) to model the hard process.
Additional radiation was modelled in the leading-logarithm approximation using \pt{}-ordered parton showers. Multiple parton--parton interactions (MPI), also referred to as the underlying event (UE), were also simulated, and modelling of the hadronization process was based on the Lund string model~\cite{Andersson:1983ia}.
 
Separate samples produced using other generators were used to derive the final jet calibration and resolution and associated uncertainties using \insitu{} techniques.
The \herwig{}~\cite{Corcella:2000bw} and \herwigpp{}~\cite{Bahr:2008pv} event generators use a $2 \to 2$ matrix element convolved with a PDF for the hard process just as \pythia{} does, but use angle-ordered parton showers and a different modelling of the UE and hadronization.
The \sherpa{} event generator~\cite{Gleisberg:2008ta} was used to produce multi-leg $2\to N$ matrix elements matched to parton showers using the CKKW~\cite{Catani:2001cc} prescription.
Fragmentation was simulated using the cluster-hadronization model~\cite{Winter:2003tt}, and the UE was modelled  using the \sherpa{} AMISIC model based on Ref.~\cite{Sjostrand:1987su}.
Samples were also produced using the \powhegbox~\cite{Nason:2004rx,Frixione:2007vw,Alioli:2010xd,Alioli:2010xa} software
that is accurate to next-to-leading order (NLO) in perturbative QCD. Parton showering and modelling of the hadronization and the UE were provided by either \pythia{} or \herwig{}, resulting in separate samples referred to as \powhegpyt{} and \powhegher{}, respectively.
Tuned values of the modelling parameters affecting the parton showering, hadronization, and the UE activity were determined for each generator set-up to match various distributions in data as summarized in Table~\ref{tab:mc} and references therein.
 
The generated stable particles, defined as those with a lifetime $\tau$ such that $c \tau > 10$~mm, were input to the detector simulation that models the particles' interactions with the detector material. Such particles are
used to build jets as explained in Section~\ref{sec:jets}. Most MC samples were generated with a full detector simulation of the ATLAS detector~\cite{SOFT-2010-01} based on \geant{}~\cite{Agostinelli:2002hh}, in which hadronic showers are simulated with the QGSP BERT model~\cite{QGS}.
Alternative samples were produced using the Atlfast-II (AFII) fast detector simulation based on a simplified modelling of particle interactions with the calorimeter, yielding a factor of ten more events produced for the same CPU time~\cite{ATLAS:2010bfa}.
The output of the detector simulation were detector signals with the same format as those from real data.

\Pileup{} events, i.e. additional $pp$ interactions that are not correlated with the hard-scatter event of interest, were simulated as minimum-bias events produced with \pythia{}  using the AM2 tuned parameter set~\cite{ATL-PHYS-PUB-2012-003} and the MSTW2008LO PDF~\cite{Martin:2009iq}. The simulated detector signals from these events were overlaid with the detector signals from the hard-scatter event based on the \pileup{} conditions of the 2012 data-taking period.
\Pileup{} events were overlaid both in the hard-scatter bunch crossing (in-time \pileup{}) and in nearby bunch crossings (out-of-time \pileup{}) with the detector signals offset in time accordingly.
These out-of-time \pileup{} signals are overlaid in such a manner as to cover the full read-out window of each of the ATLAS calorimeter sub-detectors.
The number of \pileup{} events to overlay in each bunch crossing was sampled from a Poisson distribution with a mean \avemu{} corresponding to the expected number of additional \pp{} collisions per bunch crossing.
% End of text imported from the .//sections/mc.tex input file
 
% The next lines are included from the .//sections/jets.tex input file
\section{Overview of ATLAS jet reconstruction and calibration}
\label{sec:jets}
 
\subsection{Jet reconstruction and preselection}
\label{sec:jetReco}
 
Jets are reconstructed with the \antikt{} algorithm~\cite{Cacciari:2008gp} using the
\textsc{FastJet} software package~\cite{Cacciari:2005hq,Fastjet} version 2.4.3.
Jets are formed using different inputs:
stable particles from the event generator record of simulated events resulting in \textit{\tjets}; reconstructed calorimeter clusters, producing \textit{calorimeter jets}; or inner-detector tracks to form \textit{track jets}.
 
The generated stable particles used to define \tjets{} are required to originate (either directly or via a decay chain) from the hard-scatter vertex, and hence do not include particles from \pileup{} interactions.
Muons and neutrinos are excluded to ensure that the \tjets{} are built from particles that leave significant energy deposits in the calorimeters.

Calorimeter jets are built from clusters of
adjacent calorimeter read-out cells that contain a significant energy signal above noise levels, referred to as topological clusters or \textit{\topos{}}.
Details of the formation of \topos{} are provided in Ref.~\cite{PERF-2014-07}.
In its basic definition, a \topo{} is assigned an energy equal to the sum of the associated calorimeter cell energies
calibrated at the \textit{electromagnetic scale} (EM-scale)~\cite{Aharrouche:2010zz,LArTB02uniformity, Aharrouche:2006nf, PERF-2010-04},
which is the basic signal scale accounting correctly for the energy deposited in the calorimeter by electromagnetic showers.
The direction ($\eta$ and $\phi$) of a \topo{} is defined from the centre of the ATLAS detector to the energy-weighted barycentre of the associated calorimeter cells, and the mass is set to zero.
\Topos{} can further be calibrated using the local cell signal weighting (\LCW) method~\cite{PERF-2014-07} designed to give the correct scale for charged pions produced in the interaction point.
The \LCW\ method reduces fluctuations in energy due to the non-compensating nature of the \ATLAS\ calorimeters, out-of-cluster energy depositions, and energy deposited in dead material, improving the energy resolution of the reconstructed jets in comparison with jets reconstructed using \EM-scale clusters~\cite{PERF-2011-03}.

The calorimeter jet four-momentum directly after jet finding is referred to as the \textit{constituent scale} four-momentum $p^\textrm{const}$ and is defined as the sum of the constituent \topo{} four-momenta $p^\textrm{topo}_i$:
\begin{equation}
p^\text{\,const} = \left( E^\text{\,const},\vec{p}^\text{~const} \right) = \sum_{i=1}^{N_\textrm{const}} p^\textrm{topo}_i
= \left( \sum_{i=1}^{N_\text{const}} E^\text{topo}_i, \sum_\text{i=1}^{N_\text{const}} \vec{p}^\text{~topo}_i \right).
\label{eq:jetConstScale}
\end{equation}
The constituent scales considered in this paper are EM or LCW depending on the calibration of the constituent \topos{}.
At this stage, all angular coordinates are defined from the centre of the ATLAS detector, and the
\textit{detector pseudorapidity} $\etaDet{}\equiv \eta^\textrm{const}$ and \textit{detector azimuth} $\phiDet{} \equiv \phi^\textrm{const}$ are recorded for each jet.
The most common choice in ATLAS analyses of the \antikt{} radius parameter is $R=0.4$, but $R=0.6$ is also used frequently.
Analyses that search for hadronic decays of highly boosted (high \pt{}) massive objects often use larger values of $R$ than these since the decay products of the boosted objects can then be contained within the resulting \textit{\lRjets{}}.
Due to the larger radius parameter, this class of jets spans a larger solid angle and hence are more sensitive to \pileup{} interactions than jets with $R\leq0.6$. To mitigate the influence of \pileup{} and hence improve the sensitivity of the analyses, several jet grooming algorithms have been designed and studied within ATLAS~\cite{PERF-2012-02, STDM-2011-19, PERF-2015-03, PERF-2015-04}.
In this paper,
the trimming algorithm~\cite{Krohn:2009th} (one type of grooming method) is applied to \antikt{} jets built with $R=1.0$.
This grooming procedure
starts from the constituent \topos{} of a given $R=1.0$ \antikt{} jet to create \textit{subjets} using the $k_t$ jet algorithm~\cite{KtJets} with radius parameter $R_\text{sub}=0.3$.
The \topos{} belonging to subjets with $f_\text{cut} \equiv \pt{}^\text{subjet} / \pt{}^\text{jet} < 0.05$ are discarded, and the jet four-momentum is then recalculated from the remaining \topos{}.

For each \insitu{} analysis, jets within the full calorimeter acceptance \AetaRange{4.5} with calibrated $\pt>8$~\GeV{} ($\pt>25$~\GeV\ in case of the multijet analysis) are considered.
These \pt{} thresholds do not bias the kinematic region of the derived calibration, which is $\pt\geq 17$~\GeV\ ($\pt\geq 300$~\GeV\ for the multijet analysis).
The jets are also required to satisfy ``Loose'' quality criteria, designed to reject fake jets originating from calorimeter noise bursts,  non-collision background, or cosmic rays~\cite{PERF-2012-01}, and to fulfil a requirement designed to reject jets originating from \pileup{} vertices.
The latter criterion is based on the jet vertex fraction (\JVF), computed as the scalar sum $\sum\pttrk$ of the tracks matched to the jet that are associated with the hard-scatter primary vertex divided by $\sum\pttrk$ using all tracks matched to the jet (see Ref.~\cite{PERF-2014-03} for further details).
The default hard-scatter vertex is the primary vertex with the largest $\sum_\textrm{tracks}{p_\textrm{T}^2}$, but other definitions are used for certain analyses~\cite{HIGG-2013-08}.
Each jet with $\pt<50$~\GeV\ within the tracking acceptance \AetaRange{2.4} is required to have $\JVF > 0.25$,
which effectively rejects \pileup{} jets in ATLAS 2012 $pp$ data~\cite{PERF-2014-03}.

Jets with a radius parameter of $R=0.4$ or $R=0.6$ have been built using both EM- and LCW-scale \topos{} as inputs.
These four jet reconstruction options have been studied in similar levels of detail, but for brevity the paper will focus on presenting the results
for jets built using EM-scale \topos{} with a radius parameter of $R=0.4$, which better demonstrates the importance of the GS calibration as described in Section~\ref{sec:gsc}.
Key summary plots will present the results for all four jet definitions thus showing the final performance of each of the different options.
In contrast, jets with a radius parameter of $R=1.0$ have only been studied in detail using LCW-scale \topos{} as inputs.
This choice is motivated by the common usage of such jets for tagging of hadronically-decaying particles, where the energy
and angular distribution of constituents within the jet is important.
For such a situation, LCW \topos{} are advantageous because they flatten the detector response, and thus the tagging capabilities
are less impacted by where a given energy deposit happens to be within the detector.

\subsection{Matching between jets, jet isolation, and calorimeter response}
\label{sec:jetMatch}
To derive a calibration based on MC simulation, it is necessary to match a \tjet{} to a reconstructed jet.
Two methods are used for this:
a simple, angular matching as well as a more sophisticated approach
known as \emph{jet ghost association}~\cite{areaSub}.
For the angular matching, a $\Delta R < 0.3$ requirement is used, where  $\Delta R$
is the pseudorapidity and azimuthal angle separation between the two jets added in quadrature, i.e.\
$\Delta R = \Delta\eta \,\oplus\, \Delta\phi \equiv \sqrt{(\Delta \eta)^2 + (\Delta \phi ) ^2}$.
The angular criterion $\Delta R < 0.3$ is chosen to be smaller than the jet radius parameter used for ATLAS analyses ($R=0.4$ or larger) but much larger than the jet angular resolution (Section~\ref{sec:pileupCorr}).
Jet matching using ghost association treats
each MC simulated particle as a \textit{ghost particle}, which means that they are assigned an infinitesimal \pt{}, leaving the angular coordinates unchanged.
The calorimeter jets can now be built using both the \topos{} and ghost particles as input. Since the ghost particles have infinitesimal \pt{}, the four-momenta of the reconstructed jets will be identical to the original jets built only from \topos{}, but the new jets will also have a list of associated truth particles for any given reconstructed jet.
A \tjet{} is matched to a reconstructed jet if
the sum of the energies of the \tjet{} constituents which are ghost-associated with the reconstructed jet is more than 50\% of the \tjet{} energy, i.e.\ the sum of the energies of all constituents.
This ensures that only one reconstructed jet is matched to any given \tjet{}.
If several \tjets{} fulfil the matching requirement, the \tjet{} with the largest energy is chosen as the matched jet. 
Matching via ghost association results in a unique match for each \tjet{} and hence performs better than the simple angular matching in cases where several jets have small angular separation from each other.

The simulated jet energy response is defined by
\begin{equation*}
\mathcal{R}_{E} = \left<\frac{E_\text{reco}}{E_\text{truth}}\right>,
\end{equation*}
where $E_\text{reco}$ is the reconstructed energy of the calorimeter jet, $E_\text{truth}$ is the energy of the matching \tjet{}, and the brackets denote that $\mathcal{R}_E$ is defined from the mean parameter of a Gaussian fit to the response distribution $E_\text{reco}/E_\text{truth}$.
The \pt{} and mass responses are defined analogously as the Gaussian means $\langle\pt{}_\text{,reco} / \pt{}_\text{,truth}\rangle$ and $\langle m_\text{reco}/m_\text{truth}\rangle$
of the reconstructed quantity divided by that of the matching \tjet{}.
When studying the jet response for a population of jets,
both the reconstructed and the \tjets{} are typically required to fulfil isolation requirements.
For the analyses presented in this paper, reconstructed jets are required to have no other reconstructed jet with $\pt{}>7$~\GeV\ within $\Delta R < 1.5R$, where $R$ is the \antikt{} jet radius parameter used. \Tjets{} are similarly required to have no jets with
$\pt{}>7$~\GeV\ within $\Delta R < 2.5 R$.
After requiring the particle and reconstructed jets to be isolated, the jet energy response distributions for jets with fixed $E_\text{truth}$ and $\eta$ have nearly Gaussian shapes, and
$\mathcal{R}_E$
and the jet resolution $\sigma_\mathcal{R}$ are defined as the mean and width parameters of Gaussian fits to these distributions, respectively.
For all results presented in this paper, the mean jet response is defined from the mean parameter of a fit to a jet response or momentum balance distribution as appropriate rather than the mean or median of the underlying distribution, as the fit mean is found to be significantly more robust against imperfect modelling of the tails of the underlying distribution.

\subsection{Jet calibration}
\label{sec:jetCalibOverview}
\begin{sidewaysfigure}[!ph]
\begin{center}
\includegraphics[width=1.0\textwidth]{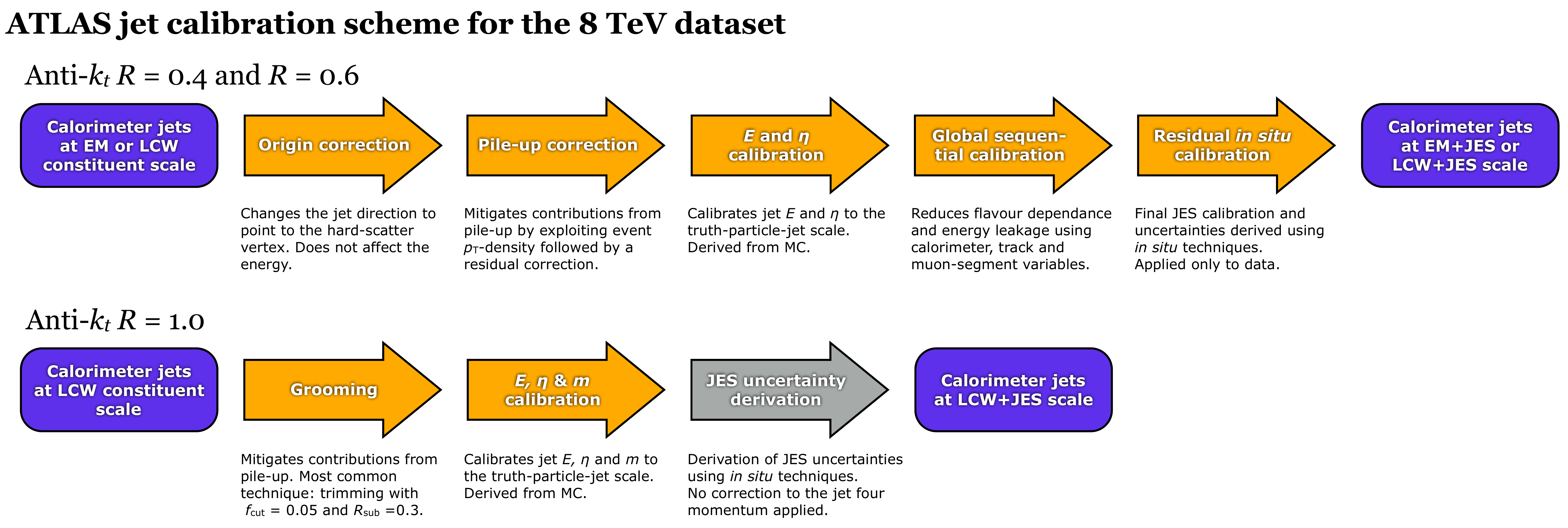}
\end{center}
\caption{Overview of the ATLAS jet calibration described in this paper. All steps are derived and applied separately for jets
built from EM-scale and LCW calibrated calorimeter clusters, except for the global sequential calibration, which is only partially applied to LCW-jets (Section~\ref{sec:gsc}).
The notations EM+JES and LCW+JES typically refer to the fully calibrated jet energy scale; however, in the sections of this paper that detail the derivations of the GS and the \insitu{} corrections, these notations refer to jets calibrated by all steps up to the correction that is being described.
}
\label{fig:JESoverview}
\end{sidewaysfigure}
 
An overview of the ATLAS jet calibration applied to the 8~\TeV\ data is presented in Figure~\ref{fig:JESoverview}.
This is an extension of the procedure detailed in Ref.~\cite{PERF-2012-01} that was applied to the 7~\TeV\ data collected in 2011.
The calibration consists of five sequential steps. The derivation and application of the first three calibration steps are described in this section,
while the global sequential calibration (GS) is detailed in Section~\ref{sec:gsc}, and the relative \insitu{} correction and the associated uncertainties
are described in Sections~\ref{sec:dijet}--\ref{sec:JEScomb}.
 
\subsubsection{Jet origin correction}
\label{sec:originCorr}
The four-momentum of the initial jet is defined according to Eq.~(\ref{eq:jetConstScale}) as the sum of the four-momenta of its constituents.
As described in Section~\ref{sec:jetReco}, the \topos{} have their angular directions $(\eta,\phi)$ defined from the centre of the ATLAS detector to the energy-weighted barycentre of the cluster. This direction can be adjusted to originate from the hard-scatter vertex of the event.
The jet origin correction first redefines the $(\eta,\phi)$ directions of the \topos{} to point to the selected hard-scatter vertex, which results in a updated set of \topo{} four-momenta.
The origin-corrected calorimeter jet four-momentum $p^\text{orig}$ is the sum of the updated \topo{} four-momenta,
\begin{eqnarray*}
p^\textrm{orig} = \sum_{i=1}^{N_\textrm{const}} p^\textrm{topo,orig}_i.
\end{eqnarray*}
Since the energies of the \topos{} are not affected, the energy of the jet also remains unchanged.
Figure~\ref{fig:etaRes} presents the impact of the jet origin correction on the jet angular resolution by comparing the axis of the
calorimeter jet ($\eta_\mathrm{reco},\phi_\mathrm{reco}$) with the axis of the matched \tjet{} ($\eta_\mathrm{truth},\phi_\mathrm{truth}$).
A clear improvement can be seen for the pseudorapidity resolution, while no change is seen for the azimuthal resolution.
This is expected as the spread of the beamspot is significantly larger along the beam axis ($\sim$50~mm) than in the transverse plane ($\ll 1$~mm).
 
\begin{figure}[!ht]
\includegraphics[width=0.48\textwidth]{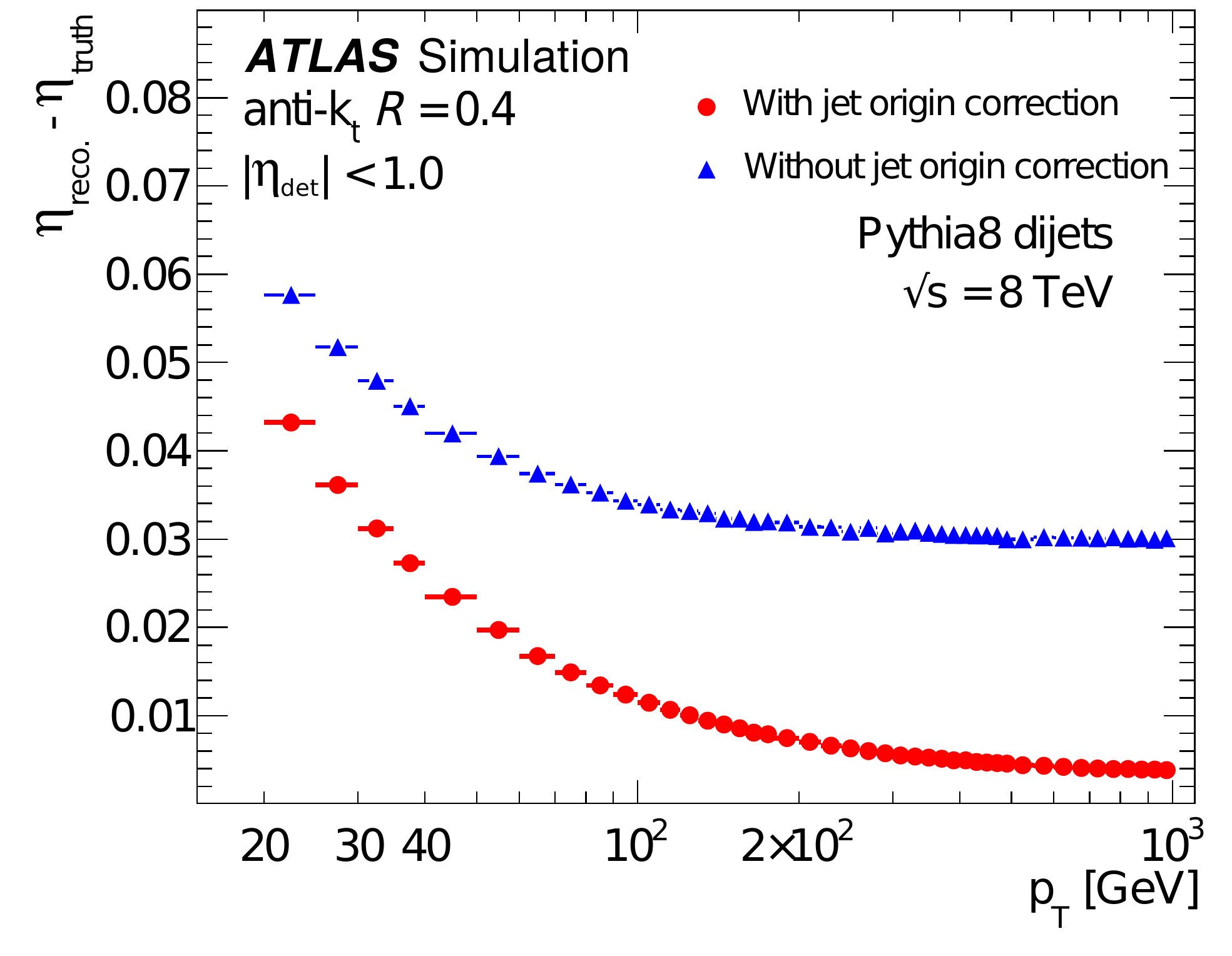}
\includegraphics[width=0.48\textwidth]{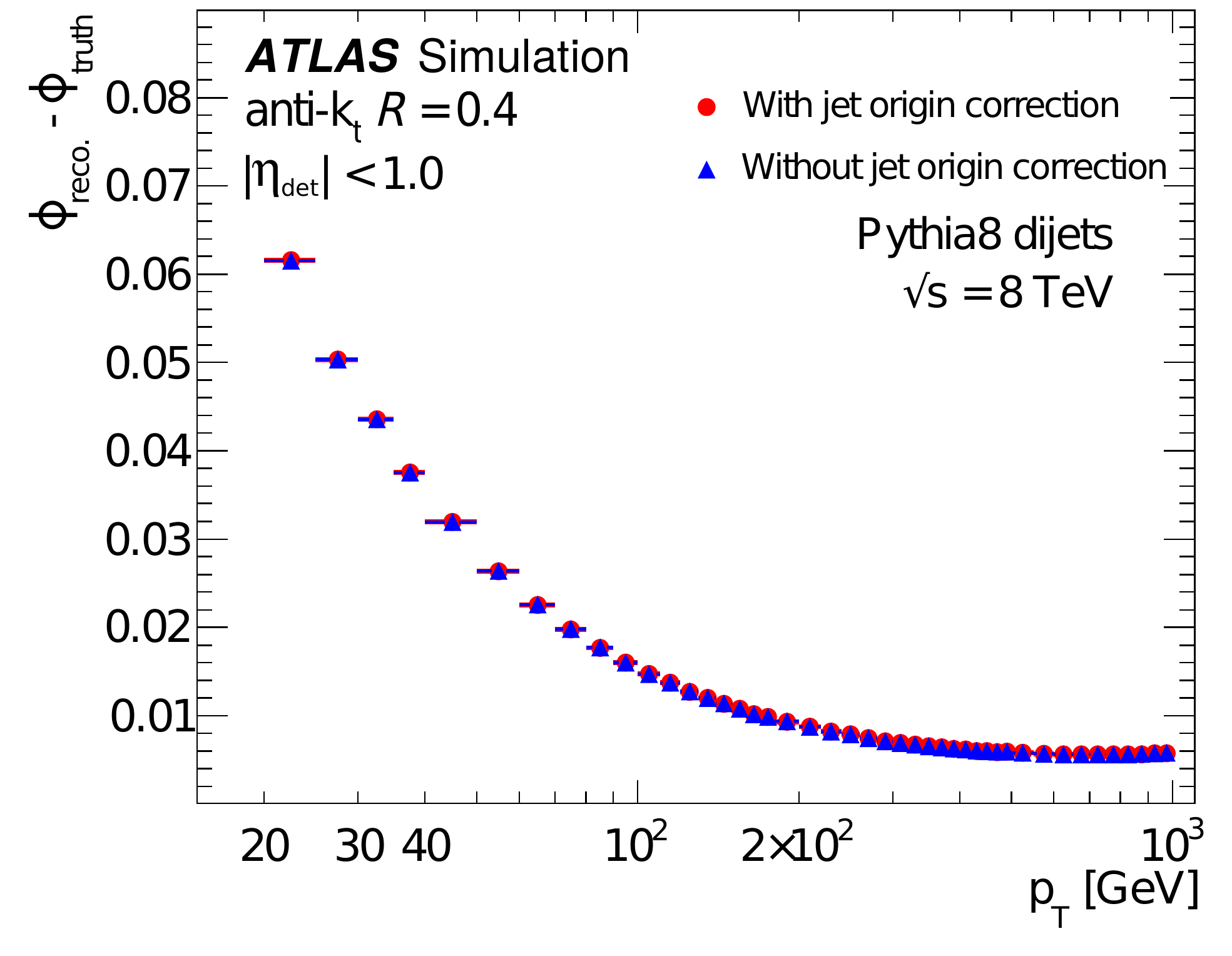}
\caption{Jet angular resolution as a function of transverse momentum for \antikt{} jets with $R = 0.4$.
The resolutions are defined by the spread of the difference between the reconstructed jet axis ($\eta_\mathrm{reco},\phi_\mathrm{reco}$) and the axis of the matched \tjet{} ($\eta_\mathrm{truth},\phi_\mathrm{truth}$) (see Section~\ref{sec:jetMatch} for matching details) in simulated events and are shown both with (circles) and without (triangles) the jet origin correction, which adjusts the direction of the reconstructed jet to point to the hard-scatter vertex instead of the geometrical centre of the detector.
}
\label{fig:etaRes}
\end{figure}

\subsubsection{\Pileup{} correction}
\label{sec:pileupCorr}
The reconstruction of the jet kinematics is affected by \pileup{} interactions.
To mitigate these effects, the contribution from \pileup{} is estimated on an event-by-event and jet-by-jet basis as the product of the event \pt{}-density $\rho$~\cite{areaSub} and the jet area $A$ in $(y,\phi)$-space, where $y$ is the rapidity of the jet~\cite{JetArea}.
The jet area is determined with the \textsc{FastJet 2.4.3} program~\cite{Cacciari:2005hq,Fastjet} using the active-area implementation, in which the jets are rebuilt after adding randomly distributed ghost particles with infinitesimal \pt{} and randomly selected $y$ and $\phi$ from uniform distributions.
The active area is estimated for each jet from the relative number of associated ghost particles (Section~\ref{sec:jetMatch}).
As can be seen in Figure~\ref{fig:jetArea}(a), the active area for a given \antikt{} jet tends to be close to $\pi R^2$.
The event \pt{}-density $\rho$ is estimated event-by-event by building jets using the $k_t$ jet-finding algorithm~\cite{KtJets} due to its tendency to naturally include uniform soft background into jets~\cite{areaSub}.
Resulting $k_t$ jets are only considered within $|\eta|<2$ to remain within the calorimeter regions with sufficient granularity~\cite{PERF-2014-03}.
No requirement is placed on the \pt{} of the jets, and the median of the $\pt{}/A$ distribution is taken as the value of $\rho$. The median is used to reduce the sensitivity of the method to the hard-scatter activity in the tails.
The $\rho$ distributions of events with average interactions per bunch crossing $\avgMu$ in the narrow range of $20<\avgMu<21$ and
several fixed numbers of primary vertices $N_\textrm{PV}$ are shown in Figure~\ref{fig:jetArea}(b).
It can be seen that $\rho$ increases with $N_\textrm{PV}$ as expected, but for a fixed $N_\textrm{PV}$, $\rho$ still has sizeable event-by-event fluctuations.
A typical value of the event \pt{}-density in the 2012 ATLAS data is $\rho = 10$~\GeV, which for a $R=0.4$ jet corresponds to a subtraction in jet \pt{} of $\rho \, A \approx 5$~\GeV.
 
\begin{figure}[!tb]
\centering
\begin{subfigure}[b]{0.49\columnwidth}\centering
\includegraphics[width=\columnwidth]{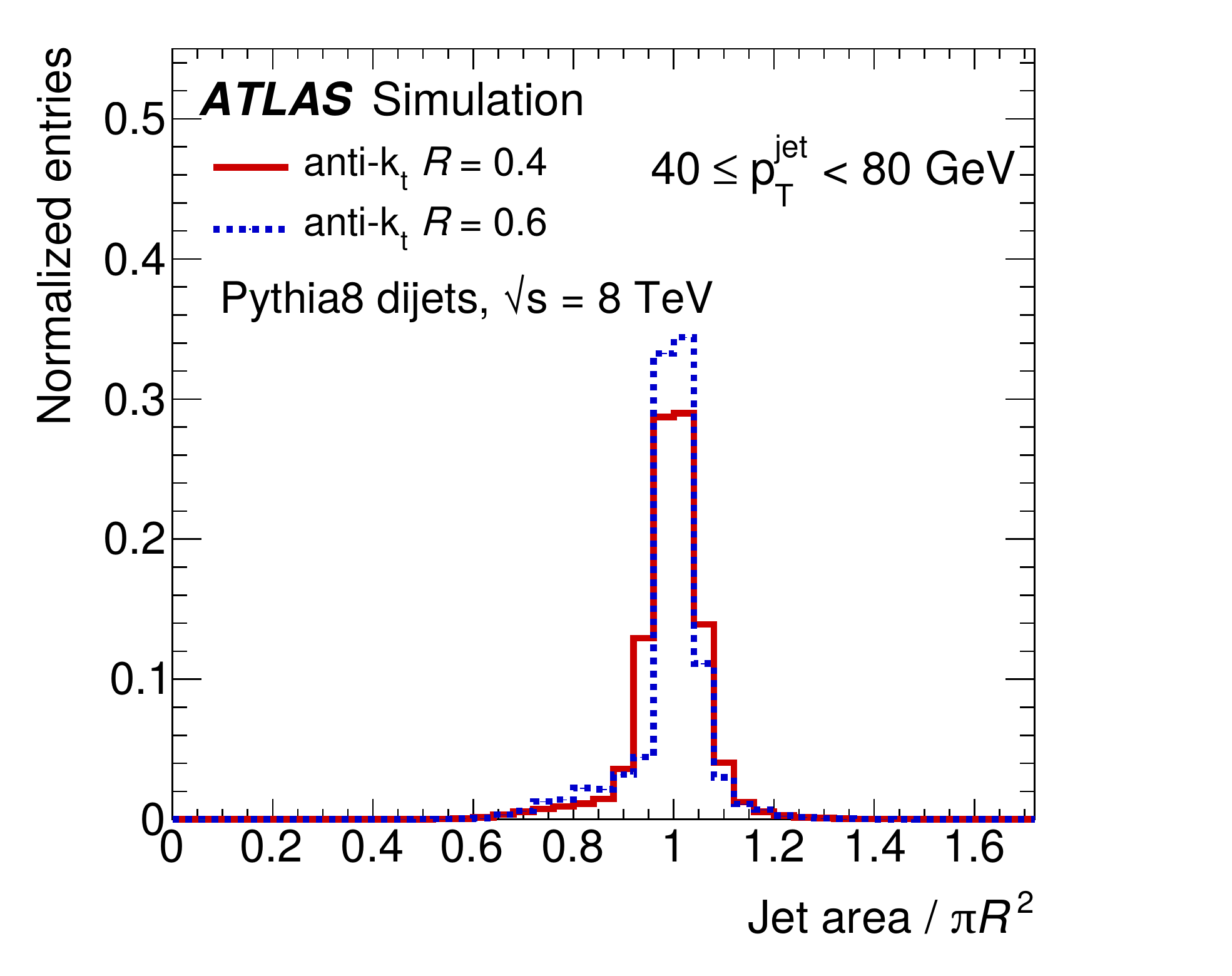}
\caption{Jet area}
\end{subfigure}
\begin{subfigure}[b]{0.475\columnwidth}\centering
\includegraphics[width=\columnwidth]{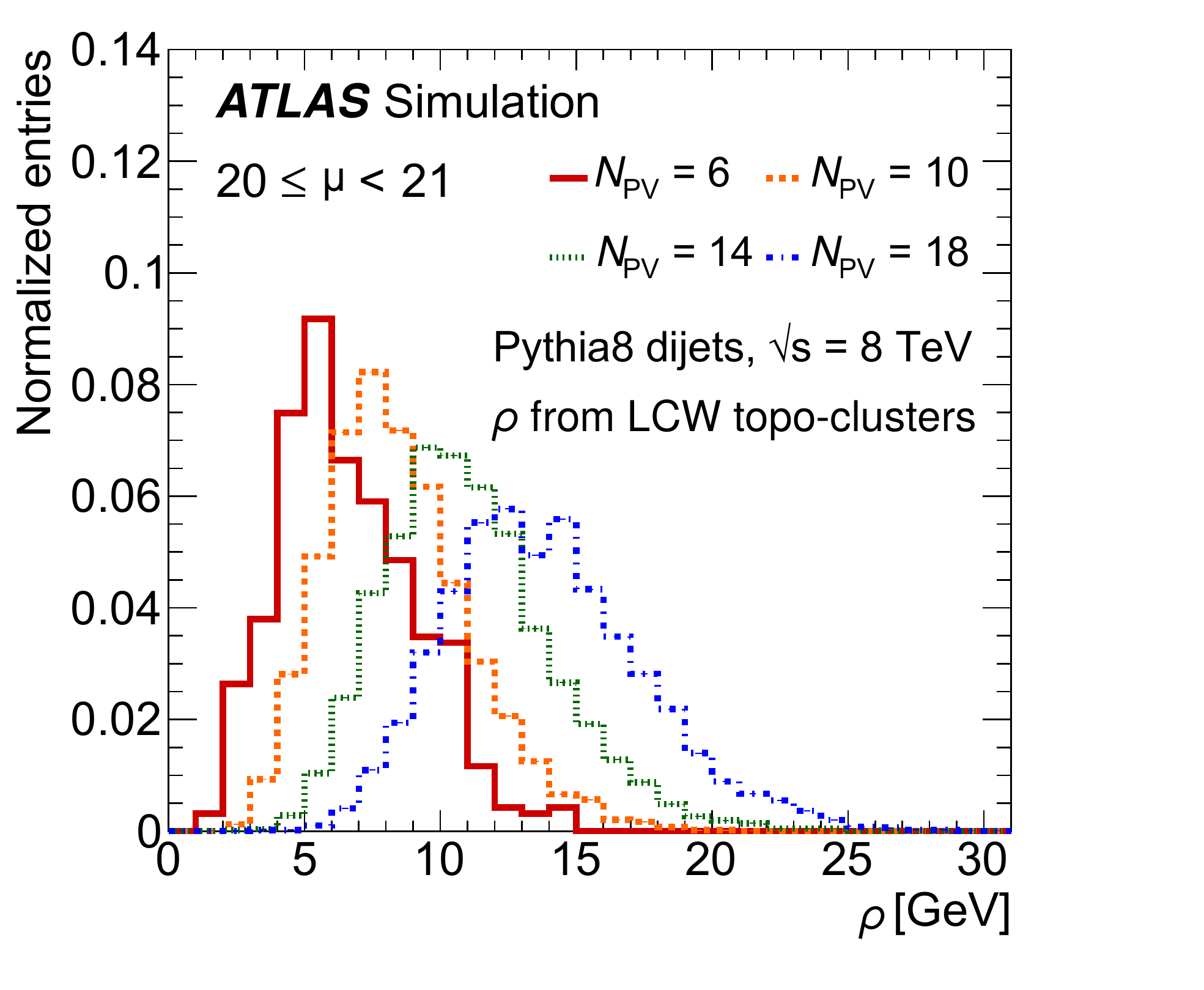}
\caption{Event \pt{}-density, $\rho$}
\end{subfigure}
\caption{
(a) Ratio of the jet active area to $\pi R^2$, where $R$ is the jet radius parameter and (b) the event \pt{}-density $\rho$.
The jet area ratio is shown separately for $R=0.4$ and $R=0.6$ jets reconstructed with the \antikt{} algorithm, and $\rho$ is shown for different numbers of reconstructed primary vertices $N_\textrm{PV}$ in events with average number of $pp$~interactions in the range $20\leq\mu{}<21$.
}
\label{fig:jetArea}
\end{figure}
 
After subtracting the \pileup{} contribution based on $\rho \, A$, the pileup dependence of $\pt^\text{jet}$ is mostly removed, especially within the region where the value of $\rho$ is derived.
However, the value of $\pt^\text{jet}$ has a small residual dependence on $N_\textrm{PV}$ and $\avgMu$, particularly in the region beyond where $\rho$ is derived and where the calorimeter granularity changes.
To mitigate this, an additional correction is derived, parameterized in terms of $N_\textrm{PV}$ and $\avgMu$, which is the same approach and parameterization as was used for the full \pileup{} correction of the ATLAS 2011 jet calibration~\cite{PERF-2012-01}. A typical value for this correction is $\pm 1$~\GeV\ for jets in the central detector region.
The full \pileup{} correction to the jet \pt{} is given by
\begin{equation}
\pt{} \, \mapsto \, \pt{} - \rho\, A - \alpha \, ( N_\textrm{PV} - 1 ) - \beta\,\avgMu,
\label{eq:PU}
\end{equation}
where the $\alpha$ and $\beta$ parameters depend on jet pseudorapidity and the jet algorithm, and are derived from MC simulation.
Further details of this calibration, including evaluation of the associated systematic uncertainties, are in Ref.~\cite{PERF-2014-03}.
No \pileup{} corrections are applied to the trimmed \lRjets{} since this is found to be unnecessary after applying the trimming procedure.
 
\subsubsection{Monte Carlo-based jet calibration}
\label{sec:EtaJES}
After the origin and \pileup{} corrections have been employed, a baseline jet energy scale calibration is applied
to correct the reconstructed jet energy to the \tjet{} energy.
This calibration is derived in MC-simulated dijet samples following the same procedure used in previous ATLAS jet calibrations~\cite{PERF-2011-03,PERF-2012-01}.
Reconstructed and \tjets{} are matched and required to fulfil the isolation criteria as described in Section~\ref{sec:jetMatch}.
The jets are then subdivided into narrow bins of $\etaDet$ of the reconstructed jet and energy of the \tjet{} $E_\text{truth}$, and $\mathcal{R}_E$ is determined for each such bin from the mean of a Gaussian fit (Section~\ref{sec:jetMatch}).
The average reconstructed jet energy $\langle E_\text{reco}\rangle$ (after
\pileup{} correction) is also recorded for each such bin.
A calibration function $c_\text{JES,1}(E_\text{reco})=1/\mathcal{R}_1(E_\text{reco})$ is determined for each \etaDet{} bin
by fitting a smooth function $\mathcal{R}_1(E_\text{reco})$ to a graph of $\mathcal{R}_E$ versus $\langle E_\text{reco}\rangle$ measurements for all $E_\text{truth}$ bins within the given \etaDet{} bin.
After applying this correction ($E_\text{reco} \mapsto c_\text{JES,1}\,E_\text{reco}$) and repeating the derivation of the calibration factor, the jet response does not close perfectly.
The derived calibration factor from the second iteration $c_\text{JES,2}$ is close to but not equal to unity.
The calibration improves after applying
three such iterative residual corrections $c_{\text{JES,}i}$ ($i\in\{2,3,4\}$) such that the final correction factor $c_\text{JES} = \prod_{i=1}^4 c_{\text{JES},i}$ achieves a jet response close to unity for each $(E_\text{truth},\etaDet{})$ bin.
 
For the \lRjets{} (trimmed \antikt{} $R=1.0$), a subsequent jet mass calibration is also applied, derived analogously to the energy calibration. Figure~\ref{fig:calibcurve} shows the energy and jet mass responses for jets with $R=0.4$ and $R=1.0$.
Jets reconstructed from LCW-calibrated \topos{} have a response closer to unity than jets built from EM-scale \topos{}.
Figure~\ref{fig:jesclosure} shows the jet $E$, \pt{}, and $m$ response plots after the application of the MC-based jet calibration.
Good closure is demonstrated across the pseudorapidity range, but there is some small non-closure
for low-\pt{} jets primarily due to
imperfect fits arising from the non-Gaussian energy response and threshold effects.
 
A small, additive correction $\Delta\eta$ is also applied to the jet pseudorapidity to account for biased reconstruction close to regions where the detector technology changes (e.g.\ the barrel--endcap transition region).
The magnitude of this correction is very similar to that of the previous calibrations (Figure~11 of Ref.~\cite{PERF-2011-03}) and can reach values as large as 0.05 near the edge of the forward calorimeters around $|\eta|=3$, but is typically much smaller in the well-instrumented detector regions.
 
\begin{figure}[!htb]
\centering
\begin{subfigure}{0.48\textwidth}\centering
\includegraphics[width=\textwidth]{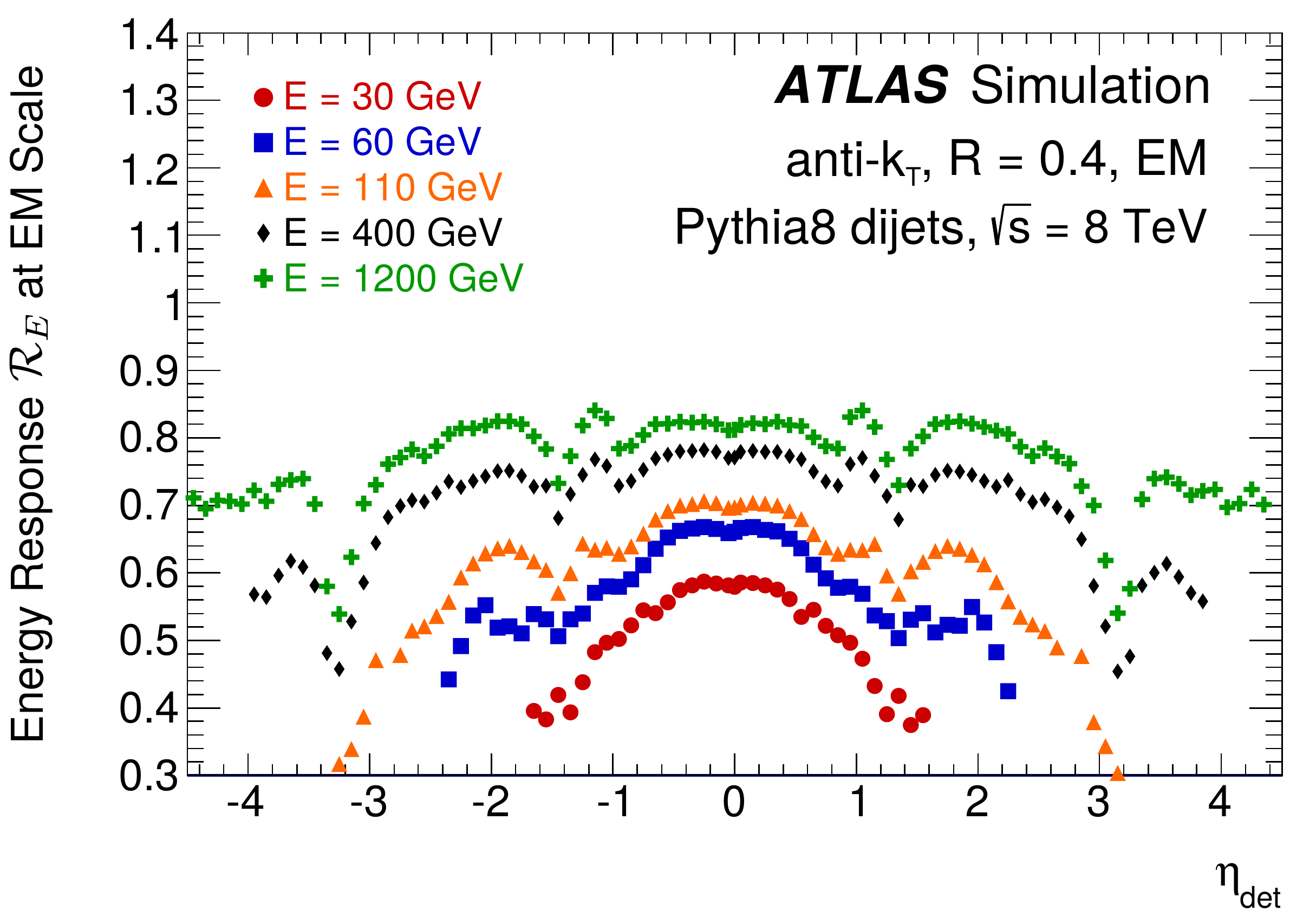}
\caption{Jet energy response $\mathcal{R}_E$, EM scale, \rfour{}}
\end{subfigure}
\begin{subfigure}{0.48\textwidth}\centering
\includegraphics[width=\textwidth]{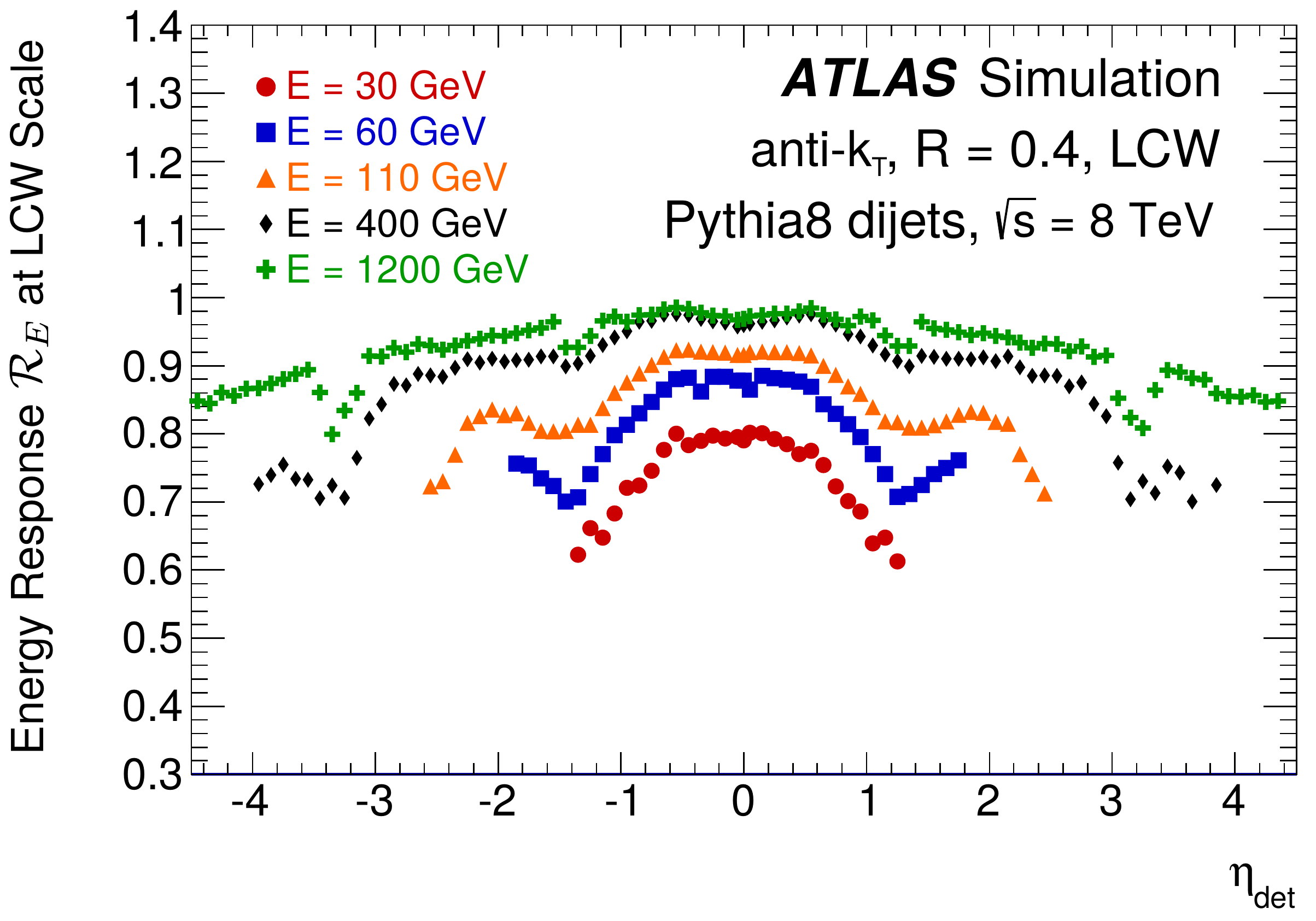}
\caption{Jet energy response $\mathcal{R}_E$, LCW scale, \rfour{}}
\end{subfigure} \\
 
\bigskip
 
\begin{subfigure}{0.48\textwidth}\centering
\includegraphics[width=\textwidth]{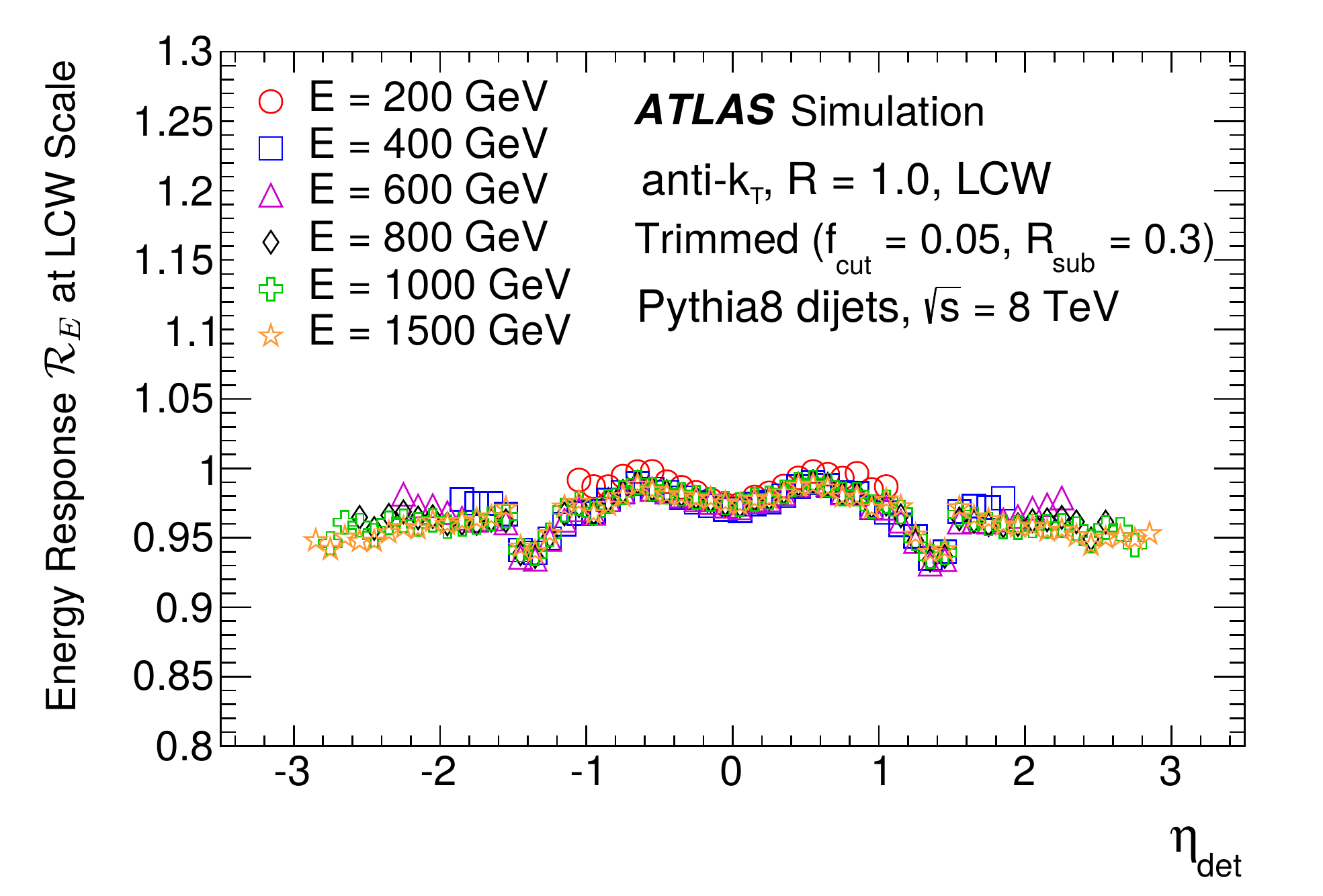}
\caption{Jet energy response $\mathcal{R}_E$, LCW scale, \rten{}}
\end{subfigure}
\begin{subfigure}{0.48\textwidth}\centering
\includegraphics[width=\textwidth]{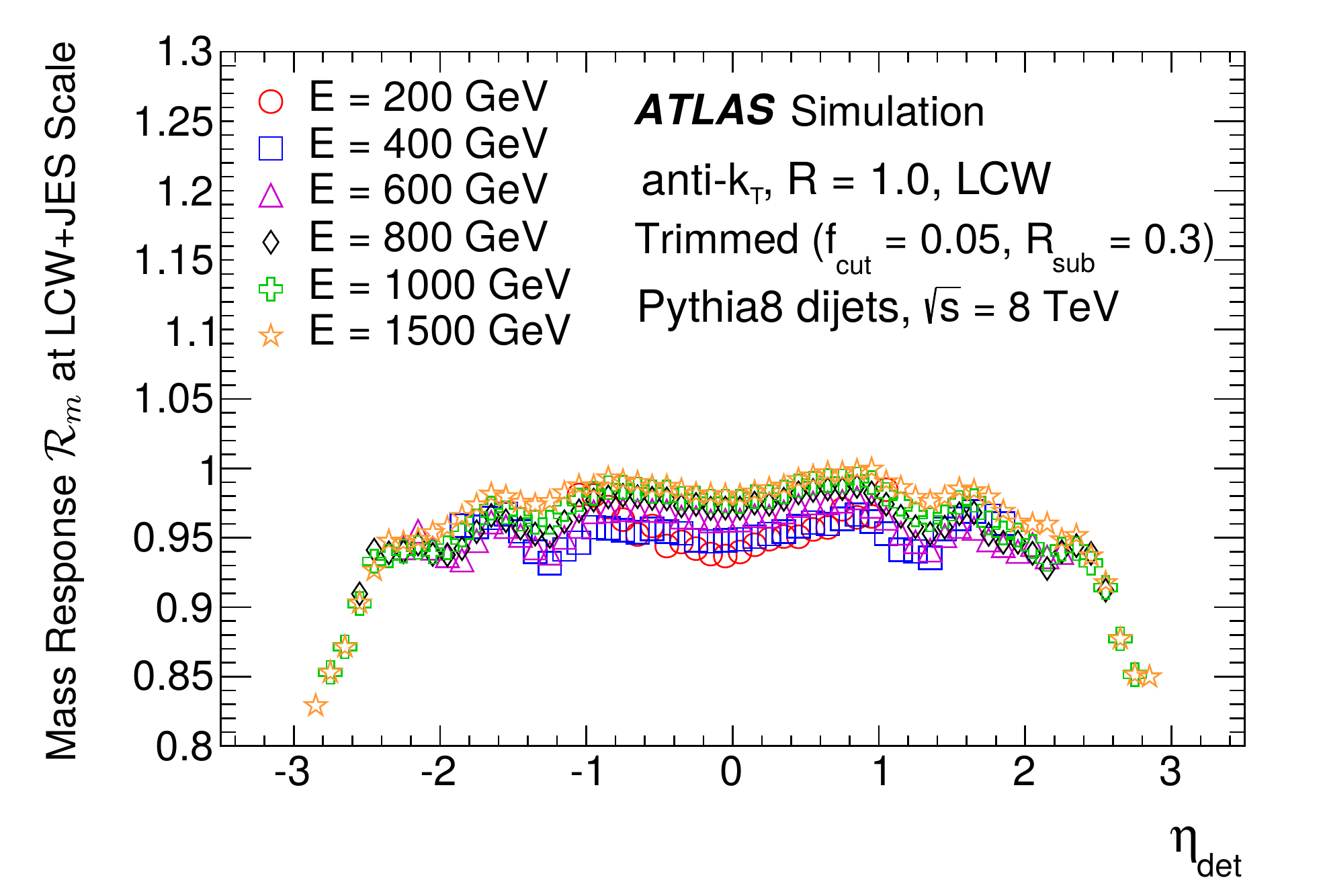}
\caption{Jet mass response $\mathcal{R}_m$, LCW scale, \rten{}}
\end{subfigure}
\caption{Jet energy and mass responses as a function of \etaDet{}
for different \tjet{} energies.
The energy responses $\mathcal{R}_E$ for \antikt{} jets with $R=0.4$ at the (a) EM scale and the (b) LCW scale and (c) for trimmed \antikt{} $R=1.0$ jets
are presented. Also, (d) the jet mass response $\mathcal{R}_m$
for the latter kind of jets is given.
\label{fig:calibcurve}
}
\end{figure}

\begin{figure}[p]
\centering
\begin{subfigure}{0.48\textwidth}\centering
\includegraphics[width=\textwidth]{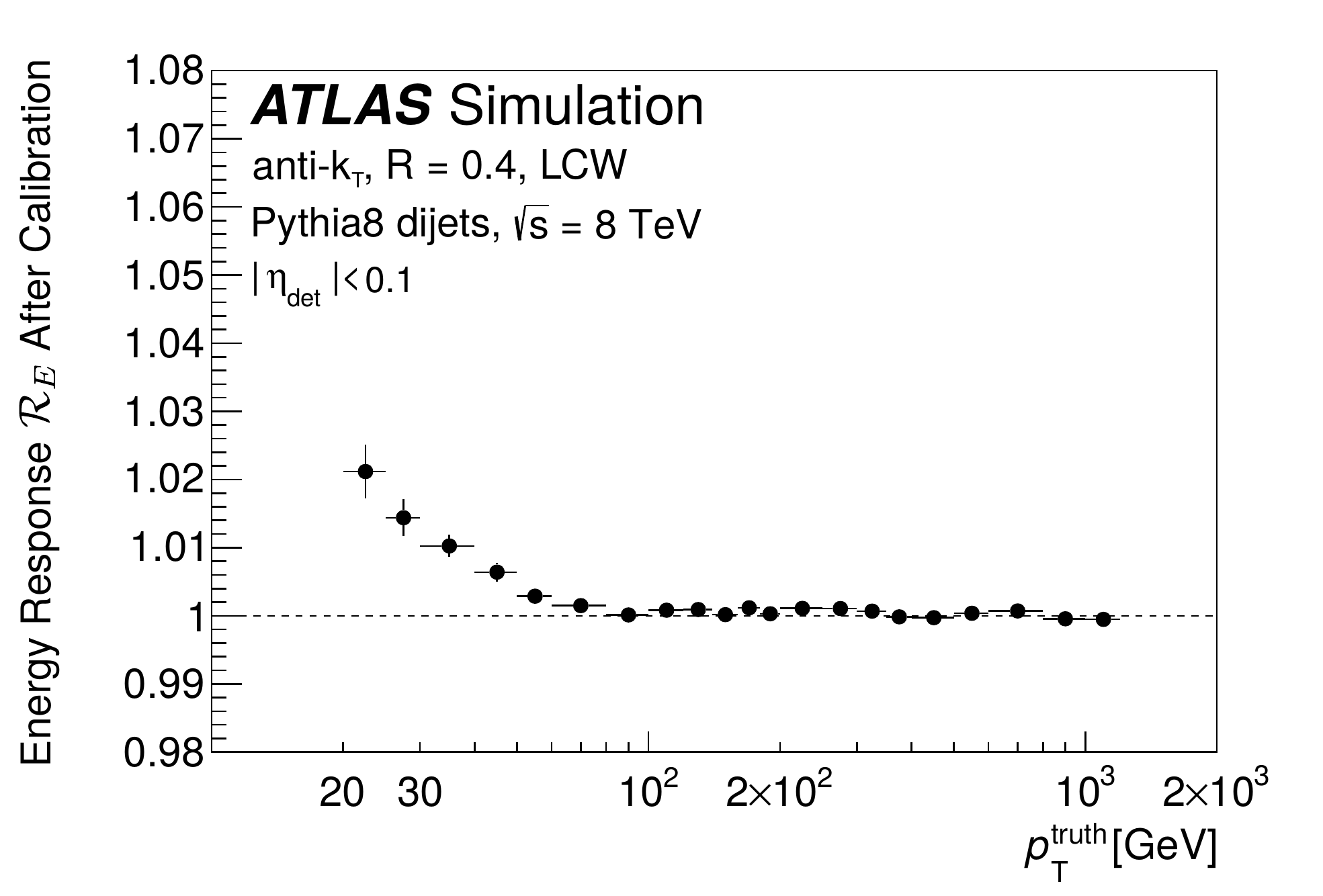}
\caption{Jet energy response $\mathcal{R}_E$ vs $\pt$, \rfour{}}
\end{subfigure}
\begin{subfigure}{0.48\textwidth}\centering
\includegraphics[width=\textwidth]{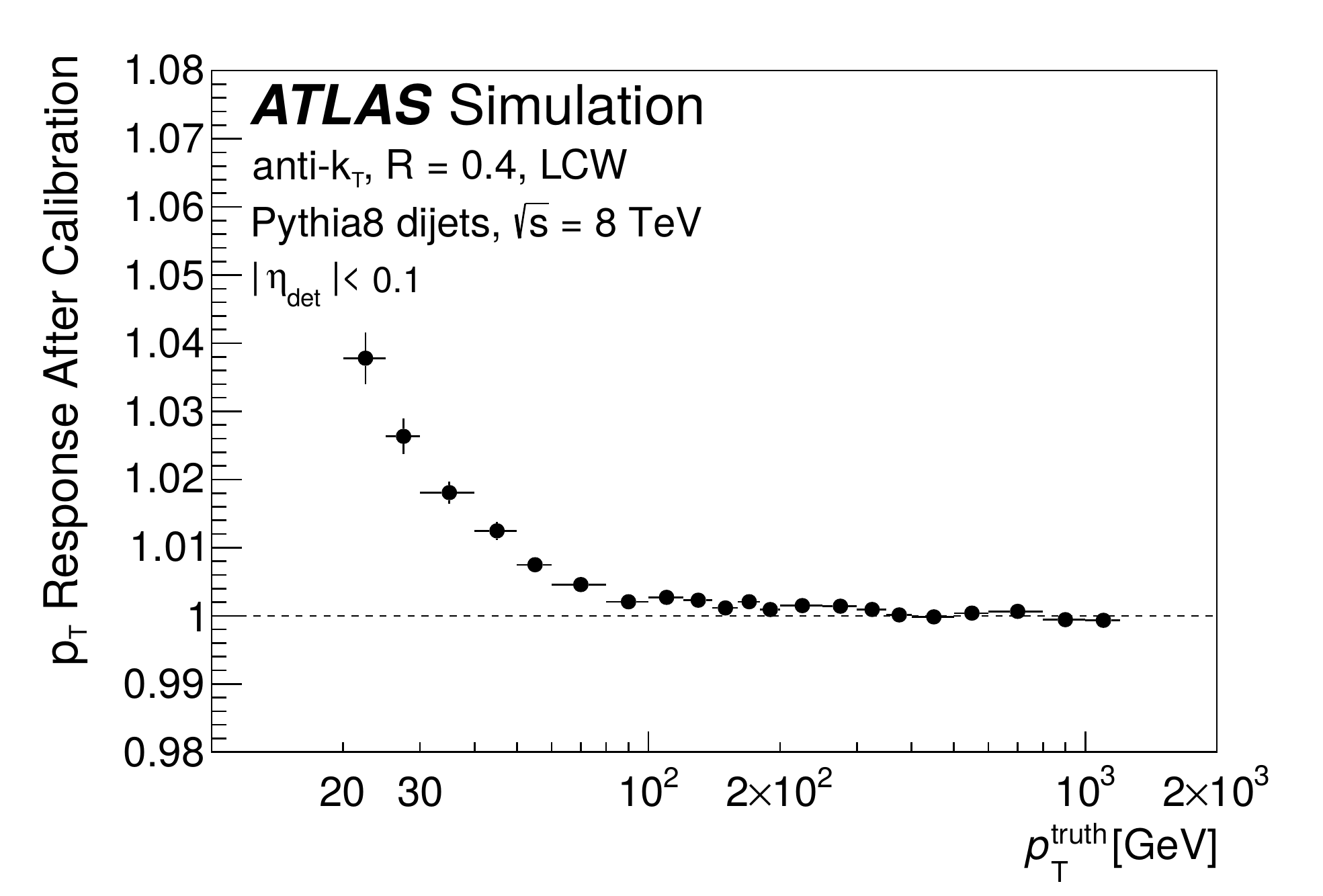}
\caption{Jet $\pt$ response vs $\pt$, \rfour{}}
\end{subfigure} \\
 
\bigskip
 
\begin{subfigure}{0.48\textwidth}\centering
\includegraphics[width=\textwidth]{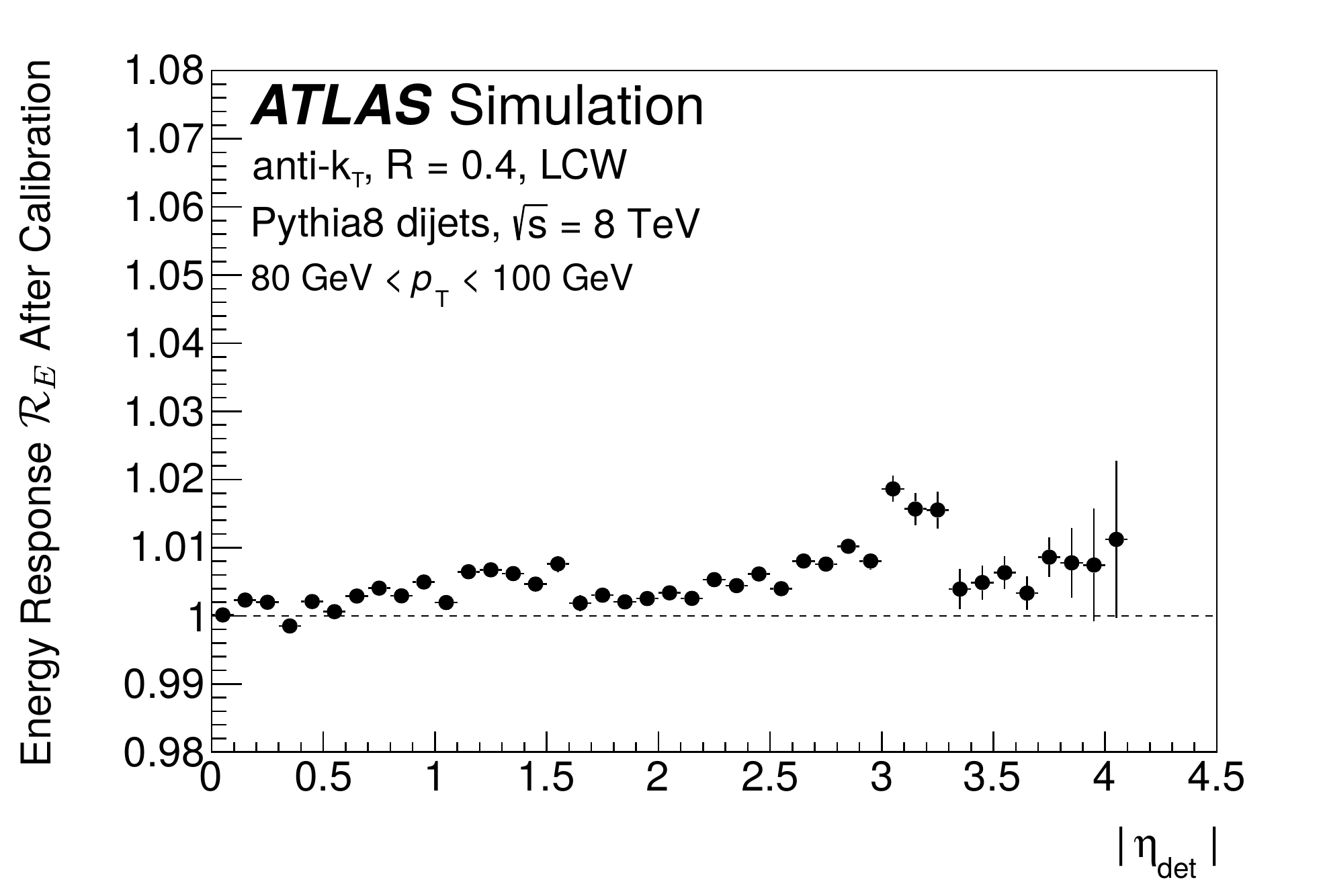}
\caption{Jet energy response $\mathcal{R}_E$ vs $|\etaDet{}|$, \rfour{}}
\end{subfigure}
\begin{subfigure}{0.48\textwidth}\centering
\includegraphics[width=\textwidth]{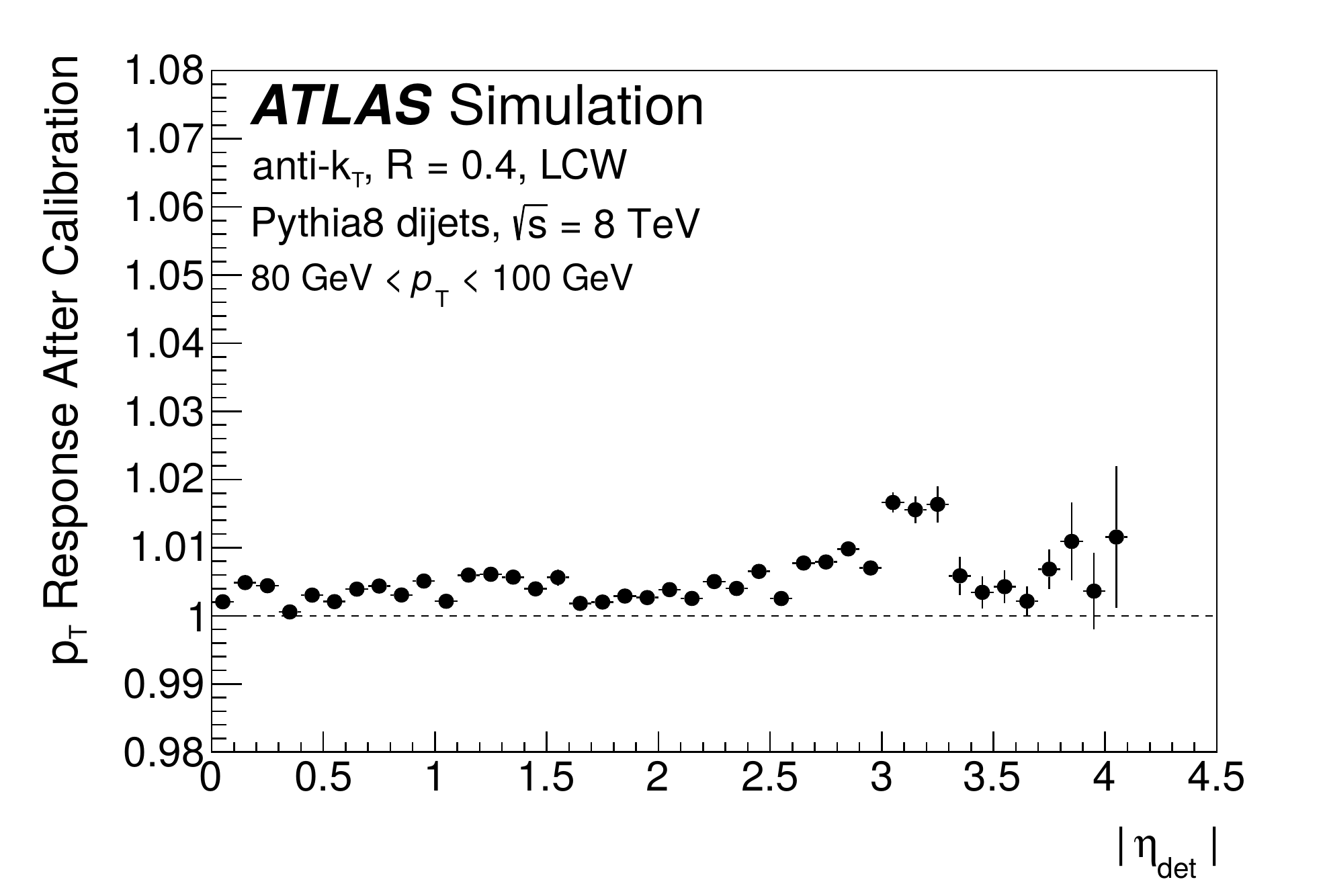}
\caption{Jet $\pt$ response vs $|\etaDet{}|$, \rfour{}}
\end{subfigure} \\
 
\bigskip
 
\begin{subfigure}{0.48\textwidth}\centering
\includegraphics[width=\textwidth]{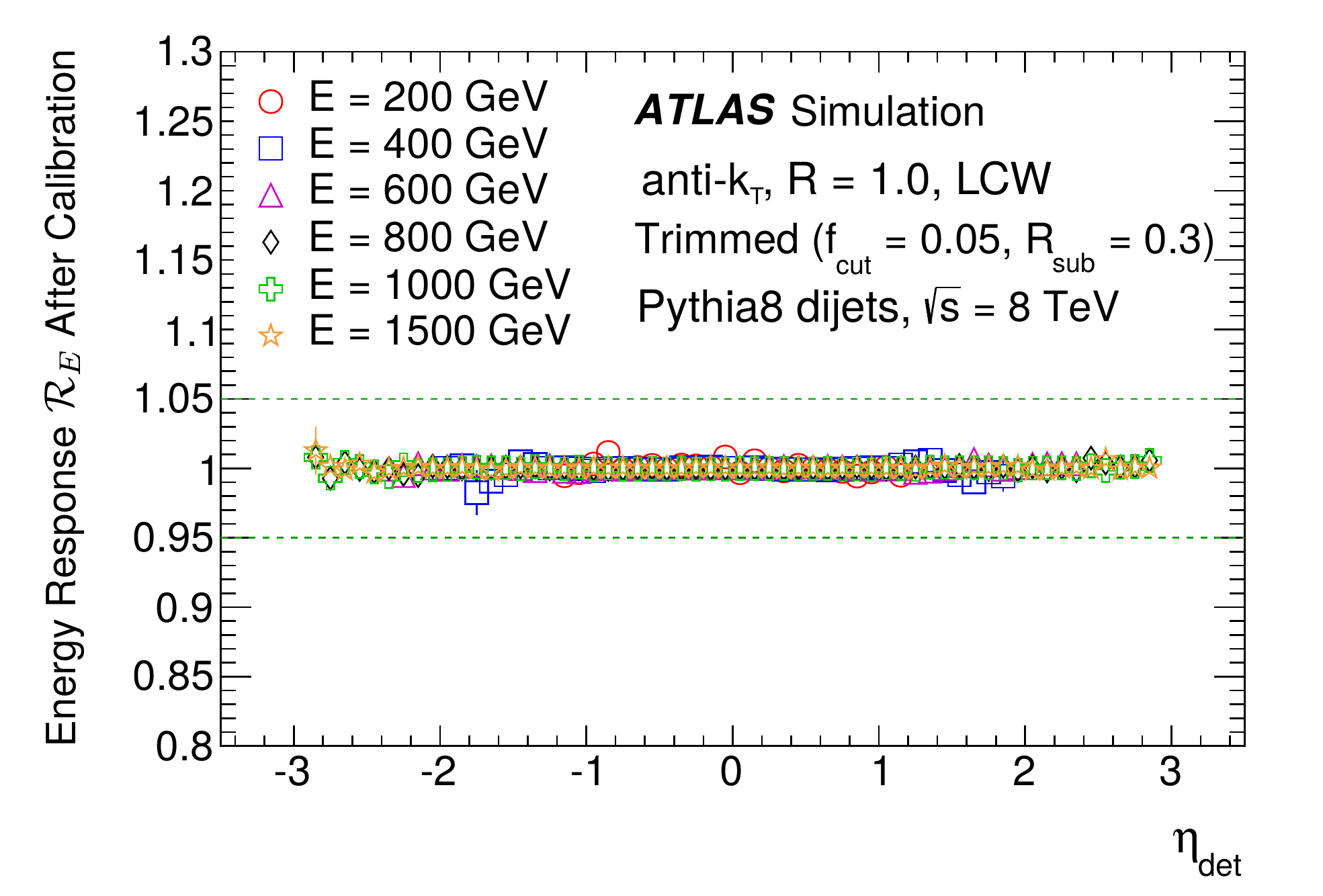}
\caption{Jet energy response $\mathcal{R}_E$ vs \etaDet{}, \rten{}}
\end{subfigure}
\begin{subfigure}{0.48\textwidth}\centering
\includegraphics[width=\textwidth]{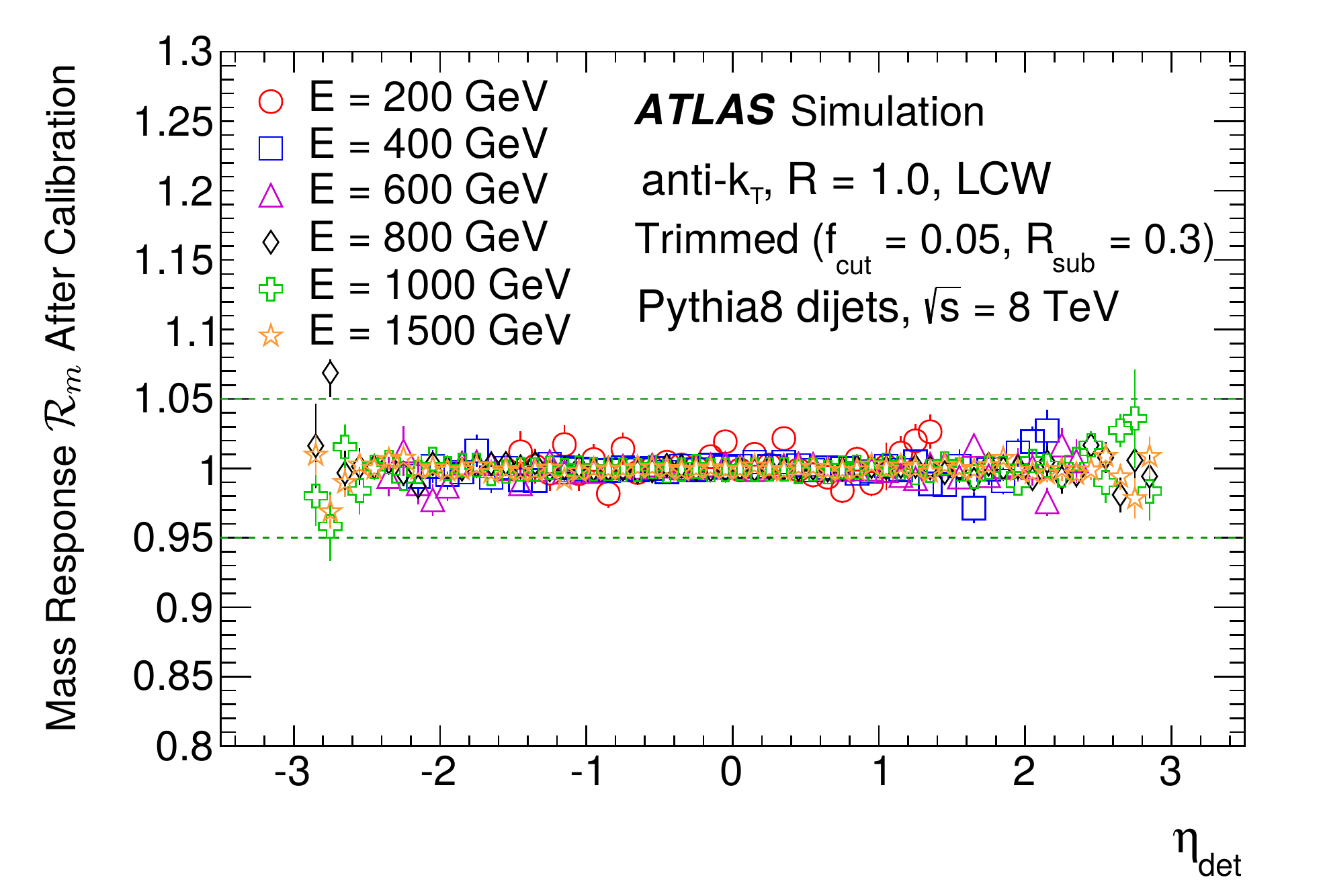}
\caption{Jet mass response $\mathcal{R}_m$ vs \etaDet{}, \rten{}}
\end{subfigure}
\caption{Jet energy, \pt{}, and mass response after the MC-based jet calibration has been applied for $R=0.4$ and $R=1.0$ \antikt{} jets reconstructed from LCW calibrated \topos{}.}
\label{fig:jesclosure}
\end{figure}

\subsection{Definition of the calibrated jet four momentum}
\label{sec:jetKinem}
 
For \sRjets{}, i.e.\ jets built with a radius parameter of $R=0.4$ or $R=0.6$, the fully calibrated jet four-momentum is specified by
\begin{equation}
\left(E,\eta,\phi,m\right) =
\left(\,c_\text{calib} \, E^\text{orig},\,\eta^\text{orig} + \Delta\eta ,\,\phi^\textrm{orig} ,\, c_\text{calib} \, m^\text{orig}\right),
\label{eq:smallRcalib}
\end{equation}
where the quantities denoted ``orig'' are the jet four-vector after the origin correction discussed in Section~\ref{sec:originCorr}, $\Delta\eta$ is the MC-based pseudorapidity calibration reported in Section~\ref{sec:EtaJES}, and $c_\text{calib}$ is a four-momentum scale factor that combines the other calibration steps:
\begin{equation}
c_\text{calib} = \begin{cases}
c_\text{PU} \cdot c_\text{JES} \cdot c_\text{GS} \cdot c_\eta \cdot c_\text{abs} & \text{for data} \\
c_\text{PU} \cdot c_\text{JES} \cdot c_\text{GS}                                & \text{for MC simulation}.
\end{cases}
\label{eq:calibFactors}
\end{equation}
Here, the \pileup{} correction factor is defined as
\begin{equation*}
c_\textrm{PU} = \frac{\pt{} - \rho\ A - \alpha(N_\textrm{PV}-1) - \beta \mu }{\pt}
\end{equation*}
in accordance with Eq.~(\ref{eq:PU}) ($\pt \mapsto c_\text{PU}\,\pt$), $c_\text{JES}$ is derived as explained in Section~\ref{sec:EtaJES}, $c_\text{GS}$ is
the global sequential calibration that is discussed in Section~\ref{sec:gsc}, and
the pseudorapidity intercalibration $c_\eta$ 
and the absolute \insitu{} calibration $c_\text{abs}$ are detailed in Sections~\ref{sec:dijet}--\ref{sec:JEScomb}.
As given in Eq.~(\ref{eq:calibFactors}), the MC-derived calibrations $c_\text{JES}$ and $c_\text{GSC}$ correct simulated jets to the \tjet{} scale,
but jets in data need the \insitu{} corrections $c_\eta$ and $c_\text{abs}$ to reach this scale.
JES systematic uncertainties are evaluated for the \insitu{} terms.
 
The calibration procedure is slightly different for the \lRjets{} used in this paper (Section~\ref{sec:jetReco}).
These jets do not receive any origin correction or global sequential calibration as the precision needs of the overall scale are not the same as for $R=0.4$ and $R=0.6$ jets.
Further, no \pileup{} correction is applied since the trimming algorithm detailed in Section~\ref{sec:jetReco} mitigates the \pileup{} dependence.
However, \lRjets{} do receive a MC-derived jet mass calibration $c_\text{mass}$. The calibrated \lRjet{} four-momentum is given by
\begin{equation}
\left(E,\eta,\phi,m\right) =
\left(\,c_\text{JES} \, E^\text{const},\,\eta^\text{const} + \Delta\eta ,\,\phi^\textrm{const} ,\,
c_\text{mass} \, m^\text{const}\right).
\label{eq:largeRcalib}
\end{equation}
By expressing the jet transverse momentum in terms of energy, mass, and pseudorapidity,
it can be seen that all calibration terms of Eqs.~(\ref{eq:smallRcalib}) and (\ref{eq:largeRcalib}) affect \pt{}, for example
\begin{equation*}
\pt = \frac{E \ominus m}{\cosh{\eta}} = \frac{c_\text{JES} \, E^\text{const} \ominus c_\text{mass} \, m^\text{const} }{\cosh \left( \eta^\text{const} + \Delta\eta \right) },
\label{eq:pt}
\end{equation*}
where the symbol $\ominus$ denotes subtraction in quadrature, i.e.\ $a \ominus b \equiv \sqrt{a^2-b^2}$.

% End of text imported from the .//sections/jets.tex input file
 
\afterpage{\clearpage}
% The next lines are included from the .//sections/gsc.tex input file
\section{Global sequential calibration}
\label{sec:gsc}
The global sequential (\GS)~calibration scheme exploits the topology of the energy deposits in the calorimeter as well as tracking information to characterize fluctuations in the jet particle content of the hadronic shower development.
Correcting for such fluctuations can improve the jet energy resolution and reduce
response dependence on the so-called ``jet flavour'', meaning dependence on the underlying physics process in which the jet was produced.
Jets produced in dijet events tend to have more constituent particles, a wider transverse profile and a lower calorimeter energy response than jets with the same \pt{} and $\eta$ produced in the decay of a $W$ boson or in association with a photon ($\gamma$+jet) or $Z$~boson ($Z$+jet). This can be attributed to differences in fragmentation between ``quark-initiated'' and ``gluon-initiated'' jets.
The GS calibration also exploits information related to the activity in the muon chamber behind uncontained calorimeter jets, for which the reconstructed energy tends to be smaller with a degraded resolution.
The calibration is applied in sequential steps, each designed to flatten the jet energy response as a function of a jet property without changing the mean jet energy.
 
\subsection{Description of the method}
\label{sec:GS-method}
 
Any variable $x$ that carries information about the jet response  can be used for the GS calibration.
A multiplicative correction to the jet energy measurement is derived by inverting the jet response as a function of this variable: $c(x) = k/R(x)$, where the constant $k$ is chosen to ensure that the average energy is not affected by the calibration, and the average jet response $R(x)$ is determined using MC simulation as described in Section~\ref{sec:jetMatch}. After a successful application, the jet response should
no longer depend on $x$.
As a result, the spread of reconstructed jet energy is reduced, thus improving the resolution.
 
Each correction is performed separately in bins of $\etaDet$, in order to account for changes in the jet \pT{} response in different detector regions and technologies.  The corrections are further parameterized as a function of {\pt} and jet property $x$: $c({\pt},x)$, except for the correction for uncontained calorimeter jets, which is constructed as a function of jet energy $E$ and
the logarithm of the number of muon segments reconstructed in the muon chambers behind the jet: $c(E,\log\Nsegments)$.
The uncontained calorimeter jet correction is constructed using the jet $E$ rather than the \pT{} to better represent the probability of a jet penetrating the full depth of the calorimeter, which depends on $\log E$.
The two-dimensional calibration function is constructed
using a two-dimensional Gaussian kernel~\cite{PERF-2012-01}
for which the kernel-width parameters are chosen to capture the shape of the response across $\etaDet$ and \pt, and at the same time provide stability against statistical fluctuations.
 
Several variables can be used sequentially to achieve the optimal resolution. The jet \pt{} after $N$ GS calibration steps is given by the initial jet \pt{} multiplied by the product of the $N$ corrections:
\begin{eqnarray}
p_\text{T}^\text{GS} = p_{\text{T},0}\,c_\mathrm{GS} = p_{\text{T},0}\,\prod\nolimits_{j=1}^{N}c_j(\,p_{\mathrm{T},j-1},\,x_j\,), &  & p_{\text{T},i} = p_{\text{T},i-1} \,c_i(p_{\text{T},i-1},x_i),
\end{eqnarray}
where $p_{\text{T},0}$ is the jet $\pt$ prior to the GS calibration.
Hence, when deriving correction $j$, one needs to start by calibrating the jets with the previous $j-1$ correction factors.
This method assumes there is little to gain from non-linear correlations of the variables used and this has been demonstrated in simulation.

\subsection{Jet observables sensitive to the jet calorimeter response}
\label{sec:jet-properties}
 
The GS calibration relies on five jet properties that were identified empirically to have a significant effect on the jet energy response.
This empirical study was conducted primarily using EM jets, while a reduced scan was performed for LCW jets given that they already exploit some of the following variables as part of the LCW procedure.
Two of the variables characterize the longitudinal shower structure of a jet, namely the fractions of energy deposited in the third electromagnetic calorimeter layer, $\fem$\footnote{
The ATLAS calorimeters have three electromagnetic layers in the pseudorapidity interval $|\eta|<2.5$, but only two in $2.5<|\eta|<3.2$. \fem{} includes energy deposits with $|\eta|<2.5$ in the third EM~layer and contributions with $2.5<|\eta|<3.2$ in the second EM~layer. Energy deposits with $|\eta|>3.2$ are not included, however a jet with $|\eta|\gtrsim 3.2$ will most often have \topos{} with $|\eta|<3.2$ that leave contributions to the second EM layer.}, and in the first hadronic Tile calorimeter layer, $\ftile$. These fractions are defined according to
\begin{eqnarray}
\fem = E^{\LAr 3}_\mathrm{EM} {\big /} E^\mathrm{jet}_\mathrm{EM}, & \mathrm{and~}
\ftile = E^{\Tile 0}_\mathrm{EM} {\big /} E^\mathrm{jet}_\mathrm{EM},
\end{eqnarray}
where the subscript EM refers to the electromagnetic scale.
The next two of the five jet properties rely on reconstructed tracks from the selected primary vertex that are matched to the calorimeter jets using ghost association (Section~\ref{sec:jetMatch}).
The tracks are required to fulfil quality criteria relating to their impact parameter and the number of
hits in the different inner-detector layers, and to have $\pt>1$~\GeV\ and $|\eta|<2.5$.
The track-based observables are the number of tracks associated with a given jet \nTrk, and the jet width $\trackWIDTH$ defined as
\begin{equation}
\trackWIDTH = \sum_{i=1}^{N_\text{trk}}p_{\text{T},i}\,\Delta R(i,\text{jet}) {\big /} \sum_{i=1}^{N_\text{trk}}p_{\text{T},i},
\end{equation}
where $N_\text{trk}$ are the number of tracks associated with the jet, $p_{\text{T},i}$ is the \pt{} of the $i^\text{th}$ track, and $\Delta R(i,\text{jet})$ is the
$\Delta R$ distance in $(\eta,\phi)$-space between the $i^\text{th}$ track and the calorimeter jet axis.
The jet width $\trackWIDTH$ quantifies the transverse structure of the jet, which is sensitive to the ``jet flavour''.
The final variable used in the GS calibration is \Nsegments{}, the number of muon segments behind the jet, which quantifies the activity in the muon chambers.
Muon segments are partial tracks constructed from hits in the muon spectrometer chambers~\cite{Aad:2014rra},
and are matched to the jet of interest in two stages.
Based on jets built using \antikt{} with $R=0.6$, \Nsegments{} is defined by the number of matching muon segments within a cone of size $\Delta R=0.4$ around the jet axis.
For \antikt{} $R=0.4$ jets, the closest $R=0.6$ jet is found (fulfilling $\Delta R<0.3$), and $\Nsegments$ is assigned to the $R=0.4$ jets according to the corresponding value for the $R=0.6$ jet.
 
Figures~\ref{fig:inputs_dataMC_comparison_LowPt} and~\ref{fig:inputs_dataMC_comparison_HighPt} show distributions comparing data with MC simulations for $\ftile$, $\fem$, $\nTrk$, $\trackWIDTH$ and $\Nsegments$ for jets with $|\etaDet|<0.6$ produced in dijet events selected as described in Section~\ref{sec:dijet-selection}. Predictions are provided using the default \pythia{} sample with full detector simulation from which the GS calibration is derived, and also using the AFII fast simulation, which is often used in physics analyses (Section~\ref{sec:mc}). For the AFII detector simulation, there is no complete implementation of the muon segments produced behind high-energy uncontained jets. Therefore, this correction is not applied to AFII samples, and no AFII prediction is provided in Figure~\ref{fig:inputs_dataMC_comparison_HighPt}(e).
It can be seen that the simulation predicts the general shapes of the data, although there are visible differences.
Similar results are found in the other \etaDet{} regions.
Disagreements in the distributions of the jet properties have little impact on the GS calibration performance as long 
as the response dependence $R(x)$ of the jet properties $x$ is well described by the simulation (Section~\ref{sec:GS-data-validation}).
 
\begin{figure}[ht]
\centering
\begin{subfigure}{0.45\textwidth}\centering
\includegraphics[width=\textwidth]{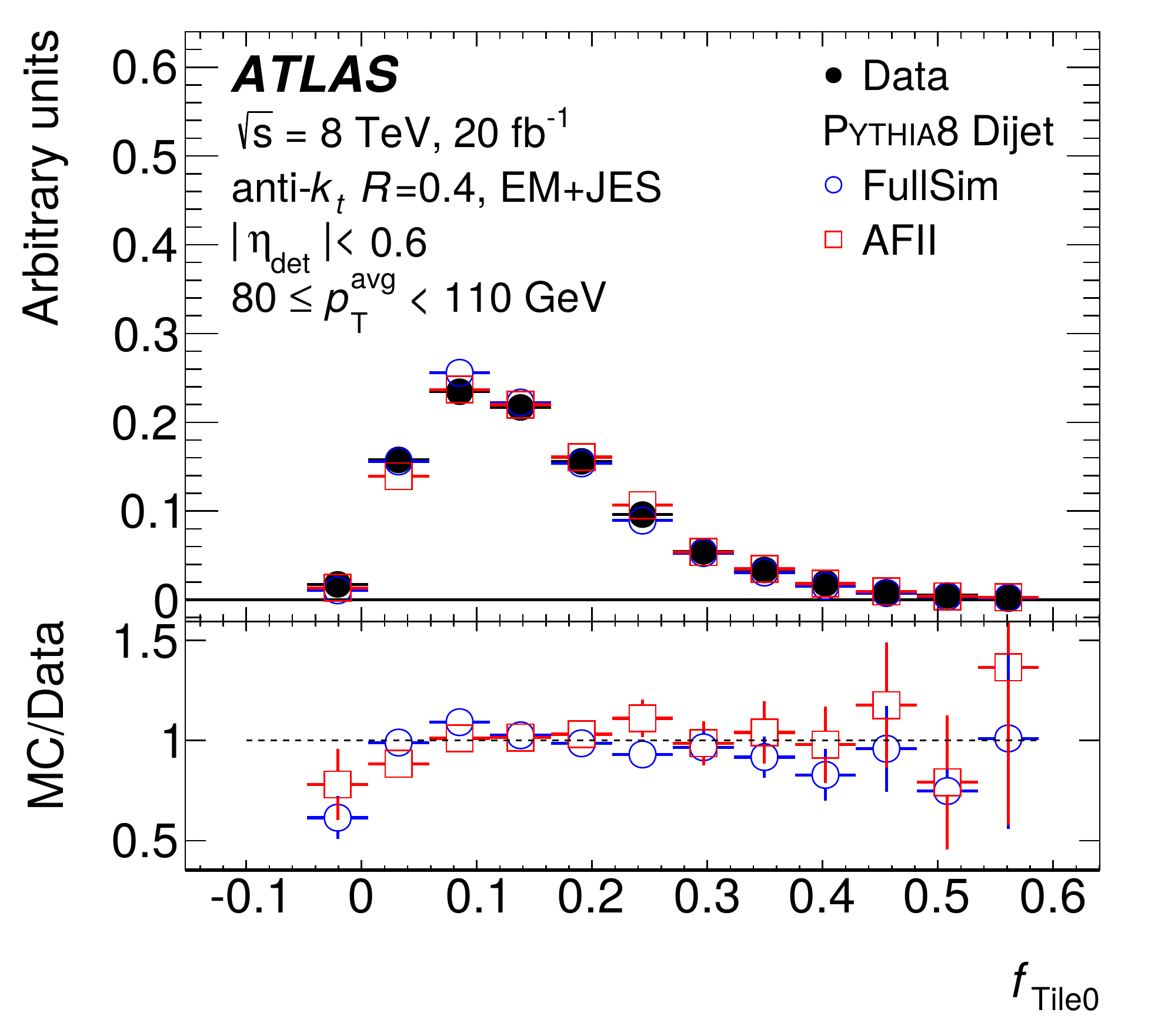}
\caption{}
\end{subfigure}
\hspace{0.03\textwidth}
\begin{subfigure}{0.45\textwidth}\centering
\includegraphics[width=\textwidth]{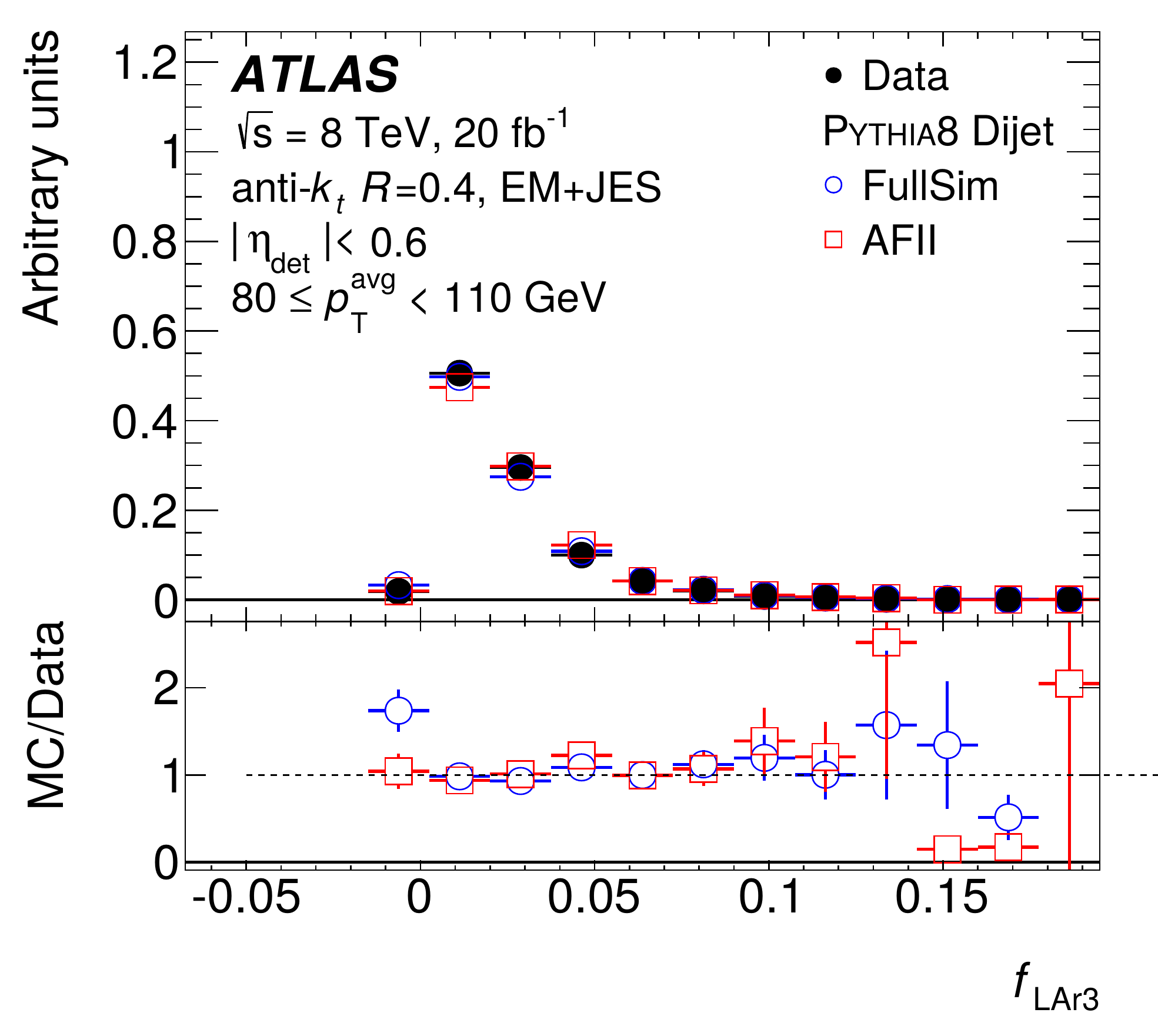}
\caption{}
\end{subfigure}\\
\begin{subfigure}{0.45\textwidth}\centering
\includegraphics[width=\textwidth]{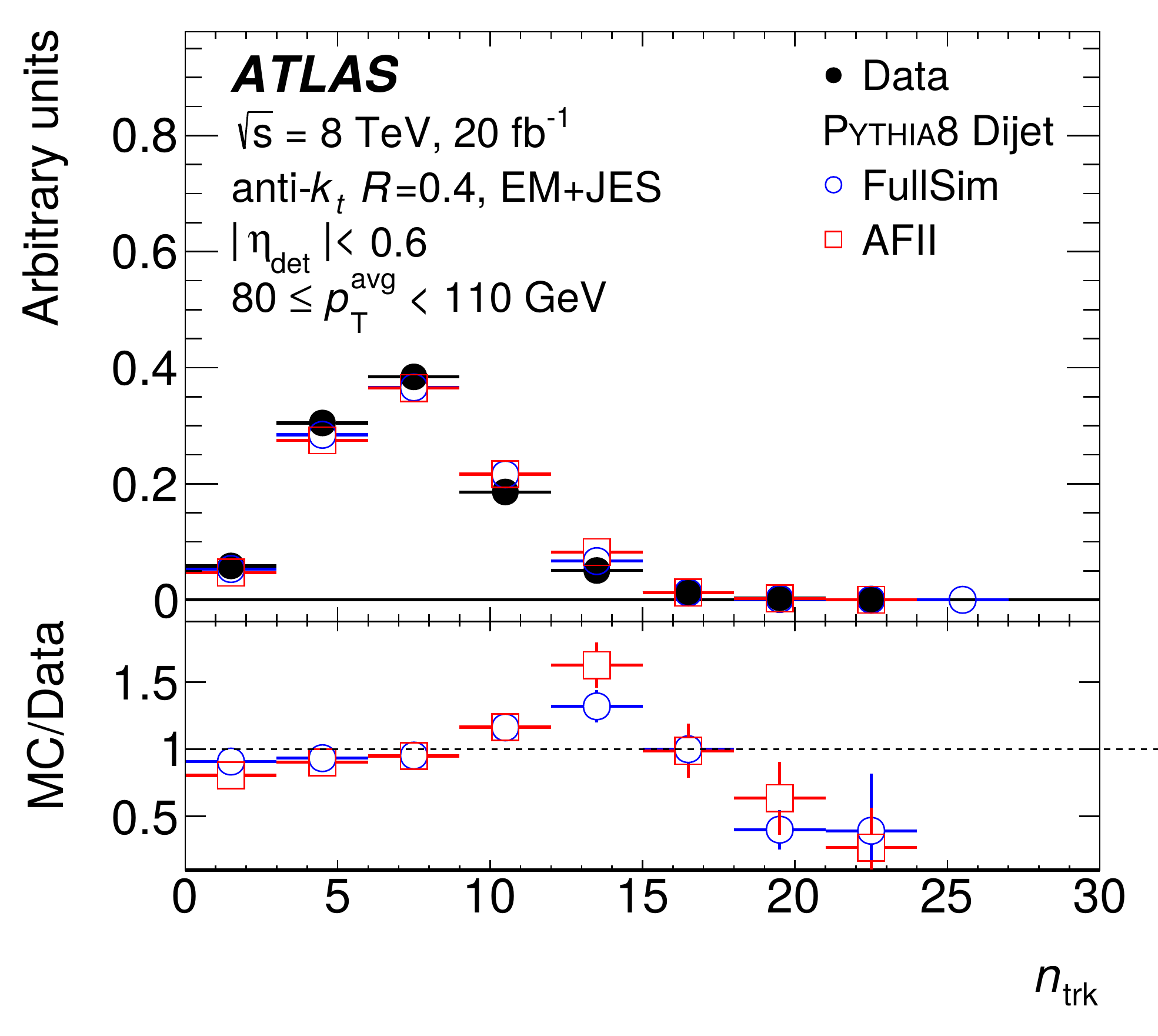}
\caption{}
\end{subfigure}
\hspace{0.03\textwidth}
\begin{subfigure}{0.45\textwidth}\centering
\includegraphics[width=\textwidth]{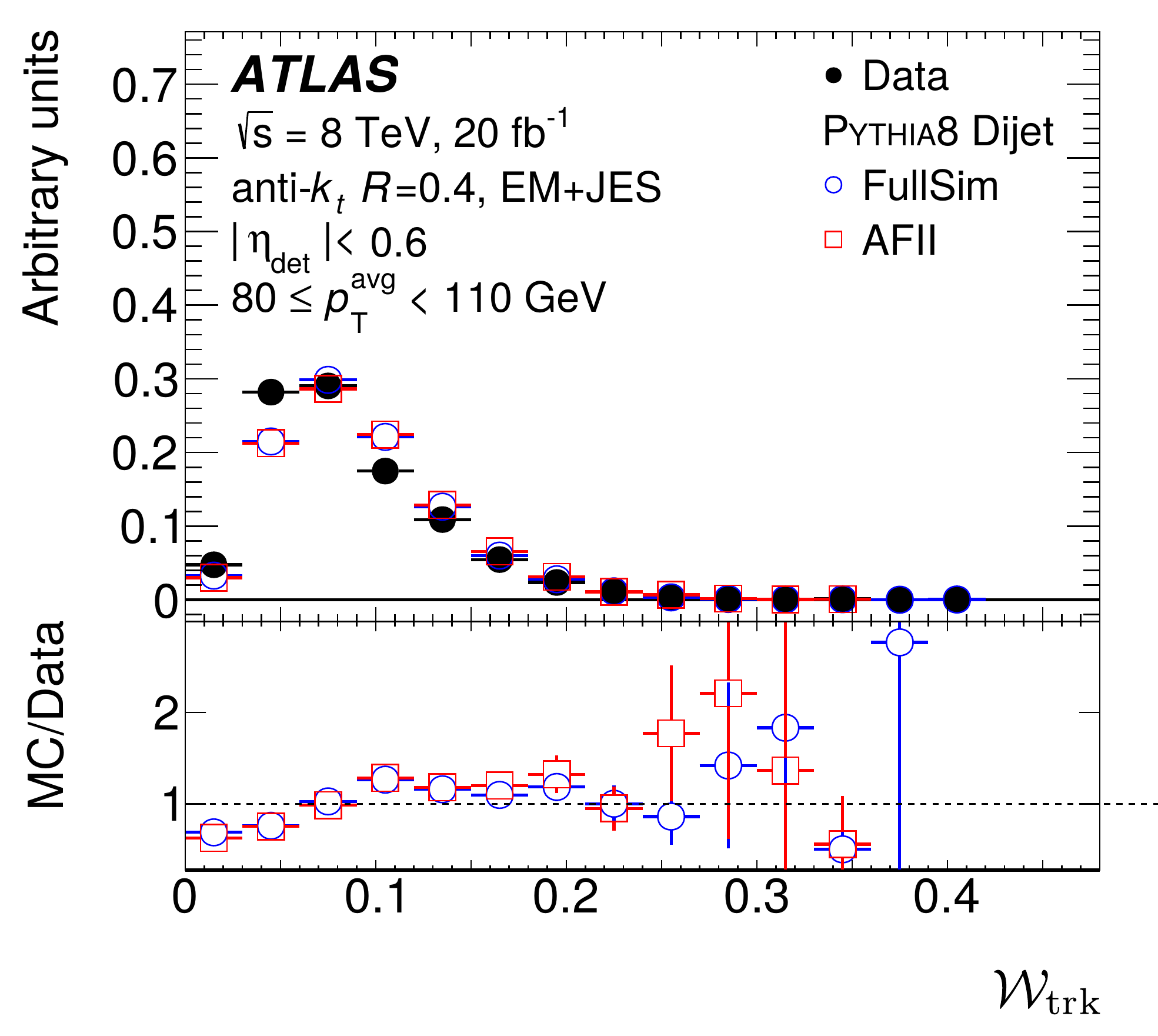}
\caption{}
\end{subfigure}
\caption{
Normalized distributions of $\ftile$, $\fem$, $\nTrk$, and $\trackWIDTH{}$ for jets $|\etaDet|<0.6$ in dijet events with $\SI{80}{\GeV}<\ptavg<\SI{110}{\GeV}$ in data~(filled circles) and \pythia{} MC simulation with both full~(empty circles) and fast~(empty squares) detector simulation. All jets are reconstructed with {\antikt} $R=0.4$ and calibrated with the \EMJES{} scheme.
The quantity $\ptavg$ is the average \pt{} of the leading two jets in an event, and hence represent the \pt{} scale of the jets being probed.
$\Nsegments$ is not shown since the vast majority of jets in this \pt{} range have $\Nsegments = 0$.
}
\label{fig:inputs_dataMC_comparison_LowPt}
\end{figure}
 
\begin{figure}[p]
\centering
\begin{subfigure}{0.45\textwidth}\centering
\includegraphics[width=\textwidth]{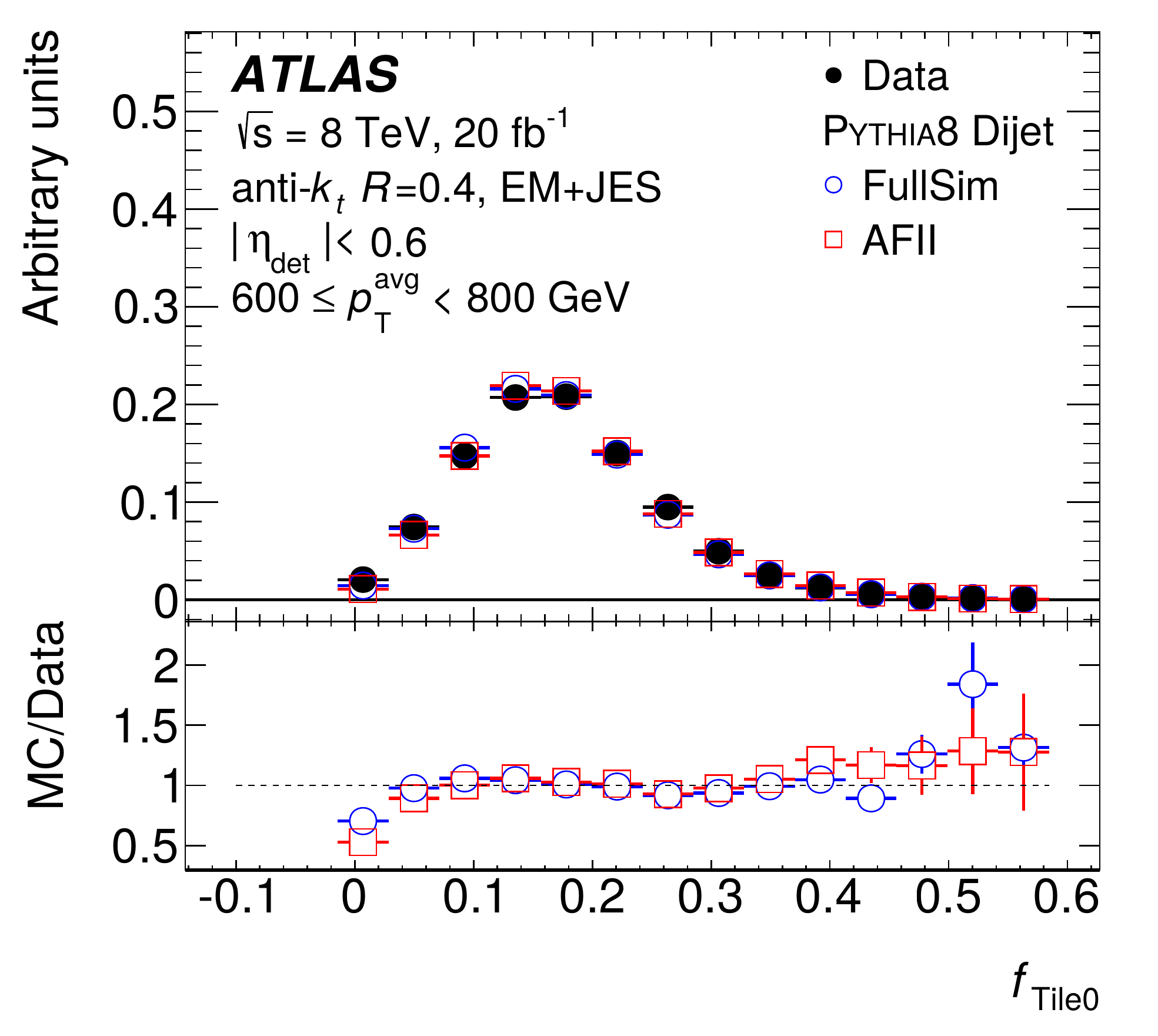}
\caption{}
\end{subfigure}
\hspace{0.03\textwidth}
\begin{subfigure}{0.45\textwidth}\centering
\includegraphics[width=\textwidth]{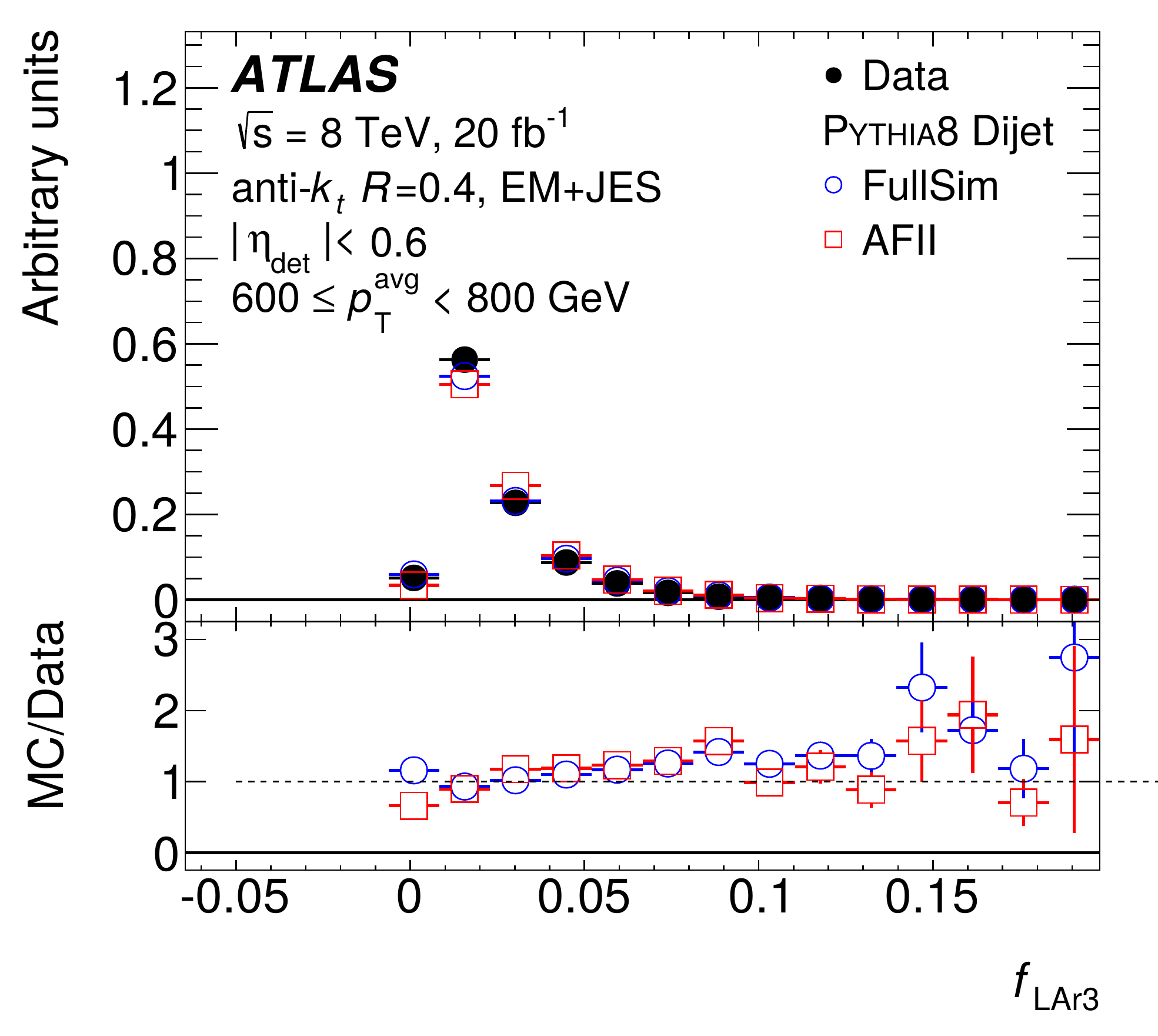}
\caption{}
\end{subfigure}
\\
\begin{subfigure}{0.45\textwidth}\centering
\includegraphics[width=\textwidth]{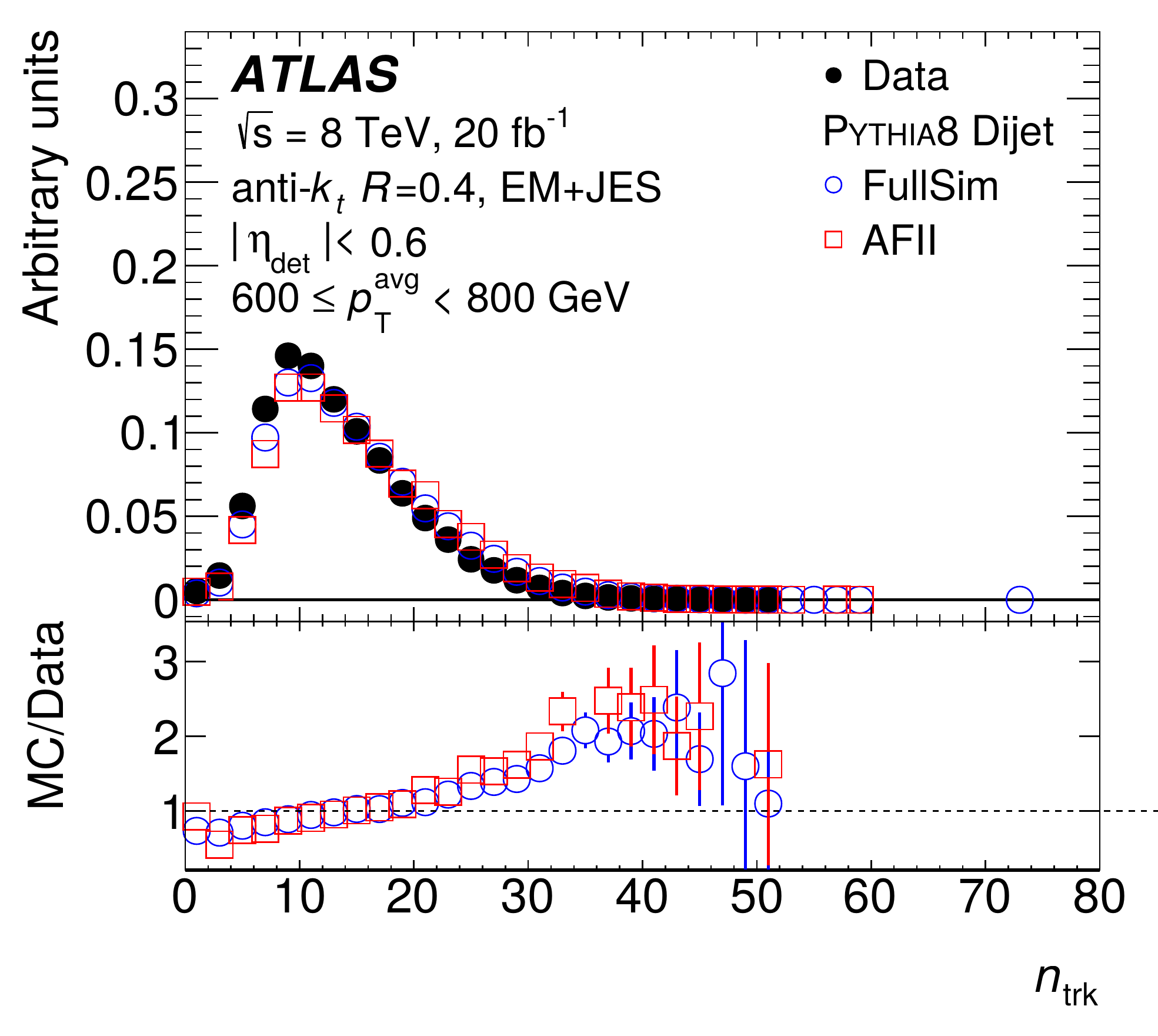}
\caption{}
\end{subfigure}
\hspace{0.03\textwidth}
\begin{subfigure}{0.45\textwidth}\centering
\includegraphics[width=\textwidth]{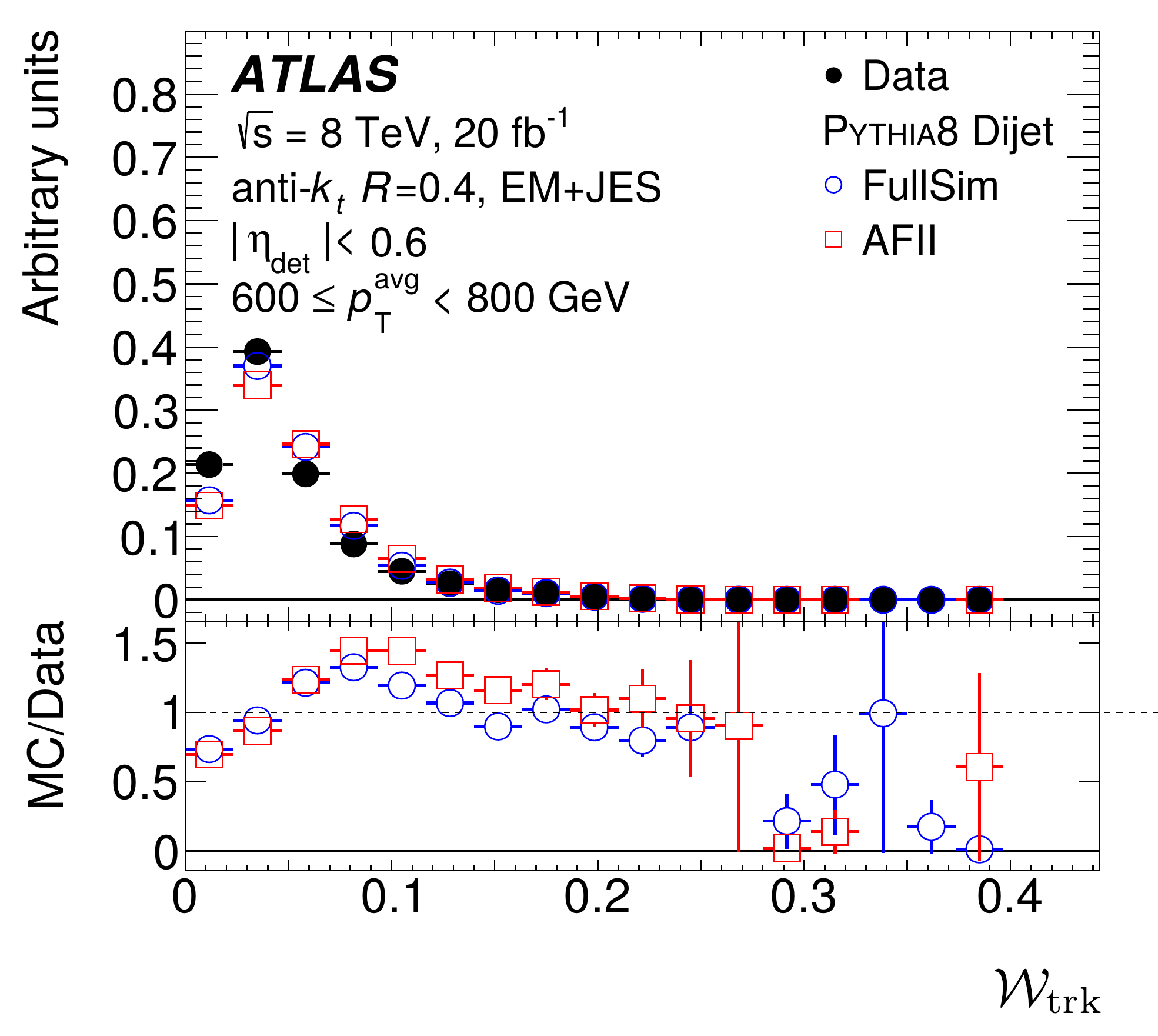}
\caption{}
\end{subfigure}
\\
\begin{subfigure}{0.45\textwidth}\centering
\includegraphics[width=\textwidth]{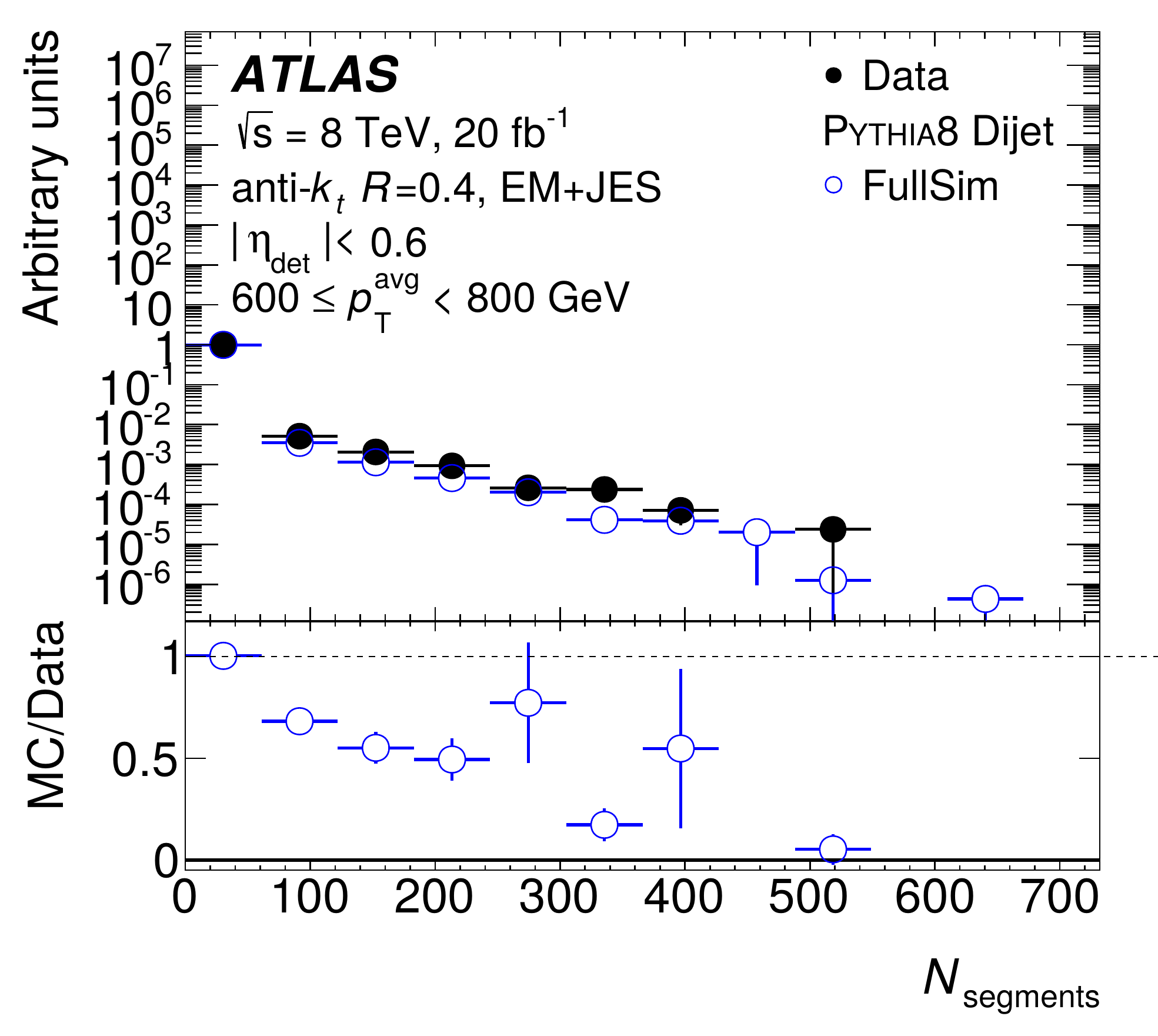}
\caption{}
\end{subfigure}
\caption{
Normalized distributions of $\ftile$, $\fem$, $\nTrk$, $\trackWIDTH{}$, and $\Nsegments$ for jets $|\etaDet|<0.6$ in dijet events with $\SI{600}{\GeV}<\ptavg<\SI{800}{\GeV}$
in the data~(filled circles) and \pythia{} MC simulation with both full~(empty circles) and fast~(empty squares) simulation.
All jets are reconstructed with {\antikt} $R=0.4$ and calibrated with the \EMJES{} scheme.
The quantity $\ptavg$ is the average \pt{} of the leading two jets in an event, and hence represent the \pt{} scale of the jets being probed.
}
\label{fig:inputs_dataMC_comparison_HighPt}
\end{figure}
 
\subsection{Derivation of the global sequential jet calibration}
\label{sec:GS-derivation}
 
The jet observables used for the GS calibration and their order of application are summarized in Table~\ref{tab:properties}. The first four corrections are determined separately in \etaDet{}-bins of width 0.1 and are parameterized down to $\pt{}=15$~\GeV.
\begin{table}[htb]
\caption{Sequence of GS corrections used to improve the jet performance in each \detEta{} region. For jets at the \LCWJES{} scale, only the
tracking and uncontained calorimeter jet corrections are applied.}
\label{tab:properties}
\centering
\begin{tabular}{c|c|c|c|c|c}
\hline\hline
$|\eta|$ region & Correction 1 & Correction 2 & Correction 3 & Correction 4 & Correction 5 \\
\hline
$[0,1.7]$ & \ftile & \fem & $\nTrk$ & $\trackWIDTH{}$ & \Nsegments \\
$[1.7,2.5]$ &  & \fem & $\nTrk$ & $\trackWIDTH{}$ & \Nsegments \\
$[2.5,2.7]$ & & \fem &  & & \Nsegments \\
$[2.7,3.5]$ & & \fem &  & & \\
\hline\hline
\end{tabular}
\end{table}
For jets at the \LCWJES{} scale, only the tracking and uncontained calorimeter jets corrections are applied since the LCW calibration already takes into account shower shape information.
No further improvement in resolution is thus achieved through the use of \ftile{} and \fem{} for LCW jets.

The calorimeter response for \EMJES{} calibrated {\antikt} $R=0.4$ jets with \pttruth{} in three representative intervals is presented as a function of the different jet property variables used by the GS calibration in
Figure~\ref{fig:resp_vs_x_beforeCorrections}.
For all properties, a strong dependence of the response as a function of the property is observed.
The $\nTrk$ and $\trackWIDTH$ show a stronger $\pt$ dependence than the other properties and this is extensible for other $\pt$ and $\etaDet$ bins and jet collections. 
The corresponding distributions after the GS calibration are shown in Figure~\ref{fig:resp_vs_x_afterCorrections}.
The jet response dependence on the jet properties is removed to within 2\% after applying the GS calibration for all observables.
Deviations from unity are expected since the correlations between the variables are not accounted for in the
GS calibration procedure.

\begin{figure}[p]
\centering
\begin{subfigure}{0.42\textwidth}\centering
\includegraphics[width=\textwidth]{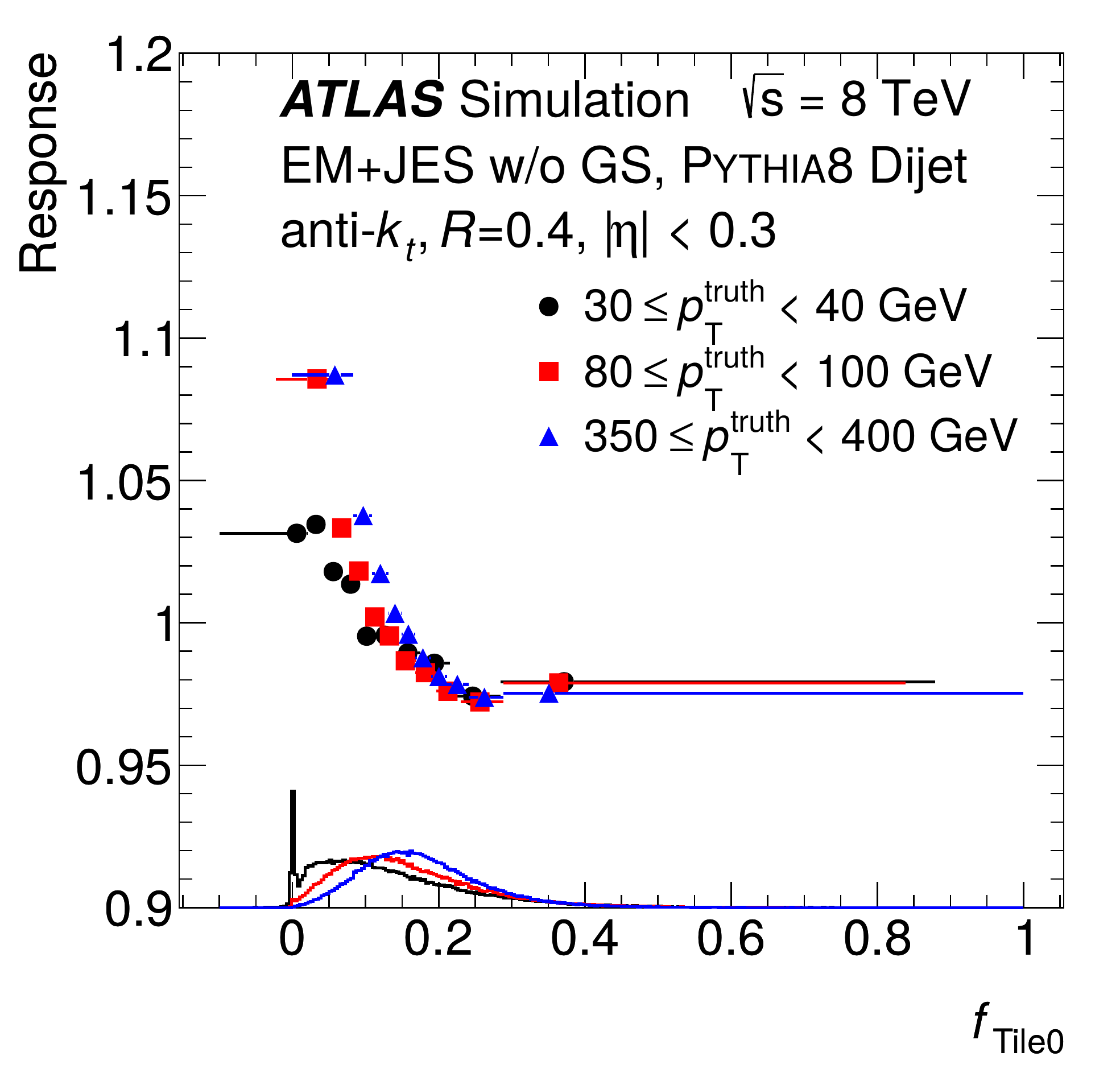}
\caption{}
\end{subfigure}
\hspace{0.03\textwidth}
\begin{subfigure}{0.42\textwidth}\centering
\includegraphics[width=\textwidth]{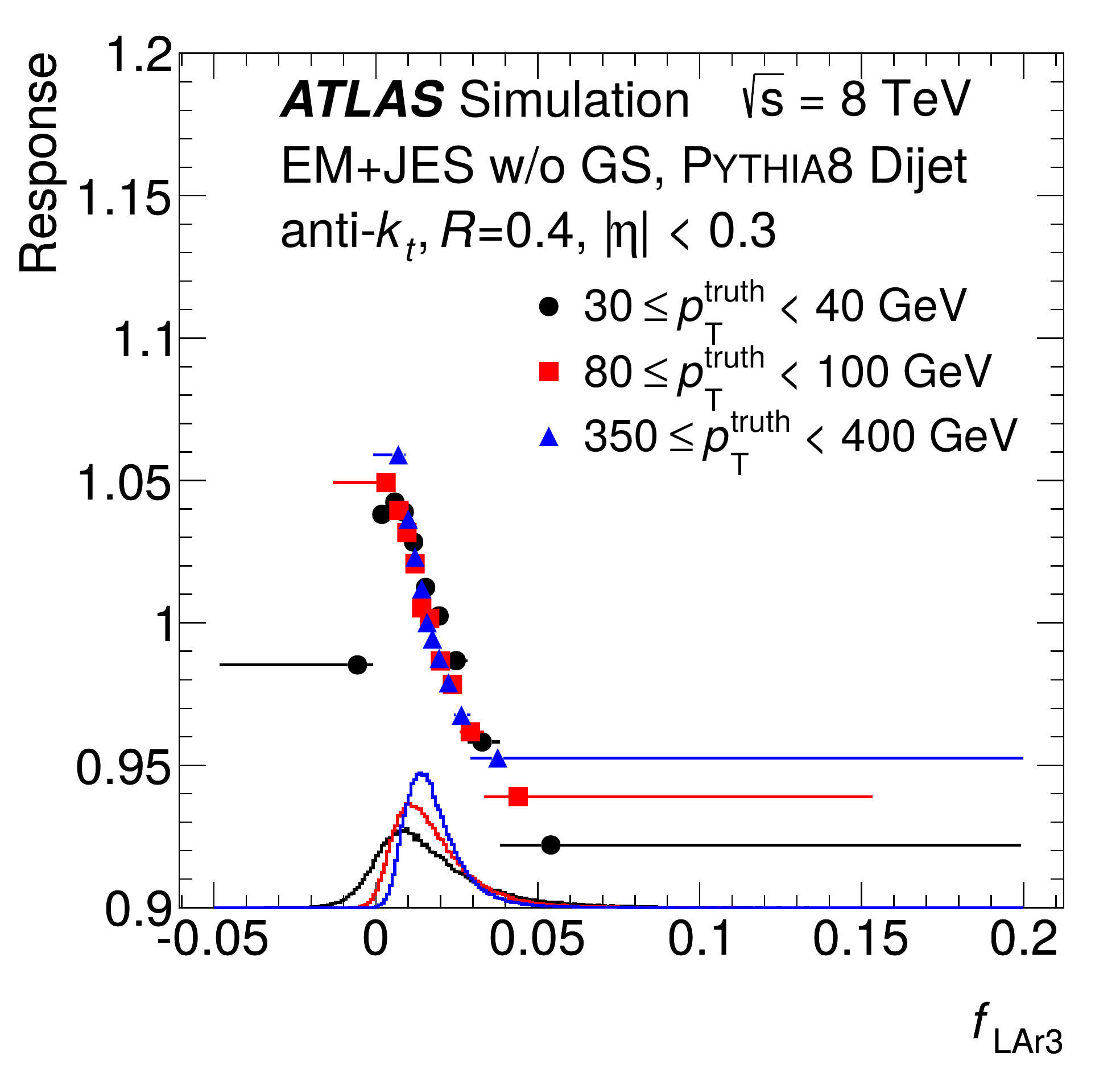}
\caption{}
\end{subfigure}
\\\vspace{-0.1cm}
\begin{subfigure}{0.42\textwidth}\centering
\includegraphics[width=\textwidth]{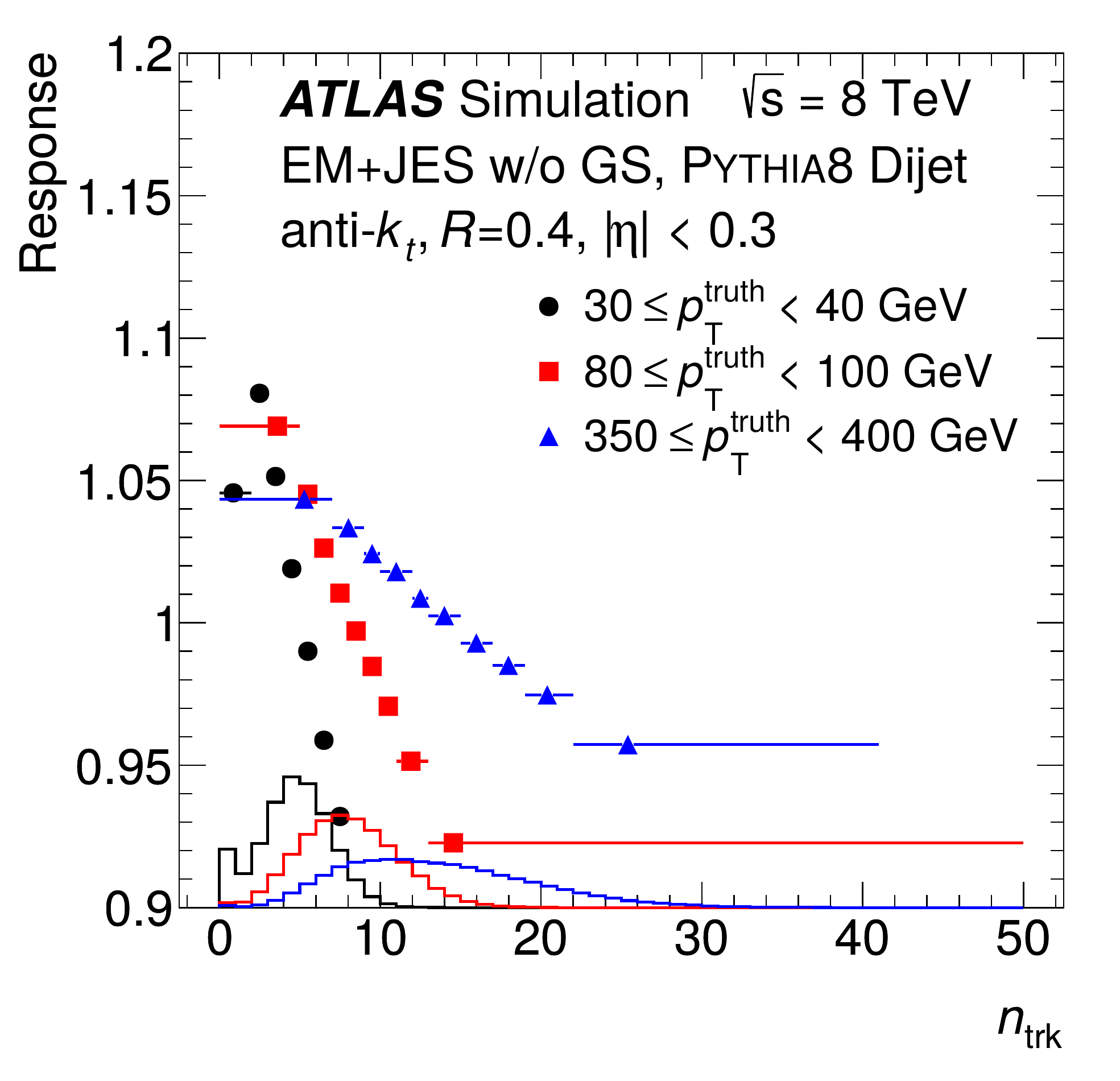}
\caption{}
\end{subfigure}
\hspace{0.03\textwidth}
\begin{subfigure}{0.42\textwidth}\centering
\includegraphics[width=\textwidth]{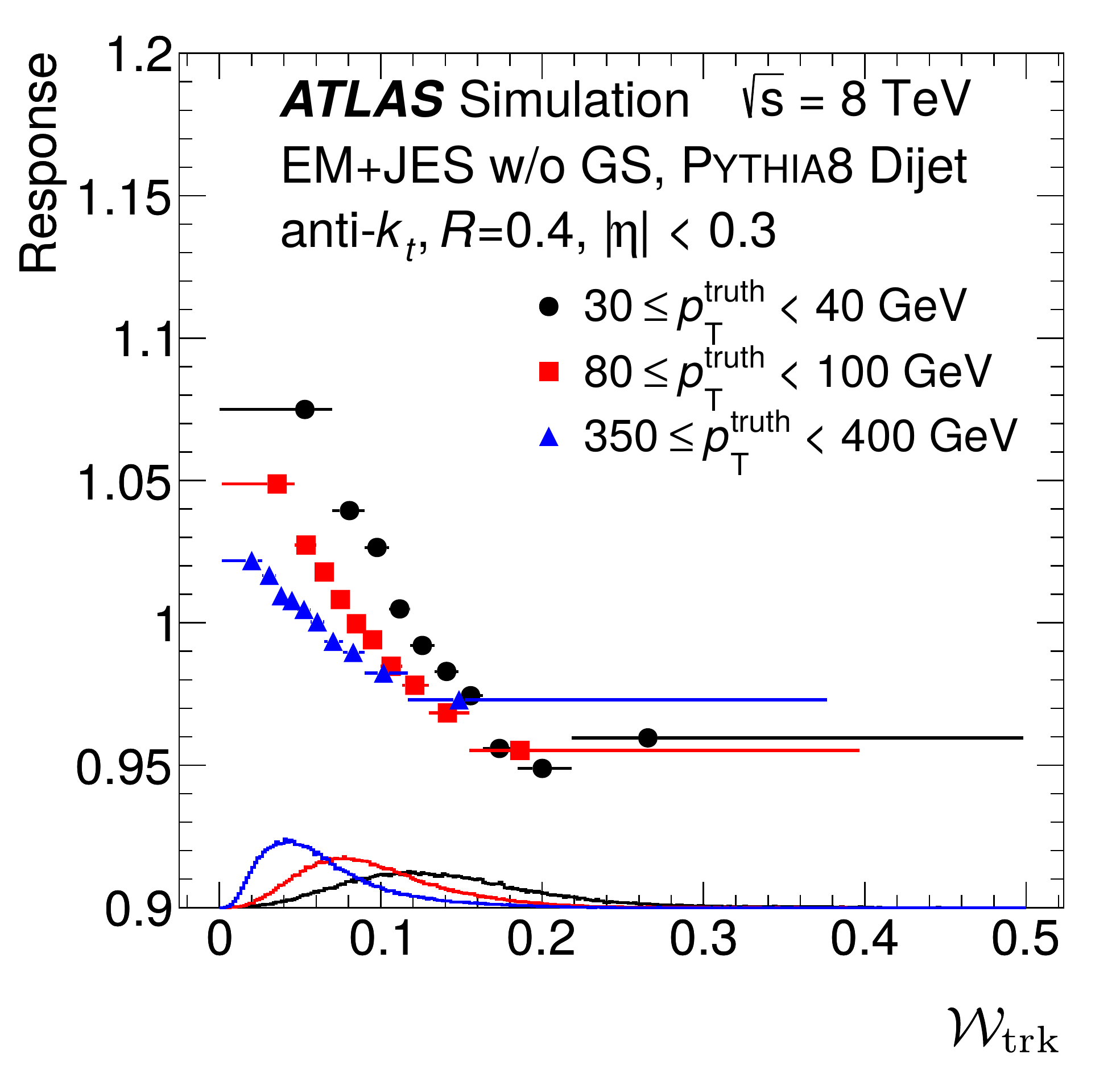}
\caption{}
\end{subfigure}
\\\vspace{-0.1cm}
\begin{subfigure}{0.42\textwidth}\centering
\includegraphics[width=\textwidth]{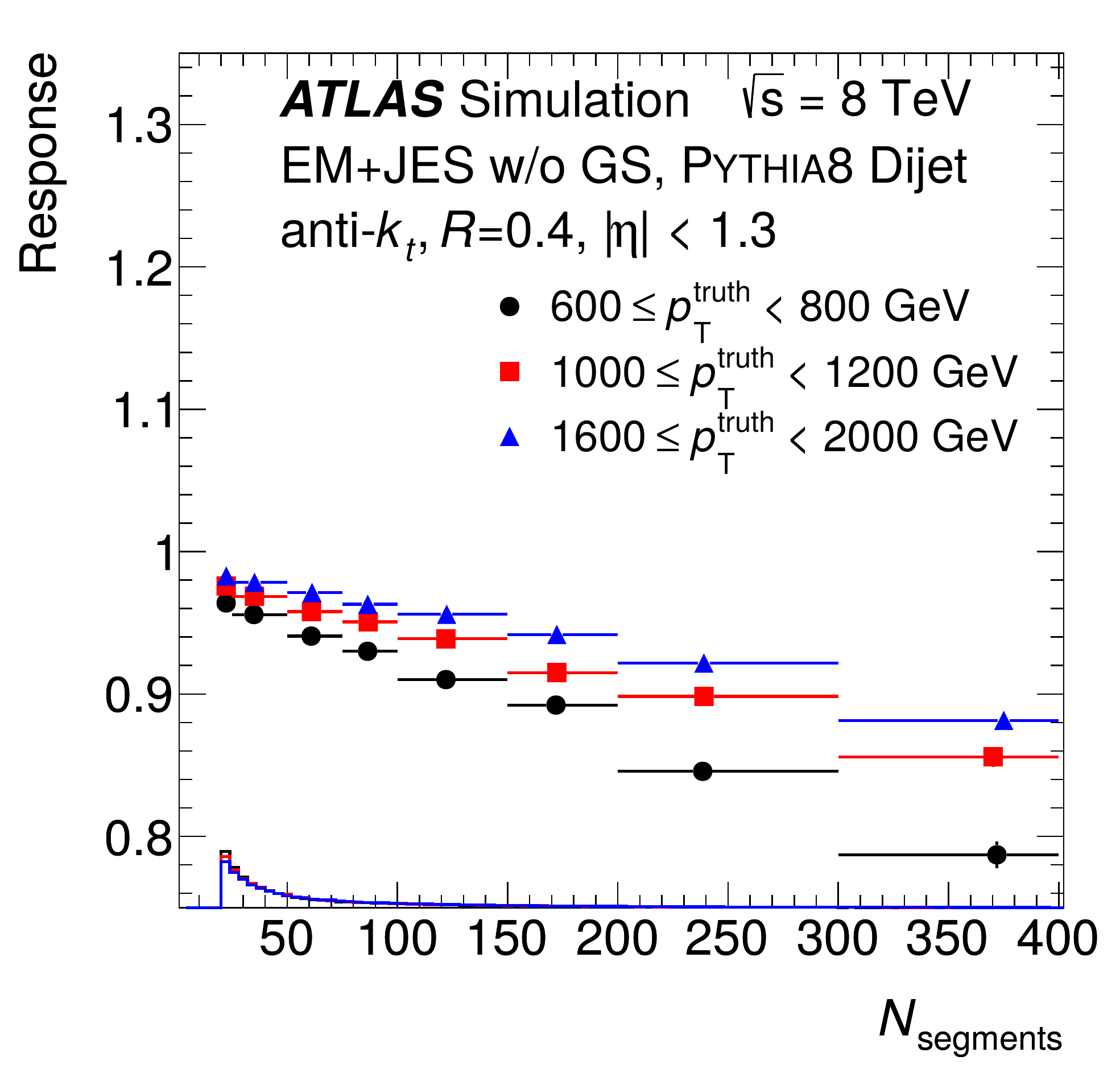}
\caption{}
\end{subfigure}
\\\vspace{-0.3cm}
\caption{Jet \pt{} response as a function of $\ftile$, $\fem$, $\nTrk$, $\trackWIDTH{}$ and $\Nsegments$ for jets with $|\etaDet|<0.3$ ($|\etaDet|<1.3$ for $\Nsegments$) in different \pttruth{} ranges. All jets are reconstructed with {\antikt} $R=0.4$ and calibrated with the \EMJES{} scheme without global sequential corrections.
The horizontal line associated with each data point indicates the bin range, and the position of the marker corresponds the centroid within this bin.
The underlying distributions of the jet properties for each \pttruth{} bin normalized to the same area are also shown as histograms at the bottom of the plots.}
\label{fig:resp_vs_x_beforeCorrections}
\end{figure}
 
\begin{figure}[p]
\centering
\begin{subfigure}{0.42\textwidth}\centering
\includegraphics[width=\textwidth]{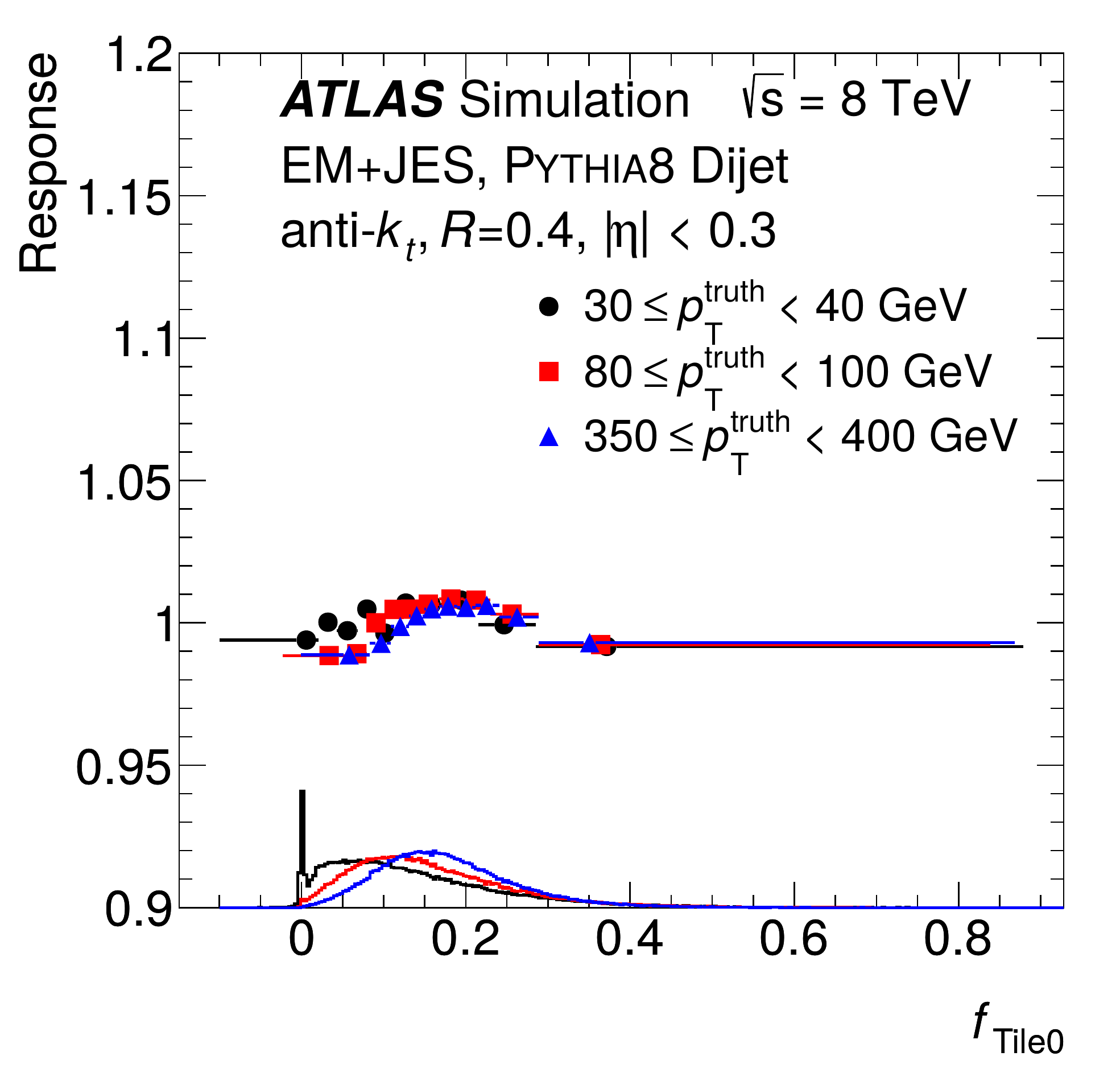}
\caption{}
\end{subfigure}
\hspace{0.03\textwidth}
\begin{subfigure}{0.42\textwidth}\centering
\includegraphics[width=\textwidth]{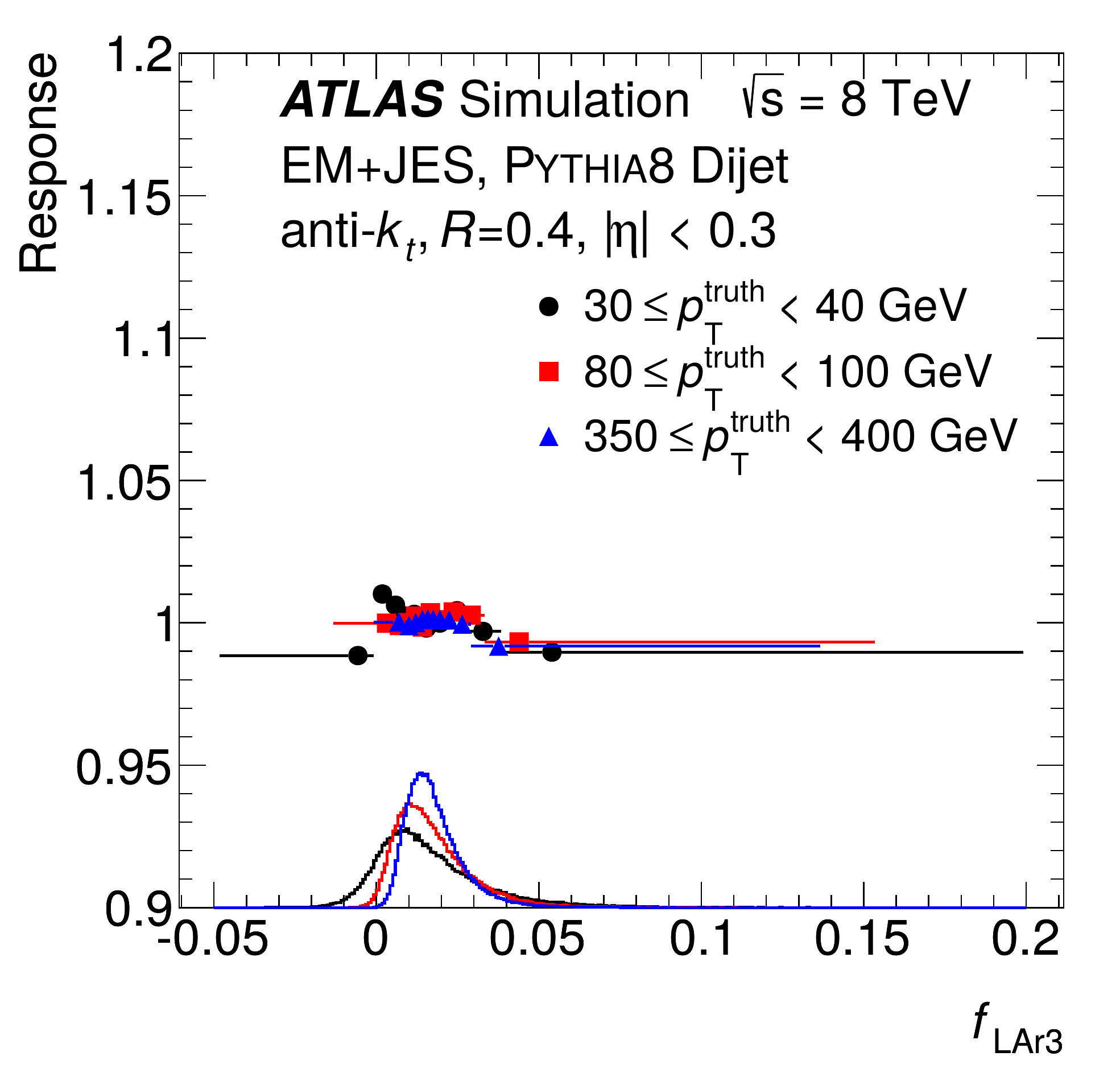}
\caption{}
\end{subfigure}
\\\vspace{-0.1cm}
\begin{subfigure}{0.42\textwidth}\centering
\includegraphics[width=\textwidth]{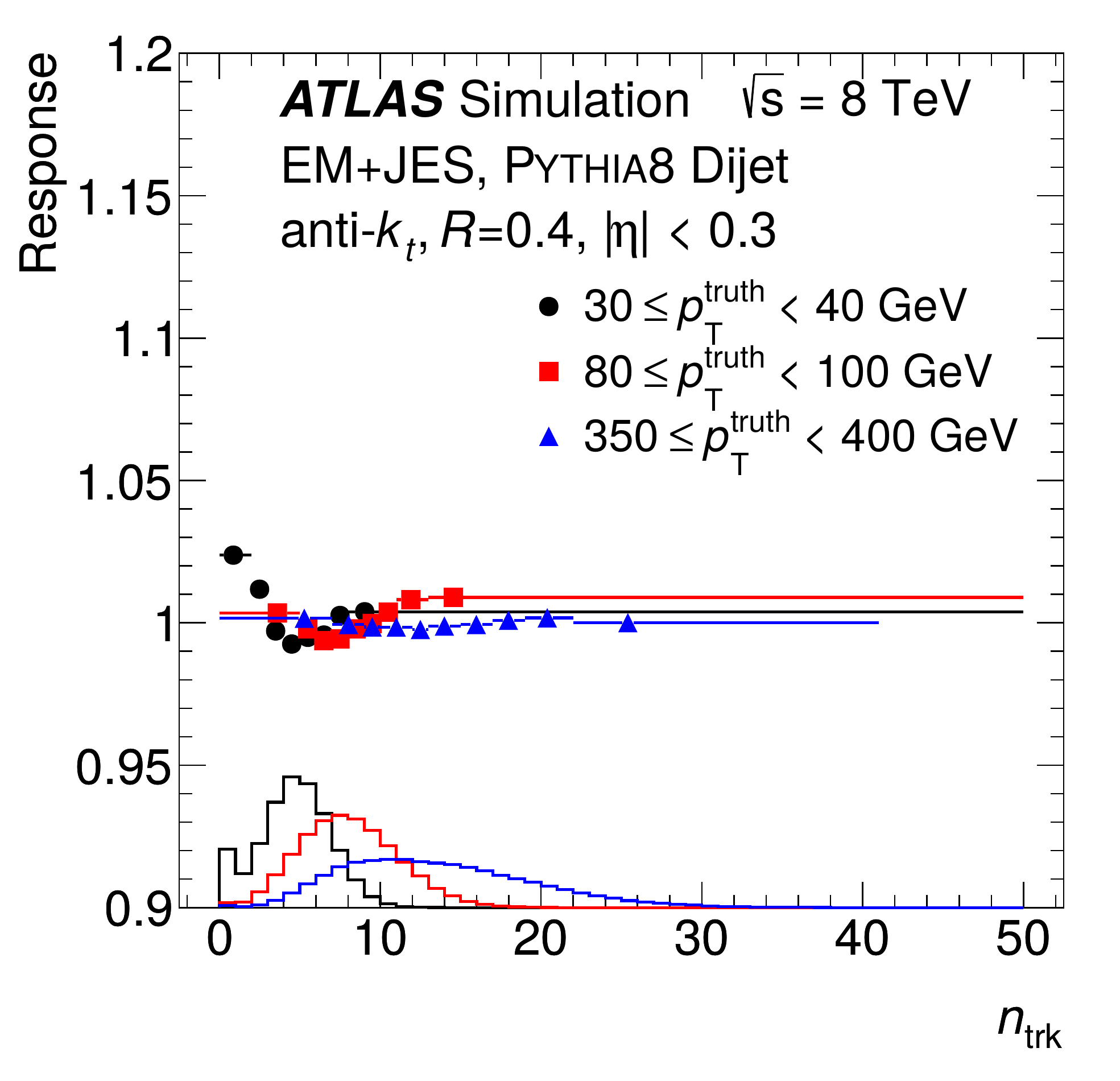}
\caption{}
\end{subfigure}
\hspace{0.03\textwidth}
\begin{subfigure}{0.42\textwidth}\centering
\includegraphics[width=\textwidth]{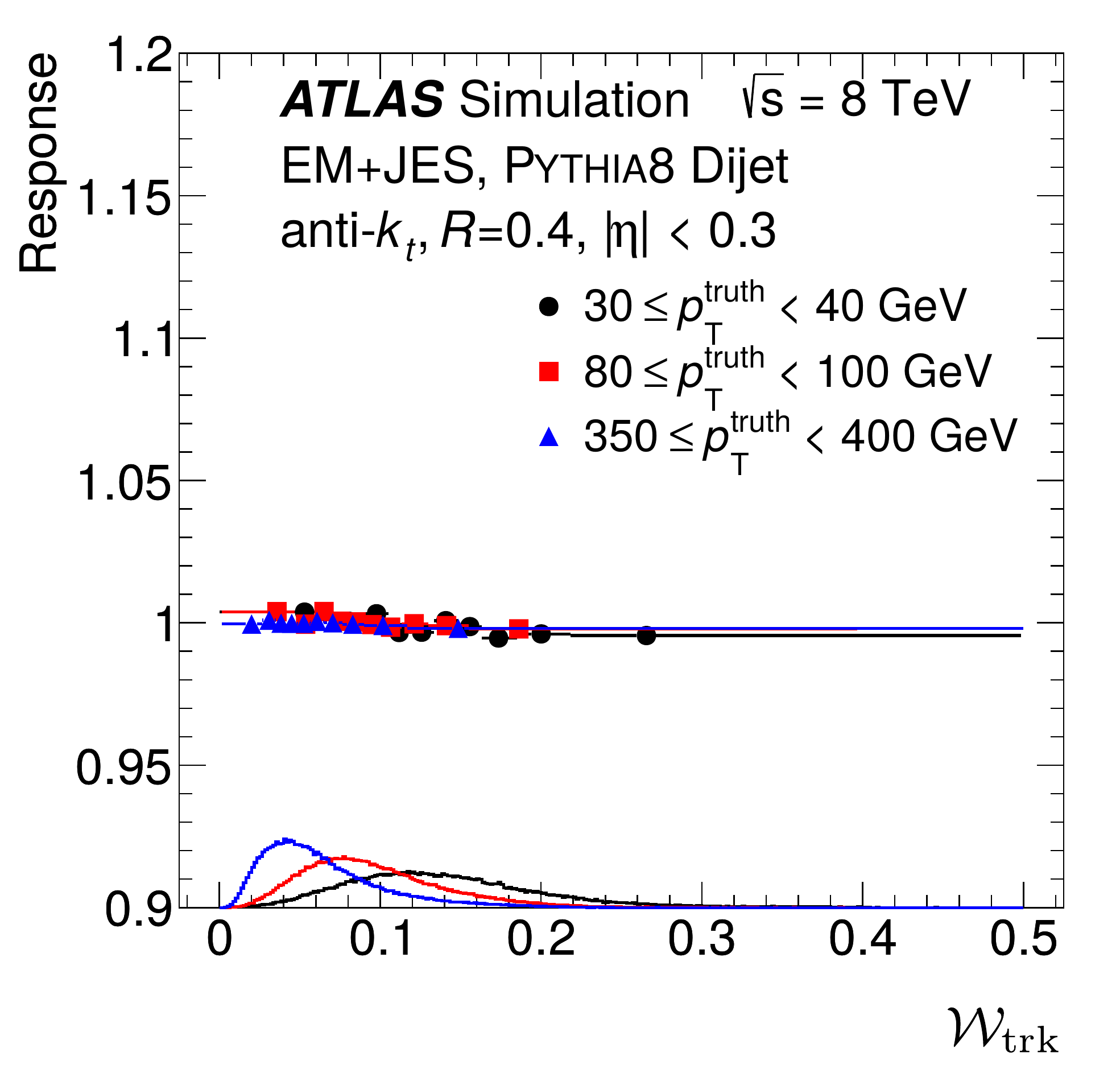}
\caption{}
\end{subfigure}
\\\vspace{-0.1cm}
\begin{subfigure}{0.42\textwidth}\centering
\includegraphics[width=\textwidth]{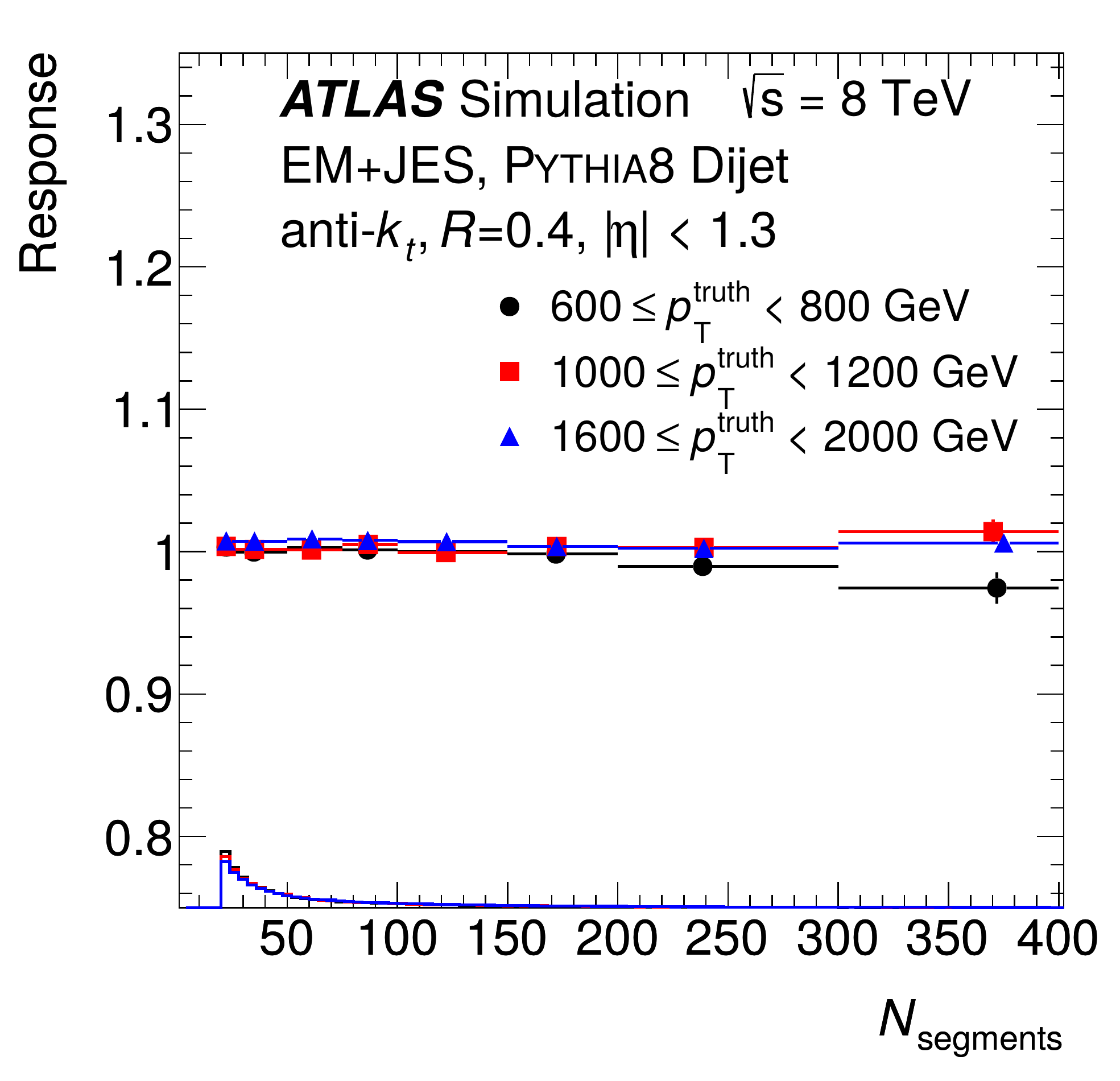}
\caption{}
\end{subfigure}
\\\vspace{-0.3cm}
\caption{Jet \pt{} response as a function of $\ftile$, $\fem$, $\nTrk$, $\trackWIDTH{}$ and $\Nsegments$ for jets with $|\etaDet|<0.3$  ($|\etaDet|<1.3$ for $\Nsegments$) in different \pttruth{} ranges. All jets are reconstructed with {\antikt} $R=0.4$ and calibrated with the \EMJES{} scheme including global sequential corrections.
The horizontal line associated with each data point indicates the bin range, and the position of the marker corresponds the centroid within this bin.
The underlying distributions of the jet properties for each \pttruth{} bin normalized to the same area are also shown as histograms at the bottom of the plots.}
\label{fig:resp_vs_x_afterCorrections}
\end{figure}

\subsection{Jet transverse momentum resolution improvement in simulation}

Figure~\ref{fig:ResponseResolutionVsPtAfterGS} shows the jet transverse momentum resolution as a function of \pttrue{} in simulated \pythia{} dijet events.
While the response remains unchanged, the jet resolution improves as more corrections are added. The relative improvement\footnote{The relative improvement in the jet
\pt{} resolution in comparison with the baseline (no-GS) calibration is calculated as $\frac{(\sigma_{\pt}/\pt)_\text{no-GS} - (\sigma_{\pt}/\pt)_\text{GS}}{ (\sigma_{\pt}/\pt)_\text{no-GS}}$, where the label no-GS refers to the jet prior to the GS calibration, i.e.\ directly after the MC-based calibration (Figure~\ref{fig:JESoverview}) and GS refers to the jet after the GS calibration.}
for \EMJES{} calibrated \antikt{} $R=0.4$ jets with central rapidity is found to be 10\% at $\pt=30$~\GeV, rising to 40\% at 400~\GeV{}.
This is equivalent to removing an absolute uncorrelated resolution source $\Delta\sigma$ of 10\% or 5\%, respectively, as can be seen in the lower part of Figure~\ref{fig:ResolutionVsPtAfterGS:a}.
The quantity $\Delta\sigma$ is calculated by subtracting in quadrature the relative jet \pt{} resolution:
\begin{equation}
\Delta\sigma = \begin{cases}
-\left(\left(\sigma_{\pt}/\pt\right)_\text{no-GS} \ominus \left(\sigma_{\pt}/\pt\right)_\text{GS}\right) & \text{if~}\left(\sigma_{\pt}/\pt\right)_\text{no-GS} > \left(\sigma_{\pt}/\pt\right)_\text{GS} \\
+\left(\left(\sigma_{\pt}/\pt\right)_\text{GS} \ominus \left(\sigma_{\pt}/\pt\right)_\text{no-GS}\right) & \text{otherwise.}
\end{cases}
\label{eq:DeltaSigma}
\end{equation}
 
\begin{figure}[b!]
\centering
\begin{subfigure}{0.48\textwidth}\centering
\includegraphics[width=\textwidth]{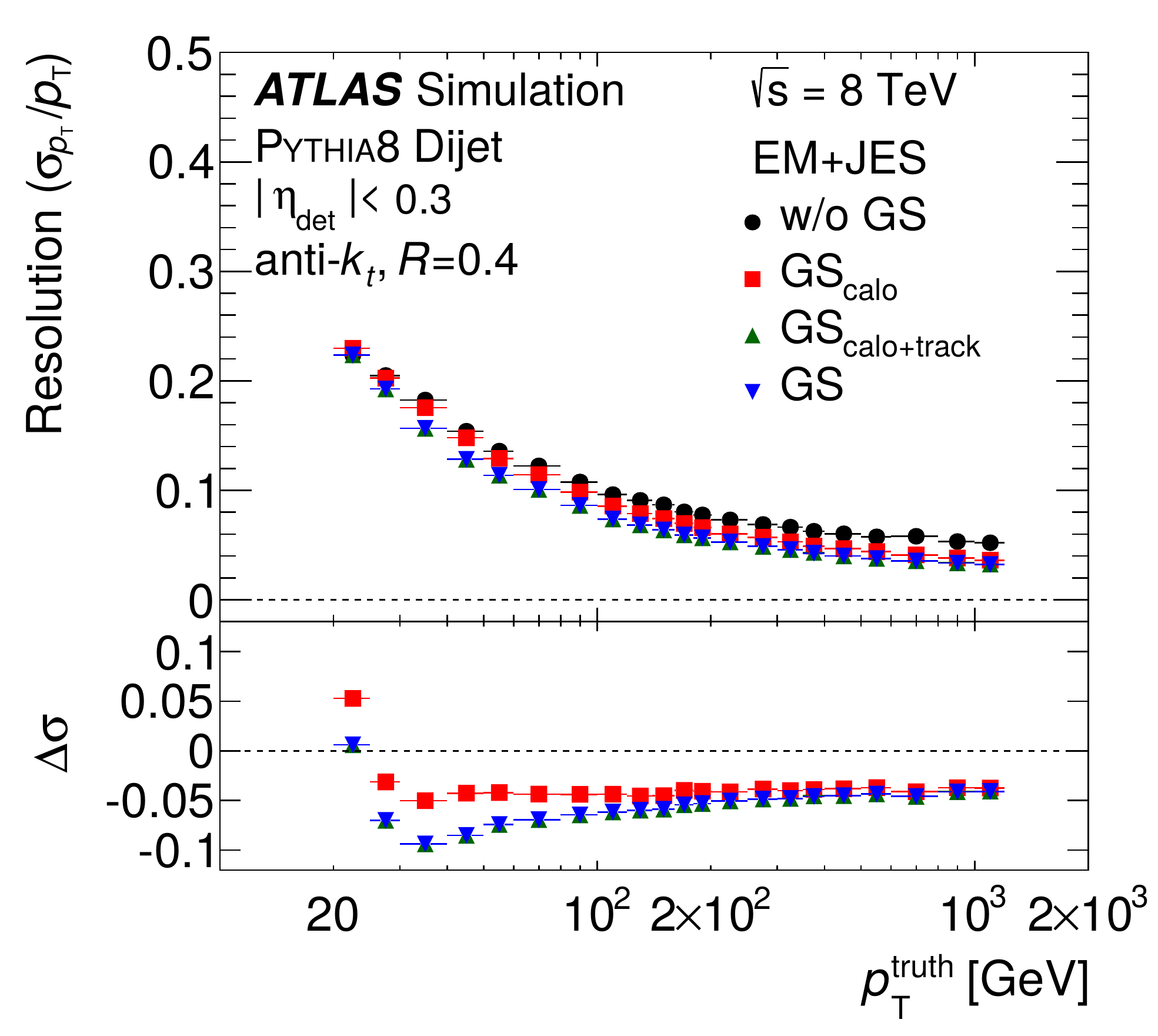}
\caption{}
\label{fig:ResolutionVsPtAfterGS:a}
\end{subfigure}
\begin{subfigure}{0.48\textwidth}\centering
\includegraphics[width=\textwidth]{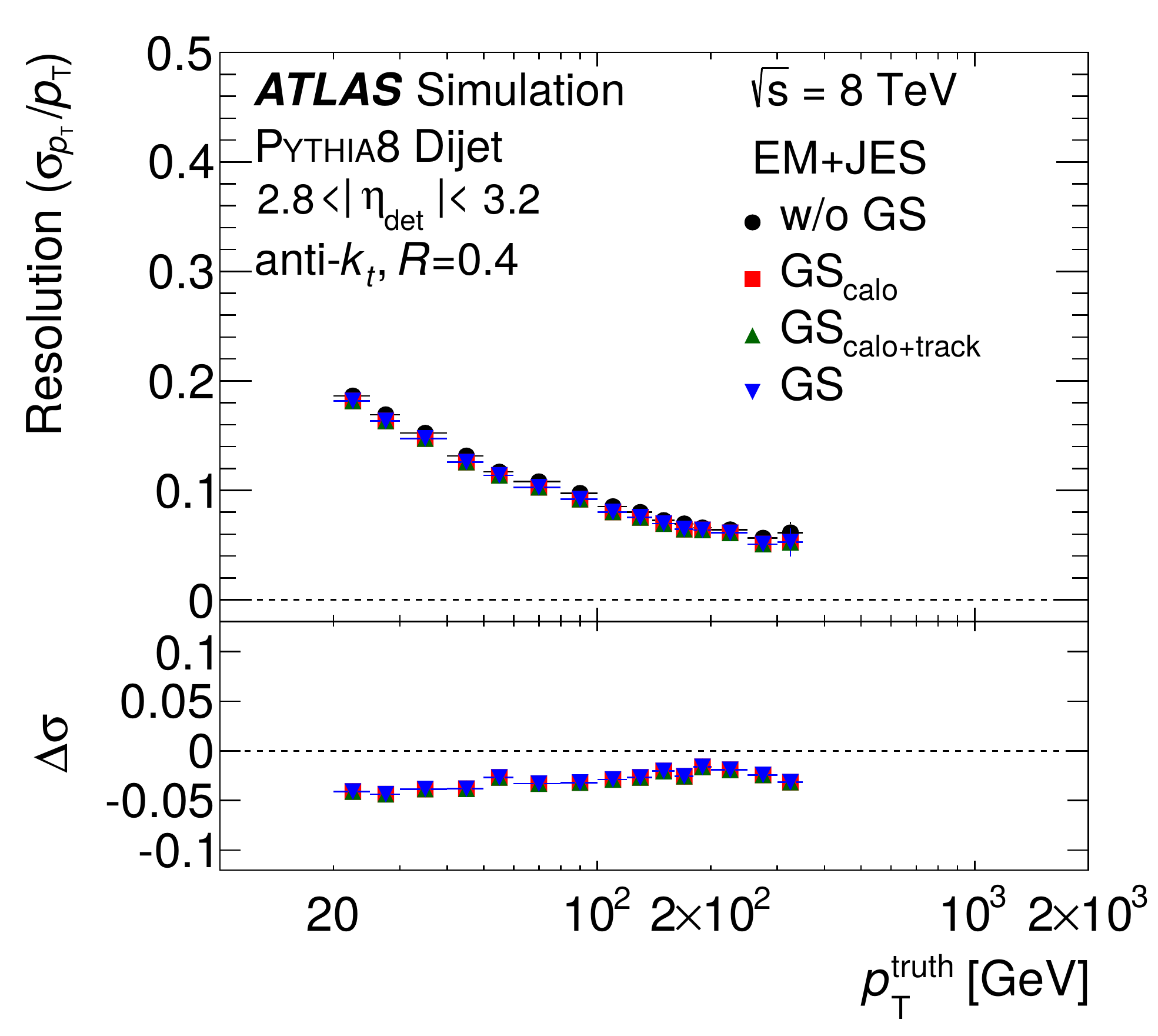}
\caption{}
\label{fig:ResolutionVsPtAfterGS:b}
\end{subfigure}
\caption{Jet \pt{} resolution as a function of \pttrue{} in the nominal \pythia{} MC sample for jets with (a) $|\etaDet|<0.3$ and (b) $2.8<|\etaDet|<3.2$.
The jets are reconstructed with {\antikt} $R=0.4$.
Curves are shown after the \EMJES{} calibration without global sequential corrections (black circles),
with calorimeter-based global sequential corrections only (red squares), with calorimeter- and track-based corrections only (green upward triangles) and including all the global sequential corrections (blue downward triangles).
The lower panels show the improvement relative to the \EMJES{} scale without global sequential corrections obtained using subtraction in quadrature (Eq.~(\ref{eq:DeltaSigma})).}
\label{fig:ResponseResolutionVsPtAfterGS}
\end{figure}
The improvement observed for jets initially calibrated with the \LCWJES{} scheme is found to be smaller, which is expected as only tracking and non-contained jet corrections are applied to these jets.
For both \EMJES{} and \LCWJES{} calibrated jets, improvements to the JER is observed across the full \pt{} range probed (\ptRange{25}{1200}).
The fact that JER reduction is observed at high jet \pt{} means that also the constant term of the calorimeter resolution (Eq.~(\ref{eq:JER})) is reduced by the GS calibration.
This improvement can be explained by considering the jet resolution distributions for different values of the jet properties.
As is evident in Figure~\ref{fig:resp_vs_x_beforeCorrections}, the mean of these distributions have a strong dependence on the jet property,
while the width of the distributions (JER) are not expected to have any such dependence at high jet \pt{}.
The GS calibration can hence be seen as aligning several similarly shaped response distributions, which each have a biased mean, towards the desired \tjet{} scale.
 
The conclusions from this section can generally be extended to the whole \etaDet{} range, although close to the calorimeter transition regions where the detector instrumentation is reduced (Figure~\ref{fig:calibcurve}), the track-based observables introduce an even stronger improvement.
The enhancement in JER due to the \GS~calibration is found to be similar for different MC generators.
 
Only a small improvement is observed after applying the last GS correction for uncontained calorimeter jets in the inclusive jet sample since only a small fraction of energetic jets are uncontained.
Figure~\ref{fig:RMSVsPtAfterPT} presents a measure of the improvement in jet energy resolution from applying the fifth GS correction both to inclusive jets and to jets with at least 20 associated
muon segments, which are less likely to be fully contained in the calorimeters.
The resolution metric is the standard deviation (RMS) of the jet response distribution divided by the arithmetic mean. This quantity is used instead of the normal resolution definition (from the $\sigma$ of a Gaussian fit as described in Section~\ref{sec:jetMatch}) since it gives information about the reduction in the low response tail.
While the improvement observed is small for an inclusive jet sample, the impact is significant for uncontained jets.  A relative resolution improvement of 10\% is seen for jets with $\pt{}\approx100$~\GeV, while the improvement is 20\% for jets with $\pt{}\approx1$~\TeV.  This corresponds to removing an absolute resolution source of 8\% or 4\%, respectively.
 
\begin{figure}[b!]
\centering
\begin{subfigure}{0.48\textwidth}\centering
\label{fig:RMSVsPtAfterPT:a}
\includegraphics[width=\textwidth]{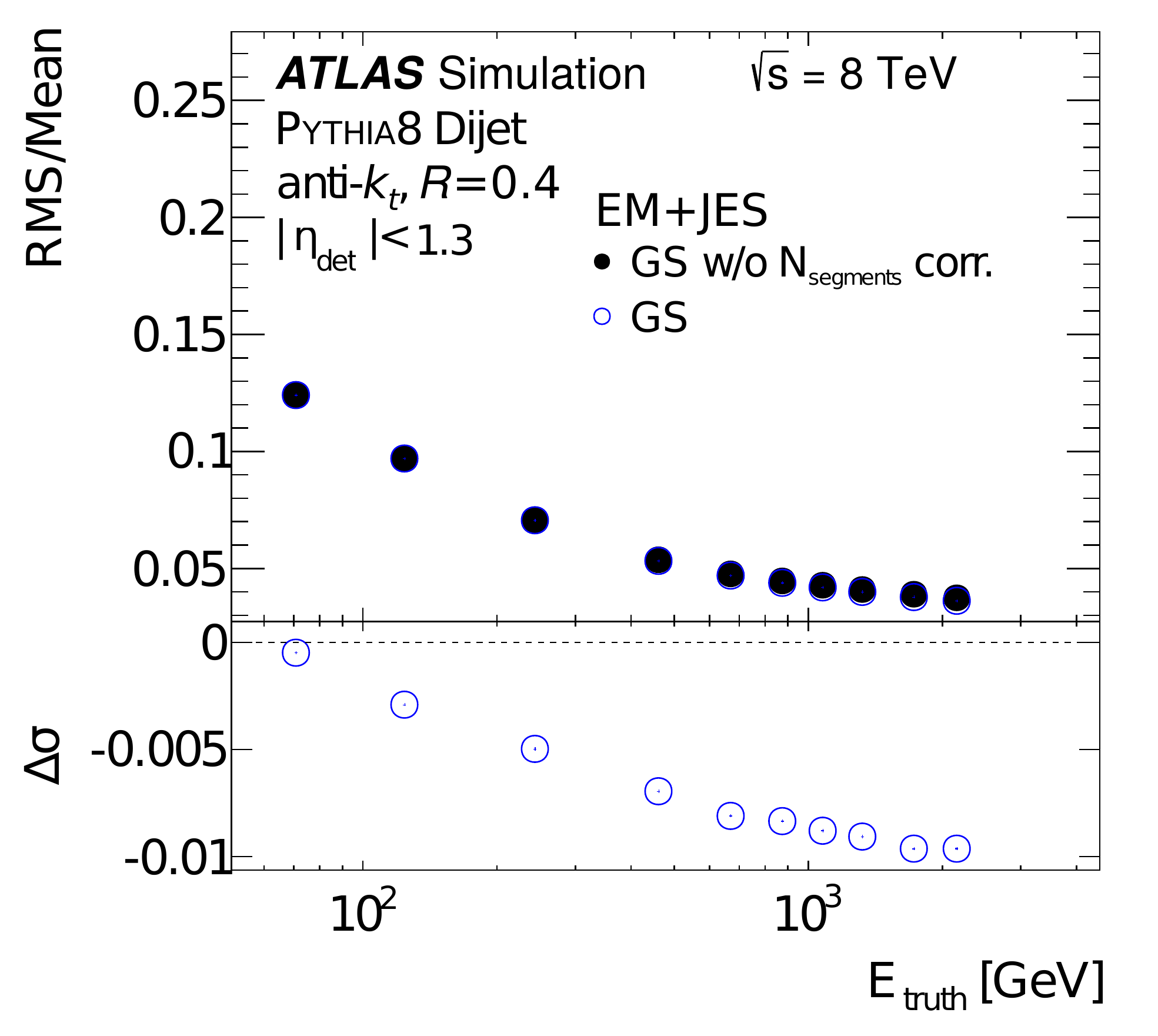}
\caption{}
\end{subfigure}
\hspace{0.001\linewidth}
\begin{subfigure}{0.48\textwidth}\centering
\label{fig:RMSVsPtAfterPT:b}
\includegraphics[width=\textwidth]{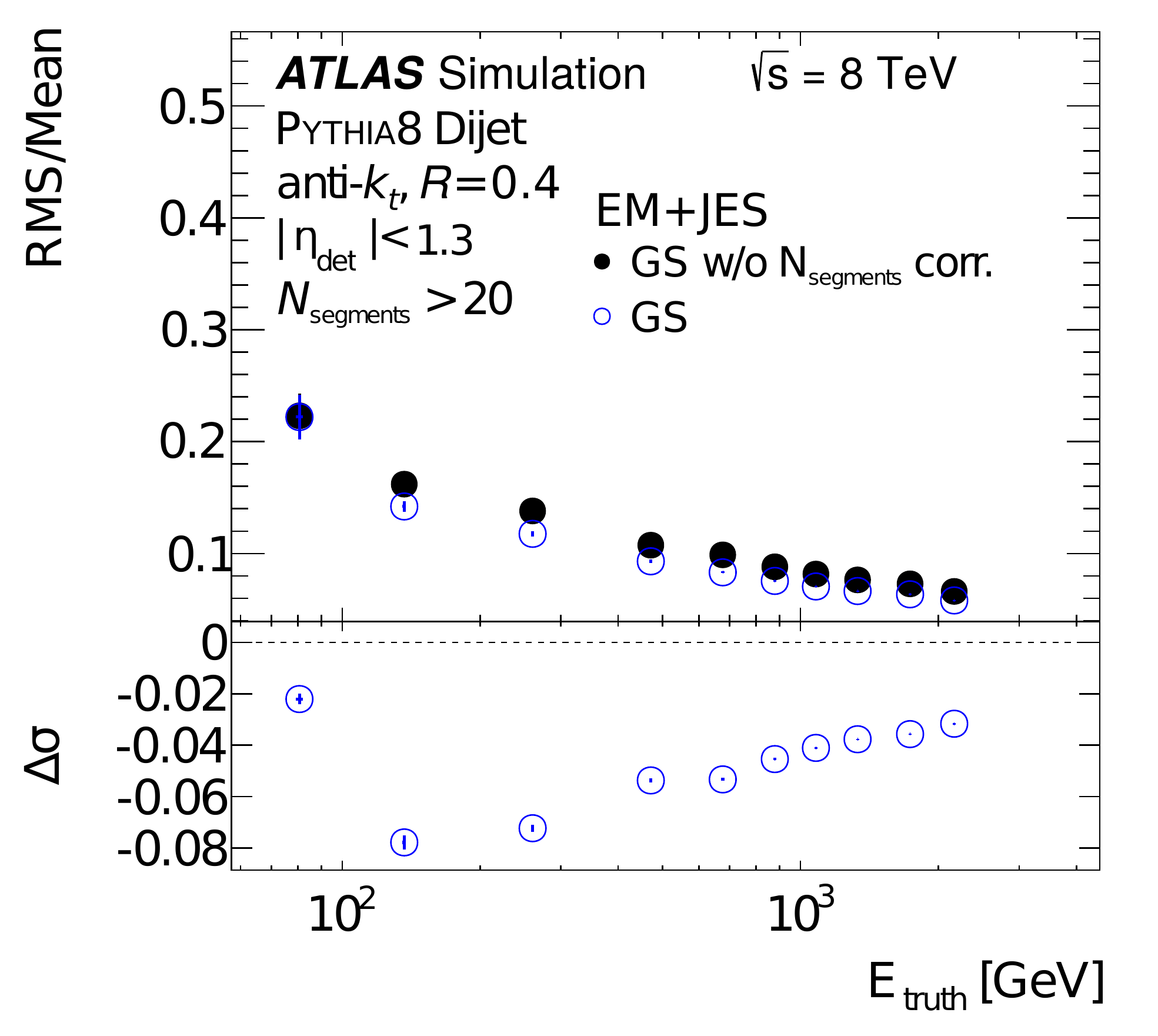}
\caption{}
\end{subfigure}
\caption{Standard deviation over arithmetic mean of the jet energy response as a function of $E^\textrm{truth}$ for $|\etaDet|<1.3$ before (filled circles) and after (empty circles) the fifth global sequential correction for (a)~all jets and (b)~calorimeter jets with $\Nsegments>20$ in the nominal \pythia{} dijet MC sample. All jets are reconstructed with {\antikt} $R=0.4$ and initially calibrated at the EM+JES scale.
The requirement $\Nsegments>20$ selects a large fraction of ``uncontained'' jets, i.e.\ jets for which some of the particles produced in the hadronic shower travel into the muon spectrometers behind the calorimeters.
The bottom panels show the improvement introduced by the corrections quantified using subtraction in quadrature (Eq.~(\ref{eq:DeltaSigma})).}
\label{fig:RMSVsPtAfterPT}
\end{figure}
 
\subsection{Flavour dependence of the jet response in simulation}
\label{sec:GS-flavour}
 
The internal structure of a jet, and thereby also its calorimeter response, depends on how the jet was produced.
Jets produced in dijet events are expected to originate from gluons more often than jets
with the same \pt{} and $\eta$ produced in the decay of a $W$ boson or in association with a photon or $Z$~boson.
The hadrons of a quark-initiated jet will tend to be of higher energy and hence penetrate further into the calorimeter, while the less energetic hadrons in a gluon-initiated jet will bend more in the magnetic field in the inner detector.
It is desirable that such flavour dependence of the calibrated jet should be as small as possible to mitigate sample-specific systematic biases in the jet energy scale
(Section~\ref{subsec:flavourUncert} for discussion of the associated uncertainty).
 
\begin{figure}[p]
\centering
\includegraphics[width=0.43\textwidth]{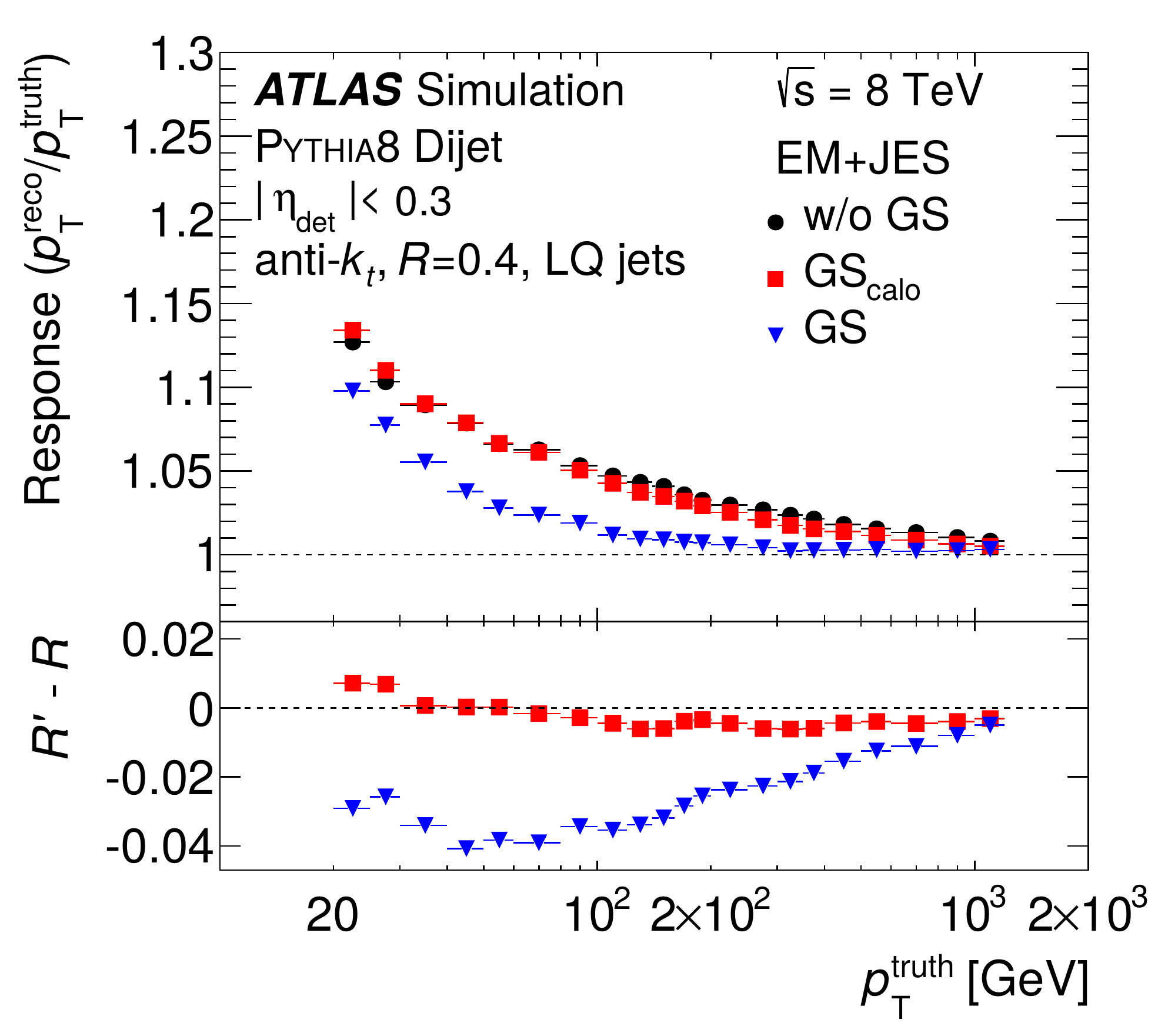}
\hspace{0.03\textwidth}
\includegraphics[width=0.43\textwidth]{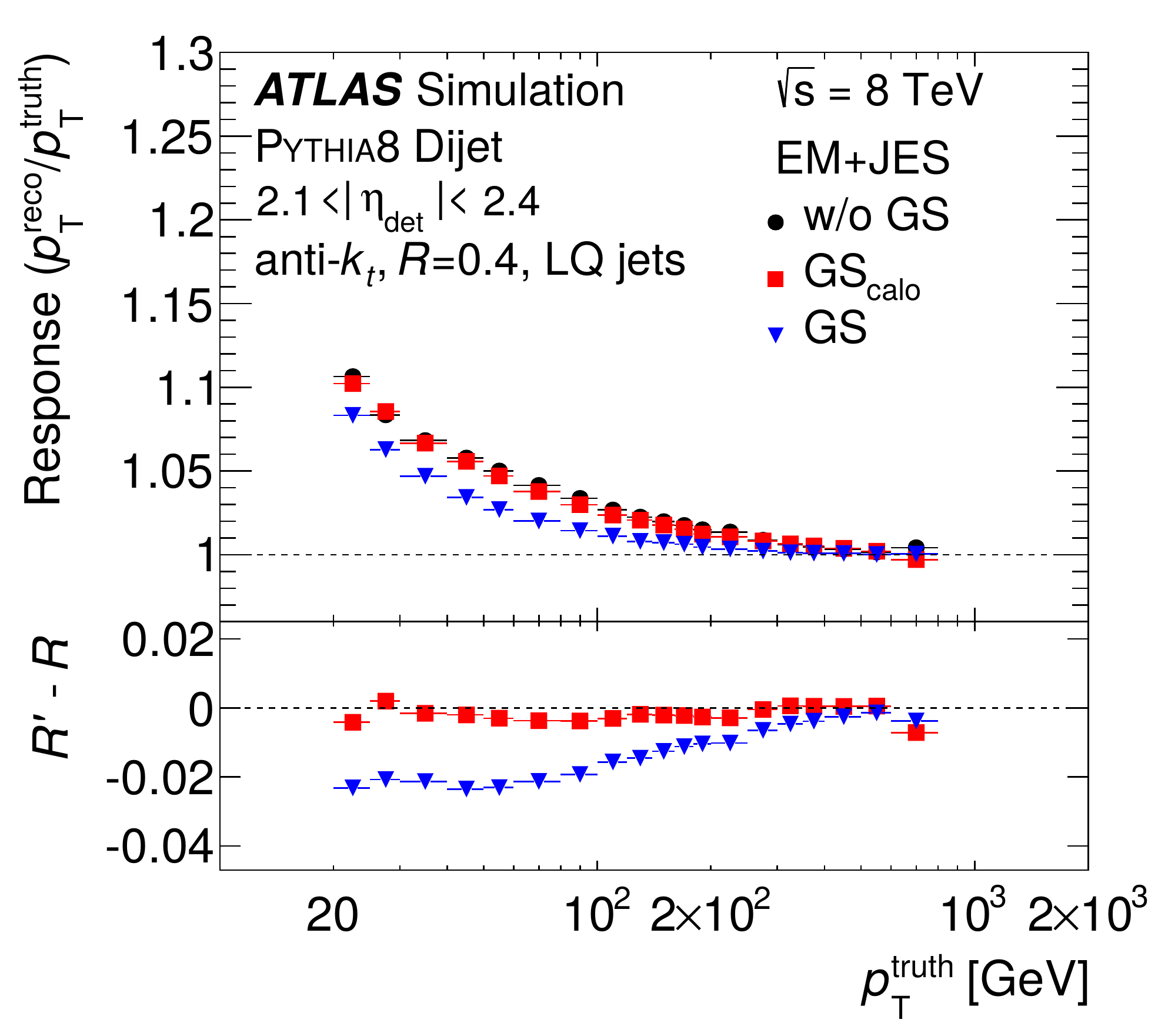}\\
\includegraphics[width=0.43\textwidth]{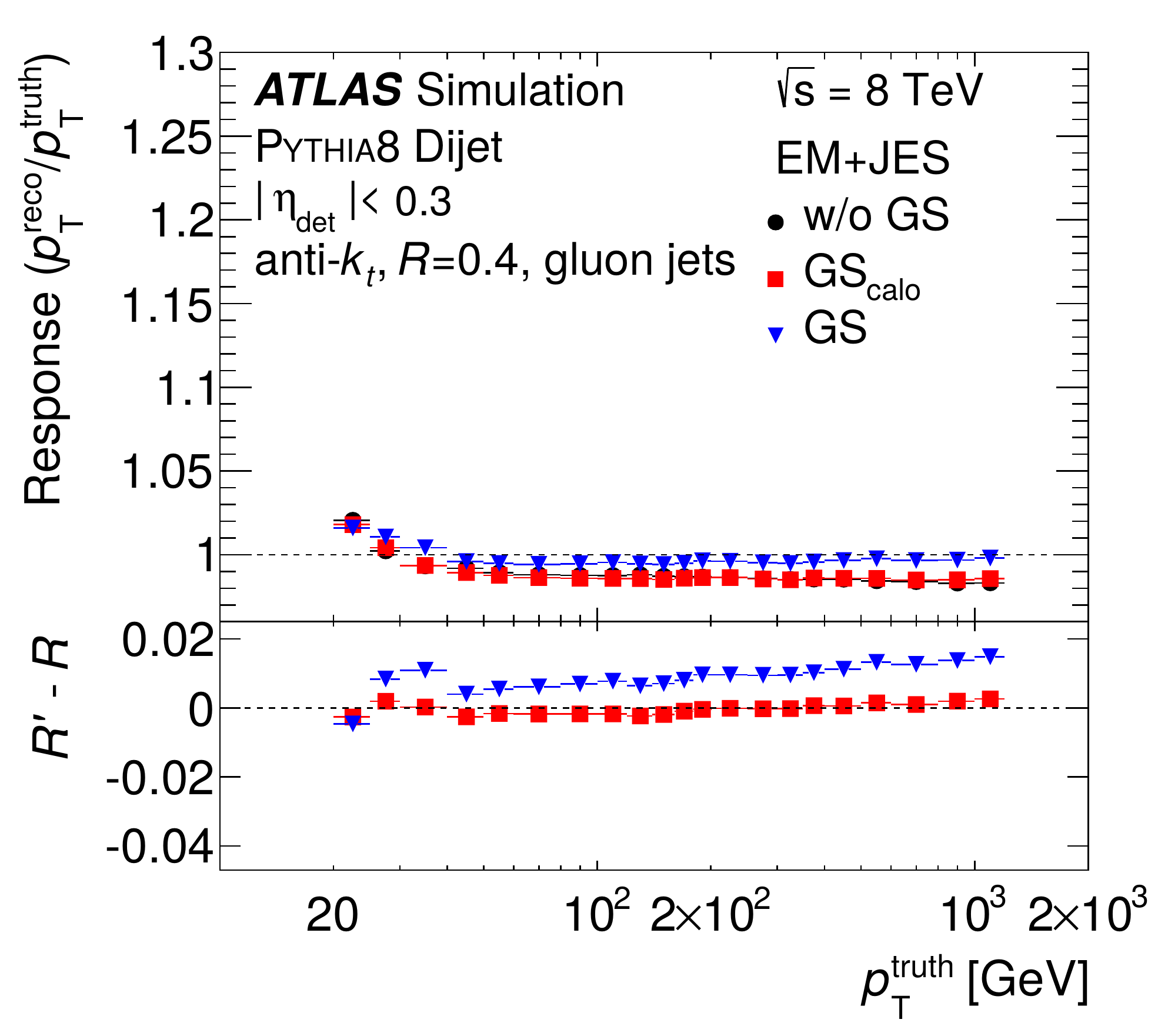}
\hspace{0.03\textwidth}
\includegraphics[width=0.43\textwidth]{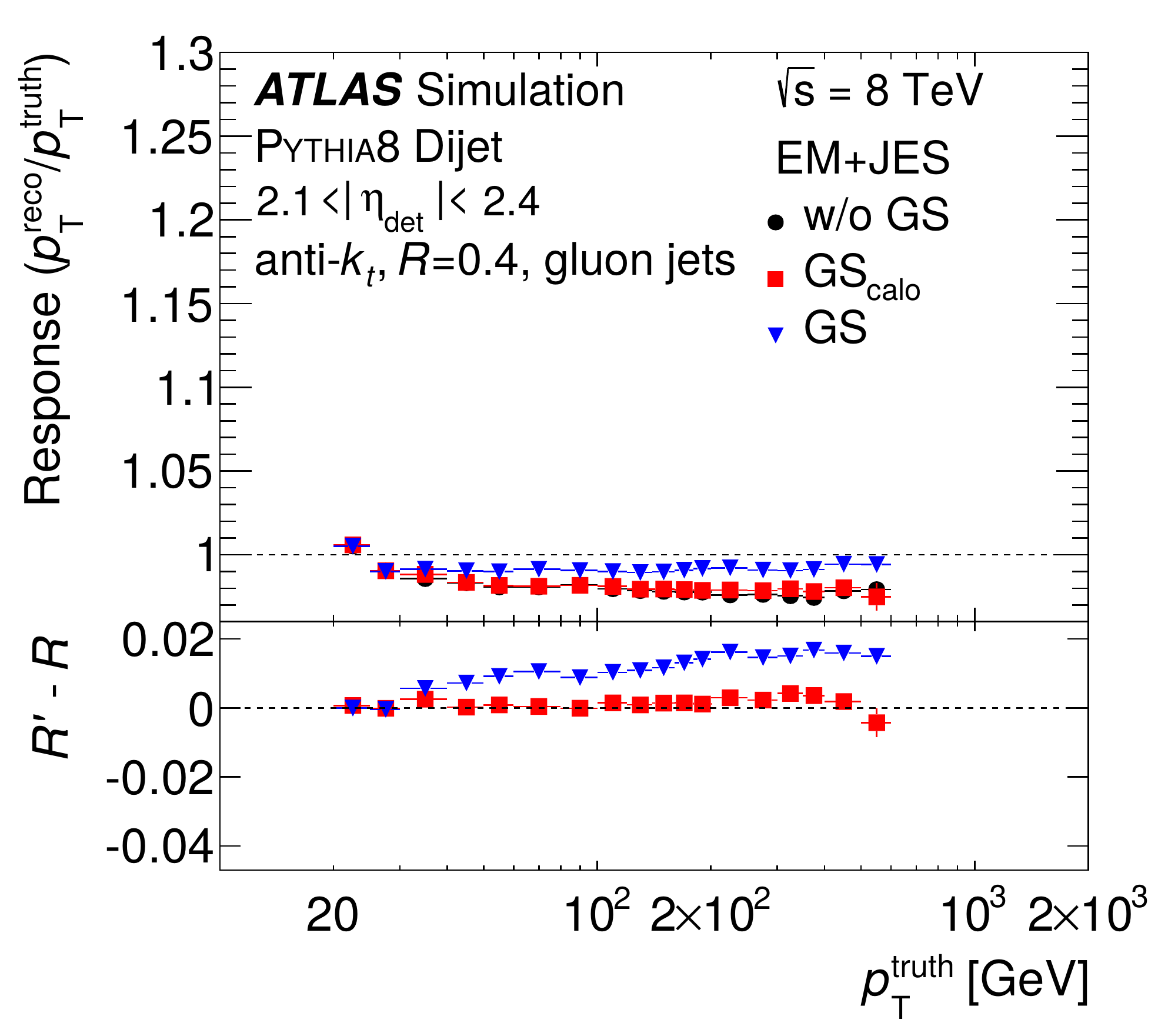}\\
\includegraphics[width=0.43\textwidth]{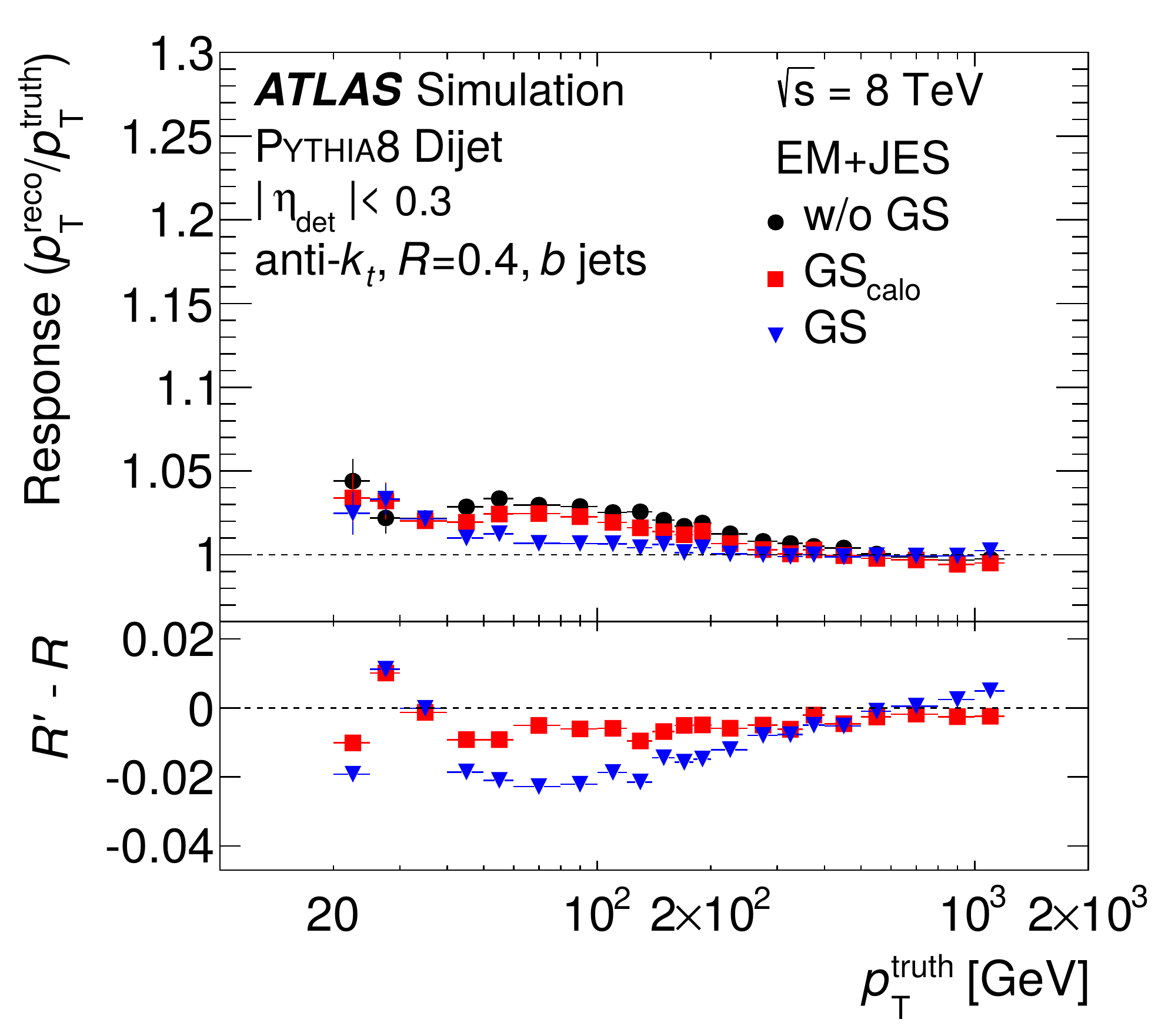}
\hspace{0.03\textwidth}
\includegraphics[width=0.43\textwidth]{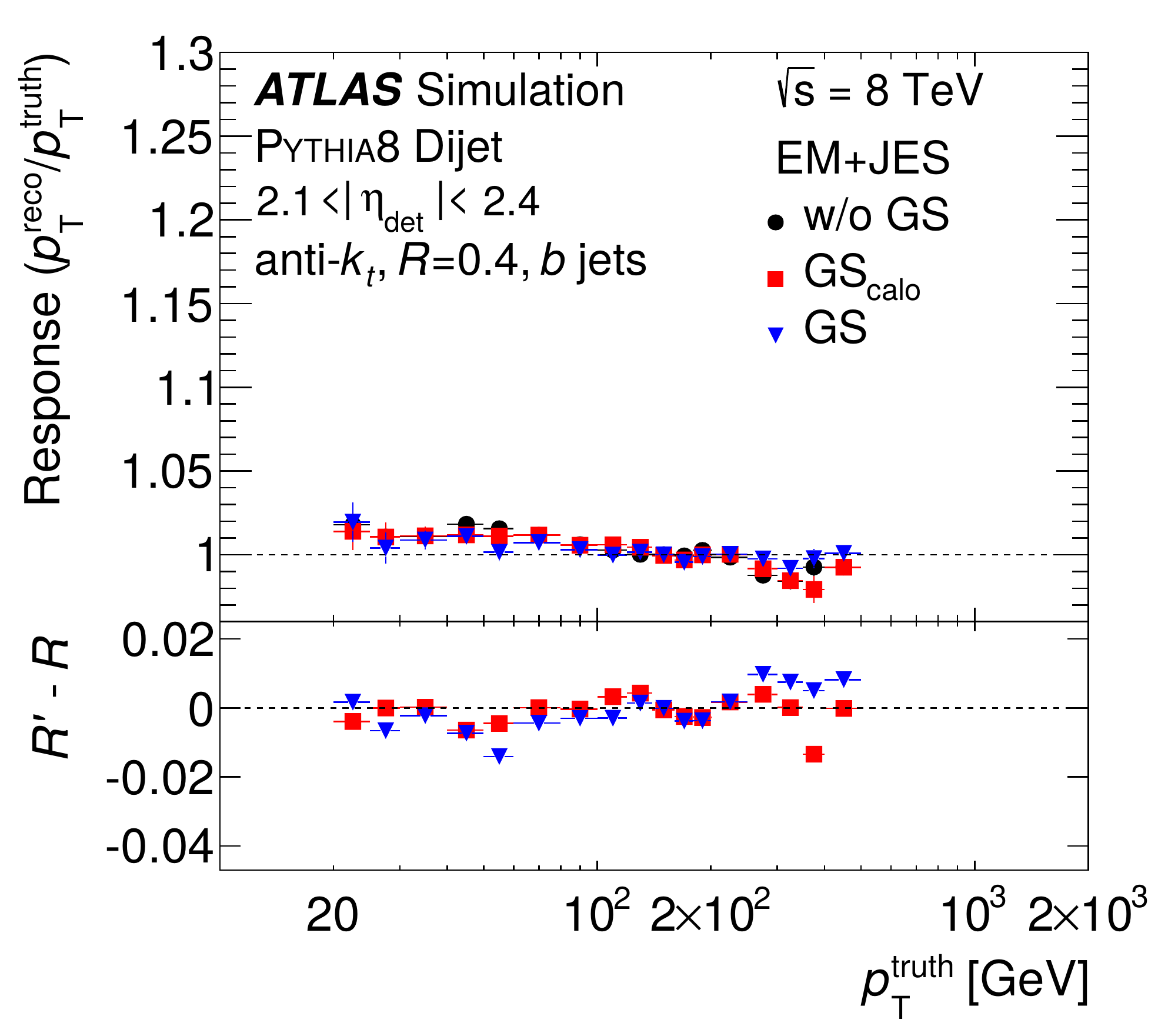}
\caption{The \pt{} response for \antikt{} \rfour{} jets as a function of \pttruth{} for light quark (LQ) jets~(top), gluon jets~(middle) and $b$-jets~(bottom)
with $|\etaDet{}|<0.3$ (left) and $2.1<|\etaDet{}|<2.4$ (right) regions
in the \pythia{} MC sample. The \pt{} response after the \EMJES{} calibration without GS corrections (circles),
with calorimeter-based GS corrections only (squares) and including all the GS corrections (triangles) are shown.
The lower box of each plot shows the impact on the jet response, subtracting the response before the GS corrections ($R$) from the
response after applying the GS corrections ($R'$).}
\label{fig:GS-flavour}
\end{figure}
 
The flavour dependence of the response is studied in simulated dijet events by assigning a flavour label to each calorimeter jet using an angular matching to the particles in the MC event record.
If the jet matches a $b$- or a $c$-hadron, it is labelled a \bjet{} or \cjet{}, respectively.
If it matches both a $b$- and a $c$-hadron, it is labelled a \bjet{}.
If it does not match any such heavy hadron, the jet is labelled ``light quark'' (LQ) or gluon initiated, based on the type of the highest-energy matching parton. The matching criterion used is $\Delta R < R$, where $R$ is the radius parameter of the jet algorithm (0.4 or 0.6).
The \pt{} responses before and after GS calibration for jets in different flavour categories are presented in Figure~\ref{fig:GS-flavour}.
For each flavour category, results are shown for two representative pseudorapidity regions.
The response for LQ jets is larger than unity since the MC-derived baseline calibration (Section~\ref{sec:jetCalibOverview}) is derived in dijet events that contain a large fraction of gluon jets.
For gluon-initiated jets the response is lower than that of LQ jets, as expected, and \bjets{} have a \pt{} response between that of LQ and gluon jets.
In all cases, the GS calibration brings the response closer to unity and hence reduces the flavour dependence, which is important as analyses do not know the flavour of each jet.
The change in \pt{} response introduced by the GS calibration for jets with $\pt{}=80$~\GeV\ with $|\eta|<0.3$ is $-4\%$, $+1\%$ and $-2\%$ for LQ jets, gluon jets and $b$-jets, respectively.

\subsection{\Insitu{} validation of the global sequential calibration}
\label{sec:GS-data-validation}
 
The GS correction is validated \insitu{} with dijet events using the tag-and-probe technique, using the event selection described in Section~\ref{sec:dijet}, with only one modification: both jets are required to be in the same $|\etaDet{}|$ region to avoid biases from any missing $\eta$-dependent calibration factors.
The jet whose response dependence is studied is referred to as the probe jet, while the other is referred to as the reference jet.
The choice of reference jet and probe jet is
arbitrary when studying the response dependence on the jet properties, and the events are always used twice, alternating the roles of reference and probe.
The response for the probe jet is measured through the dijet $\pt$ asymmetry variable $\asym$ (Eq.~(\ref{eq:asym}) and Section~\ref{sec:dijet-balance}) in bins of the average \pt{} of the probe and the reference jet \ptavg{}, and is studied as a function of the jet property of the probe jet.
\begin{figure}[!htbp]
\includegraphics[width=0.32\textwidth]{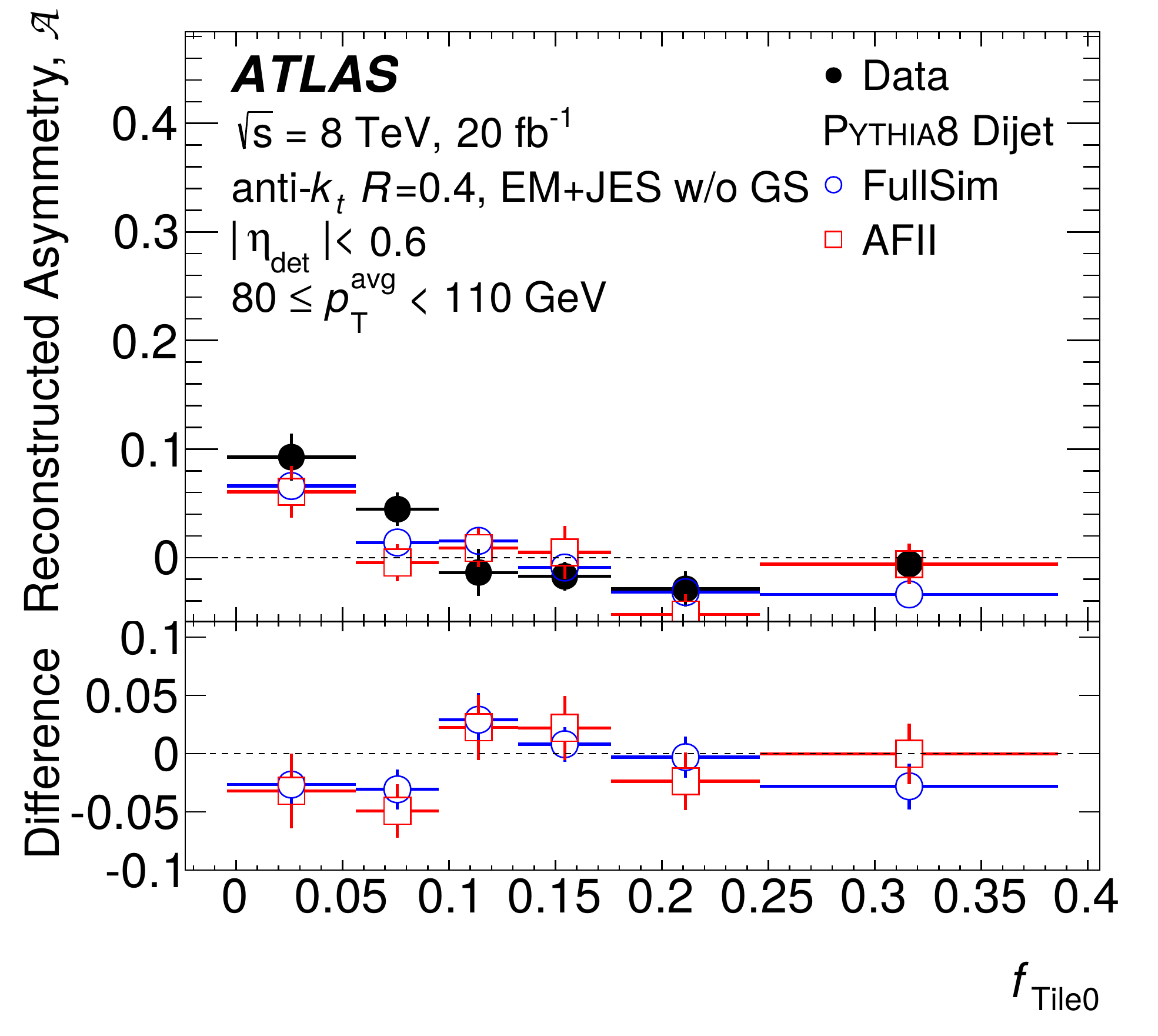}
\includegraphics[width=0.32\textwidth]{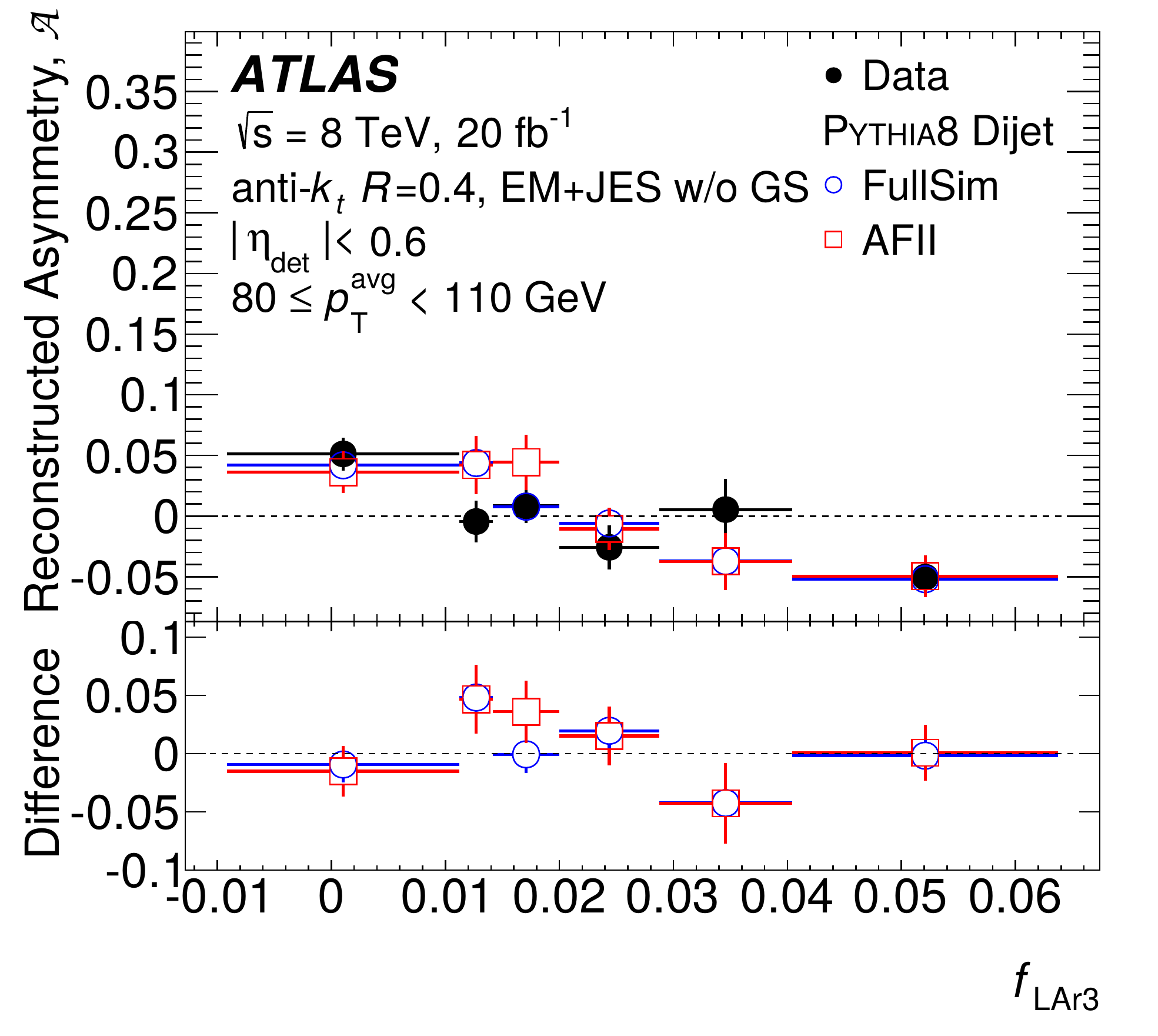}
\includegraphics[width=0.32\textwidth]{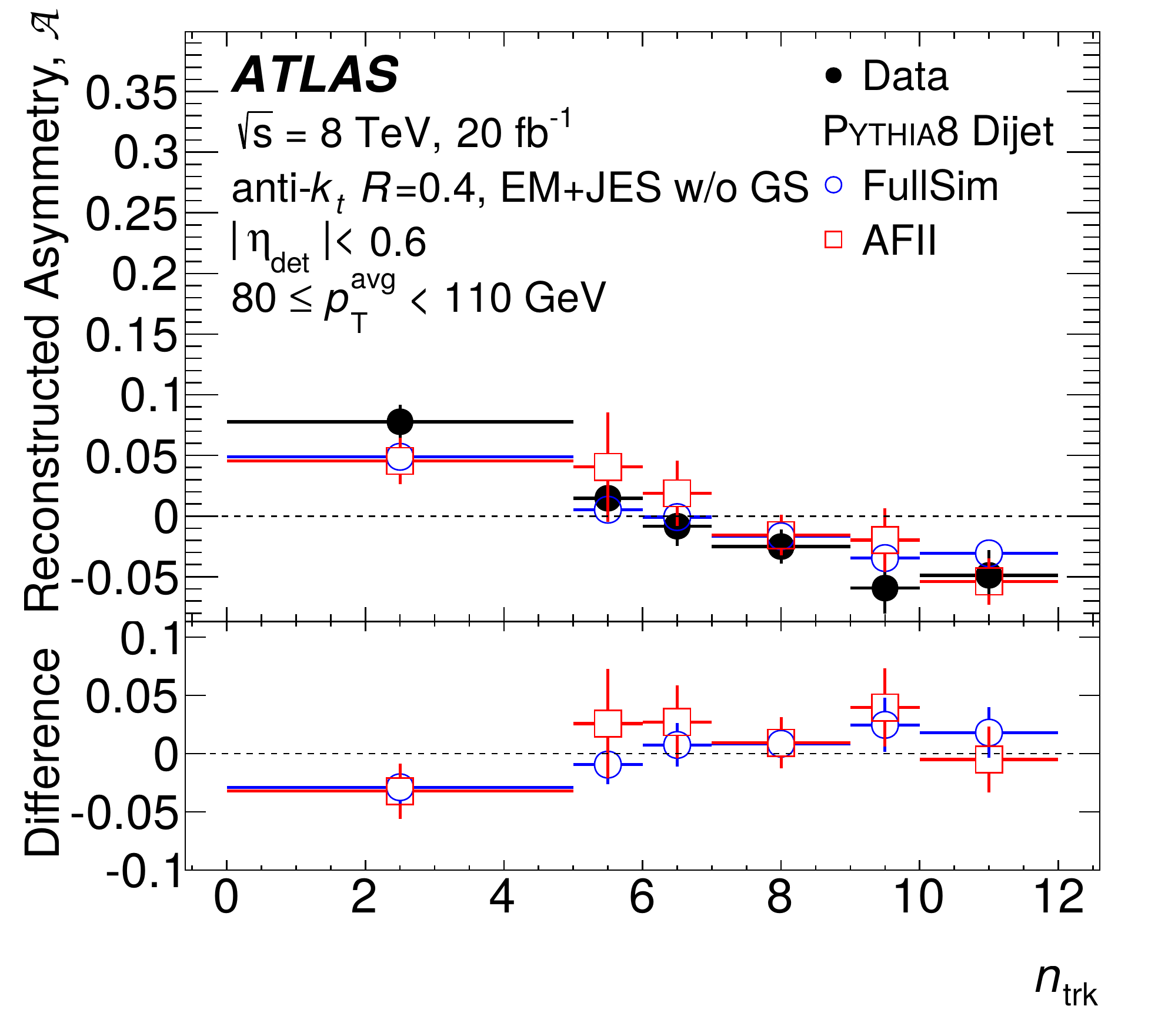}\\
\includegraphics[width=0.32\textwidth]{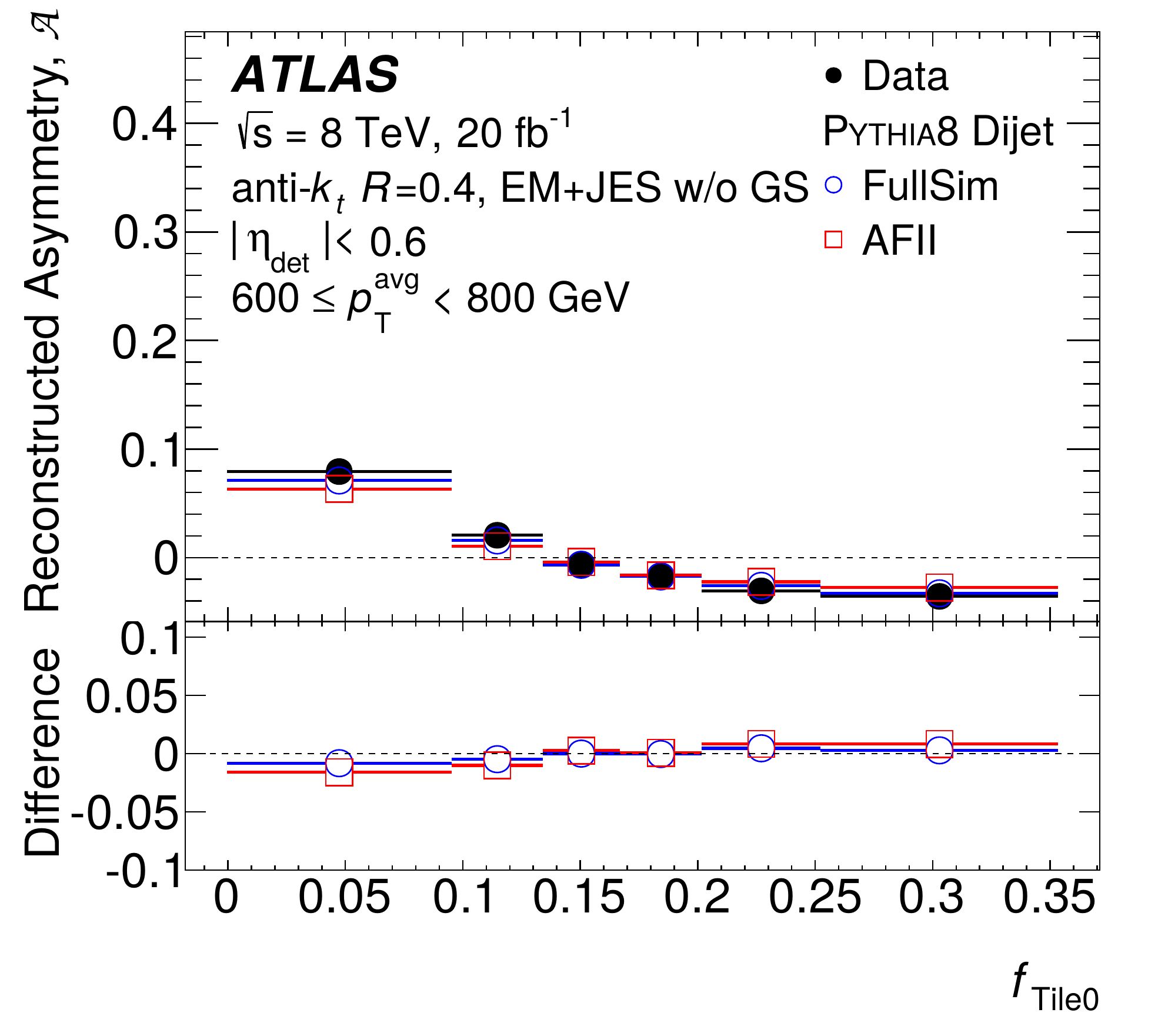}
\includegraphics[width=0.32\textwidth]{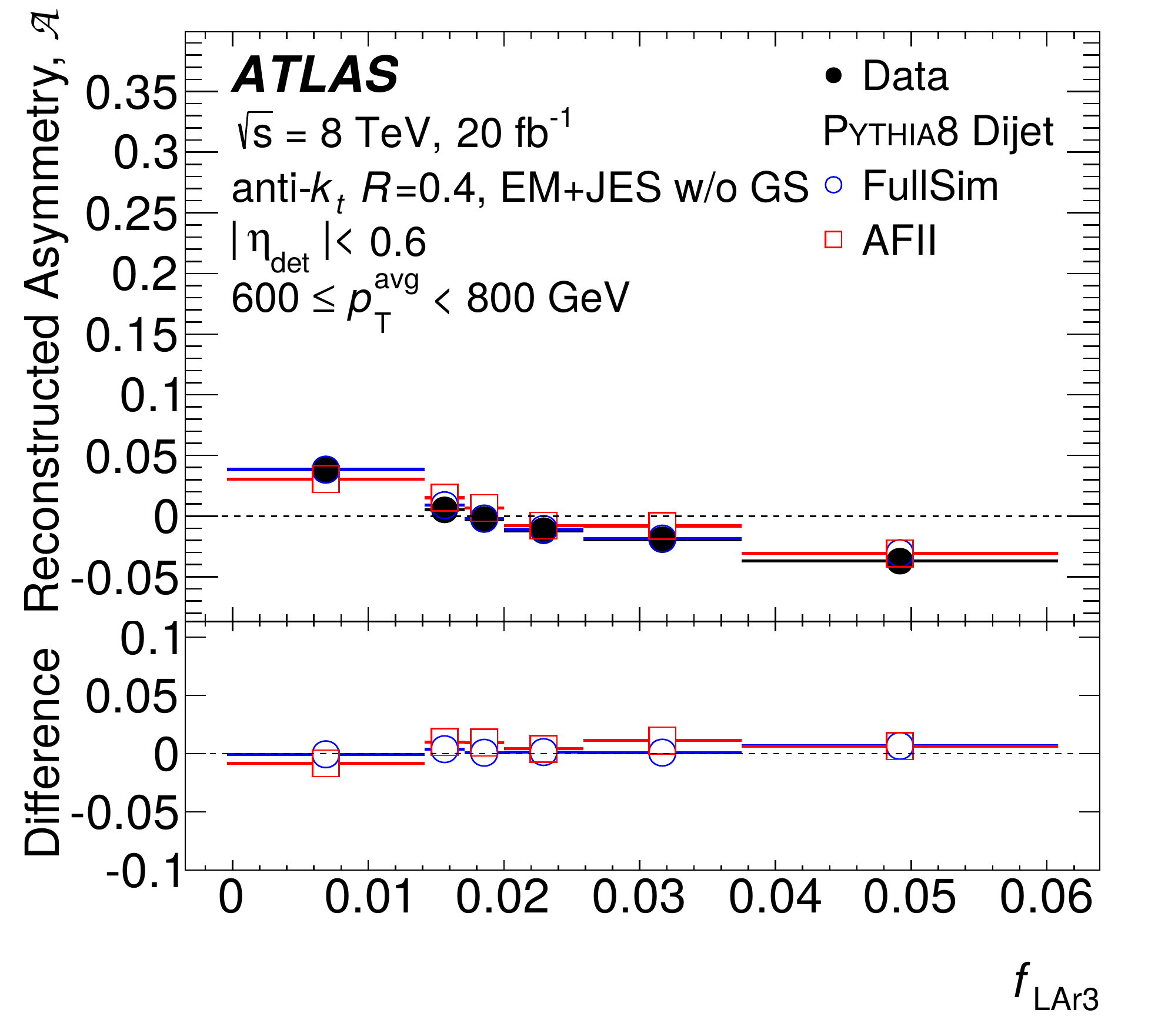}
\includegraphics[width=0.32\textwidth]{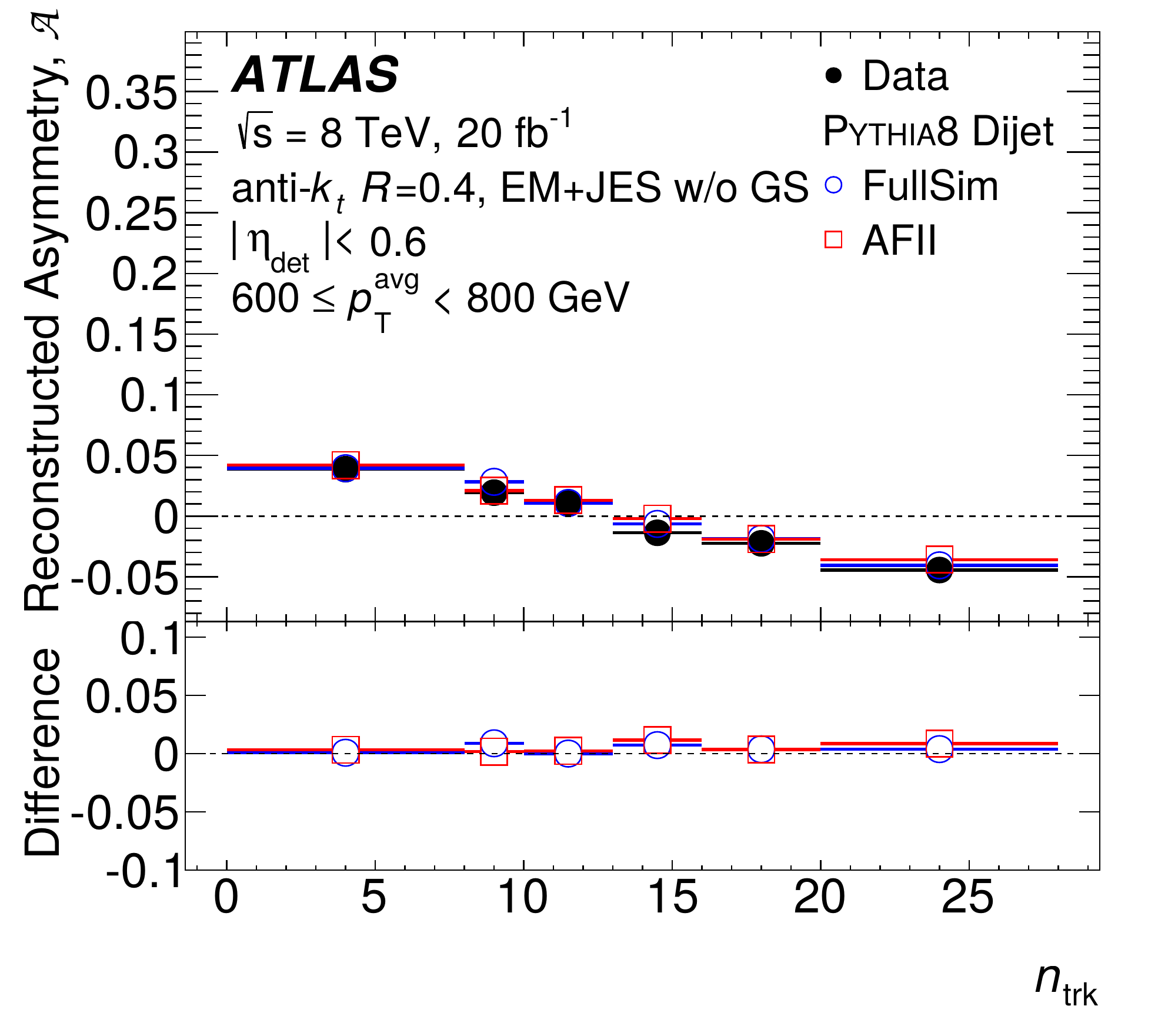}\\
\includegraphics[width=0.32\textwidth]{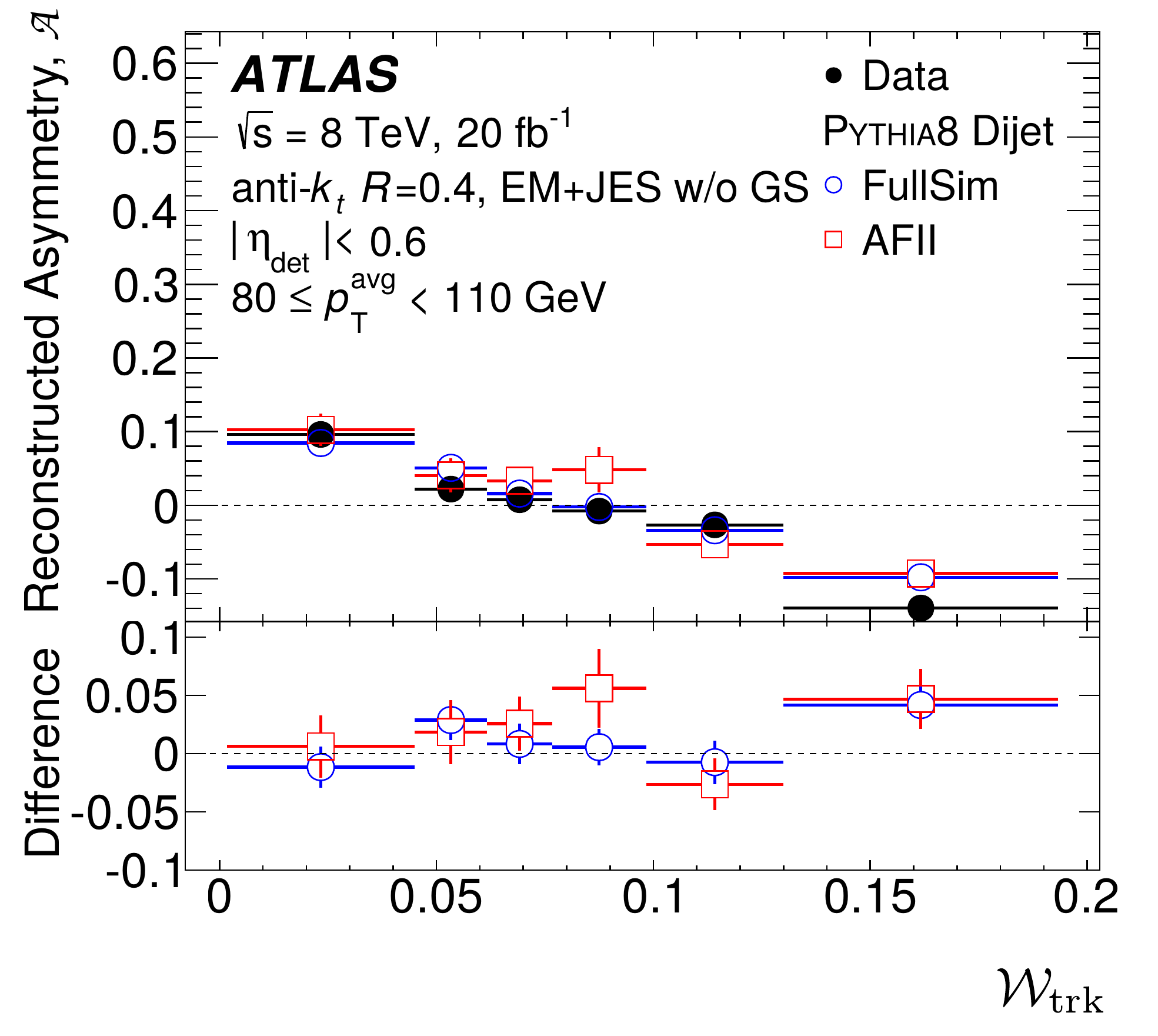}
\includegraphics[width=0.32\textwidth]{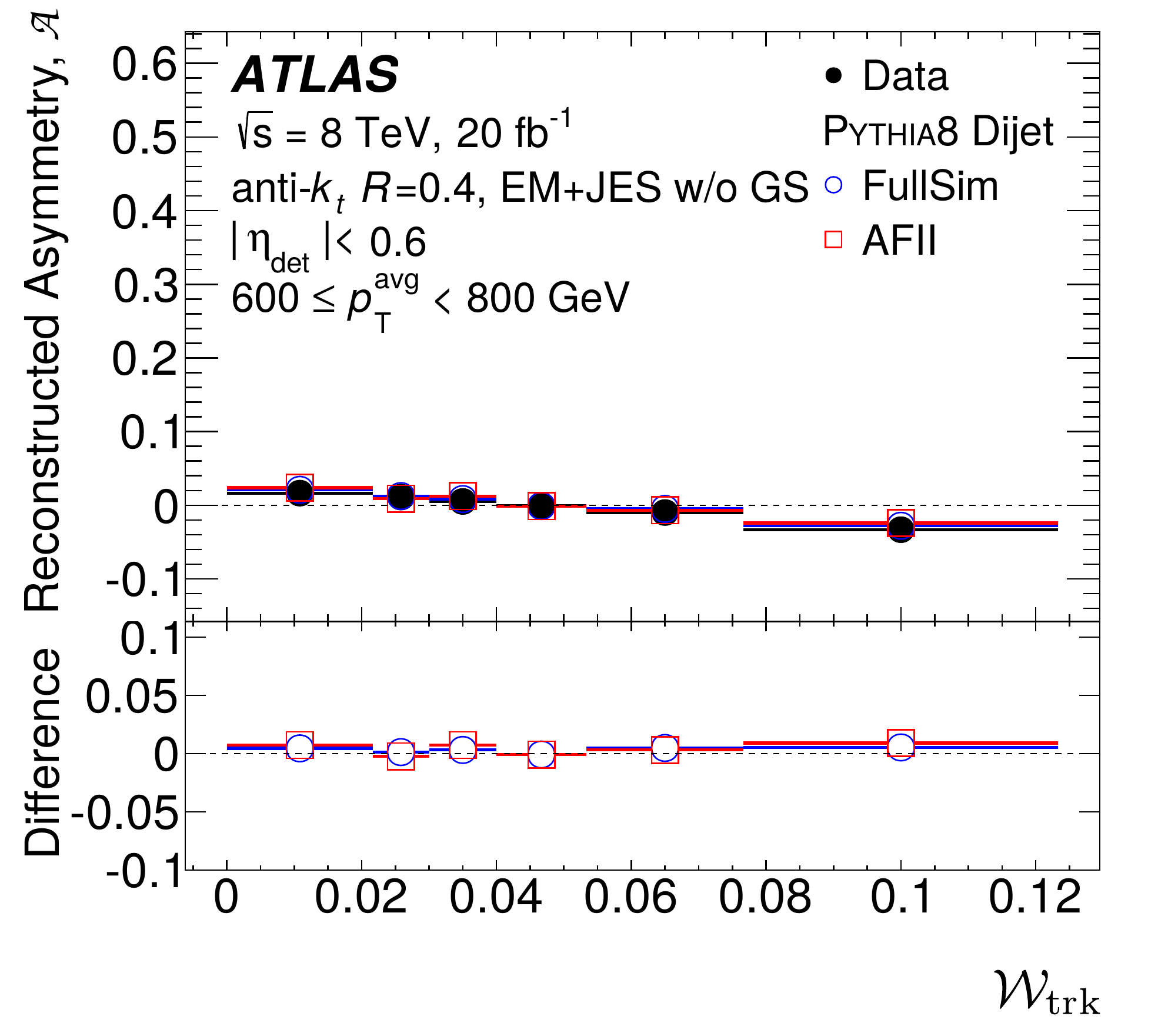}
\includegraphics[width=0.32\textwidth]{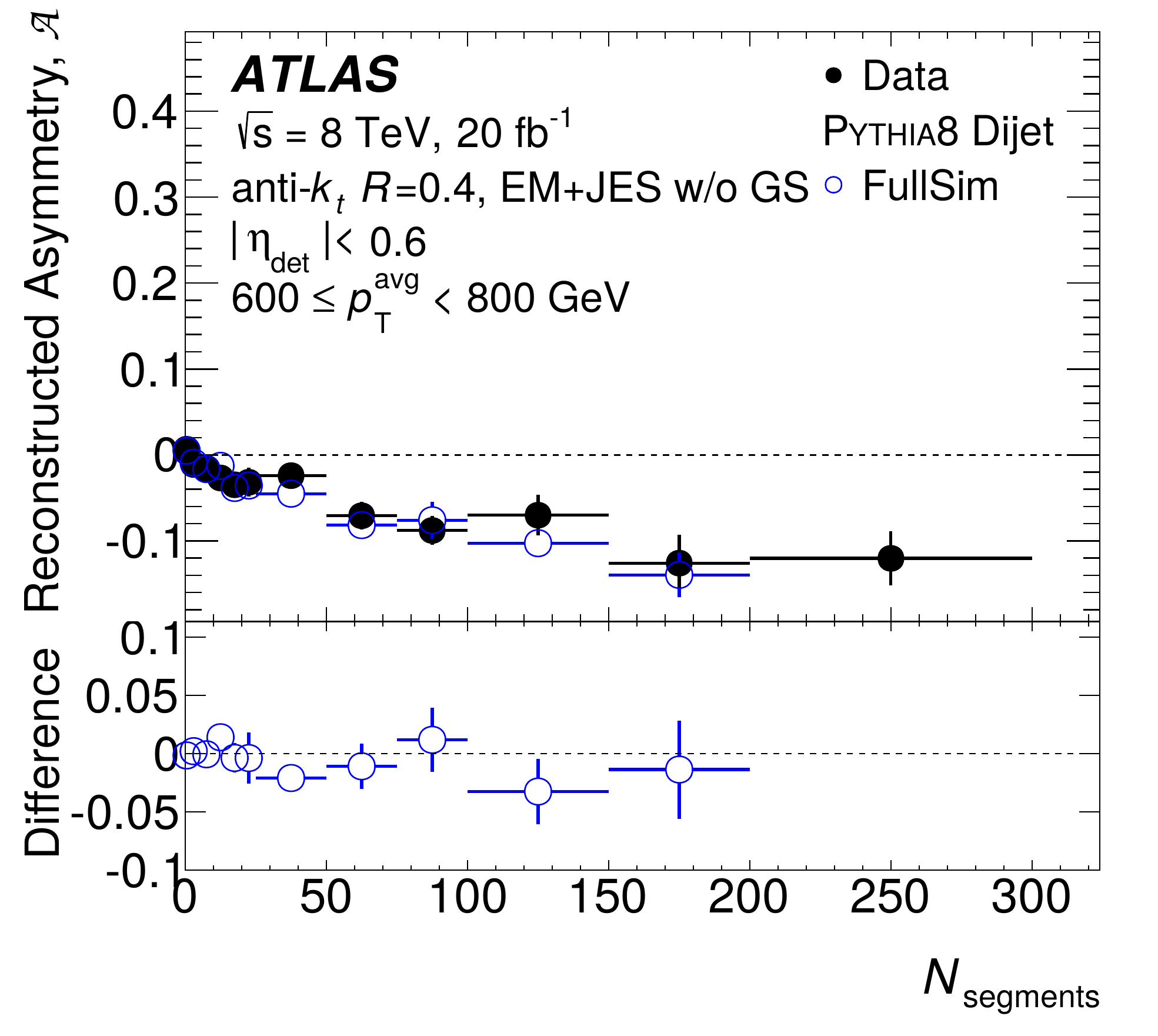}
\caption{Dijet \insitu{} validation of jet response as a function of
$\ftile$, $\fem$ and $\nTrk$ for jets with $\SI{80}{\GeV}<\ptavg<\SI{110}{\GeV}$ and $|\etaDet{}|<0.6$ (top)
and for jets with $\SI{600}{\GeV}<\ptavg<\SI{800}{\GeV}$ and $|\etaDet{}|<0.6$ (middle)
and the same quantity as a function of $\trackWIDTH$ and $\Nsegments$ (bottom).
Each set of measurements are shown for data~(filled circles) and for \pythia{} MC simulation with both full~(empty circles) and fast~(empty squares) detector modelling.
All jets are reconstructed with {\antikt} $R=0.4$ and calibrated with the \EMJES{} scheme without any global sequential corrections.}
\label{fig:GS-data-validation}
\end{figure}
Results for all variables used in the GS calibration are shown in Figure~\ref{fig:GS-data-validation} for jets with $|\etaDet|<0.6$ in two representative $\pt$ ranges.
No GS calibration is applied to either the probe or the reference jet.
It can be seen that the reference \pythia{} dijet MC sample agrees with the data within 1\% (4\%) for
$\SI{600}{\GeV}<\ptavg<\SI{800}{\GeV}$ ($\SI{80}{\GeV}<\ptavg<\SI{110}{\GeV}$)
for the calorimeter-based variables, and slightly better for the track-based observables. A similar level of agreement is seen in other jet \pt{} and \etaDet{} ranges.
These differences impact the average jet \pt{} weighted by the fraction of jets with corresponding values of the GS property in question; given that these differences occur in the tails of the distributions, the impact on the average jet \pt{} is thus minimal.
Results using MC samples produced with the AFII fast detector simulation are also shown and demonstrate similar agreement with data, although these samples have larger statistical uncertainties.
The relative data--MC agreement stays the same after the GS calibration is applied for both full and fast detector simulation.

\subsection{Comparison of jet resolution and flavour dependence between different event generators}
\label{sec:GS-MC-generators-comparison}
\begin{figure}[!htb]
\centering
\includegraphics[width=0.49\textwidth]{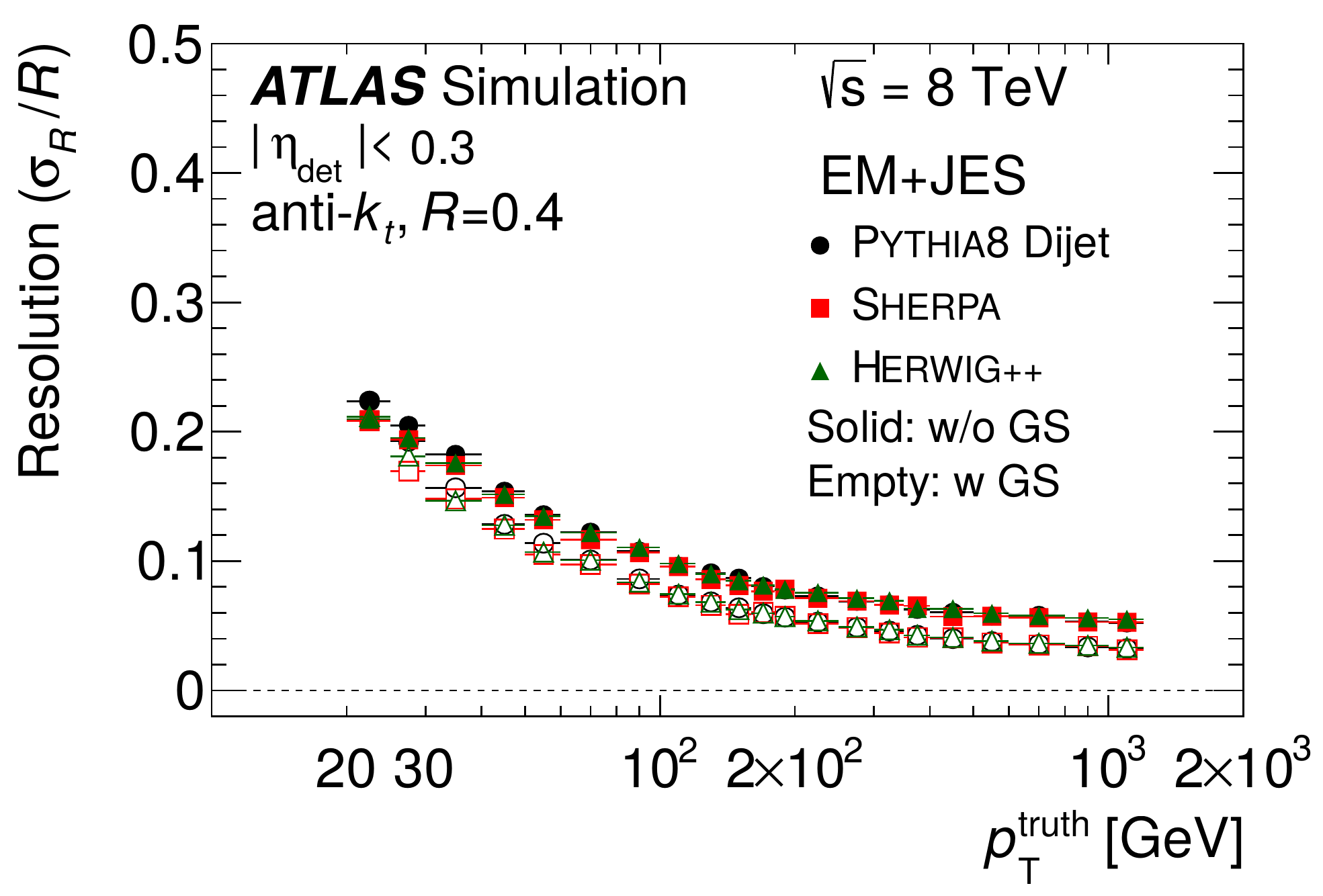}
\includegraphics[width=0.49\textwidth]{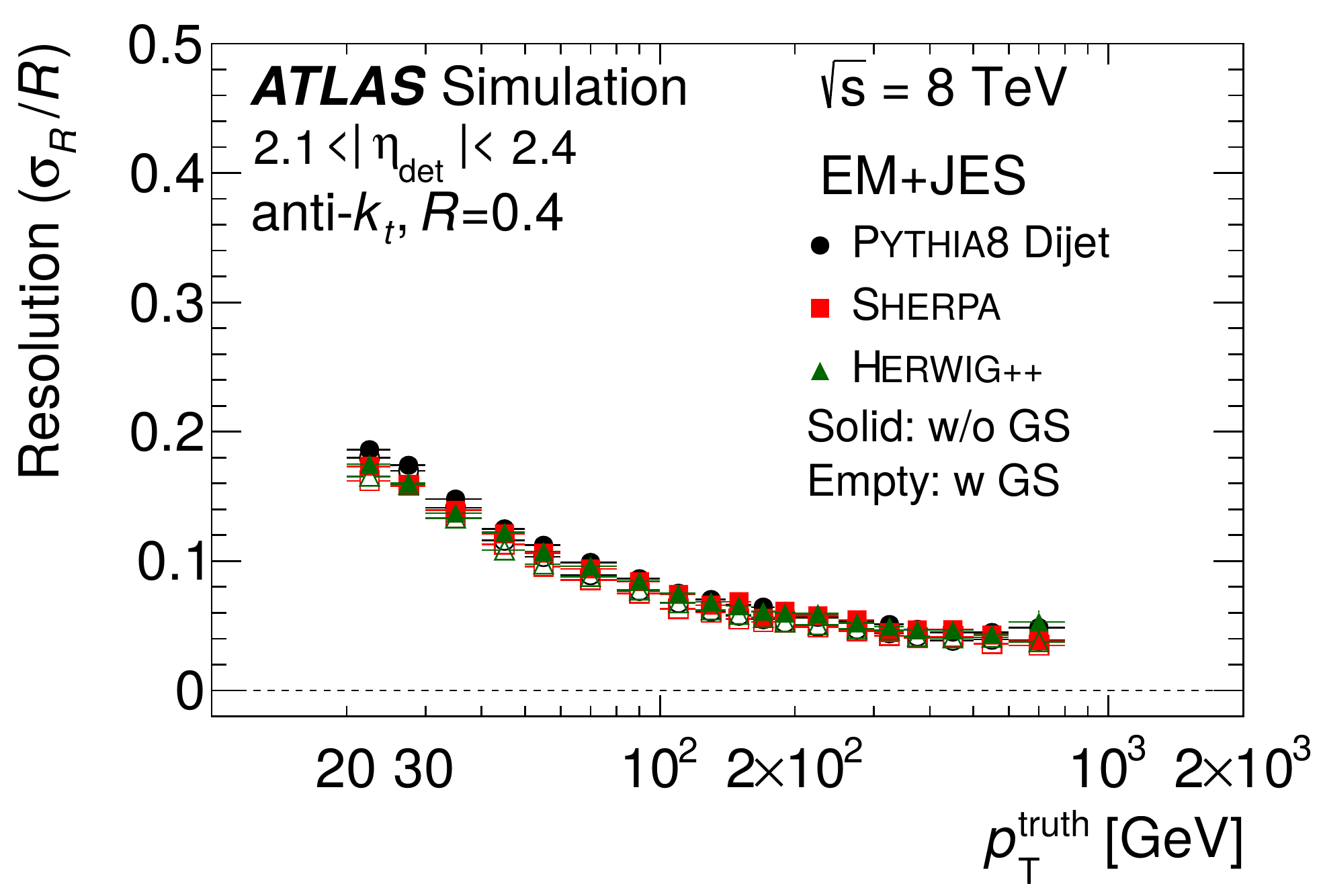}
\includegraphics[width=0.49\textwidth]{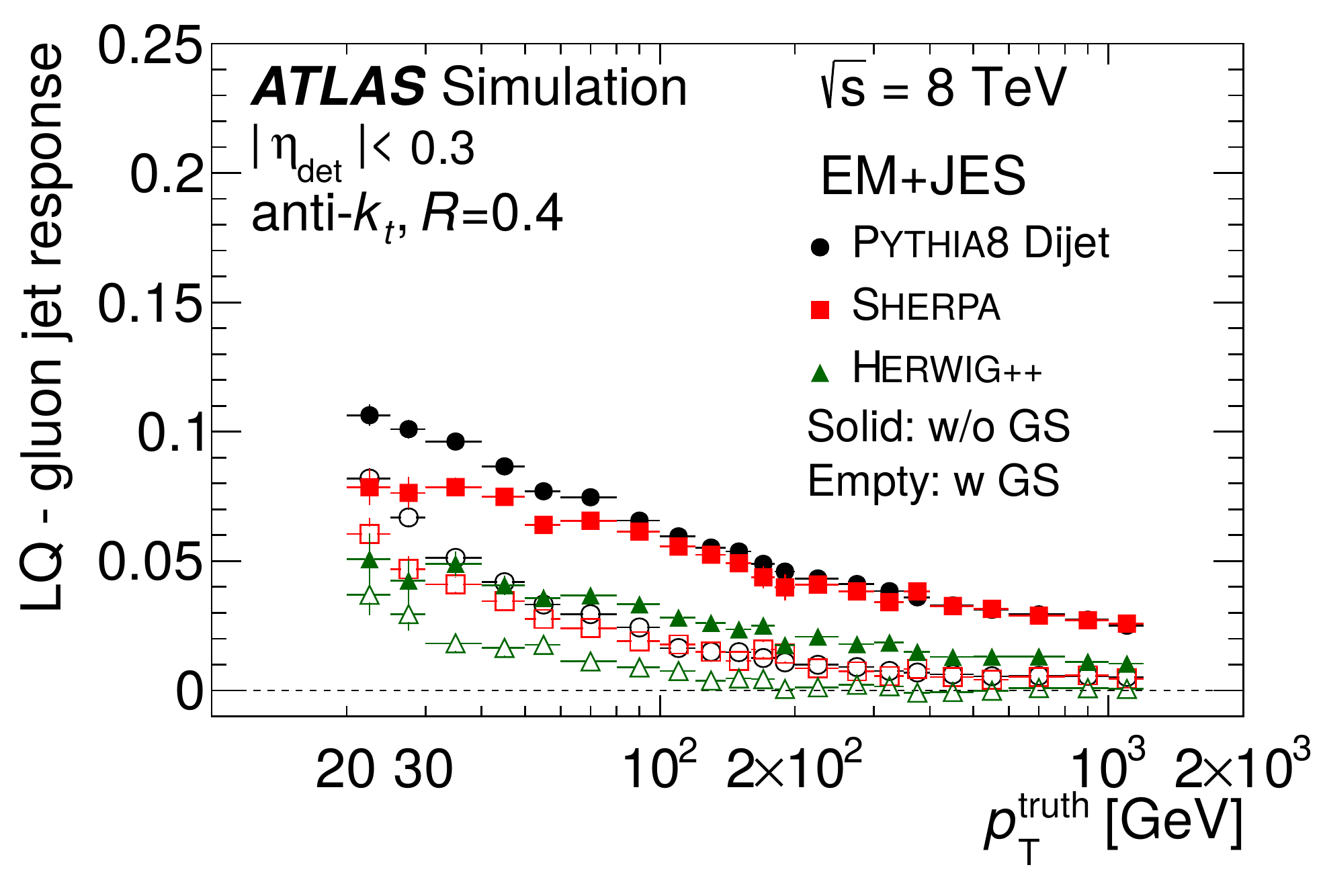}
\includegraphics[width=0.49\textwidth]{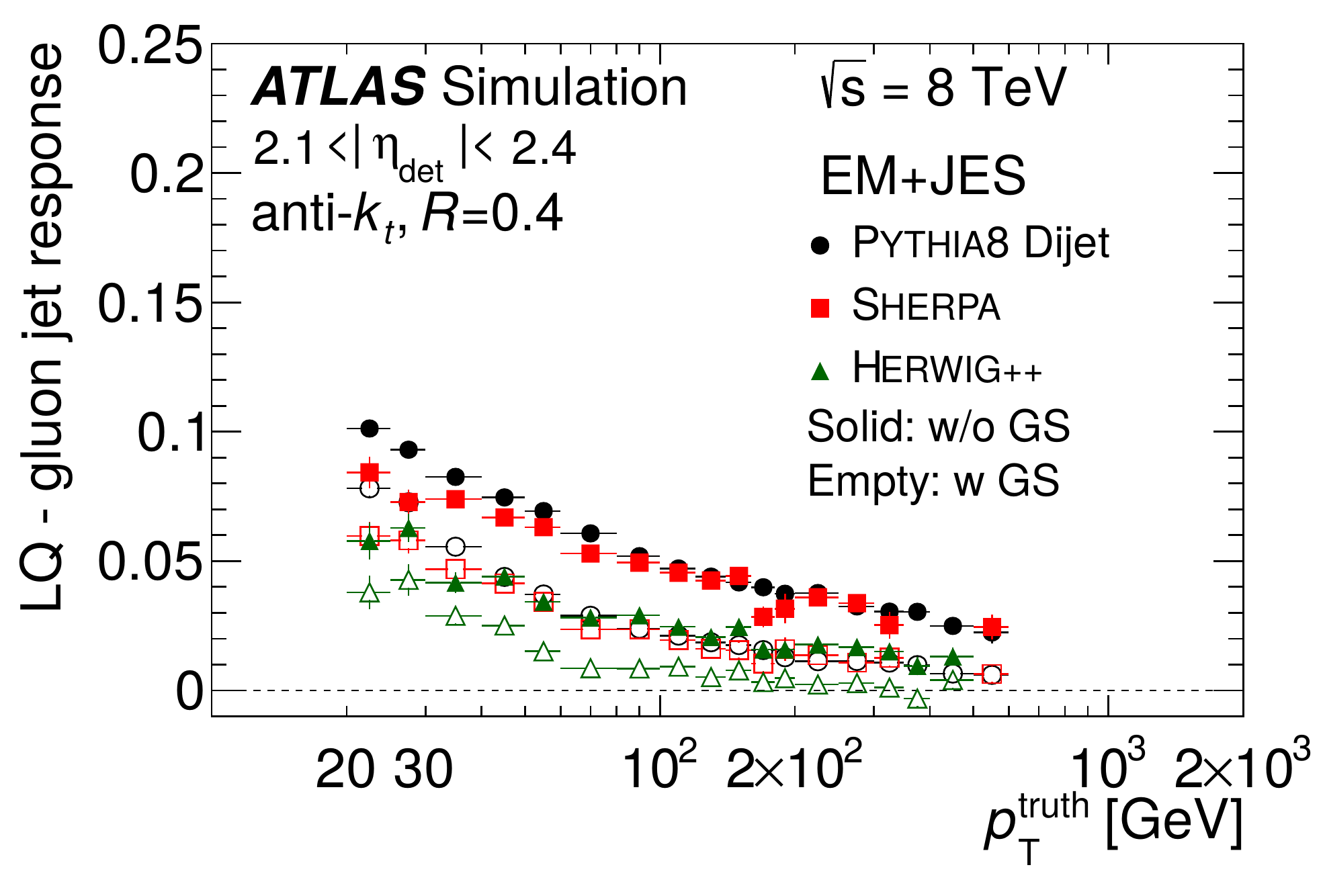}
\caption{
Jet \pt{} resolution (top) and difference in jet response between gluon and light quark (LQ) initiated jets (bottom) as a function of \pttrue{} for two representative $|\etaDet{}|$ regions. Results are shown both before (closed markers) and after (open markers) the global sequential corrections is applied, and separately for jets in the \pythia{}~(circles), \sherpa{}~(squares), and \herwigpp{}~(triangles) dijet MC samples. All jets are reconstructed with {\antikt} $R=0.4$. Jets are labelled LQ- or gluon-initiated, based on the highest-energy parton in the MC event record which fulfils an angular matching to the jet as further detailed in Section~\ref{sec:GS-flavour}.}
\label{fig:GS-generator-dependence}
\end{figure}
Figure~\ref{fig:GS-generator-dependence} presents comparisons of \pt{} resolution and response dependence on jet flavour between three MC event generators, namely \pythia{}, \herwigpp{}, and \sherpa{}, each with a different implementation of parton showering, multiple parton--parton interactions and hadronization (Section~\ref{sec:mc}).
These quantities are shown as a function of jet \pt{} both with and without GS calibration applied in two representative \etaDet{} regions.
\pythia{} tends to predict a slightly worse jet \pt{} resolution for jets with $\pt<50$~\GeV\ compared with the jet resolution in \herwigpp{} and \sherpa{},
but the improvement introduced by the GS correction is compatible between the different generators. The reduction of jet flavour dependence is studied by taking the difference between the jet responses for LQ and gluon jets, determined as discussed in Section~\ref{sec:GS-flavour} and as used for light-quark vs gluon discrimination~\cite{PERF-2013-02}.
The overall flavour dependence of the jet response is found to be smaller for \herwig{} than for \pythia{} and \sherpa{},
and in general, the LQ jet response is quite similar between the generators, while the response for gluon jets varies more.
For jets with $\pt>40$~\GeV, the response difference between LQ and gluon jets is reduced by at least a factor of two after applying the GS correction.

% End of text imported from the .//sections/gsc.tex input file

\afterpage{\clearpage}
% The next lines are included from the .//sections/dijet_calib.tex input file
\section{Intercalibration and resolution measurement using dijet events}
\label{sec:dijet}

Following the determination and application of MC-based jet calibration factors, it is important to measure the jet response and resolution \insitu{}, quantify the level of agreement between data and simulation, and correct for any discrepancy.
The first step is to investigate the jet response dependence across the detector in terms of pseudorapidity.
All results presented in this section are obtained with jets calibrated with the calibration chain up to, and including, the GS calibration (Section~\ref{sec:jetKinem}).
 
\subsection{Techniques to determine the jet calibration and resolution using dijet asymmetry}
\label{sec:dijet-balance}
 
The jet energy resolution (JER) and the relative response of the calorimeter as a function of pseudorapidity are determined using events with dijet topologies~\cite{PERF-2012-01,PERF-2011-04}.
The \pt{} balance is quantified by the dijet asymmetry
\begin{equation}
\label{eq:asym}
\asym = \frac{\ptprobe - \ptref}{\ptavg},
\end{equation}
where \ptref\ is the transverse momentum of a jet in a well-calibrated reference region, \ptprobe\ is the transverse momentum of the jet in the calorimeter region under investigation,
and $\ptavg = (\ptprobe + \ptref)/2$.
The average calorimeter response relative to the reference region, $1/c$, is then defined as
\begin{equation}
\label{eq:rr}
\frac{1}{c} \equiv \frac{2+\left<\asym\right>}{2-\left<\asym\right>} \approx \frac{\left<\ptprobe\right>}{\left<\ptref\right>},
\end{equation}
where 
$\left<\asym\right>$ is the mean of the asymmetry distribution in a given bin of \ptavg\ and \detEta, and the last equality of Eq.~(\ref{eq:rr})
can be obtained by inserting the expectation value of a first-order Taylor expansion of Eq.~(\ref{eq:asym}), giving
$\left<\asym\right> = \left(\langle\ptprobe\rangle -\langle\ptref\rangle\right)/\langle\ptavg\rangle$.
 
Two versions of the analysis are performed. In the \textit{central reference method}, the calorimeter response is measured as a function of \ptavg\ and \detEta\ relative to the region defined by $|\detEta|<0.8$.  Jets in this region are precisely calibrated using \Zjet{}, \gamjet{} and multijet data (Sections~\ref{sec:vjets} and~\ref{sec:multijet}).
In the \textit{matrix method}, multiple \etaDet{} regions are chosen and the calorimeter response in a given region is measured relative to all other regions. For a given \ptavg{} bin, $\left<\asym\right>$ is determined for each of a large number of combinations of \etaDet{} regions of the two jets involved. The calorimeter response relative to the central region is then obtained by solving a set of linear equations based on this matrix of dijet asymmetries~\cite{PERF-2012-01}. A constraint is applied that sets the average response for jets with $|\detEta|<0.8$ to unity.
The advantage of the matrix method is that a much larger fraction of events can be used, since events with both jets outside $|\detEta|<0.8$ are considered, thus reducing the statistical uncertainty of the final result.
Statistical uncertainties in the matrix method result are estimated using pseudo-experiments.
Each pseudo-experiment generates a new matrix of dijet asymmetries by sampling the average asymmetry $\left<\asym\right>$ of each bin (matrix element) according to their statistical uncertainty.
The intercalibration factors are then derived for each pseudo-experiment, and the statistical uncertainty of the calibration is obtained from the spread.
In this paper, the main results are obtained using the matrix method, and the simpler central reference method is used for validation.

The asymmetry distribution also probes the jet energy resolution.
The standard deviation of the asymmetry distribution $\Dasym ^\text{probe}$ in a given $(\ptavg{},\eta^\text{probe}_\text{det})$ bin can be expressed as
\begin{equation}
\Dasym^\text{probe} = \frac{\Dpt^\text{probe}\oplus\, \Dpt^\text{ref}}{\ptavg} =
\left<\frac{\Dpt}{\pt}\right>_\text{probe} \oplus\,   \left<\frac{\Dpt}{\pt}\right>_\text{ref} =
\left<\frac{\DE}{E}\right>_\text{probe} \oplus\,   \left<\frac{\DE}{E}\right>_\text{ref},
\label{eq:dijet-jer}
\end{equation}
where $\Dpt^\text{probe}$ and $\Dpt^\text{ref}$ are the standard deviations of \ptprobe{} and \ptref{}, respectively.
The first two equalities of Eq.~(\ref{eq:dijet-jer}) follow from error propagation of Eq.~(\ref{eq:asym}) and from the fact that after calibration $\langle \ptprobe{}\rangle = \langle \ptref{}\rangle = \langle \ptavg{} \rangle$ in a given \ptavg{} bin. The energy and \pt{} resolutions are approximately the same since contributions of the jet angular resolution are negligible (Figure~\ref{fig:etaRes}).
The standard deviation of the asymmetry distribution $\Dasym$ is obtained from a Gaussian fit to the core of the distribution.
 
The standard deviation of the probe jet \pt{} is derived from Eq.~(\ref{eq:dijet-jer}) as
\begin{equation}
\left<\frac{\Dpt}{\pt}\right>_\mathrm{probe} = \Dasym^\text{probe} \,\ominus\, \left<\frac{\Dpt}{\pt}\right>_\mathrm{ref},
\label{eq:dijet_jer}
\end{equation}
where the latter term is derived from events where both jets fall in the central reference region ($|\etaDet{}|<0.8$).
In this case, the reference region is being probed, and the first and last terms in Eq.~(\ref{eq:dijet_jer}) are hence equal, which gives
$\left<\Dpt/\pt\right>_\mathrm{ref} = \Dasym {\big/} \sqrt{2}$.
When calculating the asymmetry, the jets are fully calibrated including all data-driven correction factors.
 
The \pt{} balance strictly holds only for $2\rightarrow2$ partonic events.
In reality, the \pt{} balance between two jets is affected on an event-by-event basis by additional quark/gluon radiation outside of the jets, as well as hadronization and MPI effects that cause particle losses and additions to the jets, respectively.
To account for the impact of such effects, the dijet asymmetry standard deviation $\Dasym$ is measured separately for reconstructed and \tjets{},
and the standard deviation due to detector smearing $\Dasym^\text{det}$ is
obtained by subtracting the truth-particle quantity from the observed quantity in quadrature:
\begin{equation}
\Dasym^\text{det} = \Dasym^\text{reco} \ominus \Dasym^\text{truth}.
\label{eq:ptcl-subtraction}
\end{equation}
This final jet energy resolution measurement $\left<\Dpt/\pt\right>$ is calculated according to Eq.~(\ref{eq:dijet_jer}) after first correcting the asymmetry width \Dasym{} according to Eq.~(\ref{eq:ptcl-subtraction}).
 
\subsection{Determining the jet resolution using the dijet bisector method}
\label{sec:dijet-bisector}
 
The bisector method attempts to separate the desired part of the dijet \pt{} imbalance that is due to fluctuations in the jet calorimeter response from contributions from other effects such as soft parton radiation and the underlying event.
In the same way as for the central reference method, selection criteria are applied to select events with dijet topology, and at least one of the two jets is required to have $|\etaDet{}|<0.8$.
This jet is referred to as the ``reference jet'', while the other jet is labelled ``probe jet''. If both jets fulfil $|\etaDet{}|<0.8$, the labels are assigned randomly.
The \pt{} (imbalance) of the dijet system in the transverse plane is defined as the vectorial sum of the \pt{} vectors of the leading two jets:
$\vec{p}_\text{T}^{\,jj} = \vec{p}^\text{\,probe}_\text{T} + \vec{p}^\text{\,ref}_\text{T}$.
This vector is projected onto a Cartesian coordinate system in the transverse plane $(\psi,\upsilon)$, where the $\upsilon$-axis is defined to be along the direction that bisects the angle $\Delta\phi_{12}$ between the two jets, and the $\psi$-axis is defined to have a direction that minimizes the angle to the probe jet
as illustrated in Figure~\ref{fig:bisector}.
Both effects from the detector (response and resolution) and from physics (e.g.\ radiation) are present
in the $\psi$ component of the \pt{} balance that is oriented ``towards'' the probe jet axis, whereas
detector effects should be significantly smaller than physics effect in the
$\upsilon$ component, oriented ``away'' from both the probe and the reference jet.
As a result~\cite{PERF-2011-04}, the jet energy resolution for events in a given \ptavg{} bin where both jets are in the reference region ($|\etaDet^\text{probe}|<0.8$) is given by
\begin{equation}
\left<\frac{\Dpt}{\pt}\right>_\text{ref} \, = \,
\frac{\sigma_\psi \,\ominus\, \sigma_\upsilon}{\ptavg \sqrt{2 \,\langle \left|\cos \Delta\phi_{12}\right| \rangle}},
\label{eq:bisec}
\end{equation}
and for events where the probe jet is outside the reference region it is given by
\begin{equation}
\left<\frac{\Dpt}{\pt}\right>_\text{probe} \, = \,
\frac{\sigma_\psi \,\ominus\, \sigma_\upsilon}{\ptavg \sqrt{\langle \left|\cos \Delta\phi_{12}\right| \rangle}} \ominus  \left<\frac{\Dpt}{\pt}\right>_\text{ref}.
\label{eq:bisecFwd}
\end{equation}
The standard deviations $\sigma_\upsilon$ and $\sigma_\psi$ are evaluated as the width parameters of Gaussian fits to the $p_{\text{T}\upsilon}$ and $p_{\text{T}\psi}$ distributions, respectively.
 
Although the bisector observables in Eqs.~(\ref{eq:bisec}) and~(\ref{eq:bisecFwd}) have less dependence on soft quark or gluon emission than the asymmetry-based jet resolution measurement of Eq.~(\ref{eq:dijet-jer}),
the approach relies on the assumption that the physics effects are the same in the $\psi$ and $\upsilon$ components.
Corrections to the measured $\sigma_\psi$ and $\sigma_\upsilon$ are made by subtracting the corresponding quantities derived using \tjets{} in quadrature from the measured quantity, analogously to what is done for the central reference method (Eq.~(\ref{eq:ptcl-subtraction})).
 
\begin{figure}[t]
\centering
\includegraphics[width=0.5\textwidth]{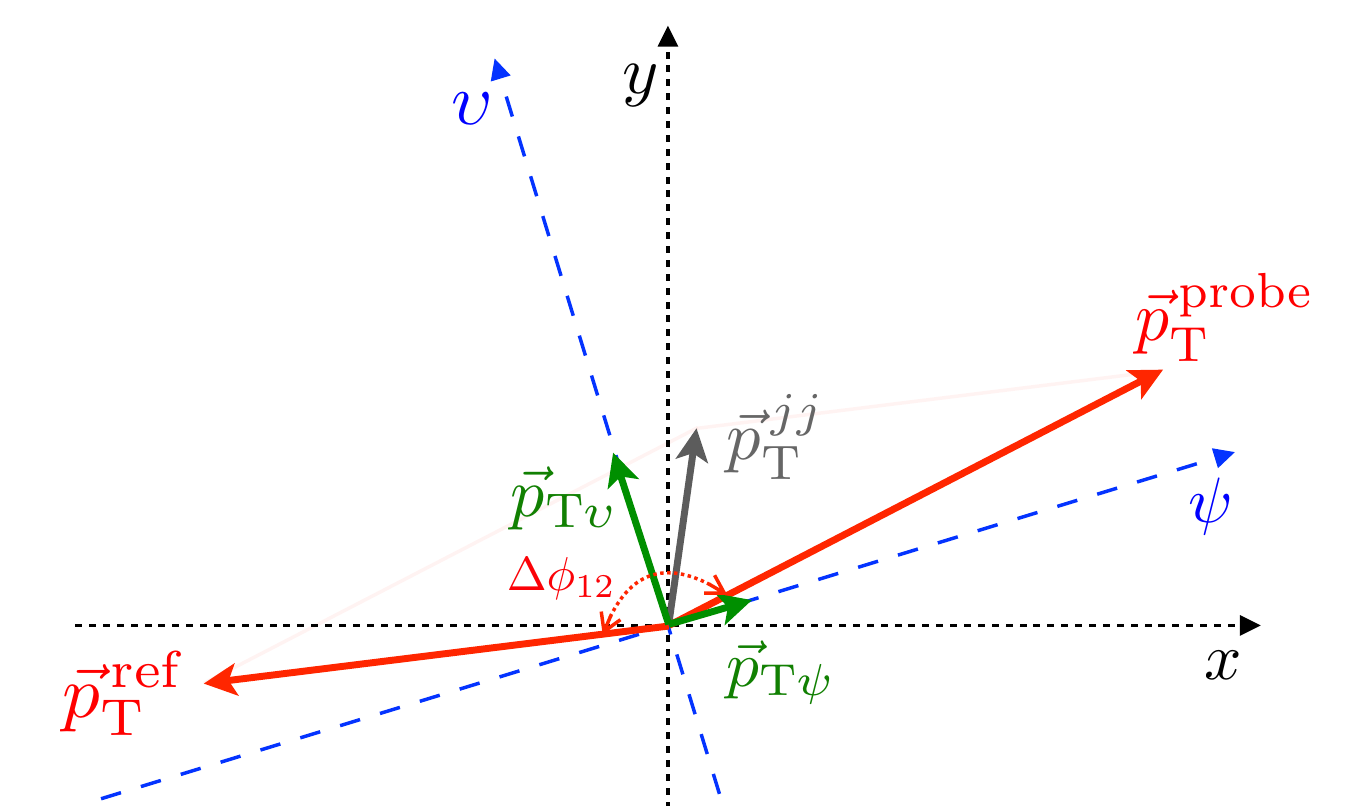}
\caption{
Illustration of observables used in the dijet bisector technique.
The $(\psi,\upsilon)$-coordinate system is defined such that the $\upsilon$-axis bisects the azimuthal angle $\Delta\phi_{12}$ between the leading two jets
while the $\psi$-axis minimizes the angle to the probe jet.
The vectorial sum of the transverse momenta of the probe and the reference jets define the dijet transverse momentum $\vec{p}_\text{T}^{\,jj}$.
Its components along the $\psi$- and $\upsilon$-axes ($p_{\text{T}\psi}$ and $p_{\text{T}\upsilon}$) are used to extract a measurement of the jet energy resolution.}
\label{fig:bisector}
\end{figure}

\subsection{Dijet selection}
\label{sec:dijet-selection}
 
Dijet events are selected using a combination of central ($|\etaDet|<3.2$) and forward ($|\etaDet|>3.2$) jet triggers.
For this selection, the trigger efficiency for each region of $\ptavg{}$ is greater than $99\%$ and approximately
independent of the pseudorapidity of the probe jet.
The jet triggers used have different \textit{prescales}, downscaling factors used to meet bandwidth constraints on the recording of data. Larger prescales are used for data recorded when the instantaneous luminosity is high or for triggers that require lower jet \pt{}. Due to the different prescales for the central and forward jet triggers, the data
collected by the different triggers correspond to different integrated luminosities. Each data event is assigned a trigger based on the \ptavg{} and \etaDet{} of the more forward jet. The data is hence split into different categories, and each is weighted according to the integrated luminosity of the dedicated trigger used following the ``exclusion method''~\cite{Lendermann:2009ah}.
Events are selected in which there are at least two jets with $\pt>25$~\GeV\ and $|\etaDet|<4.5$. To select events with a dijet topology, the azimuthal angle between the two leading jets (i.e., the reference and probe jets) is required to be $\Delta\phi_{12} > 2.5$ and events are rejected if they contain a third jet with $p_\textrm{T}^\textrm{j3} >  0.4\,\ptavg$. The jets are also required to fulfil the preselection described in Section~\ref{sec:jetReco}.

\subsection{Method for evaluating \insitu{} systematic uncertainties}
\label{sec:evalSyst}
 
The \insitu{} techniques rely on assumptions that are only approximately fulfilled, and simulation is used to account for these approximations.
For example, the momentum balance between the jet and the reference object is altered to varying degrees by the presence of additional radiation.
The impact from such radiation is reduced by event topology selection criteria.
Since the choice of the exact threshold values is arbitrary, systematic uncertainties are evaluated by rederiving the final result, which is a \DtM{} ratio, after varying these selection criteria.
Other systematic uncertainties are evaluated by altering choices used by the method, such as a parameter used in a fit or changing the MC generator.
In the case of the \gamjet{}, \Zjet{}, and multijet techniques (Sections~\ref{sec:vjets} and~\ref{sec:multijet}), uncertainties are also established by adjusting the kinematic properties (energy, \pt{}, etc.) of the reference object according to their associated uncertainties.
These variations test the ability of the MC simulation to model the physics effects since they either reduce or amplify their importance.
 
Many potential effects are considered as systematic uncertainty sources.
As explained above, each of these is evaluated by introducing a variation to the analysis.
However, due to limited statistics in both the data and MC samples, these variations have an associated statistical uncertainty (i.e.\ an ``uncertainty on the uncertainty'').
For example, an introduced variation that has no impact on the measured calibration factor (or resolution) still produces changes consistent with statistical fluctuations.
Thus, it is important to only include statistically significant variations as systematic uncertainties.
This is achieved with a two-step procedure outlined below.
 
In the first step, the statistical uncertainty of the systematic variations is evaluated in each \pt{} bin using pseudo-experiments, following the ``bootstrapping'' method~\cite{bootstrapping}.
Each such pseudo-experiment is constructed by altering the data and MC samples.  Each event
is counted $n$ times, where $n$ is sampled from a Poisson distribution with a mean of unity.
For each pseudo-experiment $i$, both the nominal $c_{\text{nom},i}$ and varied $c_{\text{var},i}$ results are extracted, and the uncertainty is evaluated as the difference between these results $\Delta c_{\text{var},i} = c_{\text{var},i}-c_{\text{nom},i}$.
If the variation is a change in the selection criteria or a change of the calibration or resolution smearing of any of the objects, the random fluctuations of the events that stay in the same bin are the same between the nominal and varied samples, while the events that migrate between bins will have independent fluctuations.
The statistical uncertainty of the systematic uncertainty amplitude is evaluated as the standard deviation of the systematic uncertainty (difference between varied and nominal result) of the pseudo-experiments.
 
In a second step, adjacent \pt{} bins might iteratively be combined until the observed variation is statistically significant. 
If the variation already is significant with the original binning, it is recorded as a systematic uncertainty. Otherwise, neighbouring bins are merged, which results in improved statistical precision.
After each bin-merging, it is checked if the systematic variation is significant, and if so, it is recorded as a systematic uncertainty.
If after all bins are merged, the variation is still not significant, the systematic effect is considered consistent with zero and is discarded.
 
For some systematic variations, there are physics reasons for the response to depend on \pt{}, such as the out-of-cone effects being relatively larger at low \pt{}.
In such cases, the bin merging step is not performed for the nominal uncertainty evaluation, but it is considered within alternative uncertainty scenarios (Section~\ref{sec:alt_unc}).

The use of the pseudo-experiments and the bin merging procedure strongly reduces the effect of statistical fluctuations when evaluating systematic uncertainties.
This procedure is used for all the \insitu{} methods discussed in this paper.
 
\subsection{Relative jet energy scale calibration using dijet events}
\label{subsec:dijetintercalibration}
 
The following subsections detail the determination of the intercalibration aimed at achieving a uniform scale for jets as a function of pseudorapidity.
 
\subsubsection{Comparison of matrix and central reference methods}
 
Figure~\ref{fig:comp} compares the relative jet response calculated using the matrix method with that obtained from the central reference method.
The relative response obtained from the matrix method differs slightly from that from the central reference method, most notably in the forward regions where the difference is up to 4\%. This is not surprising since the matrix method uses a significantly larger pool of events that have different kinematics (smaller rapidity separation) than the ones used by the central reference method.
The same shift appears in both data and MC simulation, resulting in consistent \DtM{} ratios between the two methods.
For 25~\GeV~$ \leq \ptavg{}<40$~\GeV\ the statistical precision of the matrix method generally exhibits a $40\%$ improvement compared with the precision of the central reference method.
The level of improvement decreases with increasing \ptavg{} and is typically less than $10\%$ for $\ptavg{}>400$~\GeV.
Since the final $\eta$ intercalibration is derived using \DtM{} ratios that are found to be consistent between the methods and the matrix method gives significantly smaller uncertainties, the matrix method is chosen, and hereafter all $\eta$ intercalibration results presented are derived using this method.
 
\begin{figure}
\centering
\includegraphics[width=.48\textwidth]{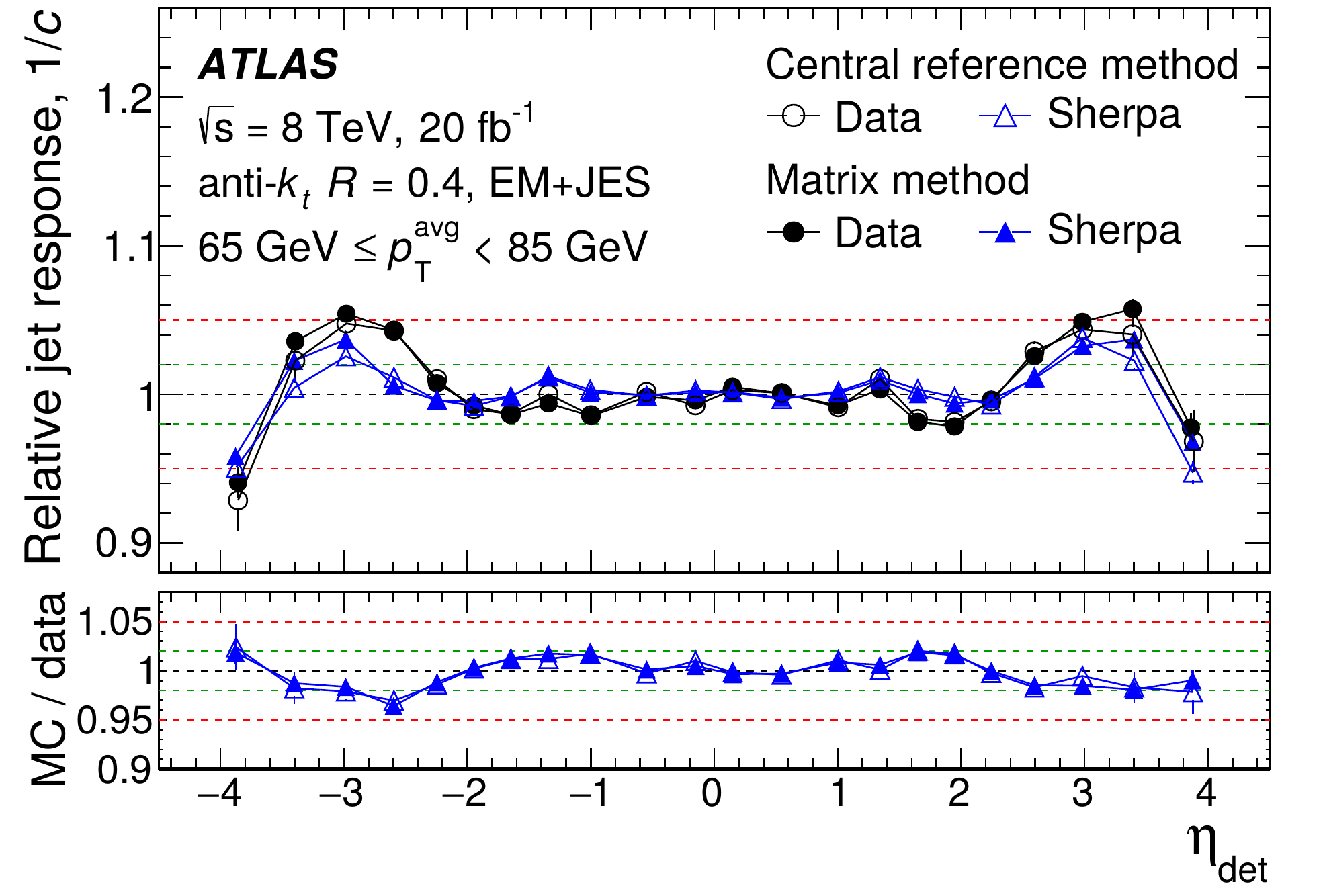}\quad
\includegraphics[width=.48\textwidth]{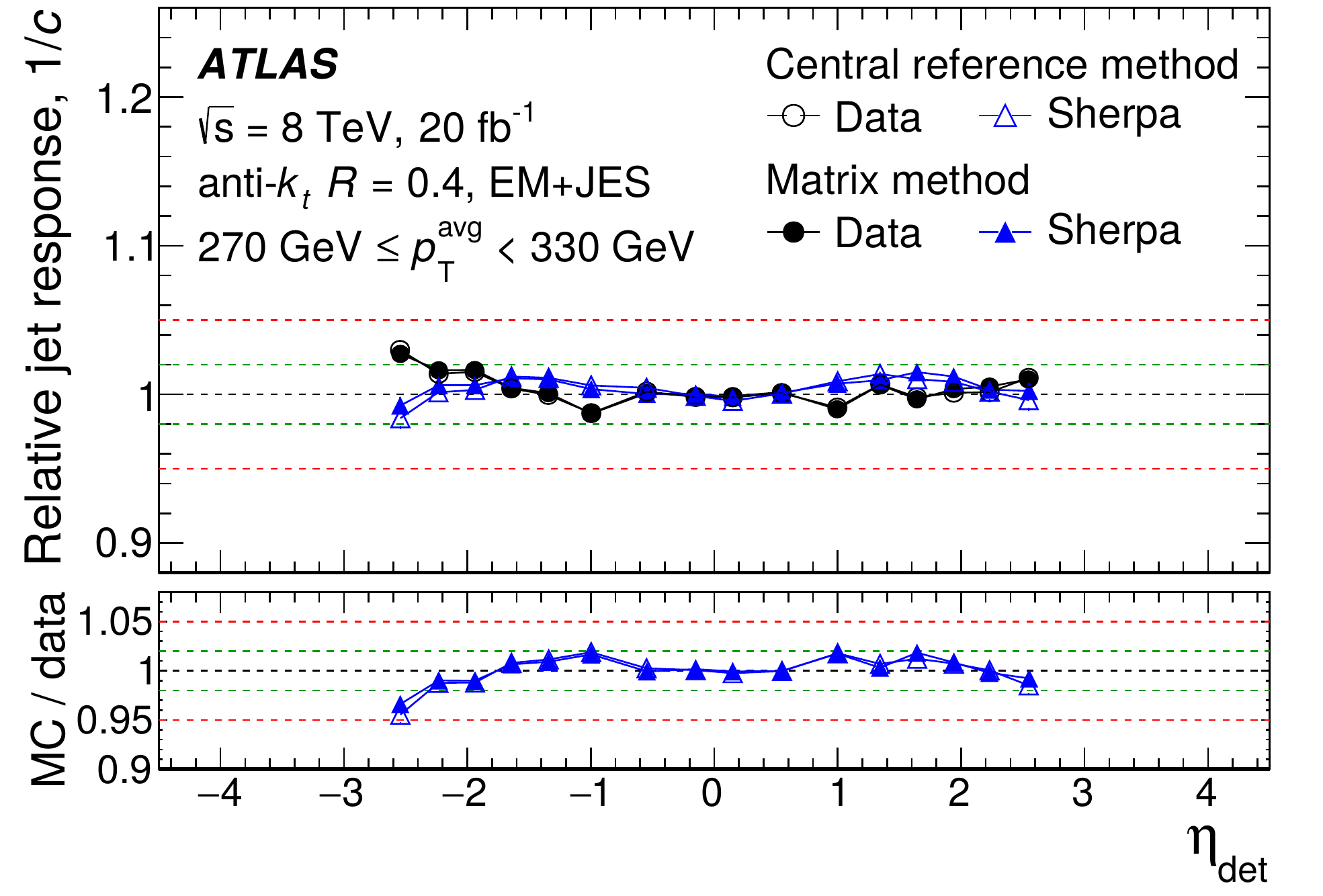}
\caption{
Relative jet response measured using the
matrix and central reference methods for \antikt{} jets with $R=0.4$ calibrated with the EM+JES scheme
as a function of the probe jet pseudorapidity.
Results are presented for 65~\GeV~$ \leq \ptavg{}<85$~\GeV\ and 270~\GeV~$ \leq \ptavg{}<330$~\GeV{} for data~(circles) and \sherpa~(triangles) using the central reference method~(empty symbols) and the matrix method~(filled symbols). Only statistical uncertainties are shown.
The dashed lines in the lower panels indicate $1\pm 0.02$ and $1\pm 0.05$.
\label{fig:comp}
}
\end{figure}
 
\subsubsection{Comparison of data with simulation}
 
Figure~\ref{fig:resp_vs_eta} shows the relative response
as a function of \etaDet{} for
data and the MC simulations for four \ptavg{} regions.
Figure~\ref{fig:resp_vs_pt} shows the relative response as a function
of $\ptavg$ for two representative $\etaDet$ bins, namely
$-1.5\leq\etaDet<-1.2$ and  $2.1\leq\etaDet<2.4$.
The general features of the response in data are reproduced reasonably well by the \sherpa\ and \powhegpyt\ predictions. Furthermore, the theoretical predictions are in good agreement with each other, with a much smaller spread than that observed in the previous studies using \pythia{} and \herwigpp{}~\cite{PERF-2012-01}, because the new theoretical predictions are accurate to leading order in perturbative QCD for variables sensitive to the third jet's activity, such as the dijet balance, whereas the \pythia{} and \herwigpp{} predictions rely on the leading-logarithm accuracy of the parton shower algorithms.
 
\begin{figure}
\centering
\includegraphics[width=.48\textwidth]{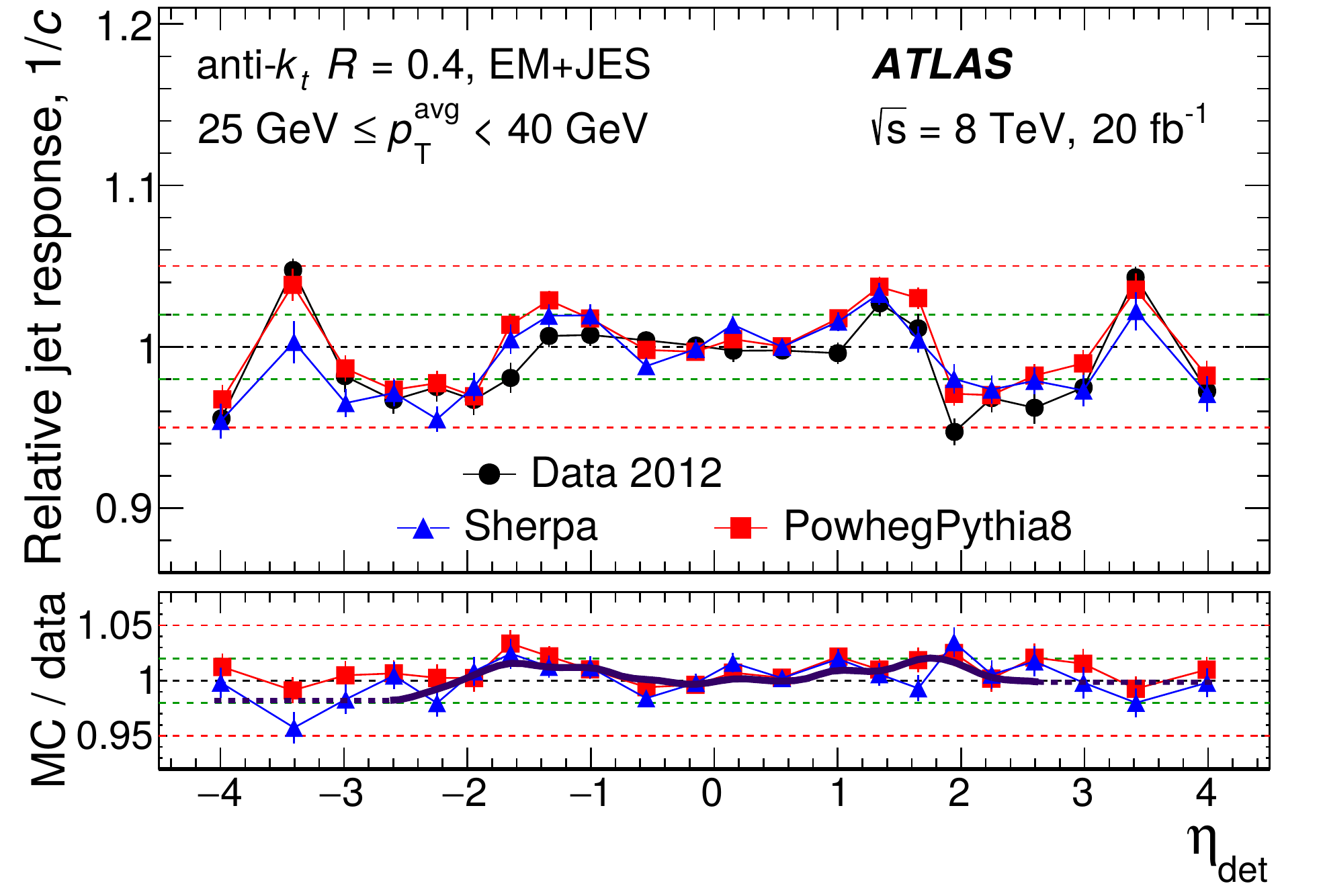}\quad
\includegraphics[width=.48\textwidth]{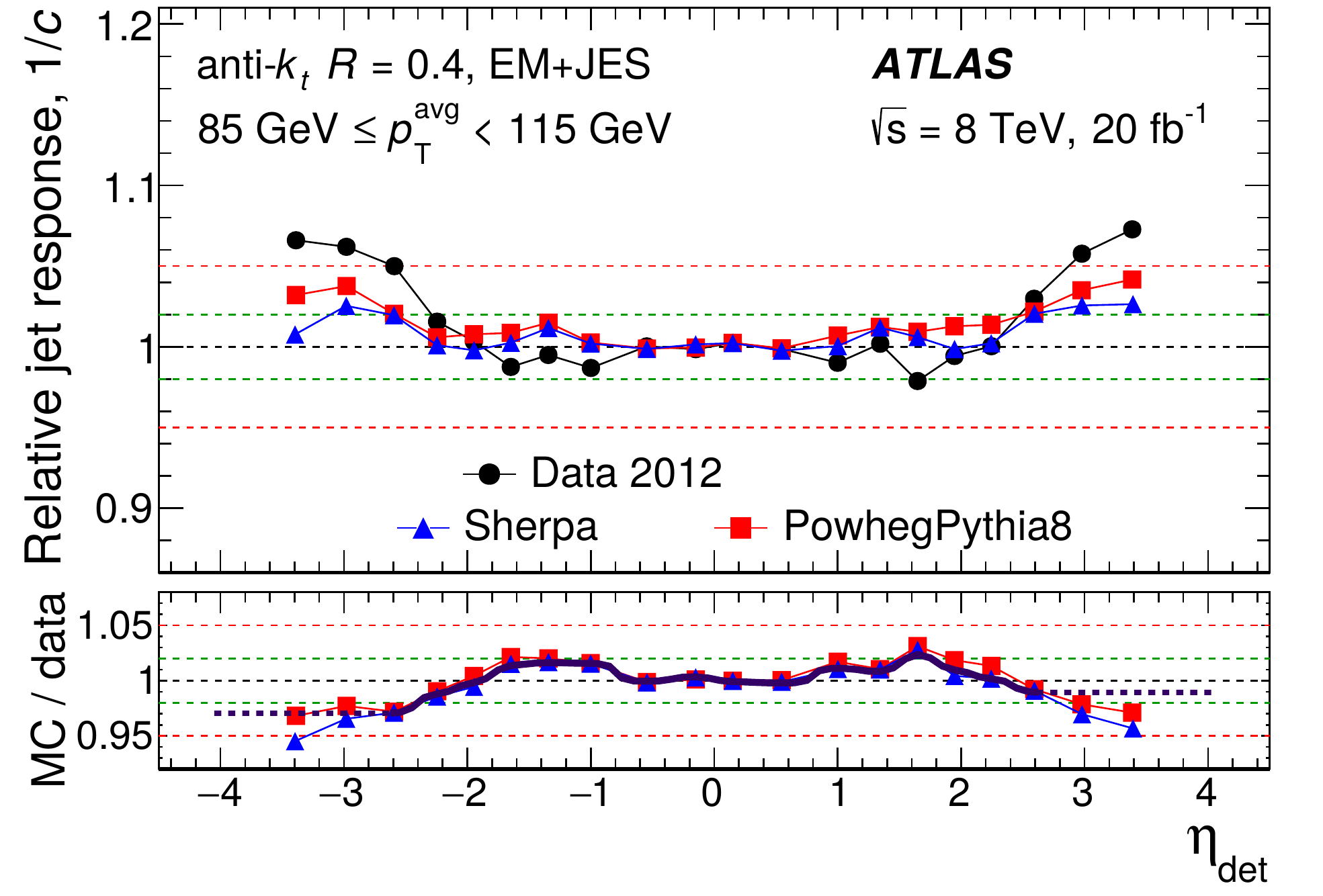}\\[1mm]
\includegraphics[width=.48\textwidth]{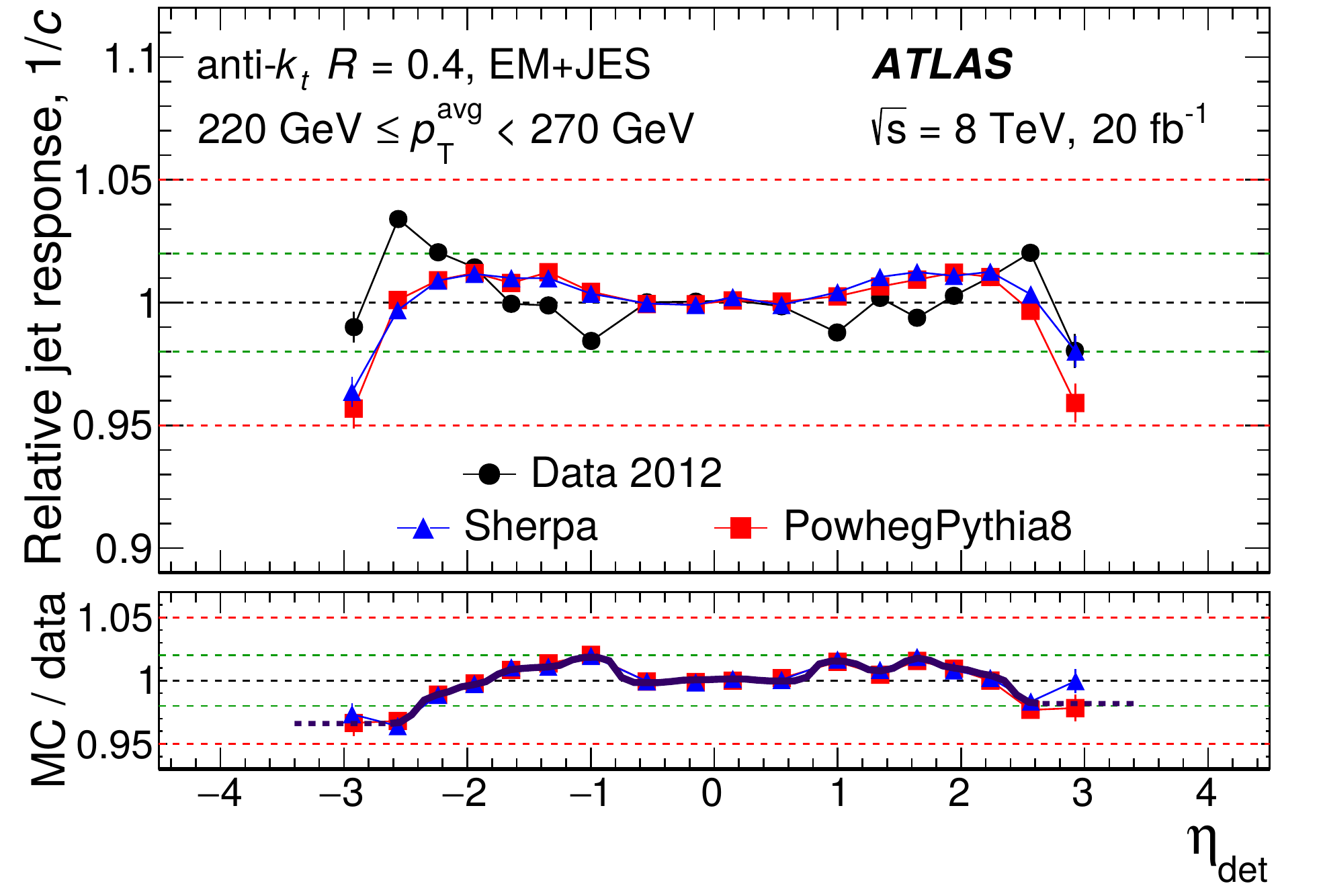}\quad
\includegraphics[width=.48\textwidth]{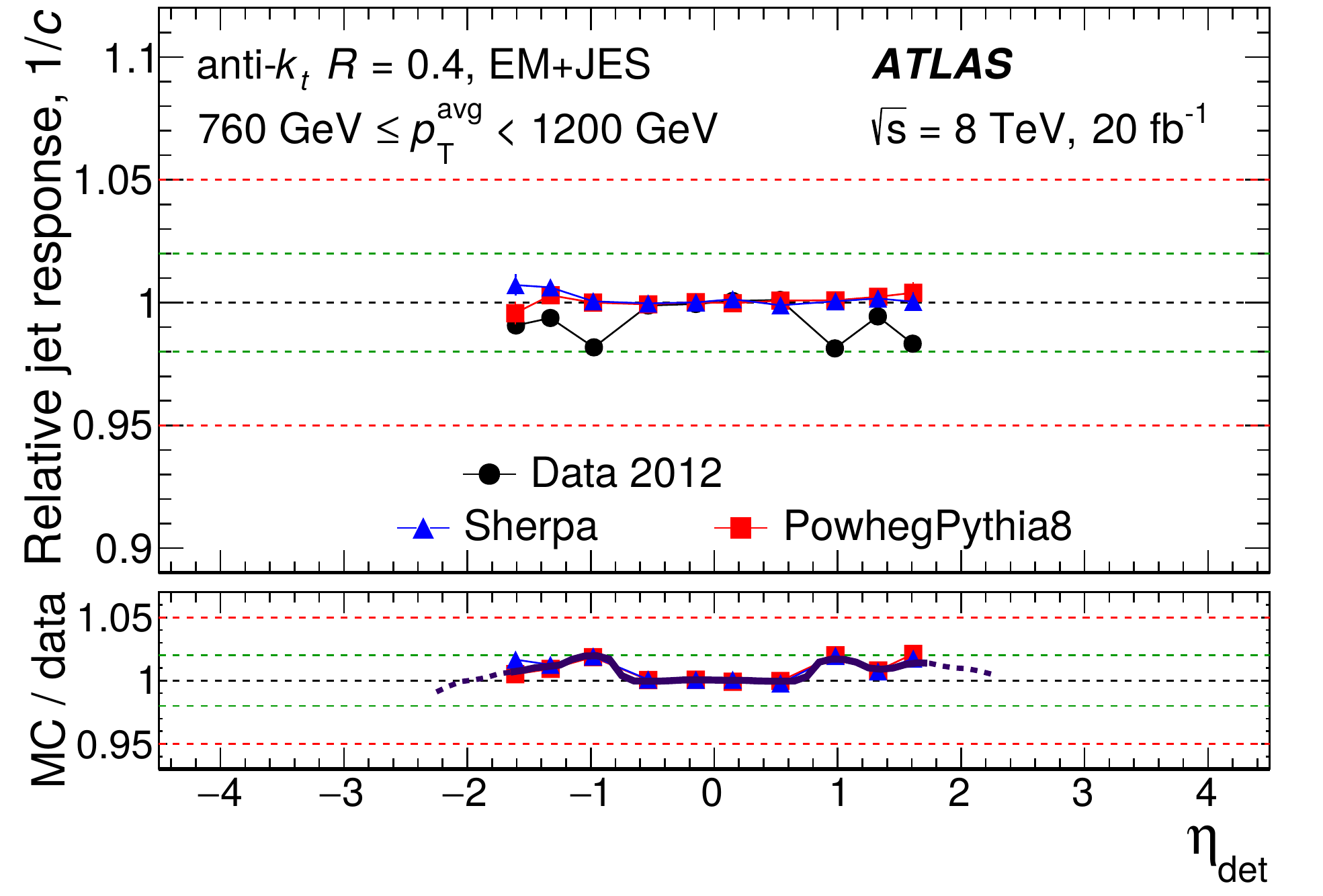}
\caption{
Relative jet response, $1/c$, as a function of the jet
pseudorapidity for \antikt{} jets with $R=0.4$ calibrated with the EM+JES scheme, for data~(black circles), \sherpa~(blue triangles) and \powhegpyt~(red squares).
Results are shown separately for
25~\GeV~$<\ptavg{}<40$~\GeV, 85~\GeV~$<\ptavg{}<115$~\GeV,
220~\GeV~$<\ptavg{}<270$~\GeV\ and 760~\GeV~$<\ptavg{}<1200$~\GeV\ with associated statistical uncertainties.
The lower part of each figure shows the ratio of relative response in MC simulation to that in data,
while the thick line indicates the resulting residual correction.
The dashed part of this line represents the extrapolation of the ratio into regions which are either statistically limited or probe $|\etaDet|>2.7$.
These measurements are performed using the matrix method. The dashed lines in the lower panels indicate $1\pm 0.02$ and $1\pm 0.05$.
\label{fig:resp_vs_eta}
}
\end{figure}
 
\begin{figure}
\centering
\includegraphics[width=.48\textwidth]{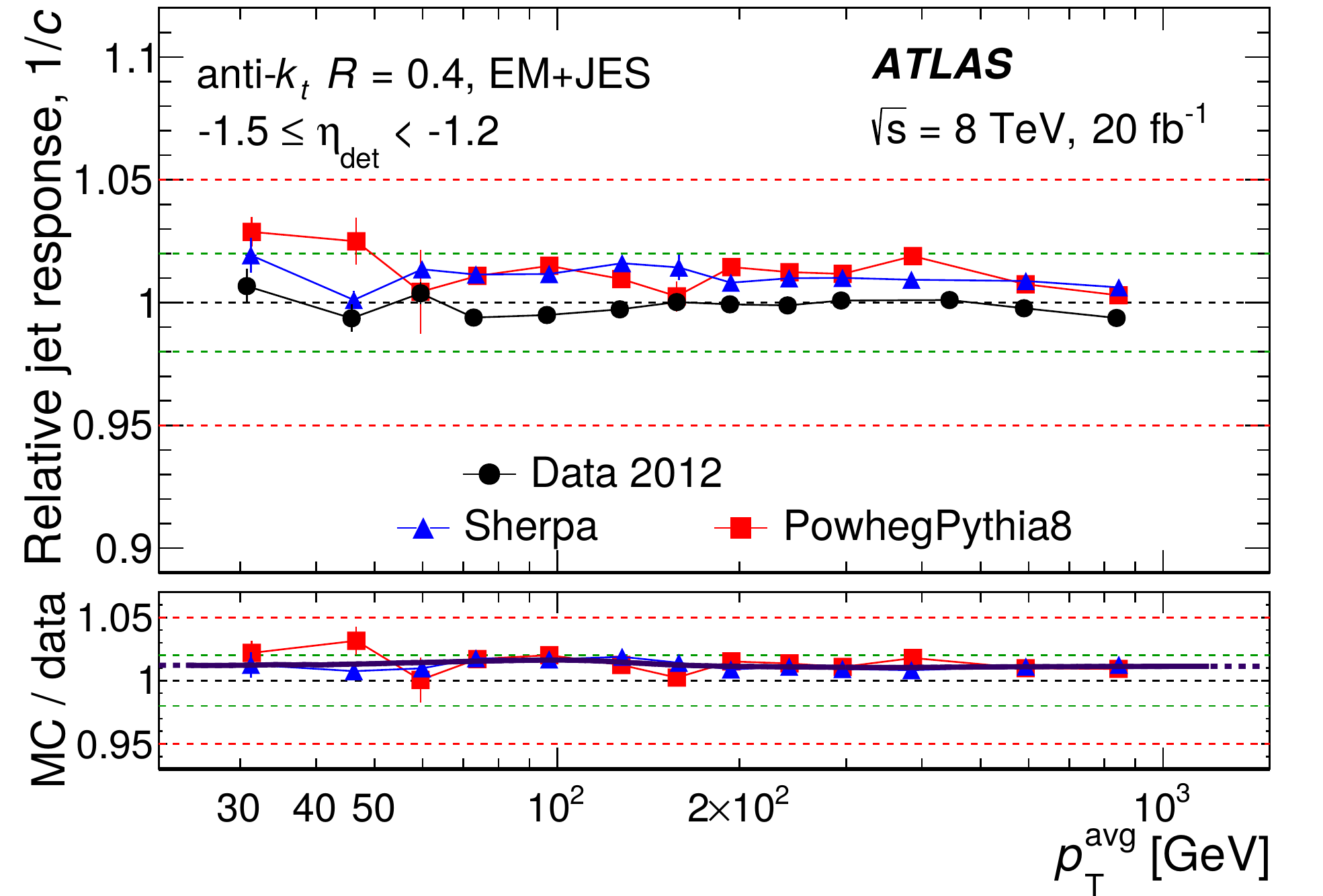}\quad
\includegraphics[width=.48\textwidth]{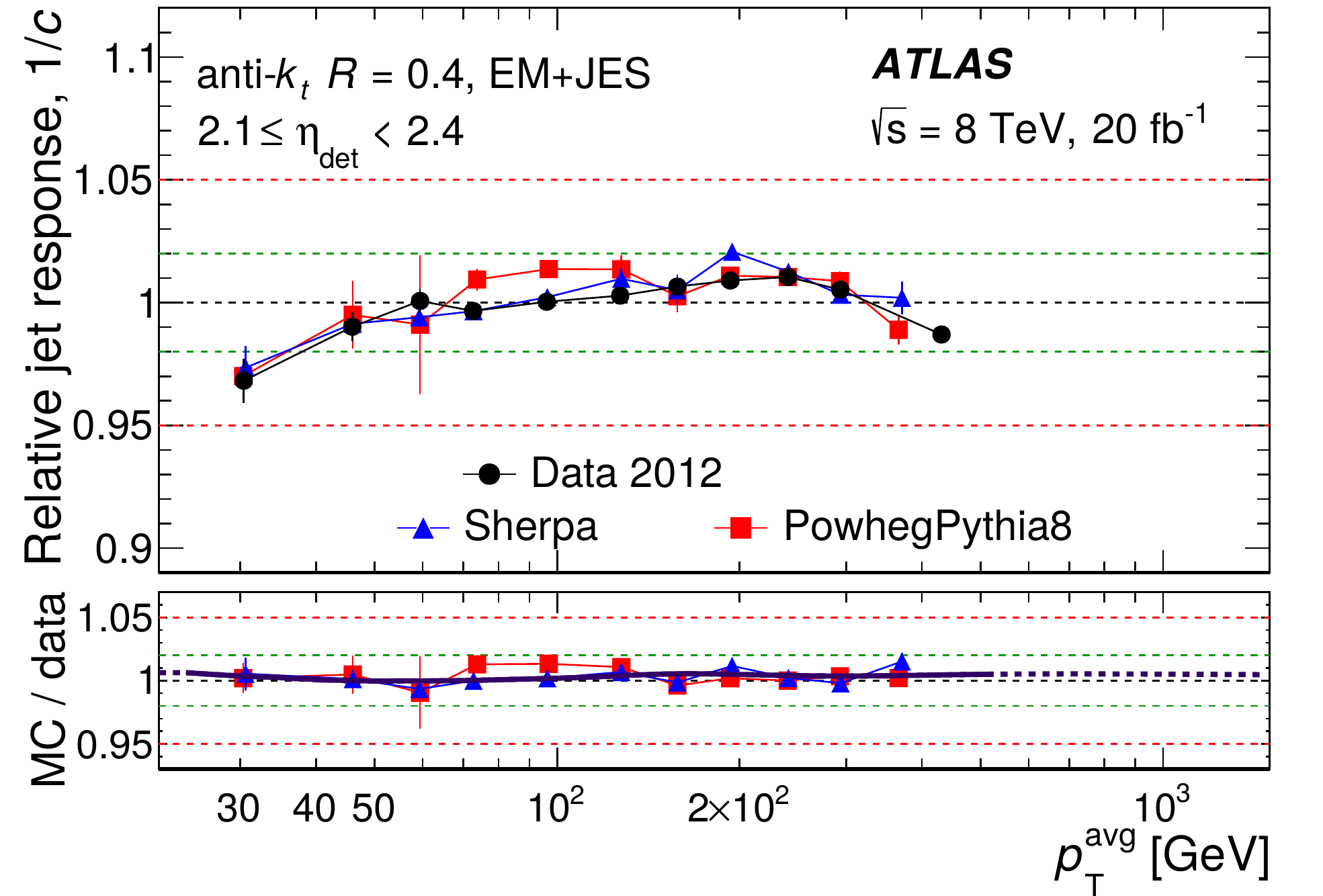}
\caption{
Relative jet response, $1/c$, as a function of the jet \pt{}
for \antikt{} jets with $R=0.4$ calibrated with the EM+JES scheme, separately for $-1.5\leq\etaDet<-1.2$ and $2.1\leq\etaDet<2.4$,
for data~(black circles), \sherpa~(blue triangles) and \powhegpyt~(red squares). The lower parts of the figures show the ratios of MC simulation to data relative response,
while the thick line indicates the resulting residual correction.
The dashed part of this line represents the extrapolation of the ratio into regions which are statistically limited.
The dashed lines in the lower panels indicate $1\pm 0.02$ and $1\pm 0.05$.
\label{fig:resp_vs_pt}
}
\end{figure}
 
\subsubsection{Derivation of residual jet energy scale correction}
\label{sec:dijetcalibration}
The residual calibration factor $c_\eta$ is derived from the ratio of data and \sherpa\ $\eta$-intercalibration factors, i.e.\ $c_{\eta,i} = c^\textrm{data}_i/c^{\textsc{Sherpa}}_i$.
The calibration factors from many bins of $\ptavg$ and $\etaDet$ are combined into a smooth function using a two-dimensional
Gaussian kernel~\cite{PERF-2012-01}.
The kernel-width parameters of this function are chosen to capture the shape
of the \MtD{} ratio across \pt\ and \etaDet, and at the same time provide stability against statistical fluctuations. The resulting residual correction $c_\eta$ is shown as a black line in the
lower panels of Figures~\ref{fig:resp_vs_eta} and~\ref{fig:resp_vs_pt}.
In these panels, it can also be seen that the calibration function is fixed for $\etaDet$ and \pt{} regions that extend beyond the data measurements. The same freezing of the calibration is also done for $|\etaDet|>2.7$ since the generator dependence becomes larger in this region. Measurements in these forward regions are not used to derive the intercalibration but are used when assessing the uncertainty.

\subsubsection{Systematic uncertainties}
\label{dijetuncertainty}
 
All intercalibration systematic uncertainties are derived as a function of \pt\ and  $|\etaDet|$ with no uncertainty assigned in the reference region ($|\etaDet|<0.8$).
No statistically significant difference is observed for positive and negative $\etaDet$ for any of the uncertainties, justifying the parameterization versus $|\etaDet|$, which increases the statistical power in the uncertainty evaluation.
 
The difference between \sherpa\ and \powhegpyt\ is used to assess the physics modelling uncertainty. Both of these generators are accurate to leading order in QCD for variables sensitive to the modelling of the third jet (such as the dijet balance).
Since there is no \textit{a priori} reason to trust one generator over
the other, the difference between the two predictions is used to estimate the modelling uncertainty.
For $0.8\leq|\etaDet|<2.7$, where data are corrected to the \sherpa\ predictions,
the full difference between \powhegpyt\ and \sherpa\ is taken as the
uncertainty.\footnote{The full difference between the generators is considered the uncertainty amplitude of a two-sided systematic uncertainty. All uncertainty components discussed in this paper are treated as two-sided uncertainties.} For $|\etaDet|\ge 2.7$, where
the calibration is frozen, the uncertainty is taken as the maximum
difference between the extrapolated calibration and the prediction from either \powhegpyt{} or \sherpa{}. The use of these event generators results in a
substantial improvement in the agreement between the theoretical predictions, thus reducing the modelling-based uncertainty by a factor of approximately two relative to the previous result~\cite{PERF-2012-01}. Despite the improvement, this modelling uncertainty remains the largest systematic uncertainty in the measurement.
 
The physics modelling uncertainty in the relative response is cross-checked using \tjets{} by varying the \powhegpyt\ predictions. The QCD renormalization and factorization scales in the \powhegbox\ are each varied by factors of 0.5 and 2.0, which has a significantly smaller impact on the relative response than the difference between the \powhegpyt\ and \sherpa{} predictions.
A comparison of the relative response between the \powhegher\ sample and the \powhegpyt\ sample
is also performed and is similar to the \tjet{} relative response between  \powhegpyt\ and  \sherpa{}. The assigned uncertainty from the difference between \sherpa\ and \powhegpyt\ is a good reflection of the underlying physics modelling uncertainty.
 
The event topology selection requires $\Delta\phi_{12}>2.5$ and $p_\textrm{T}^\textrm{j3} <  0.4\,\ptavg$.
To assess the influence of these selection criteria on the MC modelling of the \pt{} balance, the residual calibration is rederived after shifting the $\Delta\phi_{12}$ selection by $\pm 0.3$ radians and the radiation criteria based on the fractional \pt{} of a potential third jet by $\pm 0.1$. The maximum difference between the rederived calibration after the up and down shifts to the nominal is taken as uncertainty.
To assess the impact of \pileup{}, the calibration is rederived in subsets split into high and low $\mu$ ($\mu < 14 $ and $\mu \geq 17 $), and high and low $N_\textrm{PV}$ subsets ($N_\textrm{PV}<9$ and $ N_\textrm{PV} \geq 11$). The uncertainty due to \pileup{} effects is taken to be the maximum fractional difference between the varied and nominal calibrations.
Similarly, an uncertainty due to the JVF requirement is derived by redoing the calibration after tightening and loosening the JVF criteria
following the procedure defined in Ref.~\cite{PERF-2014-03}. These variations account for the
extent to which JVF is mis-modelled for jets originating from the primary interaction vertex.
An uncertainty due to imperfect modelling of the jet energy resolution is also assigned by smearing the jet four-momenta in MC simulation using Gaussian random sampling with a standard deviation calculated from the JER data-to-MC difference.
The difference between the calibrations obtained with nominal and smeared simulation is taken as the uncertainty due to JER effects.
 
The total systematic uncertainty is the sum in quadrature of the various components mentioned.
Figure~\ref{fig:uncertainty} presents a summary of the uncertainties as a function of $|\etaDet|$
for two representative values of jet transverse momentum, namely $\pt = 35$~\GeV\ and $\pt = 300$~\GeV.
The uncertainties have a strong pseudorapidity dependence, increasing with $\etaDet$,
and have a weaker \pt{}-dependence, decreasing with increasing jet \pt{}.
 
\begin{figure}
\centering
\includegraphics[width=.48\textwidth]{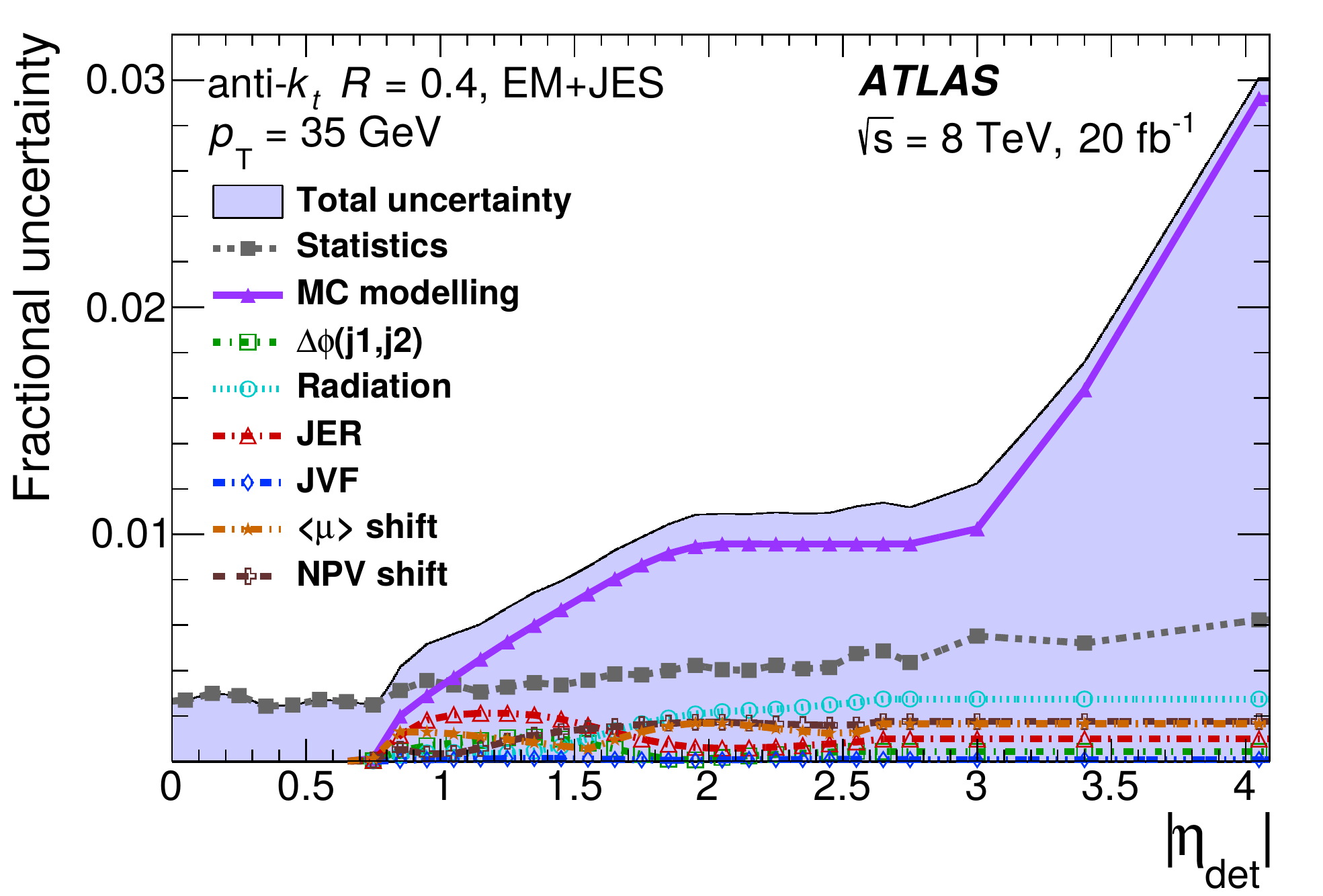}\quad
\includegraphics[width=.48\textwidth]{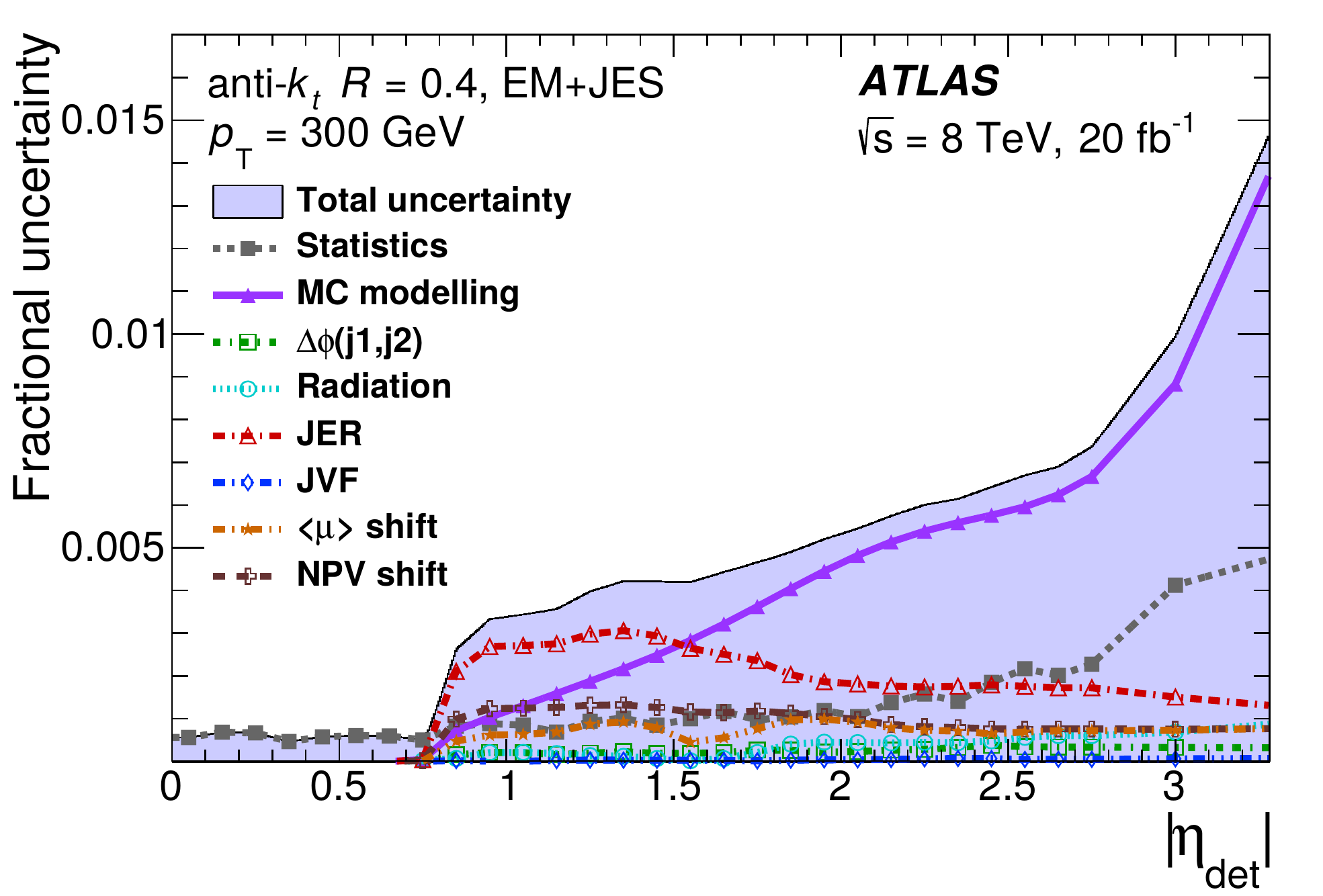}
\caption{
Summary of uncertainties in the intercalibration  as a function of the jet \etaDet{}
for \antikt{} jets with $R=0.4$ calibrated with the EM+JES scheme, separately for
$\pt{} = 35$~\GeV\ (left) and $\pt = 300$~\GeV\ (right). The individual components are added in quadrature
to obtain the total uncertainty. The MC modelling uncertainty is the dominant component for jets with $|\etaDet{}|>1.5$.
\label{fig:uncertainty}
}
\end{figure}

\subsection{Jet energy resolution determination using dijet events}
\label{sec:dijetJER}
 
Figure~\ref{fig:JER_vs_eta} shows the measured relative jet energy resolution as a function of \ptavg{} for EM+JES calibrated jets in different \etaDet{} regions of the calorimeter. The results are presented for both the dijet balance and bisector methods, and there is good agreement between the methods for all values of \ptavg{} and $|\etaDet|$. The jet energy resolution in simulated events, determined as described in Section~\ref{sec:jetMatch}, is also shown as a dotted line and is in agreement with the measured JER in data.
 
\begin{figure} [t]
\centering
\includegraphics[width=.47\textwidth]{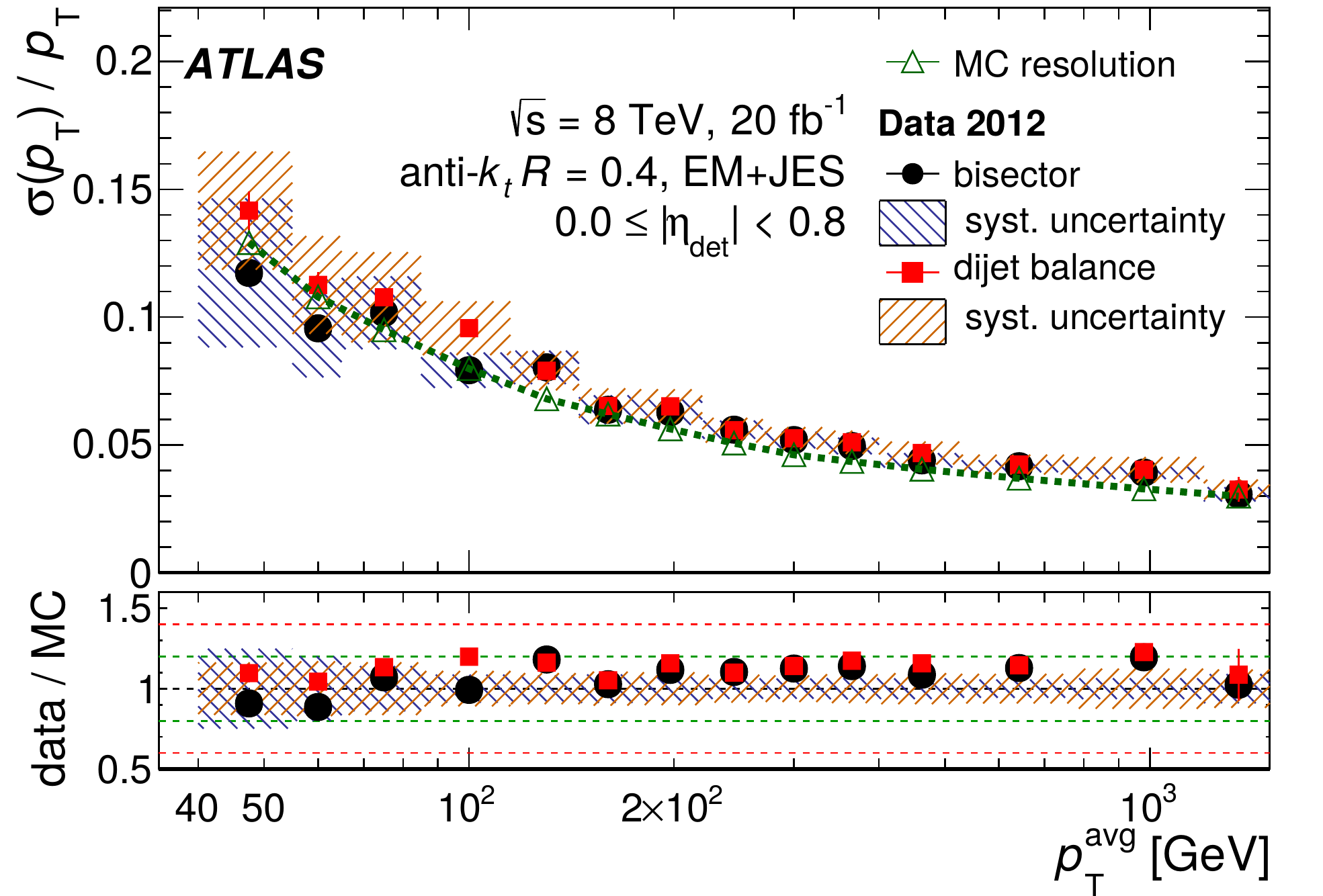}\quad
\includegraphics[width=.47\textwidth]{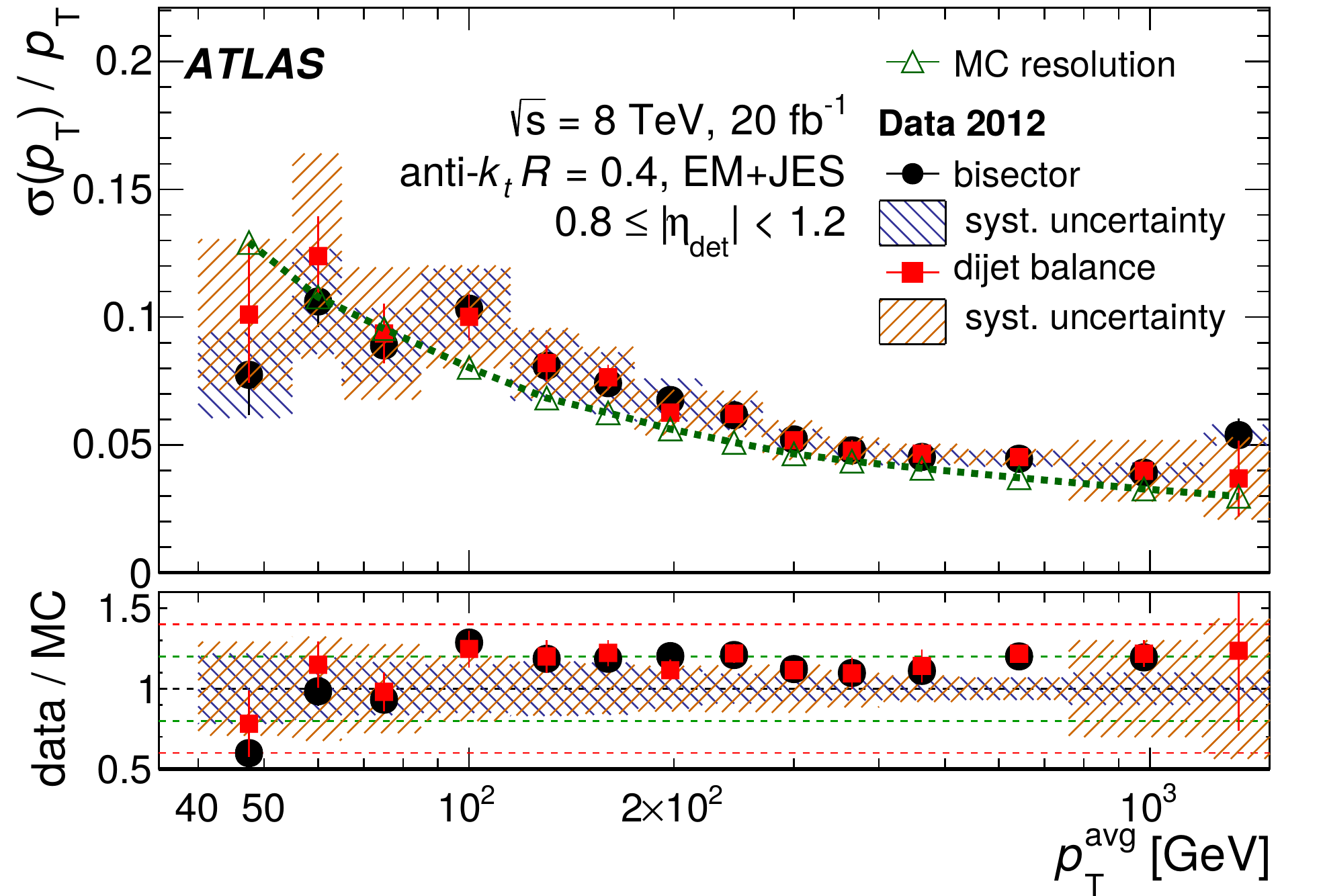}\\[1mm]
\includegraphics[width=.47\textwidth]{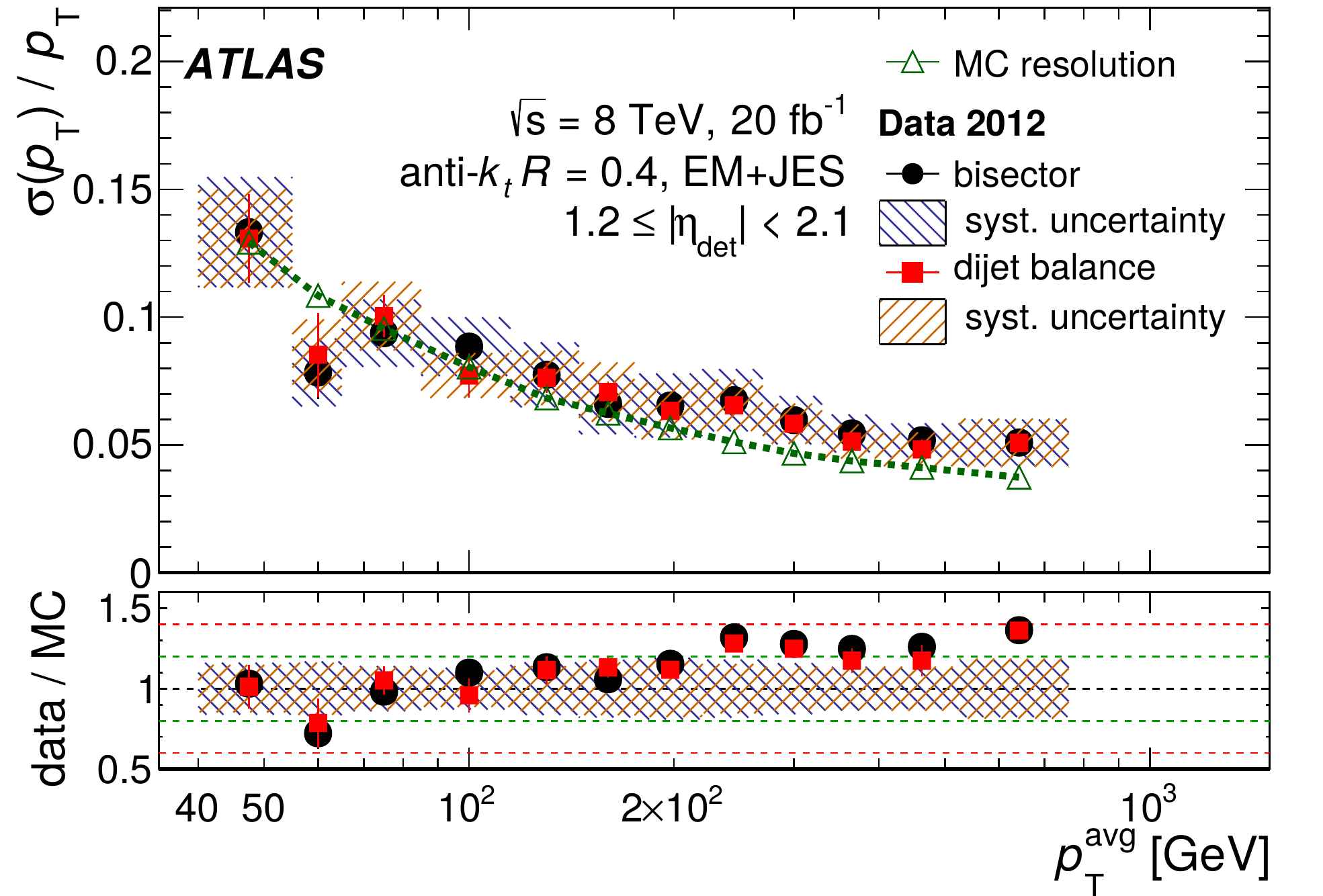}\quad
\includegraphics[width=.47\textwidth]{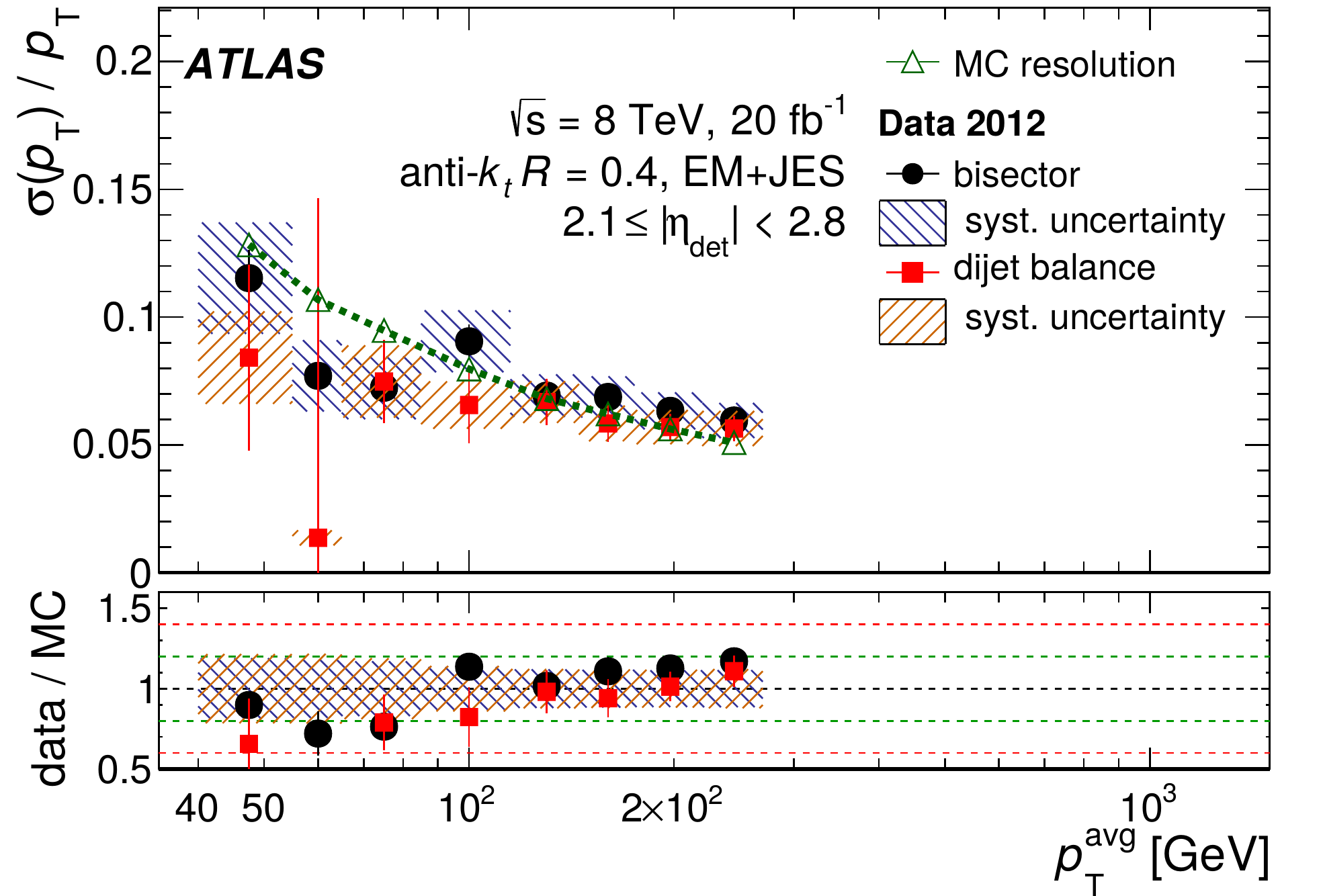}
\caption{Relative jet energy resolution obtained for EM+JES calibrated jets using the bisector~(filled circles) and dijet balance~(filled squares) methods, respectively. The MC simulated resolution derived from matching \tjets{} with calorimeter jets is presented by the open triangles connected by dashed lines. The error bars reflect the statistical uncertainty while the hashed band indicates the total systematic uncertainty.
Results are shown as a function of the jet \pt{} in four regions of detector pseudorapidity: $|\etaDet|<0.8$, $0.8 \leq |\etaDet|<1.2$, $1.2 \leq |\etaDet|< 2.1$, and $2.1 \leq |\etaDet|<2.8$.
The lower panels show the \DtM{} ratio, and the thin dashed lines indicate
$1\pm 0.2$ and $1\pm 0.4$.
}
\label{fig:JER_vs_eta}
\end{figure}
 
\subsubsection{Systematic uncertainties}
 
The JER is determined in data by subtracting the \tjet{} asymmetry from the measured asymmetry as discussed in Section~\ref{sec:dijet-balance}. The \tjet{} asymmetry is defined as the weighted average of the \tjet{} asymmetries obtained for each of the \sherpa{}, \powhegpyt{}, \pythia{}, and \herwigpp{} event samples. The uncertainty in this weighted average is taken to be the RMS deviation of the \tjet{} asymmetries obtained from the four event generators. This source of uncertainty is typically 0.02 at low \ptavg{} for both methods, falling to less than 0.01 at the highest \ptavg{}.
 
Non-closure is defined as the difference between the jet resolution measured by the \insitu{} method and the \tjet{} resolution obtained by matching truth-particle and calorimeter jets (Section~\ref{sec:jetMatch}). This is treated as a systematic uncertainty in the method.  The weighted average of the \tjet{} asymmetries predicted by \sherpa, \powhegpyt, \pythia, and \herwigpp\ is subtracted in quadrature from the weighted average of the asymmetries evaluated for reconstructed calorimeter jets. The non-closure is typically about 10\%--15\% for the bisector method, but it is larger for the dijet balance method, reaching $25\%$ in some regions.
 
Finally, there are a number of systematic uncertainties that arise from experimental sources. The uncertainty in the JES calibration is investigated by shifting the energy of the jets by the $\pm1\sigma$ uncertainty, with a typical effect between 5\% and 10\% at low \ptavg{}. The uncertainty due to the choice of JVF selection has a less than $2\%$ effect for both methods. The uncertainty due to the criterion on the azimuthal angle between the jets is investigated by changing the requirement by $\pm0.3$, with a negligible effect at high \ptavg{} for both methods, a small ($<4\%$) effect on the dijet balance results at low \ptavg{} and a larger effect (5\%--15\%) on the bisector results at low \ptavg{}. The impact of the veto on the third jet is investigated by changing the selection criteria by $\pm4$~\GeV, and is found to have a 10\%--15\% effect at low \ptavg{} for both methods, falling to a few percent at higher \ptavg{} values.
 
The total systematic uncertainty, which is taken as the sum in quadrature of all sources discussed above, is shown as a dashed band around the points in Figure~\ref{fig:JER_vs_eta}.
% End of text imported from the .//sections/dijet_calib.tex input file
 
\afterpage{\clearpage}
% The next lines are included from the .//sections/vjets_dag.tex input file
\newpage
\section{Calibration and resolution measurement using \texorpdfstring{\gamjet{}}{photon+jet} and \texorpdfstring{\zjet{}}{\Zboson+jet} events}
\label{sec:vjets}
 
This section describes the determination of the final jet calibration that corrects the absolute energy scale of the jets to achieve a \DtM{} agreement within the associated uncertainties.
The jet calibration is based on measurements conducted by \insitu{} techniques that exploit the transverse momentum balance between a well-calibrated object and the hadronic recoil (jet).
The well-calibrated object is either a photon or a \Zbos{} that decays leptonically, either \Zee{} or \Zmumu{}.
Three separate datasets are used: (\Zee{})+jet, (\Zmumu{})+jet and \gamjet{}, and two different \insitu{} techniques are used for each dataset, namely the \textit{direct balance technique} (\DB{}) and the \textit{missing projection fraction method} (\MPF).
The three independent datasets and the two analysis methods provide six separate measurements of the jet calorimeter response that can be cross-checked with each other, allowing detailed studies of systematic uncertainties. For each dataset, the method that gives the smallest overall uncertainty is chosen and is used as input to the final combination of the absolute jet calibration (Section~\ref{sec:JEScomb}).
 
Due to the steeply falling \Zbos{} \pt{} spectrum, the \Zjet{} data provide sufficient statistics to calibrate jets at lower \pt{} and are used in the range \ptRange{17}{250}.
The \Zbos{} four-momentum is reconstructed by four-vector addition of its decay products (leptons).
The \gammajet{} process has a higher cross section and covers the jet \pt{} range  \ptRange{25}{800}.
However, at low \pt{} the photon sample has a large contamination from events that do not contain any true prompt photon and hence a sizeable systematic uncertainty.
As discussed in Section~\ref{subsec:insituCombination}, a combination of both the \Zjet{} and \gammajet{} channels covers the full momentum range \ptRange{17}{800}.
 
\subsection{The direct balance and missing projection fraction methods}
\label{sec:vjets-method}
 
Both the DB and MPF methods exploit the momentum balance in events with \gamjet{} or \Zjet{} topology to study the jet calorimeter response.
Both methods benefit from accurate knowledge of the energy scale and resolution of the boson (i.e.\ the photon or the dilepton system).
The calibration of electrons and photons is accurately known through measurements using $\Zboson \to ee$ data and other final states~\cite{PERF-2010-04},
while the muon reconstruction is determined to high precision through studies of $J/\Psi \to \mu\mu$, $\Zboson \to\mu\mu$, and $\Upsilon\to\mu\mu$~\cite{PERF-2014-05}.
 
The DB response \RDB{} is
\begin{eqnarray}
\RDB = \left<\ptjet\,/\,\ptref\right>, &\mbox{ where } \ptref = \pt^{\Zgam}\,\left|\cos{\Delta\phi}\right|,
\label{eq:RDB}
\end{eqnarray}
where \ptjet{} is the \pt{} of the leading jet being probed and $\Delta\phi$ is the azimuthal angle between this jet and the boson (\Zboson\ or $\gamma$).
If the jet includes all the particles that recoil against the \Zbos{} or $\gamma$ and all particles are perfectly measured, then $\ptjet\,/\,\ptref=1$ and $\cos{\Delta\phi}=\cos{\pi}=-1$.
In reality, there is always additional QCD radiation not captured by the jet, which skews the balance.
This radiation, referred to as out-of-cone radiation (OOC), tends to be in the same hemisphere as the jet and hence biases the \DB{} to values below unity.
The reference transverse momentum \ptref{} used in the denominator of $\RDB$ is the boson momentum projected onto the jet axis in the transverse plane in order to attempt to at least partially reduce OOC effects. The \DB{} is also affected by uncertainties in the reconstructed photon, electron, or muon momenta, as well as contributions from \pileup{} and multiple parton--parton interactions (the underlying event).

The \MPF{} method~\cite{Abbott:1998xw,PERF-2011-03} is an alternative to the \DB{} technique. Rather than balancing the jet object itself against the well-measured boson, the whole hadronic recoil is used.
The \MPF{} measures the response for the full hadronic recoil, which is significantly less sensitive to OOC radiation and effects due to \pileup{} and the underlying event.
The logic of the \MPF{} method is detailed below for \gammajet{} events.  The case of \Zjet{} is the same with the \Zbos{} replacing the photon.
 
From conservation of transverse momentum, the \pt{} vector of the system of all hadrons produced in a $\gamma$+jet event, $\ptrecvec{}$, will perfectly balance the photon $\ptgammavec{}$ at the truth-particle level.
In a perfect $2\to 2$ process, \ptrecvec{} would be equal to the \ptvec{} of the parton, which in turn is that of the jet. At reconstruction level, the \pt{} of the photon (or \Zboson\ boson) is well calibrated and hence accurately reconstructed, while the hadronic response is low prior to calibration, primarily due to the non-compensating nature of the ATLAS calorimeters\footnote{The hadronic recoil is reconstructed at the constituent scale, for which the calorimeter response can have a significant energy dependance as can be seen in Figure~\ref{fig:calibcurve}.}.
There is hence a momentum imbalance, which defines the missing transverse momentum \vmet{}:
\begin{eqnarray}
\label{eq:ptclBalance}
\vec{p}^{\,\gamma}_\text{T,truth} + \vec{p}^\text{\,recoil}_\text{T,truth}  = \vec{0} & & (\text{truth-particle level})  \nonumber \\
\label{eq:recoBalance}
\ptgammavec + \ptrecvec + \vmet = \vec{0} & & (\text{detector level}).
\end{eqnarray}
Projecting the vector quantities of Eq.~(\ref{eq:recoBalance}) onto the direction of the photon $\hat{n}_{\gamma}$ and dividing the result by $p_\textrm{T}^{\gamma}$ gives the MPF observable $r_\text{MPF}$, whose mean is the MPF response $R_\text{MPF}$, where
\begin{eqnarray}
r_\text{MPF} = -\frac{\hat{n}_\gamma\cdot\vec{p}_\text{T}^\text{\,recoil}}{p_\text{T}^\gamma} = 1 + \frac{\hat{n}_{\gamma} \cdot \vmet}{\ptgamma} \qquad \textrm{ and } \qquad
R_\text{MPF} = \left<1 + \frac{\hat{n}_{\gamma} \cdot \vmet}{\ptgamma}\right>.
\label{eq:MPF}
\end{eqnarray}
The \met{} definition used in Eqs.~(\ref{eq:recoBalance}) and (\ref{eq:MPF}) is based on the calibrated momentum of the photon (or dilepton system for \Zjet\ data) using \topos{} at the constituent scale, either at the EM-scale when studying the EM+JES calibration or at the LCW-scale for the LCW+JES calibration.
Details of the ATLAS \met{} reconstruction are in Ref.~\cite{PERF-2014-04}.
 
The MPF response $R_\text{MPF}$ provides a measure of the \pt{} response of the calorimeter to the hadronic recoil for a given $p_\text{T}^\gamma$.
A feature of this method is that it is almost independent of the jet algorithm as the jet definition enters only in the event selection criteria applied (Section~\ref{sec:vjets-selection}).
Except for two relatively small corrections known as the \textit{topology} and \textit{showering corrections} (Section~\ref{sec:vjets-OOC-syst}), the $R_\text{MPF}$ determined in \gamjet{} or \Zjet{} events can be used as an estimator of the calorimeter jet response at \pileup{}-subtracted scale (Section~\ref{sec:pileupCorr}).
This is because \pileup{} is independent of both the hard interaction and the azimuth $\phi$, and so its contribution to $\hat{n}_\gamma \cdot \met{}$ will be zero on average, meaning that
$R_\text{MPF}$ already effectively subtracts the \pileup{} as is done for jets using Eq.~(\ref{eq:PU}).
Since $R_\text{MPF}$ is an approximation of the \pileup{}-subtracted jet response, it can be compared with the corresponding quantity of the MC-derived calibration in Figure~\ref{fig:calibcurve} that defines $1/c_\text{JES}$.

The $R_\text{DB}$ and $R_\text{MPF}$ parameters are determined in bins of \ptref{} (Eq.~(\ref{eq:RDB})) from the mean parameter extracted from fits to the balance distributions ($\ptjet/\ptref$ and $r_\text{MPF}$) using a \textit{Modified Poisson} distribution, which was also used in the previous ATLAS jet calibration~\cite{PERF-2012-01}.
This distribution starts from a standard Poisson distribution $f_\text{P}(n;\nu)$ and is extended to non-integer values using a Gamma function $\Gamma(n+1)$, followed by the introduction of a new parameter $s$ used to redefine the argument using $x=s^2\,n$ and defining $\mu \equiv E[x] = s^2\,\nu$, giving
\begin{equation}
f_\text{MP}(x;\mu,s) = \frac{(\mu/s^2)^{x/s^2}}{\Gamma(x/s^2+1)} \frac{\text{e}^{-\mu/s^2}}{s^2} . 
\label{eq:MP}
\end{equation}
This distribution has the same shape as a ``smoothed'' Poisson distribution with $\nu = \mu/s^2$ and has mean $\mu$ and standard deviation $\sqrt{\mu}\,s$.
For larger values of $\mu/s^2$ ($\gtrapprox 15$), it is very similar to a Gaussian distribution, while for lower values ($\mu/s^2\lessapprox 5$) the longer upper tail of a Poisson distribution is prominent.
The Modified Poisson function better describes the balance distributions and is motivated by the Poisson nature of sampling calorimeters.

The \MPF{} and \DB{} methods probe the calorimeter response to jets in a different way and are sensitive to different systematic effects.
They therefore provide complementary measurements of the jet-energy scale.
The explicit use of jets in the measurement of the jet response from \DB{} makes this technique dependent on the jet reconstruction algorithm while the \MPF{} technique is mostly independent of the jet algorithm, as explained above. Thus, in the following, when presenting \MPF{} results, no jet algorithm is explicitly mentioned.

\subsection{Event and object selection
\label{sec:vjets-selection}}
 
This section outlines the event selection used  for the \DB{} and \MPF{} analyses separately for the \gamjet{} and \Zjet{} datasets.
The two methods have similar selections, but the restriction on the subleading jet \pt{} (Section~\ref{sec:vjets-jet-sel}) is less stringent for the \MPF{} method because it is less sensitive to QCD radiation.
 
\subsubsection{Photon selection}
\label{sec:gamjet-sel}
The \gamjet{} data was collected using six different single-photon triggers, each with a different associated photon \pt{} threshold.
The five lower-threshold triggers were prescaled, while only the highest-threshold trigger was not prescaled.
A given \gamjet{} event was assigned to one of these triggers, based on the \pt{} of the leading photon reconstructed by the algorithm used in the high-level trigger.
This mapping was created such that the trigger efficiency for each \pt{} range was at least 99\%.
The lowest-threshold trigger data has the largest associated prescale factor and is used for photons between 20~\GeV\ and 40~\GeV, while the highest-threshold trigger, which was not prescaled, is used for $\pt>120$~\GeV.
 
Reconstructed photons are required to satisfy strict identification criteria ensuring that the pattern of energy deposition in the calorimeter is consistent with that expected for a
photon~\cite{Aaboud:2016yuq}.
Photons are calibrated following the procedure in Ref.~\cite{PERF-2010-04} and are required to have $\pt > 25$~\GeV\ and $|\eta|<1.37$ such that they are fully contained within the electromagnetic barrel calorimeter. In order to suppress backgrounds from calorimeter signatures of hadrons misidentified as photons, an isolation criterion is further applied to all photons.
The isolation transverse energy of a photon $E_\text{T}^\text{iso}$ is calculated from calorimeter energy deposits in a cone of size $\Delta R=0.4$ around the photon, excluding the photon itself and the expected contribution from \pileup{}~\cite{Aaboud:2016yuq}.
Photons are initially required to fulfil $E_\text{T}^\text{iso}<3.0$~\GeV. However, for events where the leading photon has \pt{} below 85~\GeV, the contamination from misidentified photons is still large. Hence, more stringent criteria are applied as follows: $E_\text{T}^\text{iso}<0.5$~\GeV\ if $\pt^\gamma<45$~\GeV, or else  $E_\text{T}^\text{iso}<1.0$~\GeV\ if $\pt^\gamma<65$~\GeV, and otherwise $E_\text{T}^\text{iso}<2.0$~\GeV\ if $\pt^\gamma<85$~\GeV.
Each event is required to have at least one photon fulfilling these criteria. In the very rare case of two such photons, the leading one is used.
 
\subsubsection{\Zbos{} selection}
\label{sec:zjet-sel}
A typical \Zbos{} selection is applied, starting by requiring dilepton triggers that were not prescaled during the data-taking period.
For the \Zee{} channel, the dielectron triggers requires two ``\verb|loose|'' electrons, defined in Ref.~\cite{ATLAS-CONF-2011-114}, with $E_{\text{T}} > 12$~\GeV{} and $|\eta|<2.5$.
For the \Zmumu{} channel, the trigger requires one ``\verb|tight|'' and one ``\verb|loose|'' muon, defined in Ref.~\cite{TRIG-2012-03}, with $\pt>18$~\GeV{} and $\pt>8$~\GeV{}, respectively.
The reconstructed electrons are required to fulfil ``\verb|medium|''  quality requirements~\cite{Aaboud:2016vfy} and are calibrated as detailed in Ref.~\cite{PERF-2010-04}.
Electrons are required to have $\pt{}>20$~\GeV\ and $|\eta|<2.37$, excluding the barrel--endcap transition region $1.37<|\eta|<1.52$.
Muons are reconstructed through the combination of trajectories and energy-loss information in several detector systems~\cite{PERF-2014-05} and are required to have $|\eta|<2.5$ and $\pt>20$~\GeV.
Each event is required to have exactly two reconstructed electrons or two muons with opposite charge. The invariant mass of the dilepton system $m_{\ell\ell}$ must then fulfil 80~\GeV $<m_{\ell\ell}<116$~\GeV, ensuring high $Z\to\ell\ell$ purity.
 
\subsubsection{Jet and boson+jet topology selection}
\label{sec:vjets-jet-sel}
Jets are reconstructed and a preselection is applied, including standard \JVF{} requirements, as described in Section~\ref{sec:jetReco}.
They are calibrated with all steps prior to the absolute \insitu{} correction (Section~\ref{sec:jetCalibOverview}).
To avoid double counting of energy depositions, jets are required to be $\Delta R>0.35$ from a photon for the \gamjet{} analysis or from any of the leptons in the \Zjet{} analysis for jets reconstructed with $R=0.4$. The corresponding criterion is $\Delta R>0.5$ for $R=0.6$ jets.
 
The leading jet is required to have $|\etaDet|<0.8$ and \pt{} greater than 10~\GeV\ for the \Zjet{} analysis and 12~\GeV\ for the \gamjet{} analysis.
To enforce a ``boson+jet'' topology and hence suppress additional QCD radiation, criteria are imposed on the azimuthal angle $\Delta\phi(Z/\gamma,\text{j1})$ between the \Zbos{} or photon and the leading jet and on the subleading jet transverse momentum $p_\text{T}^\text{j2}$, if such a jet is present.
The \DB{} analysis requires  $\Delta\phi(Z/\gamma,\text{j1})>2.8$ and $\pt^\text{j2}<\text{max}(8~\GeV,0.1\,\ptref)$ while the MPF analysis uses the criteria
$\Delta\phi(Z/\gamma,\text{j1})>2.9$ and $\pt^\text{j2}<\text{max}(8~\GeV,0.3\,\ptref)$.
The subleading jet \pt{} is always defined using the jet collection reconstructed using $R=0.4$, even when studying jets built using $R=0.6$ or $R=1.0$.
For jets built using $R=1.0$, $\pt^\text{j2}$ is defined as the \pt{} of the leading $R=0.4$ jet that fulfils $\Delta R(\text{j1},\text{j2})>0.8$, where ``j1'' refers to the leading $R=1.0$ jet, i.e.\ the jet that is being probed.
This ensures that the ``j2'' jet will have a significant proportion of its energy depositions outside of the \lRjet.

\subsection{Jet response measurements using \Zjet{} and \gamjet{} data}
\label{sec:vjets-calib}
Measurements of \RDB{} and \RMPF{} using both of the individual \Zll{} datasets and the \gammajet{} dataset are discussed below.
The subsequent combination of the \Zll{} and \gammajet{} results into the final \insitu{} calibration is detailed in Section~\ref{subsec:insituCombination}.
 
The \Zee{} and \Zmumu{} analyses probe jets over the same kinematic space and use exactly the same \ptref{} binning.
Within each bin, the balance distributions \rDB{} and \rMPF{} agree between the channels for both the cores of the distributions (including their means) and their tails (including their standard deviations). The two datasets are combined into a \Zll{} channel, which increases the statistical power of the measurement.
This combination is done consistently for data and MC simulation, and also for systematic variations.

\subsubsection{Direct balance results}
\label{sec:DB-results}
Figure~\ref{fig:DB} presents four DB \rDB{} distributions in representative \ptref{} bins from the \Zjet{} and \gammajet{} analyses using
\antikt{} jets with \rfour{} calibrated with the \EMJES{} scheme.
Good fit quality using the Modified Poisson parameterization of Eq.~(\ref{eq:MP}) is observed.
This is true for the other \ptref{} bins considered, both for data and for all MC samples considered.
 
\begin{figure}[!htb]
\begin{subfigure}{0.48\linewidth}\centering
\includegraphics[width=\linewidth]{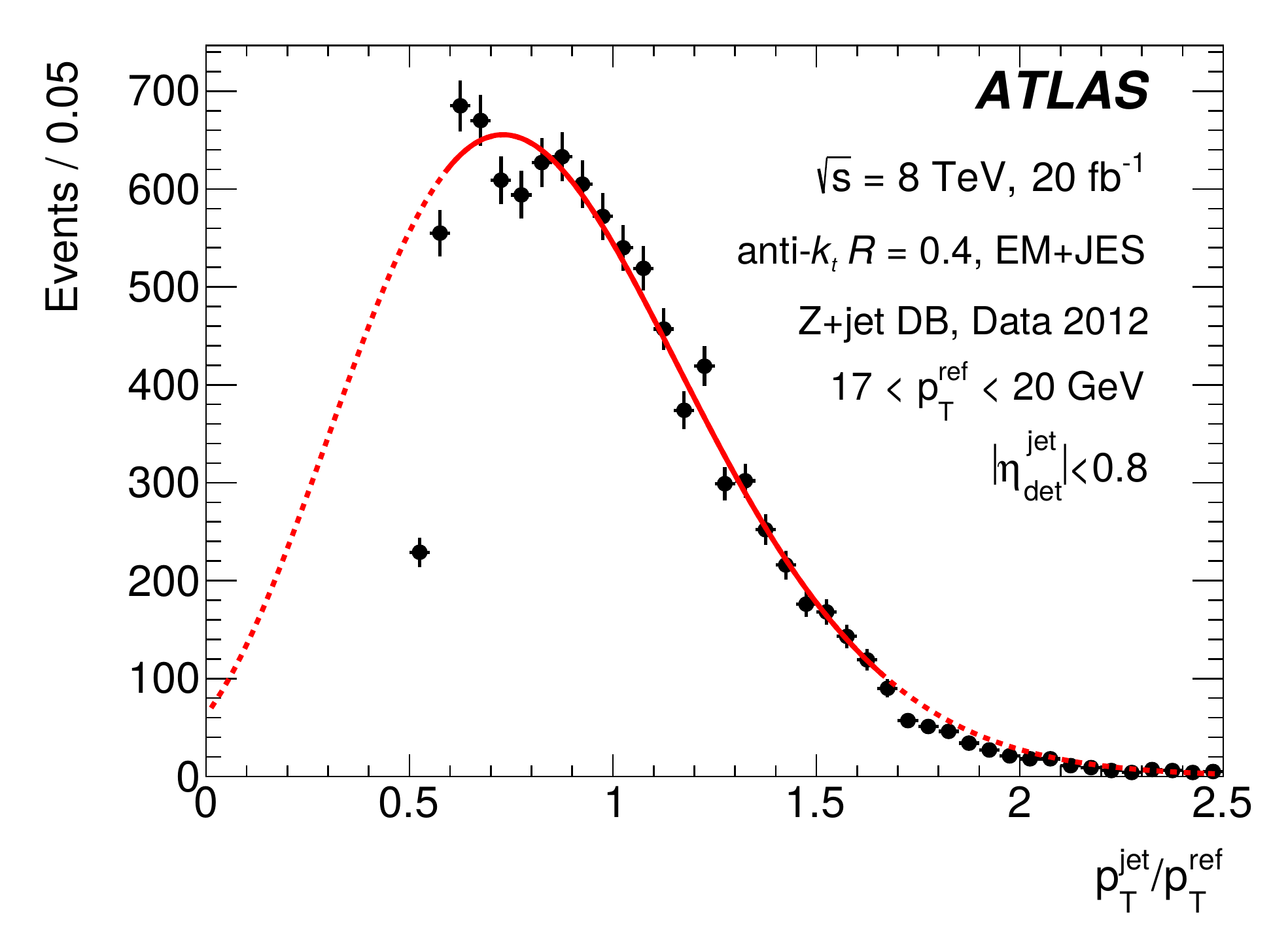}
\caption{\Zjet{}, \ptrefRange{17}{20}}
\end{subfigure}
\begin{subfigure}{0.48\linewidth}\centering
\includegraphics[width=\linewidth]{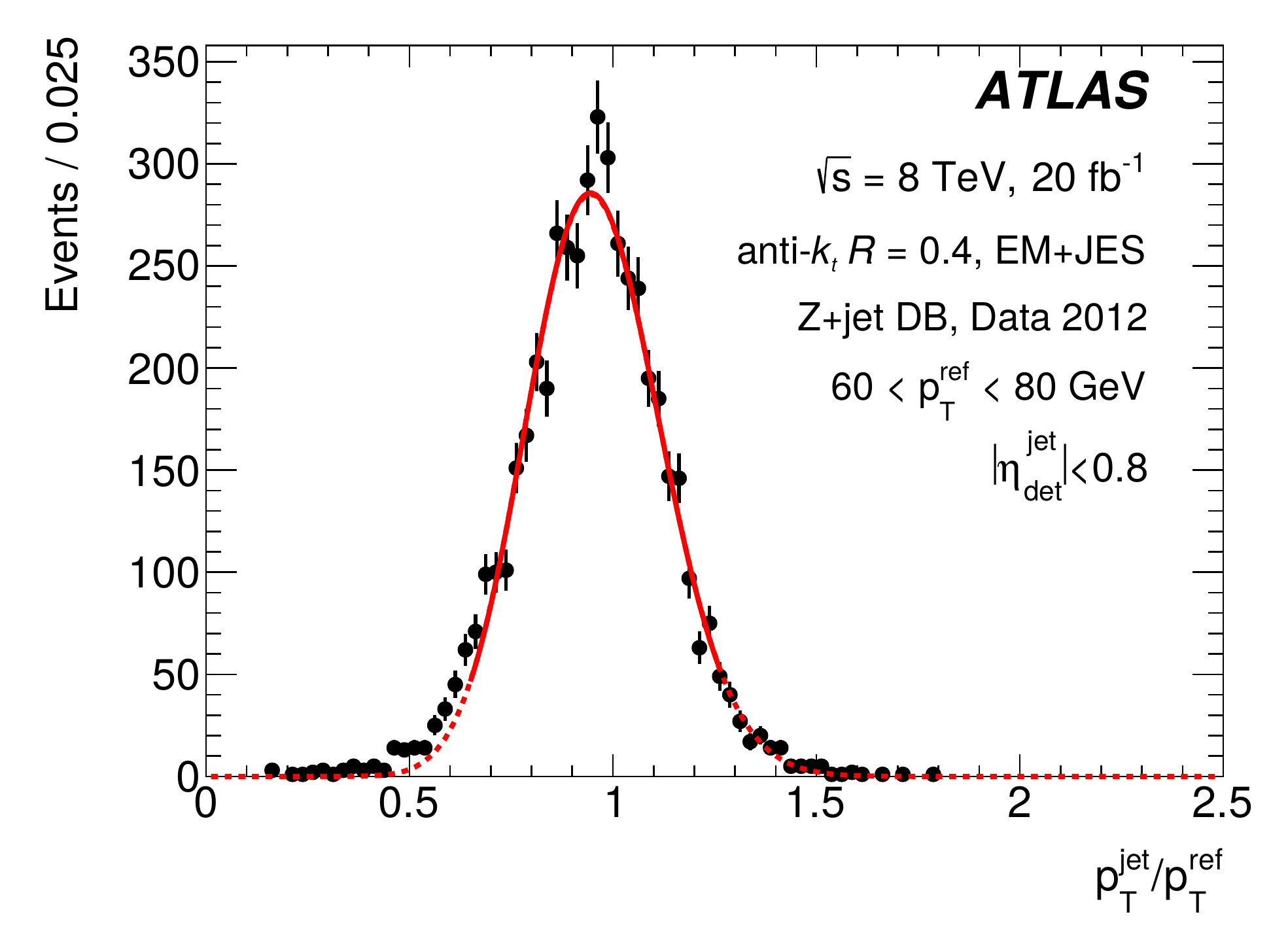}
\caption{\Zjet{}, \ptrefRange{60}{80}}
\end{subfigure} \\
 
\bigskip
 
\begin{subfigure}{0.48\linewidth}\centering
\includegraphics[width=\linewidth]{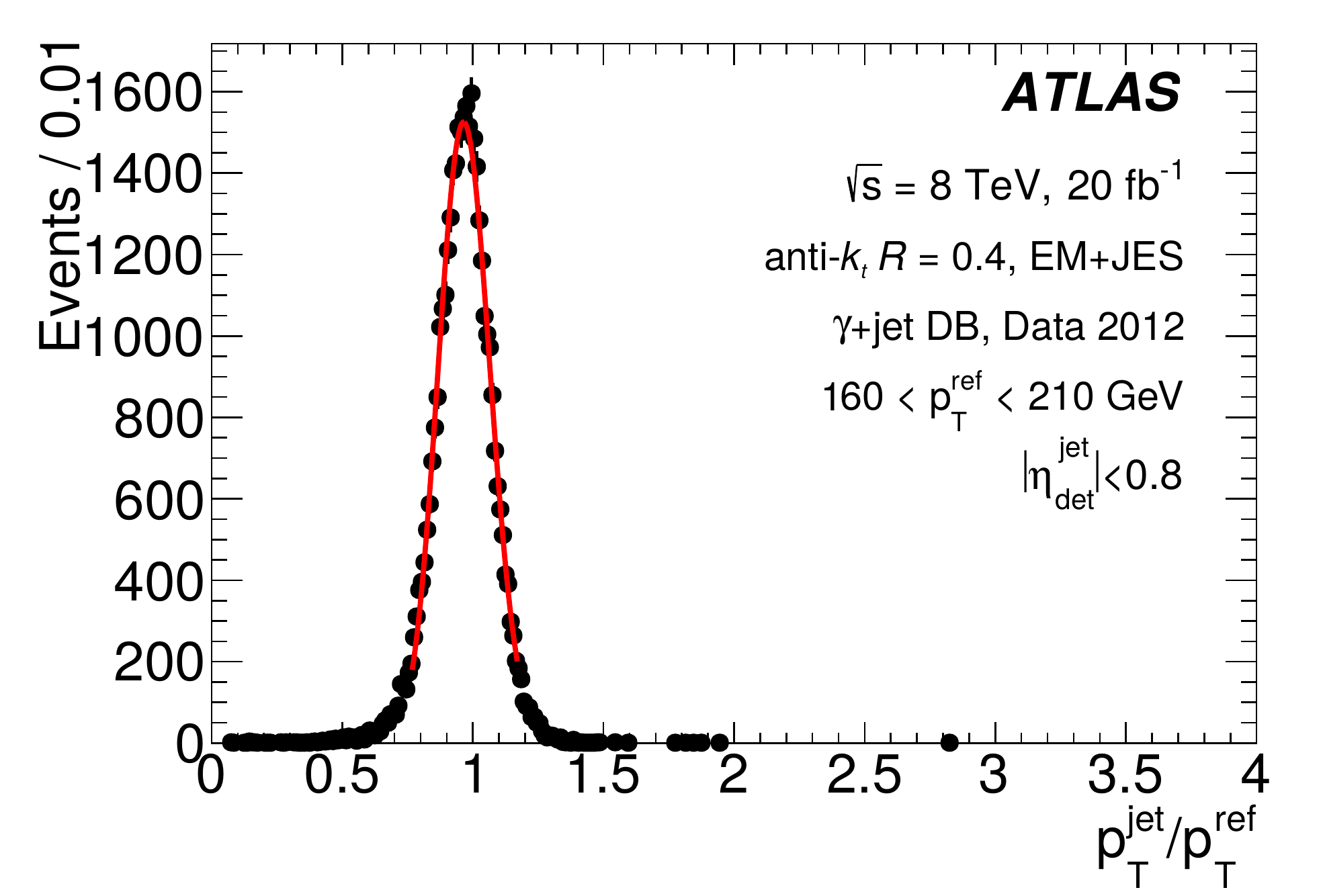}
\caption{\gammajet{}, \ptrefRange{160}{210}}
\end{subfigure}
\begin{subfigure}{0.48\linewidth}\centering
\includegraphics[width=\linewidth]{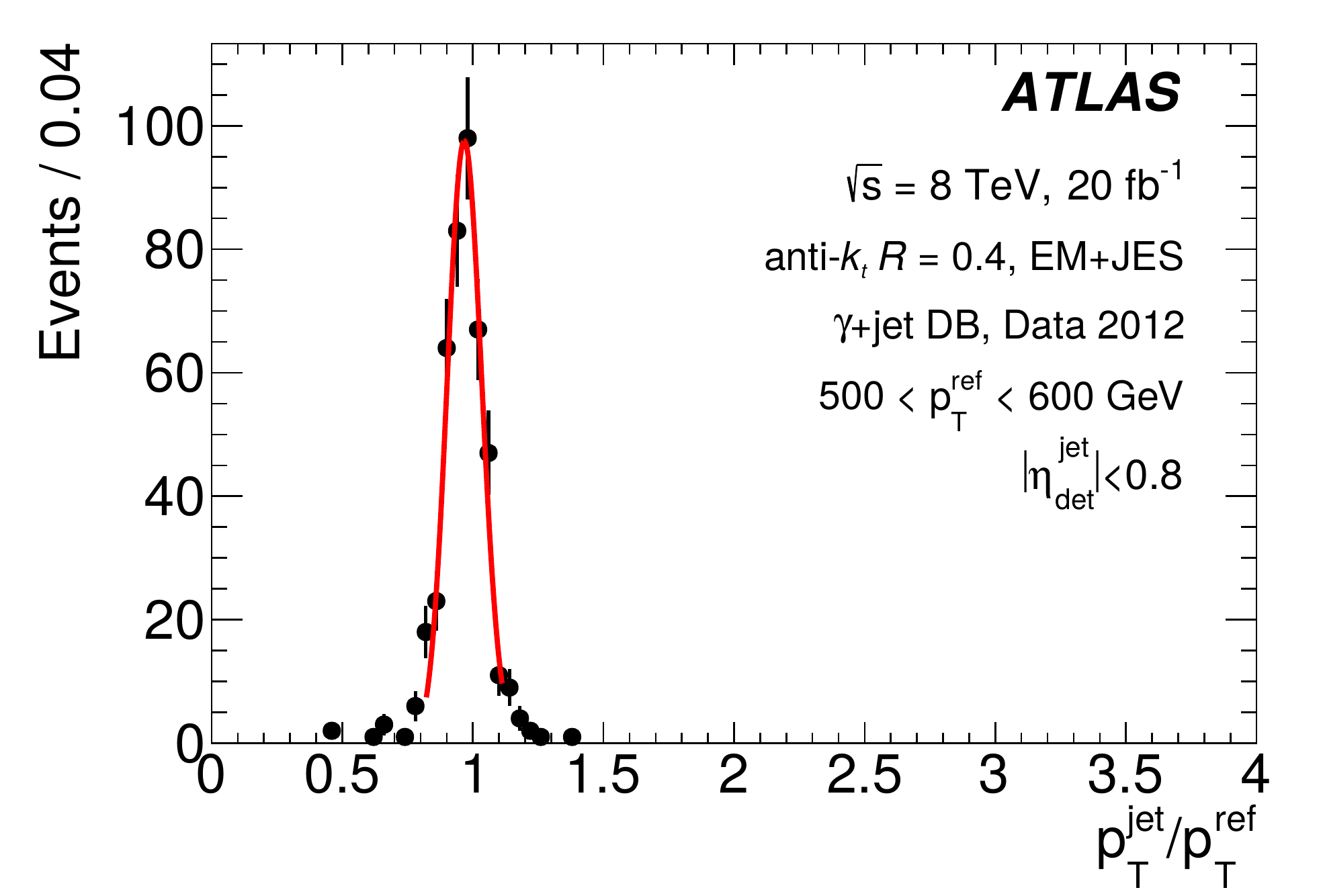}
\caption{\gammajet{}, \ptrefRange{500}{600}}
\end{subfigure}
\caption{Distributions of \rDB{} for (a)~\ptrefRange{17}{20}, (b)~\ptrefRange{60}{80}, (c)~\ptrefRange{160}{210}, and (d)~\ptrefRange{500}{600} using \antikt{} jets with \rfour{} calibrated with the \EMJES{} scheme in data from the (a,b)~\Zjet{} and (c,d)~\gammajet{} analyses. The dashed lines in (a,b) show the fitted Modified Poisson distributions of Eq.~(\ref{eq:MP}), from which the means are taken as the DB response measurements \RDB{}. The solid lines indicate the fitting ranges. The lack of data at low \rDB{} visible in (a) is due to the $\ptjet > 10$~GeV criterion.
The markers are the data counts with error bars corresponding to the statistical uncertainties.
}
\label{fig:DB}
\end{figure}
 
The value of \RDB{} is extracted for each \ptref{} bin for both data and simulation. Figure~\ref{fig:DB-vs-pTref} shows the measurements of \RDB{} in data compared with predictions from the different MC generators as a function of \ptref{} for \antikt{} \rfour{} jets calibrated with the \EMJES{} scheme.
The different MC generators agree with data within 1\% for $\pt>40$~\GeV\ with slightly worse agreement at lower \pt{}.
The worst \DtM{} agreement is for \EMJES{} calibrated \rfour{} jets in the \ptrefRange{17}{20} bin (Figure~\ref{fig:DB-vs-pTref}(a)), for which the \powheg{} MC sample predicts $\sim$5\% higher \RDB{} than what is observed in data. For LCW+JES calibrated \rfour{} jets and \rsix{} jets using both calibration schemes, the \DtM{} agreement is within 3\% across the full \ptref{} range probed.
 
\begin{figure}[!htb]
\centering
\begin{subfigure}{0.48\linewidth}\centering
\includegraphics[width=\linewidth]{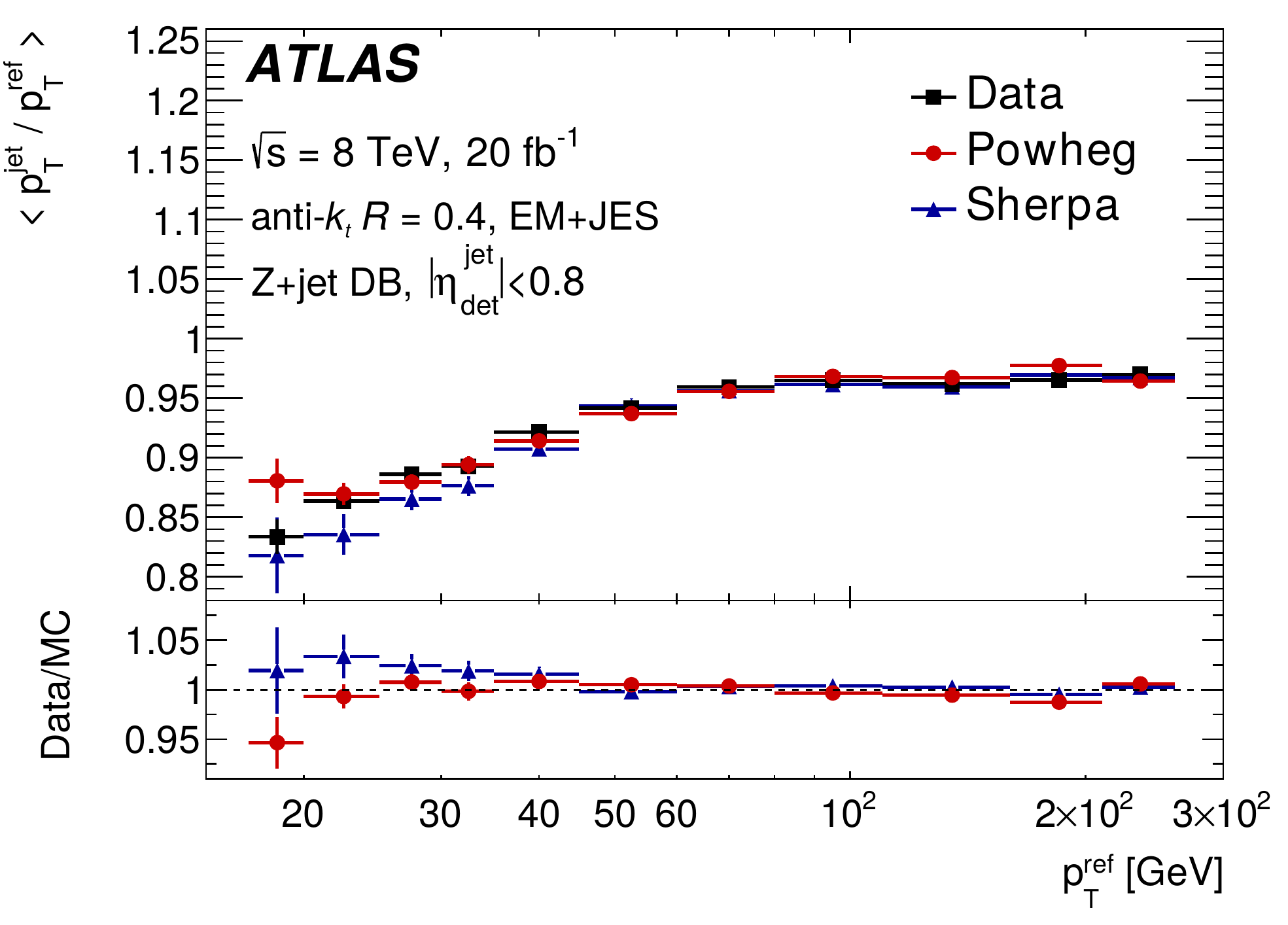}
\caption{\Zlljet{},   \antikt{} \rfour}
\end{subfigure}
\begin{subfigure}{0.48\linewidth}\centering
\includegraphics[width=\linewidth]{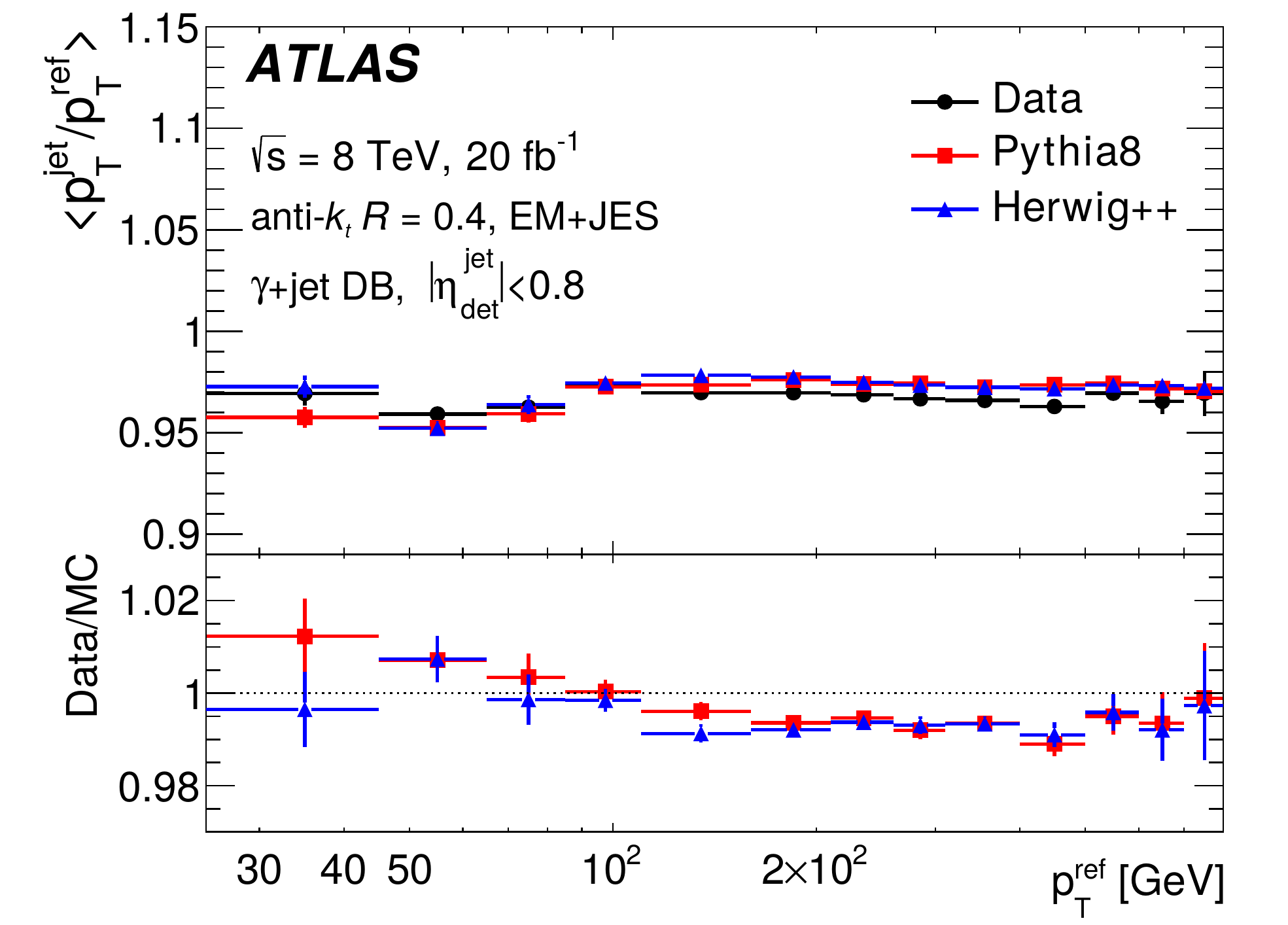}
\caption{\gammajet{}, \antikt{} \rfour}
\end{subfigure}
\caption{\RDB{} for \antikt{} jets with \rfour{} and calibrated with the \EMJES{} scheme from the (a)~\Zjet{} and (b)~\gammajet{} analyses in the data and for two MC simulations.  Only statistical uncertainties are shown.
\label{fig:DB-vs-pTref}
}
\end{figure}

For the \gammajet{} analysis, the measured responses agree within 1\% with the MC predictions for $\ptref{}<100$~\GeV\ for both \rfour{} and \rsix{} jets using both calibration schemes.
For jets with $\ptref>100$~\GeV, the MC simulation systematically tends to overestimate the measured response by approximately $1\%$.

\subsubsection{Validation of intercalibration of forward jets using \Zjet{} data}
\label{sec:vjets-fwd-jet}
 
The derived intercalibration in Section~\ref{sec:dijet} corrects jets with forward rapidities $|\etaDet|>0.8$ by about \percentRange{1}{3} (Figure~\ref{fig:resp_vs_eta}).
The total uncertainty in this calibration is presented in Figure~\ref{fig:uncertainty} and is typically below $1\%$ for jets with $|\etaDet|<3$, increasing to about $3.5\%$ for low-\pt{} jets with $|\etaDet|>4$.
 
To validate this calibration, the DB analysis is repeated for the jets with \etaRange{0.8}{4.5} using \Zjet{} events.
As in the standard analysis (Section~\ref{sec:vjets-jet-sel}), the intercalibration is applied to the jets, and hence the \DtM{} ratio of \RDB{} is expected to be uniform versus \etaDet{} within the uncertainty assigned to the intercalibration. Results of this analysis for  \EMJES{} calibrated \antikt{} \rfour{} jets are presented in Figure~\ref{fig:vjets-fwd-jets}.
The prediction of both MC generators agree with the data within the assigned uncertainties for jets with \etaRange{0.8}{2.8}.
For the region $|\etaDet|>2.8$, differences can be up to 7\% as shown in Figure~\ref{fig:vjets-fwd-jets}; however, the statistical uncertainties of the \Zll{} measurements are of similar magnitude. Hence, the results validate the derived dijet intercalibration.
 
\begin{figure}[!htb]
\centering
\begin{subfigure}{0.48\linewidth}\centering
\includegraphics[width=\linewidth]{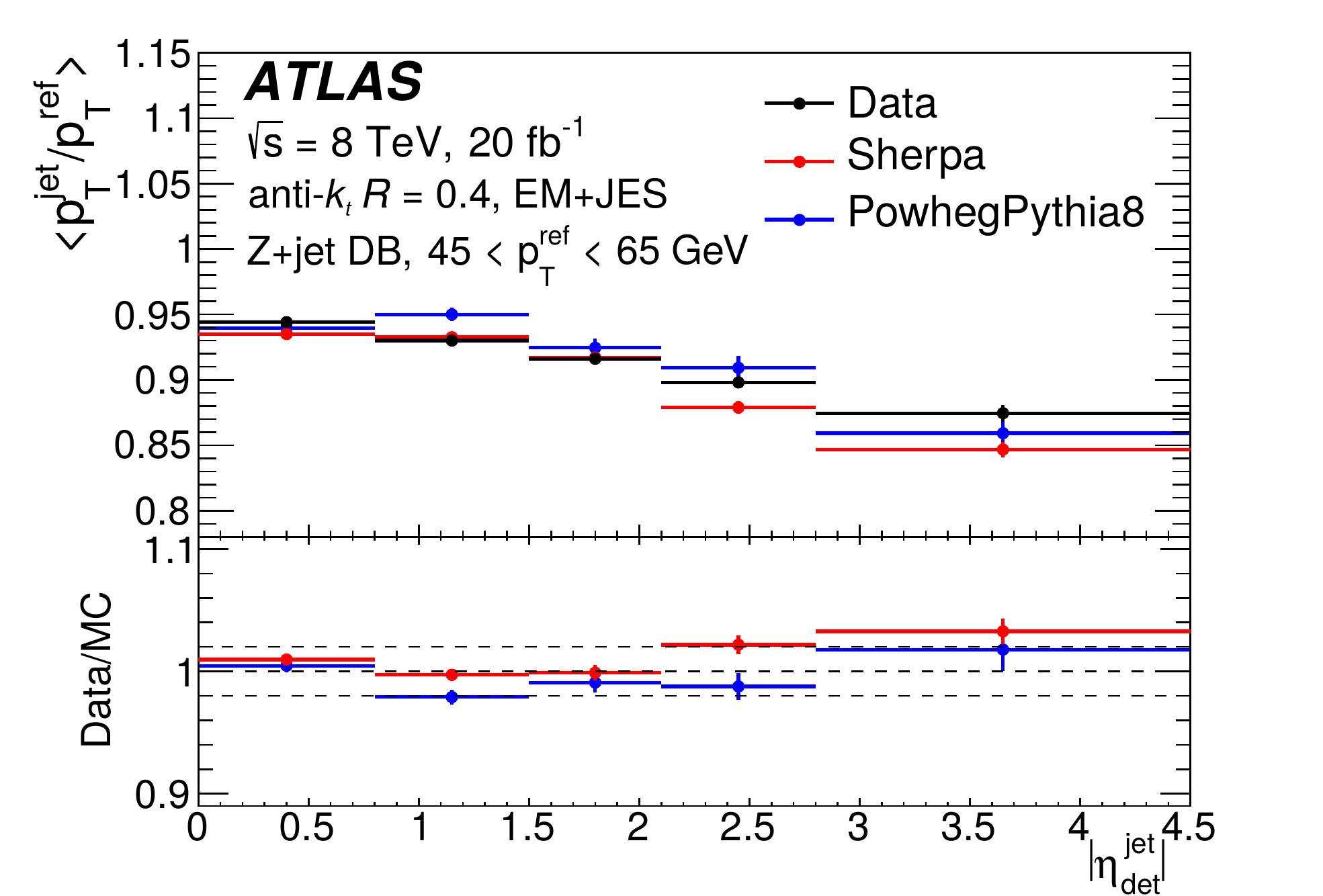}
\caption{\ptrefRange{45}{65}}
\end{subfigure}
\begin{subfigure}{0.48\linewidth}\centering
\includegraphics[width=\linewidth]{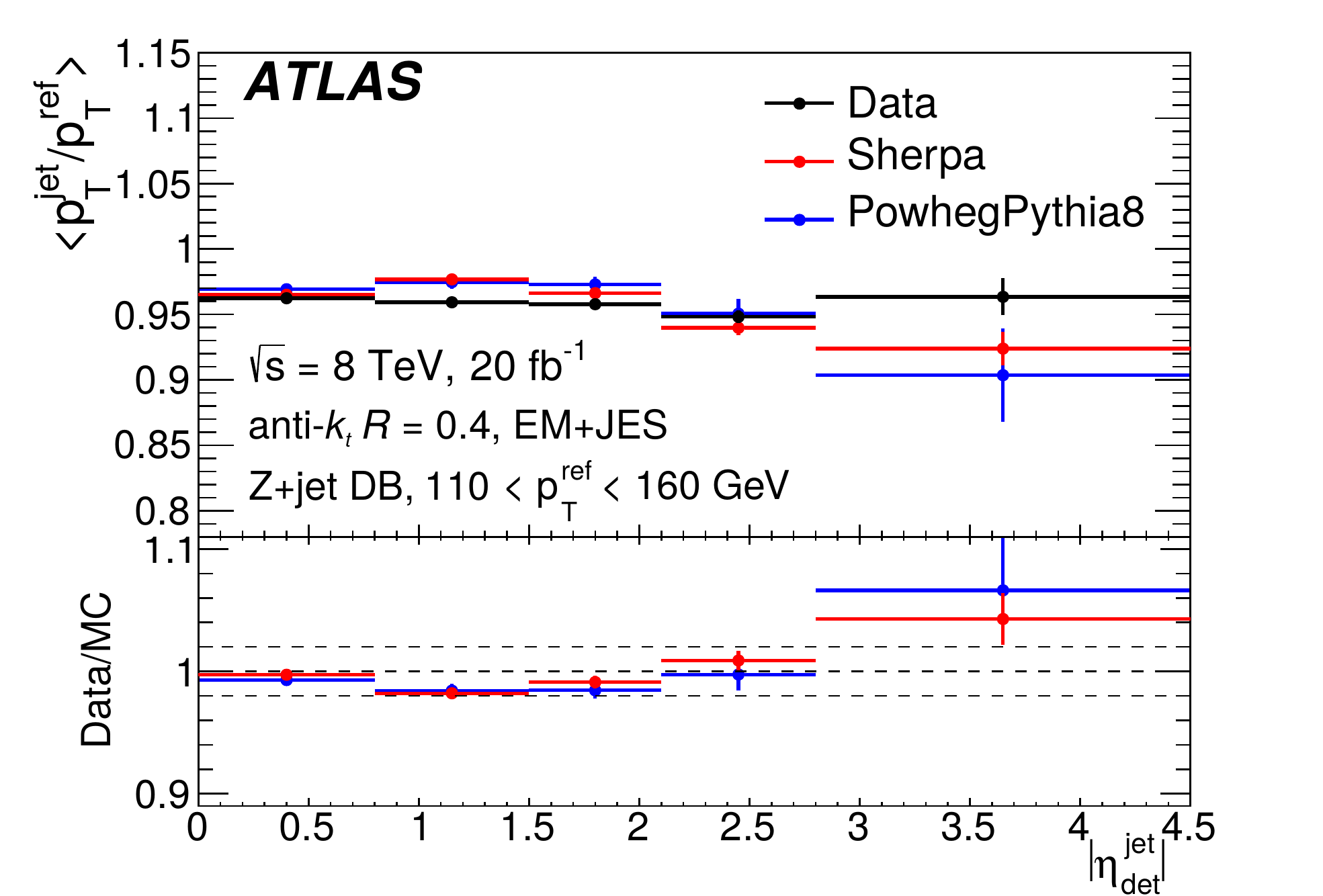}
\caption{\ptrefRange{110}{160}}
\end{subfigure}
\caption{
\RDB{} as a function of \etaDet{} for \antikt{} jets with \rfour{} calibrated with the \EMJES{} scheme in data~(black), as well as in \sherpa{}~(red) and \powhegpyt{}~(blue) simulation, for (a)~\ptrefRange{45}{65}{} and (b)~\ptrefRange{110}{160}.  Only statistical uncertainties are shown.
\label{fig:vjets-fwd-jets}
}
\end{figure}

\subsubsection{MPF results}
\label{sec:MPF-results}
Figure~\ref{fig:MPF} presents \RMPF{} calculated at both the EM and LCW scales as a function of \ptref{} extracted using both the \Zjet{} and \gammajet{} events.
As mentioned in Section~\ref{sec:vjets-method}, \RMPF{} is a measurement of the hadronic response of the calorimeter and does not include the MC-derived calibration~$c_\text{JES}$ nor the GS calibration~$c_\text{GS}$ (see Sections~\ref{sec:vjets-OOC-syst} and~\ref{sec:MPF-GSC} for further discussion on this). The ``upturn'' of \RMPF{} at low values of \ptref{} visible in Figure~\ref{fig:MPF} is an expected artefact of the jet reconstruction threshold.
Since a jet is required to be present in the event (Section~\ref{sec:vjets-jet-sel}), when this jet's \pt{} fluctuates low the event might fail the selection.
For such an event, \rMPF{} will also tend to be low. And similarly, events with jets that fluctuate high in \pt{} will have high \rMPF{} and will pass the selection.
 
For the \Zjet{} analysis, the \RMPF{} measured in data is systematically about 1\% below the prediction of \powhegpyt{}, considered the reference MC sample.
For the \gammajet{} analysis, the predictions of \RMPF{} from both MC simulations agree across the full \ptref{} range within the statistical precision.
Both simulations systematically overestimate the \RMPF{} value measured in data by about 1\% for $\ptref{}>85$~\GeV\ at the EM scale and by about 1\% for $\ptref>50$~\GeV\ at the LCW scale.
 
\begin{figure}[!htb]
\centering
\begin{subfigure}{0.48\linewidth}\centering
\includegraphics[width=\linewidth]{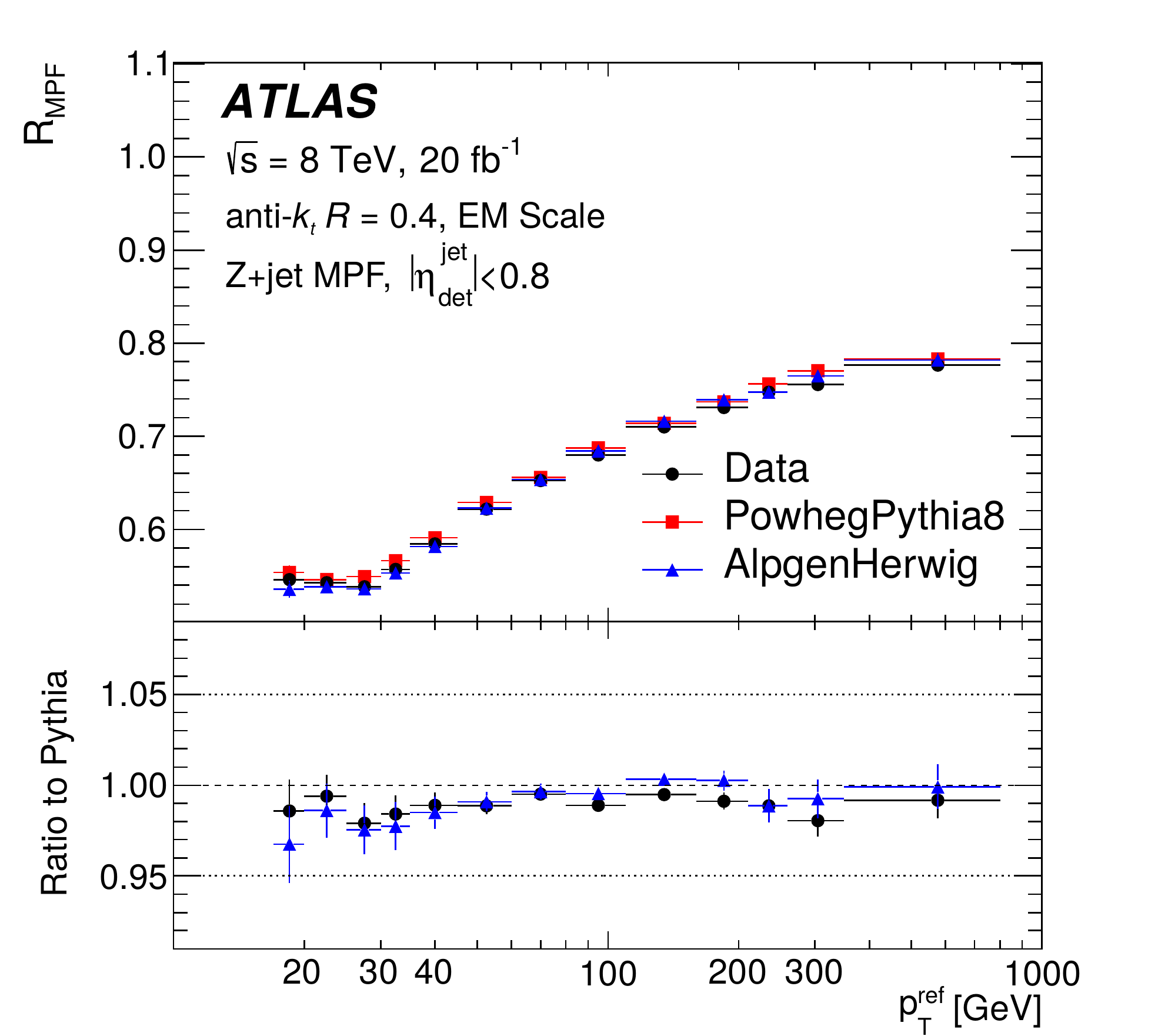}
\caption{\Zjet{},     \EM{} scale}
\end{subfigure}
\begin{subfigure}{0.48\linewidth}\centering
\includegraphics[width=\linewidth]{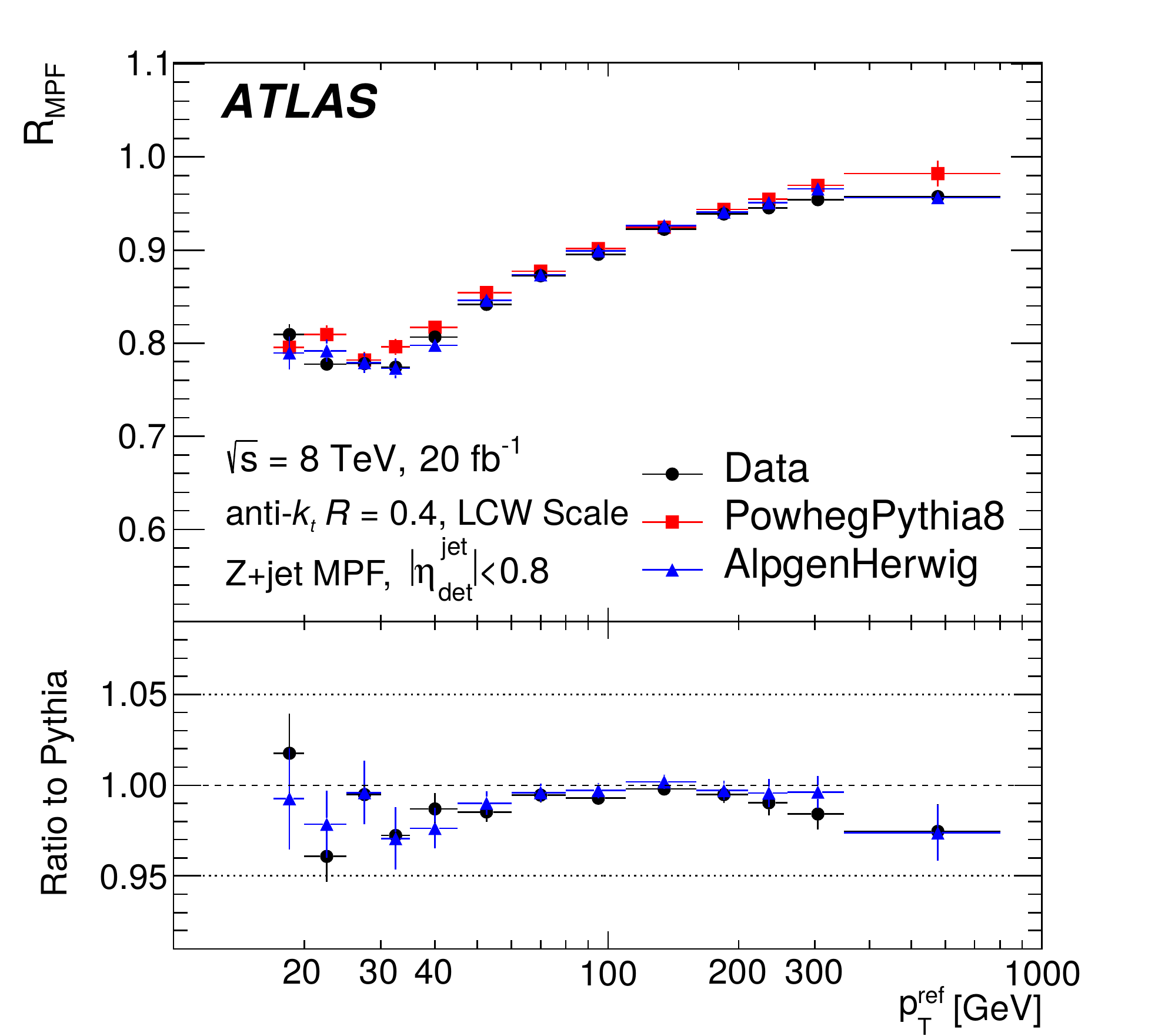}
\caption{\Zjet{},     \LCW{} scale}
\end{subfigure} \\
 
\bigskip
 
\begin{subfigure}{0.48\linewidth}\centering
\includegraphics[width=\linewidth]{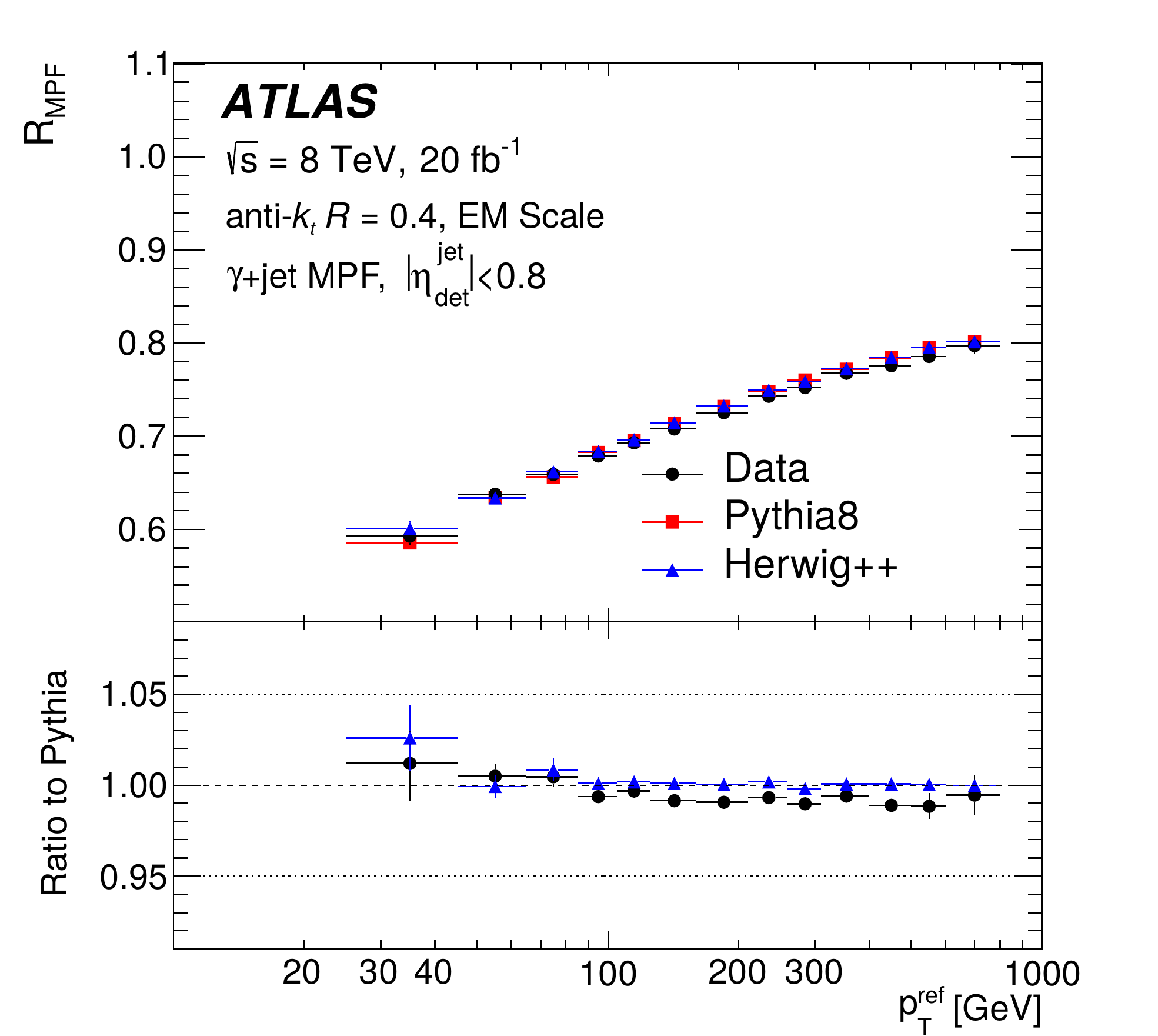}
\caption{\gammajet{}, \EM{} scale}
\end{subfigure}
\begin{subfigure}{0.48\linewidth}\centering
\includegraphics[width=\linewidth]{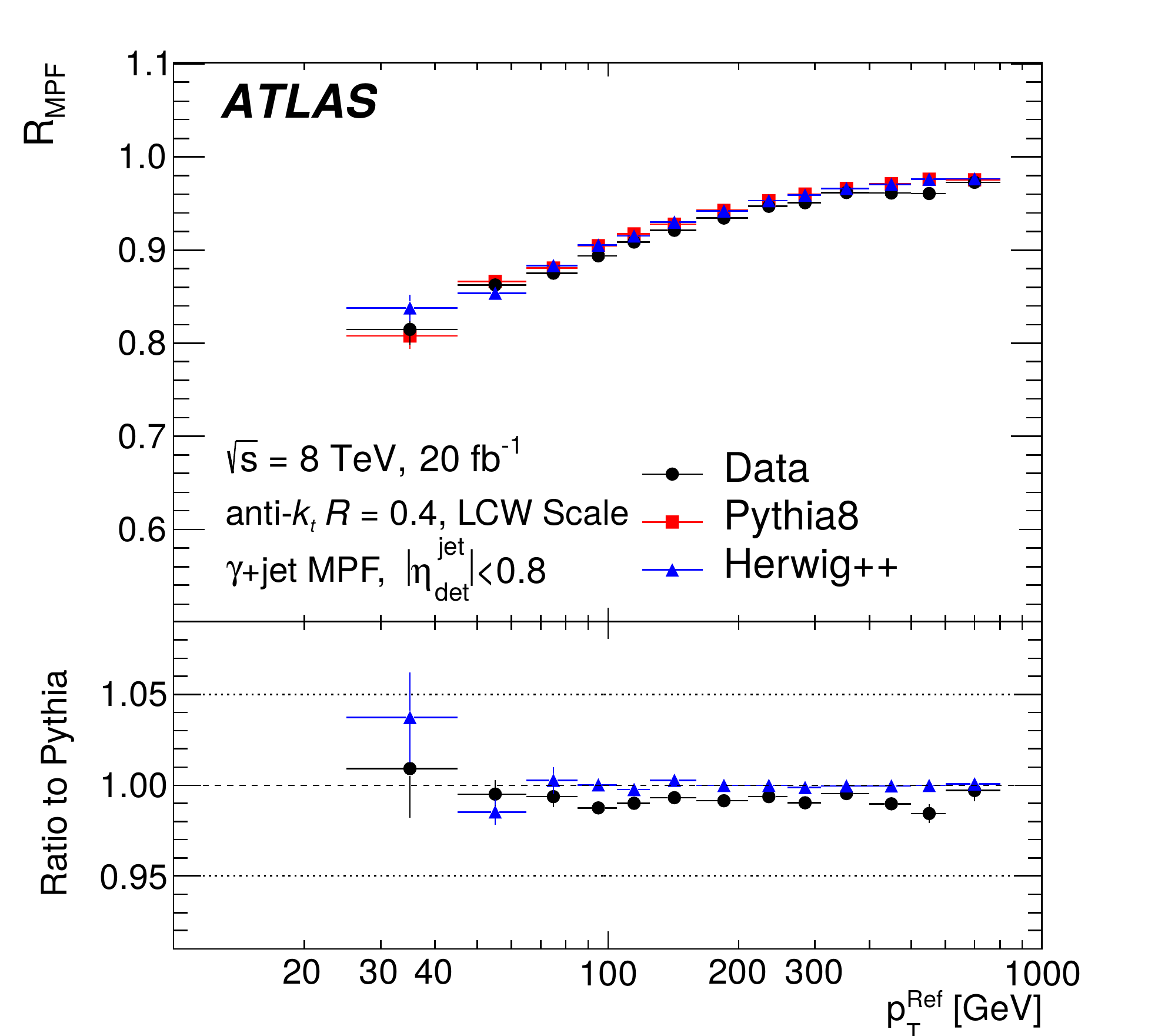}
\caption{\gammajet{}, \LCW{} scale}
\end{subfigure}
\caption{The \MPF{} response \RMPF{} in (a,b)~\Zjet{} and (c,d)~\gammajet{} events for jets calibrated (a,c)~at the \EM{} scale and (b,d)~using the \LCW{} scheme,
for both data and MC simulation, as a function of \ptref{}.  Only statistical uncertainties are shown.
}
\label{fig:MPF}
\end{figure}

\subsubsection{Systematic uncertainties}
\label{sec:vjets-syst}
 
The final JES calibration that is described in Section~\ref{sec:JEScomb} is based on the \DtM{} ratio of the response measurements.
As explained in detail in Section~\ref{sec:evalSyst}, systematic uncertainties in this quantity are evaluated by introducing variations to the analysis.
The following seven subsections present the evaluation of in total 17 uncertainty sources that affect the \DtM{} ratio of \RDB{} and \RMPF{}.
These uncertainties assess various effects that can affect the response measurement, such as impact of additional QCD radiation, choice of MC generator, effects from out-of-cone radiation and \pileup{}, and the precision of the \pt{} scale of the reference objects (photons, electrons, or muons).

\paragraph{Suppression of additional radiation}
\label{sec:vjets-syst-topo}~\\
As explained in Section~\ref{sec:vjets-jet-sel}, a ``boson+jet'' topology is selected by imposing constraints on the azimuthal angle $\Delta\phi$ between the boson and the jet and by restricting the \pt{} of any subleading jet. These criteria reduce the impact from additional QCD radiation on the momentum balance between the jet and the boson.
Systematic uncertainties from two sources are evaluated, one through varying the $\Delta\phi$ requirement and one through variations of the $\pt^\text{j2}$ selection.
Constructing a single uncertainty component from variation in simulations of the two criteria is also considered (as was done previously~\cite{PERF-2011-03});
however, the two-component approach is sufficient.
 
The $\Delta\phi$ selection is varied by $\pm 0.1$ around the nominal values of 2.9 for \MPF{} and 2.8 for \DB{} (Section~\ref{sec:vjets-jet-sel})).
The $\pt^\text{j2}$ requirement for the \DB{} analysis is similarly tightened to $\text{max}(7~\GeV,0.05\,\ptref)$ and loosened to $\text{max}(9~\GeV,0.15\,\ptref)$.
The MPF selection is varied by similar amounts around the nominal selection.
The resulting uncertainty from the $\Delta\phi$ variation is generally negligible.
The uncertainty due to the $\pt^\text{j2}$ requirement is 0.4\% or smaller for $\ptref{} < 50$~\GeV and is negligible above this value.

\paragraph{Systematic uncertainties due to out-of-cone radiation}\label{sec:vjets-OOC-syst}~\\
For the \DB{} method, the \pt{} of a jet, even if perfectly calibrated, will always tend to be smaller than that of the photon or \Zbos{} due to the out-of-cone radiation (Figure~\ref{fig:DB}). 
The impact of the out-of-cone radiation on \RDB{} is studied in both data and simulation by measuring the average \pt{} density of tracks as a function of the angular distance ($\DeltaR$) between the track direction and the leading jet axis.
Based on this \pt{} profile, the fraction of the radiation outside the jet cone is estimated (see Section 9.4 of Ref.~\cite{PERF-2012-01} for details), and an out-of-cone systematic uncertainty is evaluated on the basis of the simulation's ability to model the measured value of this quantity.
The resulting uncertainty is as large as 2\% at $\ptref=40$~\GeV\ and is smaller at higher \ptref{}.
 
In principle, the \MPF{} technique does not depend on the OOC correction because the calorimeter response is integrated over the whole detector.
However, two effects related to the OOC contribution must be considered.
The ``showering correction'' quantifies the migration of energy across the jet boundary of the calorimeter jet relative to the \tjet{} and is difficult to measure with data.
This effect is included in the DB analysis by design since it is based on reconstructed jets but is not included for the MPF method since it measures the entire hadronic recoil.
In addition, the hadronic response in the periphery of the jet is different than in the core because of the different energy densities and particle compositions.
This ``topology correction'' is also difficult to extract from data but is expected to be small
since the average \pt{} density around the jet axis decreases fairly rapidly, and
only a small fraction of the \pt{} is outside the jet radius.  MC studies have shown that the uncertainty in each of these corrections is significantly smaller than the \DB{} OOC uncertainty. 
As a conservative approach, the OOC uncertainty measured in data for the \DB{} case is used to estimate the contributions from showering and jet topology to the uncertainty in the JES determined using the \MPF{} technique. The use of this larger uncertainty does not significantly affect the total systematic uncertainty in the \JES{} from this analysis over most of the \pt{} range.
 
\paragraph{Impact of the Monte Carlo generator}~\\
For each final state, predictions of the response observables (\RDB{} and \RMPF{}) are produced with two different MC generators: \powheg{} and \sherpa{} for \Zjet{} and \pythia{} and \herwigpp{} for \gamjet{}. As detailed in Section~\ref{sec:mc}, these generators use different modelling of the parton shower, jet fragmentation, and multiple parton interactions.  The difference in the \DtM{} ratio of the response between the generators is taken as a ``generator'' systematic uncertainty source.
This is a reasonable estimate of the dependence of the $pp$ collision modelling on \RDB{} and \RMPF{}, but a possible compensation by competing modelling effects cannot be excluded. This generator modelling constitutes the largest systematic uncertainty source for \Zjet{} for $\ptref{} \lesssim 50$~\GeV, where it can be as large as 2.5\%.

\paragraph{Uncertainties associated with the reference objects}~\\
The definitions of \RDB{} and \RMPF{} 
both have the \pt{} of the reference object in the denominator and are hence sensitive to knowledge of its energy scale and resolution. For the \Zjet{} analyses, uncertainties in \ptref{} arise from the precision of the electron energy scale (EES) and energy resolution (EER) and from the muon momentum scale (MMS) and resolution (MMR), while for \gamjet{}, uncertainties arise from the precision of the photon energy scale (PES) and energy resolution (PER).
 
The EES is measured in data~\cite{PERF-2010-04} and has three uncertainty components: MC modelling of the
$Z\to ee$ decay; the material description in simulation; and the response of the calorimeter's presampler.  These are treated statistically independent of each other. The EER uncertainty is parameterized by a single component. The MMS and MMR are determined in data~\cite{PERF-2014-05} and have one and two associated uncertainty components, respectively.
Finally, the PES and PER are evaluated using extrapolation of EES and EER, and are hence affected by the same uncertainty sources.
Hence, they have the same four uncertainty sources, but these affect photons and electrons quite differently.
 
Each of the 11 uncertainty sources are propagated to the simulated samples by adjusting the four-momenta of the reconstructed electron, muon, or photon.
The uncertainties in \RDB{} and $\RMPF{}$ are then evaluated following the procedure described in Section~\ref{sec:evalSyst}.
For all objects, the resolution uncertainties are found to be negligible (0.1\% or less). For \gamjet{}, the PES uncertainties are reasonably independent of \ptref{} and their sum in quadrature amounts to 0.5\%--0.6\%. The magnitudes of the EES and MMS uncertainties are less than 0.3\%.
 
\paragraph{Impact of additional \pileup{} interactions}~\\
Jets produced in additional \pileup{} interactions are present in both data and simulation and might impact the response measurements.
To study this effect, the \JVF{} criterion is varied around its nominal value of 0.25 (Section~\ref{sec:jetReco}).
The \JVF{} criteria used for this variation are based on studies presented in Ref.~\cite{PERF-2014-03} and are 0.24 and 0.27 for \EMJES{} calibrated jets and are 0.21 and 0.28 for jets calibrated using \LCWJES{}.
 
Studies of the dependence of \RDB{} and \RMPF{} on the number of primary vertices \NPV{} in the event and on the average number of interactions per beam bunch crossing \avgmu{} were also performed. Figure~\ref{fig:mpf-pileup} presents results from these studies for the \MPF{} method for a representative selection of \ptref{} bins.
The \DtM{} ratio of \RMPF{} is found to be independent of both \NPV{} and \avgmu{} for all \ptref{} bins.
The same conclusion is reached for the \DB{} analysis.
Hence, only one \pileup{} uncertainty component is assigned, due to the \pileup{} mitigation using the \JVF{} criterion.

\begin{figure}[!htb]
\centering
\begin{subfigure}{0.45\linewidth}\centering
\includegraphics[width=\linewidth]{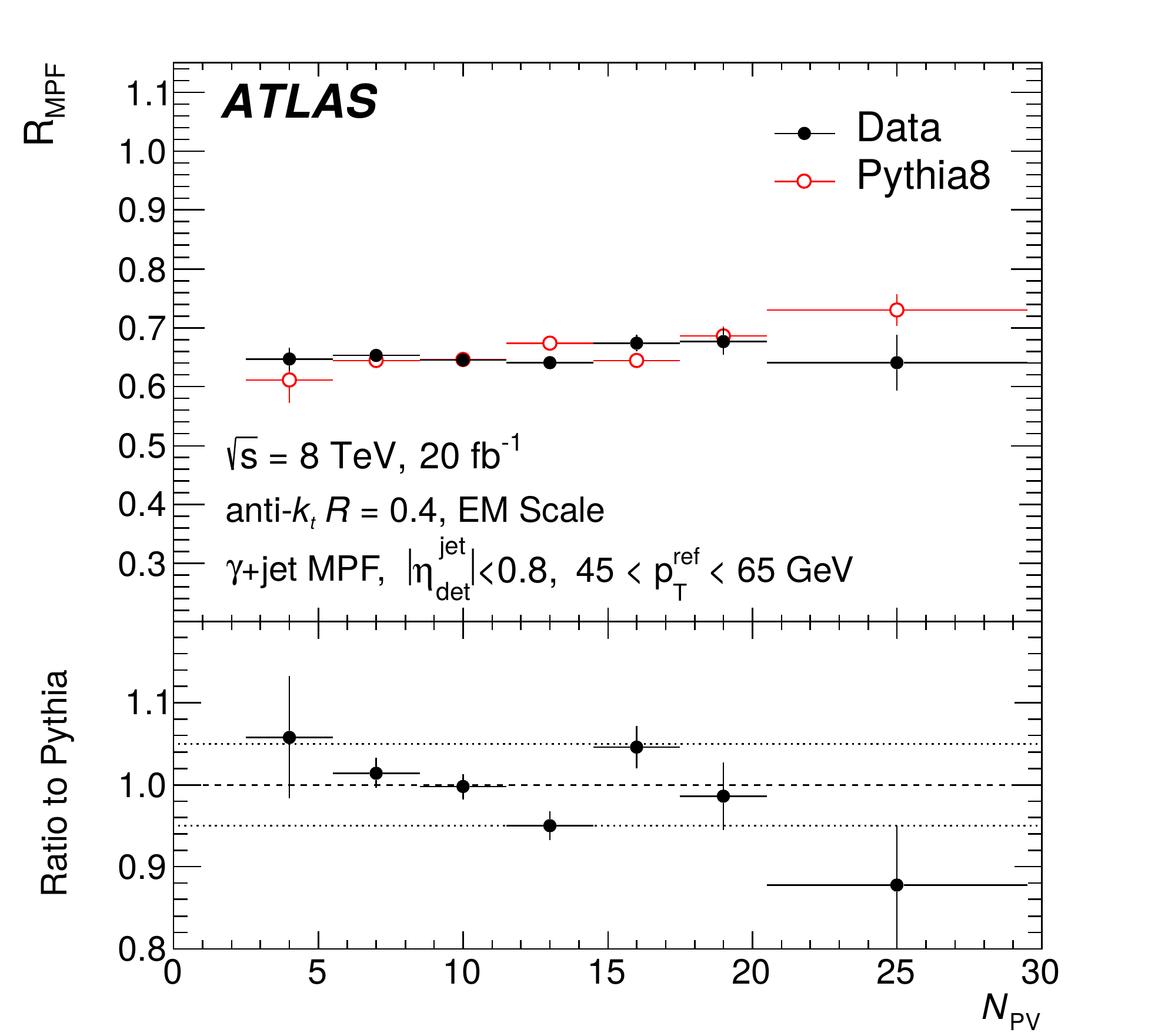}
\caption{\Npv{} dependence, \ptrefRange{45}{65}}
\end{subfigure}
\begin{subfigure}{0.45\linewidth}\centering
\includegraphics[width=\linewidth]{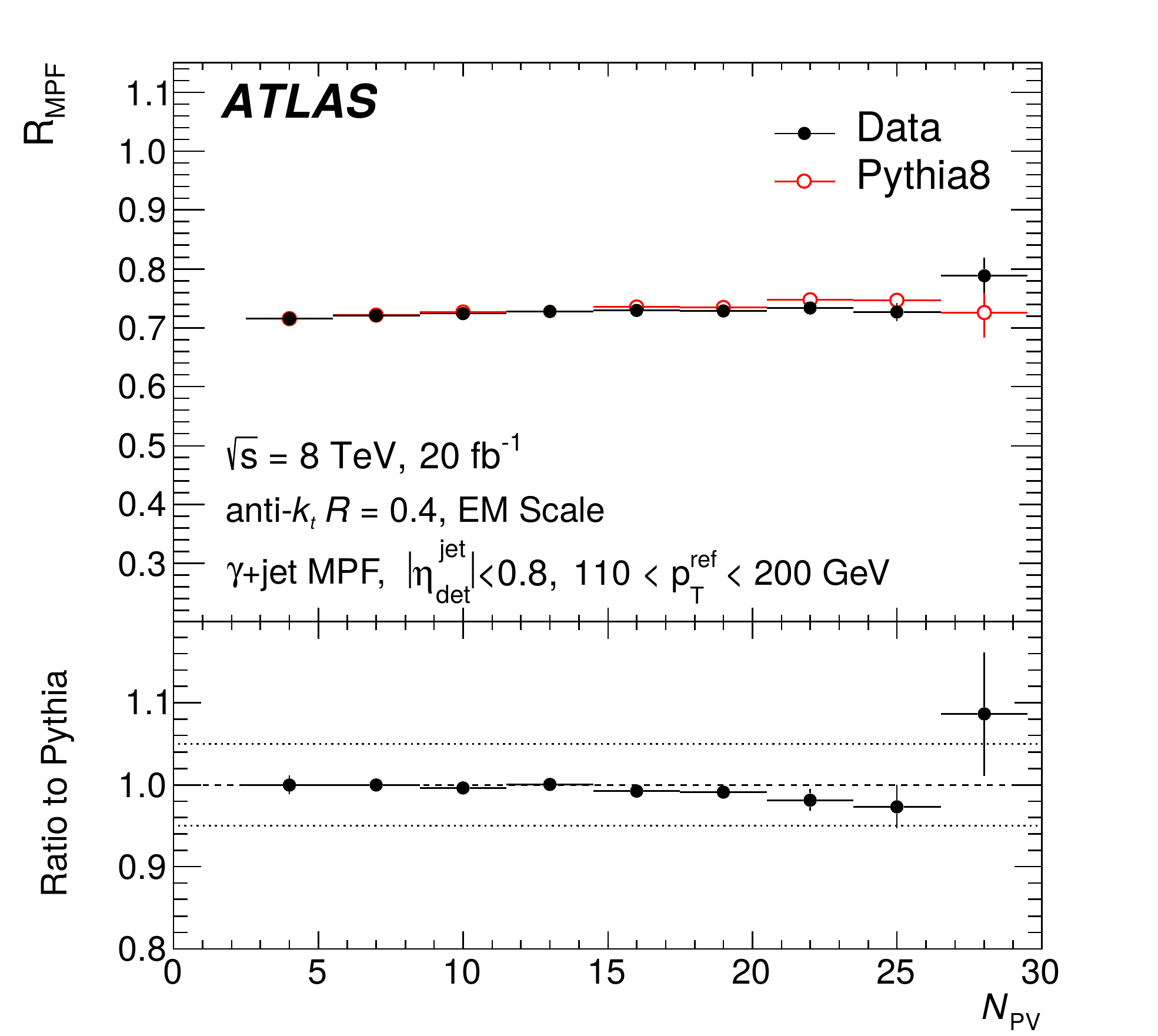}
\caption{\Npv{} dependence, \ptrefRange{110}{200}}
\end{subfigure} \\
 
\bigskip
 
\begin{subfigure}{0.45\linewidth}\centering
\includegraphics[width=\linewidth]{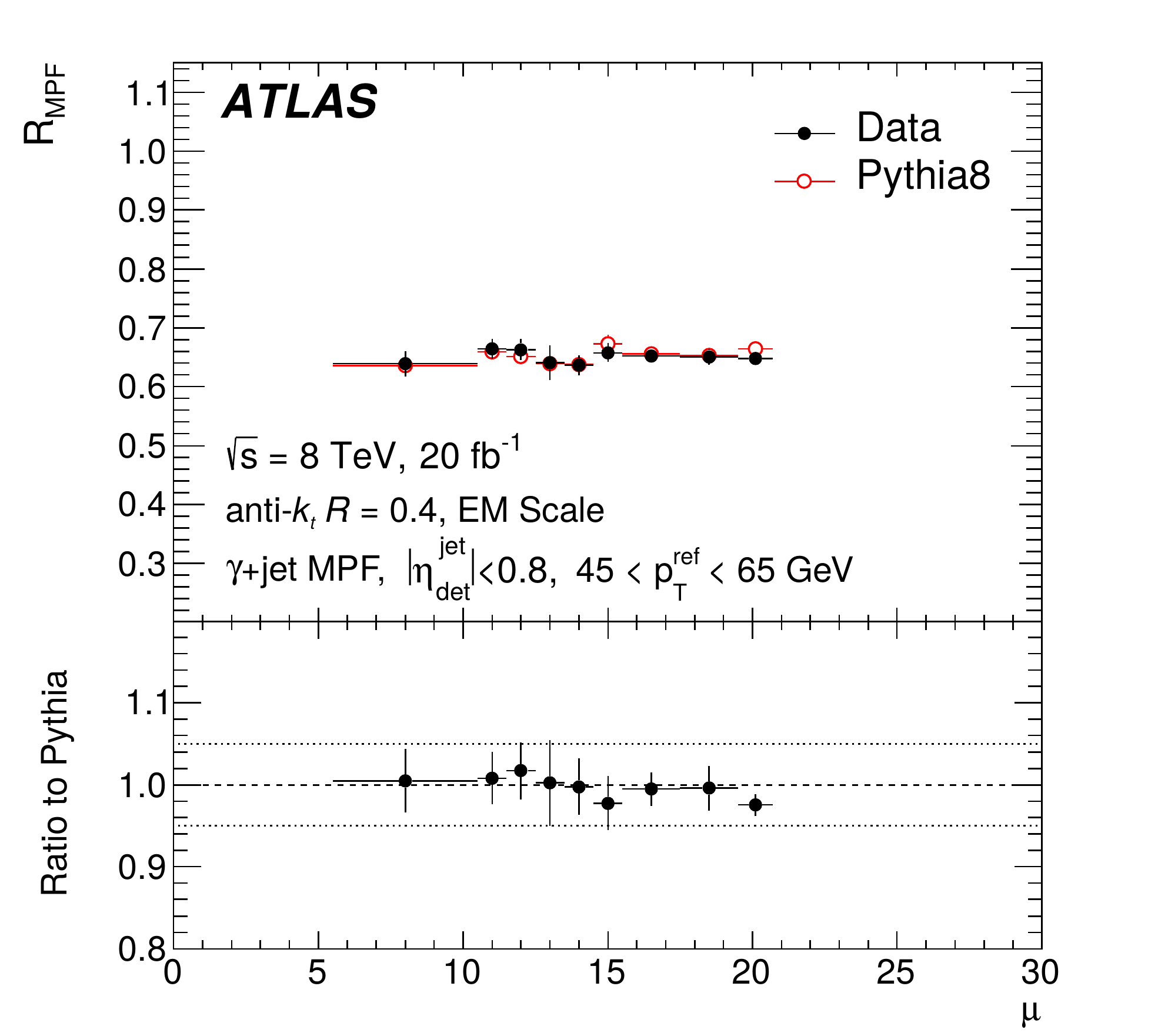}
\caption{$\mu$ dependence, \ptrefRange{45}{65}}
\end{subfigure}
\begin{subfigure}{0.45\linewidth}\centering
\includegraphics[width=\linewidth]{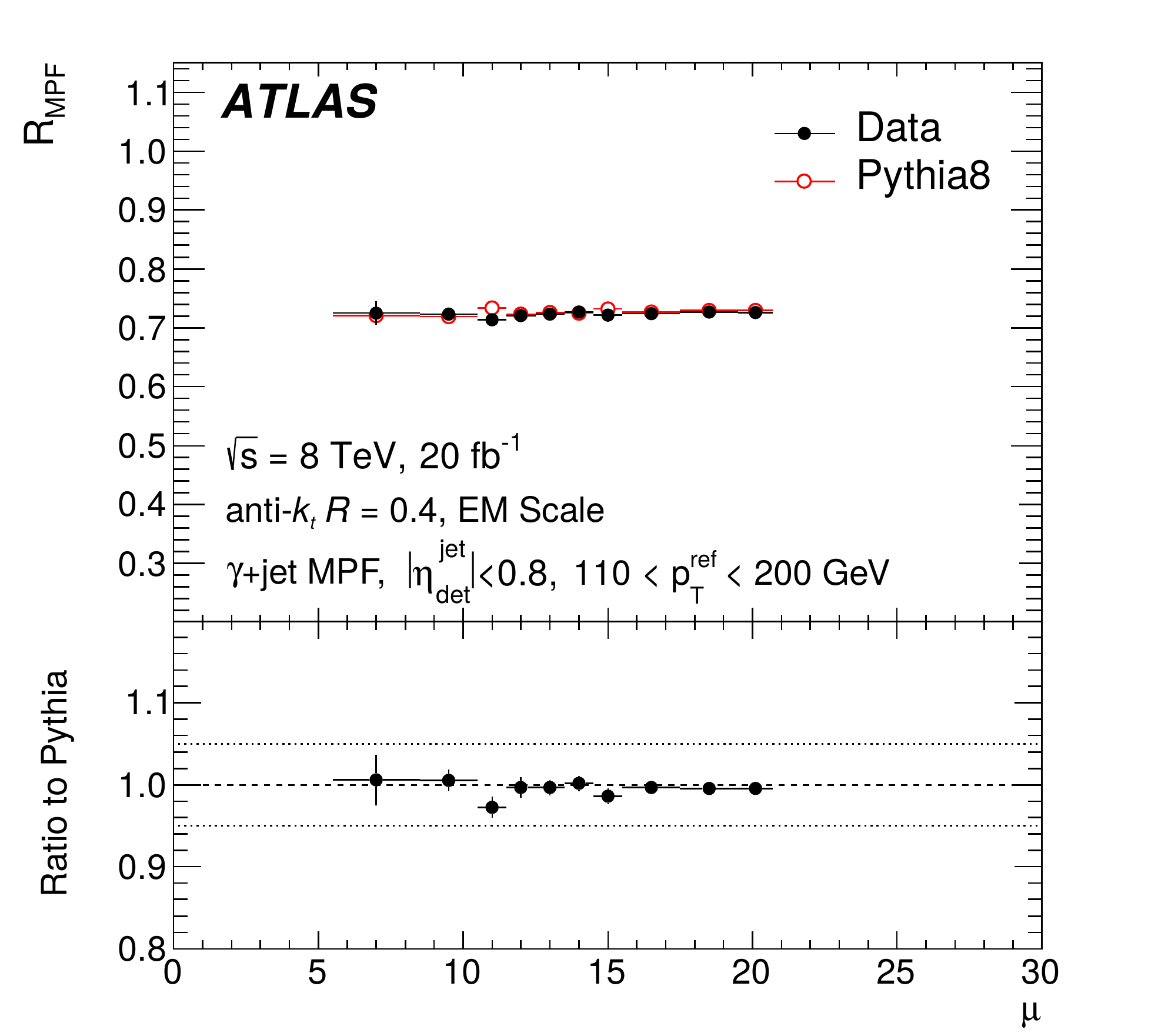}
\caption{$\mu$ dependence, \ptrefRange{110}{200}}
\end{subfigure}
\\
\caption{\MPF{} response \RMPF{} in \gammajet{} events for jets calibrated at the EM scale for both data~(black filled circles) and MC simulation~(red empty circles), as a function of (a,b)~\Npv{} and (c,d)~$\mu$ for a reference \pt{} range of (a,c)~\ptrefRange{45}{65} and (b,d)~\ptrefRange{110}{200}.  Only statistical uncertainties are shown.}
\label{fig:mpf-pileup}
\end{figure}
 
\paragraph{Impact of lack of GS correction for the \MPF{} method}\label{sec:MPF-GSC}~\\
The GS correction (Section~\ref{sec:gsc}) is based on the properties of jets.
Since the \MPF{} does not use jets directly, applying the GS correction in the standard way will have no impact on \RMPF{}.
Instead, the GS correction factor $c_\text{GSC}$ is extracted from the leading jet in each event and is used to adjust \RMPF{}.
Two methods were tested: simply scaling \RMPF{} with $c_\text{GSC}$ and recalculating \RMPF{} after adjusting \vecEtmiss{} by adding the change of the jet momentum vector due to the GS correction $(c_\text{GSC}-1)\vec{p}_\text{T}^{\,j1}$. Both methods result in a negligible change to the \DtM{} ratio of \RMPF{}, and no uncertainty is assigned for this effect.

\paragraph{Impact of background in the \gamjet{} sample}~\label{sec:vjets-photon-purity}\\
The \gamjet{} dataset suffers from non-negligible contamination from dijet events where one of the jets is misreconstructed as a photon.
The purity, i.e.\ the fraction of actual \gamjet{} events, after the nominal selection is estimated using a ``sideband'' technique based on the event yields in different control regions defined by alternative photon isolation and identification criteria. This technique is described in detail in Refs.~\cite{Aaboud:2016yuq,PERF-2011-03}. The purity increases with \ptgamma{}, being about $80\%$ at $\ptgamma = 85$~\GeV{} and rising above $90\%$ for $\ptgamma \gtrsim 200$~\GeV. The misreconstructed events tend to have higher \rDB{} and \rMPF{}. 
The difference in DB and MPF response between true \gamjet{} events and misreconstructed events is estimated by varying the photon identification criteria. The \gamjet{} MC samples used have perfect purity by definition.
The uncertainty due to the contamination from dijet events in the \gamjet{} analyses was estimated by multiplying the fraction of misreconstructed events by the relative difference in response between \gamjet{} and misreconstructed events. The resulting uncertainty decreases with \ptref{}. For the \DB{} analysis, it is $\sim$3.5\% at $\ptref{}=35$~\GeV, decreasing to $~1\%$ at $\ptref{}=100$~\GeV\ and to $<0.4\%$ for $\ptref>250$~\GeV. For MPF, this uncertainty is smaller by approximately a factor of 2.
This reduction is due to the definition of the observable, where \RMPF{} is inherently more stable against stochastically-oriented effects (pileup, fake photons, etc) compared to \RDB{} due to such contributions cancelling in \RMPF{} when averaged over many events.

\paragraph{Summary of the systematic uncertainties}~\\
Figure~\ref{fig:zjet-uncert} presents the JES uncertainties from various sources, evaluated for both the \DB{} and \MPF{} methods with
\antikt{} \rfour{} jets calibrated with the \EMJES{} using the \Zjet{} dataset. The total uncertainty is obtained by addition in quadrature of the uncertainties from different sources.
Overall, the DB and MPF methods achieve similar levels of precision.
The MC generator uncertainties dominate for $\ptref{} \lesssim 50$~\GeV\ and the out-of-cone uncertainty is also significant for $\ptref{} \lesssim 80$~\GeV.
The statistical uncertainty is the major uncertainty for the \Zjet{} analyses at $\ptref{} > 200$~\GeV.
\begin{figure}[!htb]
\centering
\begin{subfigure}{0.48\linewidth}\centering
\includegraphics[width=\linewidth]{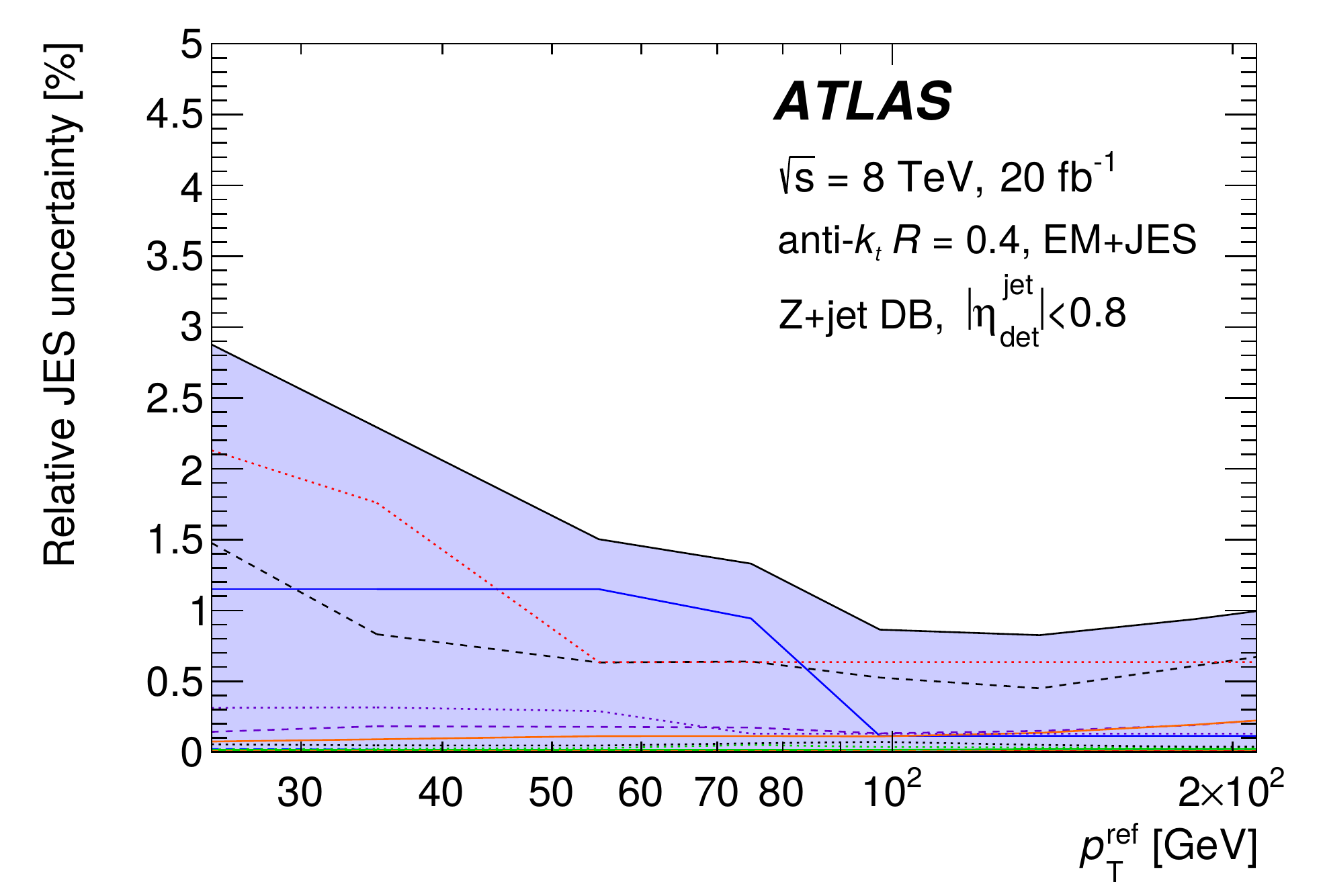}
\end{subfigure}
\begin{subfigure}{0.48\linewidth}\centering
\includegraphics[width=\linewidth]{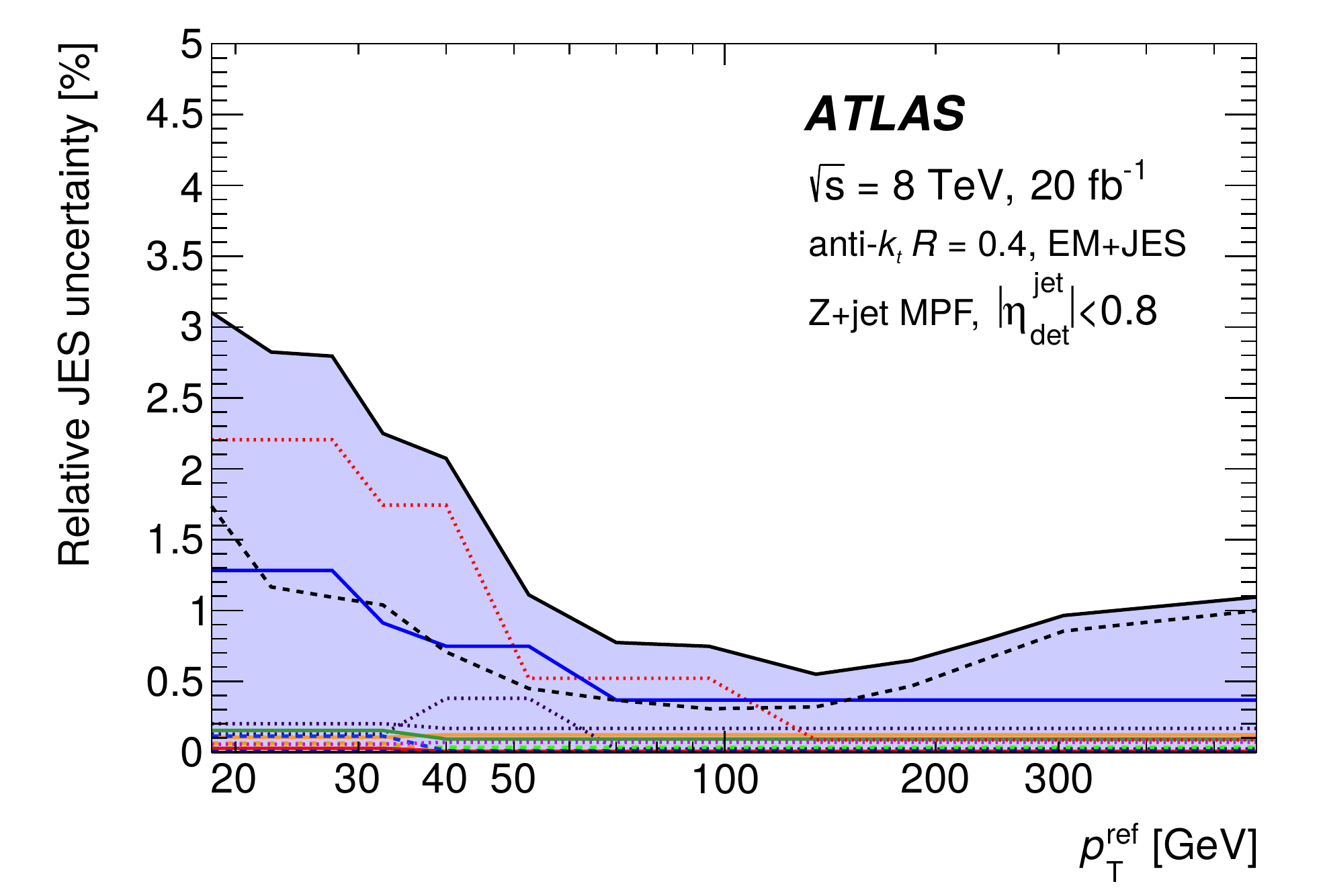}
\end{subfigure}
\begin{subfigure}[b]{0.48\linewidth}\centering
\includegraphics[width=\linewidth]{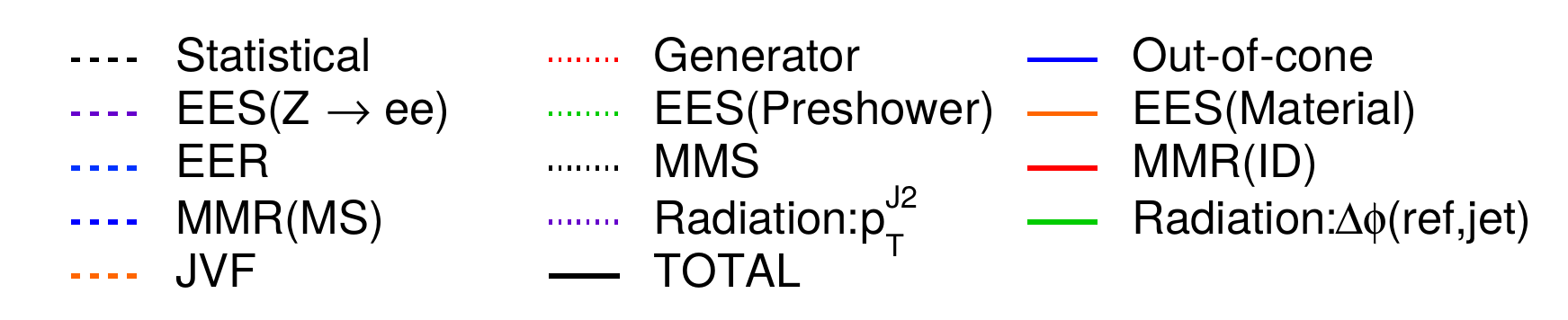}
\caption{\Zjet{} \DB{},  \antikt{} \rfour{}, \EMJES{} }
\end{subfigure}
\begin{subfigure}[b]{0.48\linewidth}\centering
\includegraphics[width=\linewidth]{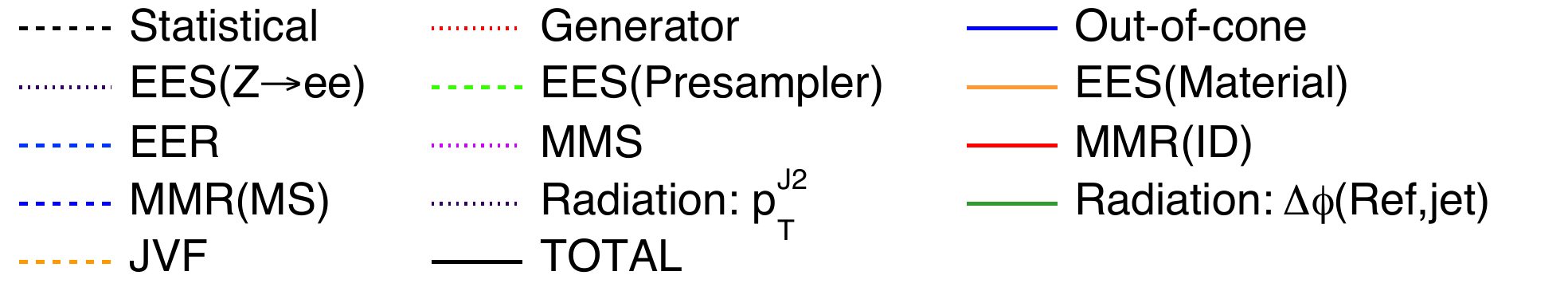}
\caption{\Zjet{} \MPF{}, \antikt{} \rfour{}, \EMJES{}}
\end{subfigure}
\caption{Summary of the JES statistical and systematic uncertainties evaluated for the \Zjet{} (a)~\DB{} and (b)~\MPF{}  analyses for \antikt{}  jets with \rfour{}  and calibrated with the \EMJES{} scheme.
The total uncertainty, shown as a shaded region, is obtained from the addition in quadrature of all uncertainty sources.
EES/EER denotes the electron energy scale/resolution, while MMS/MMR denotes the muon momentum scale/resolution.}
\label{fig:zjet-uncert}
\end{figure}
 
Figure~\ref{fig:gamjet-uncert} shows the uncertainties for the corresponding \gamjet{} analyses.
Here, the photon purity systematic uncertainty is the dominant uncertainty for the DB method for $\ptref < 100$~\GeV, while it is significantly smaller and subdominant for MPF.
The other systematic uncertainties are of similar magnitude for the two methods. For the range \ptrefRange{100}{400}, the photon energy scale contributes the dominant uncertainty.

\begin{figure}[!htb]
\centering
\begin{subfigure}{0.45\linewidth}\centering
\includegraphics[width=\linewidth]{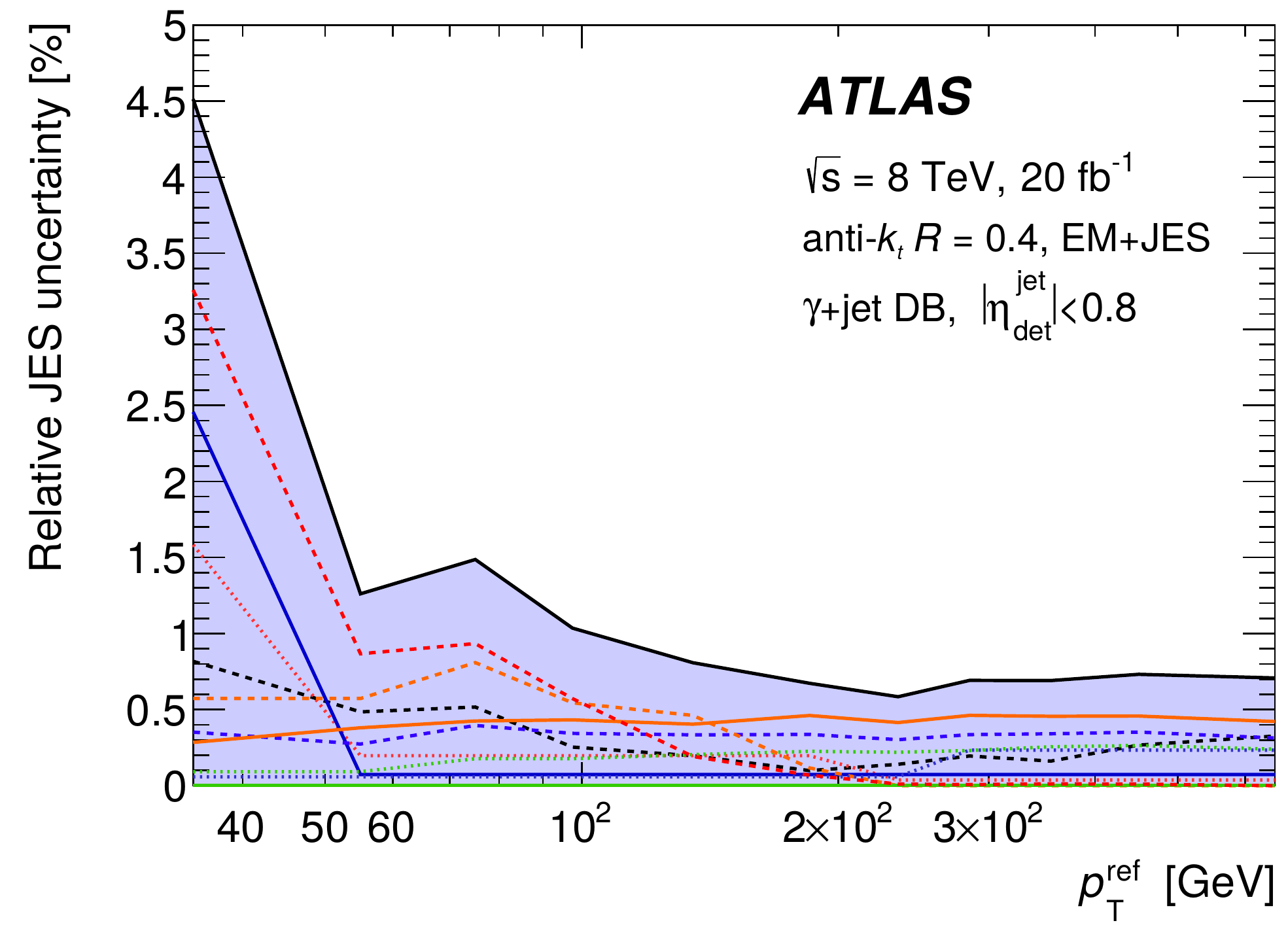}
\end{subfigure}
\begin{subfigure}{0.48\linewidth}\centering
\includegraphics[width=\linewidth]{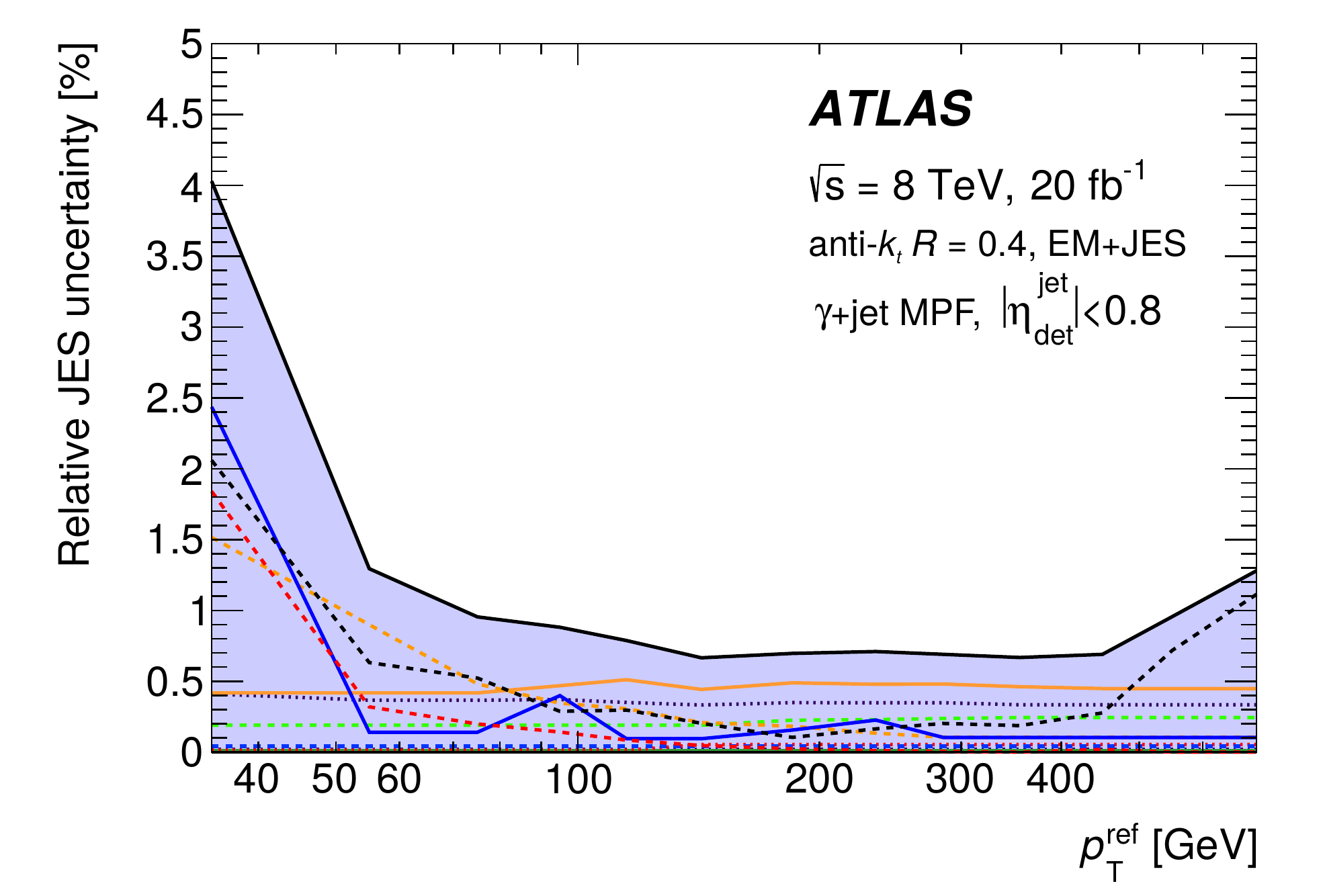}
\end{subfigure}
\begin{subfigure}[b]{0.45\linewidth}\centering
\includegraphics[width=\linewidth]{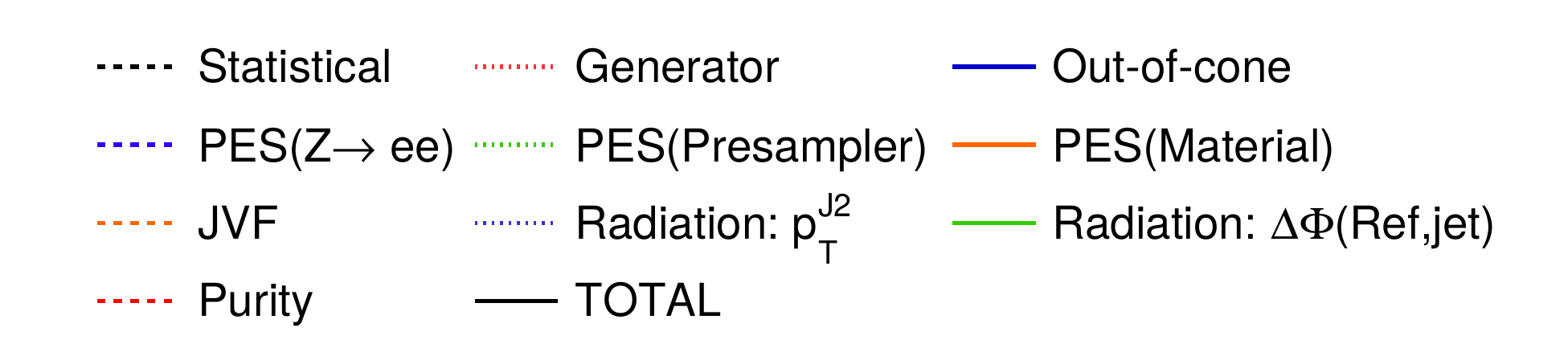}
\caption{\gammajet{} \DB{}, \antikt{} \rfour{}, \EMJES{}}
\end{subfigure}
\begin{subfigure}[b]{0.45\linewidth}\centering
\includegraphics[width=\linewidth]{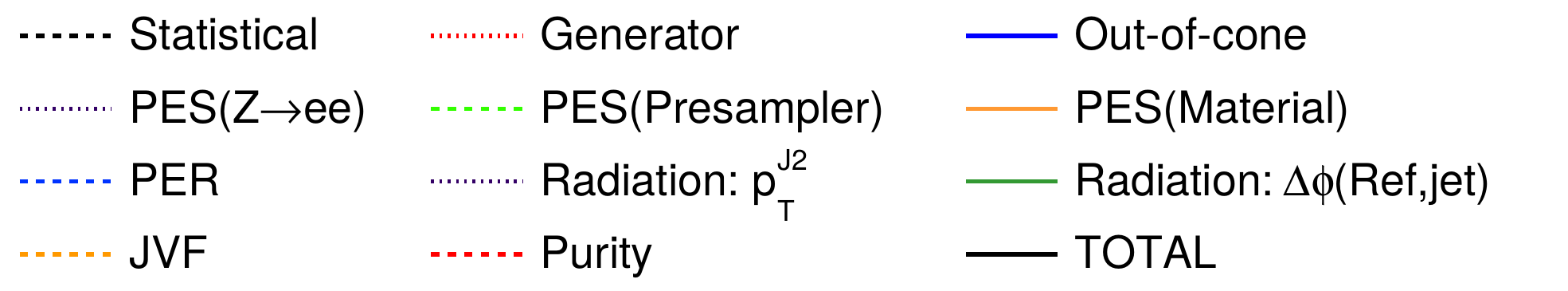}
\caption{\gammajet{} \MPF{}, \antikt{} \rfour{}, \EMJES{}}
\end{subfigure}
\vspace{-0.2cm}
\caption{Summary of the JES statistical and systematic uncertainties evaluated for the \gammajet{} (a)~\DB{} and (b)~\MPF{} analyses for \antikt{}  jets with \rfour{}  and calibrated with the ~\EMJES{} scheme.
The total uncertainty, shown as a shaded region, is obtained from the addition in quadrature of all uncertainty sources.
PES denotes the photon energy scale, while PER denotes the photon energy resolution.}
\label{fig:gamjet-uncert}
\end{figure}

\subsection{Calibration of \lRjets}
\label{sec:vjets-lRjet-calib}
 
For analyses based on pre-2012 data, the JES uncertainty of \lRjets{} has been evaluated \insitu{} using \trackjets{} (Section~\ref{sec:jetReco})~\cite{PERF-2012-02}.
This method, discussed further in Section~\ref{JES_largeR_uncert}, is limited to 2\%--7\% precision due to tracking uncertainties and the uncertainty in the charged-particle component of the jet. Furthermore, this method is restricted to the central calorimeter region \AetaRange{1.2}, since at more forward \etaDet{}, the \lRjet{} will not be fully contained in the acceptance of tracking detectors. This section presents an improved \lRjet{} JES uncertainty evaluation using \gamjet{} events.

\subsubsection{\RDB{} measurements using \gammaLRjet{} events}
The DB analysis is performed for \lRjets{} using the same approach as for \sRjets{}.
The binning in \ptref{} and \etaDet{} is different, chosen to account for the available data statistics, and \ptref{} is defined simply as $\ptref \equiv \ptgamma$ instead of projecting onto the jet axis (Eq.~(\ref{eq:RDB})).
Examples of \rDB{} distributions fitted with the Modified Poisson function are shown in Figure~\ref{fig:RDB-largeRjets}.
 
\begin{figure}[!htb]
\centering
\begin{subfigure}{0.45\linewidth}\centering
\includegraphics[width=\textwidth]{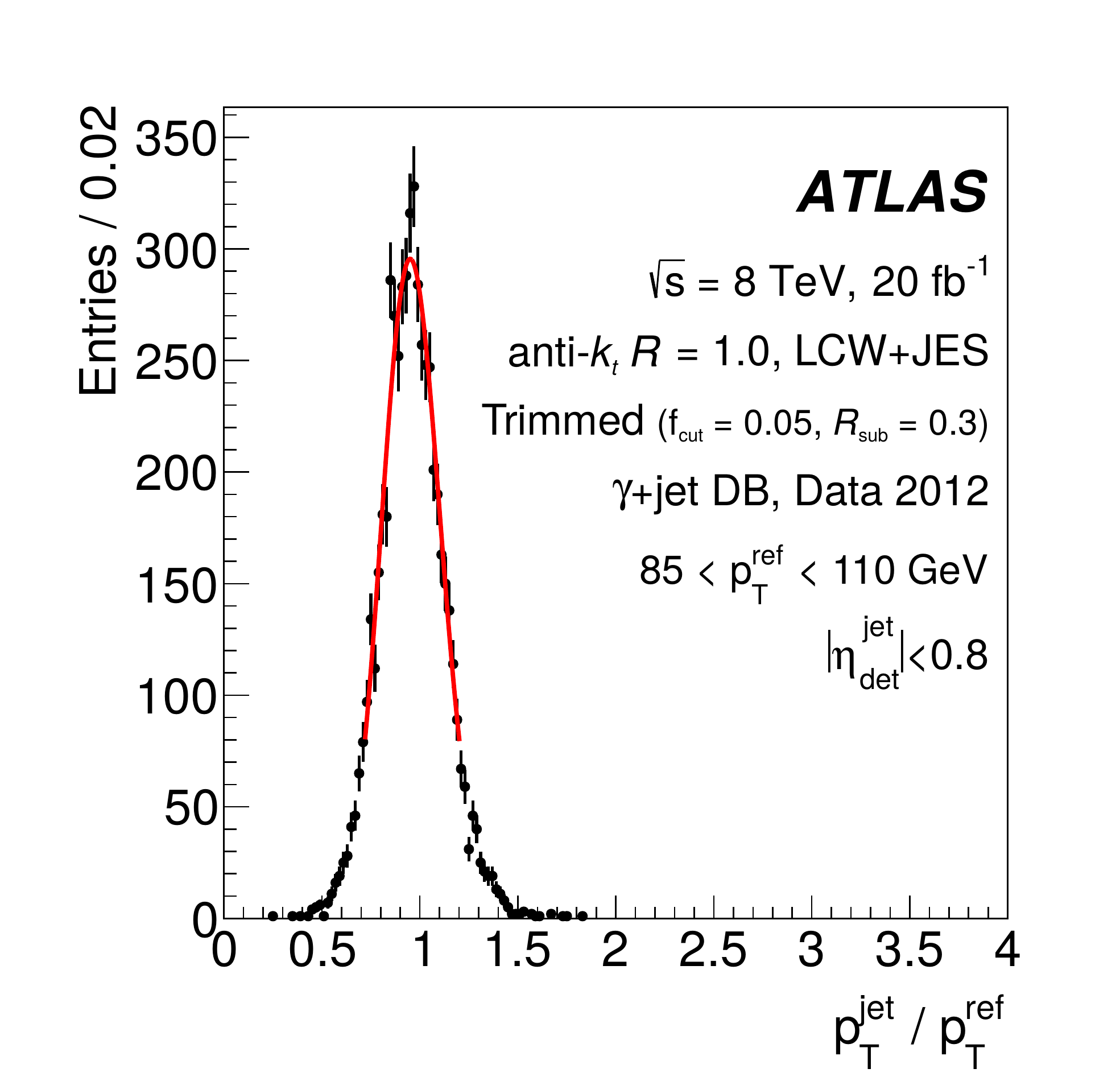}
\caption{\ptrefRange{85}{110}}
\end{subfigure}
\begin{subfigure}{0.45\linewidth}\centering
\includegraphics[width=\textwidth]{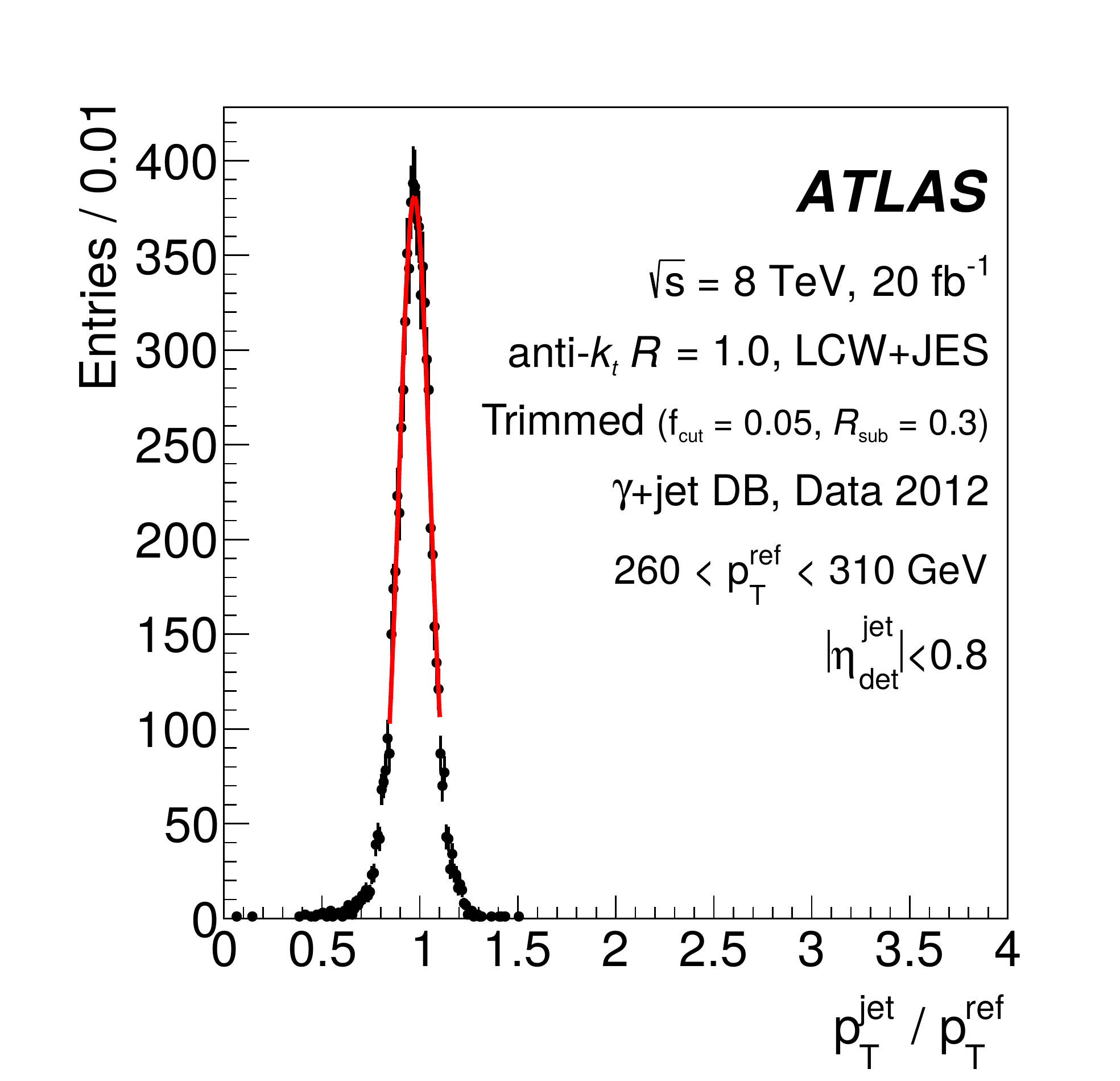}
\caption{\ptrefRange{260}{310}}
\end{subfigure}
\caption{
\rDB{} distributions for events with (a) \ptrefRange{85}{110}  and (b) \ptrefRange{260}{310} for trimmed \antikt{} jets with \rten{} calibrated with the \LCWJES{} scheme in data.
The lines present fits to the data of a Modified Poisson function (Eq.~(\ref{eq:MP})). The markers show data with error bars corresponding to the statistical uncertainties.
}
\label{fig:RDB-largeRjets}
\end{figure}
 
Figure~\ref{fig:RDB-vs-pt-largeR} presents \RDB{} as a function of \ptref{} for \lRjets{} in two \etaDet{} ranges, both for data and MC simulations.
The response in the central calorimeter region, $|\etaDet{}|<0.8$, is modelled within 1\% by the simulation, with simulations tending to overestimate the response by $\sim$0.5\%. For \lRjets{} with $0.8<|\etaDet{}|<1.2$, this deviation grows to $\sim$2\%.
Rather than using this deviation as a calibration to correct for the difference between data and MC simulation, this difference is taken as an additional uncertainty.
As detailed in Section~\ref{sec:jetKinem}, \lRjets{} do not receive any intercalibration $c_\eta$.

\begin{figure}[!htb]
\centering
\begin{subfigure}{0.48\linewidth}\centering
\includegraphics[width=\textwidth]{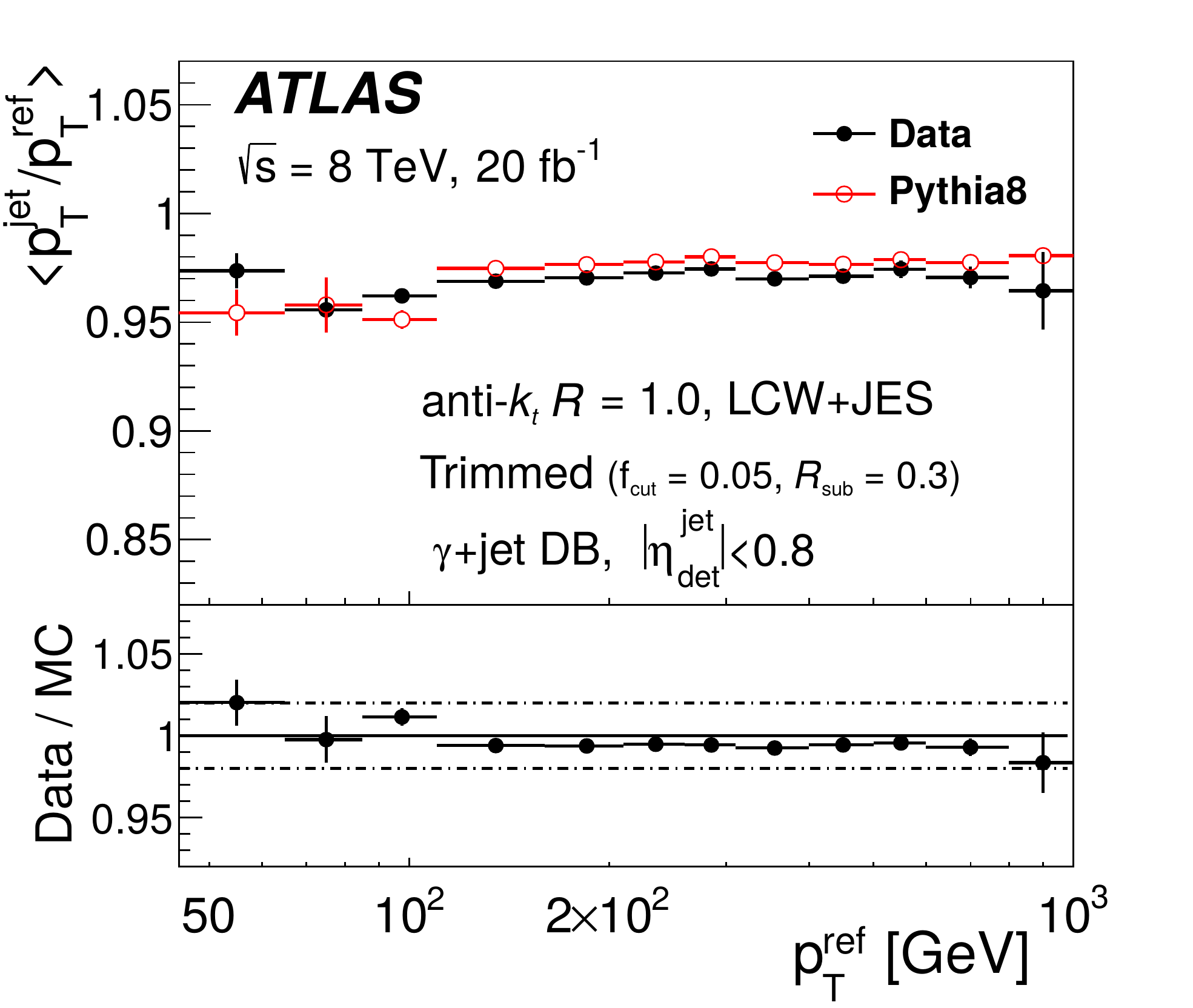}
\caption{\RDB{} vs \pt{}, \AetaRange{0.8}}
\end{subfigure}
\begin{subfigure}{0.48\linewidth}\centering
\includegraphics[width=\textwidth]{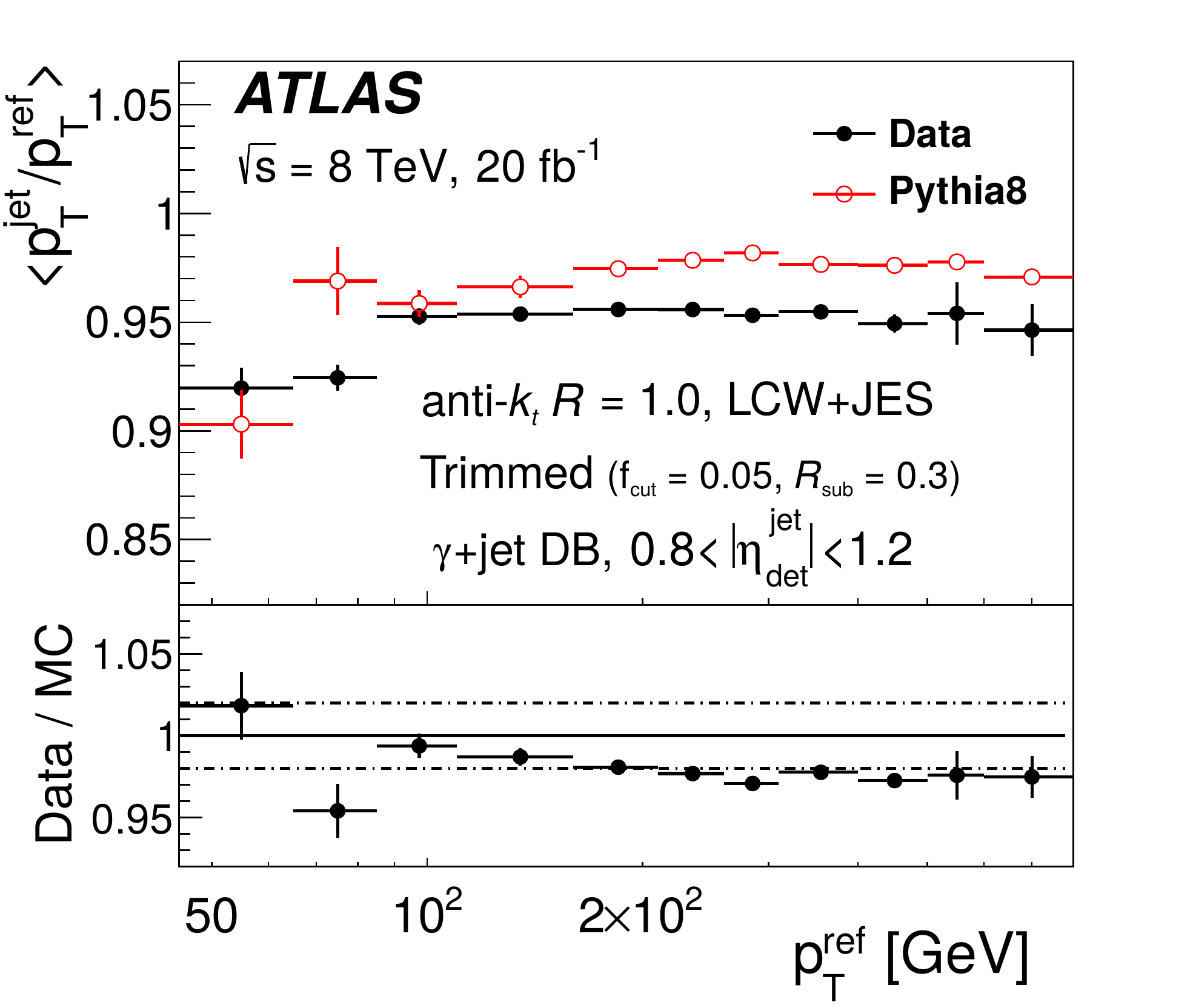}
\caption{\RDB{} vs \pt{}, \etaRange{0.8}{1.2}}
\end{subfigure}
\caption{
\label{fig:RDB-vs-pt-largeR}
\RDB{} for trimmed \antikt{} jets with \rten{} calibrated with the \LCWJES{} scheme for
(a)~\AetaRange{0.8} and (b)~\etaRange{0.8}{1.2}
for both data (filled circles) and MC simulation (empty circles), as a function of the \ptref.
Only statistical uncertainties are shown.
}
\end{figure}
 
\subsubsection{Systematic uncertainties}
\label{sec:lRjet-syst}
Most of the systematic uncertainties are evaluated in the same way as for \smallR{} jets as detailed in Section~\ref{sec:vjets-syst}.
Additional uncertainties specific to \lRjets{} and changes to the evaluation of some of the uncertainty sources are detailed below.
 
\begin{itemize}
\item
Rather than using the difference of the \DtM{} ratio of \RDB{} from unity as a calibration, it is instead taken as an uncertainty.  This allows a straightforward combination with the procedure used to derive uncertainties outside of the kinematic range for which the \gammaLRjet{} \RDB{} can be derived.  This is a significant uncertainty source, especially for jets with \etaRange{0.8}{1.2}.
\item
The OOC uncertainty is evaluated only for \lRjets{} with \AetaRange{0.8}{}, since for other \etaDet{} bins the \lRjets{} are not always fully contained within the tracking acceptance.
The uncertainty derived in this central \etaDet{} range is also applied to the more forward $|\etaDet{}|$ bins.
Due to the large radius (\rten{}), out-of-cone effects are very small, and the uncertainty is negligible for $\pT{}>100$~\GeV{}.
\item
As mentioned in Section~\ref{sec:vjets-jet-sel}, the subleading jet (labelled ``j2'') that is used to suppress additional QCD radiation is required to have an angular separation of $\Delta R(\text{j1},\text{j2})>0.8$ from the \lRjet{} (``j1''). Since the leading jet has \rten{} while the subleading jet has \rfour{}, this means that there is a significant overlap in terms of solid angle, but since the \pt{} profile of jets tend to be narrow (see the \trackWIDTH{} distribution of Figures~\ref{fig:inputs_dataMC_comparison_LowPt} and~\ref{fig:inputs_dataMC_comparison_HighPt}), the amount of energy sharing is still expected to be small.
The assigned uncertainty component is evaluated by changing the $\Delta R$ requirement from 0.8 to 1.4 to ensure that there is strictly no overlap between the two jets.
\item
Since the \sRjets{} are reconstructed independently of the \lRjets{} using the same \topos{} as input, a \lRjet{} will sometimes contain two \sRjets{} close to the \lRjet{} axis.
It is possible that events with such topologies have additional uncertainties due to the QCD modelling.
To assess this effect, an alternative subleading jet selection was applied in which ``j2'' is defined simply as the subleading $R=0.4$ jet, without any restriction based on the angle to the \lRjet{}. This means that ``j2'' will sometimes be within the \lRjet{} and sometimes not (the leading $R=0.4$ jet tends to be aligned very close to the \lRjet{} axis).
With this definition, the event selection $\ptsub{}<0.1\,\ptref$ was applied in place of the standard \ptsub{} selection, and an uncertainty component was derived from the impact of this variation.
\item
An additional dependence of the jet response for \lRjets{} on the ratio of the jet mass to the \pt{}, \MoverPT, is observed, particularly for large $|\etaDet|$. 
A systematic uncertainty is assigned to account for this dependence, derived as a triple ratio.  The data/MC ratios of the \RDB{} ratios are evaluated in the two $\MoverPT{}$ ranges shown in Figure~\ref{fig:balance_moverpt}, corresponding to $\MoverPT{}<0.15$ and $\MoverPT{}>0.15$. The systematic uncertainty is given by the ratio of the double ratios obtained for the two $\MoverPT{}$ ranges.
\end{itemize}
 
\begin{figure}[htb]
\centering
\begin{subfigure}{0.48\linewidth}\centering
\includegraphics[width=\textwidth]{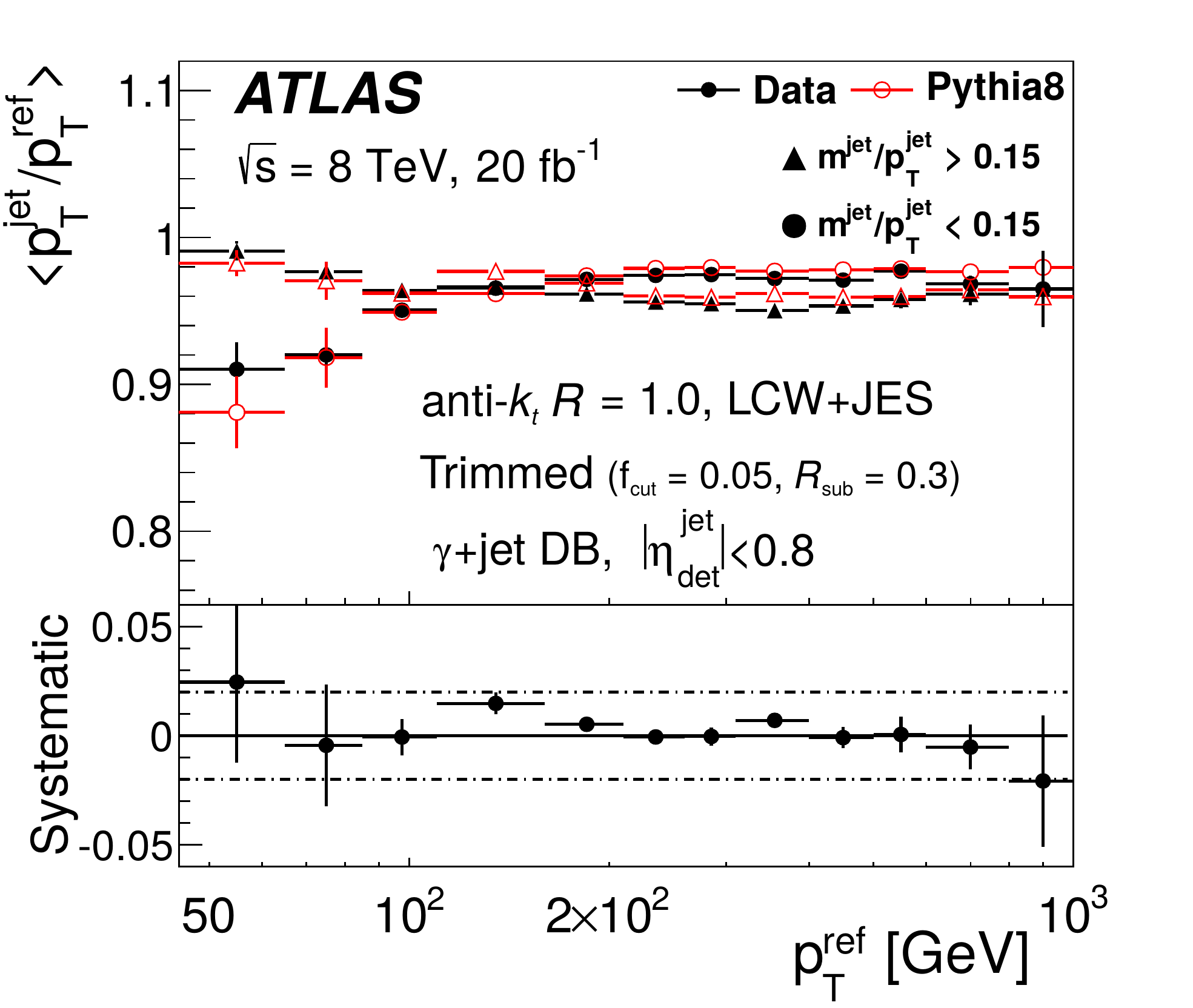}
\caption{\LRjet{} mass dependence, \AetaRange{0.8}}
\end{subfigure}
\begin{subfigure}{0.48\linewidth}\centering
\includegraphics[width=\textwidth]{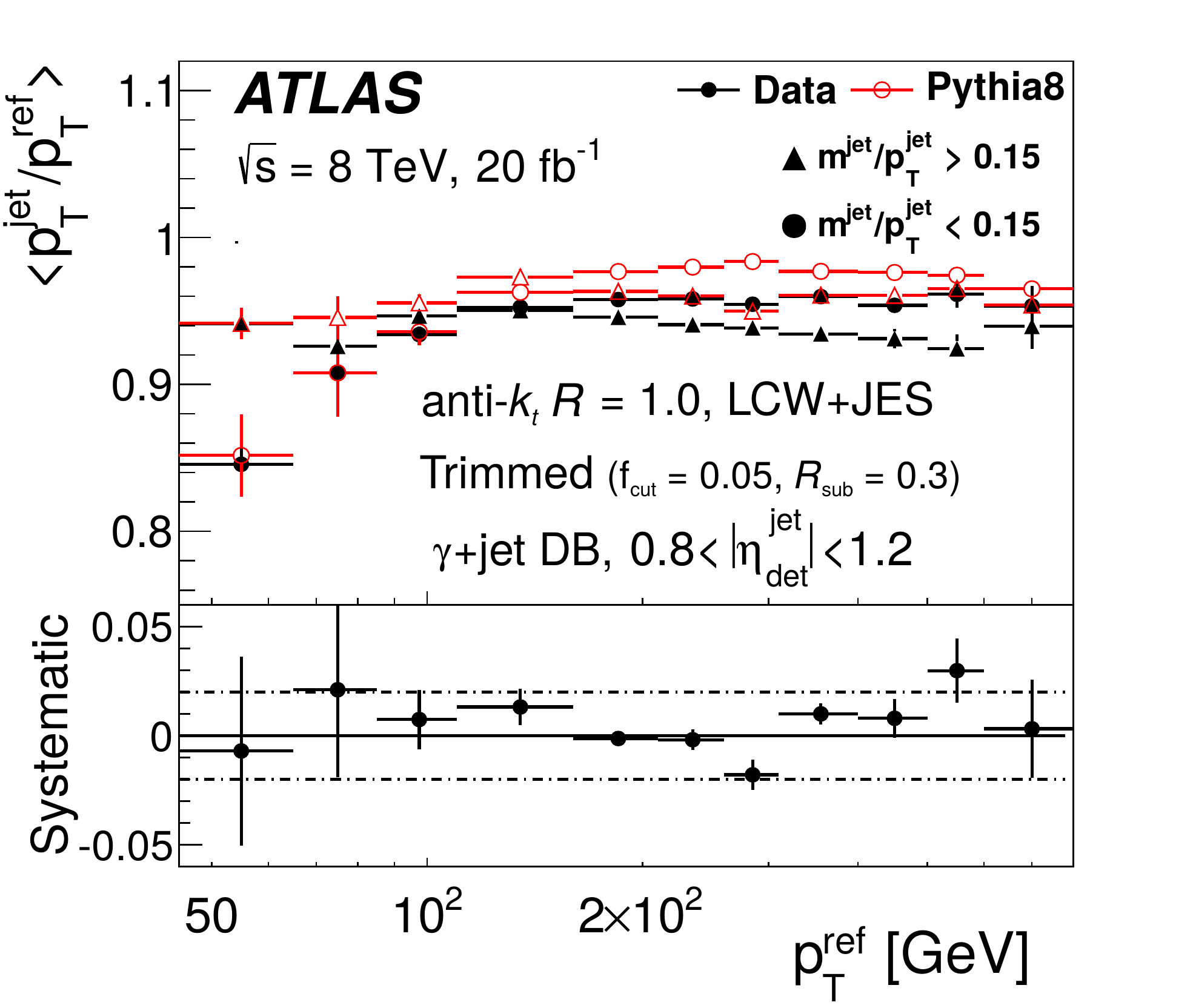}
\caption{\LRjet{} mass dependence, \etaRange{0.8}{1.2}}
\end{subfigure}
\caption{
\RDB{} as a function of \ptref{} measured for trimmed \antikt{} jets with \rten{} in \gamjet{} events shown separately for jets with (a)~\AetaRange{0.8} and (b)~\etaRange{0.8}{1.2}.
Separate results are shown for jets with $m/\pt<0.15$ and $m/\pt>0.15$, displayed with circles and triangles, respectively.
Measurements in data are shown as filled markers and MC predictions as open makers.
Statistical uncertainties are shown for each point.
The lower parts of the figures show the systematic uncertainty evaluated as the \DtM{} ratio of the ratios of \RDB{} extracted in the two \MoverPT{} ranges.
\label{fig:balance_moverpt}
}
\end{figure}
 
\subsubsection{\Pileup{} uncertainty for \lRjets}
 
As discussed in Section~\ref{sec:jetReco}, \lRjets{} do not receive any \pileup{} correction.
Due to the trimming algorithm applied, \lRjets{} are significantly less sensitive to \pileup{} than standard \sRjets{}.
To study the impact of \pileup{} on the \lRjet{} \pt{}, it is measured as a function of \Npv{} and \avgmu{} in bins of \pt{} of \trackjets{} that are matched to the \lRjets{} being probed. \Trackjets{} are resilient to \pileup{} since they are built from inner-detector tracks that are matched only to the primary vertex, and do not contain contributions (tracks) from \pileup{} vertices (in most cases). The \trackjets{} are reconstructed using the same algorithm as the calorimeter \lRjets{} (trimmed \antikt{} \rten{}).
 
\begin{figure}[!htb]
\centering
\begin{subfigure}{0.48\linewidth}\centering
\includegraphics[width=\textwidth]{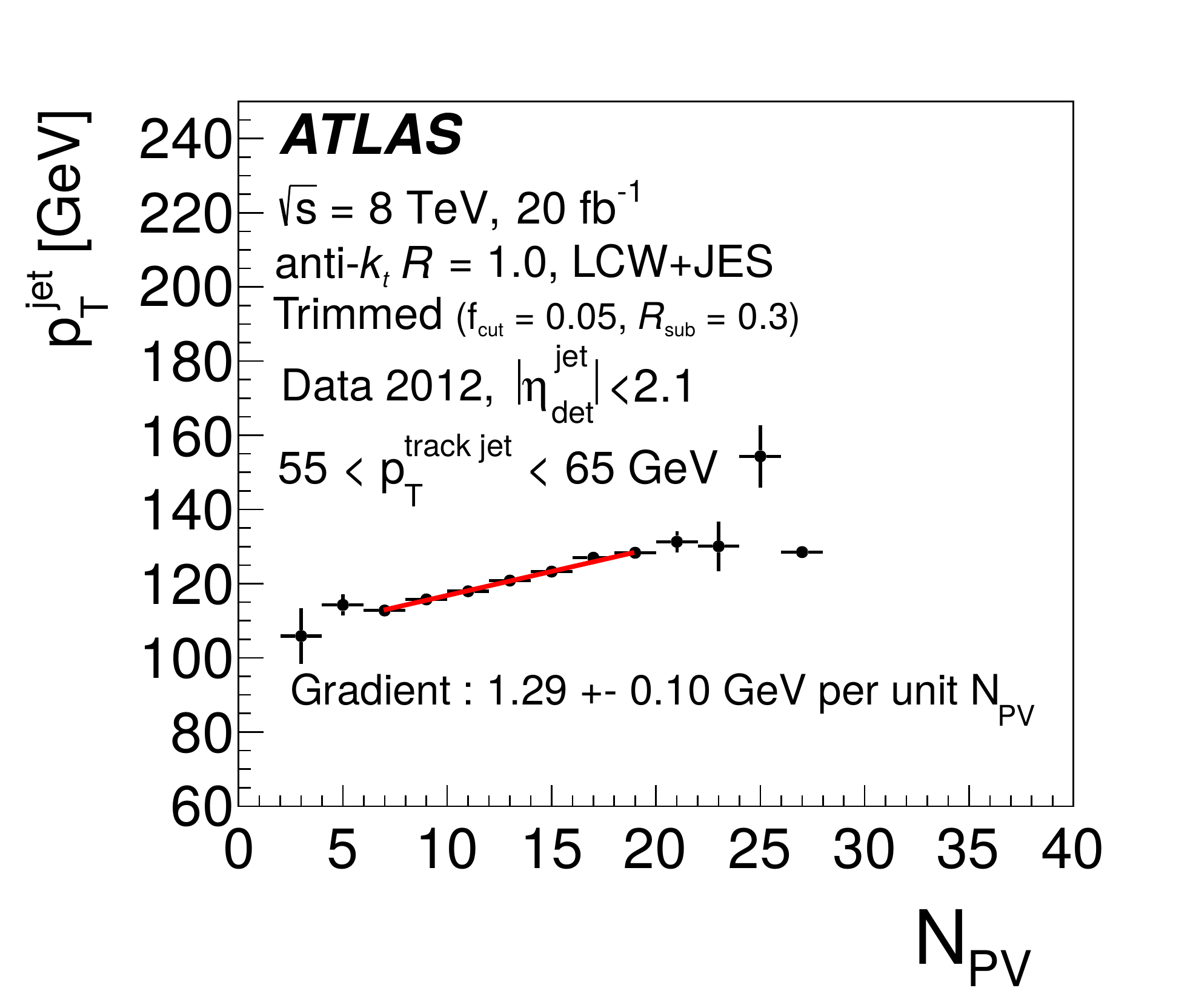}
\caption{\LRjet{} \pt{} vs \Npv{}} \label{sfig:PUa}
\end{subfigure}
\begin{subfigure}{0.48\linewidth}\centering
\includegraphics[width=\textwidth]{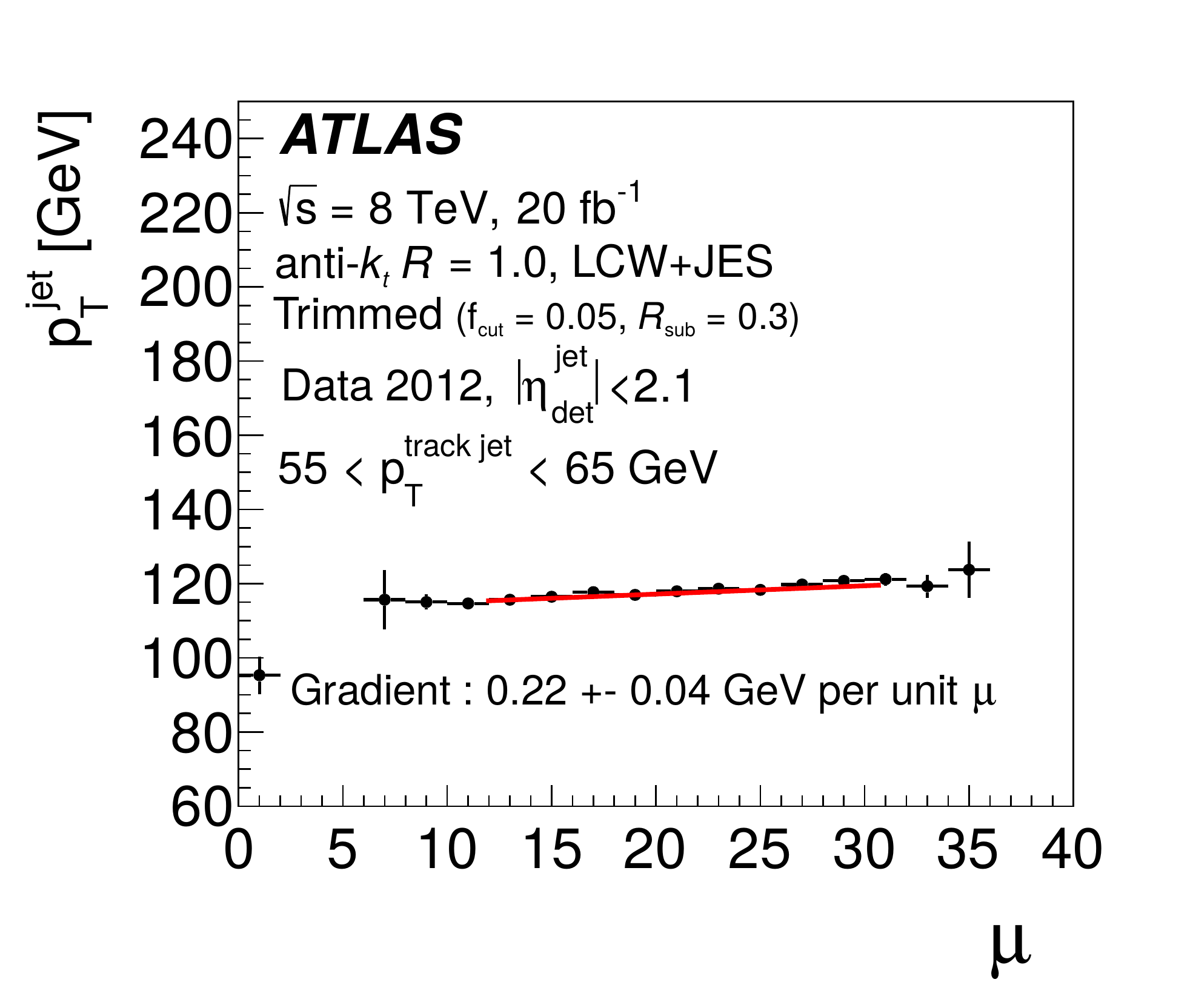}
\caption{\LRjet{} \pt{} vs \avgmu{}} \label{sfig:PUb}
\end{subfigure}
\caption{
\LRjet{} \pt{} as a function of (a)~\Npv{} in \gamjet{} events with  $20<\avemu<22$, and as a function of (b)~\avemu{} in events with $10<\Npv<12$.
The \pTtrkjet{} is required to be within  $55< \pTtrkjet <65$~\GeV{}.
The lines are linear fits. The markers show data with error bars corresponding to the statistical uncertainties.
The lowest point shown in (b) corresponds to special data-taking conditions, and is thus not expected to match the other points on the plot.
\label{fig:lRjet-pt-vs-pileup}
}
\end{figure}
 
Within a given \trackjet{} \pt{} bin, the \lRjets{} are expected to have a similar \tjet{} \pt. The reconstructed \pt{} is studied as a function of \NPV{} and \avgmu{}.
The results for a representative \trackjet{} \pt{} bin is presented in Figure~\ref{fig:lRjet-pt-vs-pileup}.
As expected (Eq.~(\ref{eq:PU}) and Ref.~\cite{PERF-2012-01}) there is a linear dependence of the jet \pt\ on both \NPV{} and \avgmu{}.
For each \trackjet{} \pt{} bin, the ``gradients'' \gradNPV{} and \gradMu{} are extracted from the slopes of a linear fits of \pt{} vs \NPV{} and \pt{} vs \avgmu{} (Figure~\ref{fig:lRjet-pt-vs-pileup}).
Figure~\ref{fig:lRjet-PUgradient-vs-pt} shows these gradients as a function of the average \pt{} of the \lRjets{}. Both of these graphs are well described with a function of the form $a+b\,\log{(\pt/p_\text{T,0})}$, where the parameters $a$ and $b$ are extracted from a fit and $p_\text{T,0}$ is a constant chosen to be 50~\GeV.
 
\begin{figure}[!htb]
\begin{center} 
\begin{subfigure}{0.48\linewidth}\centering
\includegraphics[width=\textwidth]{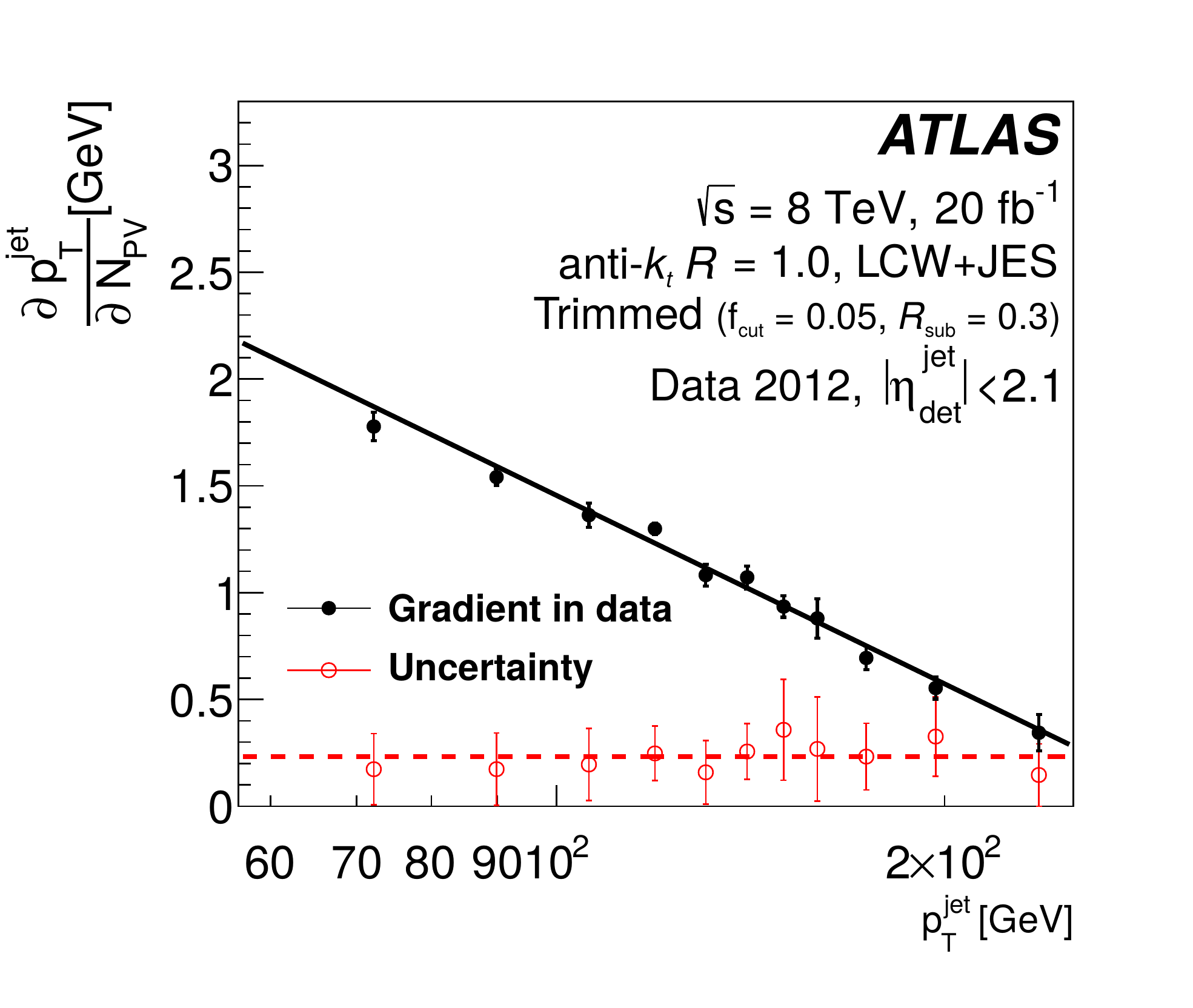}
\caption{\gradNPV{} vs \lRjet{} \pt} \label{sfig:SLa}
\end{subfigure}
\begin{subfigure}{0.48\linewidth}\centering
\includegraphics[width=\textwidth]{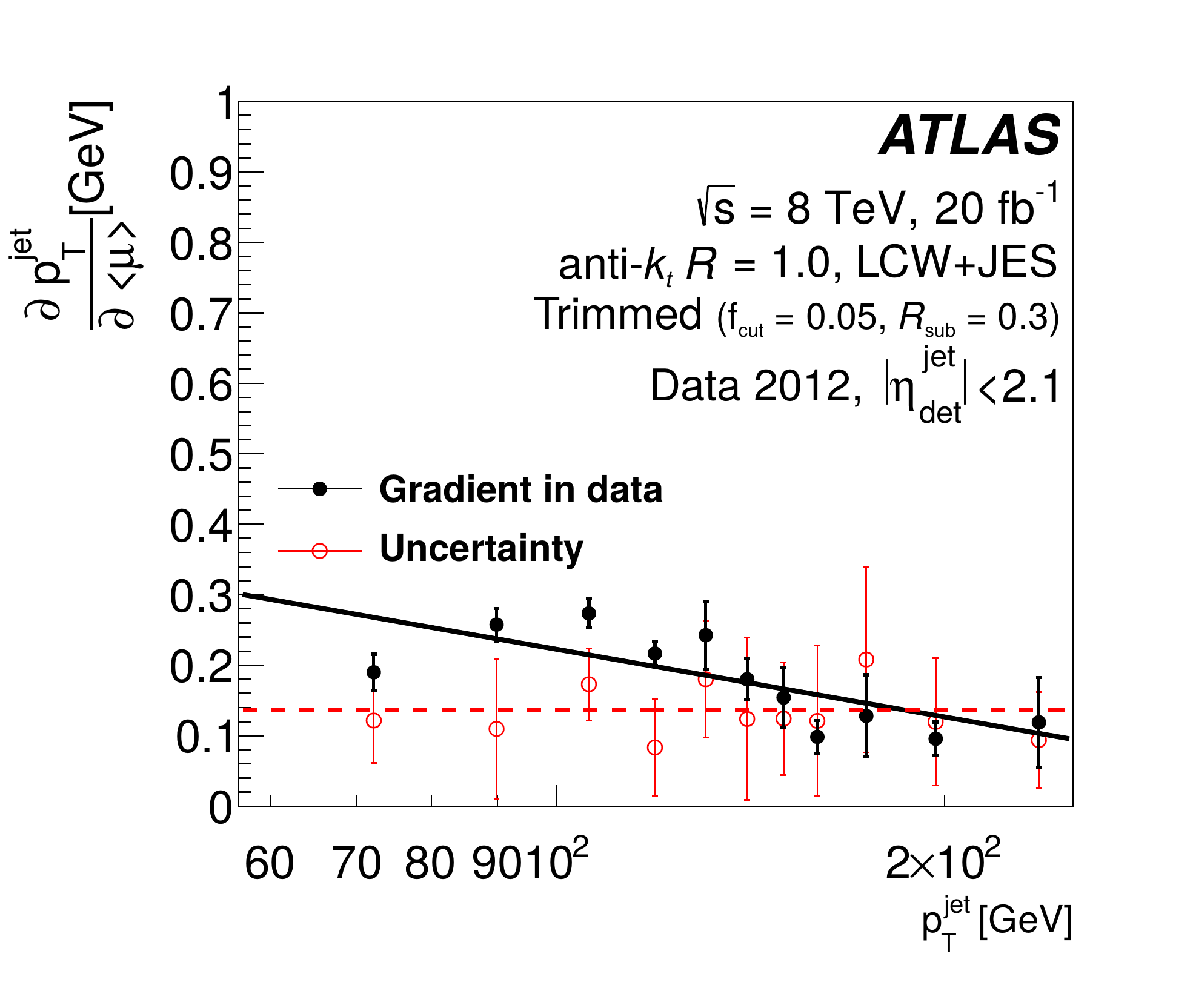}
\caption{\gradMu{} vs \lRjet{} \pt} \label{sfig:SLb}
\end{subfigure}
\end{center}
\caption{The gradients (a)~\gradNPV{} and (b)~\gradMu{} extracted in
\gamjet{} data~(filled circles) and the difference of the gradients measured in data and MC simulations (empty circles)
as functions of the \largeR{} calorimeter jet \pt{}.
The lines correspond to fits of a constant to the uncertainty (red dashed line) and a function of the form $a+b\,\log(\pt/p_\text{T,0})$ with $p_\text{T,0}=50$~\GeV\ for the gradients vs \pt{}~(black solid line). The error bars shown on the markers reflect the statistical uncertainty.
\label{fig:lRjet-PUgradient-vs-pt}
}
\end{figure}

Based on the \pt{} parameterization of the gradients from the fits to data described in the previous paragraph (Figure~\ref{fig:lRjet-PUgradient-vs-pt}), two uncertainty components are derived that have the following impact on the jet \pt{}
\begin{eqnarray}
\Delta_{\NPV} = ( \gradNPV )(\NPV - \NPV^\text{ref}) \qquad \textrm{ and } \qquad
\Delta_{\avgMu} = ( \gradMu )(\avgmu - \avgmu^\text{ref}),
\label{eq:lRjet-PU-unc}
\end{eqnarray}
where \gradNPV{} and \gradMu{} are the gradients parameterized as a function of \lRjet{} \pt{} according to the fitted functions,
\avgmu{} is the average number of interaction per bunch crossing, and \NPV{} is the number of primary vertices of the event, and
${\avemu}^\textrm{ref}=20.7$ and $\Nref=11.8$ are the average values for  the full $2012$ \gamjet{} dataset.
As can be seen from Eq.~(\ref{eq:lRjet-PU-unc}), the impact on the jet \pt{}  from the two uncertainty components can change sign depending on the amount of \pileup{}, and become zero for jets produced in events with  \pileup{} conditions matching the 2012 average values.
The resulting fractional \pt{} uncertainties are presented for two values of \lRjet{} \pt{} in Figure~\ref{fig:lRjet-pileup-unc}.
 
\begin{figure}[!htb]
\begin{center}
\begin{subfigure}{0.48\linewidth}\centering
\includegraphics[width=\textwidth]{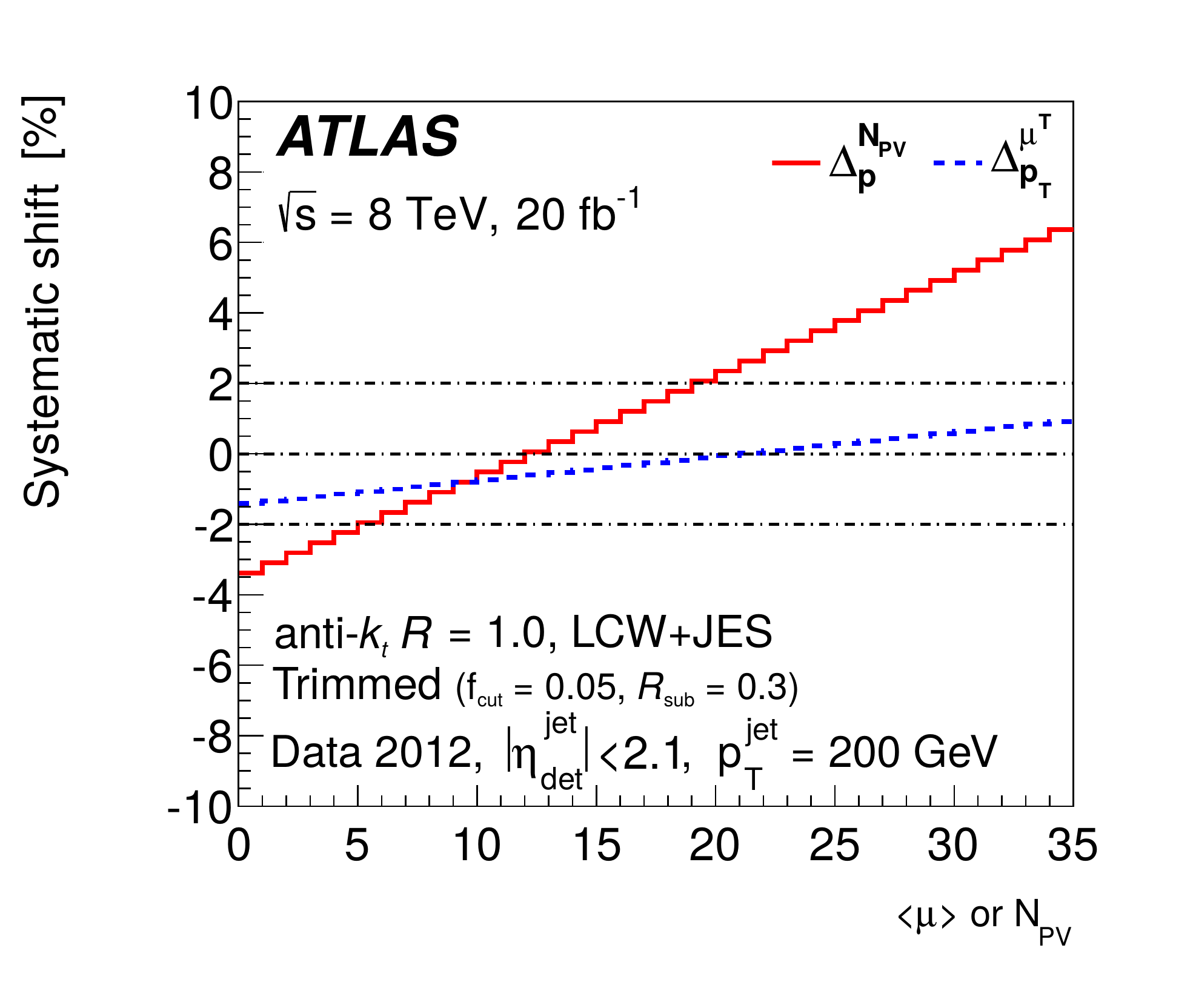}
\caption{\LRjet{} \pileup{} uncertainties, $\pt = 200$~\GeV }
\end{subfigure}
\begin{subfigure}{0.48\linewidth}\centering
\includegraphics[width=\textwidth]{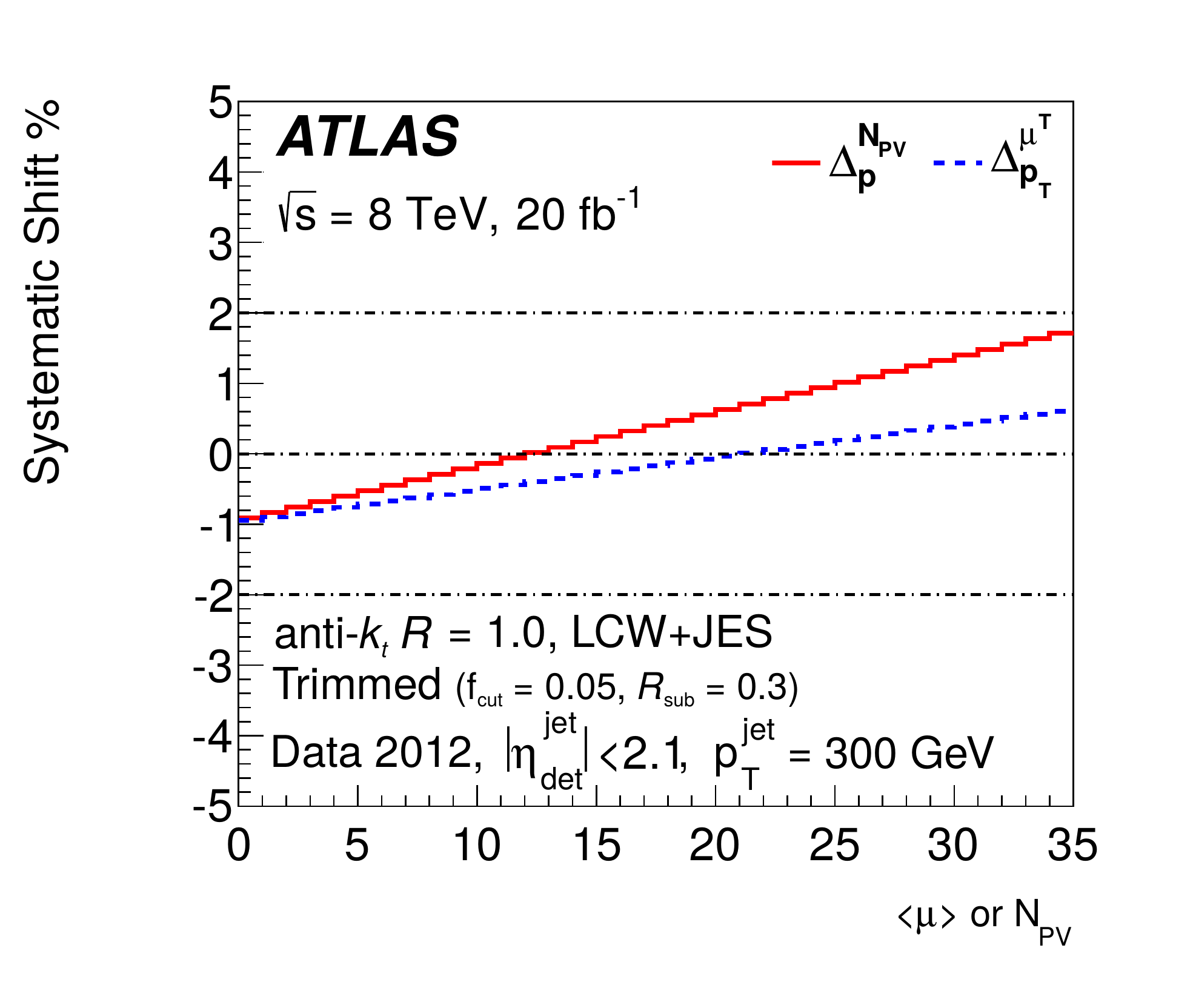}
\caption{\LRjet{} \pileup{} uncertainties, $\pt{}=300$~\GeV }
\end{subfigure}
\end{center}
\caption{
\label{fig:lRjet-pileup-unc}
Systematic uncertainties due to the \NPV{} and \avgmu{} dependence of the \lRjet{} \pt{} presented in percentage as functions of
\avemu{}~(dashed lines) and \Npv{}~(solid lines) for \lRjets{} with (a)~$\pt = 200$~\GeV{} and (b)~$\pt{}=300$~\GeV.  Horizontal lines are added at 0\% and $\pm$2\% to guide the eye.
}
\end{figure}

\begin{figure}[p]
\centering
 
\begin{subfigure}{0.48\linewidth}\centering
\includegraphics[width=\textwidth]{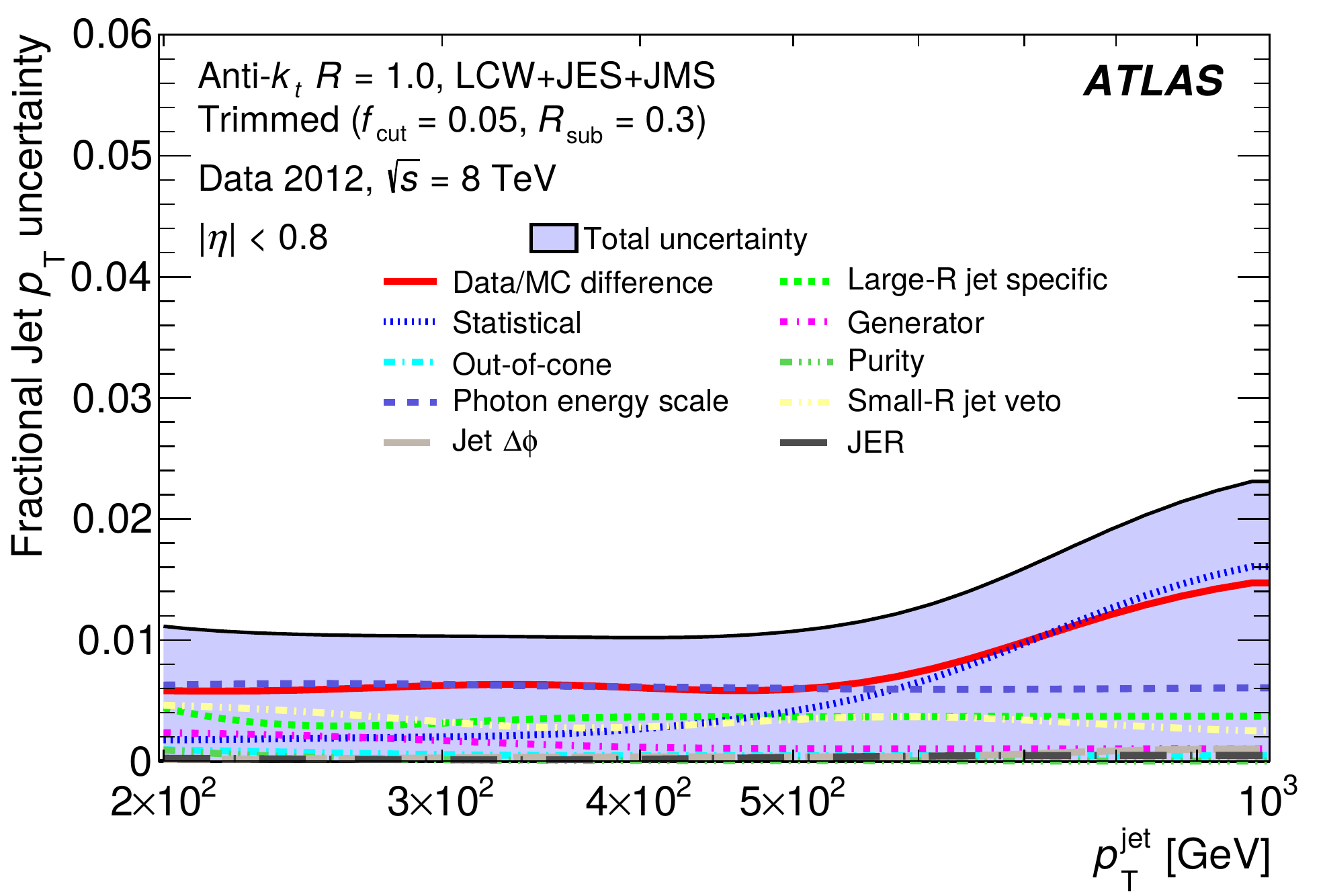}
\caption{\LRjet{} JES uncertainty, \AetaRange{0.8}}
\end{subfigure}
\begin{subfigure}{0.48\linewidth}\centering
\includegraphics[width=\textwidth]{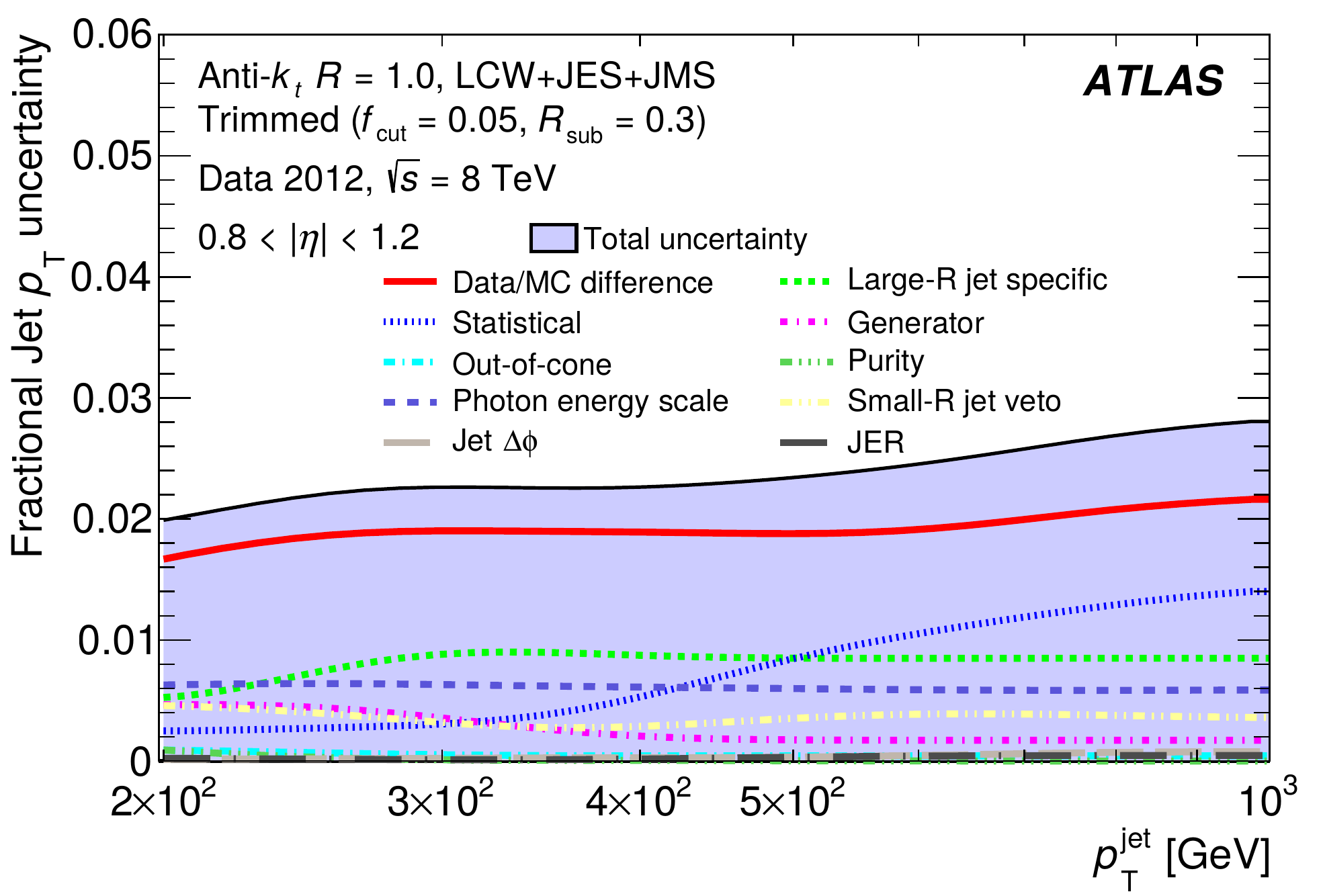}
\caption{\LRjet{} JES uncertainty, \etaRange{0.8}{1.2}}
\end{subfigure} \\
 
\bigskip
 
\begin{subfigure}{0.48\linewidth}\centering
\includegraphics[width=\textwidth]{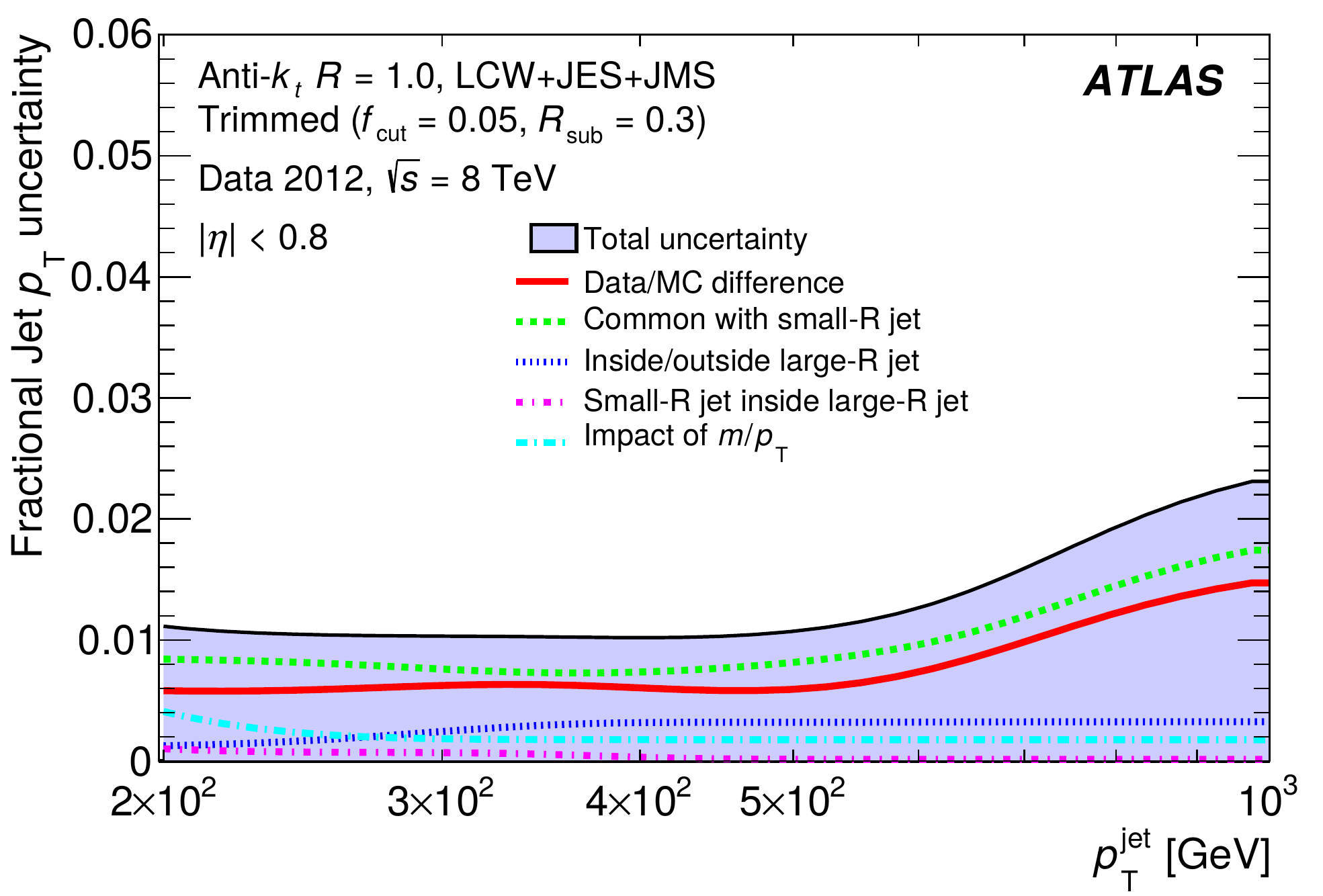}
\caption{\LRjet{} JES uncertainty, \AetaRange{0.8}}
\end{subfigure}
\begin{subfigure}{0.48\linewidth}\centering
\includegraphics[width=\textwidth]{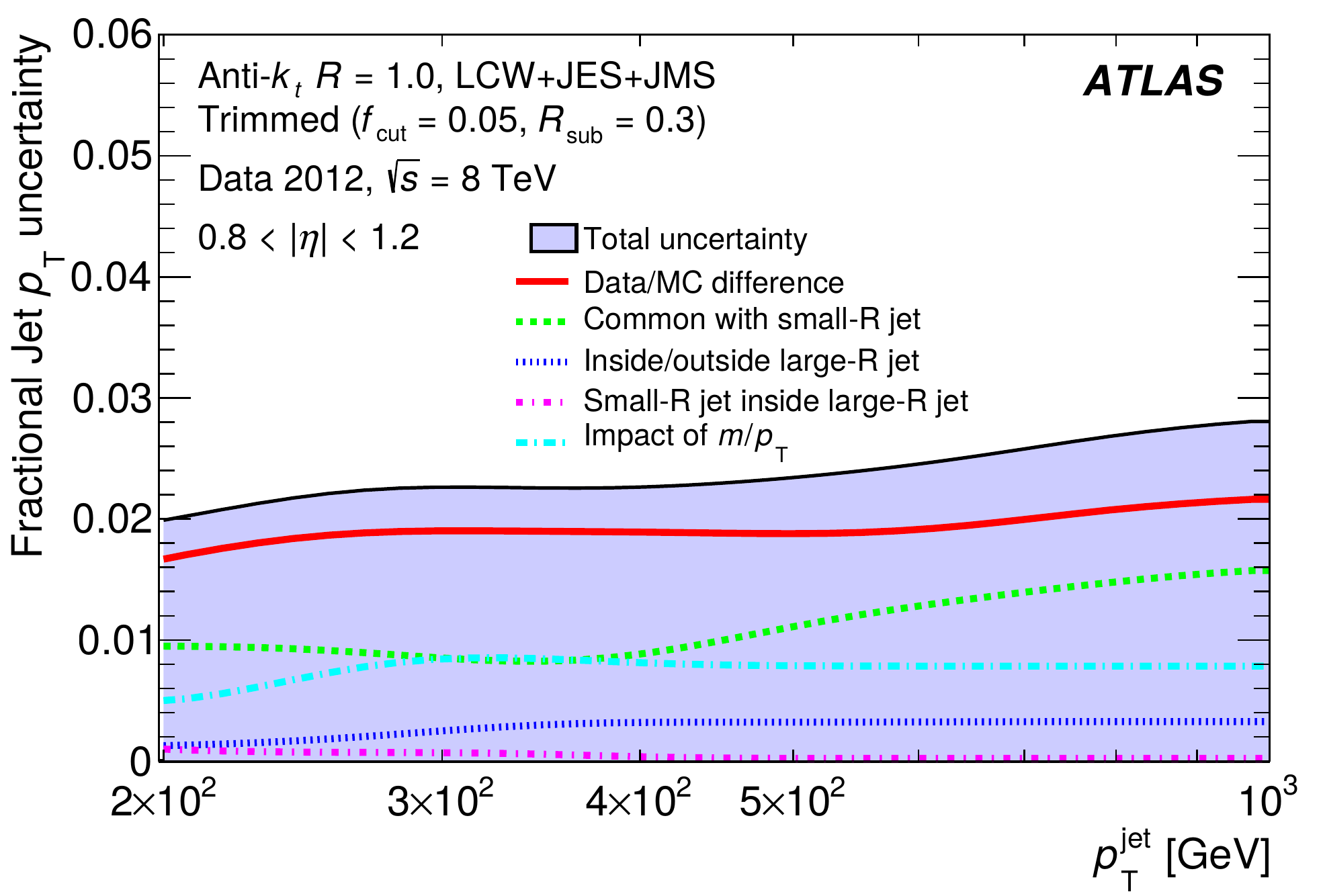}
\caption{\LRjet{} JES uncertainty, \etaRange{0.8}{1.2}}
\end{subfigure}

\caption{Statistical and systematic uncertainties in the \DtM{} ratio of \RDB{} for trimmed \rten{} \antikt{} jets calibrated with the \LCWJES{} scheme with
(a,c)~\AetaRange{0.8} and (b,d)~\etaRange{0.8}{1.2}.
The uncertainties are evaluated in bins of \ptref{}, converted to jet \pT{}, and translated into a function using a Gaussian kernel smoothing.
Figures (a) and (b) highlight the systematic uncertainty components derived analogously (and hence due to the same effects) to those for \sRjets{} (Figure~\ref{fig:gamjet-uncert}).
In Figures (c) and (d), these uncertainties are added in quadrature, and the uncertainties specific to \lRjets{} are instead displayed separately.
In all figures, the difference of the \DtM{} \RDB{} from unity is taken as an additional uncertainty rather than being applied as a calibration; this uncertainty is shown as a solid line.
\label{fig:lRjet-uncert-summary}
}
\end{figure}
\paragraph{Summary of systematic uncertainties}~\\
Figure~\ref{fig:lRjet-uncert-summary} presents a summary of the statistical and systematic uncertainties in the \lRjet{} \pt{} from the DB analysis, including a detailed breakdown of the uncertainty components that are in common with the \sRjet{} \gamjet{} measurements presented in Section~\ref{sec:vjets-syst}, while the additional uncertainty sources specific to \lRjets{} are presented in Section~\ref{sec:lRjet-syst}.
The total uncertainty for $|\eta|<0.8$ is found to be $\sim$1\% above 150~\GeV{}, rising to $\sim$2\% at 1~\TeV{}.
At larger $|\eta|$, the uncertainty increases to $\sim$2\% at low \ptref{}, rising to $\sim$3\% at 1~\TeV.
The uncertainties are dominated by the photon energy scale uncertainty, the uncertainty coming from the \lRjet{} response dependence on the ratio of \MoverPT{}, and the difference of the \DtM{} \RDB{} from unity.
The generator systematic uncertainty becomes dominant for $|\eta|>1.2$.

\subsection{Measurement of the jet energy resolution using the DB method}
\label{sec:vjets-JER}
 
The width of the DB distribution in a given \ptref{} bin is used to probe the JER.
The detector resolution of the reference object is negligible compared with that of the jet, so the method to measure the JER using \Zjet{} and \gamjet{} events is significantly simpler than that for dijets described in Section~\ref{sec:dijet}.
The event selection and binning is the same as for the \RDB{} measurements, but instead of determining the mean \RDB{} of the \rDB{} distribution within each \ptref{} bin, the width $\sigma_\text{DB}^\text{reco}$ is extracted as the standard deviation of the same Modified Poisson fit.
The relative JER $\sigma_E/E$ is then estimated using
\begin{equation}
\frac{\sigma_E}{E} = \frac{\sigma_{\pt}}{\pt} = \sigma_\text{DB}^\text{reco} \ominus \sigma_\text{DB}^\text{truth},
\label{eq:vjets-JER}
\end{equation}
where the first equality holds to a good approximation since the contribution from the angular resolution is negligible, and the second relation follows from the same reasoning as for Eq.~(\ref{eq:ptcl-subtraction}) (Section~\ref{sec:dijet-balance}).
The parameter $\sigma_\text{DB}^\text{truth}$ is obtained using a fit to the $\pttruth/\ptref$ distribution extracted using MC simulation with same selection (applied to reconstructed jets) as for the DB measurement. For each simulated event, \pttruth{} is defined from the \tjet{} that is ghost-matched (Section~\ref{sec:jetMatch}) to the leading reconstructed jet.
The simulated JER is also extracted from the MC samples with fits to $\pt^\text{reco}/\pttruth$ (Section~\ref{sec:jetMatch}).
 
\begin{figure}[!htb]
\centering
\begin{subfigure}[b]{0.48\linewidth}\centering
\includegraphics[width=\textwidth]{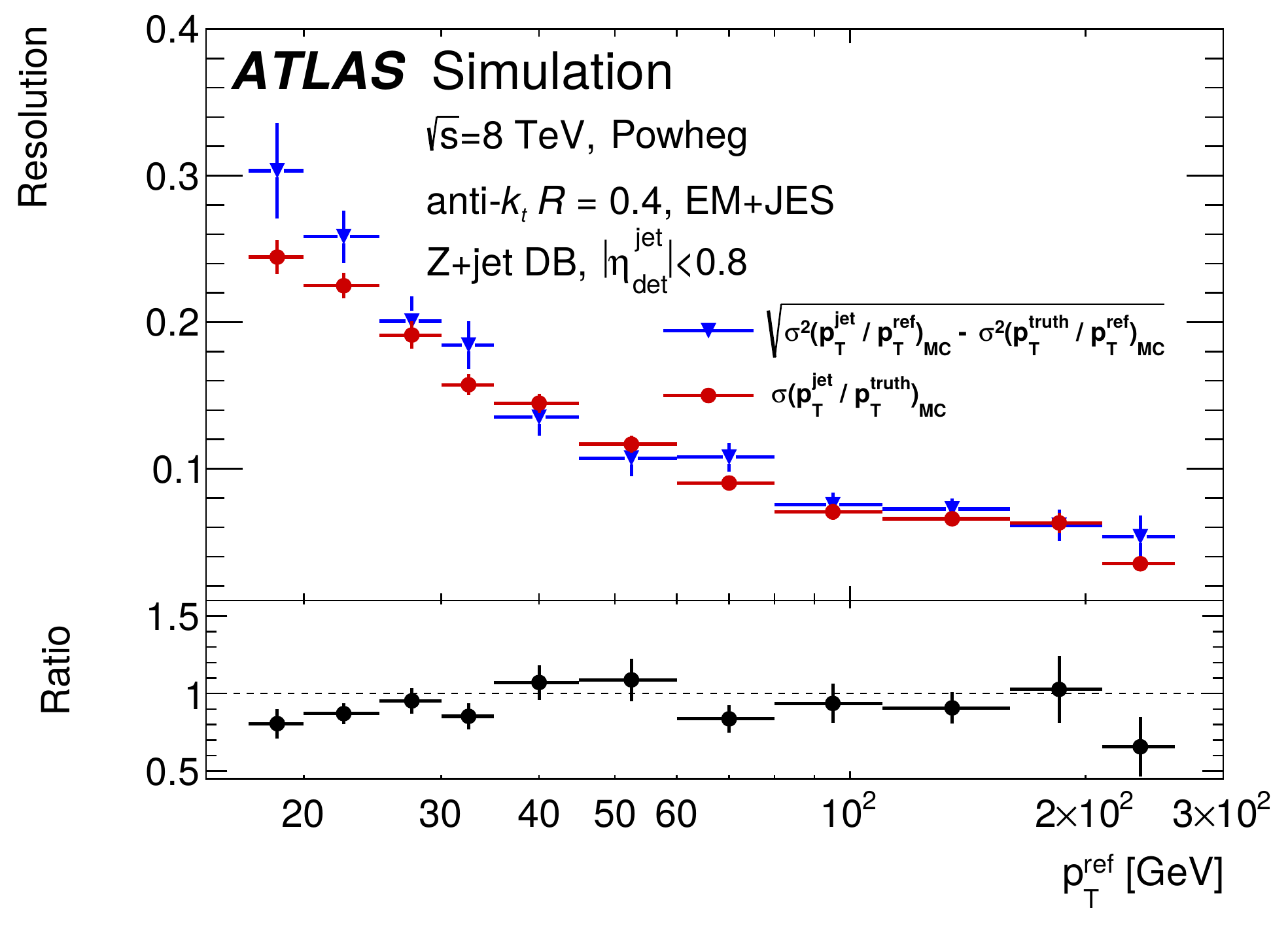}
\caption{\Zjet}
\end{subfigure}
\begin{subfigure}[b]{0.475\linewidth}\centering
\includegraphics[width=\textwidth, trim=20 0 10 0,clip]{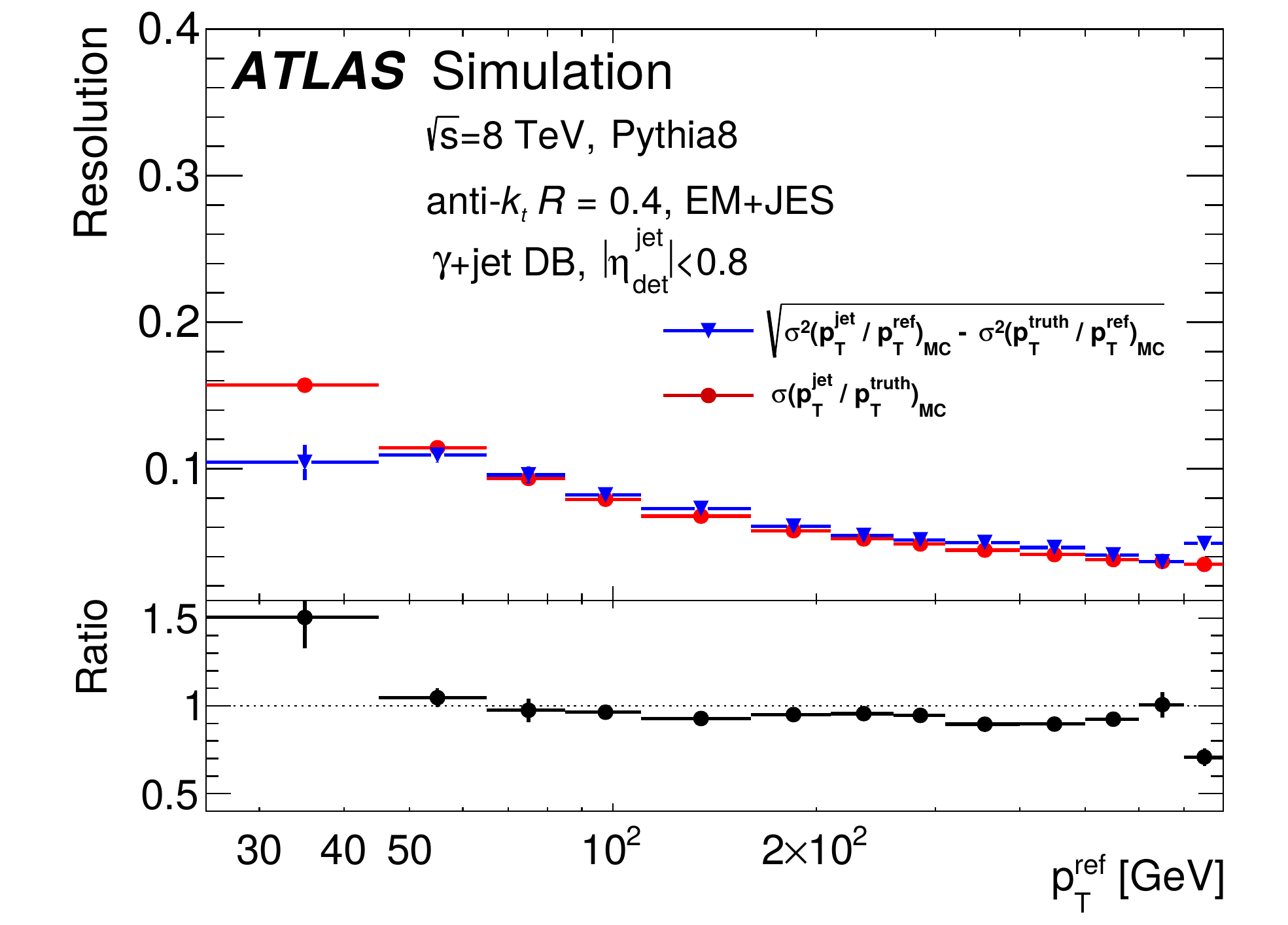}
\caption{\gammajet}
\end{subfigure}
\caption{
\label{fig:vjets-JER-closure}
Comparison of the jet energy resolution determined \insitu{}~(triangles) with the MC jet energy resolution~(circles) measured for \antikt{} \EMJES{} jets with \rfour{}, in (a) \Zjet{} and (b) \gammajet{} events. The bottom frame shows the ratio of the two.
}
\end{figure}

Figure~\ref{fig:vjets-JER-closure} presents a MC-based comparison between the relative JER
obtained using the \insitu{} technique applied to the simulated events
(Eq.~(\ref{eq:vjets-JER})) and the relative JER extracted from \tjet{} matching.
Over most of the \pt{} range probed, the \insitu{} extracted JER agrees with the expectation from simulation within 10\%, but for the first and last bins probed, the agreement is worse (20\%--40\%). The difference between the measured and expected JER is taken as a ``non-closure'' systematic uncertainty.
 
The \tjet{} DB width $\sigma_\text{DB}^\text{truth}$, used in the JER measurement (Eq.~(\ref{eq:vjets-JER})) depends on details of the physics model implemented  in the MC generator. A systematic ``generator'' uncertainty is evaluated to assess this dependence through taking the difference between the extracted JER using different MC generators.
Other systematic uncertainty sources considered for the \JER{} measurement are the same as those considered for the \JES{} measurements discussed in Section~\ref{sec:vjets-syst} and are derived for the JER analogously.
Figure~\ref{fig:vjets-JER-unc} presents the resulting JER uncertainties for \antikt{} \rfour{} jets and also the relative \DtM{} difference of the JER measurement.
Results are reported as a function of \ptref{}, separately for the measurements performed using the \Zjet{} and \gamjet{} datasets.
The total, relative \JER{} uncertainty evaluated for the \Zjet{} JER measurements varies between 20\% and 40\%, depending on the algorithm and \ptref{} values.
Dominant sources of uncertainties include the choice of MC generator for the modelling of $\sigma_\text{DB}^\text{truth}$, the non-closure, and limited statistics.
The \JER{} uncertainty of the \gamjet{} measurement is slightly smaller than that from \Zjet{} events, varying between 10\% and 30\%.
The dominant sources of systematic uncertainties are the choice of MC generator and the suppression of additional radiation.
For $\ptref{}<50$~\GeV, the \DtM{} differences can be as large as $\sim$40\%.
 
\begin{figure}[!htb]
\centering
\begin{subfigure}[b]{0.48\linewidth}\centering
\includegraphics[width=\textwidth]{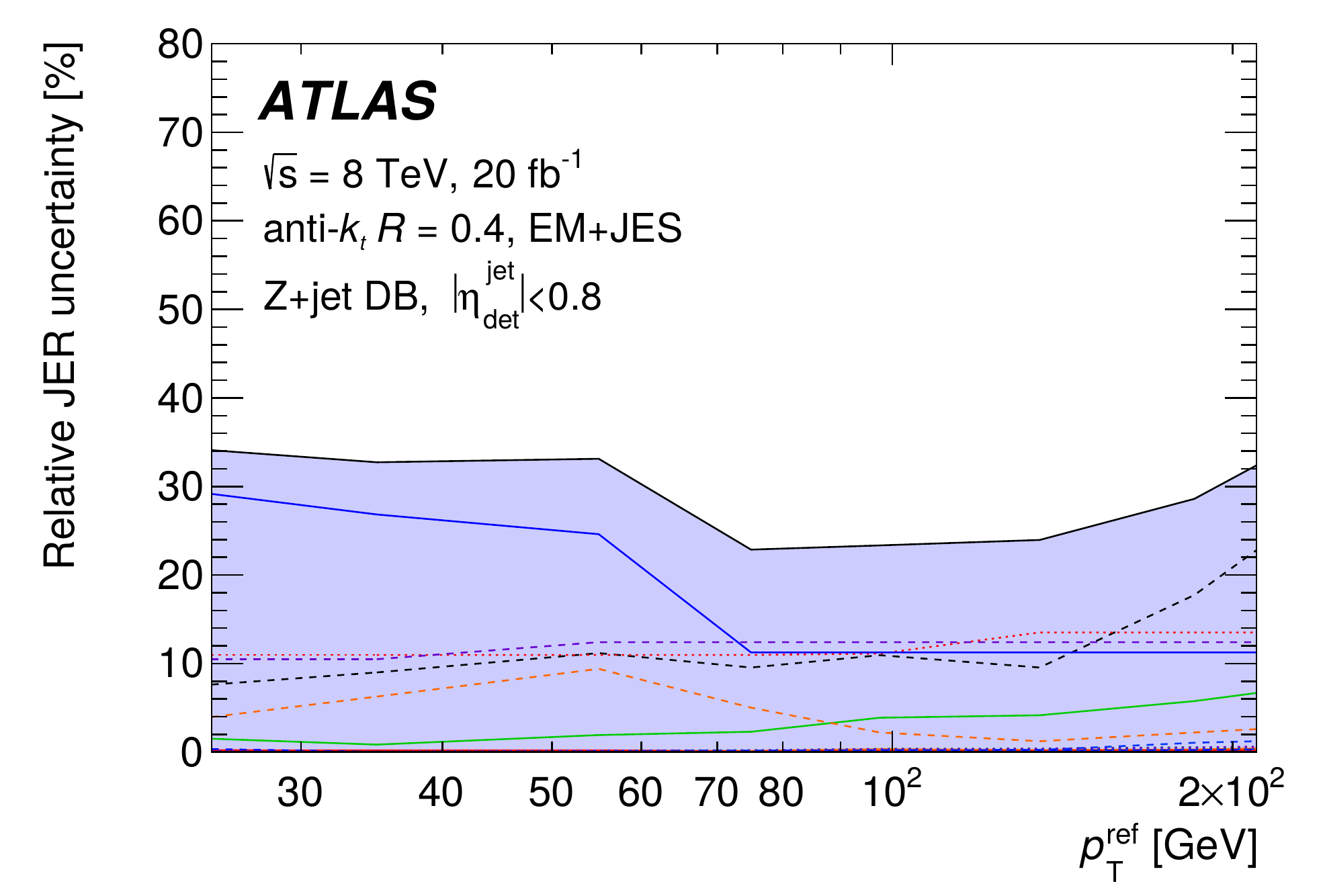}
\end{subfigure}
\begin{subfigure}[b]{0.44\linewidth}\centering
\includegraphics[width=\textwidth]{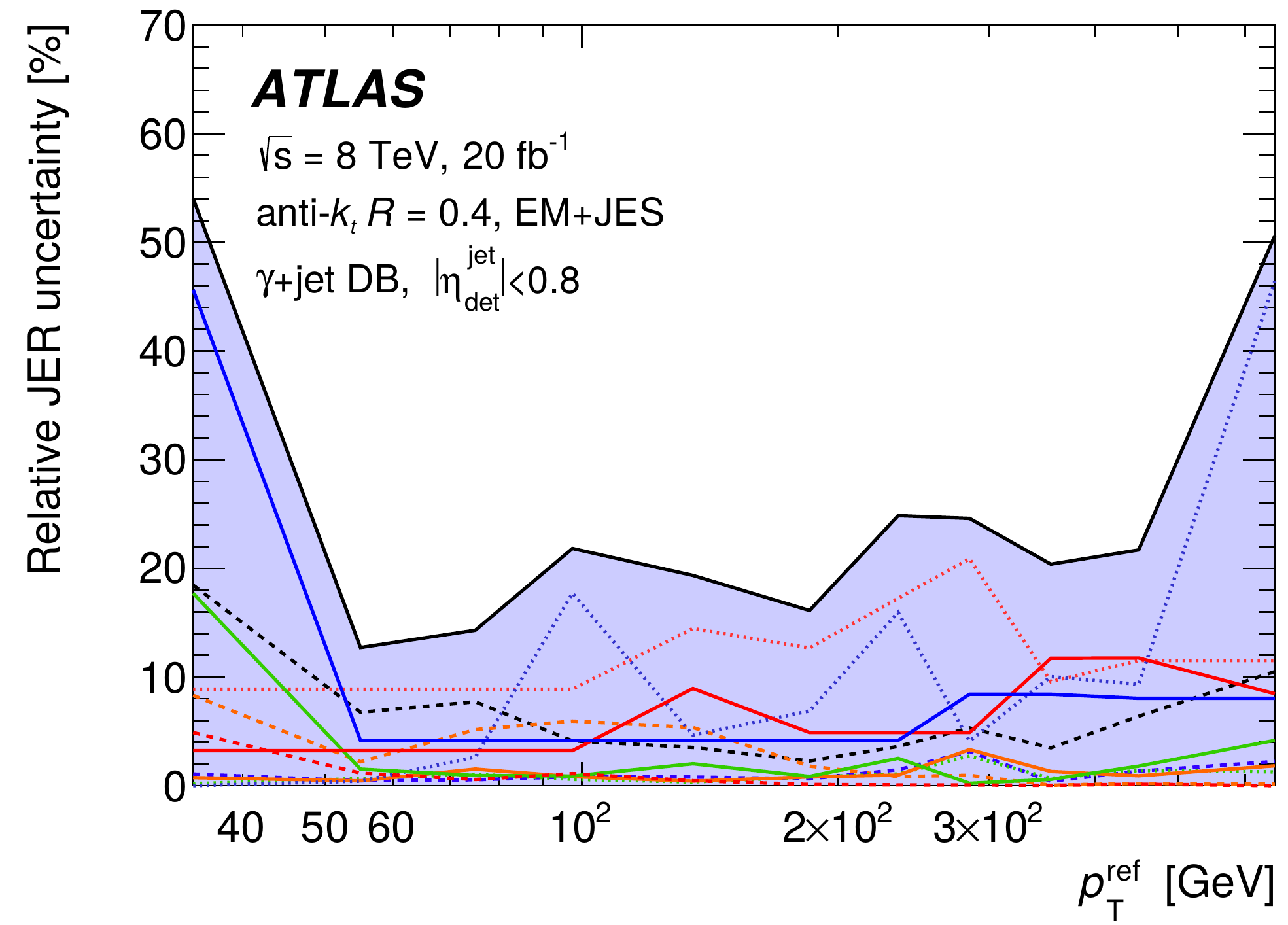}
\end{subfigure}
\begin{subfigure}{0.48\linewidth}
\includegraphics[width=\linewidth]{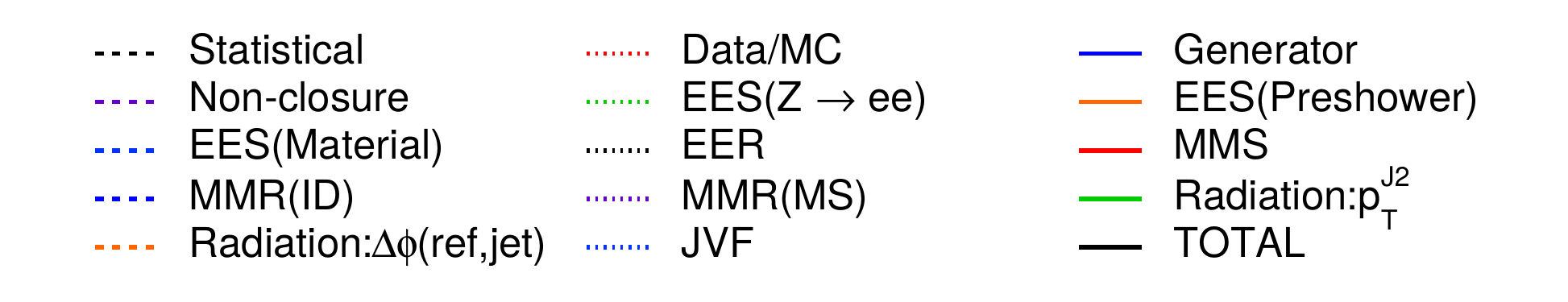}
\caption{\Zjet{},  \antikt{} \rfour{}, \EMJES{}}
\end{subfigure}
\begin{subfigure}{0.48\linewidth}
\includegraphics[width=\linewidth]{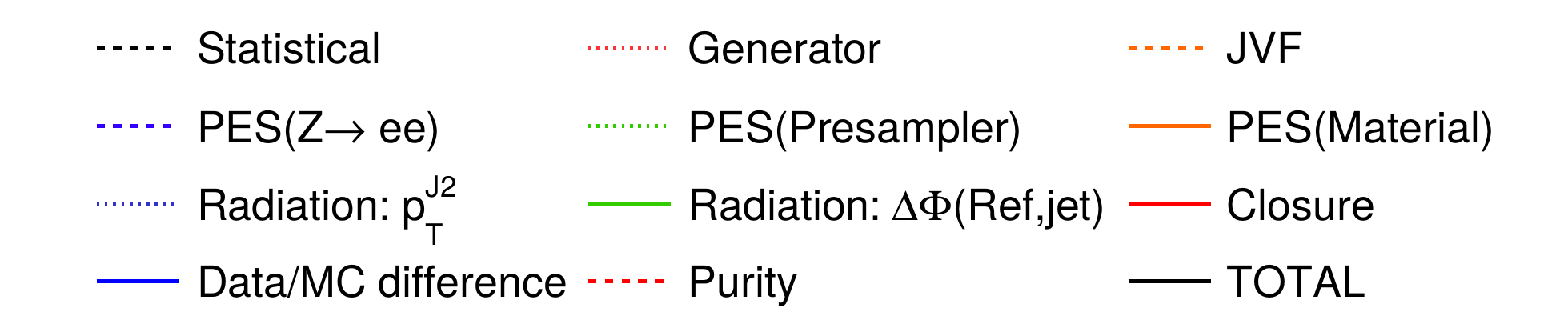}
\caption{\gammajet{},  \antikt{} \rfour{}, \EMJES{} }
\end{subfigure}
\caption{Summary of systematic uncertainties of the \JER{} measured using the \DB{} method in (a)~\Zjet{} and (b)~\gammajet{} events for \antikt{} jets with \rfour{} calibrated with the \EMJES{} scheme.  
The difference between the JER measured in data and MC simulation is presented as an uncertainty (``data/MC difference''). The total uncertainty, shown as a shaded area, is obtained by addition in quadrature of all uncertainty components (including the \DtM{} difference).
EES/EER denotes the electron energy scale/resolution, MMS/MMR denotes the muon momentum scale/resolution, and PES denotes the photon energy scale.}
\label{fig:vjets-JER-unc}
\end{figure}
 
\clearpage
% End of text imported from the .//sections/vjets_dag.tex input file
 
% The next lines are included from the .//sections/multijet.tex input file
\section{High-\texorpdfstring{\pt{}}{pT}-jet calibration using multijet balance}
\label{sec:multijet}

The \Zjet{} and \gamjet{} analyses described in the previous section probe the jet calibration in the range \ptRange{17}{800}.
For jets with \pt{} above 800~\GeV, there are an insufficient number of \Zjet{} and \gamjet{} events, and the multijet balance (MJB) technique is used instead.
This method exploits the \pt{} balance of events in which the leading (highest-\pt{}) jet is produced back-to-back with a recoil system composed of multiple lower-\pt{} jets.
The jets in the recoil system are fully calibrated including the \insitu{} corrections described in the previous sections, while the leading jet, which is being probed, is calibrated with all corrections except the absolute \insitu{} correction (Section~\ref{sec:vjets}).
 
The multijet balance observable $R_{\MJB}$ is defined as:
\begin{equation*}
R_{\MJB} = \left<\frac{\pT^\text{j1}}{\ptRecoil}\right>,
\end{equation*}
where $\pt^\text{j1}$ is the \pt{} of the leading jet and \ptRecoil{} is that from the vectorial sum of the subleading jet four-momenta.
The parameter $R_{\MJB}$~is measured in both data and MC simulations in bins of \ptRecoil.
The multijet balance observable $R_{\MJB}$ is not an unbiased estimator of the leading jet response.
It has a value below unity even at particle level due to the effects of soft quark/gluon emission outside of the jets.
The largest deviation is at low \pt{}, with data and MC simulations exhibiting similar dependence.
This underlying bias is reduced in the double ratio $R_{\MJB}^\textrm{data} / R_{\MJB}^\textrm{MC}$, allowing the response of high-\pt\ jets to be estimated. Mis-modelling in the simulation is evaluated as a systematic uncertainty of the double ratio.
 
As mentioned above,
the jets used in the construction of \ptRecoil~are fully calibrated, including all \insitu{} calibrations. However, the \insitu\ corrections from the \Zgjet\ analyses are only available for $\pt <800$~\GeV\ (Section~\ref{sec:vjets}). An iterative procedure is used to calibrate all jets that are used in the calculation of  \ptRecoil. For the first iteration of the MJB, an upper limit is imposed on the \pt{} of the recoil jets such that the second highest-\pt\ jet in the event has a $\pt<800$~\GeV. This initial selection allows corrections to $R_\MJB$ to be derived, but limits the overall statistical accuracy of the measurements at high \pt{}. To improve the statistical accuracy, $R_\MJB$ is recalculated after the application of the correction factors from the first iteration to jets in the recoil system with $\pt{}>800$~\GeV.
 
\subsection{Event selection}
\label{subsec:event_selectionMultijets}
 
Multijet events were obtained using single-jet triggers that are fully efficient for a given bin of  \ptRecoil. The triggers used for 300~\GeV~$<\ptRecoil<600$~\GeV\ were prescaled, whereas a non-prescaled jet trigger was used for \ptRecoil~$>$ 600~\GeV. Events are required to contain at least three jets with $\pt > 25$~\GeV. The leading jet is required to have $|\etaDet|<1.2$, and the subleading jets that constitute the recoil system are required to have  $|\etaDet|<2.8$.
To select non-dijet events, the leading jet in the recoil system $\pt^\text{j2}$ is required to have less than $80\%$ of the total \pt{} of the recoil system ($\ptasym < 0.8$).
Furthermore,
the angle $\alpha$ in the azimuthal plane between the leading jet three-momentum and the vector defining the recoil system is required to satisfy $ |\alpha - \pi | < 0.3$ radians, and the angle $\beta$ in the azimuthal plane between the leading jet and the nearest jet from the recoil system is required to be greater than 1 radian.

\subsection{Results}
 
Figure~\ref{fr_lab_result_emjes} shows $R_\MJB$ for data and MC simulation using the EM+JES calibration scheme.
The MJB method provides inputs for the \insitu\ jet calibration in the \pt\ range between \SI{300}{\GeV} and \SI{1900}{\GeV}.
The data and MC simulation agree to within $1\%$ across the \pt{} range probed, a feature that is reproduced by the  \Zgjet{} analyses (Section~\ref{sec:vjets}).
 
\begin{figure}[htb]
\centering{
\label{fr_lab_result_kt4emjes}
\begin{subfigure}{0.48\textwidth}\centering
\includegraphics[width=\textwidth]{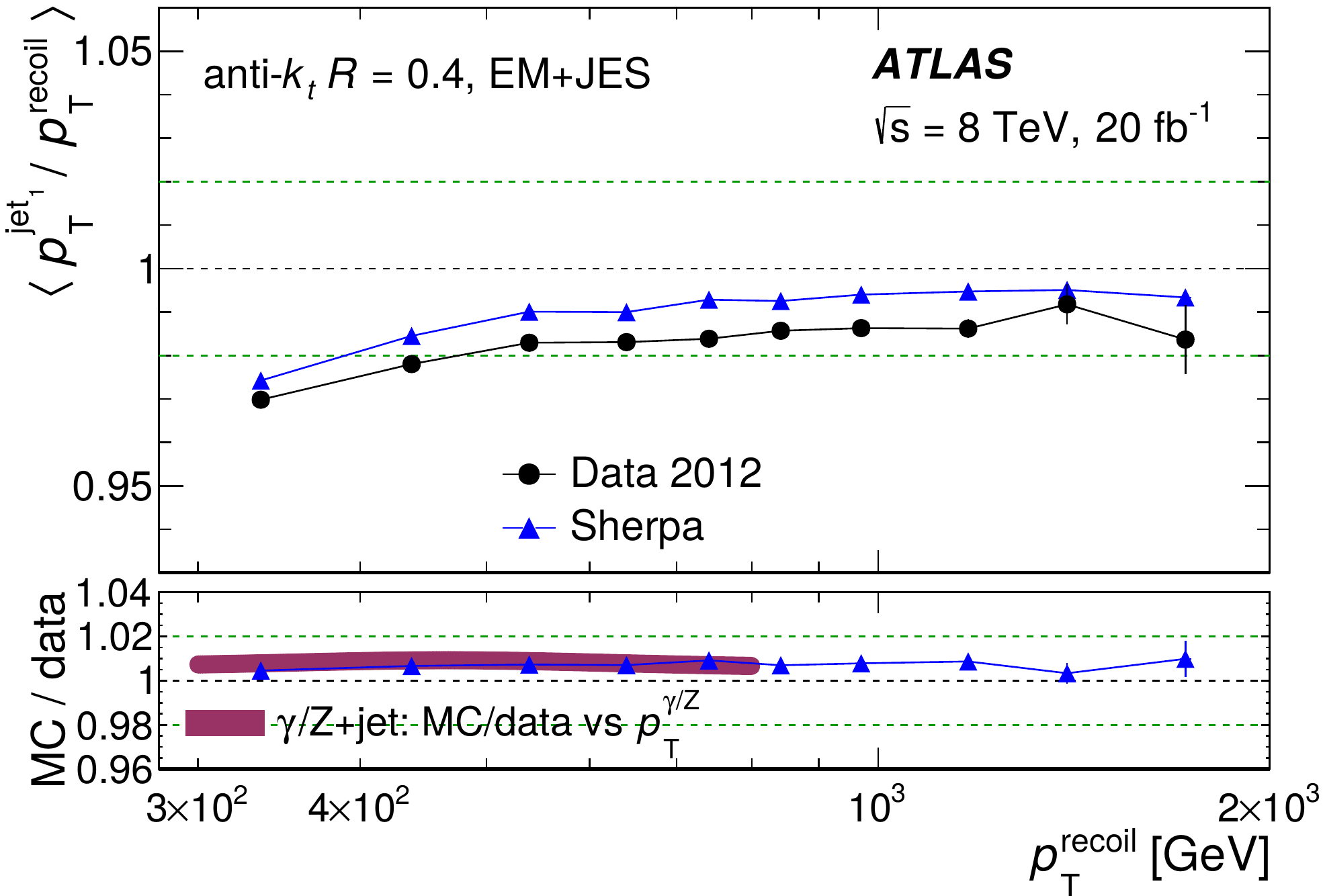}\quad
\caption{}
\end{subfigure}
\begin{subfigure}{0.48\textwidth}\centering
\includegraphics[width=\textwidth]{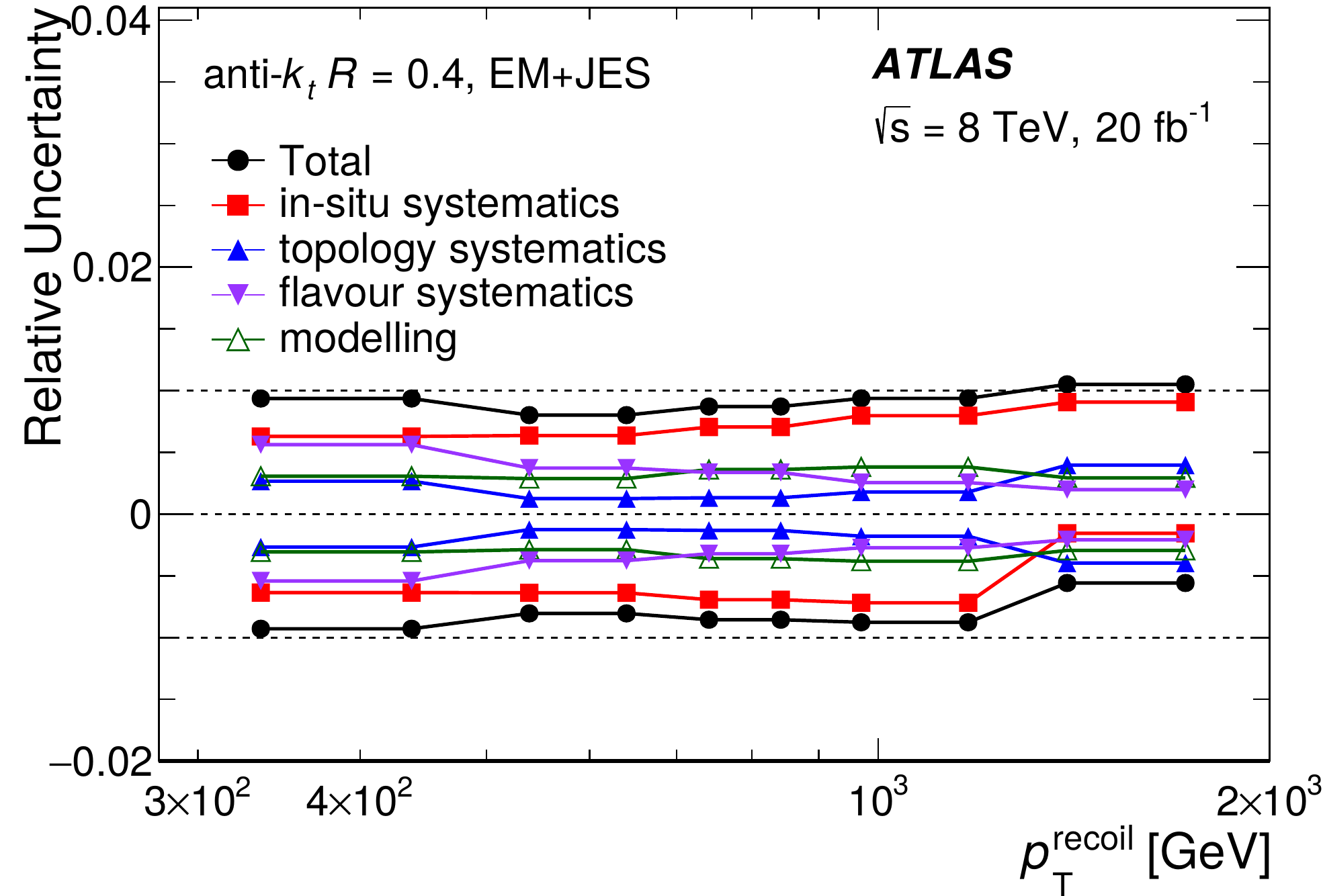}
\caption{} \label{fr_lab_result_kt4emjes_b}
\end{subfigure}
}
\caption{(a) Multijet balance $R_{\MJB}$ in data~(circles) and MC simulation~(triangles) for \antikt{} $R=0.4$ jets calibrated with the EM+JES scheme. The bottom frame compares $R_\MJB^\text{MC}/R_\MJB^\text{data}$ (triangles) with the corresponding $\gamma$/Z+jet results~(magenta solid line).
(b) The impact of \insitu, event selection (topology), physics modelling, and jet flavour systematic uncertainties on $R_\MJB$.
The error bars on the $R_\MJB$ measurements only show statistical uncertainties.}
\label{fr_lab_result_emjes}
\end{figure}
 
\subsection{Systematic uncertainties}
\label{sec:mjb-syst}
Since the jets entering  $R_\MJB$ have been calibrated using the other \insitu\ approaches, the uncertainty in the energy scale of the jets in the recoil system is defined by the systematic and statistical uncertainties of each \insitu\ procedure. To propagate the uncertainty to $R_\MJB$, all input components are individually varied by $\pm1\sigma$ and the full iterative analysis procedure is repeated for each such variation.
Changes in $R_\MJB$ due to the statistical uncertainties of the $\gamma$+jet and $Z$+jet calibrations are typically much smaller than 1\%.
 
Also, the event selection criteria and the modelling in the event generators affect the \pt~balance $R_\MJB$.
The impact of the event selection is investigated by shifting each selection criterion up and down by a specified amount and observing the change in $R_\MJB$. The \pt{} threshold for jets is shifted by $\pm 5$~\GeV, the requirement on the ratio \ptasym\ is shifted by $\pm0.1$, the angle $\alpha$ by $\pm0.1$, and the angle $\beta$ by $\pm0.5$.
The uncertainty due to MC modelling of multijet events is estimated from the symmeterized envelope of MJB corrections obtained by comparing the nominal results obtained from \sherpa\ with those obtained from \powhegpyt, \pythia{}, and \herwigpp.
 
The unknown flavour of each jet is also a source of systematic uncertainty.
The uncertainty in $R_\MJB$ due to the jet flavour response is evaluated using a correlated propagation of the jet flavour response uncertainties, i.e. all jets in the recoil system are shifted simultaneously. The jet flavour composition uncertainty is propagated to $R_\MJB$ for the first, second, and third recoil jets independently, with the final composition uncertainty obtained from the quadrature sum of the three variations. The total uncertainty due to the unknown parton flavour is taken as the sum in quadrature of the flavour response and composition uncertainties.
 
Examples of the impact of systematic uncertainties are shown in Figure~\ref{fr_lab_result_kt4emjes_b} for \antikt{} $R=0.4$ jets using the \EMJES{} calibration scheme. The uncertainties are grouped together into \insitu{}, event topology, physics modelling, and jet flavour categories. Uncertainties for \antikt{} $R=0.6$ jets or the \LCWJES{} scheme are comparable.
 
The uncertainty accounting for the difference of the jet energy resolution between data and simulation was not propagated to the recoil system of the multi-jet balance as the scale was derived before the resolution.
However, the impact of this effect on the multi-jet balance was checked after the resolution was derived, and was found to introduce per-mille level differences on the extraction of the scale.
This effect is therefore negligible compared to the existing uncertainties on the multi-jet balance shown in Figure~\ref{fr_lab_result_kt4emjes_b}.
 
% End of text imported from the .//sections/multijet.tex input file
 
\afterpage{\clearpage}
% The next lines are included from the .//sections/combination.tex input file
\section{Final jet energy calibration and its uncertainty}
\label{sec:JEScomb}
 
As detailed in Sections~\ref{sec:vjets} and~\ref{sec:multijet}, response observables that are directly proportional to the JES are constructed using \insitu{} techniques by exploiting the transverse momentum balance in $\gamma$+jet, $Z$+jet, and multijet events.
These response observables are determined in both data and MC simulations. The final residual jet calibration $c_\text{abs}$,
which accounts for effects not captured by the MC calibration, is defined through the ratio of the responses measured in data and MC simulation by
\begin{equation}
\frac{1}{c_\text{abs}} = \frac{R_\text{data}}{R_\text{MC}}.
\label{eq:doubleRatio}
\end{equation}
 
As explained in Section~\ref{sec:jetCalibOverview}, the absolute \insitu{} correction $c_\text{abs}$ is applied last in the calibration chain following the origin, \pileup{}, MC-based, and dijet \insitu{} calibrations. Just as for the dijet intercalibration (Section~\ref{sec:dijet}), the absolute correction is applied only to data to remove any residual differences in the jet response following the MC calibration.
The dijet $\eta$ intercalibration is referred to as a relative \insitu{} calibration, as it quantifies the balance between a pair of jets in different detector regions without evaluating the absolute scale of either jet.  The absolute calibration is done for the \Zjet{}, \gammajet{}, and \MJB{} techniques, which all balance the probe jet against a well-known reference quantity, thus providing a measure of the absolute scale of the jet and are known as absolute \insitu{} calibrations.

Figure~\ref{fig:insituresponse} summarizes the results of the $Z$+jet, $\gamma$+jet, and multijet balance analyses, showing the ratio of jet response in
data to jet response in MC simulations.
In the $\pt$ range 20--2000~\GeV, the response agrees between MC simulations and data at the 1\% level.  The deviation of the response from unity defines the absolute \insitu{} calibration which is applied to jets in data. There is good agreement and little tension between the three different \insitu{} methods in the regions of phase space where they overlap.
 
\begin{figure}[!ht]
\centering
\begin{subfigure}{0.48\textwidth}\centering
\includegraphics[width=\textwidth]{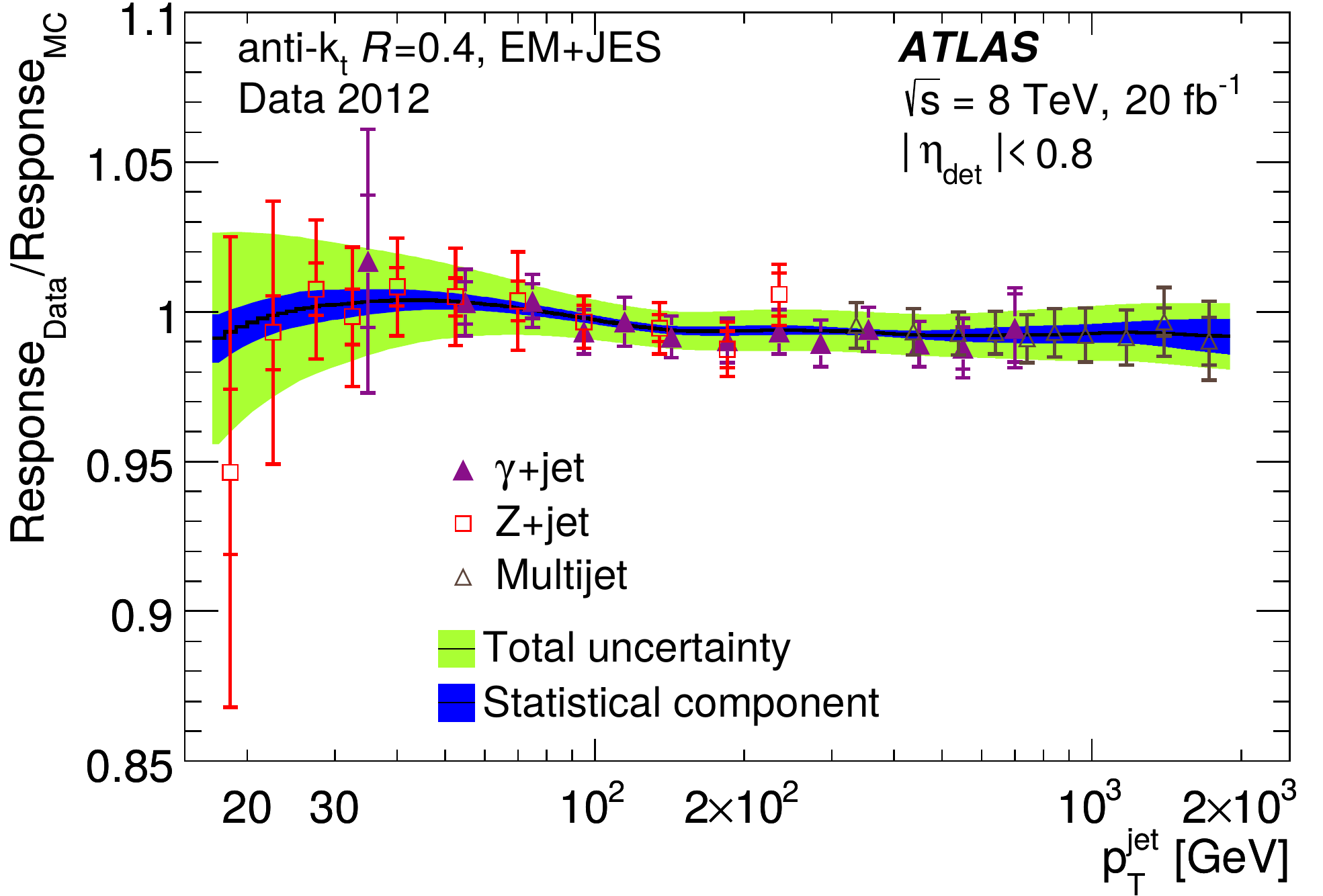}
\caption{EM+JES}
\end{subfigure}
\begin{subfigure}{0.48\textwidth}\centering
\includegraphics[width=\textwidth]{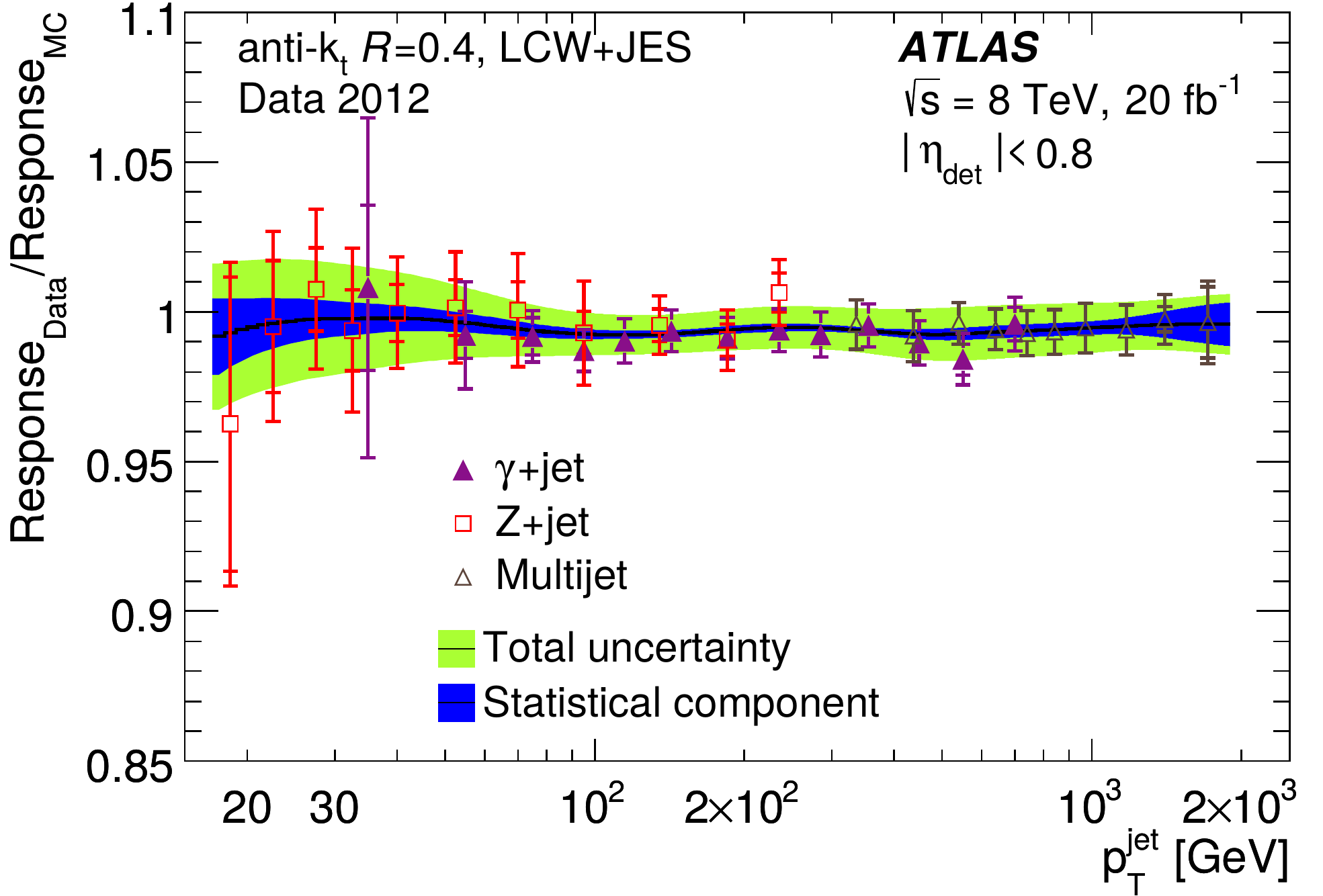}
\caption{LCW+JES}
\end{subfigure}
\caption{Ratio of response measured in data to response measured in MC samples for $Z$+jet~(empty squares), $\gamma$+jet~(filled triangles) and multijet balance~(empty triangles) \insitu{} analyses.
The method giving the smallest overall uncertainty is used, corresponding to the \DB{} approach for $Z$+jet and the \MPF{} approach for $\gamma$+jet.
Each measurement has two error bars: the smaller interval corresponds to the statistical uncertainty, while the outer error interval corresponds to the total uncertainty.
Also shown is the combined correction (line) with its associated total uncertainty (wider band) and statistical uncertainty (narrower band) as discussed in Section~\ref{subsec:insituCombination}.}
\label{fig:insituresponse}
\end{figure}

\subsection{Combination of absolute \texorpdfstring{\insitu{}}{in situ} measurements}
\label{subsec:insituCombination}
 
The separate measurements from $Z$+jet, $\gamma$+jet, and multijet balance
are combined using the procedure outlined in Ref.~\cite{PERF-2012-01}.
For the $Z$+jet and $\gamma$+jet measurements, the method giving the smallest overall uncertainty is used, corresponding to the \DB{} approach for $Z$+jet and the \MPF{} approach for $\gamma$+jet
(see Section~\ref{sec:vjets} for details on the methods).
The choice of \DB{} for $Z$+jet is a compromise between the precision of the $R=0.4$ and $R=0.6$ jet calibrations in the low \pt{} regime where the $Z$+jet final state is most relevant:
it was found that the \DB{} and \MPF{} techniques give similar uncertainties for $R=0.4$ jets, while the \DB{} technique provides improved precision for $R=0.6$ jets.
In contrast, the MPF technique is used for $\gamma$+jet events as it is found to generally provide better uncertainties across its kinematic range of relevance.
This combination uses the compatibility of the three \insitu{} measurements and their associated systematic uncertainties to produce a combined measurement
of the response ratio with associated uncertainties.
 
Table~\ref{tab:insituUncert} presents the 26 systematic uncertainty sources that affect $c_\text{abs}$.
These are evaluated as detailed in Sections~\ref{sec:vjets-syst} and~\ref{sec:mjb-syst}.
The electron and photon energy scale uncertainties are each split into four sources that are fully correlated.
These are treated as four $e/\gamma$ energy scale sources, yielding a list of 22 systematic uncertainty components.
Each source is further classified into one of the following four categories:
\begin{itemize}
\setlength\itemsep{-2mm}
\item Detector description (det.),
\item Physics modelling (model),
\item Statistics and method (stat./meth.), and
\item Mixed detector and modelling (mixed).
\end{itemize}
 
\begin{table}
\caption{Summary of the uncertainty components propagated through to the
combination of absolute \insitu{} jet energy scale measurements from \Zjet{}, \gamjet{},
and multijet balance studies. These are discussed in more detail in Sections~\ref{sec:dijet} and~\ref{sec:vjets}.}
\label{tab:insituUncert}
\begin{tabular}{l|l|l}
\hline\hline
Name                                     & Description                                                         & Category           \\
\hline
$Z$+jet                                  &                                                                     &                    \\
$e$ $E$-scale material                     & Material uncertainty in electron energy scale                       & det.               \\
$e$ $E$-scale presampler                   & Presampler uncertainty in electron energy scale                     & det.               \\
$e$ $E$-scale baseline                     & Baseline uncertainty in electron energy scale                       & mixed               \\
$e$ $E$-scale smearing                     & Uncertainty in electron energy smearing                             & mixed               \\
$\mu$ $E$-scale baseline                 & Baseline uncertainty in muon energy scale                           & det.               \\
$\mu$ $E$-scale smearing ID              & Uncertainty in muon ID momentum smearing                            & det.               \\
$\mu$ $E$-scale smearing MS              & Uncertainty in muon MS momentum smearing                            & det.               \\
MC generator                             & Difference between MC generators                                    & model              \\
JVF                                      & JVF choice                                                          & mixed              \\
$\Delta\phi$                             & Extrapolation in $\Delta\phi$                                       & model              \\
Out-of-cone                              & Contribution of particles outside the jet cone                      & model              \\
Subleading jet veto                     & Variation in subleading jet veto                                   & model              \\
Statistical components                   & Statistical uncertainty                                             & stat./meth.        \\
\hline
$\gamma$+jet                             &                                                                     &                    \\
$\gamma$ $E$-scale material              & Material uncertainty in photon energy scale                         & det.               \\
$\gamma$ $E$-scale presampler            & Presampler uncertainty in photon energy scale                       & det.               \\
$\gamma$ $E$-scale baseline              & Baseline uncertainty in photon energy scale                         & det.               \\
$\gamma$ $E$-scale smearing              & Uncertainty in photon energy smearing                               & det.               \\
MC generator                             & Difference between MC generators                                    & model              \\
$\Delta\phi$                             & Extrapolation in $\Delta\phi$                                       & model              \\
Out-of-cone                              & Contribution of particles outside the jet cone                      & model              \\
Subleading jet veto                     & Variation in subleading jet veto                                   & model              \\
Photon purity                            & Purity of sample in $\gamma$+jets                                   & det.               \\
Statistical components                   & Statistical uncertainty                                             & stat./meth.        \\
\hline
Multijet balance                         &                                                                     &                    \\
$\alpha$ selection                       & Angle between leading jet and recoil system                         & model              \\
$\beta$ selection                        & Angle between leading et and closest subleading jet                & model              \\
MC generator                             & Difference between MC generators (fragmentation)                    & mixed              \\
$\pt$ asymmetry selection          & Asymmetry selection between leading and subleading jet           & model              \\
Jet $\pt$ threshold                 & Jet \pt{} threshold                                            & mixed              \\
Statistical components                   & Statistical uncertainty                                             & stat./meth.         \\
\hline\hline
\end{tabular}
\end{table}
 
The combination is carried out using the absolute \insitu{} measurements (Eq.~(\ref{eq:doubleRatio})) in bins of $\pt^\text{ref}$
and evaluated at $\langle \pt^\textrm{ref}\rangle$.  The data-to-MC
response ratio is defined in fine $p^\textrm{ref}_\textrm{T}$ bins for each method
using interpolating second-order polynomial splines.  The combination
is then carried out using a weighted average of the absolute \insitu{} measurements
based on a $\chi^2$-minimization.  This local $\chi^2$ is used to
define the level of agreement between measurements.
 
Each uncertainty source in the combination is treated as being fully correlated across $\pt$ and $\eta$ and independent of one another.
All the uncertainty components are propagated to the combined results using pseudo-experiments~\cite{PERF-2011-03}.
To determine the correlations between different phase-space regions, it is necessary to understand the contribution of each uncertainty component to the final uncertainty.
Therefore, each individual source is propagated separately to the combined result by coherently shifting all the correction factors by one standard deviation.
Comparison of this shifted combination result with the nominal result provides an estimate of the propagated systematic uncertainty.
 
One exception is the jet flavour uncertainty of the recoil in the multijet balance method (Section~\ref{sec:multijet}). It is correlated in a non-trivial way with the additional uncertainties due to flavour composition and response considered in analyses. Including this uncertainty does not change the overall absolute \insitu{} uncertainty by a significant amount after combination with the other \insitu{} methods, so it is dropped.
 
To take tensions between measurements into account, each uncertainty source is increased by rescaling it by $\sqrt{\chi^2/n_\text{dof}}$ if $\chi^2/n_\text{dof}$ is larger than unity~\cite{Beringer:1900zz}, where $n_\text{dof}$ is the number of degrees of freedom.
The number of degrees of freedom varies with \pt{} and corresponds to the number of \insitu{} methods $n_\text{in-situ}$ that contribute to the combination minus one, i.e. $n_\text{dof} = n_\text{in-situ}-1$.
The local $\sqrt{\chi^2/n_\text{dof}}$ of the final combination (Figure~\ref{fig:chisquared}) for both jet collections is below unity for most of the $\pt$ range and barely exceeds 2 anywhere.
The combined \insitu{} factor
is the final calibration factor to be applied to data after reducing
statistical fluctuations using a sliding Gaussian kernel.
 
\begin{figure}[!ht]
\centering
\begin{subfigure}{0.48\textwidth}\centering
\includegraphics[width=\textwidth]{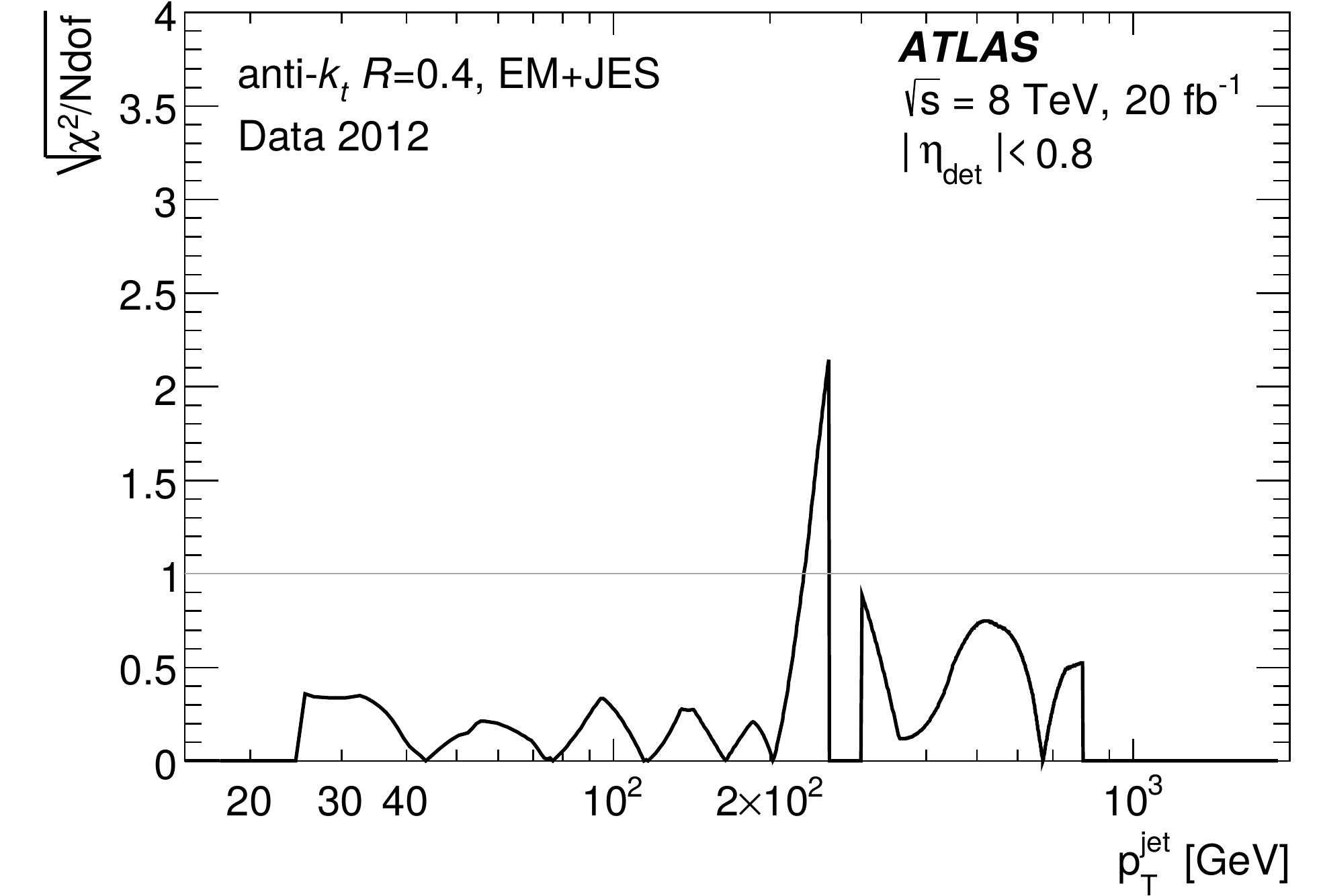}
\caption{EM+JES}
\end{subfigure}
\begin{subfigure}{0.48\textwidth}\centering
\includegraphics[width=\textwidth]{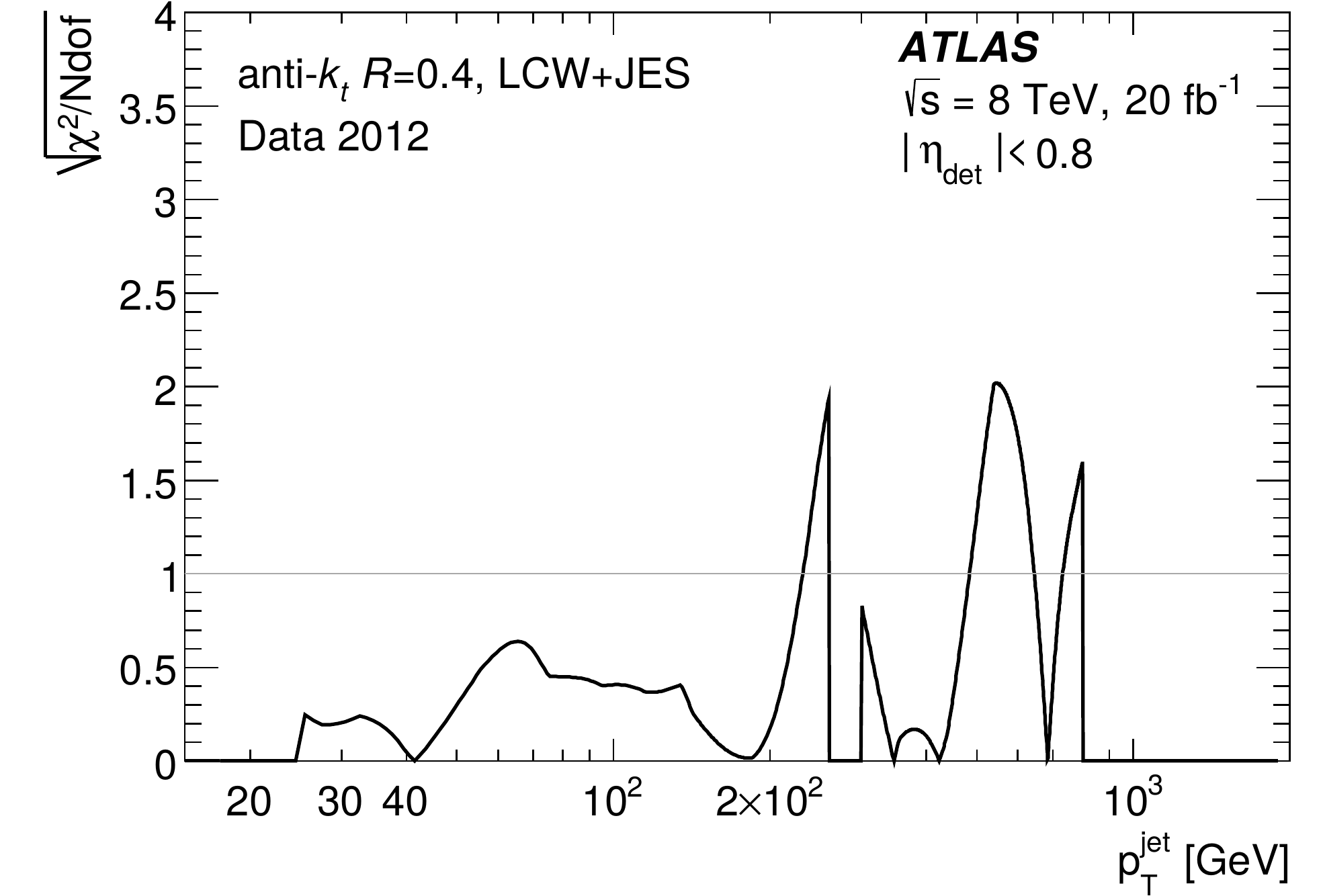}
\caption{LCW+JES}
\end{subfigure}
\caption{$\chi^2/n_\text{dof}$ for the combination of absolute
\insitu{} measurements illustrating the compatibility of the
included \insitu{} calibrations as a function of jet \pt{}.  At any
given point, there are at most two \insitu{} results being combined:
\Zjet{} and \gammajet{} at low \pT{}, or \gammajet{} and \MJB{} at
high \pT{}, which means that the number of degrees of freedom $n_\text{dof}$ is equal to one.
For a small \pt{} range near 300~\GeV, only one measurement (\gammajet{}) contributes, and there is a gap ($n_\text{dof}=0$).
The points where the curve touches zero correspond to where the two \insitu{} calibrations cross.}
\label{fig:chisquared}
\end{figure}
 
Figure~\ref{fig:allUncertainties} shows the uncertainty sources for the three absolute \insitu{} analyses used in the combination
as a function of \pt{}.  In the combination, the \Zjet{} measurement is most important at low \pt{}, the \gamjet{} measurement at medium \pt{}, and the multijet balance at high \pt{}.
 
The combined jet response, shown as a line in Figure~\ref{fig:insituresponse},
is observed to have a general offset of 0.5\% between data and MC
simulation (with data below the MC prediction).  The total uncertainty
from the combination of absolute \insitu{} techniques is shown as the wider band around the measured response and
is about 3.5\% (2.5\%) for jets with $\pt{} \approx25$~\GeV\ for EM (LCW) jets and decreases to about 1\% (1\%) for \pt{} above 200~\GeV.
 
As mentioned a spline-based combination procedure, with a local averaging within fine \pt{} intervals followed by a global smoothing, is used for the in-situ JES combination.
This method avoids assumptions on the jet energy response dependence that are implicitly present in procedures based on global fits using a functional form, which can further reduce the uncertainties (see e.g.\ Ref.~\cite{CMS-JME-13-004}).
 
\begin{figure}[!ht]
\centering
\begin{subfigure}{0.48\textwidth}\centering
\includegraphics[width=\textwidth]{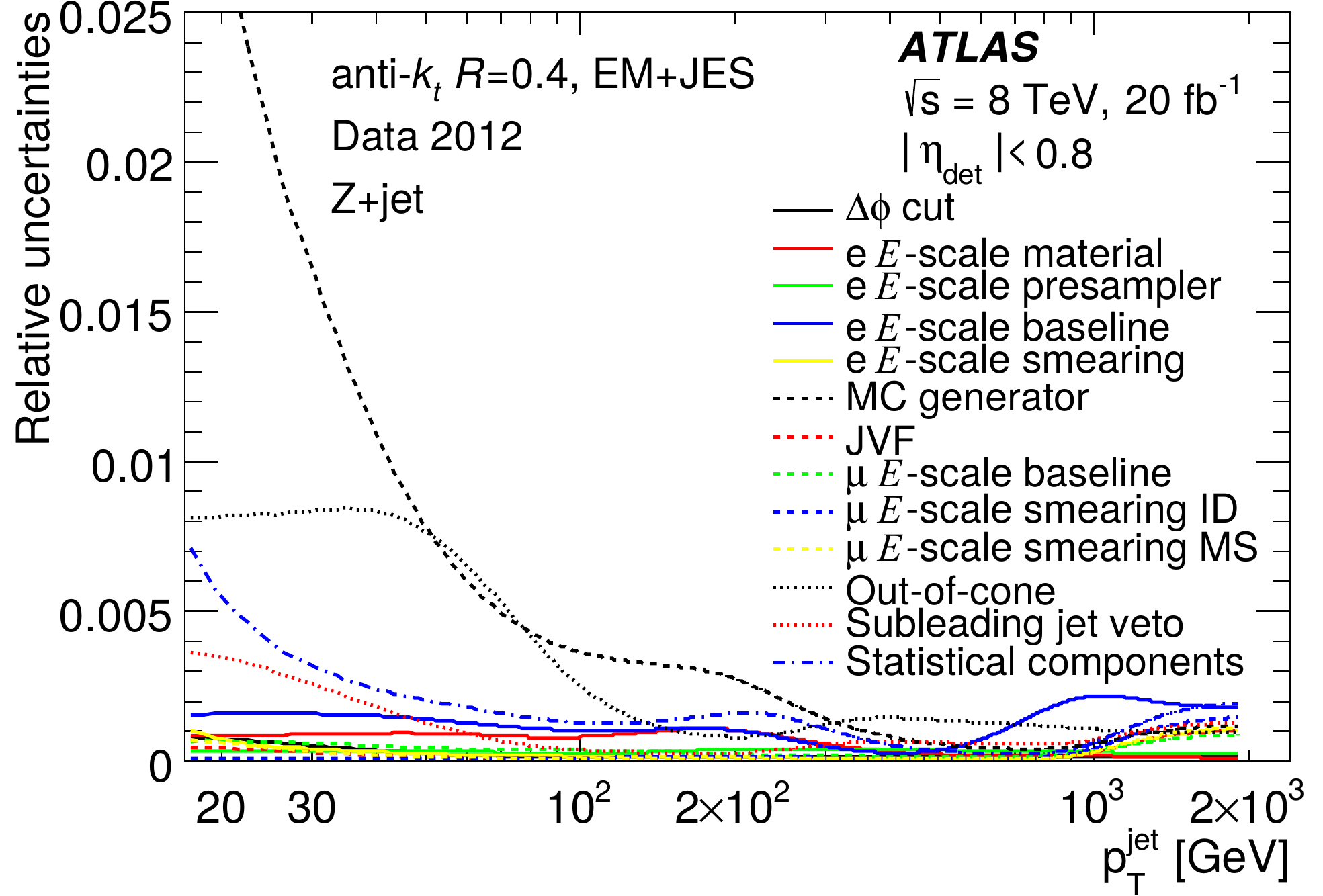}
\caption{$Z$+jets (EM+JES)}
\end{subfigure}
\begin{subfigure}{0.48\textwidth}\centering
\includegraphics[width=\textwidth]{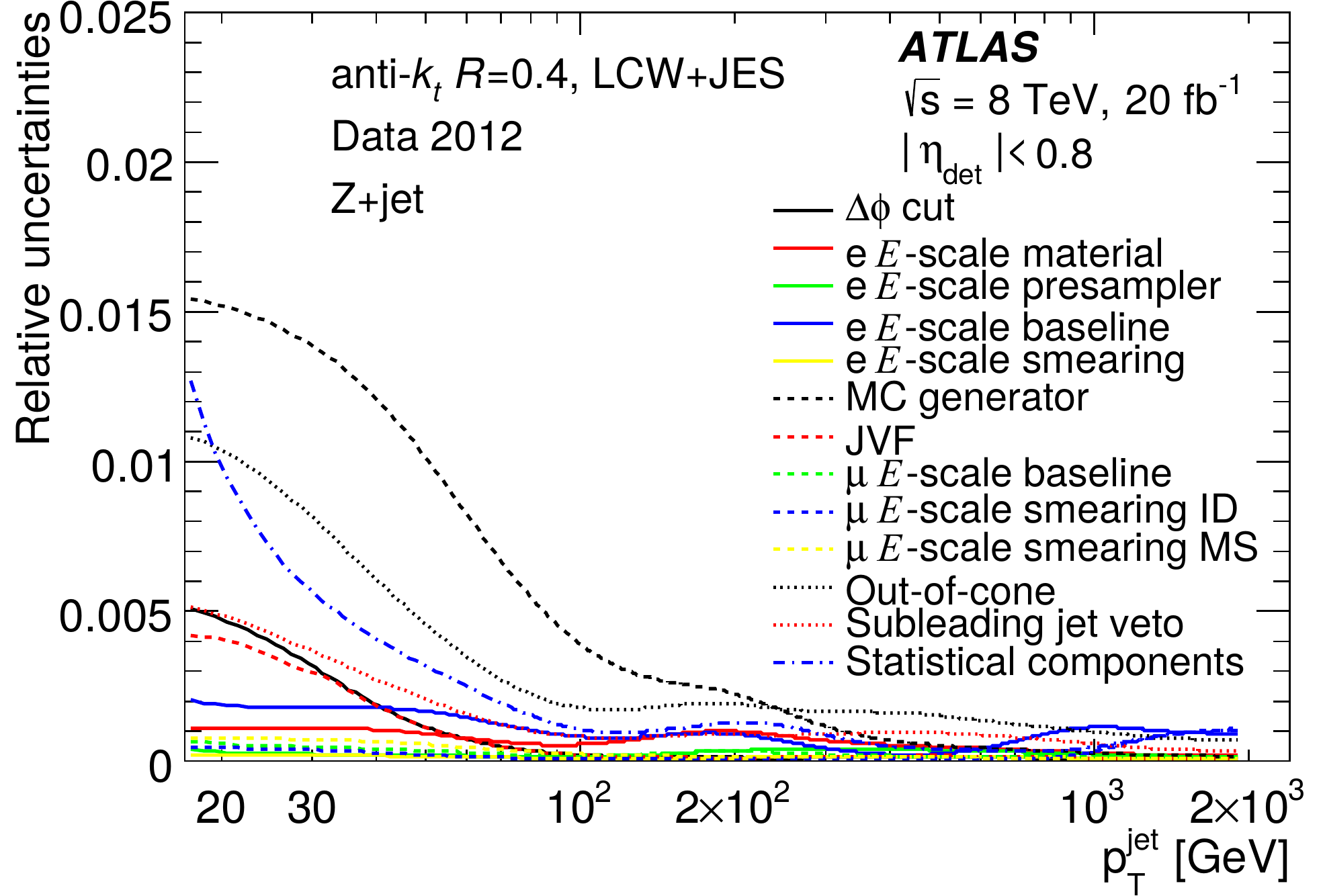}
\caption{$Z$+jets (LCW+JES)}
\end{subfigure} \\
 
\bigskip
 
\begin{subfigure}{0.48\textwidth}\centering
\includegraphics[width=\textwidth]{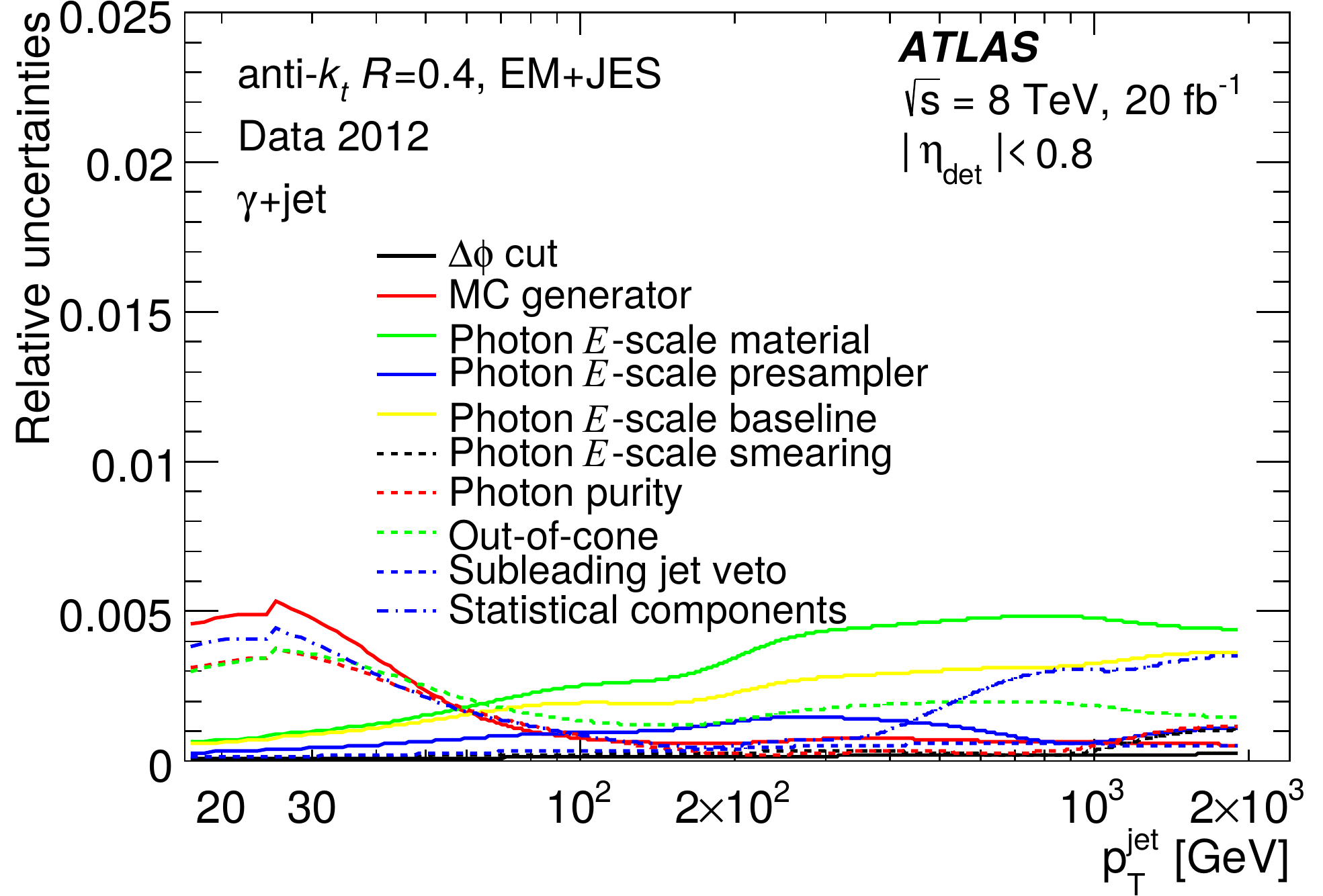}
\caption{$\gamma$+jets (EM+JES)}
\end{subfigure}
\begin{subfigure}{0.48\textwidth}\centering
\includegraphics[width=\textwidth]{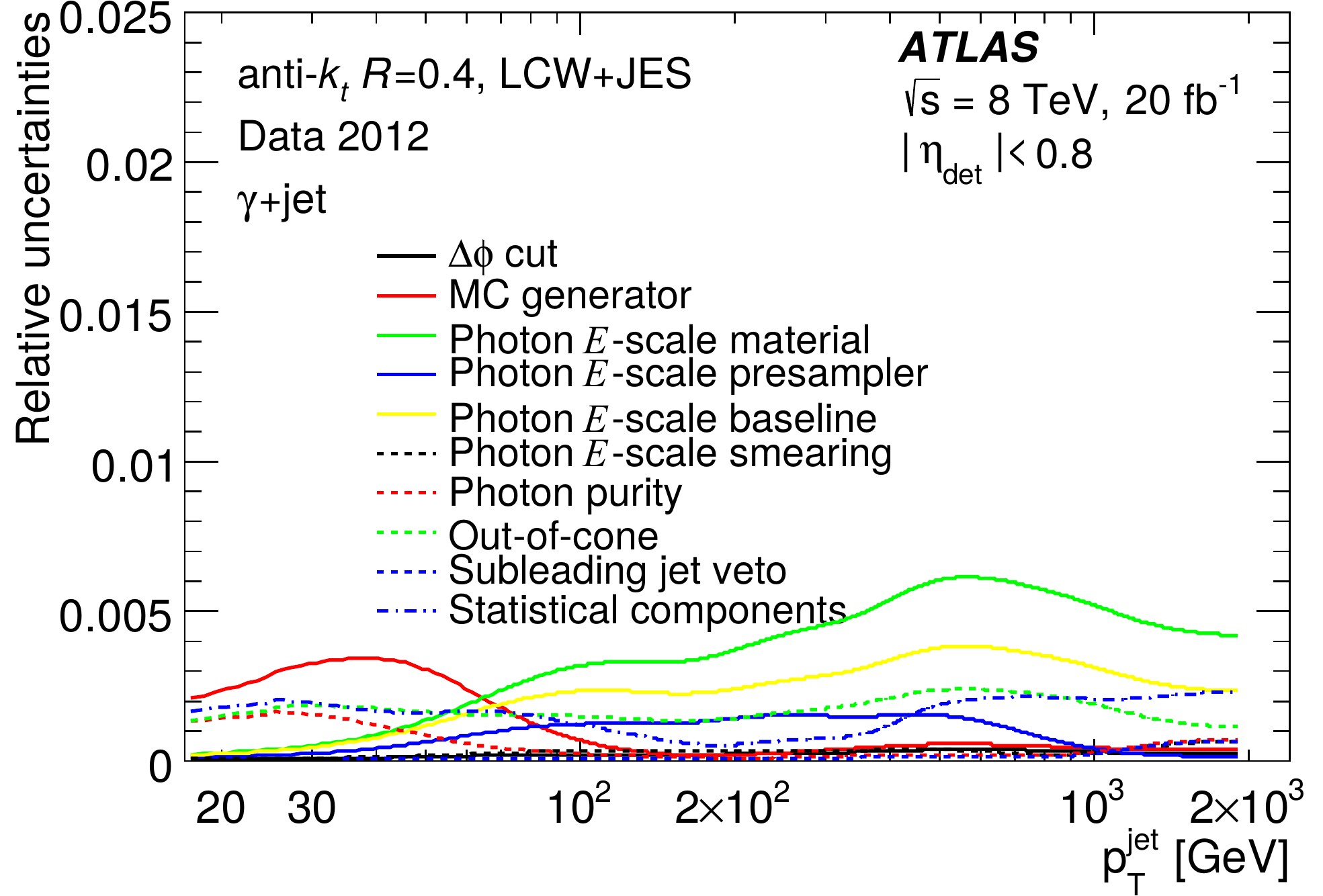}
\caption{$\gamma$+jets (LCW+JES)}
\end{subfigure} \\
 
\bigskip
 
\begin{subfigure}{0.48\textwidth}\centering
\includegraphics[width=\textwidth]{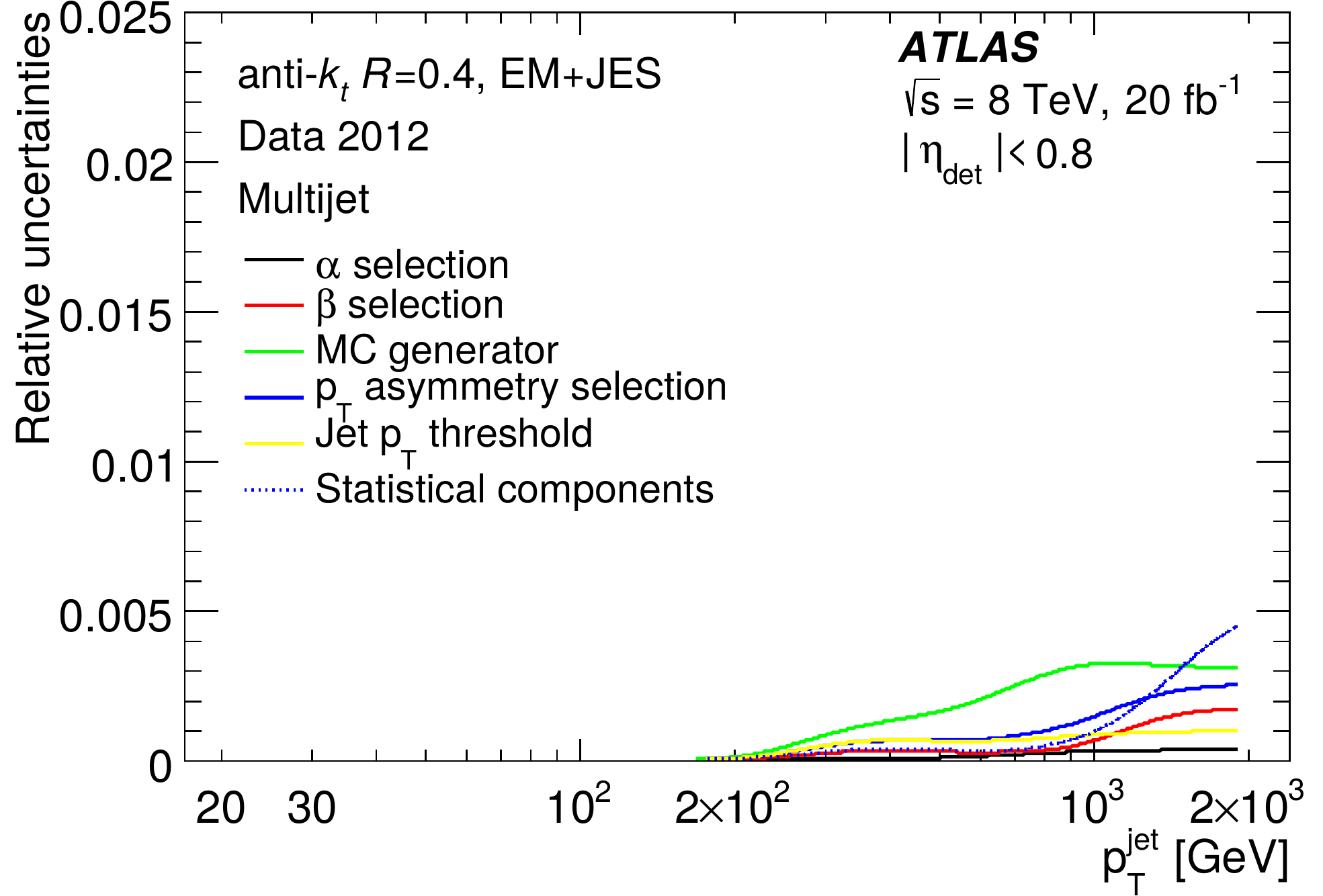}
\caption{Multijet balance (EM+JES)}
\end{subfigure}
\begin{subfigure}{0.48\textwidth}\centering
\includegraphics[width=\textwidth]{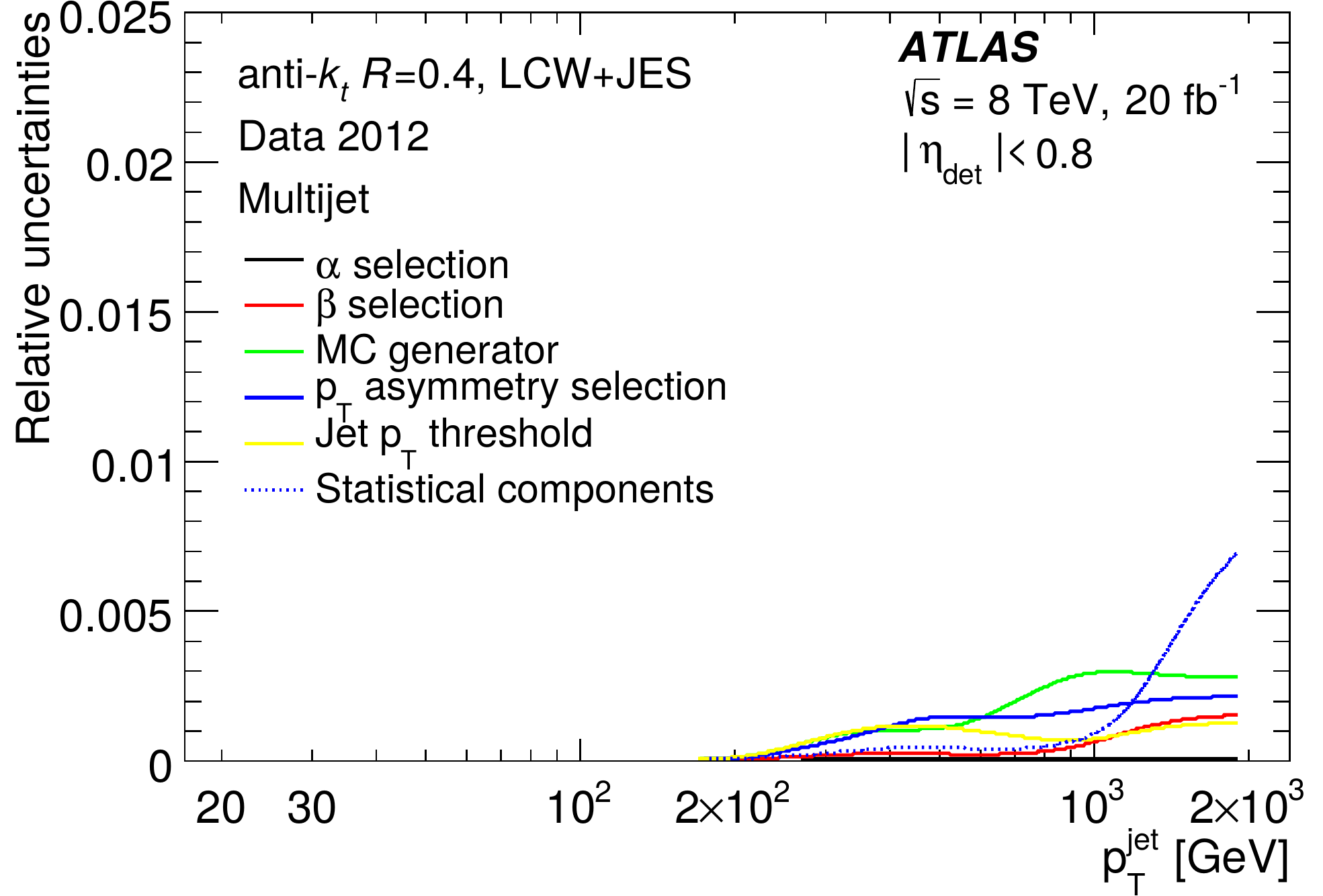}
\caption{Multijet balance (LCW+JES)}
\end{subfigure}
 
\caption{Individual uncertainty sources used in the combination for the three absolute \insitu{} calibration methods.  The systematic uncertainties displayed
correspond to those in Table~\ref{tab:insituUncert}.}
\label{fig:allUncertainties}
\end{figure}

\subsection{Jet energy scale uncertainties}
\label{JES_uncert}
\label{sec:JES_uncert}
 
In addition to the uncertainties coming from the combination of \insitu{} methods detailed above, there are several other uncertainties
that account for other potential systematic effects or expand the kinematic reach.
These additional uncertainties are described below, and summarized in Section~\ref{subsec:JESuncertSummary}

\subsubsection{Single-hadron response}
\label{subsec:singlehadron}
 
The jet energy response measured by the \insitu{} methods can also be compared with results from
a method where the jet energy scale is estimated from the calorimeter response to single hadrons measured in test beam studies.
This provides a cross-check of the direct balance \insitu{} methods, albeit with a larger uncertainty, and
also allows the extension of the \insitu{} measurements of the jet energy scale to higher energies beyond the reach of balance methods due to limited data.
In this ``single hadron'' method, jets are treated as a superposition of the individual energy deposits of their constituent particles~\cite{PERF-2011-05}.
In some cases, highly energetic jets contain constituents beyond test-beam energies.
When this occurs, a constant 10\% uncertainty is applied to each of these constituents.
 
In the previous jet energy scale measurements based on data taken in 2011~\cite{PERF-2012-01},
the absolute \insitu{} methods and the single-hadron response studies gave consistent results, indicating that MC simulation overestimated the jet response in data by approximately 2\%.
However, since the \insitu{} methods are more precise (approximately 2\% uncertainty compared to 5\%) the single-particle response method is only used at high $\pt\,$ ($>1500$~\GeV) where the statistical power of \insitu{} methods becomes limited.
The single-hadron response measurements from the 2011 data~\cite{PERF-2012-01} are propagated to high $\pt\,$ jets to provide an uncertainty where it is beyond the reach of the absolute \insitu{} analyses.

\subsubsection{\Pileup{} uncertainties}
\label{subsec:pileupUncert}
 
There are four uncertainties sources associated with the mitigation of the \pileup{} contributions to the jet momentum (Eq.~(\ref{eq:PU})) that are evaluated by comparing data with simulation using \insitu{} techniques.  Two of the uncertainties are in the values of the slope parameters $\alpha$ and $\beta$ that determine the dependence on the number of reconstructed \pileup{} vertices and the average interactions per crossing, respectively. The third uncertainty accounts for jet \pt{} dependence of the $\alpha$ and $\beta$ parameters. These uncertainties are evaluated using momentum balance in \Zjet{} events.
The fourth uncertainty is associated with a topology dependence of the event \pt{}-density $\rho$.
It is evaluated as the largest difference in measured average \pt{} density $\langle\rho\rangle$ at a given \pileup{} condition \avgmu{} between dijet, $\gamma$+jet, and \Zjet{} events. As shown in Eq.~(\ref{eq:PU}), this uncertainty is directly proportional to the jet area and is larger by approximately a factor of $0.6^2/0.4^2=2.25$ for $R=0.6$ jets compared with $R=0.4$ jets.
For $R=0.6$ jets, this tends to be the dominant uncertainty component with a typical magnitude of 2\% for jets with \pt{} around 40~\GeV. For $R=0.4$ jets in events with moderate \pileup{}, the $N_\text{PV}$-dependent uncertainty component tends to be largest for jets in the central calorimeter region while the \avgmu{} component is largest in the forward calorimeter region ($|\etaDet|>2.8$).
 
\subsubsection{Flavour-based uncertainties}
\label{subsec:flavourUncert}
 
The \insitu{} methods used to derive final corrections and uncertainties of the jet energy scale make use of event samples with particular fractions of jets initiated by quarks and gluons. The event samples in physics analyses may have jet flavour compositions which differ from that of the calibration sample.
 
The response for quark-initiated jets is considerably higher than that for gluon-initiated jets (Section~\ref{sec:GS-flavour}). Therefore, if the flavour composition of final states in a given analysis is unknown, it has an impact on the JES uncertainty. The degree to which the flavour of jets is known in an analysis can be specified in order to evaluate the corresponding uncertainty. Alternatively, analyses can be conservative and use a completely unknown flavour composition.
 
While the response for light-quark-initiated jets is found to be in good agreement between different generators, shifts are seen in the gluon jet response for different generators due to differences in the jet fragmentation. There is therefore an additional uncertainty for gluon-initiated jets, which is subdominant in the \Zjet{} and \gammajet{} regions used to constrain the uncertainty, as defined by the difference between the gluon jet response in \pythia{} and \herwigpp{}.
These differences are typically reduced using LCW \topos{} as inputs, and this is visible in the central region of the detector
when comparing to jets built using EM \topos{}.
However, this is less true in the forward region of the detector where the LCW correction is less robust due to the different properties of the more forward calorimeters.
 
Further details of this uncertainty are given in Ref.~\cite{PERF-2012-01}, and additional discussion of how the \GS{} correction reduces the jet flavour uncertainties are presented in Sections~\ref{sec:GS-flavour} and~\ref{sec:GS-MC-generators-comparison}.

\subsubsection{Summary of jet energy scale uncertainties}
\label{subsec:JESuncertSummary}
 
The total jet energy scale uncertainty is compiled from multiple sources:
\begin{itemize}
\setlength\itemsep{-2mm}
\setlength\topsep{-2mm}
\item 22 systematic sources from absolute \insitu{} methods,
\item 34 statistical sources from absolute \insitu{} methods,
\item a single-hadron response uncertainty which only affects the highest-\pT jets beyond the reach of \insitu{} techniques (Section~\ref{subsec:singlehadron}),
\item two $\eta$-intercalibration uncertainties (one systematic, one statistical),
\item four sources from uncertainties associated with the \pileup{} corrections:
\begin{itemize}
\setlength\topsep{-2mm}
\setlength\itemsep{-2mm}
\item $\avgmu$-dependent uncertainty in the \pileup{} correction,
\item \Npv{}-dependent uncertainty in the \pileup{} correction,
\item \pt{} dependence of \pileup{} corrections, and
\item $\rho$ topology dependence,
\end{itemize}
as outlined in Ref.~\cite{PERF-2014-03} (Section~\ref{subsec:pileupUncert}), and
\item two sources due to jet flavour (Section~\ref{subsec:flavourUncert}).
\end{itemize}
The last two terms are assumed to be independent, resulting in a jet energy scale uncertainty defined in terms of 65 components (nuisance parameters).
The resulting, total jet energy scale uncertainty is shown as a function of jet \pt{} in Figure~\ref{fig:JESuncertPt} and versus jet $\eta$ in Figure~\ref{fig:JESuncertEta}.
 
\begin{figure}
\begin{subfigure}{0.48\textwidth}\centering
\includegraphics[width=\textwidth]{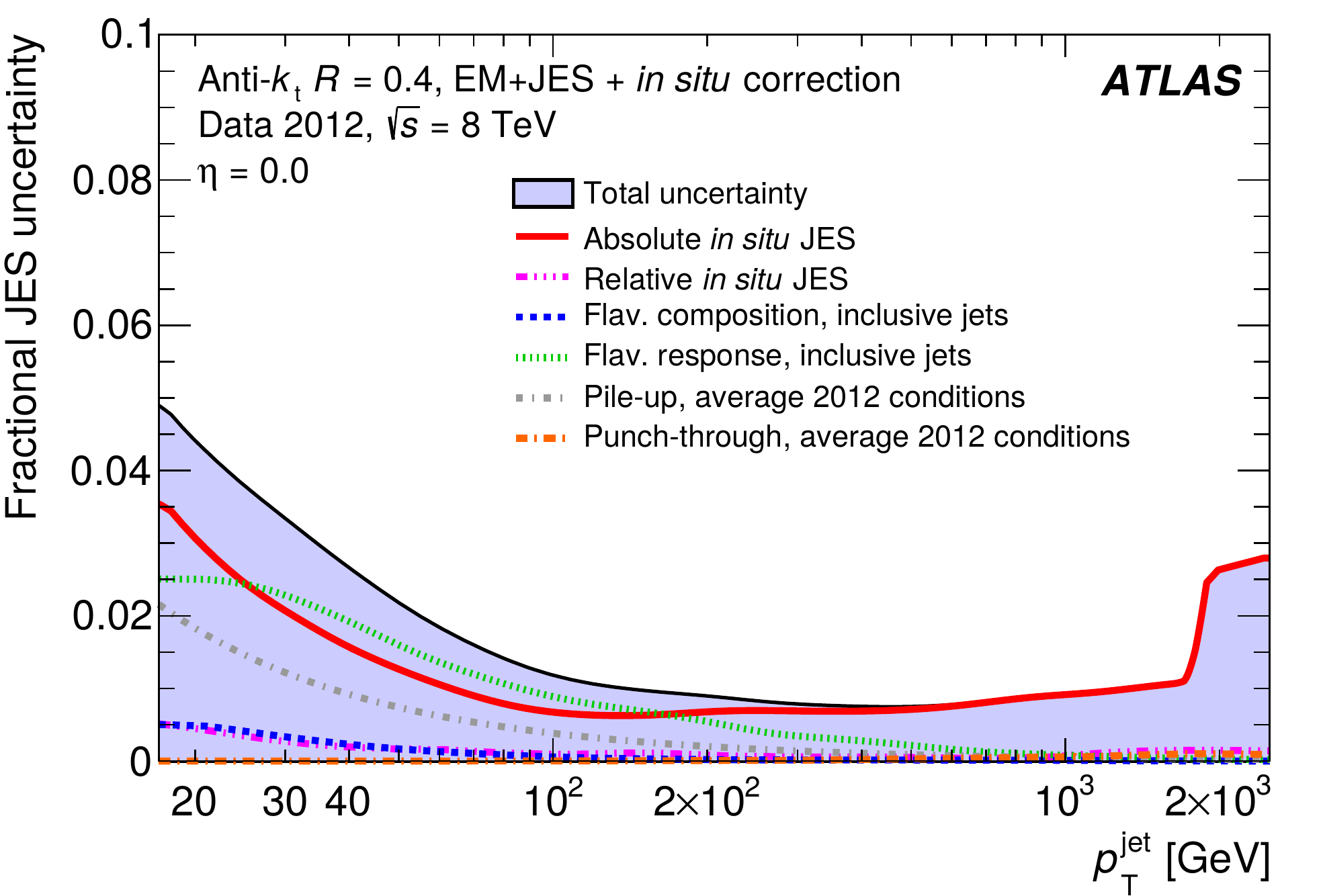}
\caption{EM in dijet events}
\end{subfigure}
\begin{subfigure}{0.48\textwidth}\centering
\includegraphics[width=\textwidth]{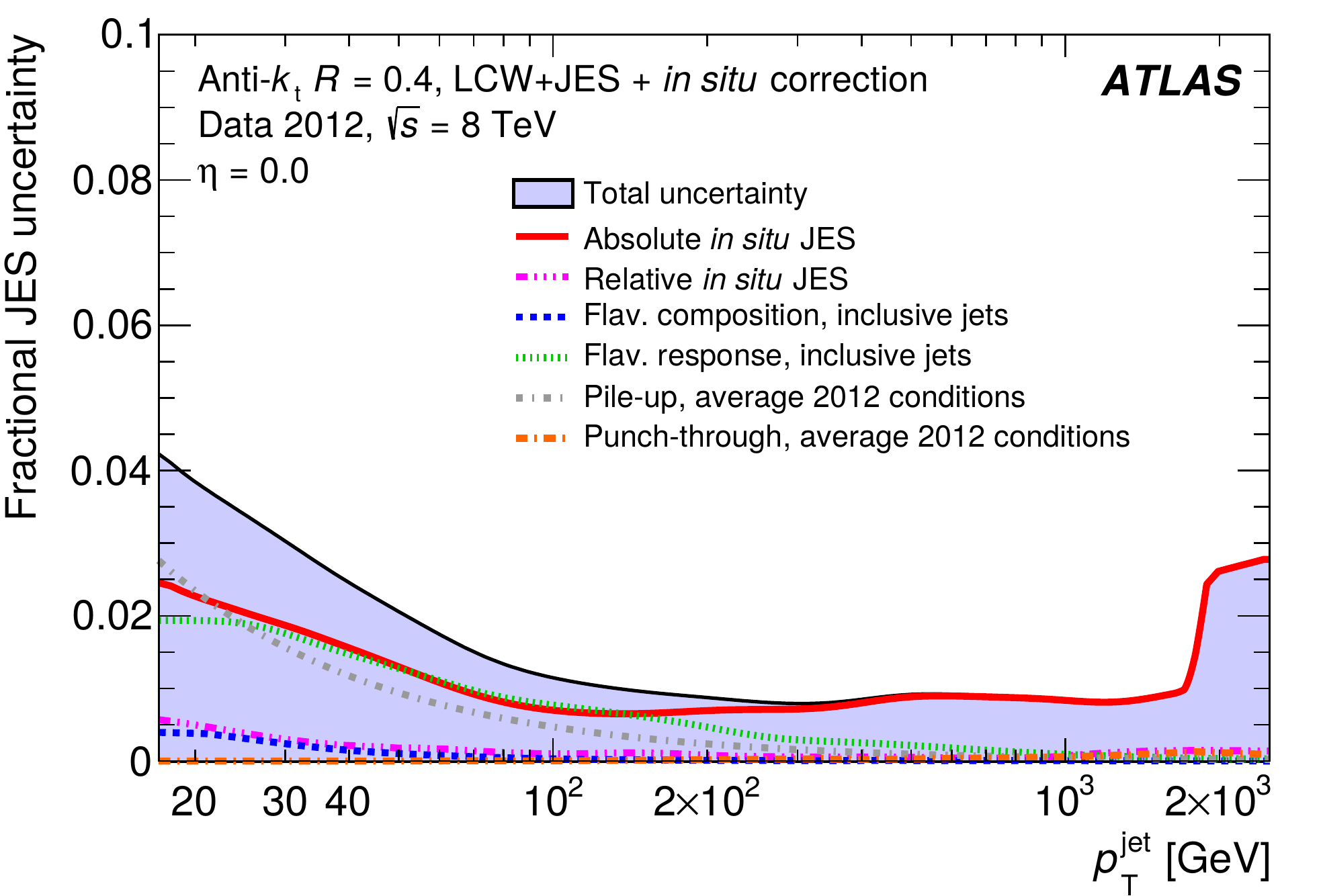}
\caption{LCW in dijet events}
\end{subfigure} \\
 
\bigskip
 
\begin{subfigure}{0.48\textwidth}\centering
\includegraphics[width=\textwidth]{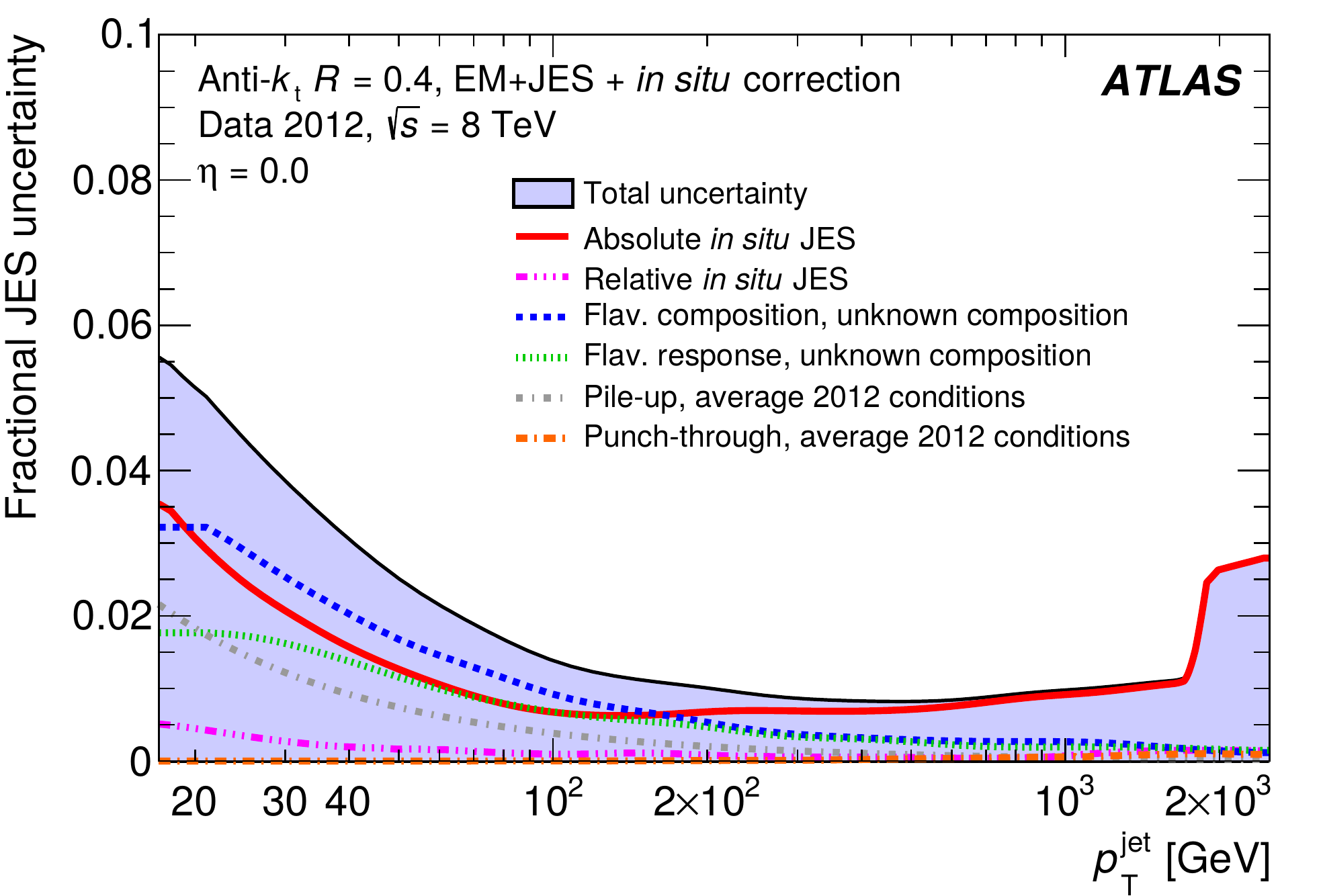}
\caption{EM in unknown composition events}
\end{subfigure}
\begin{subfigure}{0.48\textwidth}\centering
\includegraphics[width=\textwidth]{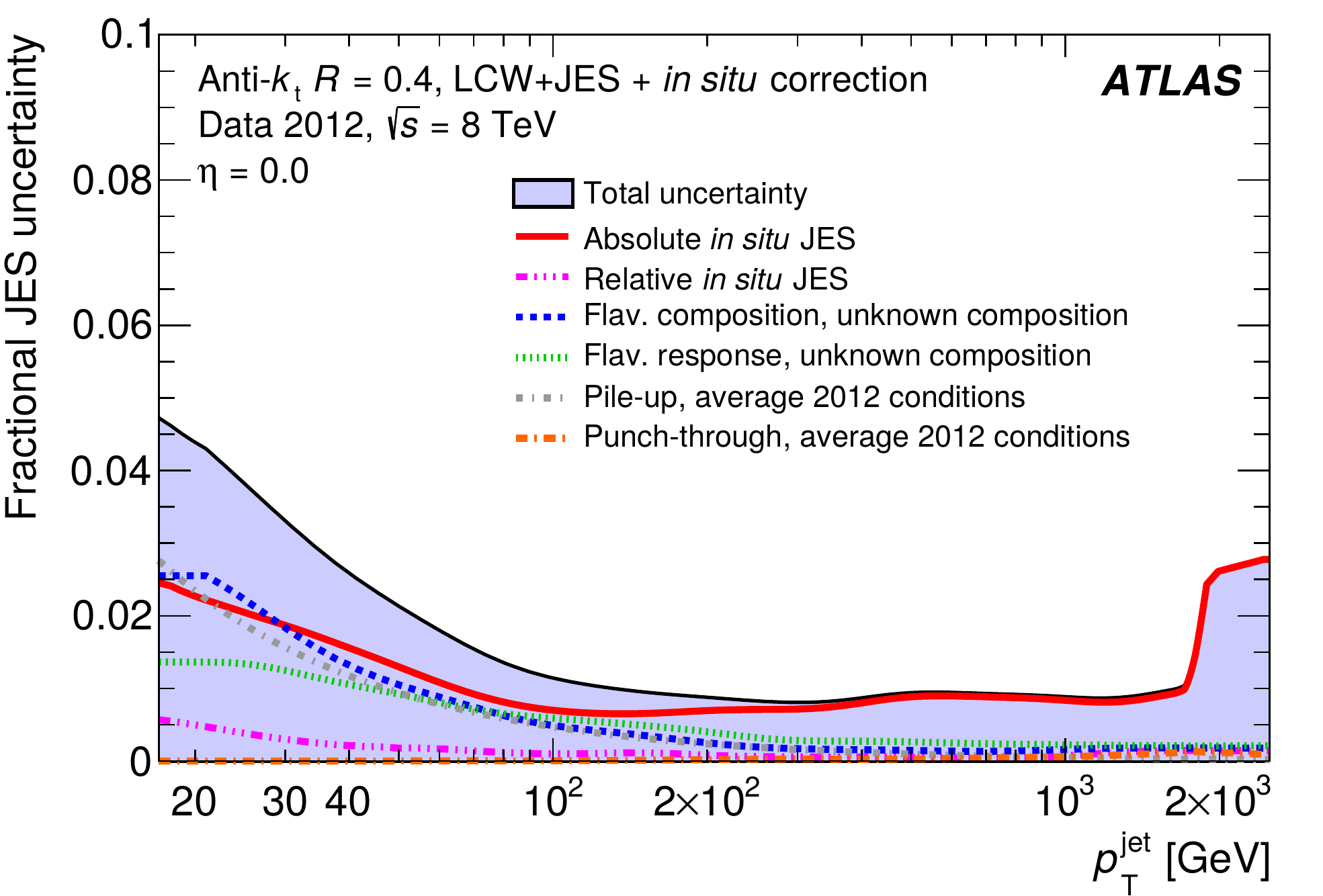}
\caption{LCW in unknown composition events}
\end{subfigure}
 
\caption{The total jet energy scale uncertainty as a function of \pT{} for central jets.  Two flavour compositions are shown, one for dijet events, where the quark/gluon composition is taken from MC simulations and an associated uncertainty from generator comparisons, and one for an unknown flavour composition (assuming 50:50 quark:gluon jets with a 100\% uncertainty).  ``Absolute \insitu{} JES'' refers to the uncertainty arising from \Zjet{}, \gammajet{}, and multijet measurements, including also the single-hadron response uncertainty at high \pT.  ``Relative \insitu{} JES'' refers to the uncertainty arising from the dijet $\eta$ intercalibration.  ``Punch-through'' refers to the uncertainty in the final (muon-based) stage of the global sequential correction.}
\label{fig:JESuncertPt}
\end{figure}
 
\begin{figure}
\begin{subfigure}{0.48\textwidth}\centering
\includegraphics[width=\textwidth]{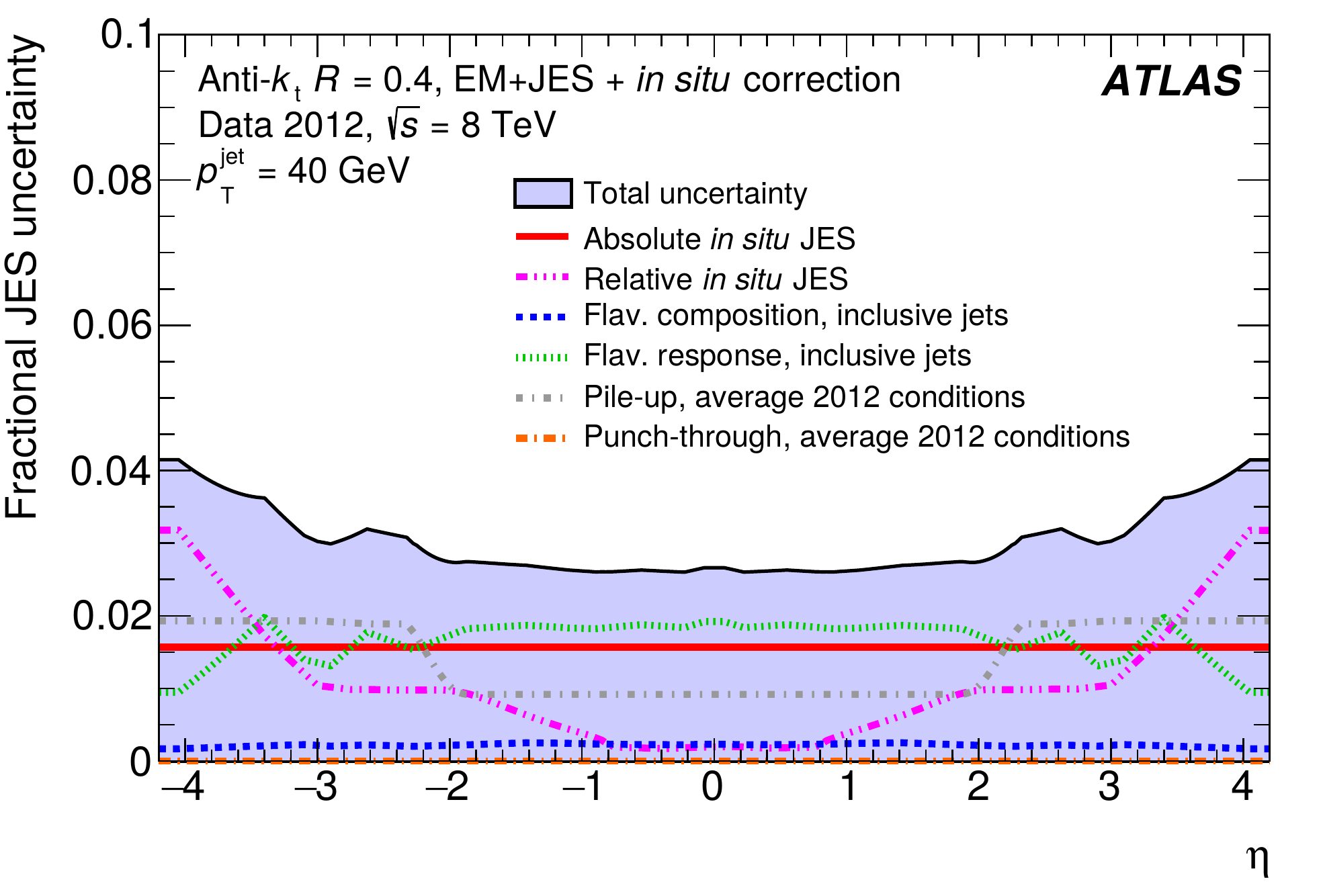}
\caption{EM in dijet events}
\end{subfigure}
\begin{subfigure}{0.48\textwidth}\centering
\includegraphics[width=\textwidth]{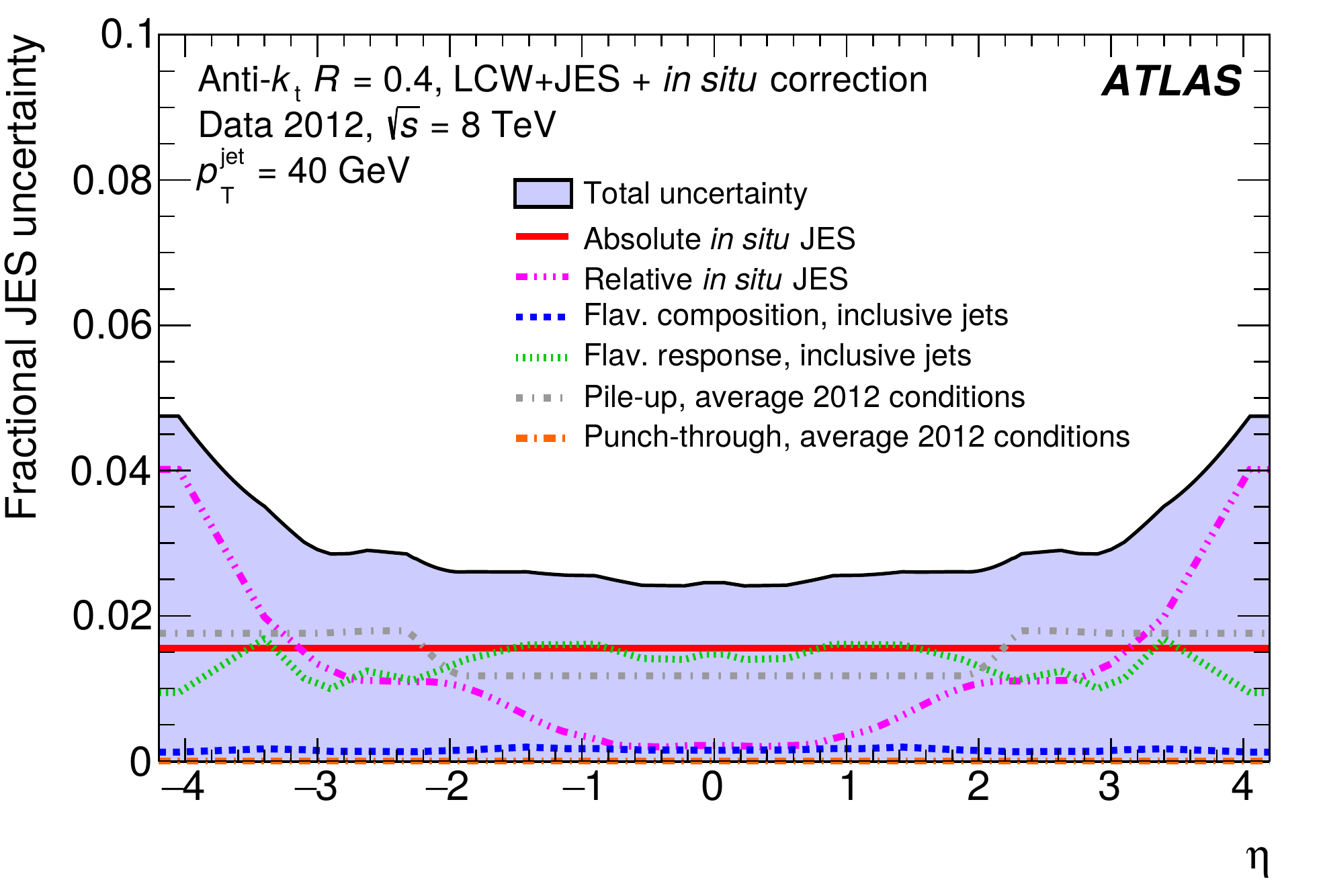}
\caption{LCW in dijet events}
\end{subfigure} \\
 
\bigskip
 
\begin{subfigure}{0.48\textwidth}\centering
\includegraphics[width=\textwidth]{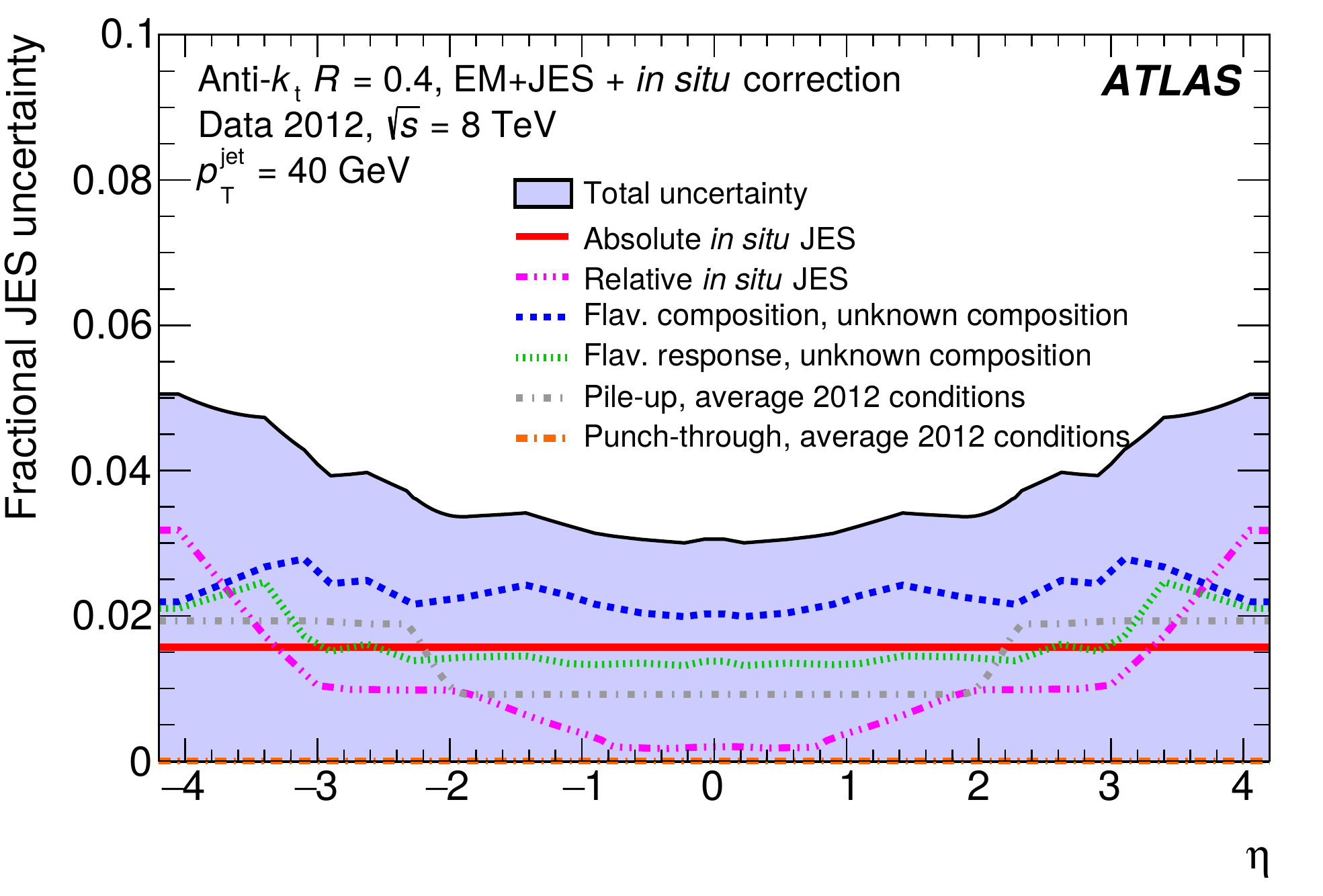}
\caption{EM in unknown composition events}
\end{subfigure}
\begin{subfigure}{0.48\textwidth}\centering
\includegraphics[width=\textwidth]{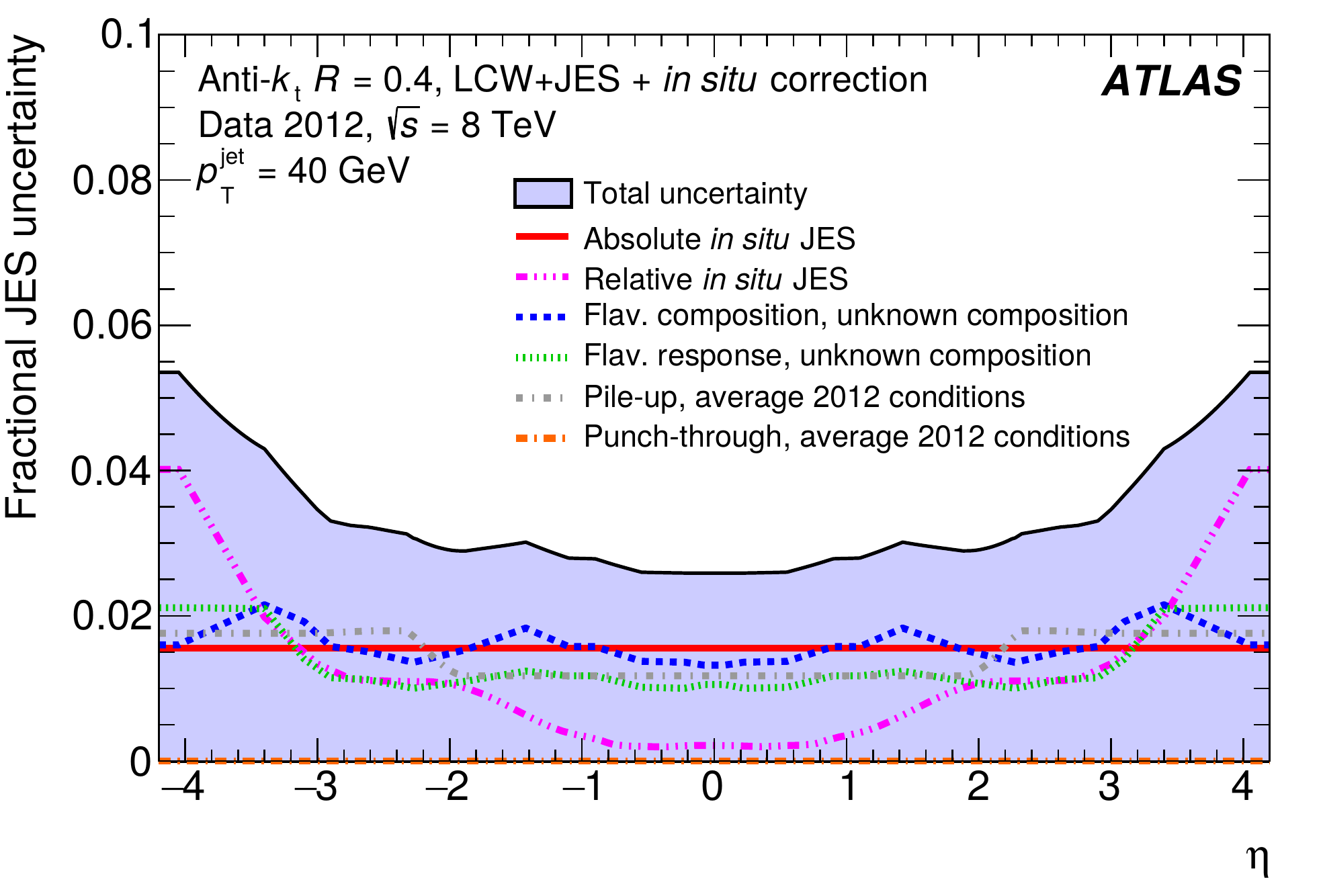}
\caption{LCW in unknown composition events}
\end{subfigure}
 
\caption{The total jet energy scale uncertainty as a function of $|\eta|$ for $\pt = 40$~\GeV\ jets.  Two flavour compositions are shown, one for dijet events, where the quark/gluon composition is taken from MC simulations and an associated uncertainty from generator comparisons, and one for an unknown flavour composition (assuming 50:50 quark:gluon jets with a 100\% uncertainty).  ``Absolute \insitu{} JES'' refers to the uncertainty arising from \Zjet{}, \gammajet{}, and multijet measurements. ``Relative \insitu{} JES'' refers to the uncertainty arising from the dijet $\eta$ intercalibration.  ``Punch-through'' refers to the uncertainty in the final (muon-based) stage of the global sequential correction.}
\label{fig:JESuncertEta}
\end{figure}

\subsubsection{Uncertainties in fast simulation}
 
All uncertainties discussed in the previous section apply to MC samples produced using either the full or fast simulation.
However, a small non-closure of the jet calibration was observed in fast simulation compared with full simulation.
To account for this, an additional systematic uncertainty must be included
in analyses using fast simulation since relative and absolute \insitu{} methods are not used to validate this simulation.
The size of this uncertainty compared with other systematic uncertainties is generally small for $R=0.4$ jets (Figure~\ref{fig:af2_uncertainties_0.4}).
However, as shown in Figure~\ref{fig:af2_uncertainties_0.6}, this uncertainty becomes sizeable for $R=0.6$ jets.
 
\begin{figure}[htb]
\begin{subfigure}{0.48\textwidth}\centering
\includegraphics[width=\textwidth]{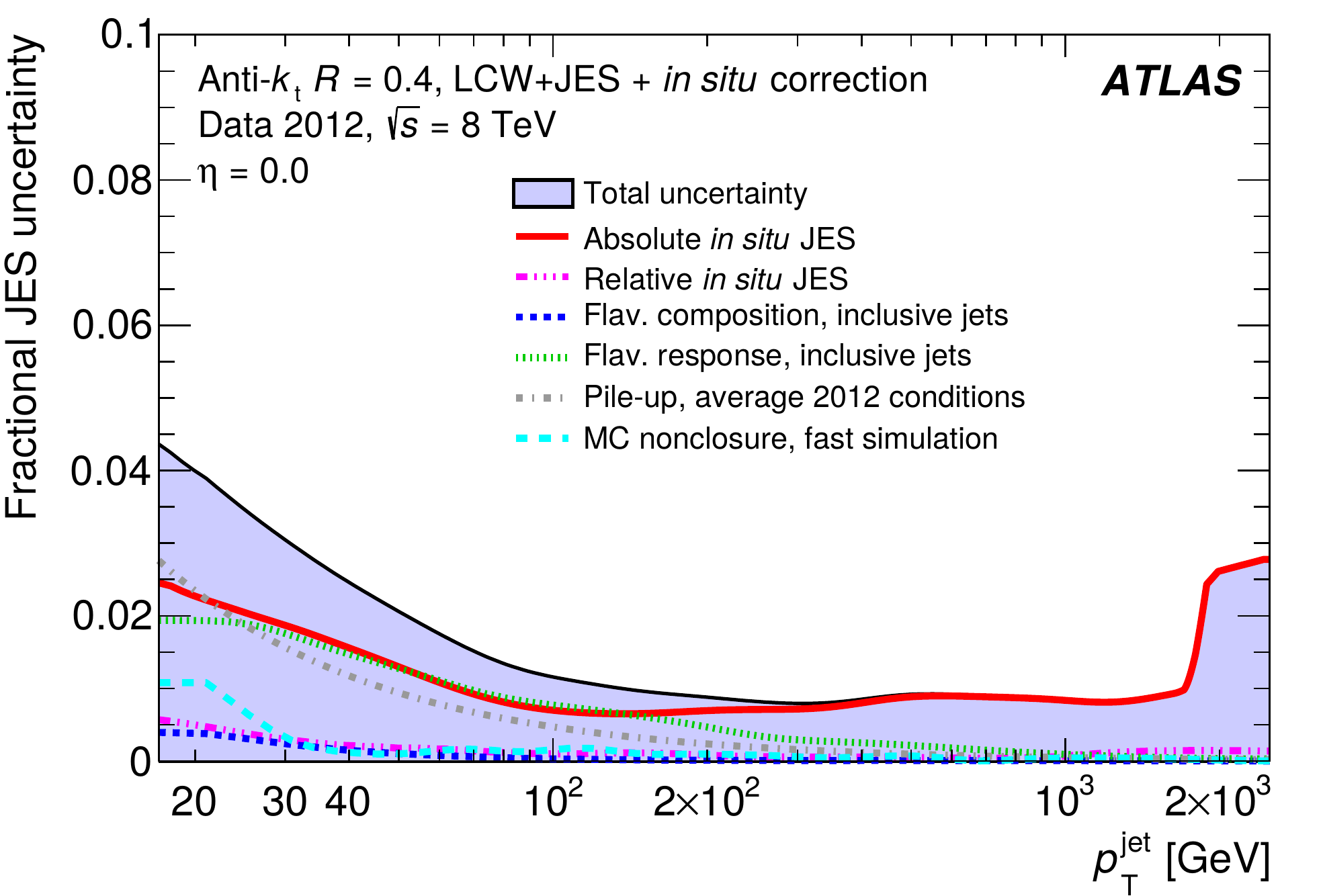}
\caption{Uncertainty vs \pT}
\end{subfigure}
\begin{subfigure}{0.48\textwidth}\centering
\includegraphics[width=\textwidth]{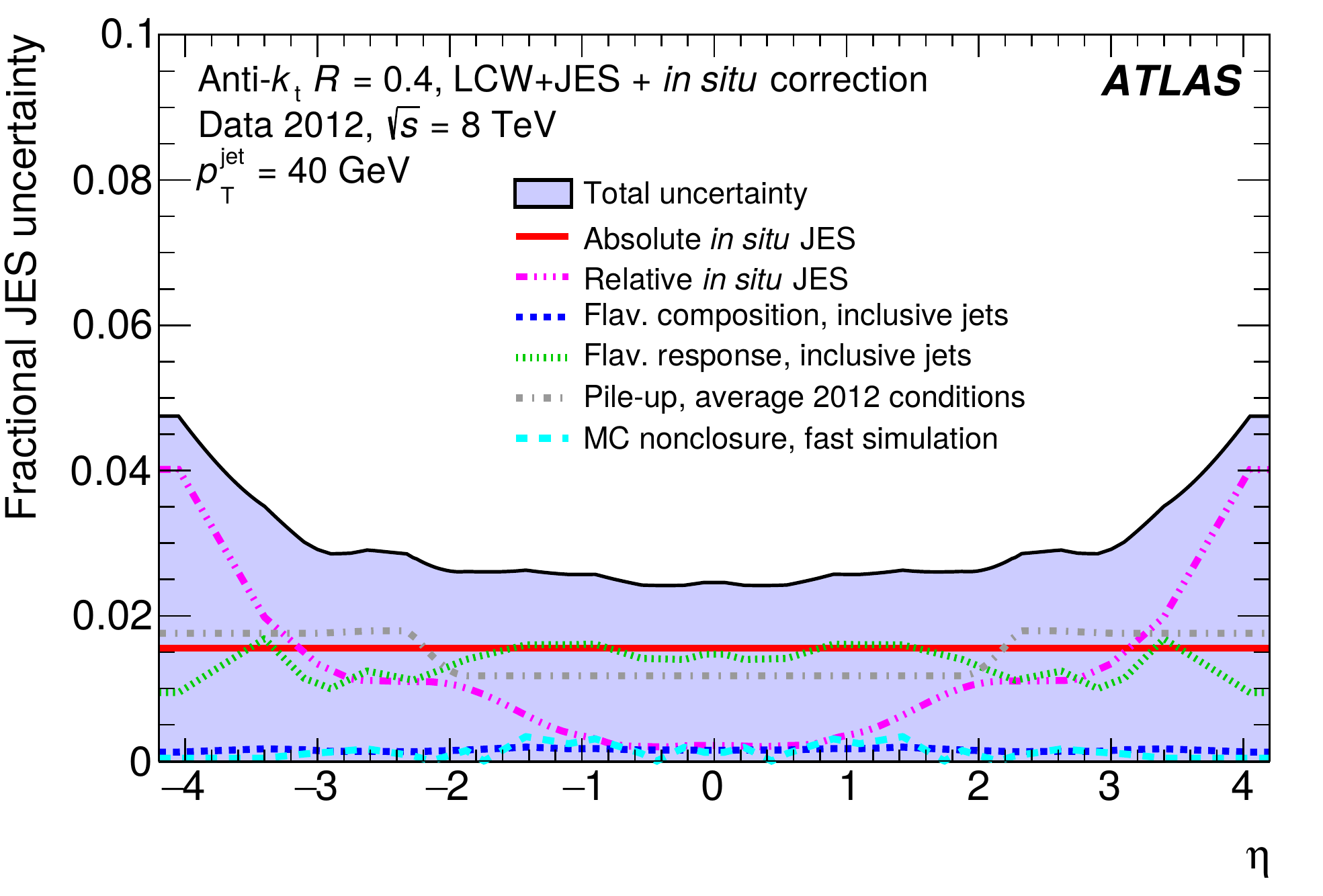}
\caption{Uncertainty vs $\eta$}
\end{subfigure}
\caption{Total uncertainty in the calibration of anti-$k_t$, $R=0.4$ jets in fast
simulation as a function of $\pt$ and $\eta$.  ``Absolute \insitu{} JES'' refers to the uncertainty arising from \Zjet, \gammajet, and multijet measurements.  ``Relative \insitu{} JES'' refers to the uncertainty arising from the dijet $\eta$ intercalibration.  ``MC non-closure, fast simulation'' refers to the additional non-closure observed in fast simulation when comparing with full simulation.}
\label{fig:af2_uncertainties_0.4}
\end{figure}
 
\begin{figure}[htb]
\begin{subfigure}{0.48\textwidth}\centering
\includegraphics[width=\textwidth]{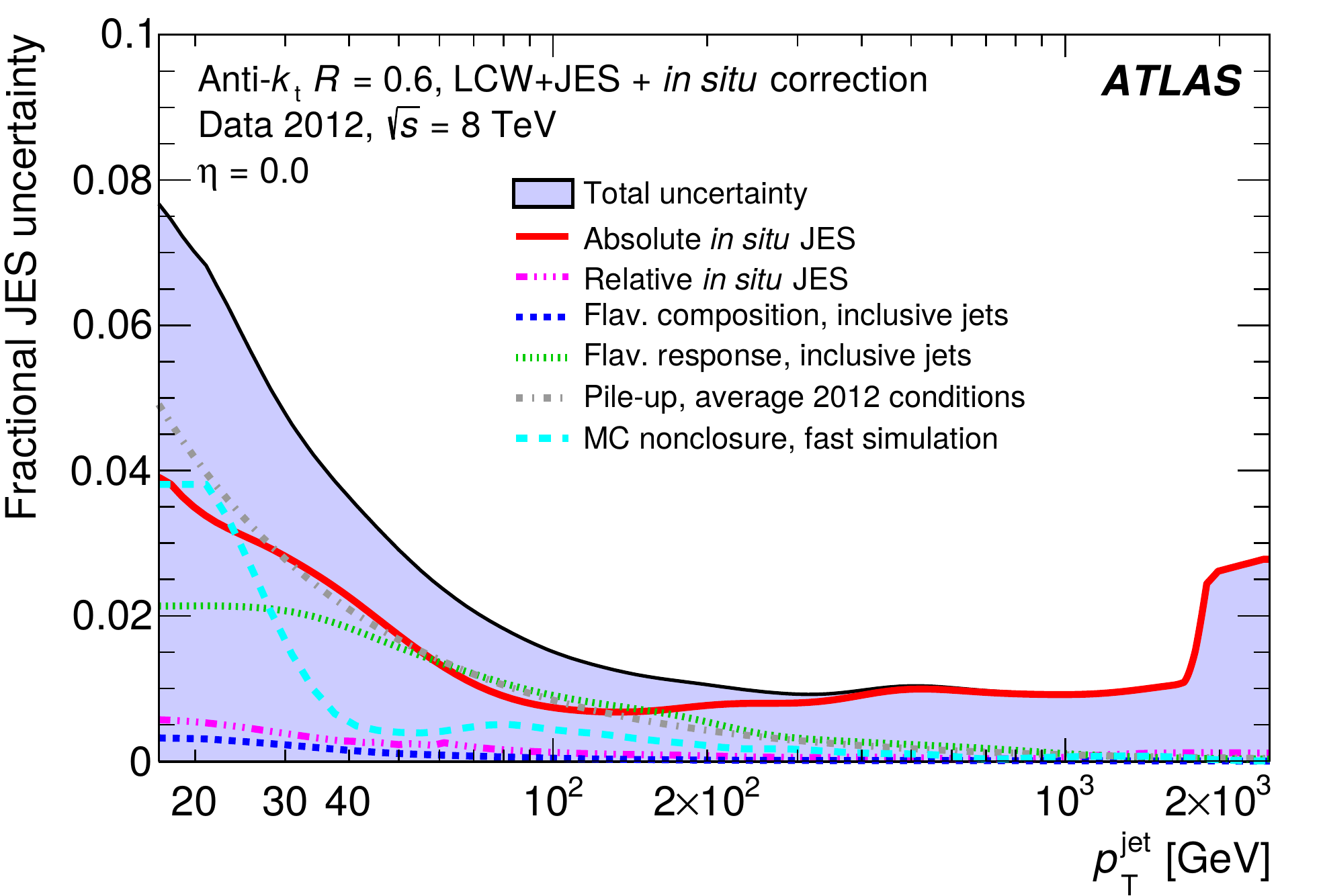}
\caption{Uncertainty vs \pT}
\end{subfigure}
\begin{subfigure}{0.48\textwidth}\centering
\includegraphics[width=\textwidth]{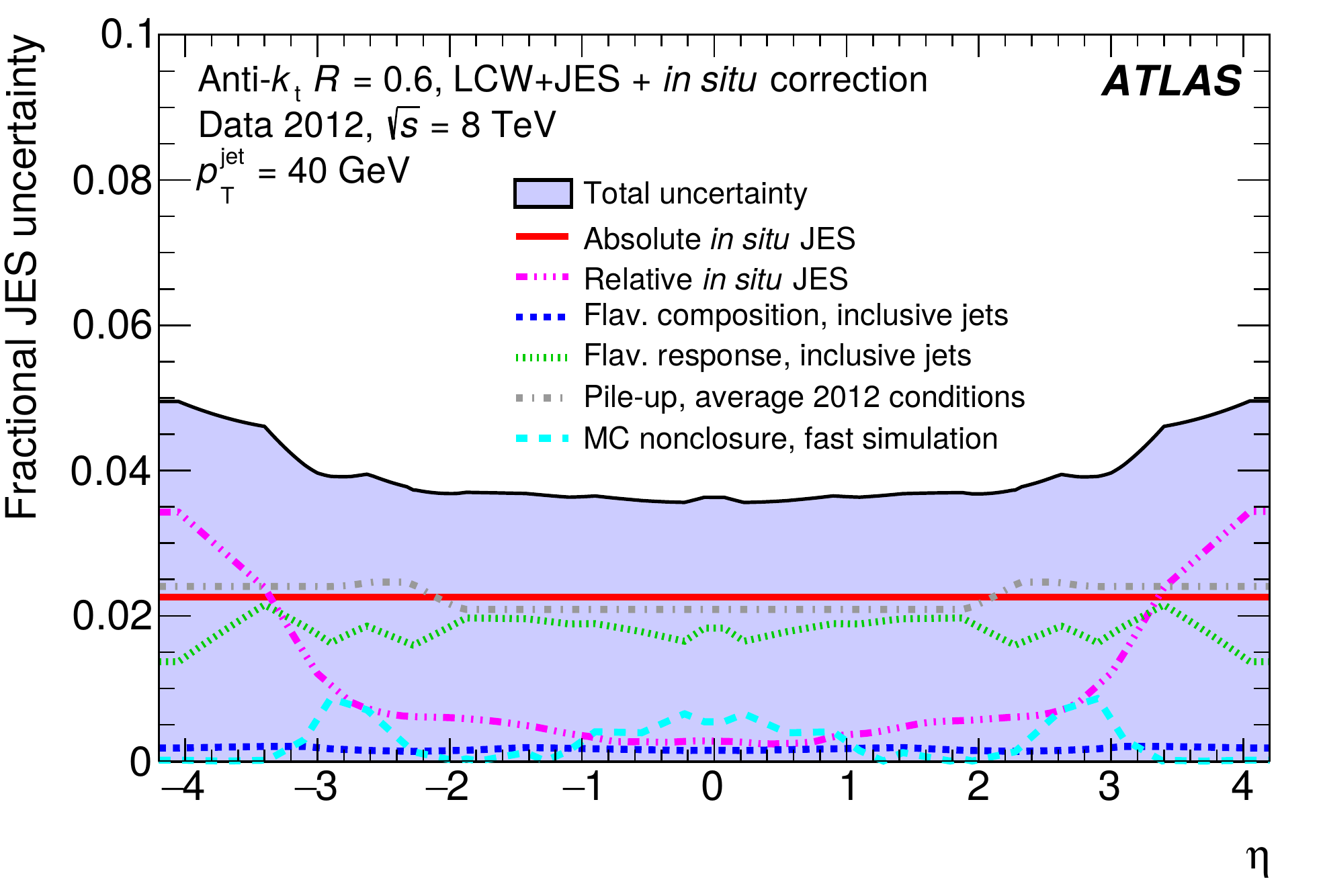}
\caption{Uncertainty vs $\eta$}
\end{subfigure}
\caption{Total uncertainty in the calibration of anti-$k_t$, $R=0.6$ jets in fast
simulation as a function of $\pt$ and $\eta$.  The large ``MC
non-closure'' term demonstrates
the limitations of using $R=0.6$ jets in fast simulation.  ``Absolute
\insitu{} JES'' refers to the uncertainty arising from \Zjet, \gammajet, and
multijet measurements. ``Relative \insitu{} JES'' refers to the
uncertainty arising from the dijet $\eta$ intercalibration.
}
\label{fig:af2_uncertainties_0.6}
\end{figure}

\subsection{Simplified description of uncertainty correlations}
\label{sec:NPreduction}
The list of uncertainties described in Section~\ref{subsec:JESuncertSummary} requires an analysis to propagate a total of 65
JES uncertainty terms to correctly account for all correlations in the jet calibration.
For many analyses it is preferable to describe such correlations using a reduced set of
uncertainty components (nuisance parameters).
 
As detailed in Ref.~\cite{PERF-2012-01}, the total covariance matrix of the JES
correction factors including all the \insitu{} sources can be diagonalized, and then a new set of
independent uncertainty sources can be derived from the eigenvectors and eigenvalues.
A good approximation of the covariance matrix is then obtained by selecting a subset
of the new uncertainty sources (those with the largest eigenvalues) and combining the remaining nuisance parameters into a
residual term.  Figure~\ref{fig:JESreduction} demonstrates this procedure, showing the nominal
correlation matrix and the difference between this and a similar matrix derived from a reduced
set of nuisance parameters.  Only uncertainties depending on a single parameter
(in this case \pT) are combined in this way and any uncertainties with dependencies on other parameters are left separate.
Including such uncertainty components with additional parameter dependencies in the combination
would not result in any significant reduction of the correlation information into fewer nuisance parameters,
as such components require additional dimensions to represent their correlations.

Two reduction schemes are provided. The first scheme reduces the number of central absolute \insitu{}
nuisance parameters, those shown in Figure~\ref{fig:allUncertainties} and the statistical components of the $\gamma+$jet, $Z+$jet, and multijet balance, from 56 to 6 (``standard'').  To preserve some knowledge of
the uncertainty source in this procedure, a second scheme is provided where the reduction is done
within categories (statistical, detector, modelling, or mixed).  This ``category based''
reduction reduces the number of central absolute \insitu{} parameters from 56 to 15.  Retaining the separation of detector, statistical, and modelling components allows the correlation between experiments and different data-taking years to be assessed in combinations of measurements.  No reduction is done for the other terms, and in addition to the 6 (15) nuisance parameters, 9 additional parameters are required, resulting in 15 (24) parameters.
This procedure gives a simpler propagation of the correlations and uncertainties associated
with the jet energy scale with very little loss of information about the correlations.
 
\begin{figure}
\begin{subfigure}{0.48\textwidth}\centering
\includegraphics[width=\textwidth]{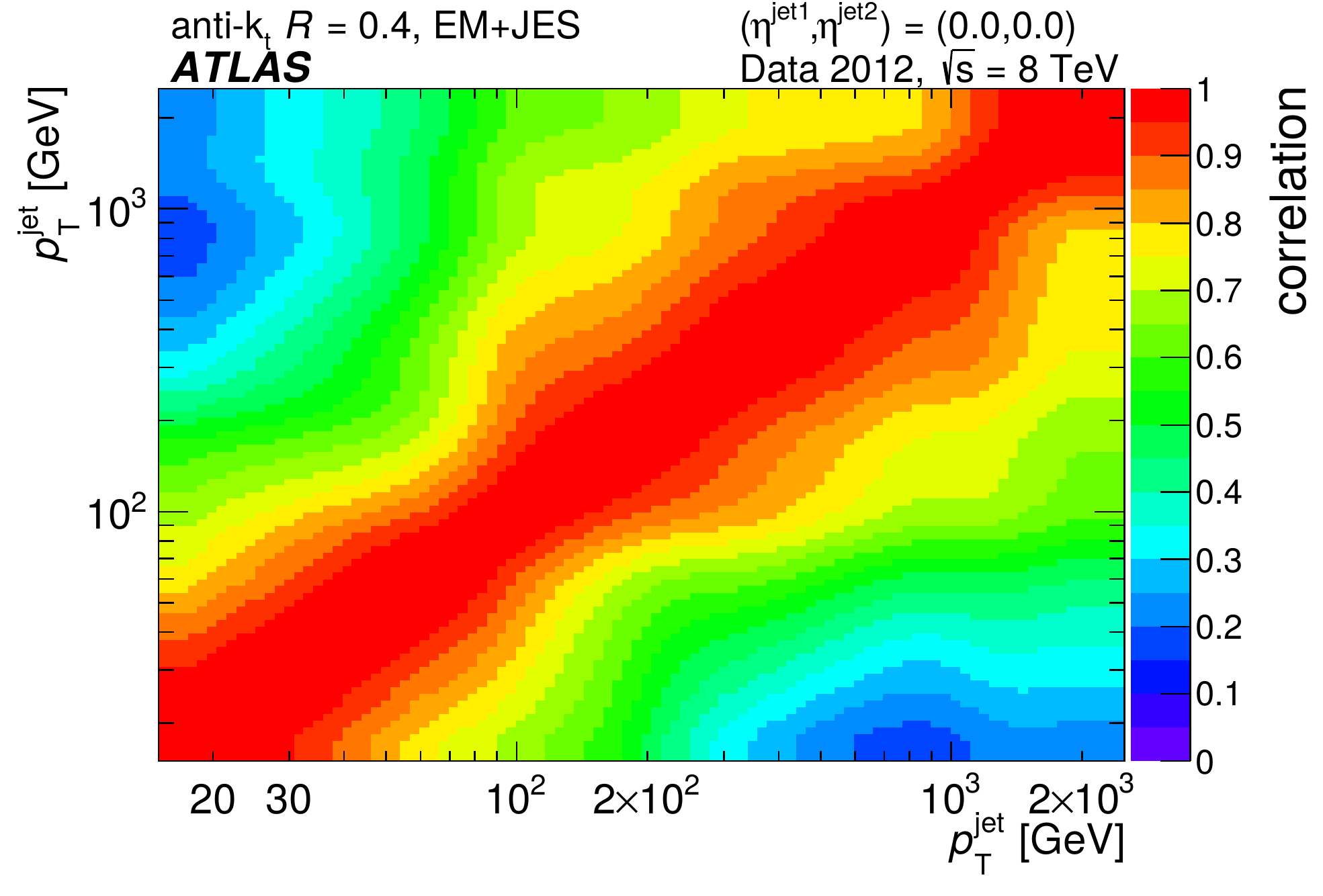}
\caption{}
\end{subfigure}
\begin{subfigure}{0.48\textwidth}\centering
\includegraphics[width=\textwidth]{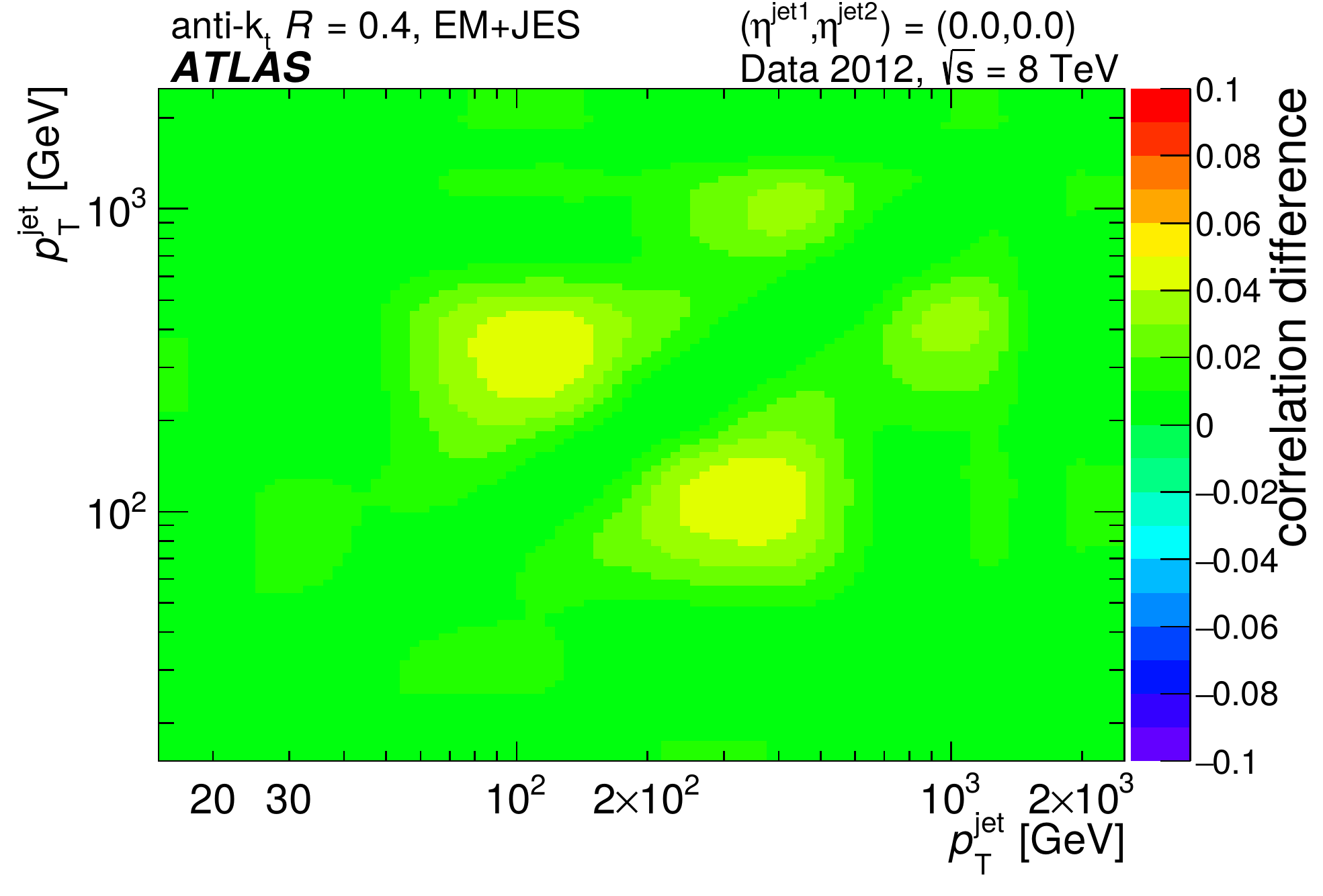}
\caption{}
\end{subfigure} \\
 
\bigskip
 
\begin{subfigure}{0.48\textwidth}\centering
\includegraphics[width=\textwidth]{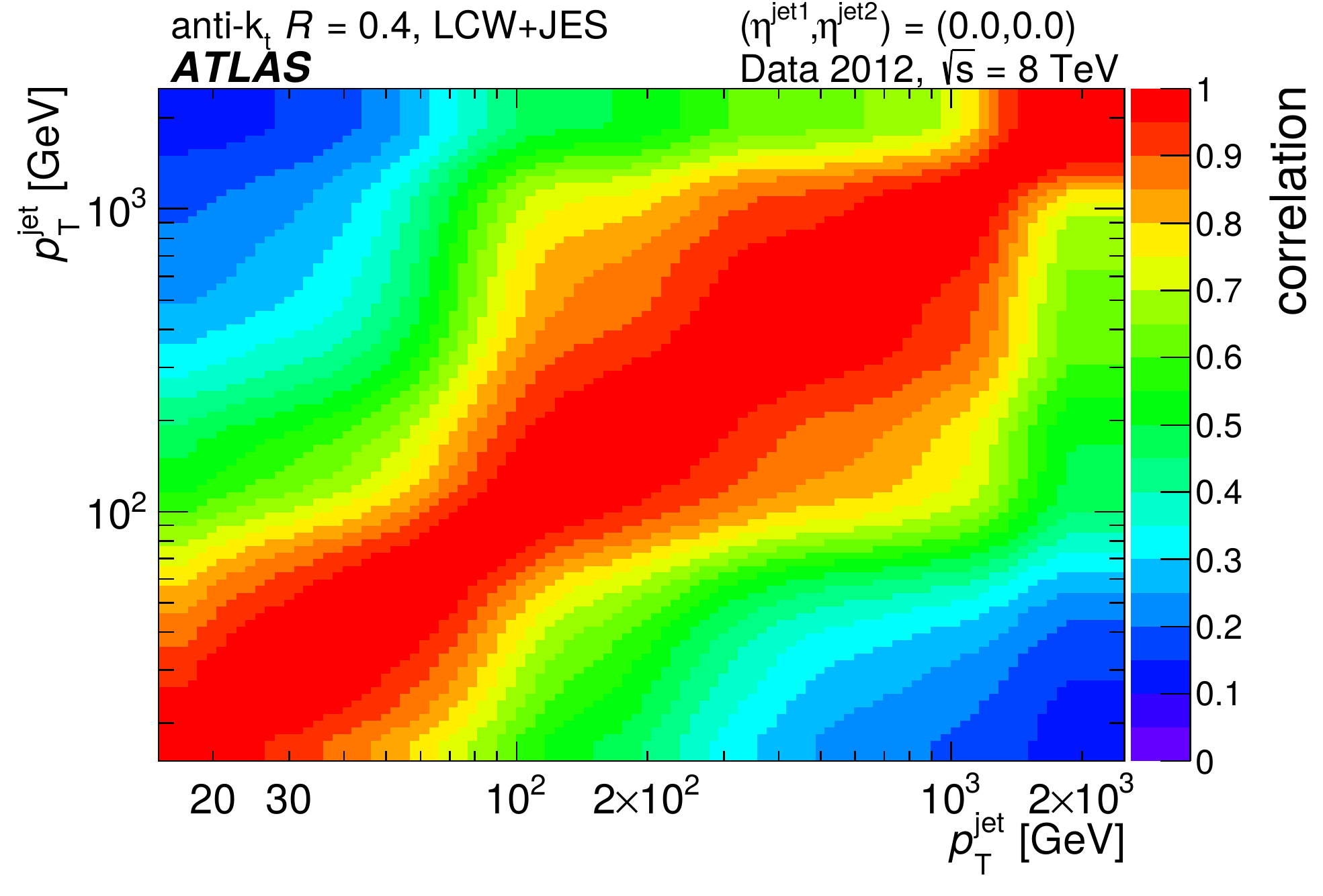}
\caption{}
\end{subfigure}
\begin{subfigure}{0.48\textwidth}\centering
\includegraphics[width=\textwidth]{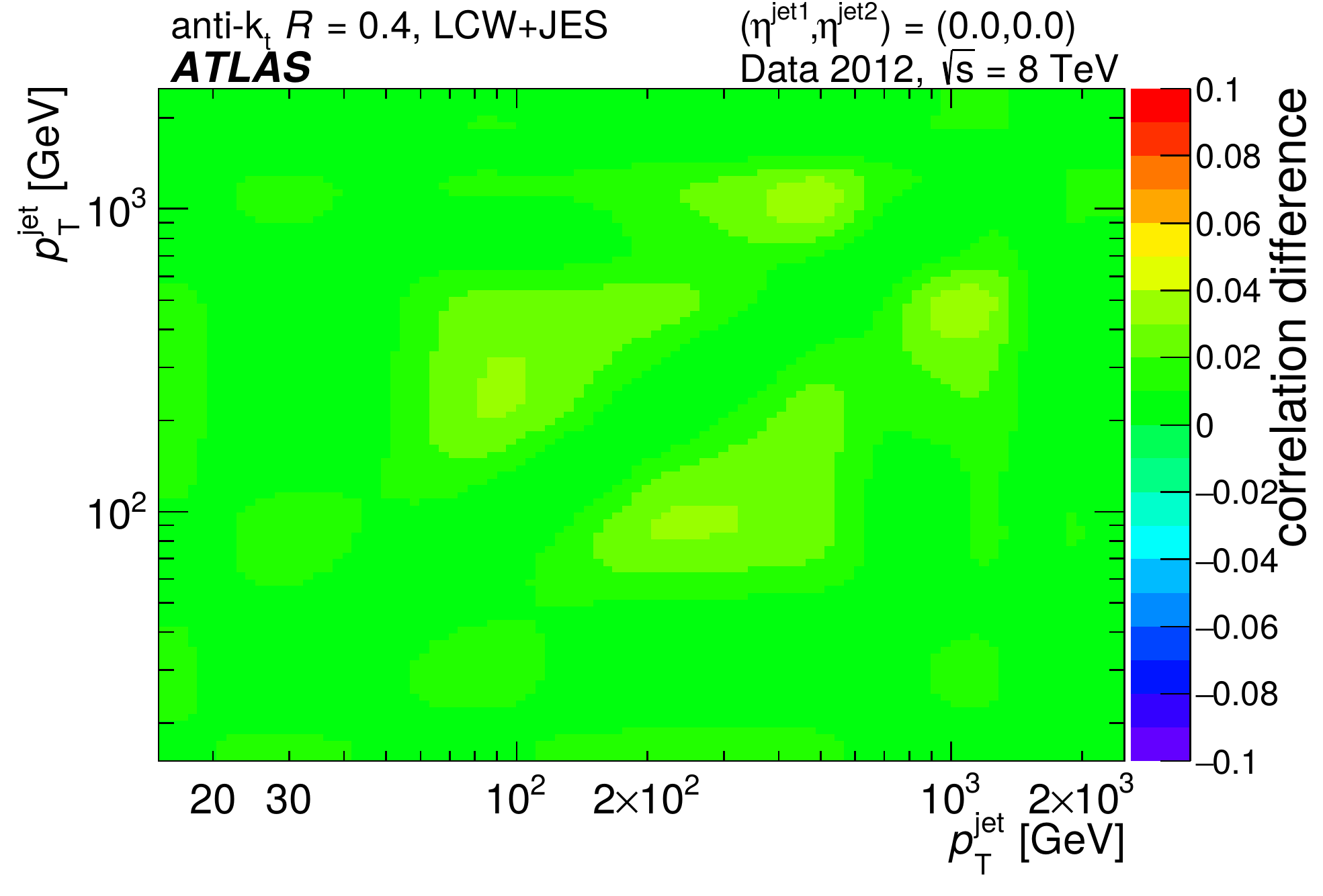}
\caption{}
\end{subfigure}
\caption{The (a,c)~JES correlation matrix and (b,d)~difference between the full correlation matrix and that derived from a reduced number (6) of absolute \insitu{} uncertainty components for \antikt{} \rfour{} jets calibrated with the (a,b)~\EMJES{} and (c,d)~\LCWJES{} schemes.}
\label{fig:JESreduction}
\end{figure}
 
A method has been developed for evaluating the correlations between the full set of 56 \insitu{} JES uncertainty terms and a reduced set.
This is especially useful for evaluating the correlations between the uncertainties obtained for two physics analyses that use different uncertainty configurations (e.g. the full set and a reduced set of JES uncertainty terms).
In this method, each JES uncertainty term in the full set is projected, in the space of uncertainties, onto the direction of each uncertainty term in the reduced set.
The corresponding projection coefficients allow expression of the uncertainties propagated by one analysis using a given configuration in terms of the components corresponding to another configuration.
Therefore, this allows correlations to be assessed between analyses using different uncertainty configurations.
 
\subsection{Alternative uncertainty configurations}
\label{sec:alt_unc}
 
Many physics analyses use ``profiling'' of uncertainties in the statistical analysis,
such as the profile log-likelihood method,
which improves the precision of the associated physics results.
These methods may make significant use of the uncertainty amplitudes and correlation in different kinematic regions, and the exact parameterization of the JES systematic uncertainties might impact the result.
Since the correlation between uncertainty sources often is unknown, the nominal uncertainty parameterization discussed in the previous sections corresponds to a ``best guess''.
Certain analyses could erroneously benefit from somewhat arbitrary choices made during the construction of this uncertainty scheme.
To allow analyses to test if their results depend on these choices,
two alternative uncertainty parameterizations are provided, one that results in stronger JES uncertainty correlations and one that gives weaker correlations.
These are constructed by making alternative assumptions about the correlation between different effects and by employing
a different rebinning prescription when propagating absolute \insitu{} derived uncertainties to the combination.

In both the strong and weak correlation scenarios, a change is made in the rebinning procedure described in Section~\ref{sec:evalSyst}.
The condition for stopping the merging of bins is altered such that the stronger (weaker) correlation scenario has more (less) bins merged.
The effect of this procedure is particularly noticeable at low $\pt$ and results in a reduction
of the absolute \insitu{} uncertainties for the stronger correlation scenario.  In addition, both alternatives use a slower turn-on of the
interpolation between multijet balance and single-particle uncertainties at $\pt\approx 1.7$~\TeV\ (Figure~\ref{fig:JESuncertPt}).
 
For the strong correlations alternative, certain uncertainty components that are treated as being uncorrelated with each other in the nominal parametrization are combined into
a correlated component. This is only done for components that are suspected to have some correlation.
The flavour composition uncertainty is also switched from using \pythia{} to derive the quark/gluon response to using \herwigpp{} to fully encompass generator dependence.
 
For the weak correlation alternative, several ``2-point'' systematic uncertainties are split into two subcomponents~\cite{ATL-PHYS-PUB-2015-014}.
The term 2-point systematic uncertainties refers to uncertainties evaluated by comparison of the nominal result with only one alternative, e.g. a comparison between the predictions from two MC generators.
The two constructed uncertainty components are defined such that their sum in quadrature
equals the original component, thus the total uncertainty is retained.
The split is performed by multiplying the original component by a factor varying linearly from 0 to 1 in either $|\eta|$ or $\log\pT{}$, forming the first subcomponent,
while the second subcomponent is formed as the quadrature complement.
Components treated this way in the alternative configurations include the $\eta$-intercalibration modelling term and flavour components.

\subsection{Large-\texorpdfstring{$R$}{R} jet uncertainties}
\label{JES_largeR_uncert}
 
Uncertainties in the large-$R$ jet calibration are determined using \insitu{} methods with the same principle as for $R=0.4$ and $R=0.6$ jets.
Jet energy scale uncertainties are derived by combining direct balance measurements (Eq.~(\ref{eq:RDB})) performed in $\gamma$+jet events
and are combined with uncertainties with \trackjets{} as reference objects. Uncertainties for the jet mass scale are derived only using \trackjets{} as reference objects.  The \trackjet{} double-ratio method is discussed below along with an additional topological uncertainty similar to the flavour composition uncertainty in small-$R$ jets.  The $\gamma$+jet studies and uncertainties are discussed in Section~\ref{sec:vjets}.
 
\subsubsection*{\Trackjet{} double-ratio method}
 
In the double-ratio method, \trackjets{} are used as reference objects since charged-particle tracks are both well measured and independent of the calorimeter and are associated with calorimeter jets using a geometrical matching in the $\eta$--$\phi$ plane.  This method assumes that energy fluctuations measured using the calorimeter are independent of the charge-to-neutral fraction of the particle-level jet's constituents.  This is only approximately true because the calorimeter response is different for charged and neutral particles.  The precision of the method requires that the \trackjet{} momentum resolution is much smaller than the calorimeter jet energy resolution, an excellent approximation for calorimeter jet momenta up to several hundred \GeV.
 
This approach was widely used in the measurement of the jet mass and substructure properties of jets in the 2011 data~\cite{PERF-2012-02}. Performance studies~\cite{STDM-2010-13} have shown that there is excellent agreement between the measured positions of clusters and tracks in data, indicating no systematic misalignment between the calorimeter and the inner detector.  However, the use of \trackjets{} as reference objects is limited to a precision in the jet mass scale of around 3\%--7\% in the central detector region due to systematic uncertainties arising from the inner-detector tracking efficiency~\cite{STDM-2010-06} and confidence in MC modelling of the charged and neutral components of jets. The \trackjet{} double ratio is compared for two different MC generators: \textsc{Pythia8} and \textsc{Herwig++}, and the larger disagreement between data and MC prediction is taken as the uncertainty. Figure~\ref{fig:JMS_track} shows the jet mass scale uncertainty for \antikt{} $R = 1.0$ trimmed jets in different detector regions. The uncertainties are derived in bins of \pt{}, $|\eta|$, and $m/\pT$, and two examples are shown.
 
\begin{figure}
\begin{subfigure}{0.49\linewidth}\centering
\includegraphics[width=\linewidth]{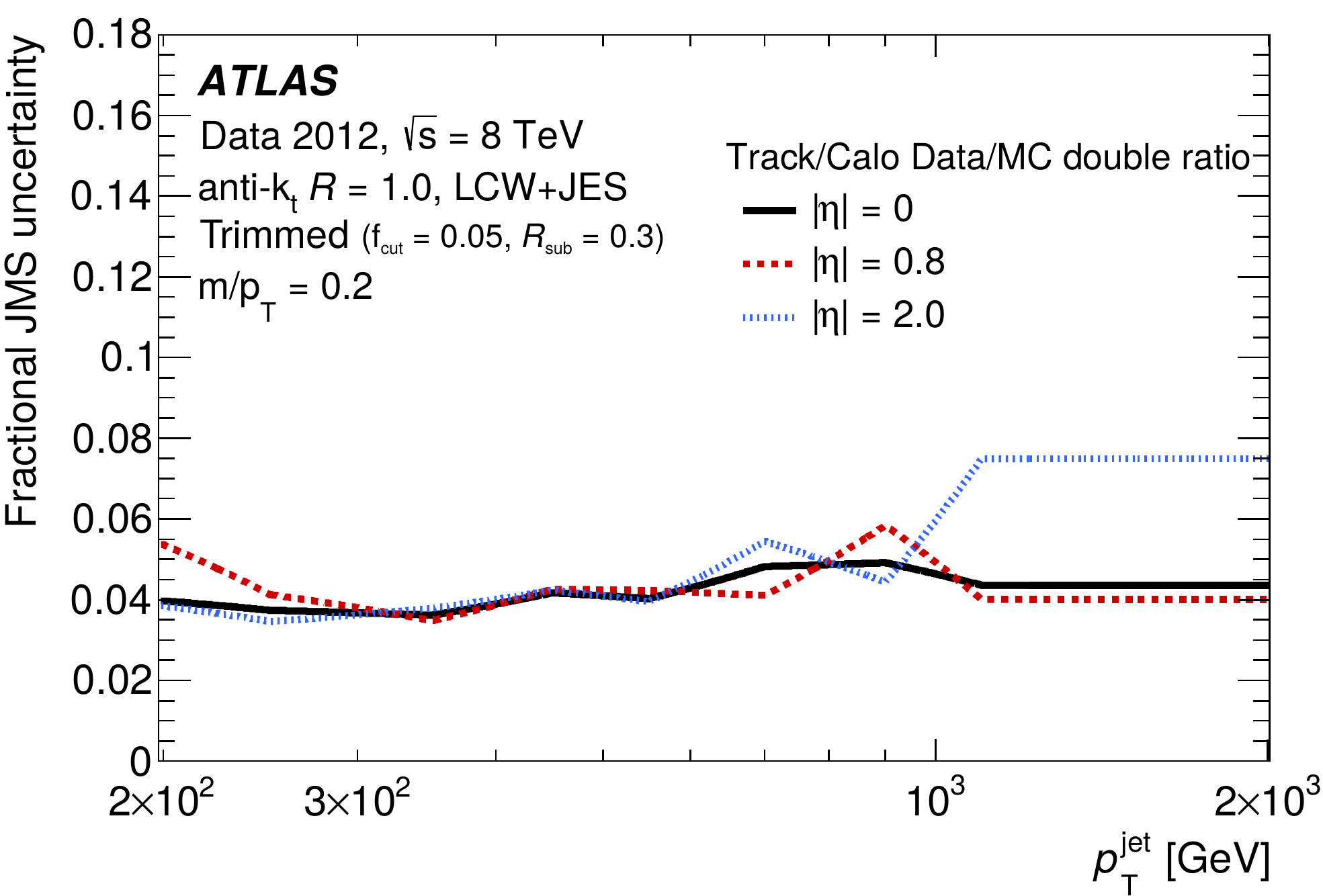}
\caption{$m/\pt$ = 0.2}\label{fig:JMS_eta000_mop020}
\end{subfigure}
\begin{subfigure}{0.49\linewidth}\centering
\includegraphics[width=\linewidth]{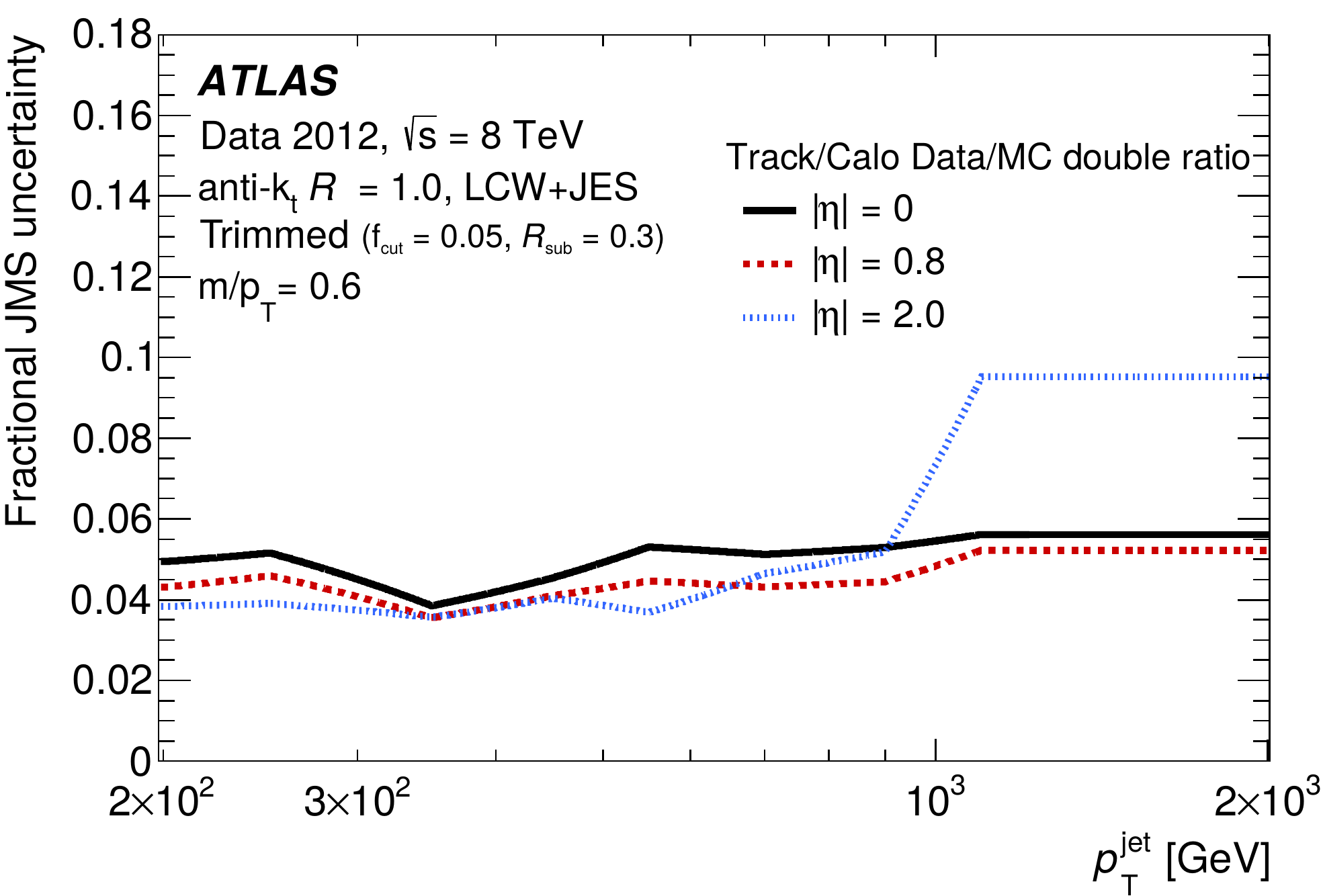}
\caption{$m/\pt$ = 0.6}\label{fig:JMS_eta000_mop060}
\end{subfigure}
\caption{Jet mass scale (JMS) uncertainties for \antikt{} $R = 1.0$ trimmed jets ($f_\mathrm{cut} = 0.05$ and $R_\mathrm{sub} = 0.3$) in different detector regions for (a)~$m/p_\mathrm{T}$ = 0.2 and (b)~$m/p_\mathrm{T}$ = 0.6.}
\label{fig:JMS_track}
\end{figure}
 
\subsubsection*{Topological uncertainty}
Similarly to the jet flavour composition uncertainty for small-$R$ jets, an uncertainty in the jet energy response for different mixtures of quark/gluon jets, boosted top jets, and $W$ jets is derived for large-$R$ jets. Simulated $t\bar{t}$ events are used to account for the different hard substructure and energy distributions within the $W$ or top jets compared with quark/gluons jets which are taken from $W$+jet samples requiring exactly one lepton. The uncertainties are derived for \antikt{} $R$ = 1.0 trimmed jets. Figure~\ref{fig:largeR_topological} shows the \pt{} dependance of the jet response in three $\eta$ regions for four different kinds of jets:
``full top'' jets have the three quarks from the top decay contained within $\Delta R = 0.8$ of the jet axis;
``$W$-only'' jets have the quarks from the $W$ decay within $\Delta R = 0.8$ of the jet axis but any $b$-quark must have $\Delta R > 1.2$; ``non-top'' jets have the top quark separated from the jet axis by $\Delta R > 2.0$; and, ``QCD jets'' are jets from a leptonically decaying $W$+jets sample. The topological uncertainty (Figure~\ref{fig:largeR_topological}) is determined by the envelope of the responses of these different types of jets.
 
\begin{figure}
\centering
\begin{subfigure}{0.47\linewidth}\centering
\includegraphics[width=\linewidth]{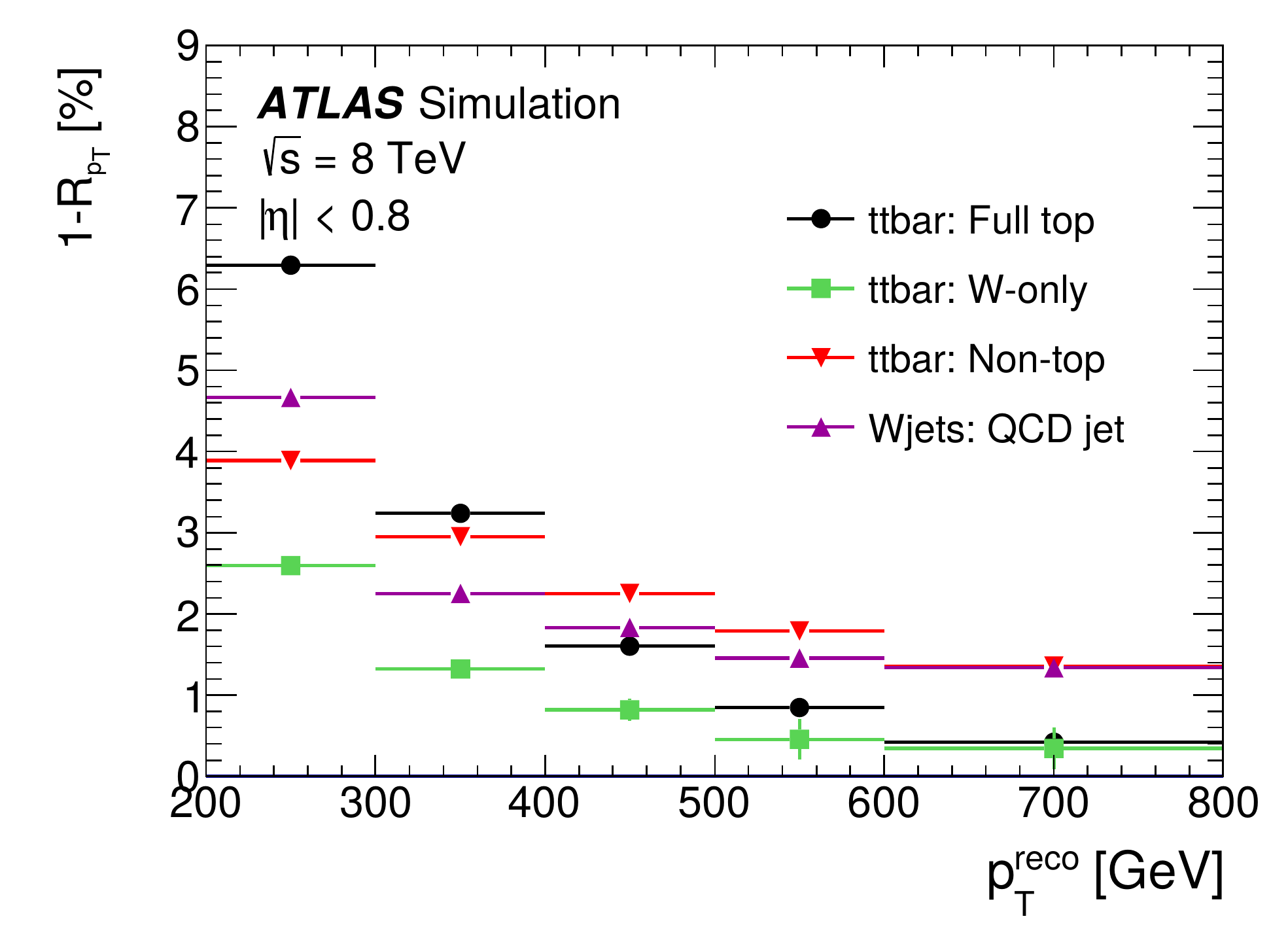}
\caption{$0.0 < |\eta| < 0.8$}\label{fig:JpTS_topo_eta0}
\end{subfigure}
\begin{subfigure}{0.49\textwidth}\centering
\includegraphics[width=\textwidth]{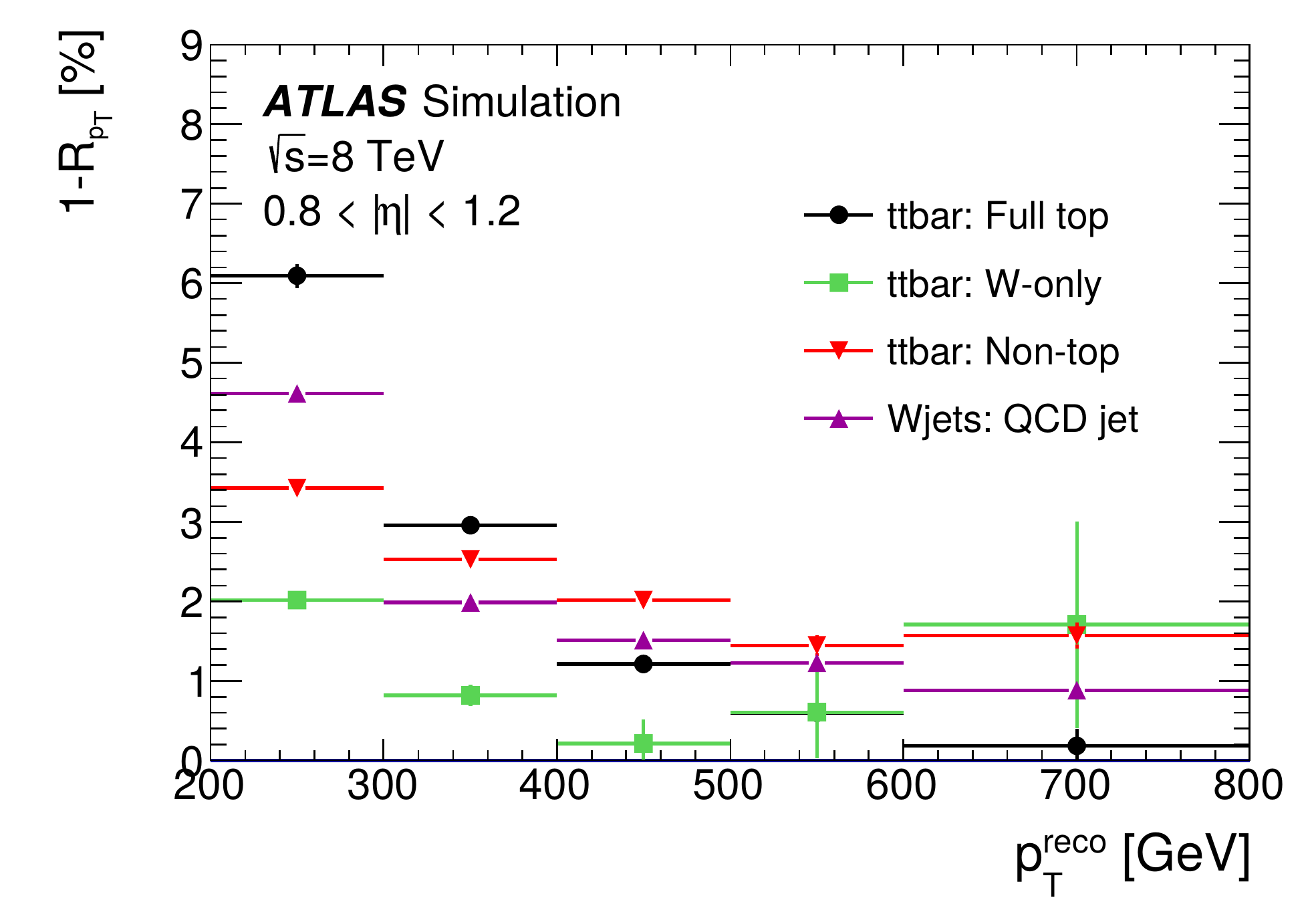}
\caption{$0.8 < |\eta| < 1.2$}\label{fig:JpTS_topo_eta8}
\end{subfigure} \\
 
\bigskip
 
\begin{subfigure}{0.49\textwidth}\centering
\includegraphics[width=\textwidth]{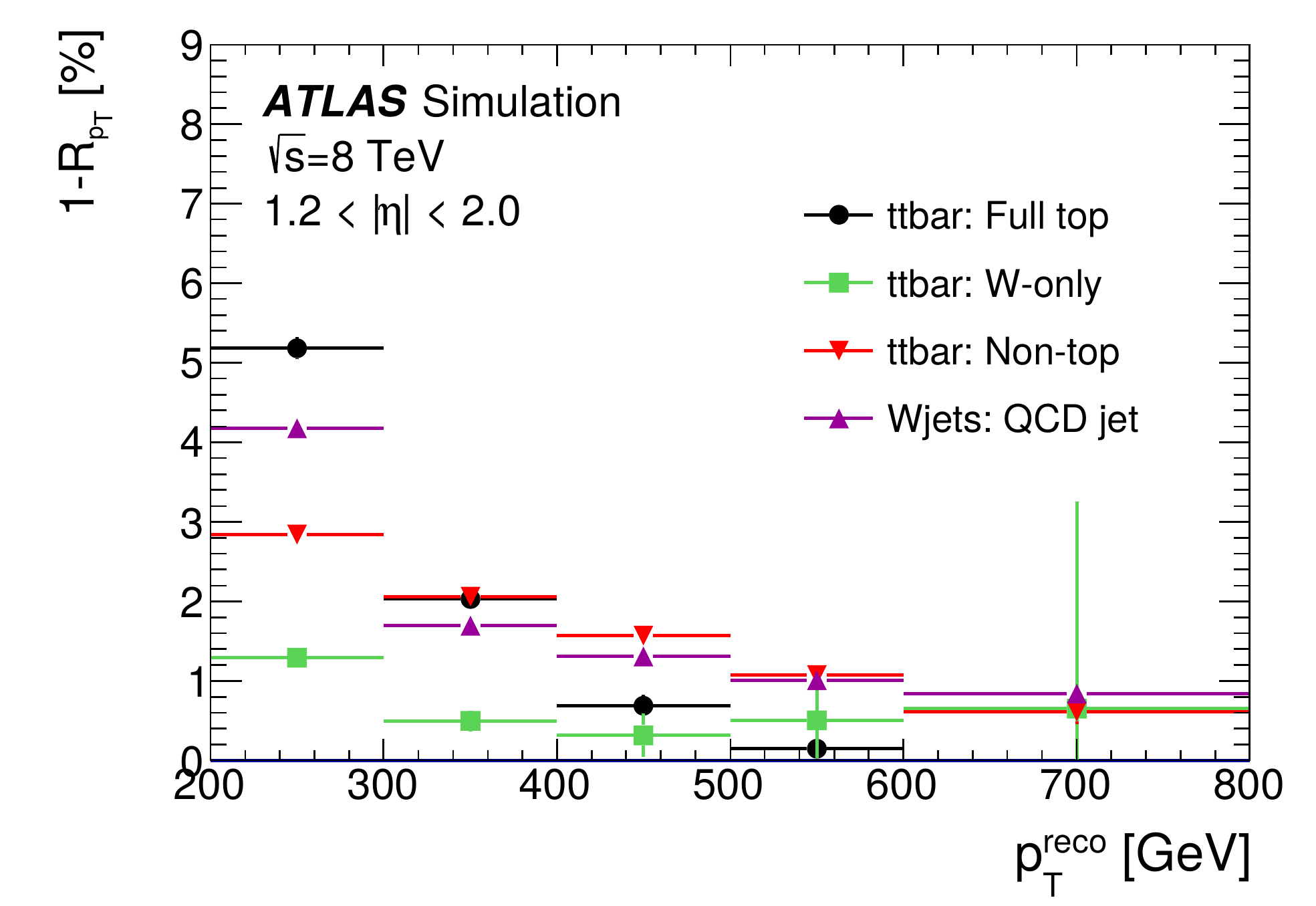}
\caption{$1.2 < |\eta| < 2.0$}\label{fig:JpTS_topo_eta12}
\end{subfigure}
\caption{One minus the jet \pt{} response ($\langle \pttruth - \pt^\text{reco}\rangle/\pttruth = 1-\mathcal{R}_{\pt}$) for \antikt{} $R = 1.0$ trimmed jets with different flavour composition for (a) $0.0 < |\eta| < 0.8$, (b) $0.8 < |\eta| < 1.2$, (c) $1.2 < |\eta| < 2.0$. The categories in the plot are defined by (1)~``$t\bar t$ full top'' jets~(circles) that represent jets for which the three quarks from a hadronic top quark decay are contained within $\Delta R = 0.8$ of the jet axis; (2)~``$t\bar t$ W-only'' jets~(squares), for which the quarks from the $W$~boson decay are within $\Delta R = 0.8$ of the jet axis while the $b$-quark fulfils $\Delta R > 1.2$; (3)~``$t\bar t$ non-top'' jets~(lower triangles) that represent jets for which the top quark is $\Delta R > 2.0$ from the jet; and, (4)~``Wjets QCD'' jets~(upper triangles) representing jets from a leptonically decaying $W$ boson in a $W$+jets MC sample. These are plotted as a function of reconstructed jet \pt{} $(\pt^\text{reco})$, but due to the large bin size compared with the \pT resolution, the choice of plotting $\pt^\textrm{reco}$ or \pttruth{} is of little significance.}
\label{fig:largeR_topological}
\end{figure}
 
\subsubsection*{Combination}
 
The jet \pt{} scale uncertainties are available within $|\eta| < 2.0$ but the available data at high \pt{} ($\pt > 800$~\GeV) is limited for the direct \gammajet{} \pt{} balance method.  By contrast, the uncertainties from the \trackjet{} double ratios cover $\pt > 800$~\GeV. To benefit from the drastically reduced \pt{} scale uncertainties derived with $\gamma$+jet events, a linear interpolation is performed around \pt{} = 800~\GeV\ between the two methods. The uncertainty arising from the topological composition of the jet is added in quadrature to form the total uncertainty. This total uncertainty and its components are shown as a function of \pT in Figure~\ref{fig:JpTS_Combination}.
 
\begin{figure}
\begin{subfigure}{0.48\textwidth}\centering
\includegraphics[width=\textwidth]{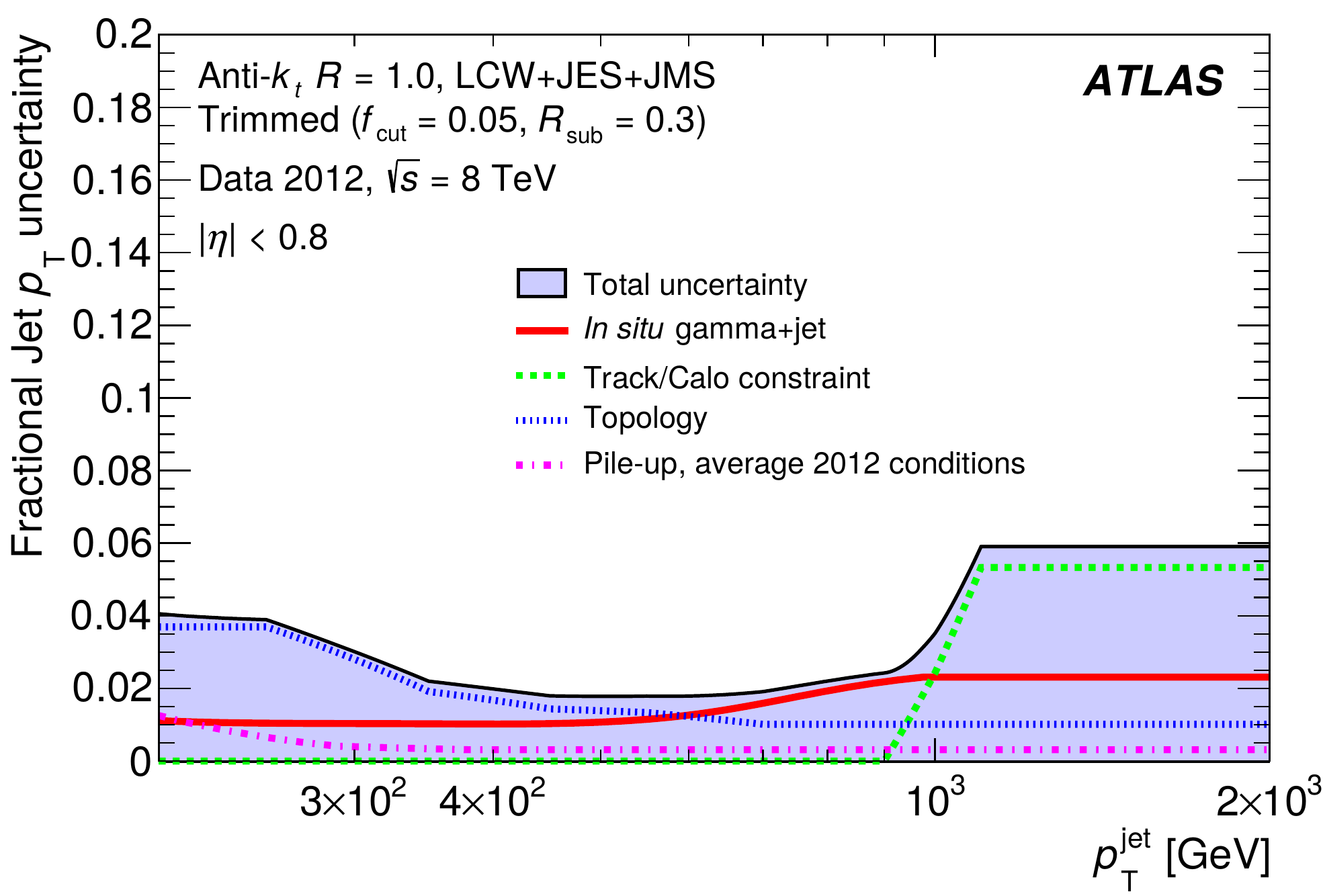}
\caption{$m/\pt$ = 0.2}\label{fig:JPTS_eta080_mop020}
\end{subfigure}
\begin{subfigure}{0.48\textwidth}\centering
\includegraphics[width=\textwidth]{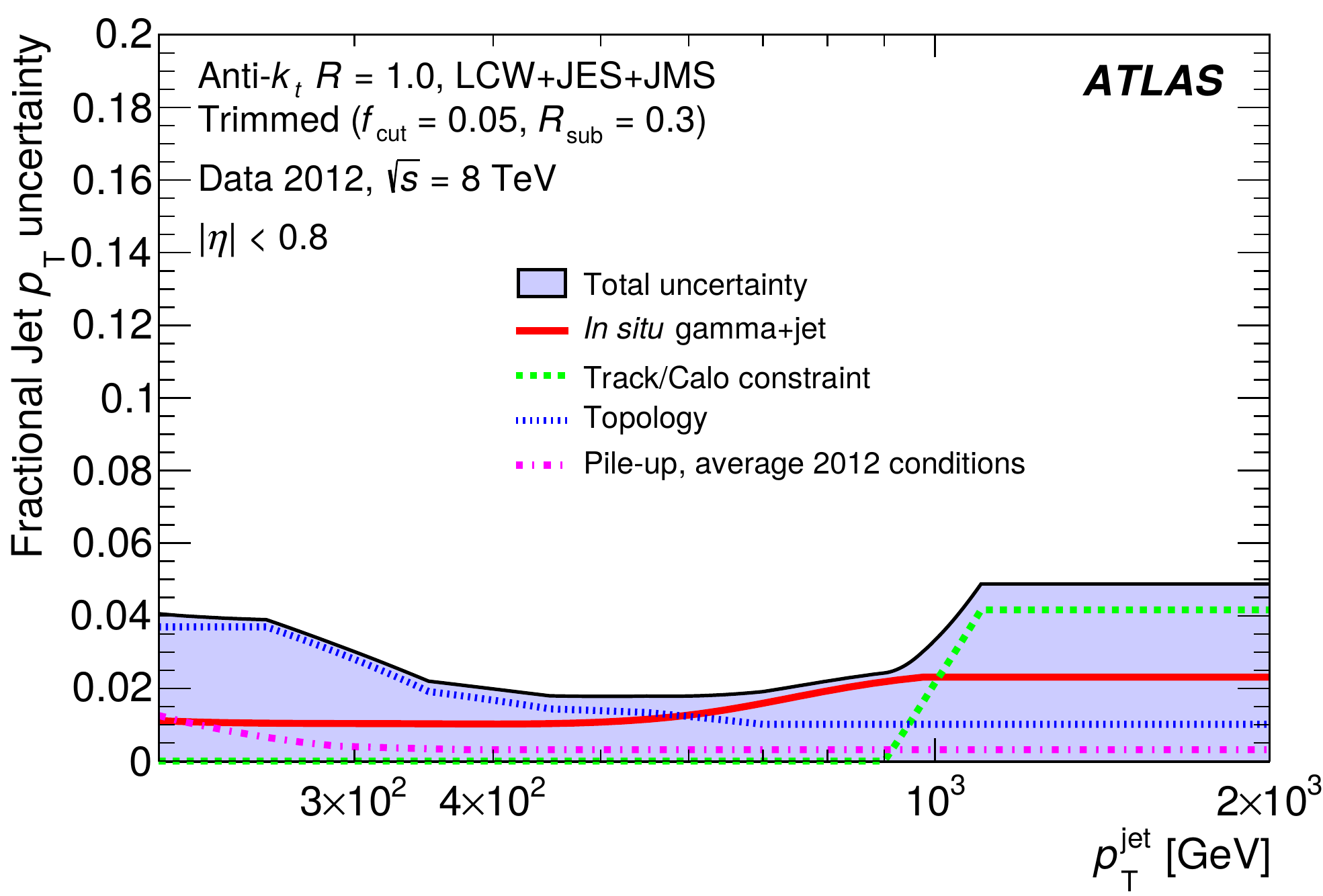}
\caption{$m/\pt$ = 0.6}\label{fig:JPTS_eta080_mop060}
\end{subfigure}
\caption{Combination of the uncertainties in the jet \pt{} scale for \antikt{} $R$ = 1.0 trimmed jets for $|\eta| = 0$ and two values of $m/\pt$: (a) $m/\pt$ = 0.20 and (b) $m/\pt$ = 0.60.}
\label{fig:JpTS_Combination}
\end{figure}

\section{Final jet energy resolution and its uncertainty}
\label{sec:JERcomb}
 
The measurement of the jet energy resolution (JER) in data is a multi-step process.
As detailed in Sections~\ref{sec:dijet}--\ref{sec:vjets}, the analyses employed to measure the JER are essentially the same as for the jet calibration, but the observable of interest is not the mean of the response observable but is its width.
For the central rapidity region, the JER is measured with good precision using $\gamma$+jet and $Z$+jet events.
In the forward pseudorapidity region and for high \pT{}, dijet events provide the most precise determination of the JER.
For very low \pT jets there is a significant contribution to the jet energy resolution from \pileup{} particles and electronic noise.
Using the data taken in 2012, new methods have been developed to measure the \pileup{} component.
 
The jet energy resolution is parameterized as a function of three terms~\cite{PERF-2011-04},
\begin{equation}
\frac{\sigma_{p_\text{T}}}{\pT} = \frac{N}{\pT} \oplus \frac{S}{\sqrt{\pT}} \oplus C \label{eq:JER} \ ,
\end{equation}
where $N$ parameterizes the effect of noise (electronic and \pileup{}), $S$ parameterizes the
stochastic effect arising from the sampling nature of the calorimeters, and $C$ is a $\pT$-independent constant term.
It is the determination of these terms in data that is the subject of this section.
 
In Section~\ref{subsec:MCJER}, the MC simulated jet energy resolution is discussed, followed by the determination of the noise term in data in Section~\ref{subsec:JER:Noise}. The combination of the measurements of the noise term and the $Z$+jet, $\gamma$+jet, and dijet measurements, described in Sections~\ref{sec:vjets} and \ref{sec:dijet}, respectively, is detailed in Section~\ref{subsec:jercombination}. The uncertainty in the measurement of the jet energy resolution arising from the various \insitu{} methods is propagated through the fit to the $\pt$ dependence of the jet energy resolution.
 
\subsection{JER in simulation}
\label{subsec:MCJER}
 
The jet energy resolution is measured in simulated event samples as described in Section~\ref{sec:jetMatch}, \ie{}
it is defined as the width parameter $\sigma$ of a Gaussian fit to the jet energy response distribution restricted to the range $\pm 1.5\,\sigma$ around the mean. Figure~\ref{fig:mcjer} shows the resolution determined using \pythia{} dijet MC samples both with full \geant{} detector simulation and with fast simulation. The two simulations generally agree very well, although there are some discrepancies in the very forward regions. The distribution is shown both with and without the \GS{} correction, which significantly improves the resolution (decreasing the resolution of $R=0.4$ EM+JES jets from 10\% to 7\% at 100 \GeV), particularly for jets built from EM-scale clusters.  The resolution is shown as functions of $\pT^\textrm{truth}$ and $|\eta_\textrm{det}|$.  As expected, the resolution improves quickly with increasing $\pT^\textrm{truth}$.  The resolution for a fixed value of \pt{} gets better towards more forward regions (this is not the case for constant jet energy).
 
\begin{figure}[!ht]
\centering
\begin{subfigure}{0.49\textwidth}\centering
\includegraphics[width=\textwidth]{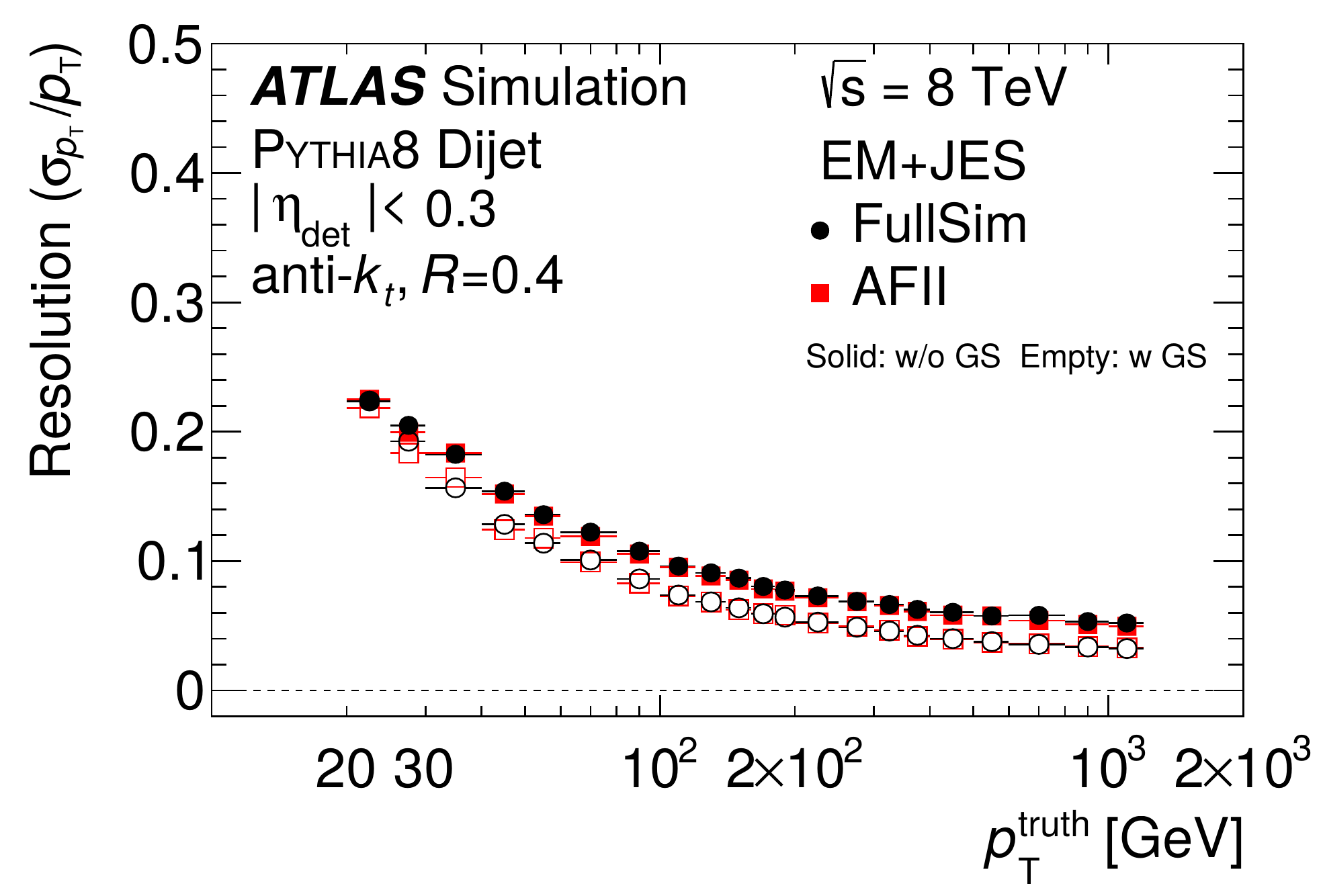}
\caption{}
\end{subfigure}
\begin{subfigure}{0.49\textwidth}\centering
\includegraphics[width=\textwidth]{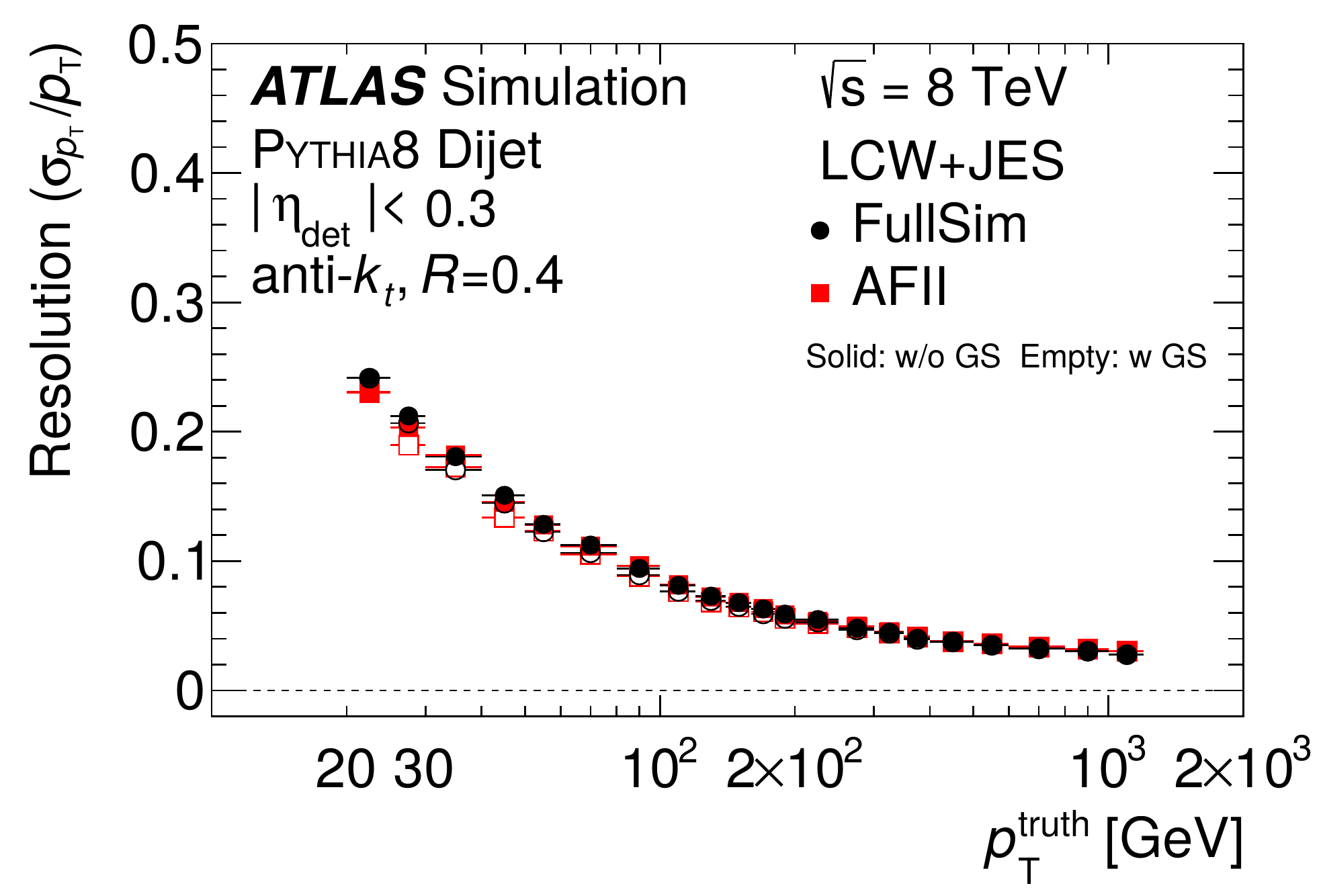}
\caption{}
\end{subfigure} \\
 
\bigskip
 
\begin{subfigure}{0.49\textwidth}\centering
\includegraphics[width=\textwidth]{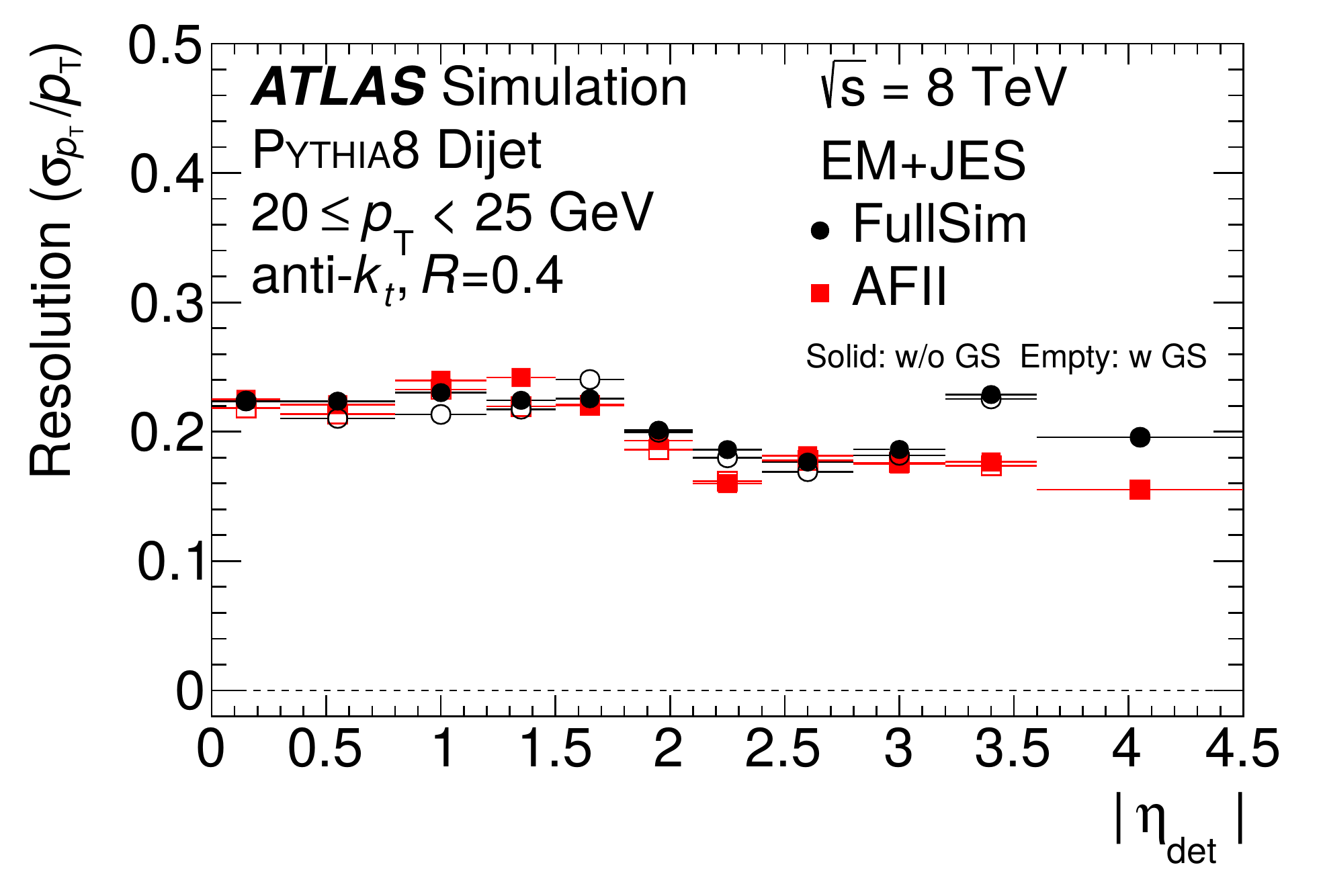}
\caption{}
\end{subfigure}
\begin{subfigure}{0.49\textwidth}\centering
\includegraphics[width=\textwidth]{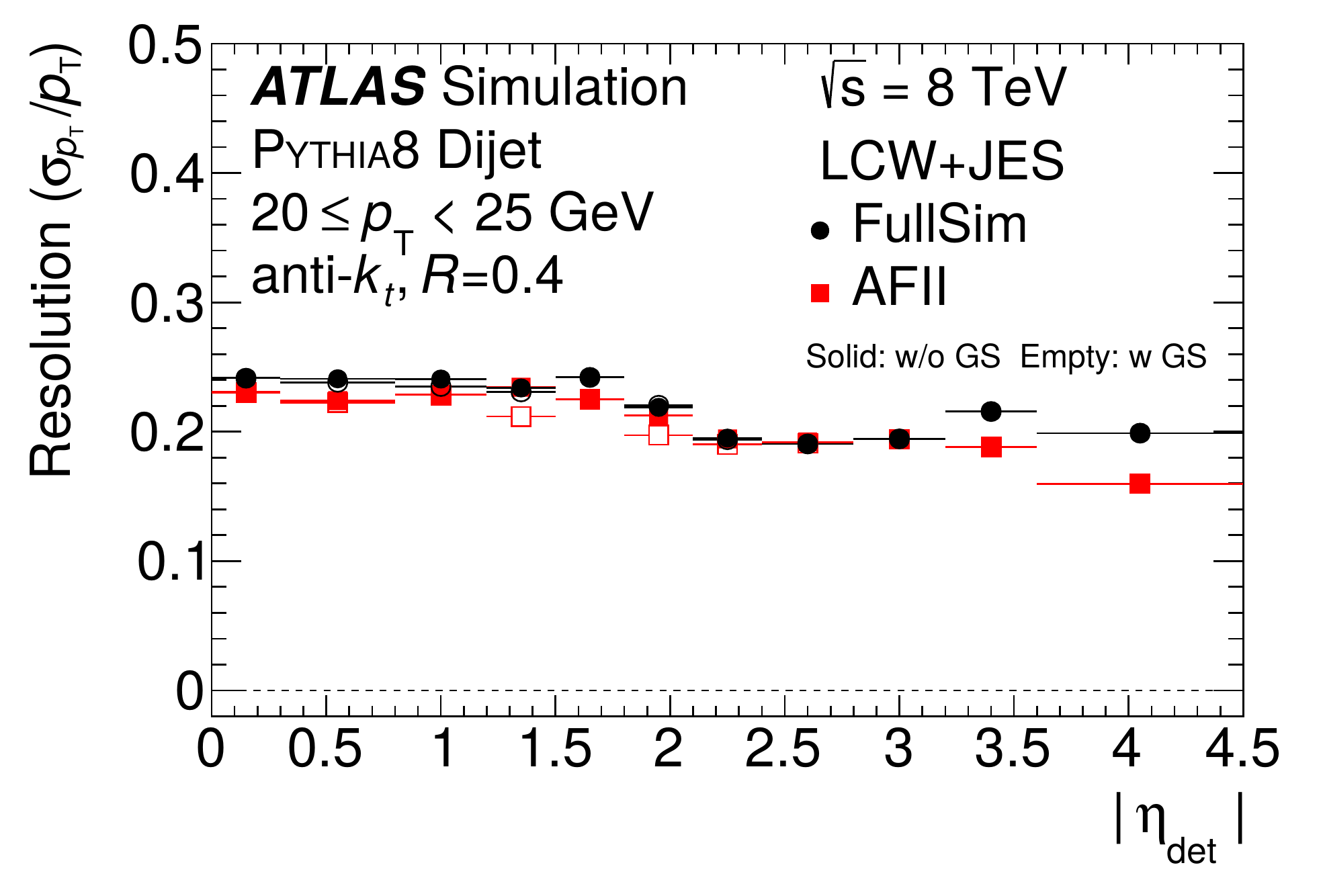}
\caption{}
\end{subfigure}
\caption{Jet energy resolution measured in dijet MC samples as a function of $\pT^\textrm{truth}$ for (a) EM+JES and (b) LCW+JES jets~(filled markers). The resolution in both events simulated with the full \geant{} toolkit~(circles) and with fast simulation~(squares) are shown. Additionally the improvement from the global sequential correction is shown~(empty markers). Figures (c) and (d) show the dependence of the resolution on $|\eta_\textrm{det}|\,$ for low-\pT{} (20--25~\GeV) jets and the level of agreement between full simulation (circles) and fast simulation (squares).}
\label{fig:mcjer}
\end{figure}
 
\subsection{Determination of the noise term in data}
\label{subsec:JER:Noise}
 
Noise, both from the calorimeter electronics and from \pileup{}, forms a significant component of the JER at low \pT{}.
The noise term is not evaluated for $R=1.0$ trimmed jets, as they are only used for $\pT{}>200~\GeV{}$ at which point the noise term is negligible.
It is quite challenging to measure the JER at low \pt{} with
\insitu{} techniques (Section~\ref{subsec:jercombination}) as uncertainties increase at low \pt{} and the stochastic and noise terms are correlated at intermediate \pt{}. Two alternative methods have hence been developed to target the noise term. These attempt to
extract the noise at the constituent scale (the scale of the input \topos{}) as explained in Sections~\ref{subsubsec:ZeroBias} and~\ref{subsubsec:rho}. They are translated into the effect on the jet resolution at the calibrated scale in Section~\ref{subsubsec:noise}. Good agreement is found between the methods, and a closure test is performed using MC simulations in Section~\ref{subsubsec:noiseClose}, leading to a final value for the noise term in the jet energy resolution.
 
The JER noise term receives contributions from the cells inside the \topos{} created by the actual \tjet{} as well as from
\pileup{}. The noise term is significantly affected by the \topo{} formation threshold as jets will contain a varying fraction of particles that have enough energy to form a \topo{}.
The noise term in data without \pileup{} is denoted $N^{\mu=0}$. As just mentioned, this term will be affected by a contribution corresponding to the number of constituent particles produced without enough energy to produce \topos{} or that have been swept out of the cone by the magnetic field, and also by the electronic noise from the cells inside the \topos{}. \Pileup{} particles can result in increased noise of \topos{} seeded by the \tjet{} particles, and also create new \topos{} that are included in the jet.
The latter effects is assumed to dominate, and its contribution to the JER noise term is denoted $N^\mathrm{PU}$.
A third source of noise are \topos{} created solely from electronic noise in the entire calorimeter.
This is assumed to be a negligible effect as the \topos{} require a calorimeter cell with $4\,\sigma$ energy over noise, which is also confirmed in data from events without collisions.
The following subsections present two different measurements of $N^\mathrm{PU}$.
 
\subsubsection{\Pileup{} noise measured using random cones in zero-bias data}\label{subsubsec:ZeroBias}
 
In the random cone method, a cone of given size is formed at a random values of $\eta$ and $\phi$ in zero-bias data, and the energies of all clusters (at either EM or LCW scale) that fall within this cone are combined.
The data was collected using a zero-bias trigger that records events occurring one LHC revolution after an event is accepted by a L1 electron/photon trigger. 
The total \pt{} of a random cone is hence expected to only capture contributions from \pileup{} interactions.
Since jets formed with the \antikt{} algorithm tend to be circular (Figure~\ref{fig:jetArea}(a)), fluctuations of the \pt{} in a random cone can be considered
a measure of the expected \pileup{} fluctuations that are captured by an \antikt{} jet with a radius parameter equal to the cone size.
 
The $\eta$ of the cone is randomly sampled within the range for which the noise is being probed, and the random cone method proceeds by forming a second cone at $\phi+\pi$ (``back-to-back'' in azimuth to the first cone) but at a new random $\eta$, also restricted to the $\eta$ range probed.
The effect of the noise in these cones is expected to be the same on average\footnote{The noise is $\eta{}$ dependent, but since both $\eta{}$ values are sampled randomly within the probe region, the noise will be the same on average.}, and the difference in the random cone $\pT{}$, $\Delta \pT{}$, is plotted.
The difference between two cones is used to remove any absolute offset present as the jet calibration would remove any absolute bias affecting the jets.
The noise is studied as a function of $\eta$ by restricting the $|\eta|$ values that can be chosen for the random cones as previously mentioned.
Since the \topos{} that enter the random cone have no origin correction applied (Section~\ref{sec:originCorr}), the $\eta$ of the random cone corresponds to $\etaDet$ of a jet.
An example of the distribution of this noise in data is shown in Figure~\ref{JER:Noise:coneBal}.
Due to the random nature of the \pileup{} energy deposits with significant energy over noise, the $\Delta \pT{}$ distribution is not expected to be Gaussian.
The 68\% confidence interval of this distribution is defined as the width.
Since $\Delta \pT{}$ gives the fluctuations of two cones, this value is divided by $\sqrt{2}\,$ to give an estimate of the noise term due to \pileup{} $N^\text{PU}$ at the constituent scale for a given jet.
 
The growth of this noise term at the constituent scale as a function of the average number of interactions per bunch crossing is shown in Figure~\ref{JER:Noise:coneR} separately for $|\eta|<0.8$ and $3.2<|\eta|<4.5$. From these results, it is clear that the MC simulations overestimate the influence of \pileup{} events, and this effect is increased in the forward region.
Also, the noise term at constituent scale is larger for LCW than EM \topos{},
because the LCW weighting acts to increase the energy scale of the \topos{}, which also increases the constituent-level noise term.
The EM- and LCW-scale noise terms can only be fairly compared after applying the jet calibration factor, which is done later in Section~\ref{subsubsec:noise}.
Figure~\ref{JER:Noise:coneEta} shows the average \pileup{} noise fluctuations expected in different jets in 2012 for the different $|\eta|\,$ regions.  The data--MC agreement deteriorates in the more forward regions of the detector.  This is likely to arise from poor modelling of the \pileup{} being exacerbated in this region due to the change in detector granularity and noise thresholds.
 
To extract the \pileup{} noise term for average 2012 conditions, the noise term in random cones is extracted from the total 2012 zero-bias dataset. To ensure that the $\avgmu$ distribution used in other \insitu{} measurements (dijet, \Zjet, and \gammajet) is identical to that in the zero-bias dataset, a reweighting is applied dependent on the $\avgmu$ distribution.  This reweighting has a very small effect as the zero-bias trigger and prescales are designed to produce a dataset which mimics the $\avgmu$ distribution of the full dataset used for physics.  In addition, to enable a direct comparison between data and MC simulations, the simulated $\avgmu$ distribution is reweighted to that of the data.
 
\begin{figure}[!ht]
\centering
\begin{subfigure}{0.48\textwidth}\centering
\includegraphics[width=\textwidth]{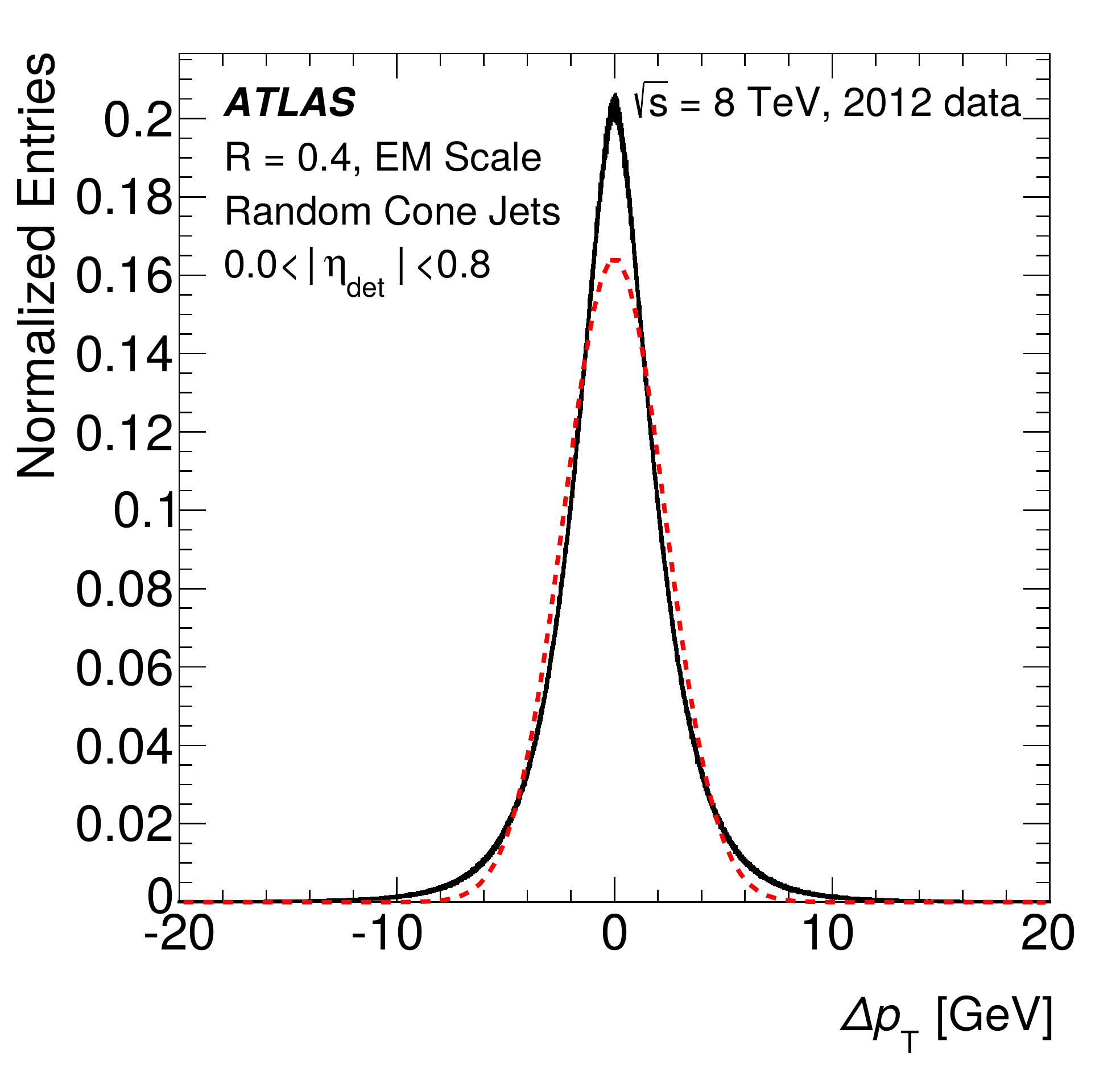}
\caption{EM}
\end{subfigure}
\begin{subfigure}{0.48\textwidth}\centering
\includegraphics[width=\textwidth]{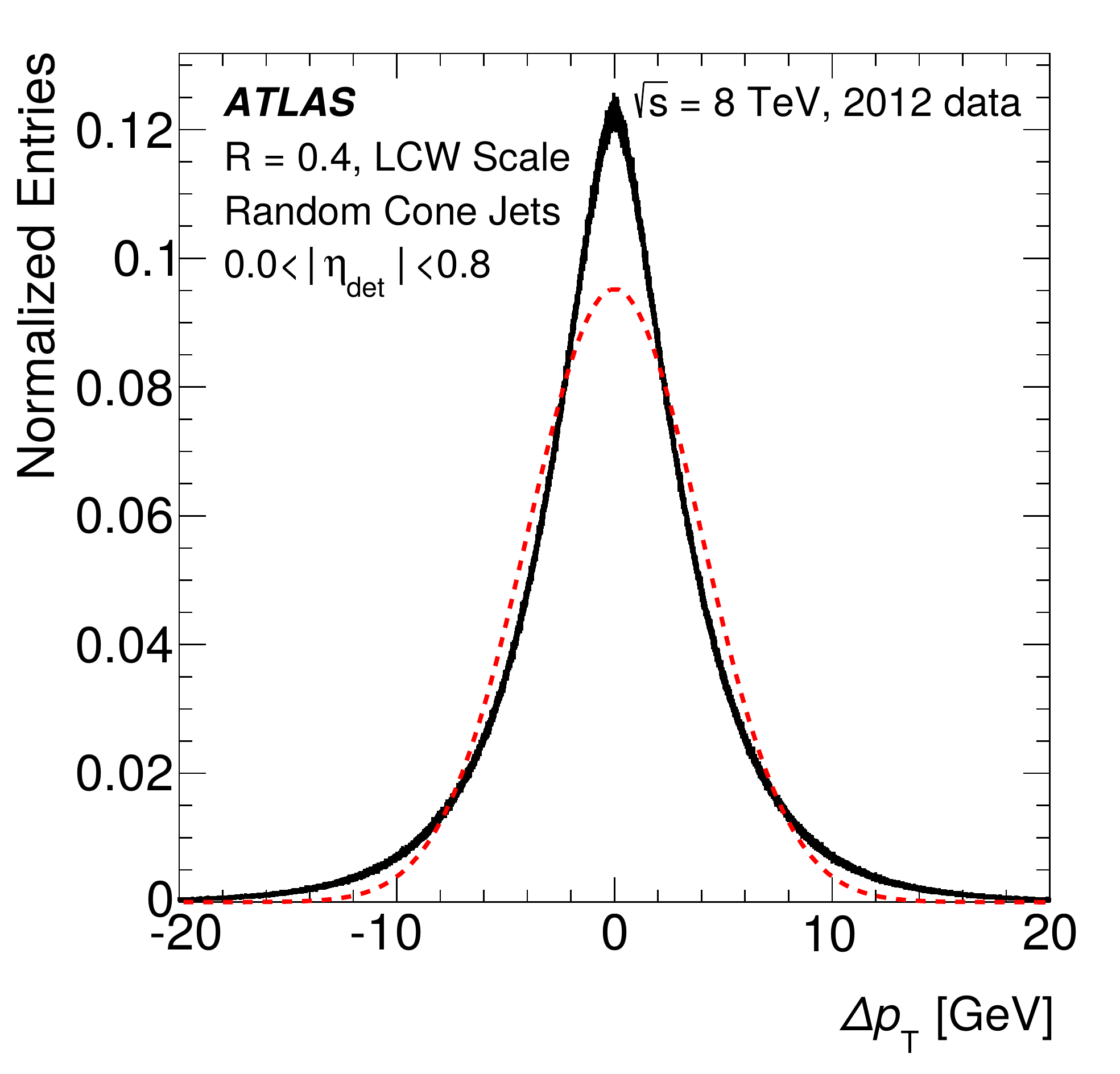}
\caption{LCW}
\end{subfigure}
\caption{The balance of random cones of size 0.4 ($\Delta\pT{}=\pT^1-\pT^2$) in the central region $|\eta|<0.8$ in 2012 zero-bias data using EM and LCW clusters.  The non-Gaussian shape of this distribution is demonstrated by the inclusion of a Gaussian fit~(dashed lines) to the data~(solid lines).}
\label{JER:Noise:coneBal}
\end{figure}
 
\begin{figure}[t]
\centering
\begin{subfigure}{0.48\textwidth}\centering
\includegraphics[width=\textwidth]{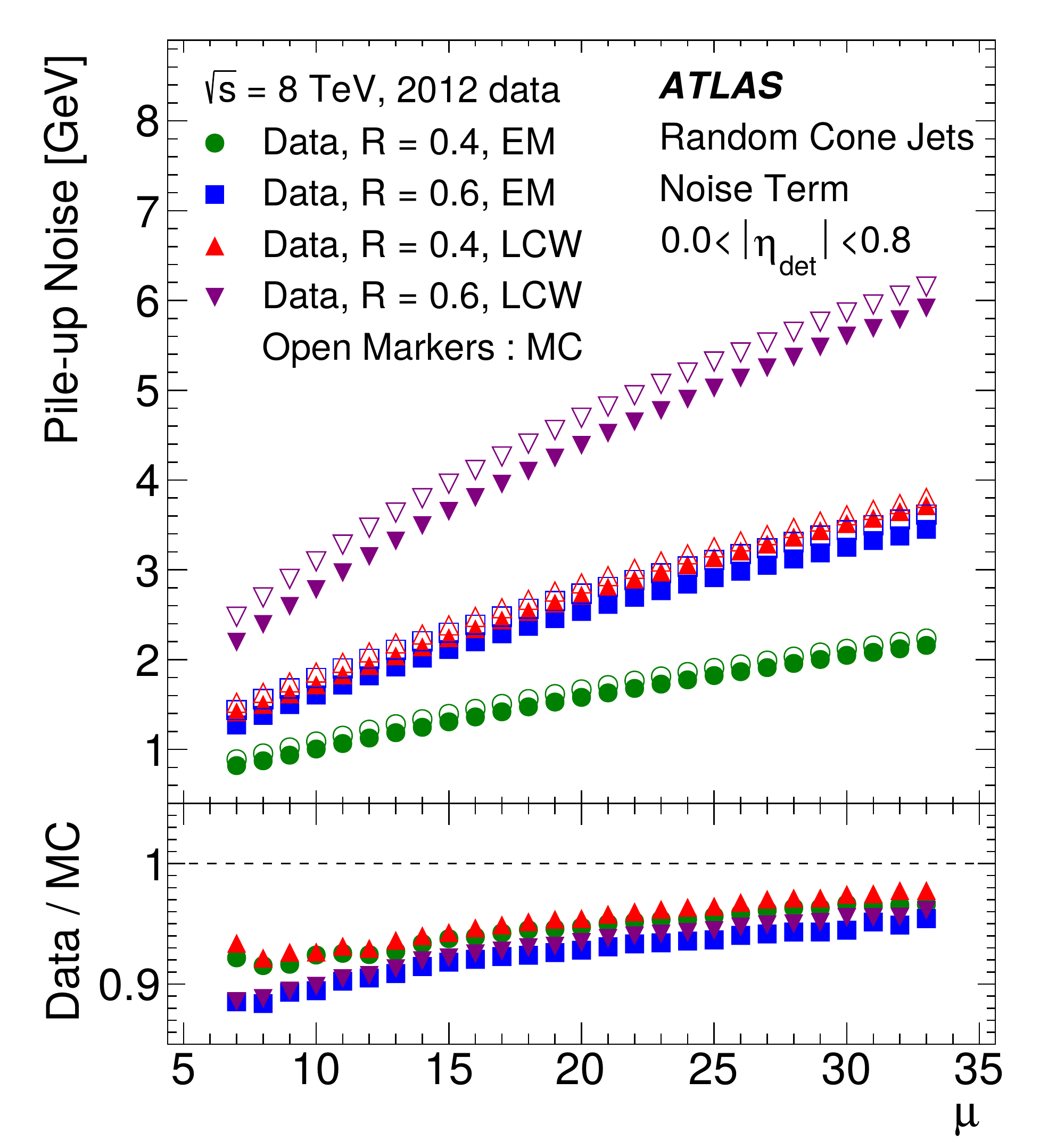}
\caption{$0.0<|\eta|<0.8$}
\end{subfigure}
\begin{subfigure}{0.48\textwidth}\centering
\includegraphics[width=\textwidth]{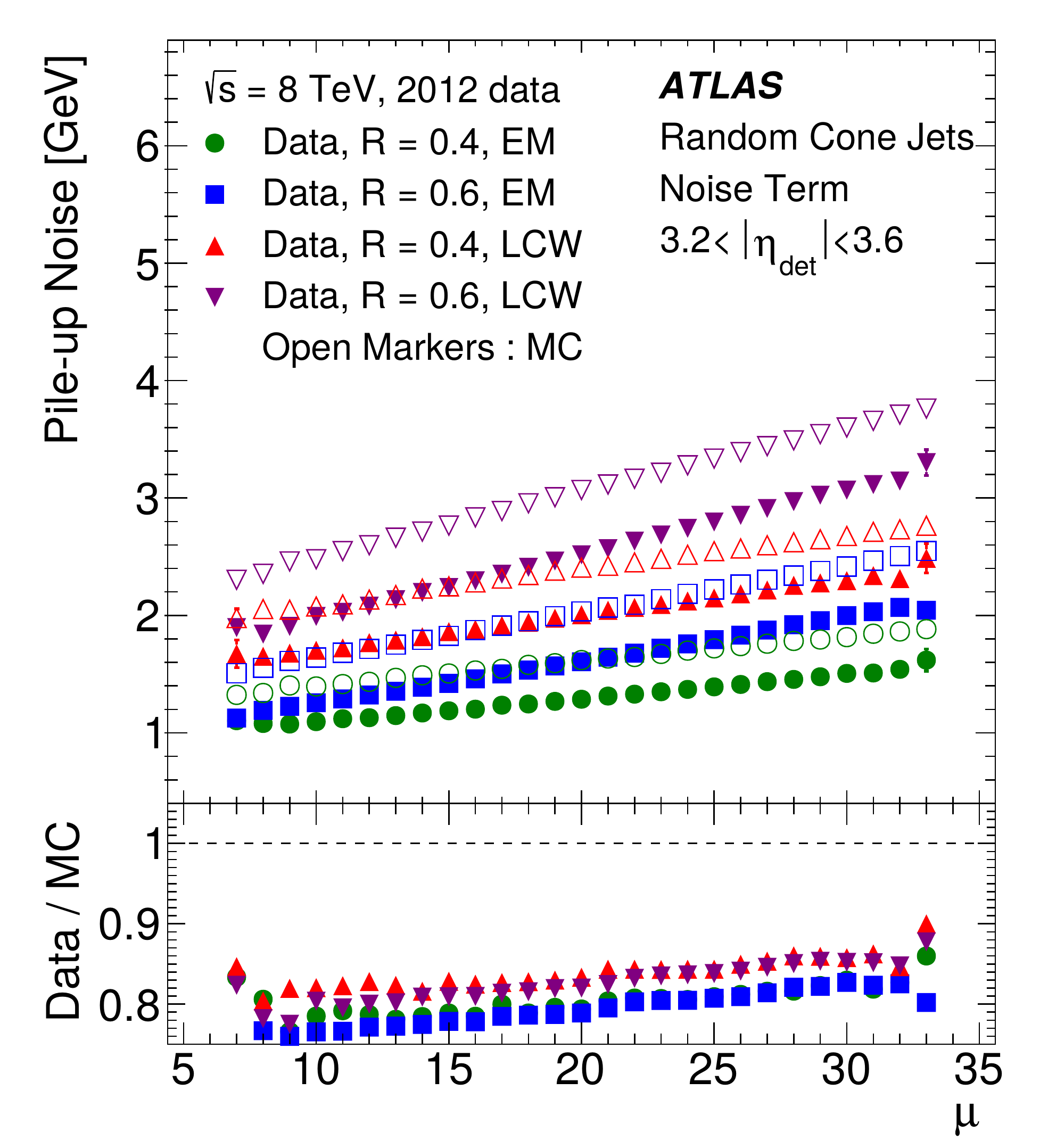}
\caption{$3.2<|\eta|<3.6$}
\end{subfigure}
\caption{The magnitude of the expected fluctuations in different jet radii at the constituent scale derived using the random cone procedure as a function of $\avgmu$, for data (filled markers) and MC simulations (open markers).  The results are shown for $R=0.4$ and $R=0.6$ cones and at both the EM and LCW scales probing two calorimeter $|\eta|$ regions, one central ($|\eta|<0.8$) and one forward ($3.2<|\eta|<3.6$).  A scale factor of 1.09 has been applied to \avgMu{} in the MC simulations to correct for extra activity observed in the minimum-bias tune.}
\label{JER:Noise:coneR}
\end{figure}

\begin{figure}[!ht]
\centering
\includegraphics[width=0.7\textwidth]{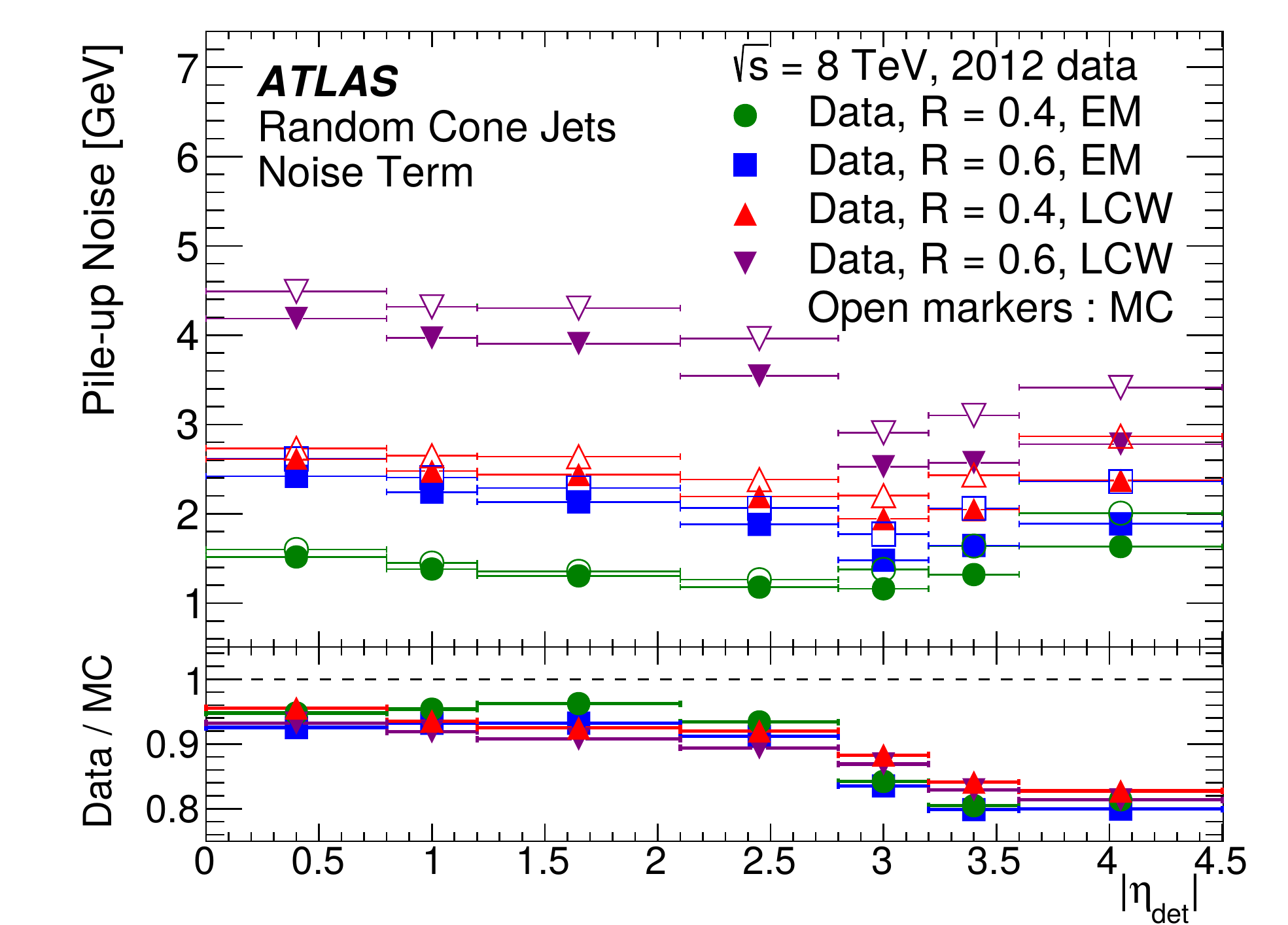}
\caption{The magnitude of the expected fluctuations within different jet radii at the constituent scale as a function of $|\eta|$ for data (filled markers) and MC simulations (open markers).  The results are shown for $R=0.4$ and $R=0.6$ cones and at both the EM and LCW scales.}
\label{JER:Noise:coneEta}
\end{figure}
 
\subsubsection{\Pileup{} noise term measurements using the soft jet momentum method}\label{subsubsec:rho}
 
As explained in Section~\ref{sec:pileupCorr}, the event \pt{}-density $\rho$ is obtained by reconstructing jets using the $k_t$ algorithm without applying any jet \pT{} threshold and defining $\rho$ to be the median of the jet \pt{}-density $\pt/A$, where $A$ is the area of the jet.
Starting from this quantity, the noise term of the JER due to \pileup{} $N^\text{PU}$ is extracted by defining a new observable $\sigma_\rho$ that is a measure of the fluctuations in \pt{} per unit area assuming a stochastic model of noise.
Due to using the median (rather than the mean) in its definition, $\rho$ is to first order insensitive to the hard process.
Any type of data can in principle be used for the measurement.
The results presented in this section are based on $Z\to\mu\mu$ data.
The following steps are performed:
\begin{itemize}
\item
Jets are reconstructed using the Cambridge--Aachen algorithm~\cite{Salam:2009jx} with $R=0.6$ and required to have $|\eta|<2.1$. No $\pt$ threshold is applied, and the jet \pt{} extends down to zero.
\item
For each jet, the quantity $r=(\pT-\rho\, A)/\sqrt{A}$ is calculated, where $A$ is the jet area defined using the Voronoi procedure~\cite{JetArea}.
Since no jet \pt{} threshold is applied, many jets will be built from noise only.
The distribution of $r$ is expected to be centred at zero since after subtracting $\rho\, A$ there should be as many jets above the \pT\ density as below.
\item
The observable $\sigma_\rho$ is defined event-by-event from the width of the $r$ distribution of all jets in the event.
To avoid complications of non-Gaussianity and the hard-scatter event biasing the upper side tail, $\sigma_\rho$ is defined by half the difference between the 84\% and 16\% quantile points.
\end{itemize}
 
The size of the expected fluctuations at the constituent scale of a given jet is given by $\sigma_\rho\,\sqrt{A}$.
The distributions of $\sigma_\rho\,$ for EM-scale and LCW-scale clusters in $Z\rightarrow\mu\mu$ 
data and \pythia{} samples are shown in Figure~\ref{JER:SigmaRho}.
$Z\rightarrow\mu\mu$ events are used to select an unbiased set of events for data-to-MC comparison, thus avoiding the use of any jet-based trigger which would bias the jet distributions.
As in the random cone method (Figure~\ref{JER:Noise:coneR}), the \pileup{} noise is overestimated in the MC simulations.
An estimate of the noise term due to \pileup{} is obtained by scaling the mean value of the $\sigma_\rho$ distribution by $\sqrt{\pi R^2}$.
 
\begin{figure}[!ht]
\centering
\includegraphics[width=0.49\textwidth]{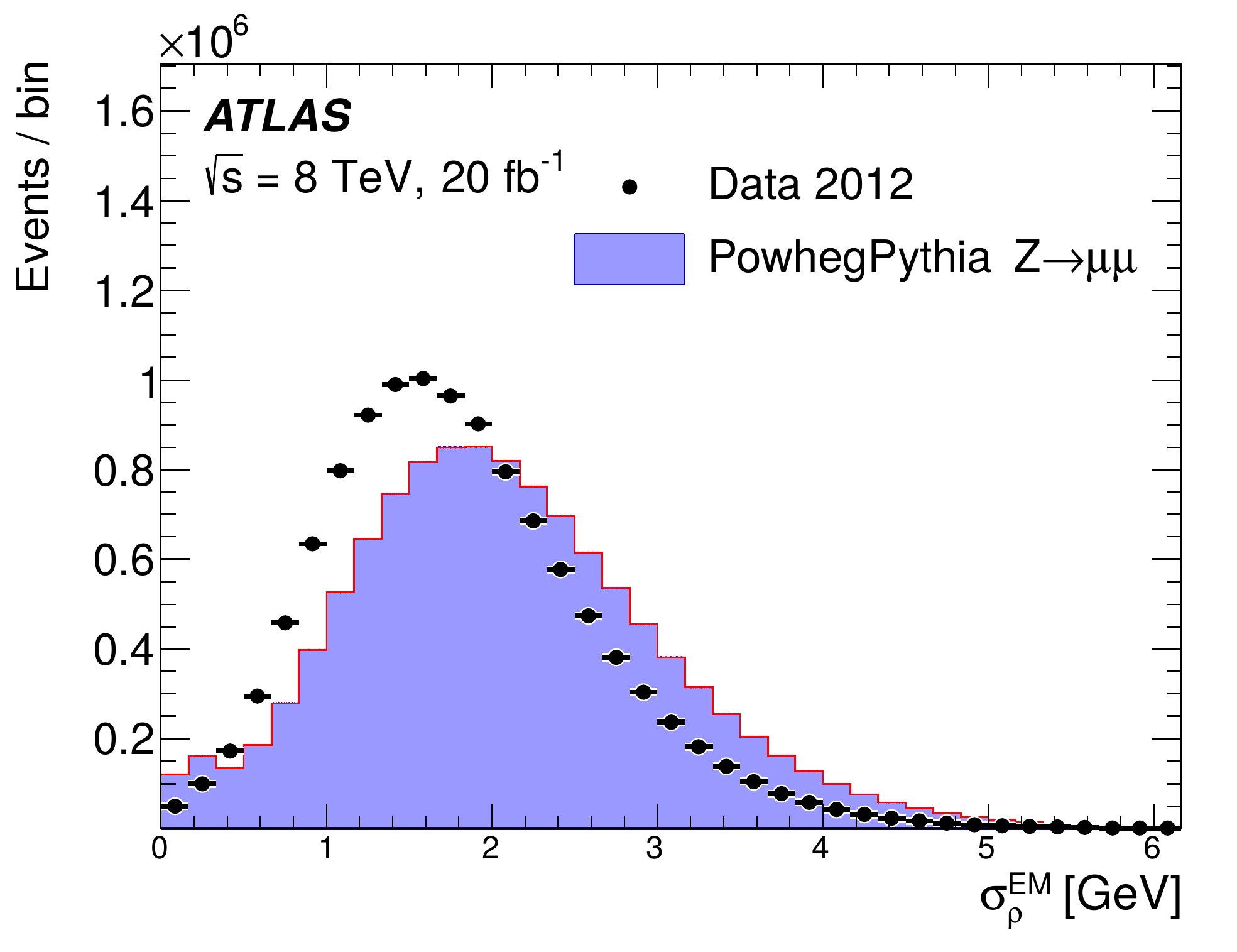}
\includegraphics[width=0.49\textwidth]{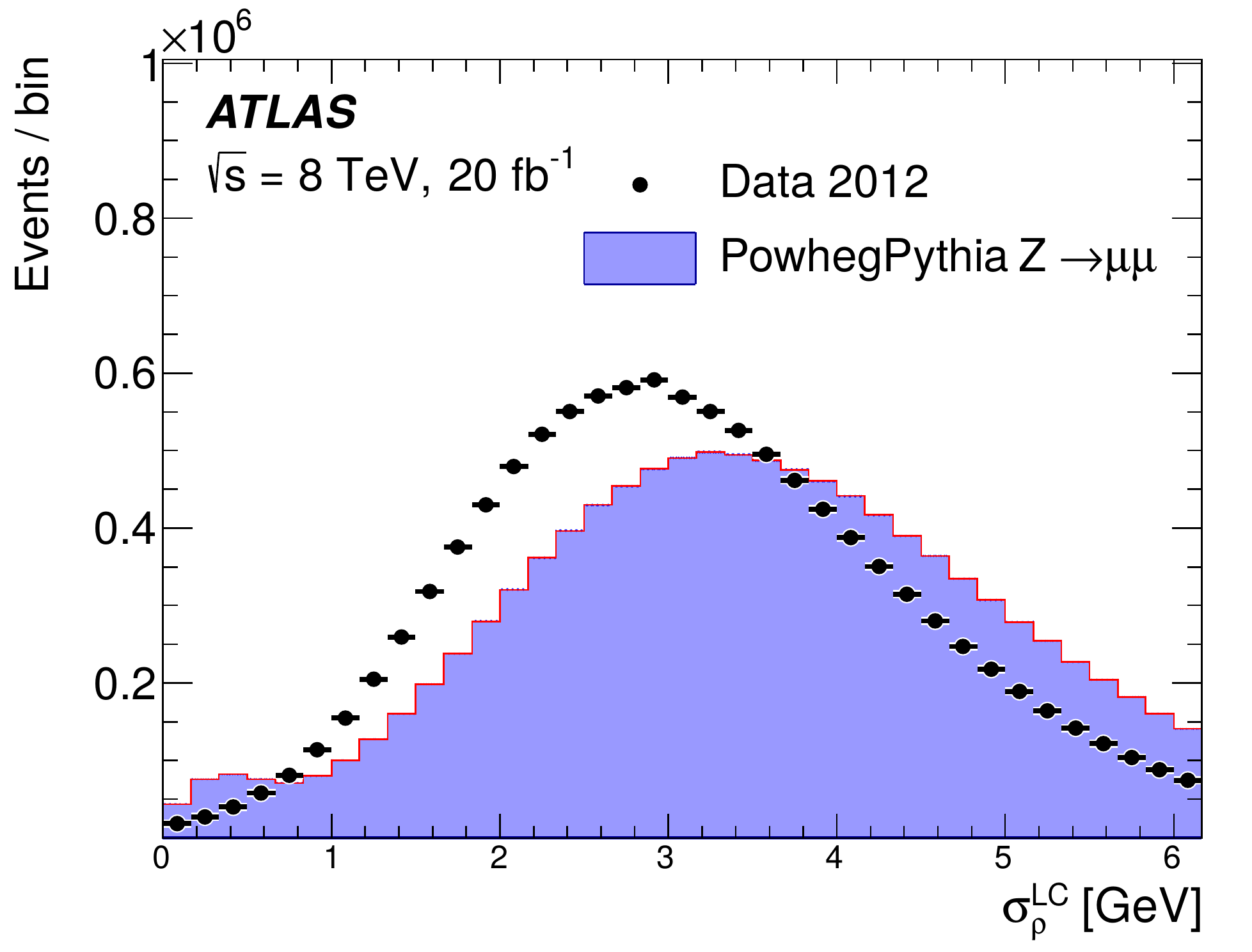}
\caption{Extracted values of $\sigma_\rho$ in data (points) and MC simulations (histogram) for a sample of $Z\rightarrow\mu\mu\,$ events at the EM scale (left) and the LCW scale (right).
This observable quantifies the fluctuations of the \pt{} density $\rho$, i.e.\ the \pt{} per area in $(y,\phi)$-space.}
\label{JER:SigmaRho}
\end{figure}
 
\begin{table}
\caption{Measurements of $\langle\sigma_\rho\rangle$ and $\langle\sigma_\rho\rangle\,\sqrt{A}\,$, where $\langle\sigma_\rho\rangle$ is the mean of the $\sigma_\rho$ distribution, and the random cone results, both using data and MC simulations.
The area is defined by $A=\pi R^2$, where $R$ is the radius parameter.
The $\sigma_\rho\, \sqrt{A}$ results, which is a noise term measurement from the soft jet momentum method, is extracted using the region $|\eta|<2.1$ while the noise term measurement using the random cone method is extracted for jet $|\eta|<0.8$.
Statistical uncertainties of both measurements are negligible.}
\label{tab:sigmaRhoResults}
\centering
\begin{tabular}{l|c|c|c|c}
\hline\hline
& EM  & LCW & EM  & LCW \\
& $R=0.4$ & $R=0.4$ & $R=0.6$ & $R=0.6$ \\ \hline
$\langle\sigma_\rho\rangle\,$ ($Z\rightarrow\mu\mu$, data) [\GeV] & 1.81 & 3.25 & 1.81 & 3.25 \\
$\langle\sigma_\rho\rangle\,$ ($Z\rightarrow\mu\mu$, MC) [\GeV] & 2.09 & 3.72 & 2.09 & 3.72 \\ \hline
$\langle\sigma_\rho\rangle\, \sqrt{A}\,$ ($Z\rightarrow\mu\mu$, data) [\GeV] & 1.28 & 2.30 & 1.92 & 3.46 \\
Random cone, data [\GeV] & 1.52 & 2.61 & 2.42 & 4.19 \\
Difference [\%] & 16 & 12 & 21 & 17 \\ \hline
$\langle\sigma_\rho\rangle\, \sqrt{A}\,$ ($Z\rightarrow\mu\mu$, MC) [\GeV] & 1.48 & 2.64 & 2.22 & 3.96 \\
Random cone, MC [\GeV] & 1.60 & 2.73 & 2.61 & 4.49 \\
Difference [\%] & 7.5 & 4.4 & 15 & 12 \\ \hline\hline
\end{tabular}
\end{table}
 
\subsubsection{Comparison of methods and construction of the noise term}
\label{subsubsec:noise}
 
As described in the previous two subsections, the random cone and the soft jet momentum methods can both be used to measure the
noise term of the jet energy resolution.
It is useful to compare their results and to contrast the two methods. As well as using different data samples,
these methods make quite different assumptions about the underlying physics:
\begin{itemize}
\item The soft jet momentum method implicitly assumes the \pileup{} noise is stochastic (such that it grows with $\sqrt{A}$).
\item The random cone method measures the noise in several $\eta$-bins, while the soft jet momentum method does not consider any $\eta$-dependence of the noise within the probed detector region $|\eta|<2.1$.
\item The symmetry assumption of the two cones back-to-back in azimuth in the zero bias events is not required by the soft jet momentum method.
\end{itemize}
Further, while the soft jet momentum method gives an estimate of the noise term in each event (as is done for the calculation of $\rho$), the random cone method gives the noise term over an event sample. Table~\ref{tab:sigmaRhoResults} compares the measured noise term at the constituent scale using the two methods.
The two sets of measurements agree at the level of 20\%.
 
\subsubsection{Closure test of the \pileup{} noise measurement in MC simulation}\label{subsubsec:noiseClose}
 
A closure test is performed on the \pileup{} noise measurements by comparing the random cone result with the \pileup{} noise extracted using \tjets{} in MC simulation. The \pileup{} noise in MC simulations is extracted by measuring the MC JER (Section~\ref{sec:jetMatch}) in two \pythia{} dijet samples: one without \pileup{} and one sample with 2012 \pileup{} conditions. By subtracting the JER measured in the sample without \pileup{} from the JER measured in the sample with \pileup{}, the contribution from the \pileup{} noise is isolated and can be compared with the measurement of the noise term using the random cone method. However, this comparison cannot be done directly since the random cone measures the noise at constituent scale (EM or LCW), while the JER is measured at the fully calibrated scale (EM+JES or LCW+JES). To account for this mismatch in scale, the random cone measurements are scaled by the average MC calibration factor $\langle c_\text{JES}\rangle$ evaluated for the jets in the kinematic region of interest.
The results of these tests are shown in Figure~\ref{JER:Noise:closurePt} as a function of \pT for both EM and LCW jets.
The relevant comparison is that of the estimated noise term $N^\text{PU}$ and the quadrature difference of the MC JER measurements with and without \pileup{}.
In the central region $|\eta|<0.8$, good closure is observed, both for EM+JES and LCW+JES.
In this region, the calorimeters have high granularity, and as a consequence energy clusters from \pileup{} and from the hard-scatter signal tend to form separately with little overlap.
Slightly larger non-closure is observed towards the more forward regions, which is expected due to the coarser angular granularity and higher noise thresholds, which result in a larger overlap between energy deposits from \pileup{} and the hard scatter.
 
The same closure test was performed for the $N^\text{PU}$ measured with the soft jet momentum method, and the difference between the results is taken as a systematic uncertainty due to the arbitrariness of the selection of method.
Additionally, the degree of non-closure of the method is taken as a systematic uncertainty.
 
\begin{figure}[!ht]
\centering
\includegraphics[width=0.45\textwidth]{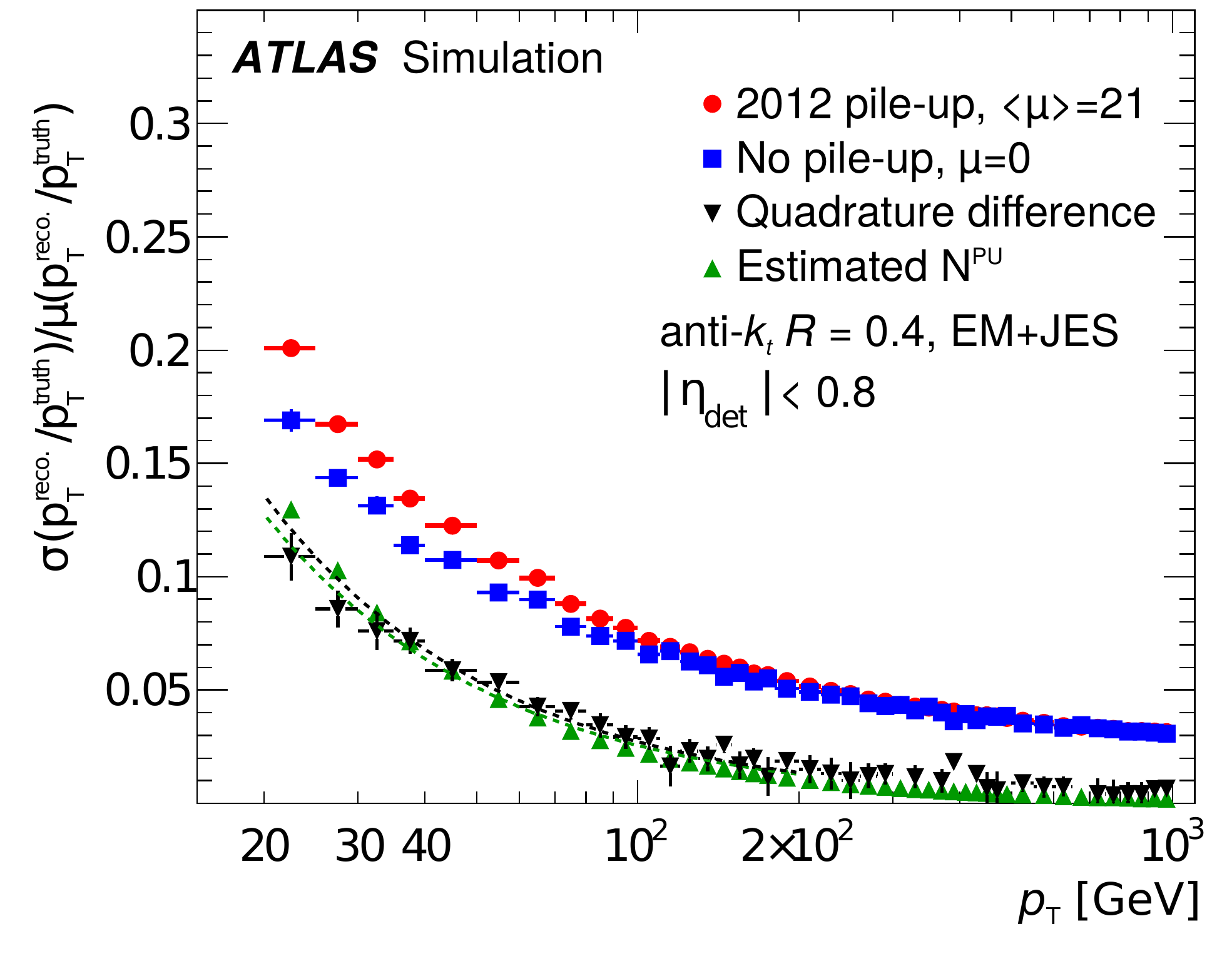}
\includegraphics[width=0.45\textwidth]{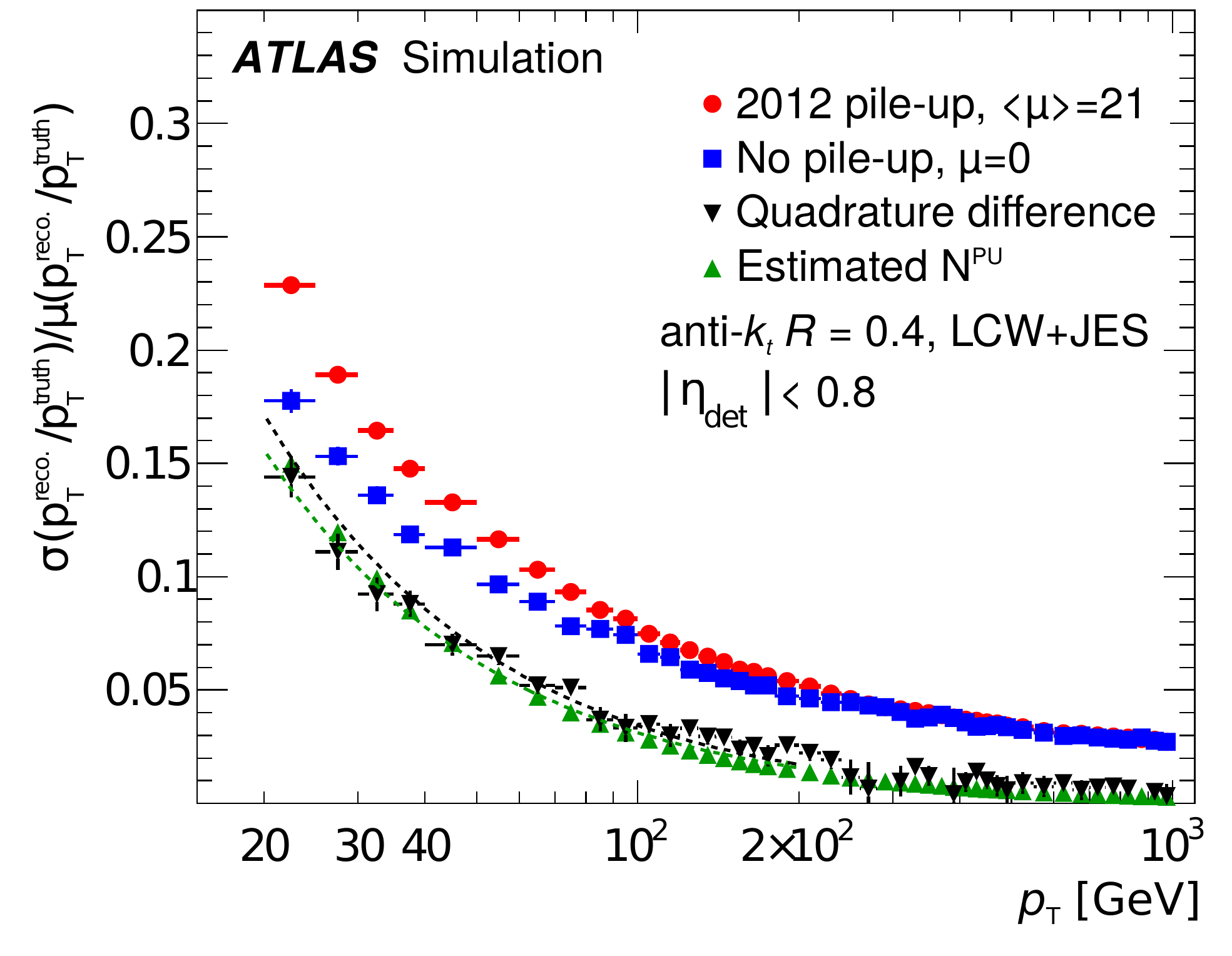}\\
\includegraphics[width=0.45\textwidth]{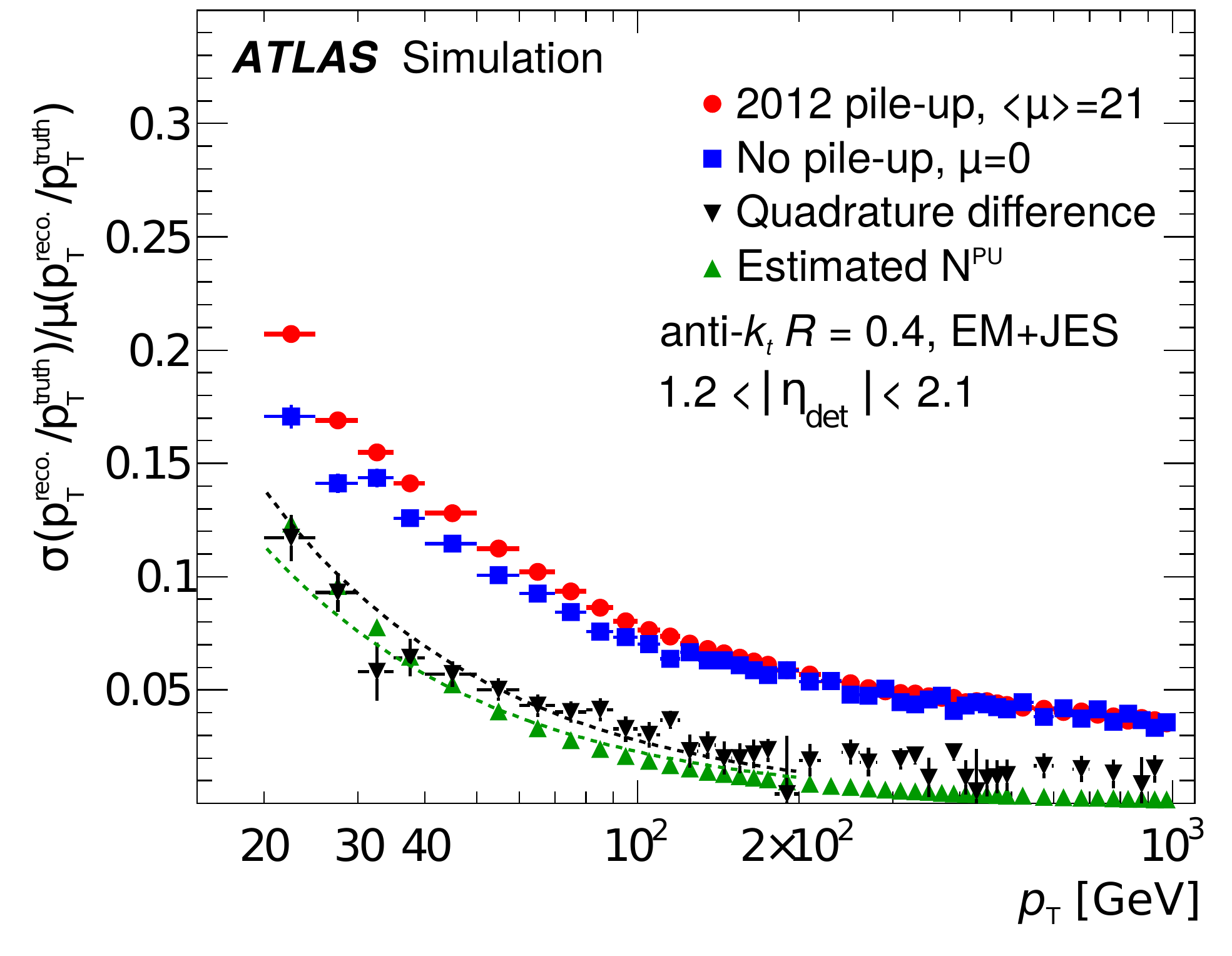}
\includegraphics[width=0.45\textwidth]{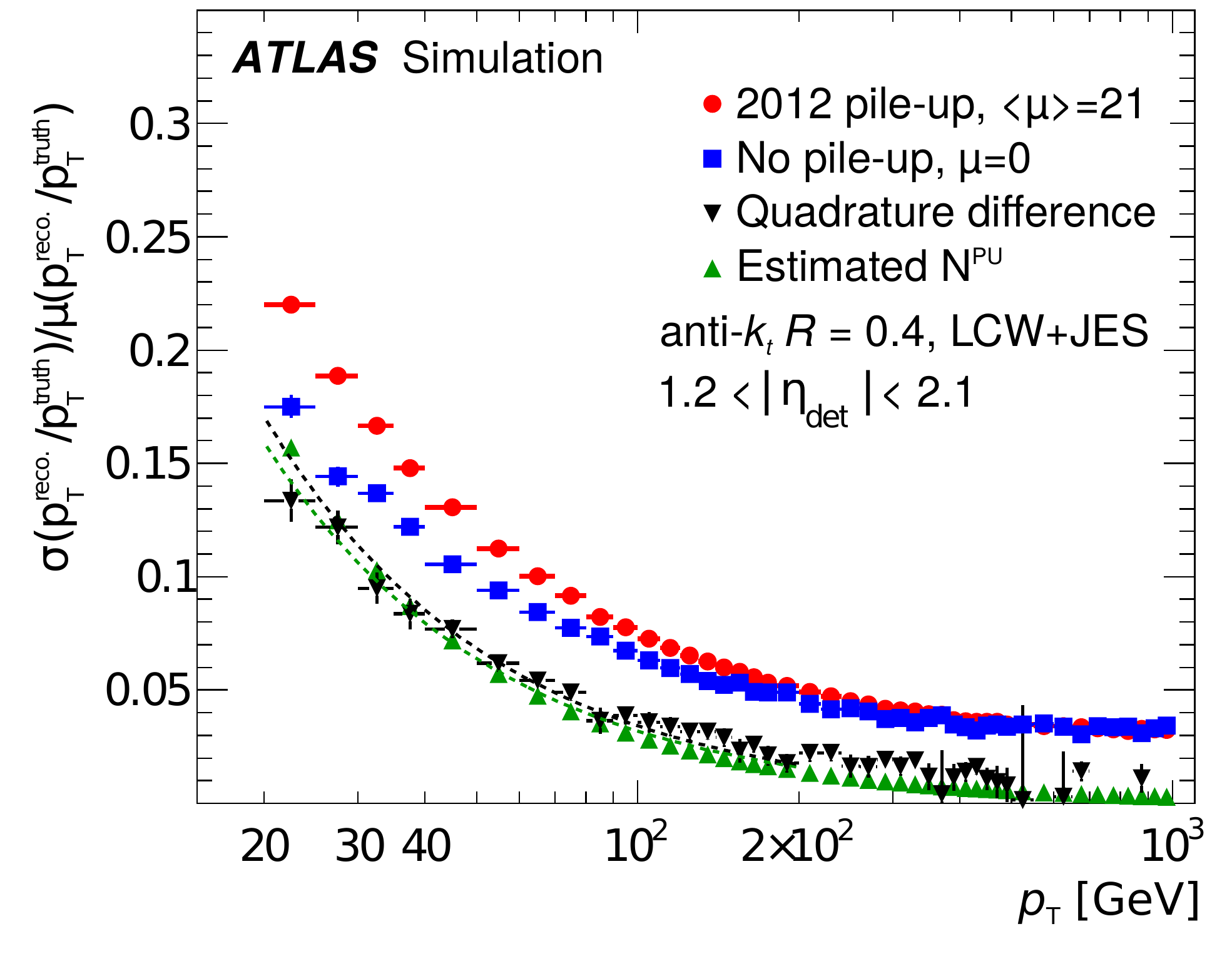}\\
\caption{
Comparison between the \pileup{} noise term $N^\text{PU}$ extracted using the random cone method~(upward triangles) with the expectation from MC simulation~(downward triangles).
Results are shown for jets built from EM~(left) and LC~(right) \topos{}, for jets with $|\etaDet{}|<0.8$~(top) and $1.2<|\etaDet{}|<2.1$~(bottom). The expected $N^\text{PU}$ is obtained by quadrature subtraction of the JER obtained from MC simulation of events with nominal \pileup{} (circles) from that of events with no \pileup{} (squares).
Fits performed to the measured and expected \pileup{} noise data are
displayed as dotted curves.
Quadrature differences corresponding to points where, due to statistical fluctuations, the resolution is worse in the no \pileup{} scenario
are not displayed.}
\label{JER:Noise:closurePt}
\end{figure}

\subsubsection{Noise term in the no \pileup{} scenario}
 
The random cone and soft jet momentum methods provide measurements of the part of the noise term arising from \pileup{} activity $N^{\textrm{PU}}$.
In the dijet MC sample without \pileup{}, for which $\mu = 0$, the noise term does not have any \pileup{} contribution but does include other effects such as electronic noise on the signal clusters and threshold effects.  To get a handle on the additional noise terms not included in the random cone or soft jet methods, the $\mu=0$ MC simulated resolution is fitted with the standard $N$, $S$ and $C$ parameterization of Eq.~(\ref{eq:JER}) to extract the no \pileup{} noise term $N^{\mu=0}$. The result of such fits are presented in Table~\ref{tab:NoPUnoise}.
 
\begin{table}
\centering
\caption{The noise term $N^{\mu=0}$ in \GeV\ extracted in a dijet MC sample without \pileup{}.
The values and uncertainties are extracted from a fit.
For data, an additional 20\% uncertainty is assigned, based on the 2010 measurements~\cite{PERF-2011-04}.}
\label{tab:NoPUnoise}
\begin{tabular}{c|c|c|c|c}
\hline\hline
& EM+JES R$=0.4$ & LCW+JES R$=0.4$ & EM+JES R$=0.6$ & LCW+JES R$=0.6$ \\ \hline
$|\eta|<0.8$ & $2.28\pm0.13$ & $2.66\pm0.09$ & $1.83\pm0.12$ & $2.54\pm0.09$ \\
$0.8<|\eta|<1.2$ & $1.95\pm0.25$ & $2.14\pm0.17$ & $1.29\pm0.25$ & $2.34\pm0.15$ \\
$1.2<|\eta|<2.1$ & $2.52\pm0.18$ & $2.99\pm0.09$ & $0.90\pm0.28$ & $2.94\pm0.09$ \\
$2.1<|\eta|<2.8$ & $2.25\pm0.30$ & $2.19\pm0.13$ & $0\pm0.95$ & $2.24\pm0.11$ \\ \hline\hline
\end{tabular}
\end{table}
 
The total jet energy resolution (Eq.~(\ref{eq:JER})) was measured in 2010 and agreed between data and MC simulations within 10\% for jet \pt{} in the range 30~$\GeV{}<\pt<500~\GeV{}$~\cite{PERF-2011-04}.  For $\pt=30$~\GeV\ in the central region, the noise term is responsible for more than half of the total resolution.  Given that the dominant resolution source leads to a total resolution modelled to the level of 10\%, this implies that the noise term itself agrees between data and MC simulation to the level of 20\% in simulated samples without \pileup{}.
This conclusion is also supported by single-particle measurements~\cite{PERF-2011-05}. This extrapolation includes some additional assumptions in the MC modelling of the detector as several settings changed between 2010 and 2012, most notably the \topo{} noise thresholds; however, 20\% is considered a conservative estimate of the uncertainty in this component.
 
The total JER noise term $N$ is defined by combining the noise term extracted in the no \pileup{} sample with that originating from \pileup{} (measured above) using a sum in quadrature, i.e. $N = N^\text{PU}\oplus N^{\mu=0}$.

\subsection{Combined \texorpdfstring{\insitu{}}{in situ} jet energy resolution measurement}
\label{subsec:jercombination}
 
The JER measurements based on the bisector method in dijet events reported in Section~\ref{sec:dijetJER} and the vector boson plus jet balance reported in Section~\ref{sec:vjets-JER} are statistically combined using a chi-squared minimization of the function in Eq.~(\ref{eq:JER}).
In this fit, the noise term is held at the central value found in the previous section, while measurements of the $S\,$ and $C\,$ terms are extracted.
The uncertainties in each term are evaluated in the same way they were in the JES determination in Section~\ref{subsec:insituCombination}, \ie{} by re-evaluating the JER measurement after a 1$\sigma$ shifts of each individual uncertainty source. The degree of agreement between the three \insitu{} measurements is in Figure~\ref{fig:jercombination:chi2}, which shows the $\chi^2$ per degree of freedom as a function of \pT.  The low values of the $\chi^2\,$ per degree of freedom across the \pT\ range demonstrates that the \insitu{} methods agree well.
As expected, there is a large anti-correlation between the $S$ and $C$ parameters of $-0.25$ ($-0.44$) for \EMJES{} (\LCWJES) calibrated jets, and
the $\chi^2\,$ per degree of freedom for the fit to find $N$, $S$ and $C$ is 8/35 (15/35) when correlations are not considered, and
71/35 (58/35) when correlations are considered.
The relatively large size of the $\chi^2$ per degree of freedom when correlations are considered indicates a limitation in the fitting function used.
It is a possible indication of the need for higher-order terms in the series to better describe the resolution dependence on \pT.
A similar effect is seen when looking at the fit to these three parameters in MC simulations.
 
When propagating the uncertainty in the noise term to the fit the resulting changes in the fitted values of $N$, $S\,$ and $C\,$ for anti-$k_t$ $R=0.4\,$ EM+JES (LCW+JES) jets are $\genfrac{}{}{0pt}{}{+0.63}{-0.63}$, $\genfrac{}{}{0pt}{}{-0.038}{+0.030}$, $\genfrac{}{}{0pt}{}{+0.001}{-0.001}$ ($\genfrac{}{}{0pt}{}{+0.74}{-0.74}$, $\genfrac{}{}{0pt}{}{-0.048}{+0.039}$, $\genfrac{}{}{0pt}{}{+0.002}{-0.002}$). Again, correlations between the different components are observed, namely increasing $N\,$ results in a reduced $S\,$ and increased $C$.
 
To reduce the number of parameters which need to be propagated, the full set of eigenvectors is built according to the total effect on the JER measurement of each uncertainty component (rather than the effect of each component on the $N$, $S\,$ and $C$ terms individually).  These uncertainty sources can then be reduced in number by using an eigenvector decomposition (diagonalization) as was done for the JES. This allows the full correlations to be retained and propagated to analyses.  Figure~\ref{fig:jereigenvectors} shows the three eigenvectors after this diagonalization. Combining in quadrature the results from varying $N\,$ and propagating the \insitu{} uncertainties gives $N=3.33\pm0.63~(4.12\pm0.74)$~\GeV, $S=0.71\pm0.07~(0.74\pm0.10)$~$\sqrt{\GeV }$, and $C=0.030\pm0.003~(0.023\pm0.003)$ for anti-$k_t$ $R=0.4$ EM+JES (LCW+JES) jets. Figure~\ref{fig:jercombination:central} shows the individual measurements of the resolution in the central region, the result of the combination, and the associated uncertainty. The uncertainty in the jet energy resolution for anti-$k_t\,$ $R=0.4\,$ jets is less than 0.03 at 20~\GeV\ and below 0.01 above 100~\GeV.
 
\begin{figure}
\centering
\begin{subfigure}{0.45\textwidth}\centering
\includegraphics[width=\textwidth]{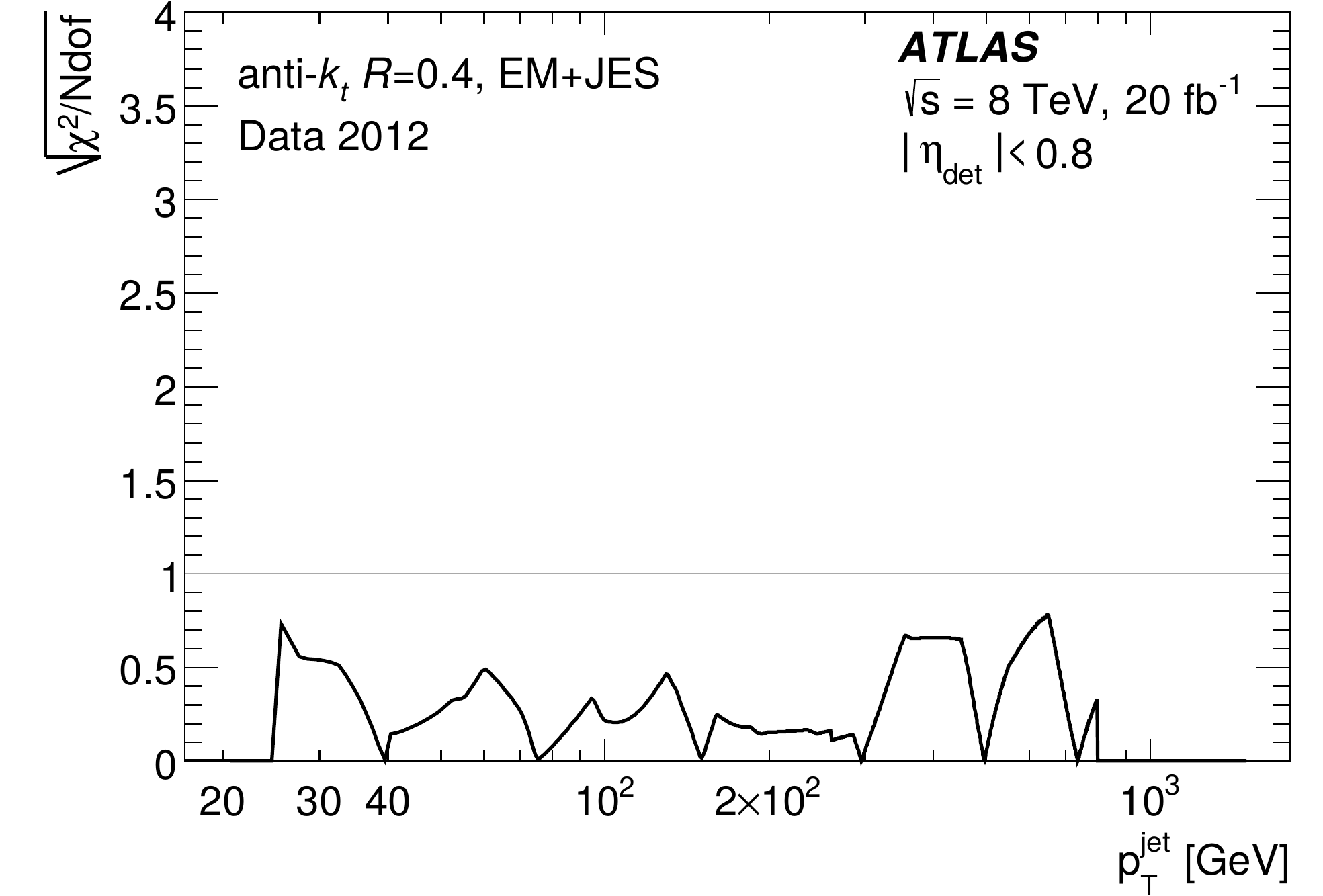}
\caption{\EMJES{}}
\end{subfigure}
\begin{subfigure}{0.45\textwidth}\centering
\includegraphics[width=\textwidth]{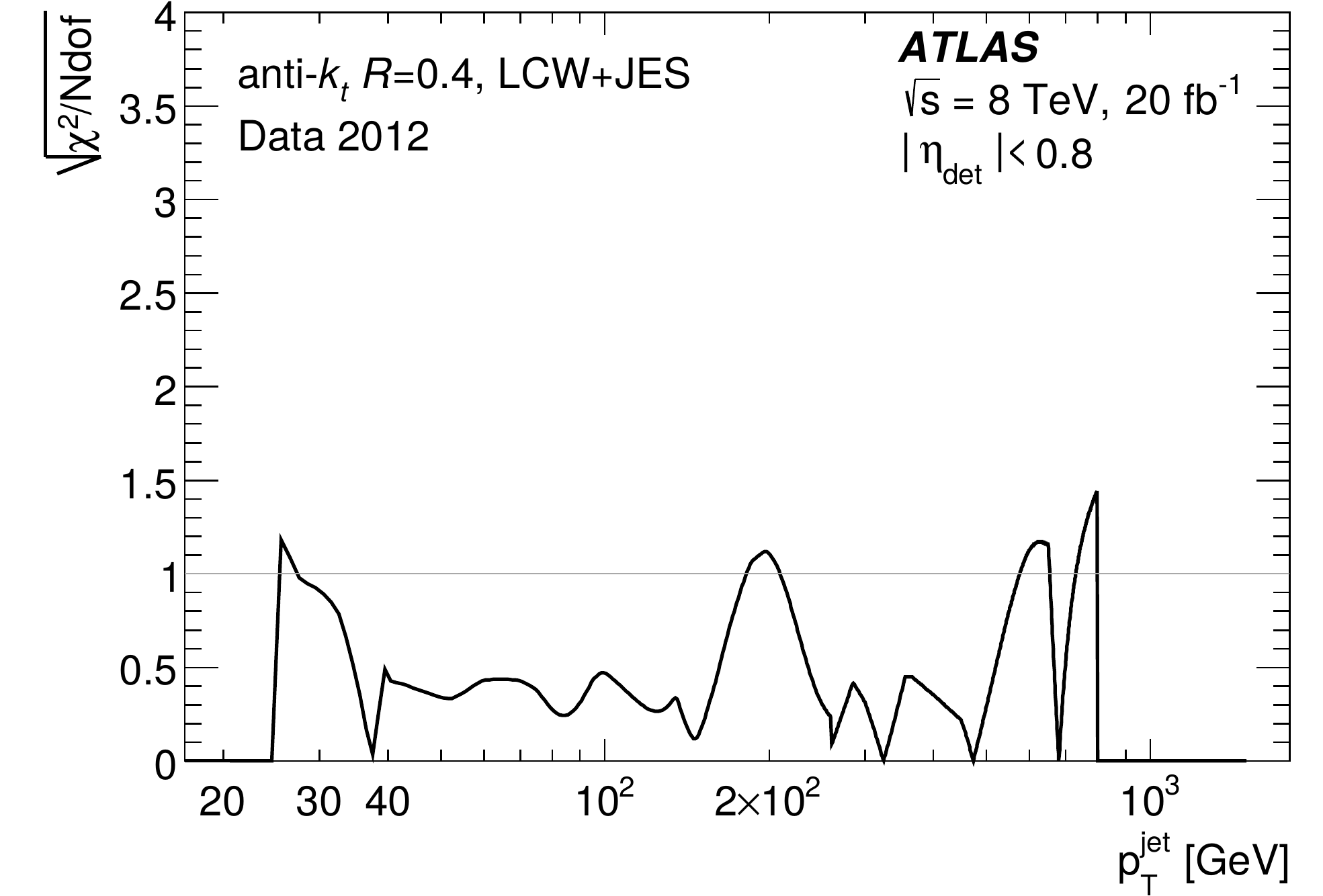}
\caption{\LCWJES{}}
\end{subfigure}
\caption{The $\chi^2\,$ per degree of freedom for the jet energy resolution showing the compatibility of the three \insitu{} measurements of the jet energy resolution (dijet asymmetry, $Z$+jet balance, and $\gamma$+jet balance) for jets calibrated with the (a)~\EMJES{} and (b)~\LCWJES{} schemes.}
\label{fig:jercombination:chi2}
\end{figure}
 
\begin{figure}
\centering
\begin{subfigure}{0.45\textwidth}\centering
\includegraphics[width=\textwidth]{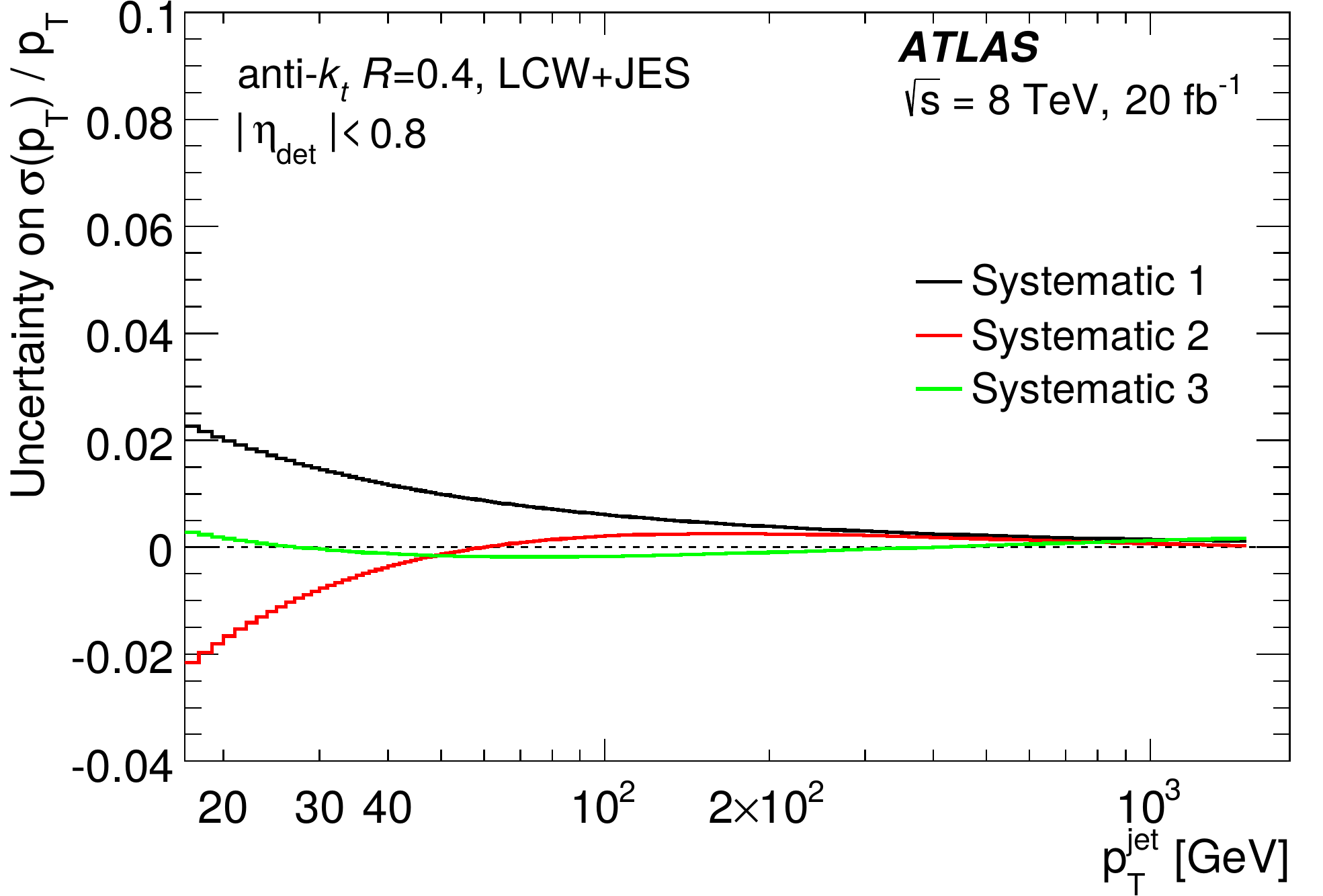}
\caption{$|\eta|<0.8$}
\end{subfigure}
\begin{subfigure}{0.45\textwidth}\centering
\includegraphics[width=\textwidth]{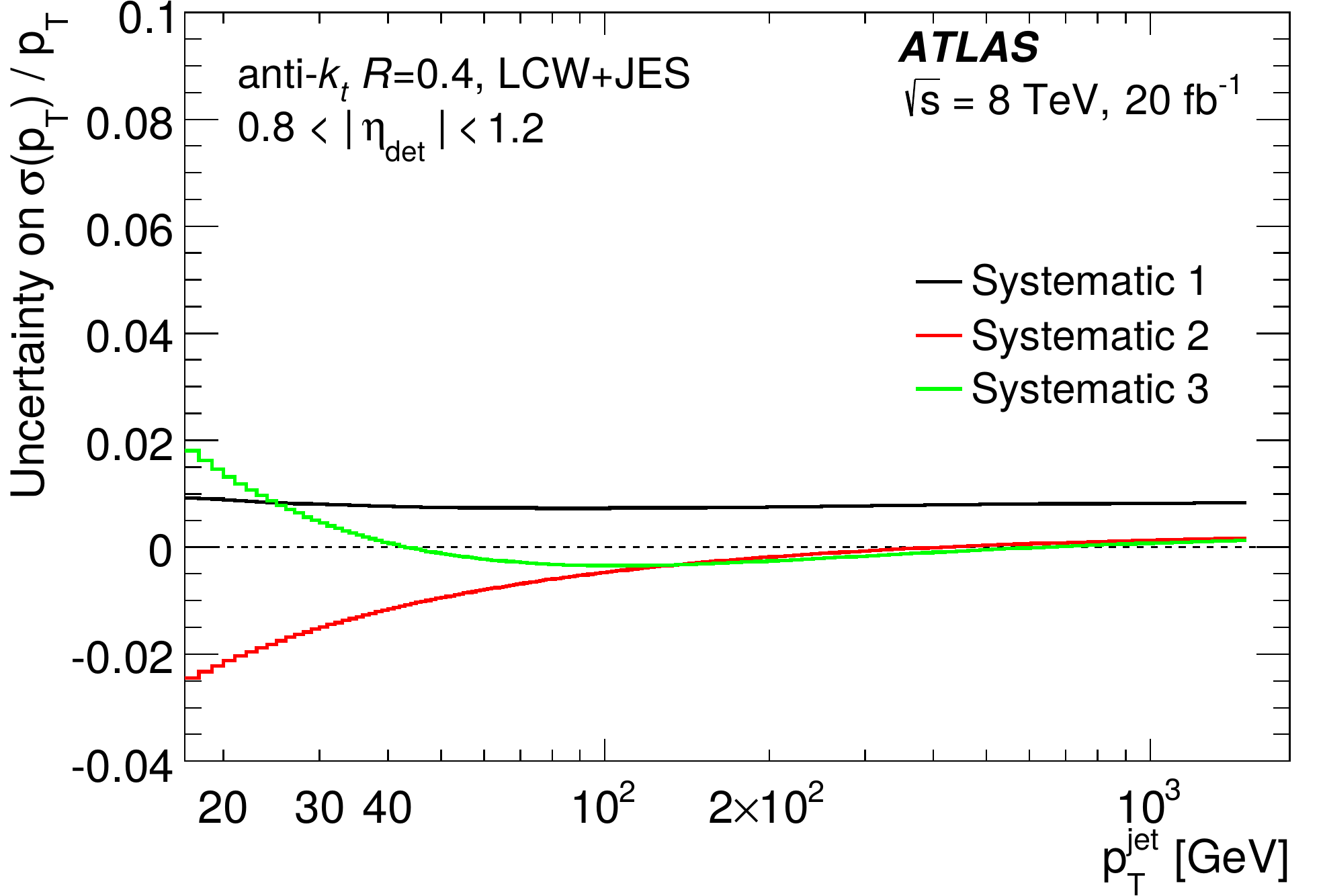}
\caption{$|\eta|<0.8$}
\end{subfigure} \\
 
\bigskip
 
\begin{subfigure}{0.45\textwidth}\centering
\includegraphics[width=\textwidth]{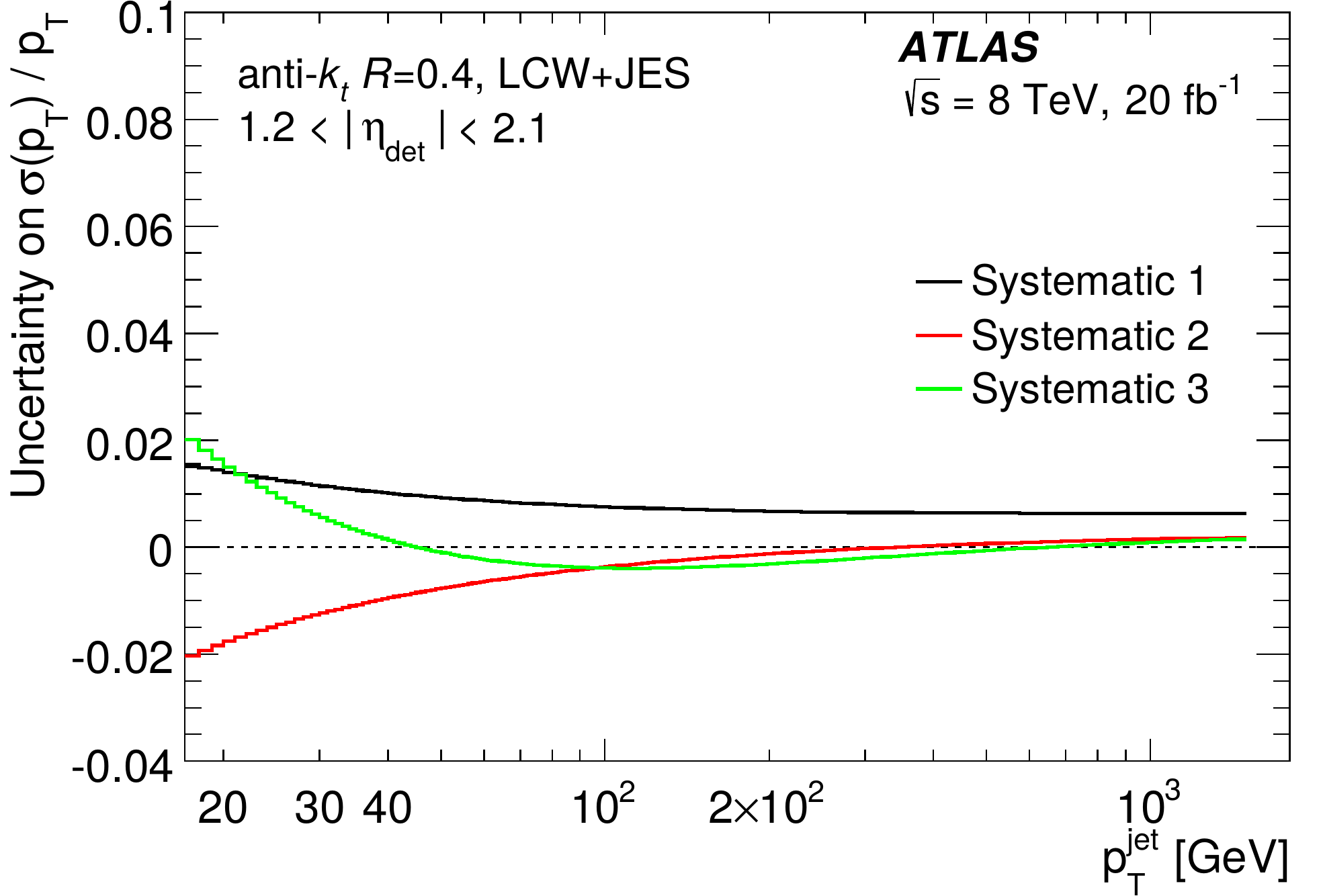}
\caption{$1.2<|\eta|<2.1$}
\end{subfigure}
\begin{subfigure}{0.45\textwidth}\centering
\includegraphics[width=\textwidth]{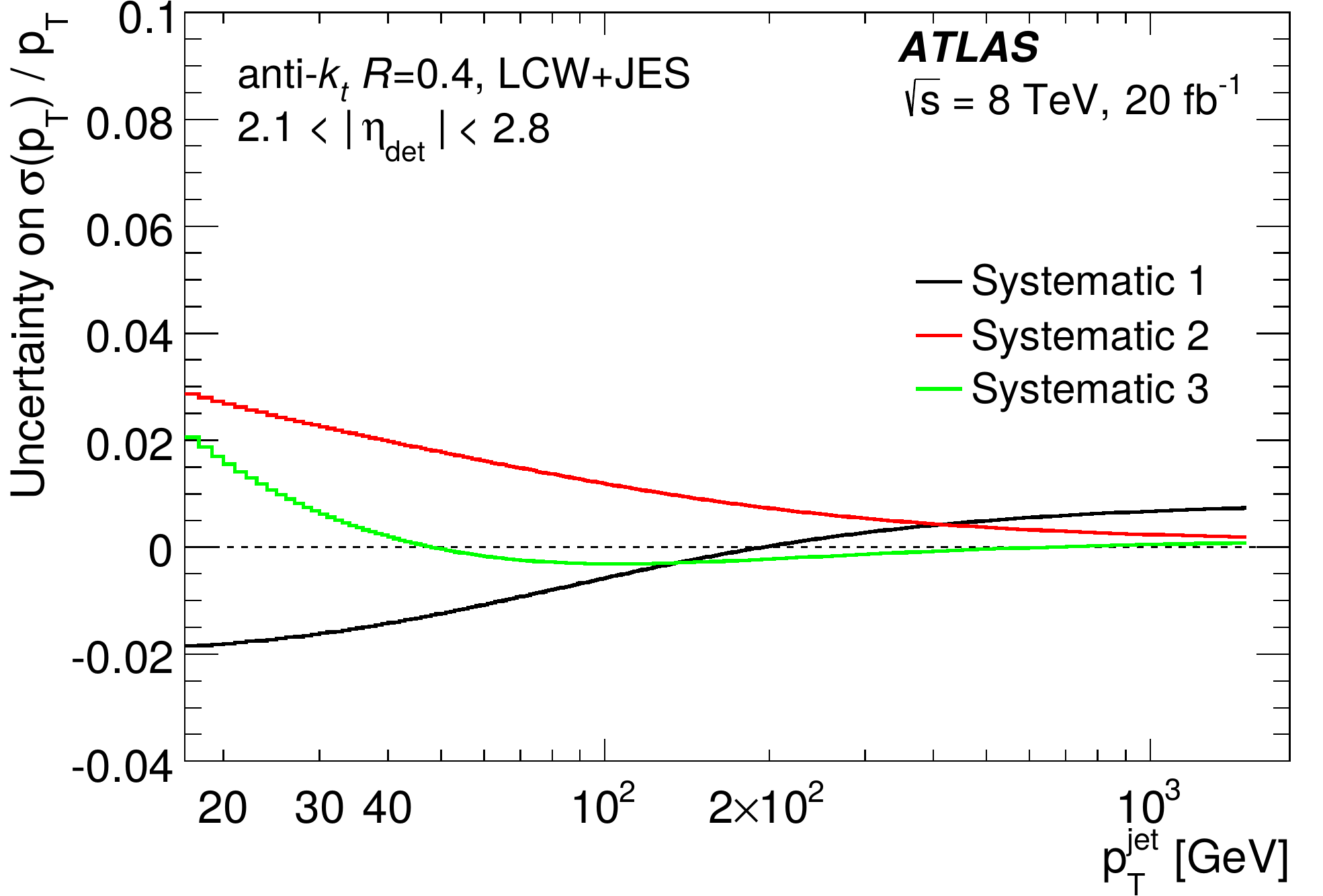}
\caption{$2.1<|\eta|<2.8$}
\end{subfigure}
\caption{The three eigenvectors after the eigenvector reduction of the nuisance parameters in the jet energy resolution measurement for LCW+JES $R=0.4$ jets for four different $|\eta|$ bins.
These nuisance parameters fully describe the correlations.}
\label{fig:jereigenvectors}
\end{figure}
 
When considering the more forward $|\eta|\,$ bins, the large statistical uncertainty in \Zjet\ and \gammajet\ events means that only dijet measurements are useful. These are combined with the measured noise term in data in the same way as in the central region. For LCW+JES anti-$k_t\,$ $R=0.4\,$ jets all the different regions are shown in Figure~\ref{fig:jercombination:forward} and the extracted $N$, $S$, and $C$ parameters for all jet collections are shown in Tables~\ref{tab:jerResults_EM} and~\ref{tab:jerResults_LC}.
 
Finally, to account for correlations between the measurements at different $|\eta|\,$ a correlation matrix as a function of $\pT$ and $|\eta|\,$ is built.
The systematic uncertainties of the noise term and dijet balance results are assumed to be correlated between $|\eta|\,$ regions.
The eigenvector reduction is performed, which results in, at most, 12 uncertainty components required to capture all the correlations between the \pt and $|\eta|\,$ regions covered by the \insitu{} studies.
 
\begin{table}[p]
\caption{Extracted values of the $N$, $S$, and $C$ terms from a combined fit to the jet energy resolution measurements for $R=0.4$ and $R=0.6$ jets, both calibrated with the EM+JES scheme. 
The quoted uncertainties of the $N$, $S$, and $C$ terms are highly correlated with each other.}
\label{tab:jerResults_EM}
\centering
\begin{tabular}{c|ccc|ccc}
\hline\hline
& \multicolumn{3}{c|}{EM+JES, $R=0.4$} & \multicolumn{3}{c}{EM+JES, $R=0.6$} \\
\hline
$|\eta|$ range& $N$ [\GeV] & $S$ [\GeV$^{0.5}$] & $C$ & $N$  [\GeV] & $S$ [\GeV$^{0.5}$] & $C$ \\
\hline
$(0,0.8)$ & $3.33\pm0.63$ & $0.71\pm0.07$ & $0.030\pm0.003$ & $4.34\pm0.93$ & $0.67\pm0.08$ & $0.030\pm0.003$ \\
$(0.8,1.2)$ & $3.04\pm0.70$ & $0.69\pm0.13$ & $0.036\pm0.003$ & $4.06\pm0.93$ & $0.76\pm0.10$ & $0.031\pm0.003$ \\
$(1.2,2.1)$ & $3.34\pm0.80$ & $0.61\pm0.16$ & $0.044\pm0.008$ & $3.96\pm0.91$ & $0.56\pm0.14$ & $0.042\pm0.007$ \\
$(2.1,2.8)$ & $2.9 \pm1.0 $ & $0.46\pm0.30$ & $0.053\pm0.011$ & $3.41\pm0.84$ & $0.48\pm0.27$ & $0.049\pm0.012$ \\
\hline\hline
\end{tabular}
\end{table}
 
\begin{table}[p]
\caption{
Extracted values of the $N$, $S$, and $C$ terms from a combined fit to the jet energy resolution measurements for $R=0.4$ and $R=0.6$ jets, both calibrated with the LCW+JES scheme. 
The uncertainties shown are highly correlated between the $N$, $S$, and $C$ terms.}
\label{tab:jerResults_LC}
\centering
\begin{tabular}{c|ccc|ccc}
\hline\hline
& \multicolumn{3}{c|}{LCW+JES, $R=0.4$} & \multicolumn{3}{c}{LCW+JES, $R=0.6$} \\
\hline
$|\eta|$ range& $N$ [\GeV] & $S$ [GeV$^{0.5}$] & $C$ & $N$  [\GeV] & $S$ [\GeV$^{0.5}$] & $C$ \\
\hline
$(0,0.8)$ & $4.12\pm0.74$ & $0.74\pm0.10$ & $0.023\pm0.003$ & $5.50\pm0.99$ & $0.66\pm0.12$ & $0.026\pm0.004$ \\
$(0.8,1.2)$ & $3.66\pm0.75$ & $0.64\pm0.13$ & $0.039\pm0.009$ & $5.40\pm0.98$ & $0.78\pm0.15$ & $0.032\pm0.005$ \\
$(1.2,2.1)$ & $4.27\pm0.75$ & $0.58\pm0.15$ & $0.034\pm0.007$ & $5.7 \pm 1.0$ & $0.62\pm0.16$ & $0.031\pm0.006$ \\
$(2.1,2.8)$ & $3.38\pm0.65$ & $0.26\pm0.36$ & $0.050\pm0.010$ & $5.2 \pm 1.0$ & $0.51\pm0.38$ & $0.028\pm0.019$ \\
\hline\hline
\end{tabular}
\end{table}
 
\begin{figure}
\centering
\includegraphics[width=0.45\textwidth]{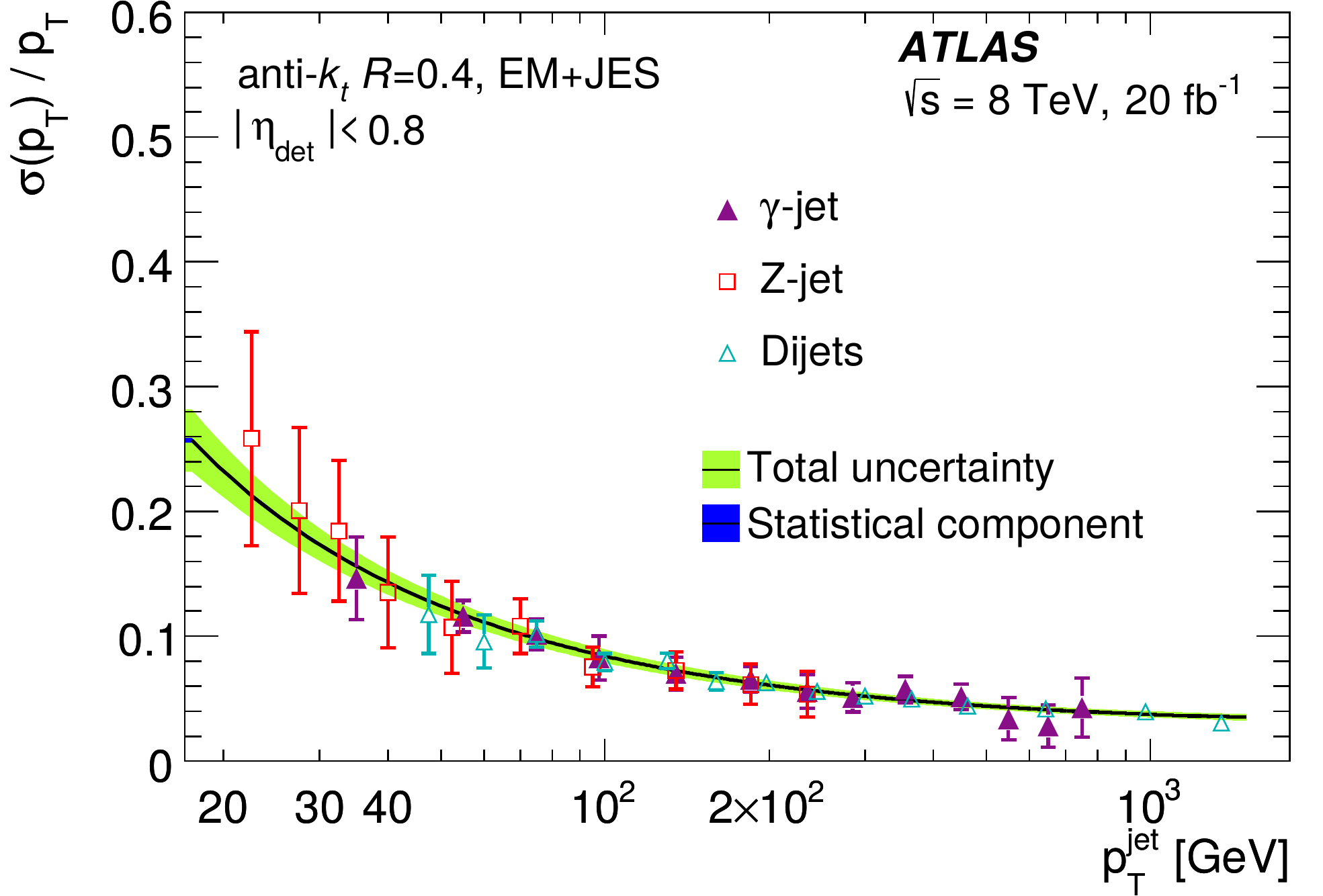}
\includegraphics[width=0.45\textwidth]{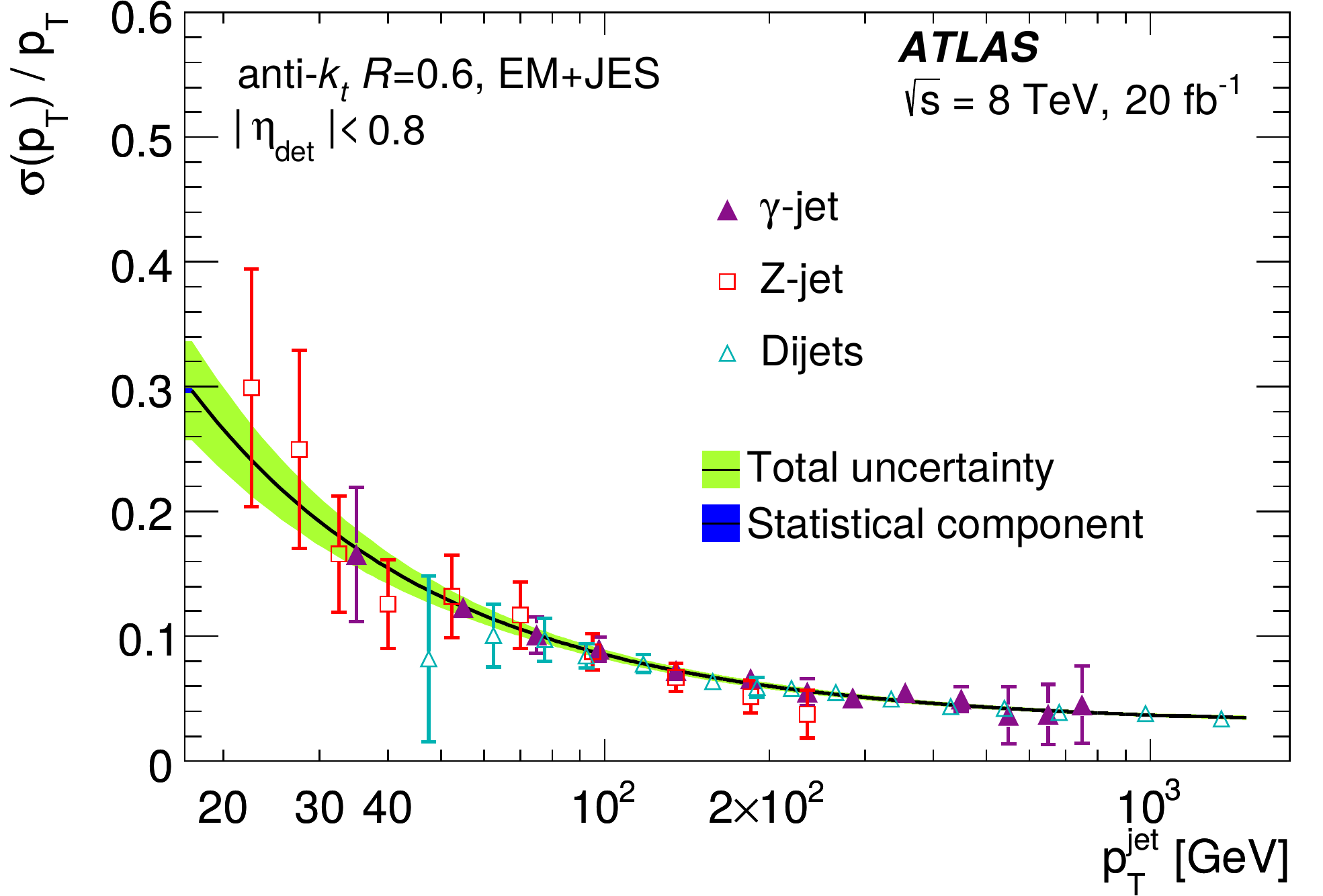}\\
\includegraphics[width=0.45\textwidth]{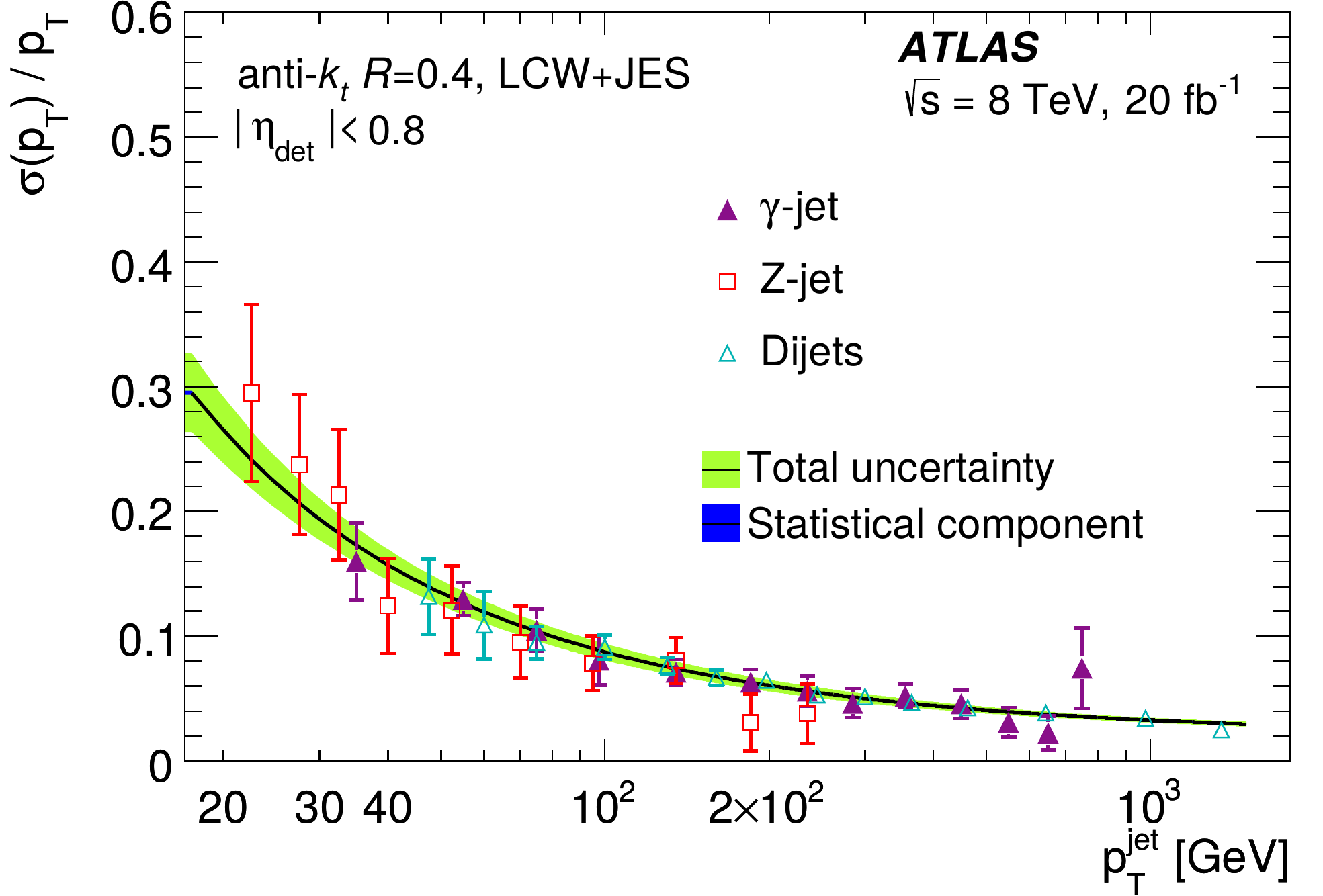}
\includegraphics[width=0.45\textwidth]{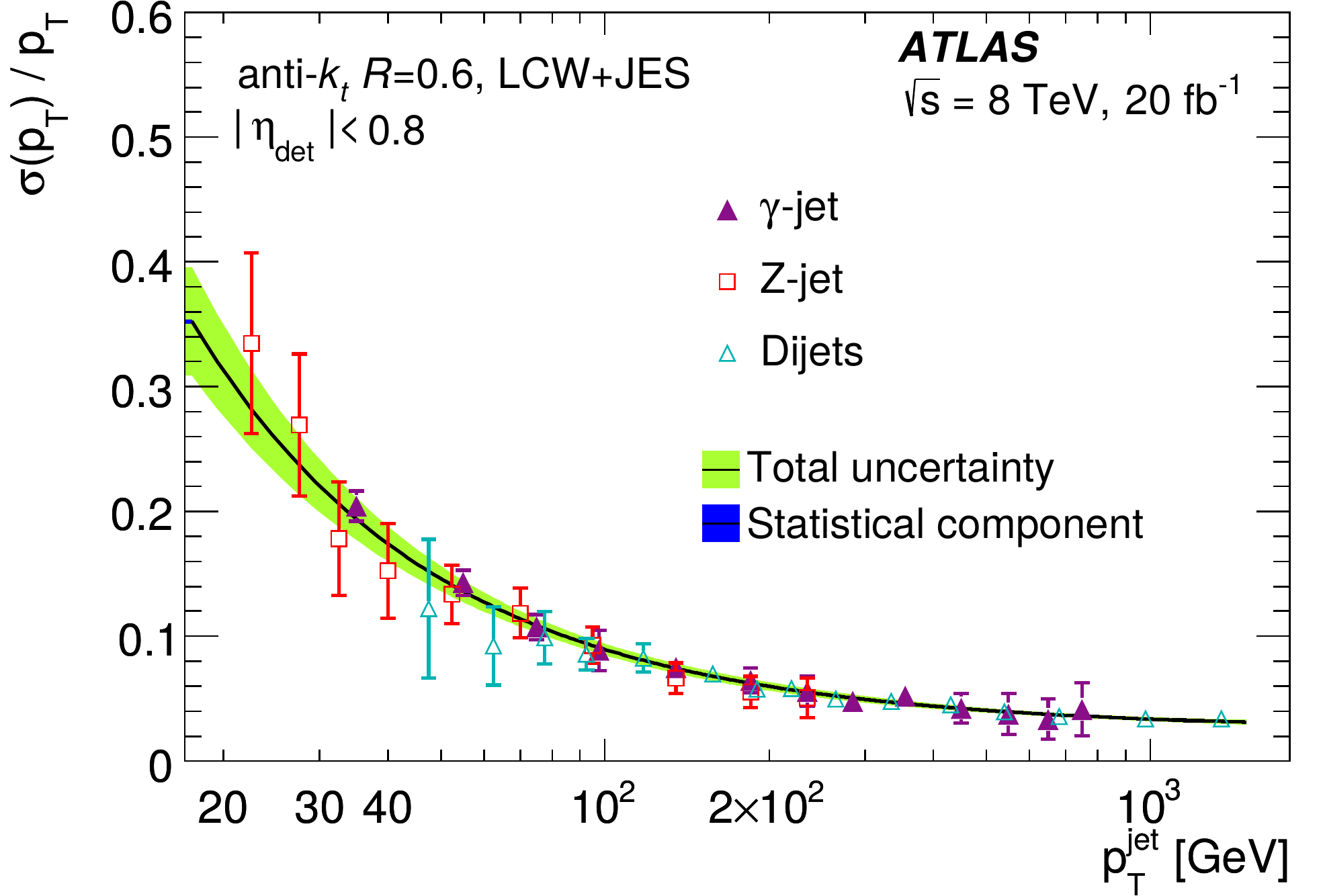}
\caption{The jet resolution as a function of \pT for the four different jet collections in the central region. The three \insitu{} inputs to the measurement, namely $Z$+jet~(empty squares), $\gamma$+jet~(filled triangles), and dijet balance~(empty triangles) are shown displaying the compatibility of the measurements.  The final fit using the function in Eq.~(\ref{eq:JER}) is included with its associated statistical and total uncertainty.}
\label{fig:jercombination:central}
\end{figure}
 
\begin{figure}
\centering
\includegraphics[width=0.45\textwidth]{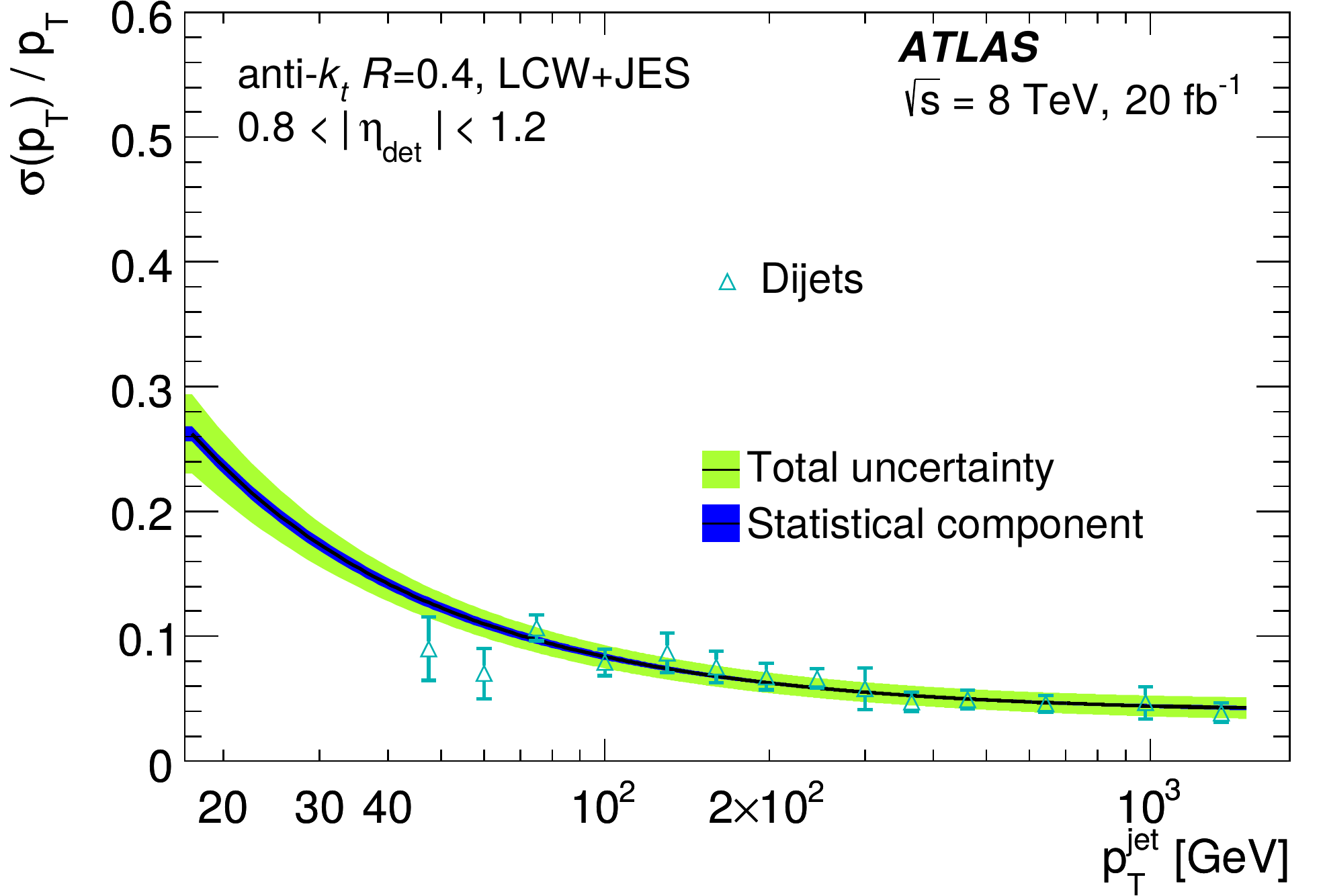}
\includegraphics[width=0.45\textwidth]{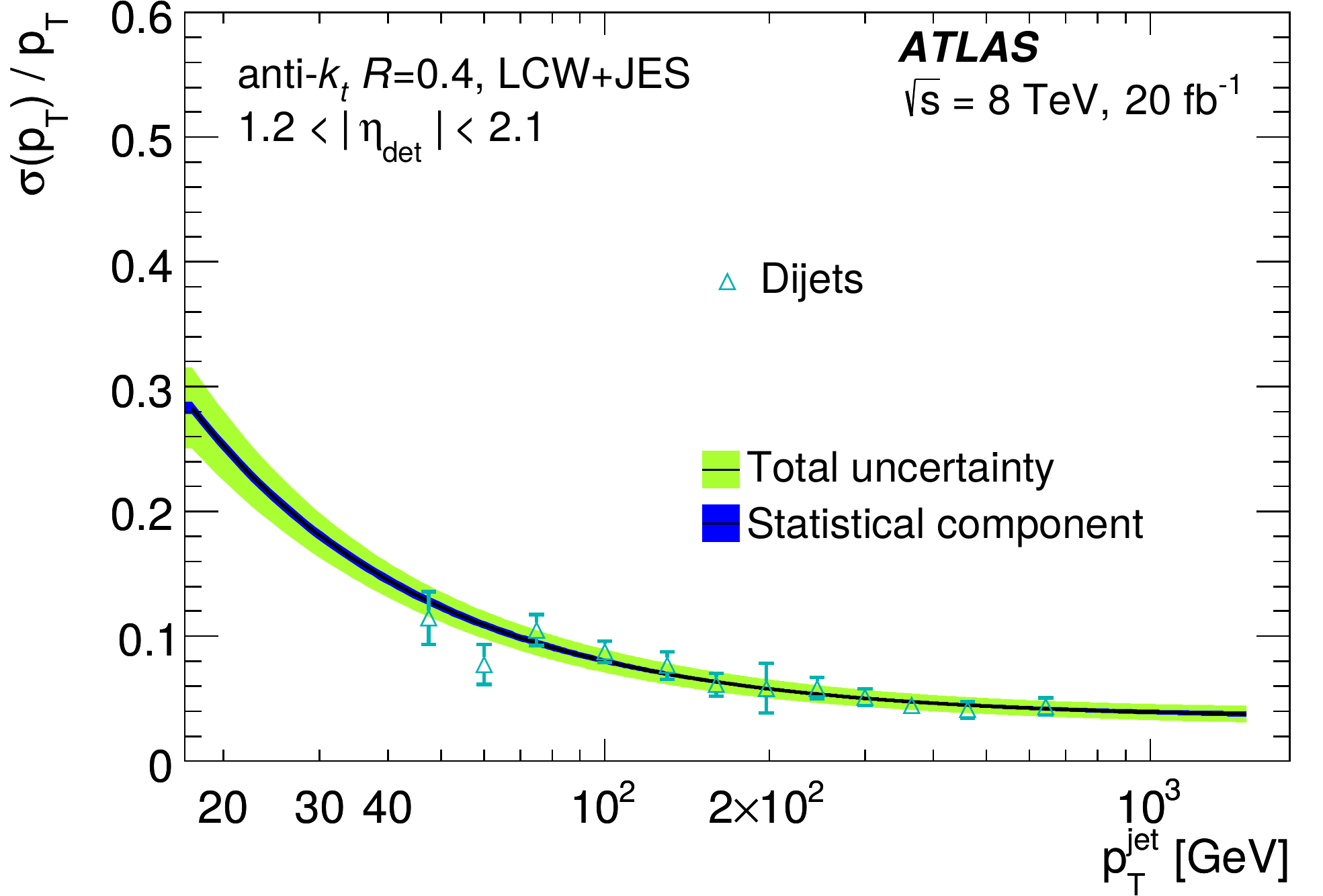}\\
\includegraphics[width=0.45\textwidth]{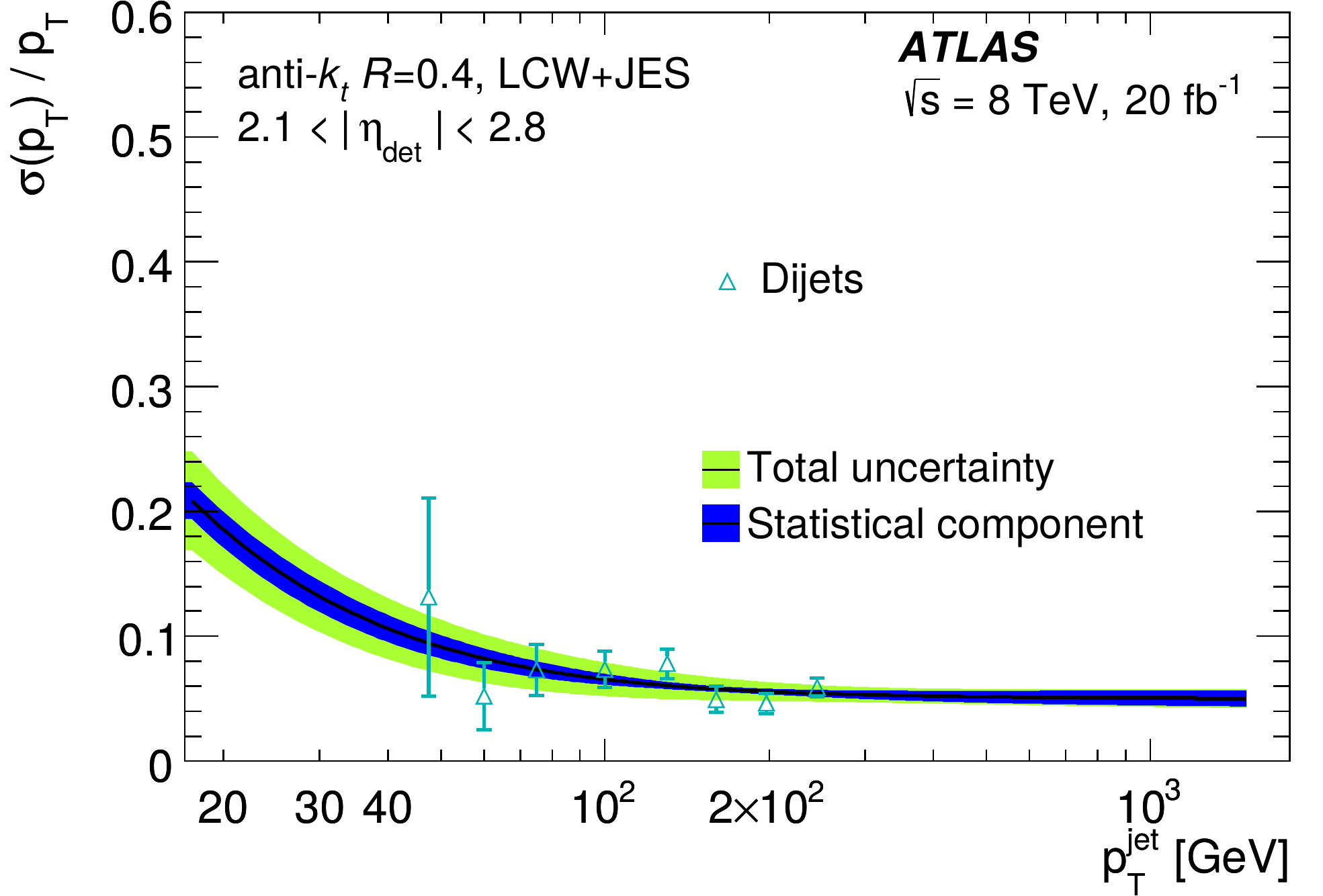}
\caption{The jet resolution as a function of \pT for LCW+JES anti-$k_t$ $R=0.4$ jets in three more-forward regions. The dijet \insitu{} inputs~(empty triangles) to the measurement are shown. The final fit using the function in Eq.~(\ref{eq:JER}) is included with its associated statistical and total uncertainty.}
\label{fig:jercombination:forward}
\end{figure}
% End of text imported from the .//sections/combination.tex input file
 
\FloatBarrier
% The next lines are included from the .//sections/conclusions.tex input file
\section{Conclusions}
\label{sec:conclusions}
 
This article describes the determination of the jet energy scale (JES) and jet energy resolution (JER) for data recorded by the ATLAS experiment in 2012 at $\sqrt{s} = 8$~\TeV.
 
The calibration scheme used for \antikt{} jets reconstructed using radius parameter $R=0.4$ or $R=0.6$ corrects for \pileup{} and the location of the primary interaction point before performing a calibration based on MC simulation.
These initial steps in the calibration provide stability of the calibration as a function of \pileup{} and improve the angular resolution of jets.
Following the MC-simulation-derived baseline calibration, a global sequential correction is performed.
It is also derived from MC simulations using information about how the jet deposits energy in the calorimeter, the tracks associated with the jet,
and the activity in the muon chambers behind the jet (particularly important for high-$\pt$ jets).
This improves the resolution of jets and reduces the difference in energy scale between quark- and gluon-initiated jets.
 
Following these MC-based calibration steps, the data taken in 2012 are used to perform a residual calibration that constrains
the uncertainties. This is performed for \antikt{} jets with \rfour{} and \rsix{} calibrated with both the \EMJES{} and \LCWJES{} schemes.
Dijet events are used to calibrate jets in the forward region relative to the central region as a function of jet transverse momentum and pseudorapidity.
The uncertainties of this calibration step have been significantly reduced compared with previous results primarily though the use of event generators
with improved modelling of multijet production. The total uncertainties are typically below $1\%$ for central jets, rising to $3.5\%$ for low-\pt{} jets
at high absolute pseudorapidity.
Central jets are calibrated by exploiting the balance between jets recoiling against either a photon or a $Z$~boson.
In the pseudorapidity region \etaRange{0.8}{2.8}, the jet energy scale is validated with \Zjet{} events using the direct \pt{} balance technique.
The jet energy scale calibration for central jets with high \pt{} is determined using events in which an isolated high-\pt{}
jet recoils against a system of lower-\pt{} jets.
The final calibration is obtained through a statistical combination of the different measurements.
This results in a correction at the level of 0.5\% to the JES in data with an associated uncertainty of less than 1\% for central anti-$k_t$ $R=0.4$ jets with $100<\pT<1500$~GeV.
At higher \pT, the uncertainty is about 3\% as \insitu{} measurements become statistically limited, and instead the calibration relies on single-hadron response studies.
 
The jet energy scale of trimmed \antikt{} jets with \rten{} is derived using MC simulation in the same way as for $R=0.4$ and $R=0.6$ jets,
thus calibrating the jets to the \LCWJES{} scale.  In an additional step, a dedicated calibration of the jet mass for the \rten{} jets is derived.
The MC-derived calibration is tested \insitu{} using the direct balance method with $\gamma$+jet events.
These studies are used to evaluate uncertainties in the calibration.
The total uncertainty for \AetaRange{0.8} is found to be around $3\%$ for jets with low \pt{}, falling to about $1\%$ for jets with $\pt \geq 150$~\GeV.
At larger $|\etaDet|$, the uncertainty increases to \percentRange{4}{5} at low jet \pt{}, decreasing to \percentRange{1}{2} for $\pt > 150$~\GeV.
 
The JER is measured in 2012 data using several \insitu{} methods.
The JER \pileup{} noise term is determined using novel techniques that exploit the increased level of \pileup{} interactions in the 2012 data. Three measurements of the JER as a function of jet \pt{} and \etaDet{} are performed using $\gamma$+jet, $Z$+jet and dijet data.
A final measurement of the JER is obtained using a statistical combination of these measurements, using a methodology similar to that used for the JES calibration.
The different \insitu{} inputs are found to be consistent with each other over the kinematic regions where they overlap.
For \antikt{} $R=0.4$ jets in the central calorimeter region $|\eta|<0.8$ calibrated with the EM+JES calibration scheme, the JER resolution parameters are measured to be $N=3.33\pm0.63$~\GeV, $S=0.71\pm0.07$~$\sqrt{\text{\GeV}}$, and $C=0.030\pm 0.003$, which corresponds to a relative JER of $\sigma_{p_\text{T}} / \pt = (23 \pm 2)\%$ for $\pt = 20$~\GeV\ and $\sigma_{p_\text{T}} / \pt = (8.4 \pm 0.6)\%$ for $\pt = 100$~\GeV.
The jet energy resolution in data is generally well reproduced by the MC simulation.
In certain kinematic regions, the simulated jets have a slightly smaller resolution than jets in data.
In physics analyses, the \pt{} of the simulated jets is corrected by random smearing to match the resolution observed in data.
The required amount of smearing is of similar order of magnitude as the jet energy resolution uncertainties.
 
% End of text imported from the .//sections/conclusions.tex input file
 
\section*{Acknowledgements}
 
% The next lines are included from the .//acknowledgements/Acknowledgements.tex input file

We thank CERN for the very successful operation of the LHC, as well as the
support staff from our institutions without whom ATLAS could not be
operated efficiently.
 
We acknowledge the support of ANPCyT, Argentina; YerPhI, Armenia; ARC, Australia; BMWFW and FWF, Austria; ANAS, Azerbaijan; SSTC, Belarus; CNPq and FAPESP, Brazil; NSERC, NRC and CFI, Canada; CERN; CONICYT, Chile; CAS, MOST and NSFC, China; COLCIENCIAS, Colombia; MSMT CR, MPO CR and VSC CR, Czech Republic; DNRF and DNSRC, Denmark; IN2P3-CNRS and CEA-DRF/IRFU, France; SRNSFG, Georgia; BMBF, HGF and MPG, Germany; GSRT, Greece; RGC and Hong Kong SAR, China; ISF and Benoziyo Center, Israel; INFN, Italy; MEXT and JSPS, Japan; CNRST, Morocco; NWO, Netherlands; RCN, Norway; MNiSW and NCN, Poland; FCT, Portugal; MNE/IFA, Romania; MES of Russia and NRC KI, Russia Federation; JINR; MESTD, Serbia; MSSR, Slovakia; ARRS and MIZ\v{S}, Slovenia; DST/NRF, South Africa; MINECO, Spain; SRC and Wallenberg Foundation, Sweden; SERI, SNSF and Cantons of Bern and Geneva, Switzerland; MOST, Taiwan; TAEK, Turkey; STFC, United Kingdom; DOE and NSF, United States of America. In addition, individual groups and members have received support from BCKDF, CANARIE, Compute Canada and CRC, Canada; ERC, ERDF, Horizon 2020, Marie Sk{\l}odowska-Curie Actions and COST, European Union; Investissements d'Avenir Labex, Investissements d'Avenir Idex and ANR, France; DFG and AvH Foundation, Germany; Herakleitos, Thales and Aristeia programmes co-financed by EU-ESF and the Greek NSRF, Greece; BSF-NSF and GIF, Israel; CERCA Programme Generalitat de Catalunya and PROMETEO Programme Generalitat Valenciana, Spain; G\"{o}ran Gustafssons Stiftelse, Sweden; The Royal Society and Leverhulme Trust, United Kingdom.
 
The crucial computing support from all WLCG partners is acknowledged gratefully, in particular from CERN, the ATLAS Tier-1 facilities at TRIUMF (Canada), NDGF (Denmark, Norway, Sweden), CC-IN2P3 (France), KIT/GridKA (Germany), INFN-CNAF (Italy), NL-T1 (Netherlands), PIC (Spain), ASGC (Taiwan), RAL (UK) and BNL (USA), the Tier-2 facilities worldwide and large non-WLCG resource providers. Major contributors of computing resources are listed in Ref.~\cite{ATL-GEN-PUB-2016-002}.
 
% End of text imported from the .//acknowledgements/Acknowledgements.tex input file

\printbibliography
\clearpage
% ATLAS Collaboration author list
% Reference date of PERF-2014-02 is 2017-12-18
% Author list last updated on date 08-OCT-19
% Data extracted on 28-Feb-2020 for paper reference PERF-2014-02
% at 5:28pm
 
\begin{flushleft}
{\Large The ATLAS Collaboration}

\bigskip

M.~Aaboud$^\textrm{\scriptsize 35d}$,    
G.~Aad$^\textrm{\scriptsize 101}$,    
B.~Abbott$^\textrm{\scriptsize 128}$,    
O.~Abdinov$^\textrm{\scriptsize 13,*}$,    
B.~Abeloos$^\textrm{\scriptsize 64}$,    
S.H.~Abidi$^\textrm{\scriptsize 167}$,    
O.S.~AbouZeid$^\textrm{\scriptsize 145}$,    
N.L.~Abraham$^\textrm{\scriptsize 155}$,    
H.~Abramowicz$^\textrm{\scriptsize 161}$,    
H.~Abreu$^\textrm{\scriptsize 160}$
% ...
\bigskip
\\

% ...
$^{13}$Institute of Physics, Azerbaijan Academy of Sciences, Baku; Azerbaijan.\\
% ...
$^{35}$$^{(a)}$Facult\'e des Sciences Ain Chock, R\'eseau Universitaire de Physique des Hautes Energies - Universit\'e Hassan II, Casablanca;$^{(b)}$Facult\'{e} des Sciences, Universit\'{e} Ibn-Tofail, K\'{e}nitra;$^{(c)}$Facult\'e des Sciences Semlalia, Universit\'e Cadi Ayyad, LPHEA-Marrakech;$^{(d)}$Facult\'e des Sciences, Universit\'e Mohamed Premier and LPTPM, Oujda;$^{(e)}$Facult\'e des sciences, Universit\'e Mohammed V, Rabat; Morocco.\\
% ...
$^{64}$IJCLab, Universit\'e Paris-Saclay, CNRS/IN2P3, 91405, Orsay; France.\\
% ...
$^{101}$CPPM, Aix-Marseille Universit\'e, CNRS/IN2P3, Marseille; France.\\
% ...
$^{128}$Homer L. Dodge Department of Physics and Astronomy, University of Oklahoma, Norman OK; United States of America.\\
% ...
$^{145}$Santa Cruz Institute for Particle Physics, University of California Santa Cruz, Santa Cruz CA; United States of America.\\
% ...
$^{155}$Department of Physics and Astronomy, University of Sussex, Brighton; United Kingdom.\\
% ...
$^{160}$Department of Physics, Technion, Israel Institute of Technology, Haifa; Israel.\\
$^{161}$Raymond and Beverly Sackler School of Physics and Astronomy, Tel Aviv University, Tel Aviv; Israel.\\
% ...
$^{167}$Department of Physics, University of Toronto, Toronto ON; Canada.\\
% ...
$^{a}$ Also at Borough of Manhattan Community College, City University of New York, New York NY; United States of America.\\
$^{b}$ Also at Centre for High Performance Computing, CSIR Campus, Rosebank, Cape Town; South Africa.\\
$^{c}$ Also at CERN, Geneva; Switzerland.\\
$^{d}$ Also at CPPM, Aix-Marseille Universit\'e, CNRS/IN2P3, Marseille; France.\\
$^{e}$ Also at D\'epartement de Physique Nucl\'eaire et Corpusculaire, Universit\'e de Gen\`eve, Gen\`eve; Switzerland.\\
% ...
$^{*}$ Deceased

\end{flushleft}

% Created with Glance <Atlas.Glance@cern.ch>
\clearpage
\appendix

\end{document}